\documentclass{article}

\usepackage{xcolor}
\usepackage{arxiv}
\usepackage{graphicx}
\usepackage{subfig}

\usepackage{lmodern}
\usepackage{siunitx}
\usepackage{booktabs}
\usepackage{etoolbox}
\usepackage{soul}

\usepackage[utf8]{inputenc} 
\usepackage[T1]{fontenc}    
\usepackage{hyperref}       
\usepackage{url}            
\usepackage{booktabs}       
\usepackage{amsfonts}       
\usepackage{nicefrac}       
\usepackage{microtype}      
\usepackage[toc,page,titletoc]{appendix}
\usepackage{lineno}
\usepackage{amsmath,bm}
\usepackage{amsfonts,amssymb}
\usepackage{amsmath}

\usepackage{soul}

\usepackage{algorithm}
\usepackage[noend]{algpseudocode} 
\usepackage{tabularx}
\usepackage{array}

\algrenewcommand\algorithmicforall{\textbf{foreach}}
\algrenewcommand\algorithmicindent{.8em}
\newcolumntype{K}[1]{>{\centering\arraybackslash}p{#1}}
\algnewcommand{\Inputs}[1]{%
  \State \textbf{Inputs:}\hspace*{\algorithmicindent}\parbox[t]{.8\linewidth}{\raggedright #1}
}
\algnewcommand{\Initialize}[1]{%
  \State \textbf{Initialize:}\hspace*{\algorithmicindent}\parbox[t]{.8\linewidth}{\raggedright #1}
}
\algnewcommand{\Outputs}[1]{%
  \State \textbf{Outputs:}\hspace*{\algorithmicindent}\parbox[t]{.8\linewidth}{\raggedright #1}
}
\makeatletter
\newcommand{\multiline}[1]{%
  \begin{tabularx}{\dimexpr\linewidth-\ALG@thistlm}[t]{@{}X@{}}
    #1
  \end{tabularx}
}

\title{COVID-19 and Influenza Joint Forecasts Using Internet Search Information in the United States}

\author{Simin Ma\thanks{H. Milton Stewart School of Industrial and Systems Engineering, Georgia Institute of Technology, Atlanta, Georgia, USA.} \\ \And Shaoyang Ning \thanks{Department of Mathematics and Statistics, Williams College, Williamstown, MA 01267, USA.}\\ \And Shihao Yang \footnotemark[1] }

\begin{document}
\maketitle

\begin{abstract}
As COVID-19 pandemic progresses, severe flu seasons may happen alongside an increase in cases in cases and death of COVID-19, causing severe burdens on health care resources and public safety. A consequence of a twindemic may be a mixture of two different infections in the same person at the same time, ``flurona''. Admist the raising trend of ``flurona'', forecasting both influenza outbreaks and COVID-19 waves in a timely manner is more urgent than ever, as accurate joint real-time tracking of the twindemic aids health organizations and policymakers in adequate preparation and decision making. Under the current pandemic, state-of-art influenza and COVID-19 forecasting models carry valuable domain information but face shortcomings under current complex disease dynamics, such as similarities in symptoms and public healthcare seeking patterns of the two diseases. Inspired by the inner-connection between influenza and COVID-19 activities, we propose ARGOX-Joint-Ensemble which allows us to combine historical influenza and COVID-19 disease forecasting models to a new ensemble framework that handles scenarios where flu and COVID co-exist. Our framework is able to emphasize learning from COVID-related or influenza signals, through a winner-takes-all ensemble fashion. Moreover, our experiments demonstrate that our approach is successful in adapting past influenza forecasting models to the current pandemic, while improving upon previous COVID-19 forecasting models, by steadily outperforming alternative benchmark methods, and remaining competitive with publicly available models.
\end{abstract}


\section{Introduction}
\begin{figure}[htbp]
\centering
\subfloat[GA COVID-19 Cases (thick black curve) and the \%ILI of itself (red curve) and its neighbours]
{\includegraphics[width=0.4\textwidth]{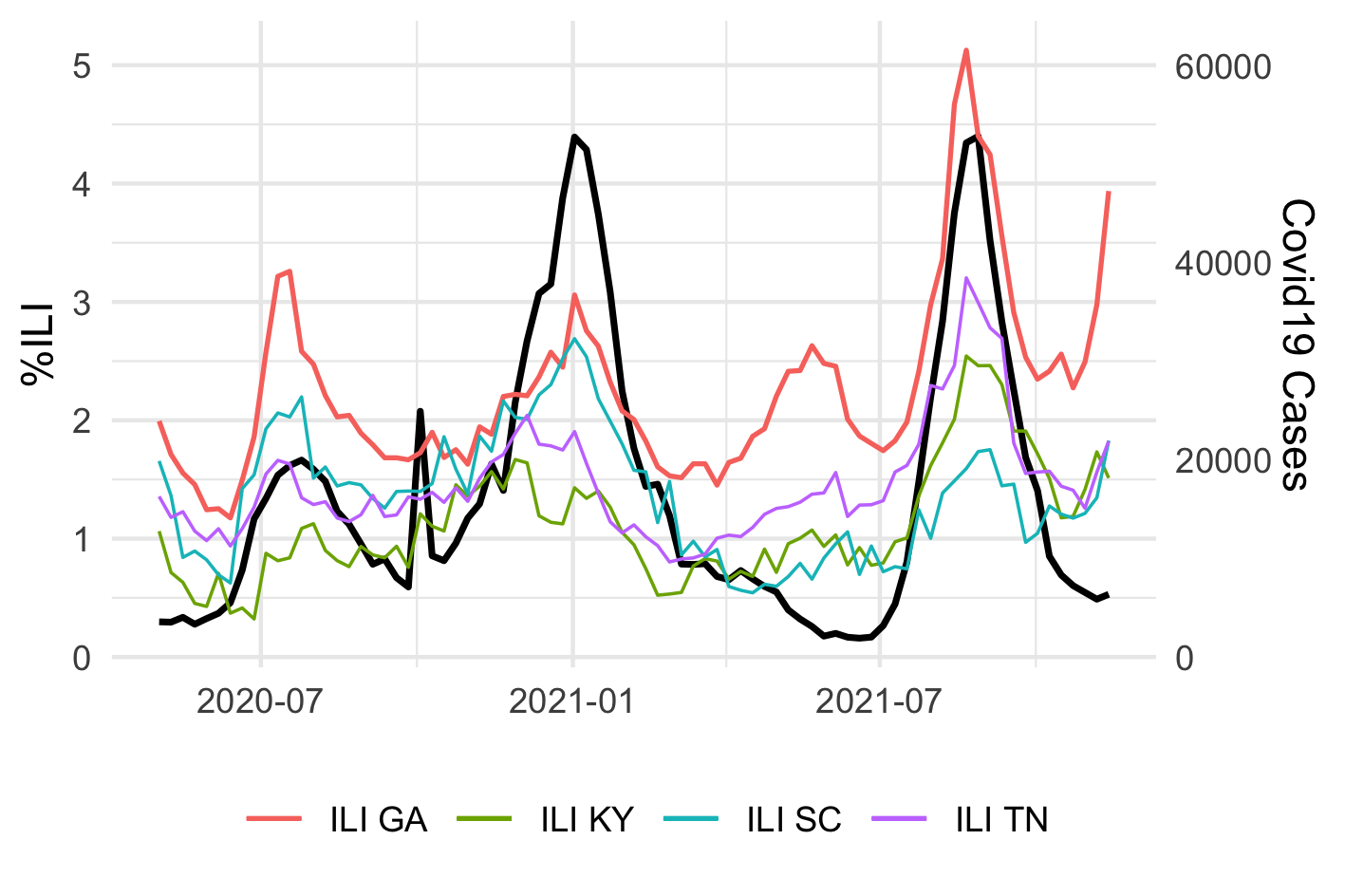}\label{fig:GA_Case_ILI}}\hfill
\subfloat[GA COVID-19 Death (thick black curve) and the \%ILI of itself (red curve) and its neighbours]
{\includegraphics[width=0.4\textwidth]{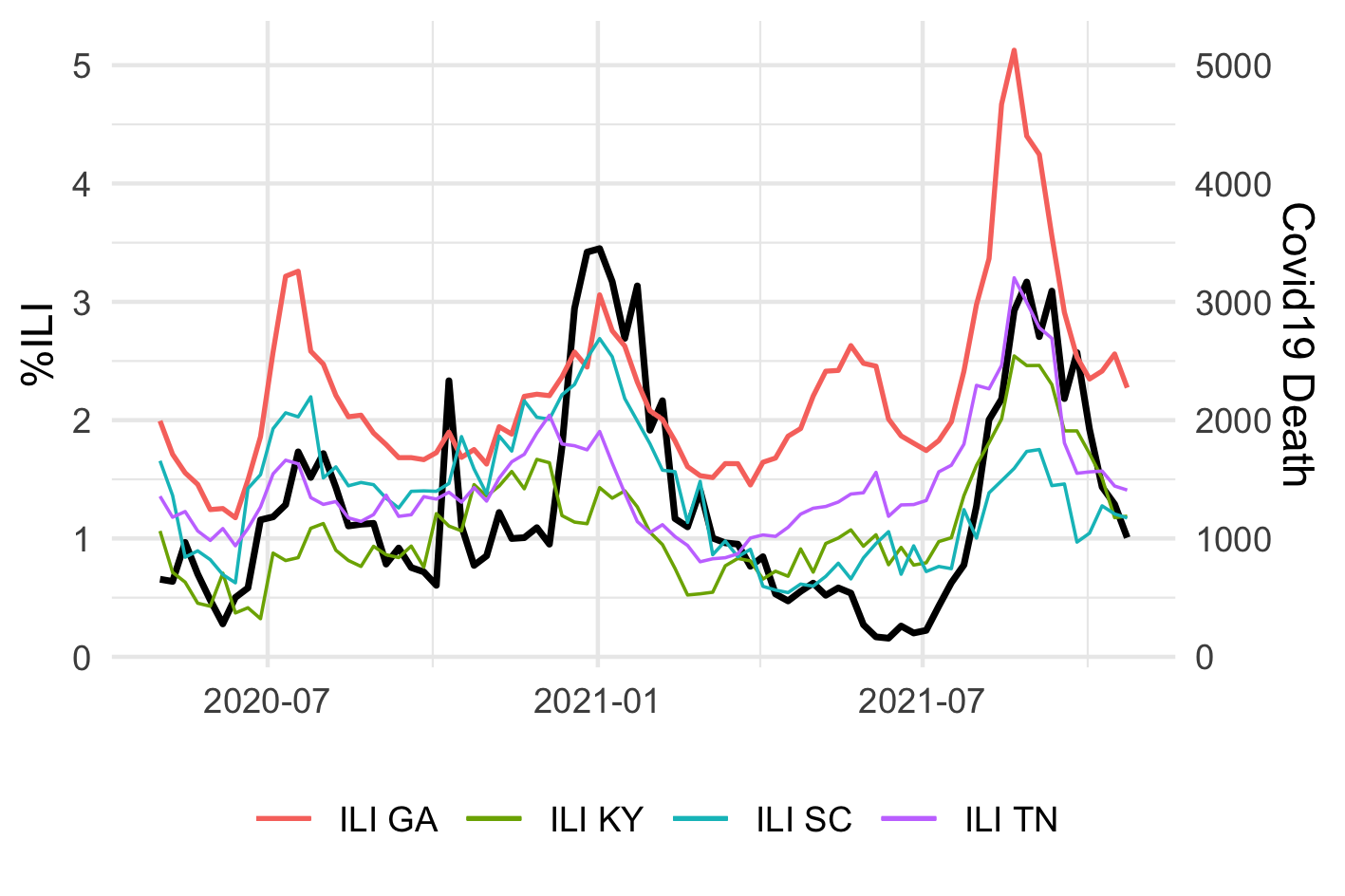}\label{fig:GA_Death_ILI}}
\caption{GA and its neighbours' \%ILI and GA COVID-19 cases (top) and death (bottom) Illustration of GA COVID-19 cases/death (black) growth relationships with itself (red) and its neighbouring states' \%ILI. Y-axis are adjusted accordingly.}
\label{fig:GA_COVID_ILI}
\end{figure}

The rising numbers of newly detected ``flurona'' cases \cite{hassan_2022}, i.e., individuals infected by both influenza (flu) and COVID-19, have raised serious concerns about the potential of a "twindemic" of flu and COVID-19 among the general public \cite{berger2022twindemic}. The recent epidemic trends of flu and COVID-19 have shown remarkable similarity (figure \ref{fig:GA_COVID_ILI}). The fast-developing COVID-19 pandemic, coupled with a severe flu season, would overwhelm the already heavily-burdened health care systems, causing further inconceivable losses \cite{rubin2020happens}. This calls for an urgent need to establish an accurate and robust bi-disease tracking/forecasting system to provide public health officials with reliable, timely information to make informed decisions to control and prevent the onset of a ``twindemic''. To this end, we propose ARGOX-Joint-Ensemble, a principled framework that utilizes the connectivity between flu and COVID-19 to integrate previously-proposed forecasting models and adapt to a new era where flu and COVID co-evolve.

Accurate tracking of flu outbreaks and trends is important but non-trivial. In fact, flu affects 9–41 million people annually between 2010-2020 seasons in the United States and resulting in between 12-52 thousands of deaths \cite{CDC_Flu_Stat}. For decades, the U.S. Centers for Disease Control and Prevention (CDC) monitors flu activities through Influenza-like Illness Surveillance Network (ILINet), which collects the weekly reported number of outpatients with Influenza-like Illness (ILI) from thousands of healthcare providers, and publishes the weekly ILI percentages (\%ILI, i.e., the percentages of outpatients with ILI) at the national, regional levels (10 Health and Human Services (HHS) regions in the US), and state levels. However, due to the time required for data collection and administrative processing, The ILI reports from CDC lag behind real time by 1-2 weeks, thus unable to provide most accurate and timely information on the disease development. Numerous flu tracking approaches have therefore been proposed,  utilizing statistical models \cite{brooks2018nonmechanistic, osthus2019dynamic}, mechanistic models such as compartmental models \cite{shaman2012forecasting, tizzoni2012real, yang2014comparison, yang2015inference}, ensemble approaches \cite{reich2019collaborative}, and deep learning models \cite{wang2019defsi, venna2018novel}. Several approaches rely on external signals such as environmental conditions and weather reports \cite{shaman2010absolute, tamerius2013environmental}; social media, such as Twitter posts \cite{paul2014twitter,signorini2011use} and Wikipedia article views \cite{mciver2014wikipedia,generous2014global}; search engine data, such as: Google \cite{GFT_2008, yang2015accurate, ARGO2_Regional, ARGOX}, Yahoo \cite{polgreen2008using}, and Baidu internet searches \cite{yuan2013monitoring}. However, this new forecasting problem of adapting to a new emerging pandemic scenario (COVID-19) is complex and cannot be addressed by traditional historical wILI methods alone. 

Similarly, many influenza forecasting approaches are adapted and modified to predict the newly emerged COVID-19 pandemic \cite{Delphi_KF,shaman2012forecasting}. In particular, machine learning (data-driven) methods \cite{DeepCOVID_GT, UCSB_attention, Delphi_KF} and compartmental models \cite{COVID19Simulator, UCLA_sueir, yang2021estimating} are the most popular and prevailing approaches for the publicly-available COVID-19 spread forecasts, according to the weekly forecast reports compiled by CDC \cite{CDC_Ensemble}. Yet, they also do not capture the inner-correlation between the two diseases, which could be a crucial factor as both infectious diseases co-evolve.

Despite the development in the methodology tracking individual diseases, joint tracking of flu and COVID-19 remains challenging. In the midst of the on-going COVID-19 pandemic, \%ILI collected by CDC may get ``contaminated'' in the current season, due to symptomatic similarities with COVID-19 as well as various biological and demographic factors, which brings more challenges to the forecasting tasks. Additionally, influenza outbreaks and trends can also be used to help with indirect COVID-19 surveillance, and cases and death predictions, due to the proximity between the two diseases. Finally, it is widely believed that COVID-19 may be in circulation for a long period of time and co-evolve with influenza, especially when COVID-19 variant continue to evolve \cite{hassan_2022}. Hence, a unified robust forecasting framework for both diseases is eminent indispensable.

Inspired by the affinity between influenza and COVID-19's growth trends (figure \ref{fig:GA_COVID_ILI}), we propose to leverage external COVID-related signals (confirmed cases) for influenza (wILI) forecasts, and vise-versa for COVID-19 cases and death predictions, along with relevant public search information. Yet, to build COVID-ILI joint prediction model with online search data, many challenges remain to be addressed. For instance, such inner-correlation between COVID-19 and influenza is latent and varies across geographical areas, which can be challenging to capture and utilize for forecasts. Furthermore, the COVID-ILI is a new phenomanon, with limited external signals, while relevant internet search information can be noisy and unstable; hence, it would be a significant challenge to efficiently learn the model under data paucity, and data instability.


Unfortunately, few attempts have been made to study the connection between COVID-19 and influenza trends, and to incorporate their simultaneous growths for forecasting, while considering the geographical dependence structure (in state-level). Most of the existing works adapted influenza forecasting model framework and applied towards COVID-19 predictions, or vise-versa. For example, Ref \cite{arokiaraj2020correlation} studies the influenza vaccination rates' correlation with COVID-19 deaths, and states its potential prediction power of death trends. Ref \cite{wang2021association} extends this study to identify association between vaccination rates and COVID-19 infection, death and hospitalization, as well as arguing for their forecasting potentials. Ref \cite{huang2021universal} uses incidence patterns from  past influenza seasons, COVID-19 time series information, and demographic covariates in a Generalized Linear Model to forecast next week's country-level case counts, under mild assumptions on the similarity of the transmission mechanisms between COVID-19 and flu. Ref \cite{rodriguez2021steering}, on the other hand, explores seasonal similarities between historical influenza seasons and current COVID-19 related signals using a deep
clustering module (learn lower-dimensional representation of the signals and reconstruct for forecasting using attention), and produces 1 week ahead independent state-level influenza forecasts.



Here we propose ARGOX-Joint-Ensemble, a principled way to combine and guide previously-proposed flu and COVID-19 forecasting models, to adapt to new scenarios where flu and COVID co-exist. In particular, we modified upon previously-proposed forecasting models by incorporating COVID-19 signals for flu predictions and vise-versa for COVID-19 forecasts, in a spatial-temporal fashion to efficiently capture and incorporate COVID-ILI signals for state-level forecasts, while manipulating model features for national-level forecasts. Finally, we employ an ensemble approach to efficiently combine COVID and flu forecasting methods into one joint framework, which is able to effectively shift focuses between COVID and influenza signals for both diseases' forecasts, and produce robust forecasts despite unstable search information signals as inputs.

\section{Data Acquisition and Pre-processing}
This paper focuses on the 50 states of the United States, plus Washington D.C for COVID-19 cases and death forecasting, while excluding Florida (whose ILI data is not available from CDC) and including New York City for \%ILI forecasting. For COVID-19 cases and death forecasting, we use confirmed cases, confirmed death, ILI and Google search query frequencies as inputs. For \%ILI forecasting, we use delayed \%ILI, COVID-19 cases, and Google search query frequencies as inputs.

\subsection{COVID-19 Reporting Data}\label{sec:COVID-Data}
We use reported COVID-19 confirmed cases and death of United States from New York Times (NYT) \cite{NYT_COVID} as features in our model. When comparing against other benchmark methods published in CDC COVID Forecast-hub \cite{CDC_ForecastHub}, we use COVID-19 confirmed cases and death from JHU CSSE COVID-19 dataset \cite{JHU_Data}, a curated dataset used by the CDC at their official website, as the groundtruth. We do not use JHU COVID-19 dataset as input features in our model because JHU COVID-19 dataset retrospectively corrects past confirmed cases and death due to reporting error and federal and state policies, while NYT dataset does not revise past data, which gives more realistic forecasts. Both data sources are collected from January 21, 2020 to October 9, 2021. 

\subsection{CDC's ILINet data}\label{sec:CDCILINet-Data}
CDC releases a report of \%ILI for the previous week every Friday, which contains the percent of outpatient visits with influenza-like illness for the whole nation, 10 HHS regions, 50 states (except Florida), Washington DC, and New York City (separated from New York State) \footnote{\url{https://www.cdc.gov/flu/weekly/overview.htm}}. CDC’s \%ILI data for this study are collected from January 21, 2020 to October 9, 2021. 

\subsection{Google Search Data}
The online search data used in this paper is obtained from Google Trends \cite{GoogleTrends}, where one can obtain the search frequencies of a term of interest in a specific region, time frame, and time frequency by typing in the search query on the website. With Google Trends API, we are able to obtain a daily time series of the search frequencies for the term of interest, including all searches that contain all of its words (un-normalized) \cite{GoogleTrends}.

We use 23 highly correlated COVID-19 related Google search queries discovered in prior study \cite{ma2021covid} (in daily frequency) for COVID-19 cases and death forecasts, while using influenza related queries (weekly frequency) from previous study \cite{ARGO, ARGOX} for \%ILI forecasts. We obtain the search queries for national, regional (summation from states) and state level. For COVID-19 forecasts, we follow the prior work's data cleaning procedures \cite{ma2021covid}, and find the optimal delay of each Google search query from COVID-19 cases/death \cite{ma2021covid} (shown in table \ref{tab:23query_optimallag} in supplementary materials), serving as inputs to the forecasting models. Figure \ref{fig:Nat_cases_lossoftaste} and \ref{fig:Nat_death_lossoftaste} show that the peak of COVID-19 search volume for query ``loss of taste'' ahead of the peak in reported cases and deaths, confirming strong connections between people's search behaviors and COVID-19 trends.

\subsection{\%ILI Data Imputation}
\%ILI is weekly indexed while COVID-19 cases and deaths are daily indexed. As we propose a joint forecast framework for both COVID-19 cases/deaths and \%ILI in this study, the discrepancy in time stamps between the two needs to be resolved. For this study, we impute daily \%ILI from weekly \%ILI, since imputing daily data will enable us with larger training sets. 

Hypothesizing that CDC published daily data will exhibit in-week seasonality due to reporting bias, similar to COVID-19 incremental cases' testing biases (some days in a week have more testing capacities than others), we impute daily ILI from weekly ILI by adapting COVID-19 daily incremental cases. Here, we will work with ILI (re-scaled from \%ILI), the number of outpatients with Influenza-Like-Illnesses, throughout the data imputation as well as serving as inputs to COVID-19 forecasts. For each imputation iteration, we randomly select COVID-19 incremental cases from a week in the past (prior to the week currently working) and normalize them to sum up to the current imputing week's ILI. This will be our daily imputed ILI for the current week. We impute for $N=100$ iterations and use them throughout this study.

\section{Methods}
\subsection{National Level}\label{sec:ARGO}
Here, we propose a joint framework for national level COVID-19 cases and death, by additionally incorporating flu information in the previously-proposed national COVID-19 forecast model, ``ARGO-Inspired'' \cite{ma2021covid}, while adapting COVID-19 cases information for \%ILI predictions in the ``ARGO'' method \cite{ARGO}.

Motivated by the robust performance of ``ARGO-Inspired'' method \cite{ma2021covid} and the connection between COVID-19 cases/deaths and \%ILI (figure \ref{fig:GA_COVID_ILI}), we simply add delayed daily imputed ILI information in the $L_1$ penalized LASSO regression as extra exogenous variables to produce future 4-week COVID-19 incremental cases and death predictions (where we take the median across all predictions generated by $100$ imputation iterations' fitting procedures as the final daily predictions and aggregate into weekly).

Meanwhile, we obtain an accurate estimate of 1-week ahead national \%ILI using the ``ARGO'' method \cite{ARGO}, by additionally incorporating national COVID-19 incremental cases (weekly aggregated) as exogenous variables.

Detailed regression formulations are included in the Supplementary Materials. We denote this method as ``ARGO-Joint'' method.

\subsection{State Level}\label{sec:state_model}
\begin{figure}[htbp]
\centering
\includegraphics[width=0.7\textwidth]{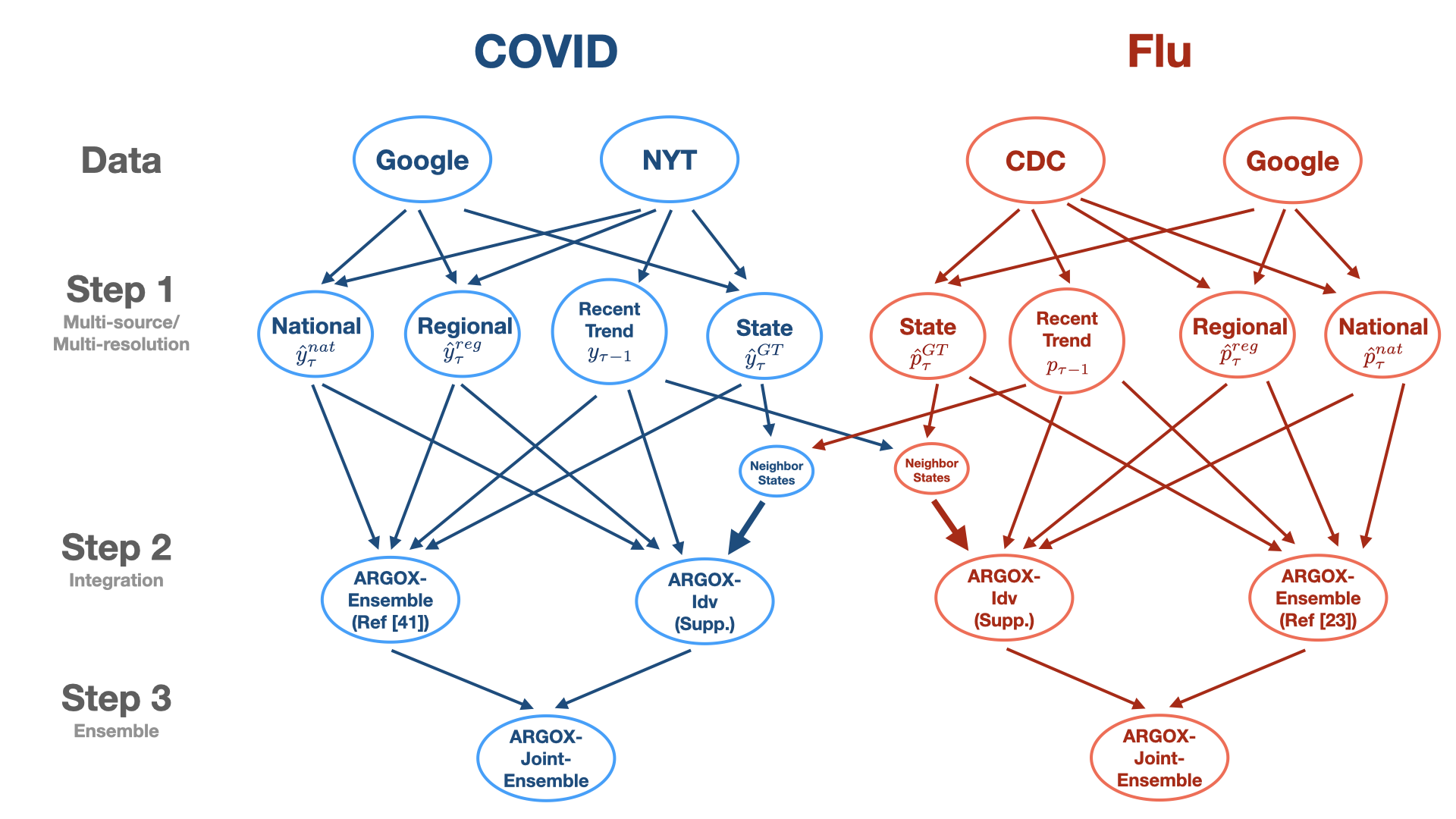}
\caption{Flow Chart of the State-Level forecast framework for COVID-19 cases/death and \%ILI}
\label{fig:flowchart}
\end{figure}

To handle the complicated disease dynamic when COVID-ILI co-evolves, we propose a new ensemble framework, ``ARGOX-Joint-Ensemble'', which uses COVID-ILI joint information to guide previously-proposed disease forecasting methods for unified COVID and \%ILI state-level forecasting.

A high-level illustration of our propose method is shown in figure \ref{fig:flowchart}, where ARGOX-Joint-Ensemble operates in 3 steps. 

\textbf{In the first step}, we gather the raw estimates of COVID-19 cases/death (left of figure \ref{fig:flowchart}) and raw estimates of \%ILI (right of figure \ref{fig:flowchart}) in different geographical resolution. For COVID-19, our raw estimates for state $m$ week $\tau$ incremental cases/deaths $y_{\tau,m}$ are $\hat{y}^{GT}_{\tau,m}$, $\hat{y}_{\tau, r_m}^{reg}$, $\hat{y}^{nat}_{\tau}$, and $y_{\tau,m}$, where $r_m$ is the region number for state $m$. Here, we denote $GT$ and $reg$ to be state/regional estimates with internet search information only, and $nat$ to be national estimates (same as prior study \cite{ma2021covid}). Similarly, we obtain the raw estimates for state $m$ weekly \%ILI $p_{\tau,m}$: $\hat{p}^{GT}_{\tau,m}$ \cite{ARGOX}, $\hat{p}_{\tau, r_m}^{reg}$ \cite{ARGO2_Regional}, $\hat{p}^{nat}_{\tau}$ \cite{ARGO}, $p_{\tau-1,m}$. 

\textbf{In the second step}, we fit two models separately using the raw estimates from step 1 as inputs. Motivated by the connection between neighbouring states' \%ILI and COVID-19 trends (figure \ref{fig:GA_COVID_ILI}), we first propose the ``ARGOX state-by-state'' method (denoted as ARGOX-Idv), which produces individual state-level COVID-19 cases/death predictions by utilizing neighbouring state's \%ILI information and internet search information, while adopting neighbouring state's COVID-19 cases and internet search information for \%ILI predictions. We also fit previously-proposed disease forecasting models for COVID-19 \cite{ma2021covid}, and \%ILI \cite{ARGOX}, in the second step, since they have already shown robustness prior to the newly emerged disease dynamics. Specifically, for COVID predictions, we fit the previously-proposed COVID-19 forecasting model (ARGOX-Ensemble) \cite{ma2021covid} and the newly-proposed model (ARGOX-Idv), which additionally incorporates ILI information ($p_{\tau-1,m}$) and state-level internet information ($\hat{y}^{GT}_{\tau,m}$) from neighbouring states ($\forall m\in \mathcal{M}$) along with inputs from step 1. Similarly, we fit the previously-propose influenza forecasting model \cite{ARGOX} and ARGOX-Idv, which additionally incorporates neighbouring states' COVID-19 incremental cases ($y^C_{\tau,m}$ where $C$ denotes cases) and internet search information ($\hat{p}_{\tau,m}^{GT}$). 

\textbf{In the third (last) step}, we gather the two methods in step 2, to produce the final winner-takes-all ensemble predictions for future 4 weeks COVID-19 cases/deaths and future 1 week \%ILI. Particularly, for a training period of 15 weeks, we evaluate both predictor with mean squared error (MSE) and select the one with lowest MSE as the ensemble predictor for future weeks. 

ARGOX-Idv and minor implementation modifications on previously-proposed method \cite{ma2021covid, ARGOX} are presented in the supplementary materials.

\section{Retrospective Evaluation}
\subsection{Evaluation Metrics}\label{sec:Error_Metric}
We use three metrics to evaluate the accuracy of an estimate of COVID-19 cases/death and \%ILI against the actual COVID-19 cases/death (published by JHU) and \%ILI (published by CDC): the root mean squared error (RMSE), the mean absolute error (MAE), and the Pearson correlation (Correlation). RMSE between an estimate $\hat{y}_t$ and the true value $y_t$ over period $t=1,\ldots, T$ is $\sqrt{\frac{1}{T}\sum_{t=1}^T \left(\hat{y}_t - y_t\right)^2}$. MAE between an estimate $\hat{y}_t$ and the true value $y_t$ over period $t=1,\ldots, T$ is $\frac{1}{T}\sum_{t=1}^T \left|\hat{y}_t - y_t\right|$. 
Correlation is the Pearson correlation coefficient between $\hat{\bm{y}}=(\hat{y}_1, \dots, \hat{y}_T)$ and $\bm{y}=(y_1,\dots, y_T)$.

\subsection{Retrospective Comparisons}
In this section, we conduct retrospective estimation of the 1-4 weeks ahead COVID-19 cases and death, and 1 week ahead \%ILI, at the US national and state level for the period of July 4, 2020 to Oct 9, 2021. We analyze our Joint framework's performances by conducting sensitivity analysis (comparing with our own methods), as well as comparing against other publicly available methods from CDC \cite{UMASS,CDC_Ensemble,MOBS_GLEAM,LANL_GrowthRate,UA_EpiGro,epiforecasts_MIT,USC-SI_kJalpha,JHU_CSSE-DECOM,UVA-Ensemble,CU-Select,COVIDAnalytics-DELPHI} for COVID-19 forecasts. CDC official predictions are a compilation of predictions from all teams that submit their weekly predictions every Monday since January 15th 2020, contributed by different research groups or individuals and providing robust and consistent predictions. We consider top 6 CDC published teams for COVID-19 cases and deaths prediction comparisons, after filtering out teams having missing values in their reporting over the period and states we are considering, among over 100 teams submitted to CDC. The teams considered in the COVID-19 cases' and deaths' retrospective comparisons are slightly different, subject to the teams forecasting targets.

For COVID-19 cases and death forecast, we focus on United State's 51 states/districts (including Washington DC). For \%ILI forecast, we focus on United State's 51 states/districts/city (including Washing DC and excluding Florida) and New York City. 

\subsubsection*{COVID-19 Death}
For national level sensitivity analysis, we compare with two other models (i) persistence (Naive) and (ii) ARGO inspired prediction \cite{ma2021covid}, and (iii) ARGO Joint prediction, with the actual COVID-19 weekly incremental death released by JHU dataset \cite{JHU_Data} in multiple error metrics (section \ref{sec:Error_Metric}). ARGO inspired predictions for national level are obtained from a $L_1$ penalized regression, utilizing the lagged COVID-19 cases, death and optimal lagged Google search terms, while the ARGO Joint prediction further added lagged national (imputed) \%ILI as exogenous variables. The Naive (persistence) predictions use current week's incremental death counts from New York Times (NYT) as next 1 to 4 weeks' estimation. For fair comparisons, both Ref \cite{ma2021covid} and ARGO Joint prediction use 56 days training period. 

Fig. \ref{fig:Death_Nat_Our} displays the estimates against the observed COVID-19 weekly incremental death. Table \ref{tab:Nat_Death_Our} (in the supplementary materials) further summarizes the accuracy metrics for all estimation methods, which shows that ARGO Joint estimations improves upon the previous-proposed method (Ref \cite{ma2021covid}) in almost all future time periods with the help of delayed \%ILI information serving as extra exogenous features (see figure \ref{fig:Nat_Coef_Death} for the coefficient heatmap). Delayed flu information is able to help the previous ARGO framework not overshoot and quickly recover from estimation spikes, especially over the period between December 2020 to March 2021 (see figure \ref{fig:Death_Nat_Our}). However, the connection between \%ILI information and COVID-19 deaths (target) degenerates as prediction horizon extends to 3 and 4 weeks ahead, since it is reasonable to believe that \%ILI information's influence is in a cascading fashion on COVID-19 case first and then death. Also, the improvement is limited from Ref \cite{ma2021covid}, as the search terms and time series' signals are saturated and the previous ARGO Inspired method is already doing a good job on national level (see figure \ref{fig:Death_Nat_Our}. This is also shown in table \ref{tab:Nat_Death_Other}, when comparing ARGO Joint estimations with other publicly available methods released by CDC. Similar to the previous ARGO Inspired method, ARGO Joint estimations is also able to achieve competitive performance on national level with the subtle improvement. 

For state level sensitivity analysis, we compare with two other methods: (i) persistence (Naive) predictions, (ii) ARGOX Ensemble prediction (Ref \cite{ma2021covid}), and (iii) ARGOX Joint Ensemble. Ref \cite{ma2021covid} predictions for each state are obtained from a three-methods ensemble, whereas ARGOX-Joint-Ensemble adds the additional ARGOX-Idv method (section \ref{sec:state_model}) to produce a four-methods ensemble. For fairness, Ref \cite{ma2021covid} and ARGOX-Joint-Ensemble methods both use 15 consecutive and overlapping weeks as training period. Table \ref{tab:State_Death_Our} summarizes the overall results of the comparing methods, averaging over the 51 states for the whole period of July 4th 2020 to Oct 9th 2021. Our ARGOX-Joint-Ensemble method improves upon Ref \cite{ma2021covid} thanks to ARGOX-Idv, which is a strong predictor by gathering neighbouring states' death and \%ILI information to serve in its spatial-temporal framework. Unlike the other predictors in the ensemble framework, ARGOX-Idv focuses on each individual states separately utilizing its neighbouring states' information, which shows its power in the ensemble framework as the pandemic progresses. This is also shown in the detail break down of methods contributing to ARGOX-Joint-Ensemble in table \ref{tab:Ensemble_Selection}, in which ARGOX-Idv is being selected the most on average (around 34\%). Other predictors also contribute heavily to the ensemble approach and together they provide us a Flu+COVID unified framework with improved robustness and accuracy. This is also evident in table \ref{tab:State_Death_Other} when comparing against other CDC published teams, as the ARGOX-Joint-Ensemble produces competitive accuracy and is among the top three models in term of all three error metrics.

Detailed state by state sensitivity comparisons are shown in in figure \ref{fig:State_Ours_Death_AK}-\ref{fig:State_Ours_Death_WY} and table \ref{tab:State_Ours_Death_AK}- \ref{tab:State_Ours_Death_WY}. 

\begin{table}[ht]
\sisetup{detect-weight,mode=text}
\renewrobustcmd{\bfseries}{\fontseries{b}\selectfont}
\renewrobustcmd{\boldmath}{}
\newrobustcmd{\B}{\bfseries}
\addtolength{\tabcolsep}{-4.1pt}
\footnotesize
\centering
\begin{tabular}{lrrrr}
  \hline
& 1 Week Ahead & 2 Weeks Ahead & 3 Weeks Ahead & 4 Weeks Ahead \\ 
    \hline \multicolumn{1}{l}{RMSE} \\
 \hspace{1em} Ref \cite{ma2021covid} &  76.27 & 87.78 & 102.79 & \B124.84 \\ 
   \hspace{1em} ARGOX Joint Ensemble & \B70.04 & \B77.62 & \B95.71 & 125.17 \\ 
   \hspace{1em} Naive & 83.39 & 107.14 & 127.51 & 147.05 \\ 
   \multicolumn{1}{l}{MAE} \\
  \hspace{1em}  Ref \cite{ma2021covid} & 42.34 & 51.38 & 62.41 & 77.14 \\ 
   \hspace{1em} ARGOX Joint Ensemble & \B39.56 & \B46.81 & \B56.93 & \B75.39 \\ 
   \hspace{1em} Naive & 46.14 & 62.26 & 78.66 & 94.56 \\ 
   \multicolumn{1}{l}{Correlation} \\
 \hspace{1em} Ref \cite{ma2021covid} & 0.82 & 0.80 & 0.77 & 0.71 \\ 
   \hspace{1em} ARGOX Joint Ensemble  & \B0.83 & \B0.82 & \B0.79 & \B0.74 \\ 
   \hspace{1em} Naive& 0.79 & 0.71 & 0.63 & 0.54 \\ 
   \hline
\end{tabular}
\caption{Comparison of different methods for state-level COVID-19 1 to 4 weeks ahead incremental death in 51 U.S. states. The averaged RMSE, MAE, and correlation are reported and best performed method is highlighted in boldface. In average, ARGOX-Joint-Ensemble achieves round 6.6\% RMSE, 10\% MAE, and 2.6\% Correlation improvements for 1-4 weeks ahead prediction.} 
\label{tab:State_Death_Our}
\end{table}

\begin{table}[ht]
\sisetup{detect-weight,mode=text}
\renewrobustcmd{\bfseries}{\fontseries{b}\selectfont}
\renewrobustcmd{\boldmath}{}
\newrobustcmd{\B}{\bfseries}
\addtolength{\tabcolsep}{-4.1pt}
\footnotesize
\centering
\begin{tabular}{lrrrrr}
 \hline
 & 1 Week Ahead & 2 Weeks Ahead & 3 Weeks Ahead & 4 Weeks Ahead & Average \\ 
  \hline \multicolumn{1}{l}{RMSE} \\
\hspace{1em} COVIDhub-ensemble\cite{CDC_Ensemble} & 62.19 & 71.39 & 80.59 & 92.81 & 76.74 \\ 
 \hspace{1em} UMass-MechBayes\cite{UMASS} & 63.02 & 74.29 & 87.13 & 107.05 & 82.87 \\ 
 \hspace{1em} MOBS-GLEAM\_COVID\cite{MOBS_GLEAM} & 70.13 & 77.62 & 95.61 & 124.11 & 91.86 \\ 
 \hspace{1em} ARGOX\_Joint\_Ensemble & (\#3)70.04 & (\#4)77.62 & (\#4)95.71 & (\#3)125.17 & (\#4)92.12 \\ 
  \hspace{1em} UA-EpiCovDA\cite{UA_EpiGro} & 80.13 & 100.85 & 115.18 & 121.61 & 104.44 \\ 
 \hspace{1em} LANL-GrowthRate\cite{LANL_GrowthRate} & 77.83 & 97.79 & 114.21 & 131.67 & 105.37 \\ 
\hspace{1em} Naive & 83.39 & 107.14 & 127.51 & 147.05 & 116.27 \\ 
\hspace{1em} epiforecasts-ensemble1\cite{epiforecasts_MIT} & 111.41 & 157.79 & 167.75 & 197.94 & 158.72 \\ 
 \hline \multicolumn{1}{l}{MAE} \\
\hspace{1em} COVIDhub-ensemble\cite{CDC_Ensemble} & 36.04 & 42.14 & 49.24 & 57.20 & 46.16 \\ 
\hspace{1em}  UMass-MechBayes\cite{UMASS} & 36.09 & 43.74 & 51.61 & 62.15 & 48.40 \\ 
\hspace{1em}  ARGOX\_Joint\_Ensemble & (\#3)39.56 & (\#3)46.81 & (\#3)56.93 & (\#4)75.39 & (\#3)54.64 \\ 
 \hspace{1em} MOBS-GLEAM\_COVID\cite{MOBS_GLEAM} & 43.01 & 50.70 & 60.32 & 70.73 & 56.19 \\ 
\hspace{1em}  LANL-GrowthRate\cite{LANL_GrowthRate} & 46.81 & 58.62 & 70.43 & 83.33 & 64.80 \\ 
\hspace{1em}  UA-EpiCovDA\cite{UA_EpiGro} & 48.01 & 62.61 & 71.97 & 78.51 & 65.28 \\ 
\hspace{1em}  Naive & 46.14 & 62.26 & 78.66 & 94.56 & 70.41 \\ 
\hspace{1em}  epiforecasts-ensemble1\cite{epiforecasts_MIT} & 51.84 & 65.12 & 76.77 & 92.41 & 71.53 \\ 
  \hline \multicolumn{1}{l}{Correlation} \\
\hspace{1em} COVIDhub-ensemble\cite{CDC_Ensemble} & 0.86 & 0.83 & 0.81 & 0.77 & 0.82 \\ 
 \hspace{1em} UMass-MechBayes\cite{UMASS} & 0.86 & 0.82 & 0.79 & 0.76 & 0.81 \\ 
 \hspace{1em} ARGOX\_Joint\_Ensemble & (\#4)0.83 & (\#3)0.82 & (\#3)0.79 & (\#3)0.74 & (\#3)0.80 \\ 
\hspace{1em}  MOBS-GLEAM\_COVID\cite{MOBS_GLEAM} & 0.84 & 0.80 & 0.76 & 0.72 & 0.78 \\ 
\hspace{1em}  LANL-GrowthRate\cite{LANL_GrowthRate} & 0.82 & 0.75 & 0.71 & 0.65 & 0.73 \\ 
 \hspace{1em} UA-EpiCovDA\cite{UA_EpiGro} & 0.78 & 0.70 & 0.64 & 0.63 & 0.69 \\ 
 \hspace{1em} epiforecasts-ensemble1\cite{epiforecasts_MIT} & 0.78 & 0.71 & 0.65 & 0.58 & 0.68 \\ 
\hspace{1em}  Naive & 0.79 & 0.71 & 0.64 & 0.54 & 0.67 \\ 
   \hline
\end{tabular}
\caption{Comparison among different models' 1 to 4 weeks ahead U.S. states level weekly incremental death predictions (from 2020-07-04 to 2021-10-09). The RMSE, MAE, Pearson correlation and their averages are reported. Methods are sorted based on their average. Our ARGOX-Joint-Ensemble's ranking for each error metric are included in parenthesis.}
\label{tab:State_Death_Other}
\end{table}

\subsubsection*{COVID-19 Cases}
For COVID-19 cases sensitivity analysis, we compare with the same models stated in the COVID-19 death retrospective comparison section above for both national and state levels. Fig. \ref{fig:Case_Nat_Our} displays the estimates against the observed COVID-19 weekly incremental cases, while table \ref{tab:Nat_Case_Our} (in the supplementary materials) further summarizes all estimations' performances in the three error metrics. Ref \cite{ma2021covid} and ARGO Joint estimations both steadily outperforms the naive estimations for 1 and 2 weeks ahead, demonstrating that optimally delayed Google search terms and time series information employed from death forecasting framework have strong predictive power on case trends as well. However, since Google search queries' optimal delays from COVID-19 cases are much shorter than from death (table \ref{tab:23query_optimallag}), when forecast horizon extends beyond 2 weeks ahead, majority of Google search terms' optimal lags from cases are smaller than the forecast horizon, resulting in Google search terms' signal deterioration, and worse prediction performances of Ref \cite{ma2021covid} and ARGOX-Joint. Yet, by incorporating \%ILI information as additional features, ARGO-Joint is able to rely more on the time series information and delayed \%ILI when Google search queries' signals deteriorates (see figure \ref{fig:Nat_Coef_Case}). Indeed, Ref \cite{ma2021covid} fall short against the naive estimations in 3-4 weeks ahead predictions, while ARGO-Joint is still able to produce steady estimations and recover from overshooting. This is also shown in table \ref{tab:Nat_Case_Other}, when comparing ARGO Joint estimations with other publicly available methods released by CDC. Nevertheless, ARGO-Joint's long term predictions are still impacted by the short term search information signals shown in  table \ref{tab:Nat_Case_Our}, as it ``degenerates'' to the naive method and only barely beats it for 3 and 4 weeks ahead predictions. 

Table \ref{tab:State_Case_Our} summarizes the overall results of the comparing methods (in sensitivity analysis) averaging over the 51 states, while table \ref{tab:State_Case_Other} summarizes the overall results when comparing with top CDC published teams (that report cases estimations). By incorporating neighbouring states' flu and case information, our ARGOX-Joint-Ensemble method improves upon Ref \cite{ma2021covid} and remains competitive among the top performers released by CDC, demonstrating the robustness of Flu+COVID joint framework on the cases predictions. Meanwhile signal deterioration still persists and affects state level long term forecasting (see table \ref{tab:State_Case_Our} and \ref{tab:State_Case_Other}), as the four ensembled methods are obtained by extracting different Geographical levels' Internet search information (national, regional and state-levels). Overall, our Flu+COVID joint framework is consistent with the top 3 CDC released teams in all error metrics, exhibiting strong short term national and state level cases forecasting while maintains its accuracy alongside the naive method in long term forecasting. Detailed state by state sensitivity analysis are shown in figures \ref{fig:State_Ours_Case_AK}-\ref{fig:State_Ours_Case_WY} and tables \ref{tab:State_Ours_Case_AK}-\ref{tab:State_Ours_Case_WY}.

\begin{table}[ht]
\sisetup{detect-weight,mode=text}
\renewrobustcmd{\bfseries}{\fontseries{b}\selectfont}
\renewrobustcmd{\boldmath}{}
\newrobustcmd{\B}{\bfseries}
\addtolength{\tabcolsep}{-4.1pt}
\footnotesize
\centering
\begin{tabular}{lrrrr}
  \hline
& 1 Week Ahead & 2 Weeks Ahead & 3 Weeks Ahead & 4 Weeks Ahead \\ 
    \hline \multicolumn{1}{l}{RMSE} \\
 \hspace{1em} Ref \cite{ma2021covid} &  3510.69 & 5942.41 & 7768.19 & 11008.92 \\ 
   \hspace{1em} ARGOX Joint Ensemble & \B3097.00 & \B5311.69 & \B7535.77 & 10725.59 \\ 
   \hspace{1em} Naive & 3700.11 & 5901.32 & 7683.88 & \B9706.93 \\ 
   \multicolumn{1}{l}{MAE} \\
  \hspace{1em}  Ref \cite{ma2021covid} & 2309.33 & 3785.76 & 4925.58 & 6635.53 \\ 
   \hspace{1em} ARGOX Joint Ensemble & \B2042.83 & \B3393.36 & \B4830.95 & \B6660.19 \\ 
   \hspace{1em} Naive & 2368.96 & 3949.65 & 5230.68 & 6797.63 \\ 
   \multicolumn{1}{l}{Correlation} \\
 \hspace{1em} Ref \cite{ma2021covid} &0.94 & 0.88 & 0.80 & \B0.70 \\ 
   \hspace{1em} ARGOX Joint Ensemble  & \B0.95 & \B0.89 & \B0.82 & 0.64 \\ 
   \hspace{1em} Naive& 0.94 & 0.84 & 0.74 & 0.55 \\
   \hline
\end{tabular}
\caption{Comparison of different methods for state-level COVID-19 1-4 weeks ahead incremental case in 51 U.S. states. The averaged RMSE, MAE, and correlation are reported and best performed method is highlighted in boldface. \%7 average RMSE improvement, 6\% average MAE improvement from Ref \cite{ma2021covid}.} 
\label{tab:State_Case_Our}
\end{table}

\begin{table}[ht]
\sisetup{detect-weight,mode=text}
\renewrobustcmd{\bfseries}{\fontseries{b}\selectfont}
\renewrobustcmd{\boldmath}{}
\newrobustcmd{\B}{\bfseries}
\addtolength{\tabcolsep}{-4.1pt}
\footnotesize
\centering
\begin{tabular}{lrrrrr}
 \hline
 & 1 Week Ahead & 2 Weeks Ahead & 3 Weeks Ahead & 4 Weeks Ahead & Average \\ 
  \hline \multicolumn{1}{l}{RMSE} \\
\hspace{1em}COVIDhub-ensemble\cite{CDC_Ensemble} & 3235.33 & 5041.92 & 6946.03 & 8608.85 & 5958.03 \\ 
  \hspace{1em}USC-SI\_kJalpha\cite{USC-SI_kJalpha} & 3685.74 & 5371.76 & 7505.12 & 10054.56 & 6654.30 \\ \hspace{1em}ARGOX\_Joint\_Ensemble & (\#1)3097.00 & (\#2)5311.69 & (\#2)7535.77 & (\#5)10725.59 & (\#3)6667.51 \\ 
  \hspace{1em}Naive & 3700.11 & 5901.32 & 7683.88 & 9706.93 & 6748.06 \\ 
  \hspace{1em}LANL-GrowthRate\cite{LANL_GrowthRate} & 4400.60 & 6321.68 & 8132.45 & 9796.18 & 7162.72 \\ 
  \hspace{1em}JHU\_CSSE-DECOM\cite{JHU_CSSE-DECOM} & 3723.61 & 6036.30 & 8253.63 & 11393.54 & 7351.77 \\
  \hspace{1em}CU-selec\cite{CU-Select} & 3821.30 & 6620.17 & 10550.82 & 12701.43 & 8423.43 \\ 
  \hspace{1em}Karlen-pypm\cite{karlen2020characterizing} & 4370.43 & 7931.34 & 13174.26 & 20084.94 & 11390.24 \\ 
 \hline \multicolumn{1}{l}{MAE} \\
\hspace{1em}COVIDhub-ensemble\cite{CDC_Ensemble} & 2070.99 & 3267.65 & 4626.24 & 5867.19 & 3958.01 \\ 
  \hspace{1em}USC-SI\_kJalpha\cite{USC-SI_kJalpha} & 2248.76 & 3324.88 & 4790.20 & 6344.28 & 4177.03 \\
  \hspace{1em}ARGOX\_Joint\_Ensemble & (\#3)2042.83 & (\#3)3393.36 & (\#3)4830.95 & (\#3)6660.19 & (\#3)4231.83 \\ 
  \hspace{1em}LANL-GrowthRate\cite{LANL_GrowthRate} & 2758.55 & 3861.20 & 5101.73 & 6515.04 & 4559.13 \\
  \hspace{1em}JHU\_CSSE-DECOM\cite{JHU_CSSE-DECOM} & 2273.46 & 3741.02 & 5202.54 & 7086.21 & 4575.81 \\
  \hspace{1em}Naive & 2368.96 & 3949.65 & 5230.68 & 6797.64 & 4586.733 \\ 
  \hspace{1em}CU-select\cite{CU-Select} & 2451.64 & 3908.73 & 5857.80 & 7166.84 & 4846.25 \\ 
  \hspace{1em}Karlen-pypm\cite{karlen2020characterizing} & 2556.34 & 4358.50 & 6797.77 & 9902.27 & 5903.72 \\ 
  \hline \multicolumn{1}{l}{Correlation} \\
\hspace{1em}COVIDhub-ensemble\cite{CDC_Ensemble} & 0.95 & 0.89 & 0.80 & 0.70 & 0.83 \\ 
  \hspace{1em}JHU\_CSSE-DECOM\cite{JHU_CSSE-DECOM} & 0.94 & 0.87 & 0.79 & 0.69 & 0.82 \\ 
  \hspace{1em}ARGOX\_Joint\_Ensemble  & (\#2)0.95 & (\#2)0.89 & (\#2)0.82 & (\#5)0.64 & (\#3)0.82 \\ 
  \hspace{1em}Karlen-pypm\cite{karlen2020characterizing} & 0.93 & 0.87 & 0.78 & 0.70 & 0.82 \\ 
  \hspace{1em}LANL-GrowthRate\cite{LANL_GrowthRate} & 0.92 & 0.86 & 0.78 & 0.66 & 0.81 \\ 
  \hspace{1em}Naive &0.93 & 0.84 & 0.73 & 0.61 & 0.78 \\ 
  \hspace{1em}CU-select\cite{CU-Select} & 0.93 & 0.84 & 0.68 & 0.57 & 0.75 \\ 
  \hspace{1em}USC-SI\_kJalpha\cite{USC-SI_kJalpha} & 0.87 & 0.81 & 0.72 & 0.61 & 0.75 \\ 
  \hline
\end{tabular}
\caption{Comparison among different models' 1 to 4 weeks ahead U.S. states level weekly incremental death predictions (from 2020-07-04 to 2021-10-09). The RMSE, MAE, Pearson correlation and their averages are reported. Methods are sorted based on their average. Our ARGOX-Joint-Ensemble's ranking for each error metric are included in parenthesis.}
\label{tab:State_Case_Other}
\end{table}

\subsection*{\%ILI}
To evaluate the accuracy of our \%ILI estimations, we compared the estimates with the actual \%ILI released by CDC weeks later, and different benchmark methods for national and state level forecasts. 

For national level, we compare (a) the persistence (Naive) estimates, which simply use CDC’s reported \%ILI of the previous week as the estimate for the current week, (b) estimates by the lag-3 autoregressive model (AR-3 model), (c) ARGO (Ref \cite{ARGO}), and (d) ARGO-Joint (section \ref{sec:ARGO}). ARGO \cite{ARGO} predictions are obtained from a $L_1$ penalized regression, utilizing delayed \%ILI and (flu related) Google search queries, while ARGO-Joint additionally uses weekly aggregated COVID-19 cases as exogenous variables. For fair comparisons, AR-3, Ref \cite{ARGO} and ARGO-Joint all use 52 weeks training period. Figure \ref{fig:ILI_Nat_Our} displays the estimates against actual CDC-reported \%ILI, whereas table \ref{tab:Nat_ILI_Our} (in the supplementary materials) further summarizes all estimations' performances in three error metrics. Ref \cite{ARGO} outperforms the naive and AR-3 (time series alternative) estimates in all error metrics.  ARGO-Joint further improves from Ref \cite{ARGO} by capturing the COVID-19 cases' evolvement during the same period, being the best performer shown in table \ref{tab:State_ILI_Our}, with 6.1\% RMSE, 11\% MAE and 0.2\% Pearson Correlation improvements from Ref \cite{ARGO}. By utilizing people’s search behavior to foresee future trends, Ref \cite{ARGO} and ARGO-Joint are able to overcome delaying effect in the \%ILI predictions, and predict almost perfectly from July 2021 to October 2021 (as the 52 weeks training period covers the previous testing horizons). By integrating COVID-19 cases information, ARGO-Joint is able to further prevent undesired over-predictions comparing to Ref \cite{ARGO}.

For state level, we compare (a) the persistence (Naive) estimates, (b) estimates by the lag-1 vector autoregressive model (VAR-1 model), (c) ARGOX model (Ref \cite{ARGOX}), (d) ARGOX-Idv (section \ref{sec:state_model}), (e) ARGOX-Joint-Ensemble (section \ref{sec:state_model}). Table \ref{tab:State_ILI_Our} shows the comparing methods' performance averaging across the 50 states and NYC. ARGOX-Joint-Ensemble gives the leading performance uniformly in all metrics, which achieves around 17\% error reduction in RMSE, around 24\% error reduction in MAE and around 13\% increase in Pearson correlation compared to the best alternative in the whole period. Among all the methods we compare in this section, Ref \cite{ARGOX}, ARGO-Idv, and ARGOX-Joint-Ensemble all uniformly outperforms the naive and VAR predictions, while ARGOX-Joint-Ensemble is the only method consistently beats the naive estimates in all the states (except NYC) in all error metric shown in figures \ref{fig:State_Ours_ILI_AL}-\ref{fig:State_Ours_ILI_NYC} and tables \ref{tab:State_Ours_ILI_AL}-\ref{tab:State_Ours_ILI_NYC} (detailed state-by-state comparisons). Figure \ref{fig:ILI_Our_Violine} further shows all states' RMSE, MAE and Pearson Correlation in the violin charts, with mean and standard deviations, where the joint Flu+COVID ensemble framework reveals its robustness over geographical variability and extracts a strong combination from the other two ARGOX alternatives, shown in table \ref{tab:Ensemble_Selection}.

\begin{table}[ht]
\sisetup{detect-weight,mode=text}
\renewrobustcmd{\bfseries}{\fontseries{b}\selectfont}
\renewrobustcmd{\boldmath}{}
\newrobustcmd{\B}{\bfseries}
\addtolength{\tabcolsep}{-4.1pt}
\centering
\begin{tabular}{|c|c|c|c|}
  \hline
Methods & RMSE & MAE & Correlation \\ \hline\hline
Naive &0.273 &0.183 &0.812 \\ 
    \hline VAR &0.380 &0.264 &0.704 \\ 
    \hline Ref \cite{ARGOX} & 0.260 &0.178  & 0.823 \\ 
    \hline ARGOX-Idv & 0.248 & 0.165 & 0.846 \\ 
    \hline ARGOX-Joint-Ensemble & \B0.205 &\B0.125  & \B0.960 \\ 
   \hline
\end{tabular}
\caption{Comparison of different methods for state-level \%ILI 1 week ahead \%ILI in 51 U.S. states. The best performed method ishighlighted in boldface.} 
\label{tab:State_ILI_Our}
\end{table}

\section{Discussion}
ARGO-Joint extends from previously-proposed ARGO \cite{ma2021covid} method by incorporating COVID-19 cases and \%ILI as exogenous variables for national level \%ILI and cases/death predictions, respectively, serving as a consistent framework for Flu+COVID national-level forecates. Meanwhile, ARGOX-Joint-Ensemble is a unified Flu+COVID forecasting framework that efficiently combines multi-source, multi-resolution information, and provides accurate, reliable, real-time COVID-19 cases, death and \%ILI tracking at states level. Previously-proposed ARGOX and its alternative methods \cite{ARGO,ARGOX,ma2021covid} effectively gathers publicly available search data from Google at different resolutions (national, regional and state-level) and time series information of the forecasting target (COVID-19 or \%ILI). Motivated by the connections between \%ILI and COVID-19 cases/death trends (see figure \ref{fig:GA_COVID_ILI}) among each individual states and their neighbouring states, this study further proposed ARGOX-Idv which additionally gathers neighbouring states' \%ILI information for each individual states' COVID-19 cases/death forecast and utilizes COVID-19 cases for \%ILI forecast. Lastly, by aggregating ARGOX-Idv and the previously-proposed COVID-19/ \%ILI forecasting methods in an winner-takes-all ensemble fashion, ARGOX-Joint-Ensemble provides accurate, reliable real-time Flu+COVID joint tracking at the state level. By incorporating cross-regional and temporal correlation of both COVID-19 and influenza activities for accurate estimation, ARGOX-Joint-Ensemble outperforms most benchmark methods and stays competitive against other publicly available models. 

Benefited from the COVID-19 \cite{ma2021covid} and influenza tracking \cite{ARGO} frameworks that have already shown their strength by recognizing and utilizing strong temporal auto-correlation and dependence between people search behaviors and time series information (from forecast targets), ARGO-Joint takes a step further by complementing COVID-19 and influenza information interchangeably between the COVID-19 and influenza forecasting frameworks. By simply incorporating delayed influenza and COVID-19 cases activities for COVID-19 cases/death and \%ILI forecasts respectively, ARGO-Joint shows performances improvements for COVID-19 cases/death and \%ILI future predictions, with fast recovery from over-estimations and smoother forecasts (especially when forecasting 3-4 weeks ahead for COVID-19 cases and deaths).

For state level, we effectively combine neighbouring states' influenza activity, publicly available data from Google searches and delayed COVID-19 cases/death to produce ARGOX-Idv state level COVID-19 cases/death estimations (vise-versa for \%ILI estimates). This proposed ARGOX-Idv is an interactive method that connects influenza and COVID-19 information in geographical proximity, and is a joint Flu+COVID method that use each other's growth trend and help each other. To further improve accuracy and robustness, we efficiently combine ARGOX-idv with previously-proposed COVID-19 and \%ILI forecasting methods \cite{ARGOX, ma2021covid} to produce winner-takes-all ensemble (ARGOX-Joint-Ensemble) forecast for 1 to 4 weeks ahead state level cases/deaths, and 1 week ahead \%ILI. The combined methods are adapted directly from influenza and COVID-19 predictions with minimal changes, which demonstrates their accuracy, and general applicability to boost the joint Flu+COVID ensemble's performances together with ARGOX-Idv. Furthermore, the ARGO-Joint-Ensemble is able to outperform the constituent models for all states in all 1 to 4 weeks ahead COVID-19 cases/death predictions, and 1 week ahead \%ILI predictions. Our national and state-level models are also competitive to other state-of-arts models from CDC for COVID-19 forecasts, demonstrating the strength of the unifying framework to tract and enables public health officials to make timely decisions regarding both diseases simultaneously.

Like all big-data-based models, our result has certain limitations. ARGO-Joint and ARGOX-Joint-Ensemble's accuracy depends on the reliability and stability of its inputs—Google Trends data, historical \%ILI data from CDC, and COVID-19 cases/death data from NYT.  Since the optimal delay between COVID-19 cases/death and Google search queries have short time-span (table \ref{tab:23query_optimallag}), information in Google search data deteriorates as forecast horizons expand, which could potentially impact the robustness and accuracy of our 3 and 4 weeks ahead predictions. Fortunately, by recognizing the correlation between influenza activities and COVID-19 growth trend in both national and state level, \%ILI and COVID-19 cases/death are able to rely more on each other when Google search queries' signal degenerates, which enables robust estimations and fast recoveries from overshooting to steadily outperform benchmark methods. Nevertheless, models to further capture long-term COVID-19 trends, and boost long-term forecasting performances could be an interesting future direction.

In light of recurrent influenza and COVID-19 waves, accurate joint localized and national tracking of epidemic activity has become more important than ever before.Such high-precision, robust national and state-level surveillance information, provided by ARGO-Joint and ARGOX-Joint-Ensemble, enables public health officials to make timely decisions and optimal resource reallocation regarding the changes in COVID-19 and flu epidemics simultaneously and jointly. The reliable estimations by this joint Flu+COVID framework gives public more insights into both diseases and can serve valuable resources for government officials.

\clearpage


\newpage
\setcounter{page}{1}
\rfoot{\thepage}
\section*{Supporting Information for
Use Internet Search Data to Accurately Track State Level Influenza Epidemics}
\subsection*{Simin Ma, Shaoyang Ning, Shihao Yang}
\noindent Correspondence to:  shihao.yang@isye.gatech.edu
\noindent This PDF file includes:
\begin{itemize}
\item Supplementary Text
\item Supplementary Figs.
\item Supplementary Tables
\end{itemize}

\setcounter{table}{0}
\renewcommand{\thetable}{S\arabic{table}}%
\setcounter{figure}{0}
\renewcommand{\thefigure}{S\arabic{figure}}%

This Supplementary Material is organized as following:

\subsection*{Imputation of Daily ILI}
Here we impute daily number of outpatients with Influenza like-illness (ILI) from weekly ILI (\%ILI scaled by $100$), by adapting COVID-19 daily incremental cases. 

Here we denote $\tau$ as weekly indicator and $t$ as daily indicator. Let $y^C_{t,m}$ be the New York Times COVID-19 cases increment at day $t$ of area $m$, where the area $m$ can refer to the entire nation, one specific HHS region such as New England, or one specific state such as Georgia). Let $p^*_{\tau,m}$ be the CDC published \%ILI for week $\tau$ (with ending date on Saturday for each week), and $p_{\tau,m}$ be the re-scaled ILI for week $\tau$. Lastly, let $p^{(n)}_{t,m}$ be the imputed CDC ILI at day $t$ for $n^{\text{th}}$ imputation iteration of area $m$. Then, for all $n=1,\ldots,N$ imputation iteration, $p_{\tau,m}=\sum_{i=0}^6p^{(n)}_{t_{\tau,sat}-i,m}$, where $t_{\tau,sat}$ denote the Saturday of week $\tau$. This means that sum of imputed daily ILI of week $\tau$ equals to the weekly ILI of week $\tau$.

Detailed imputation steps are shown in Algorithm \ref{Alg:ImputeMethod} below, where we randomly select  COVID-19 incremental cases from a week in the past (after April 1st 2020) and normalize them to sum up to the \%ILI of the current imputing week as the final imputed values for $\{p^{(n)}_{t-i,m}\}_{j=0,\ldots,6}$. We impute for total of $N=100$ iterations and use them throughout this study.

\begin{algorithm}[htbp]
    \caption{Impute Daily CDC \%ILI for area $m$} 
	\begin{algorithmic}[1]
	    \Inputs{$y^C_{t,m},\,p_{\tau,m}$ for days $t\in\{1,\ldots,T\}$ and weeks $\tau\in\{1,\ldots,\mathcal{T}\}$ for area $m$. $N$ total imputation iterations.}
	    \For{$n=1:N$}
		\For{$\tau=1:\mathcal{T}$}
		\State Uniformly select $\Tilde{\tau}$ from $\{1,2,\ldots, \tau\}$.
		\State \multiline{Obtain normalized and scaled week $\Tilde{\tau}$ COVID-19 cases $\{\frac{y^C_{\Tilde{t}-6,m}p_{\tau,m}}{\sum_{i=0}^6y^C_{\Tilde{t}-i,m}}, \frac{y^C_{\Tilde{t}-5,m}p_{\tau,m}}{\sum_{i=0}^6y^C_{\Tilde{t}-i,m}},\ldots, \frac{y^C_{\Tilde{t},m}p_{\tau,m}}{\sum_{i=0}^6y^C_{\Tilde{t}-i,m}}\}$, such that sum to week $\tau$ \%ILI, where $\Tilde{t}$ is the ending Saturday of week $\Tilde{\tau}$.}
		\State \multiline{The imputed ILI for week $\tau$ are $p^{(n)}_{{t}-j,m}=\frac{y^C_{\Tilde{t}-j,m}p_{\tau,m}}{\sum_{i=0}^6y^C_{\Tilde{t}-i,m}}$ for $j\in\{0,1,\ldots,6\}$, where $\Tilde{t}$ is the ending Saturday of week $\Tilde{\tau}$, and ${t}$ is the ending Saturday of week ${\tau}$.}
		\EndFor
		\EndFor
		\Outputs{$p^{(n)}_{t,m}$ the $n^\text{th}$ imputed CDC ILI sequence at day $t\in\{1,\ldots,T\}$ of area $m$ for $n=1,\ldots,100$.}
	\end{algorithmic} 
	\label{Alg:ImputeMethod}
\end{algorithm}

\clearpage
\subsection*{ARGO-Joint Prediction}
\subsubsection*{\textit{COVID-19 Cases and Death}}
Here, we discuss the national-level COVID-19 cases and death prediction model in detail. On the high-level, we employ delayed (daily imputed) ILI information into ``ARGO-Inspired'' method \cite{ma2021covid} for both cases and death predictions, but with slight exogenous variable adjustments for cases predictions. Since the COVID-19 case trend leads the death trend not vise-versa, we include only COVID-19 delayed cases as time-series exogenous variables in cases prediction model, while using the same important Google search queries (with different optimal lags in the LASSO regression shown in table \ref{tab:23query_optimallag}) and including delayed ILI information. As we use the imputed ILI used as exogenous variables in both cases and death forecasting framework, the final predictions will be the median taken across the imputation iterations. 

Let $X^C_{i,t,m}$ be the COVID-19 related Google Trends data of search term $i$ day $t$ of area $m$; $X^F_{i,\tau,m}$ be the Flu related Google Trends data of search term $i$ week $\tau$ of area $m$; $y^D_{t,m}$ be the New York Times COVID-19 death increment at day $t$ of area $m$; $y^C_{t,m}$ be the New York Times COVID-19 confirmed case increment at day $t$ of area $m$; $p^*_{\tau,m}$ be the CDC published \%ILI for week $\tau$ of area m; $p^{(n)}_{t,m}$ be the $n^\text{th}$ imputed \%ILI for day $t$ of area m, where the area $m$ can refer to the entire nation, one specific HHS region (such as New England), or one specific state (such as Georgia). Let $O^D_k$ be the optimal lag for the $k$th Google search term with respect to (w.r.p) COVID-19 incremental death, and $O^C_k$ be the optimal lag for the $k$th Google search term w.r.p to COVID-19 incremental death, which are the same for all area $m$ respectively. Let $\mathbb{I}_{\{t, r\}}$ be the weekday $r$ indicator for $t$ (i.e., $\mathbb{I}_{\{t, 1\}}$ indicates day $t$ being Monday, and $\mathbb{I}_{\{t, 6\}}$ indicates day $t$ being Saturday), which accounts for the weekday seasonality in COVID-19 incremental death time series.

Inspired by ARGO method \cite{ARGO}, with information available as of time $T$, to estimate $y^C_{T+l,m}$ and $y^D_{T+l,m}$ for $l>0$, the incremental COVID-19 cases and death on day $T+l$ of area $m$, $L_1$ regularized linear estimators are used:
\begin{equation}\label{eqn:argo_case_yhat}
\begin{aligned}
    \hat{y}^{C,(n)}_{T+l,m} &= \hat{\mu}^{C,(n)}_{y,m} + \sum^{I}_{i=0}\hat{\alpha}^{C,(n)}_{i,m}y^C_{T-i,m} + \sum_{j\in \mathcal{J}}\hat{\beta}^{C,(n)}_{j,m}y^C_{T+l-j,m} + \sum^{K}_{k=1}\hat{\delta}^{C,(n)}_{k,m}X_{k,T+l-\hat{O}^C_k,m}  + \sum^6_{r=1}\hat{\gamma}^{C,(n)}_{r,m}\mathbb{I}_{\{T+l, r\}} \\
    & + \sum_{h\in\mathcal{H}}\eta_{h,m}^{C,(n)} p^{(n)}_{\mathcal{T}-h,m}
\end{aligned}
\end{equation}
\begin{equation}\label{eqn:argo_death_yhat}
\begin{aligned}
    \hat{y}^{D,(n)}_{T+l,m} &= \hat{\mu}^{D,(n)}_{y,m} + \sum^{I}_{i=0}\hat{\alpha}^D_{i,m}y^D_{T-i,m} + \sum_{j\in \mathcal{J}}\hat{\beta}^{D,(n)}_{j,m}y^C_{T+l-j,m} + \sum^{K}_{k=1}\hat{\delta}^{D,(n)}_{k,m}X_{k,T+l-\hat{O}^D_k,m}  + \sum^6_{r=1}\hat{\gamma}^{D,(n)}_{r,m}\mathbb{I}_{\{T+l, r\}} \\
    &+\sum_{h\in\mathcal{H}}\eta_{h,m}^{D,(n)} p^{(n)}_{\mathcal{T}-h,m}
\end{aligned}
\end{equation}
where we use lagged confirmed cases, cases' optimal lagged Google search terms and previous weeks' ILI for incremental cases predictions, and use lagged death, lagged confirmed cases, deaths' optimal lagged Google search terms and previous weeks' ILI for death prediction, in each imputation iterations. We then take the median over the estimates cross the imputation iterations above as our final estimations, i.e. final incremental case estimation is $\hat{y}^C_{T+l,m}=\mathrm{Med}_{n=1,\ldots,100}\left(\hat{y}^{C,(n)}_{T+l,m}\right)$ and final incremental death estimation is $\hat{y}^D_{T+l,m}=\mathrm{Med}_{n=1,\ldots,100}\left(\hat{y}^{D,(n)}_{T+l,m}\right)$. For $l^{\text{th}}$ day ahead incremental cases predictions at area $m$, the coefficients of the case forecasting model $\bm{\theta}^C=\{\mu_{y,m}^C,\bm{\alpha}^C=(\alpha^C_{1,m},\ldots,\alpha^C_{I,m}), \bm{\beta}^C=(\beta_{1,m}^C,\ldots,\beta_{|\mathcal{J}|,m}^C), \bm{\delta}^C=(\delta_{1,m}^C,\ldots,\delta_{K,m}^C), \bm{\gamma}^C=(\gamma_{1,m}^C, \ldots, \gamma_{6,m}^C),\bm{\eta}^C = \{\eta_{1,m}^C,\ldots,\eta_{|\mathcal{H}|,m}^C\}\}$ are obtained via
\begin{equation}\label{eqn:argo_case_obj}
    \begin{aligned}
    \underset{\bm{\theta}^C, \bm{\lambda}^C}{\mathrm{argmin}} 
    \sum_{t=T-M-l+1}^{T-l} &  \Bigg( y^C_{t+l,m}-\mu^C_{y,m} - \sum^{6}_{i=0}{\alpha}^C_{i,m}y^C_{t-i,m} - \sum_{j\in \mathcal{J}}{\beta}^C_{j,m}y^C_{t+l-j,m} \\
    & -  \sum^{27}_{k=1}{\delta}^C_{k,m}X_{k,t+l-\hat{O}^C_k,m} -\sum^6_{r=1}\gamma_{r,m}^C\mathbb{I}_{\{t+l,r\}}-\sum_{h\in\mathcal{H}}\eta_{h,m}^Cp^{(n)}_{\tau-1,m} \Bigg)^2 + \bm{\lambda}^C\|\bm{\theta}^C\|_1
    \end{aligned}
\end{equation}
For $l^{\text{th}}$ day ahead incremental deaths predictions at area $m$, The coefficients of the death forecasting model $\bm{\theta}^D=\{\mu_{y,m}^D,\bm{\alpha}^D=(\alpha^D_{1,m},\ldots,\alpha^D_{I,m}), \bm{\beta}^D=(\beta_{1,m}^D,\ldots,\beta_{|\mathcal{J}|,m}^D), \bm{\delta}^C=(\delta_{1,m}^D,\ldots,\delta_{K,m}^D), \bm{\gamma}^D=(\gamma_{1,m}^D, \ldots, \gamma_{6,m}^D),\bm{\eta}^D = \{\eta_{1,m}^D,\ldots,\eta_{|\mathcal{H}|,m}^D\}\}$ are obtained via
\begin{equation}\label{eqn:argo_death_obj}
    \begin{aligned}
    \underset{\bm{\theta}^D, \bm{\lambda}^D}{\mathrm{argmin}} 
    \sum_{t=T-M-l+1}^{T-l} &\Bigg( y_{t+l,m}^D-\mu_{y,m}^D - \sum^{6}_{i=0}{\alpha}^D_{i,m}y_{t-i,m}^D - \sum_{j\in \mathcal{J}}{\beta}^D_{j,m}y^C_{t+l-j,m} \\
    \;\;\;&-  \sum^{27}_{k=1}{\delta}^D_{k,m}X_{k,t+l-\hat{O}^D_k,m} -\sum^6_{r=1}\gamma^D_{r,m}\mathbb{I}_{\{t+l,r\}}- \sum_{h\in\mathcal{H}}\eta_{h,m}^Dp^{(n)}_{\tau-1,m} \Bigg)^2+ \bm{\lambda}^D\|\bm{\theta}^D\|_1
    \end{aligned}
\end{equation}

Here, we dropped the imputation iteration $(n)$ for simplicity. We set $M=56$, i.e. 56 days as training period; $I=6$ considering consecutive 1 week lagged target time series; $\mathcal{J}=\max\left(\{7,14,21,28\},l\right)$ considering weekly lagged confirmed cases; $K=23$ highly correlated COVID-19 related Google search terms; $\hat{O}^C_k=\max\left(O^C_k,l\right)$ be the adjusted optimal lag w.r.p to COVID-19 cases, and $\hat{O}^D_k=\max\left(O^D_k,l\right)$ be the adjusted optimal lag w.r.p to COVID-19 deaths of $k$th Google search term subject to $l^{\text{th}}$ day ahead prediction; $\mathcal{H}=\max\left(\{7,14,21,28\},l\right)$ considering weekly lagged ILI. We set hyperparameters $\bm{\lambda}^C$ and $\bm{\lambda}^D$ through cross-validation separately.

To further impose smoothness into our predictions, we use the three-day moving average of the coefficients for predicting day $T+l$, which slightly boosts our prediction accuracy.

Using the above formulation, we forecast future 4 weeks of daily incremental COVID-19 cases and death of area $m$, i.e. $\{\hat{y}^C_{T+1,m},\ldots,\hat{y}^C_{T+28,m}\}$ and $\{\hat{y}^D_{T+1,m},\ldots,\hat{y}^D_{T+28,m}\}$, and aggregate them into weekly predictions. In other words, $\hat{y}^C_{T+1:T+7,m}=\sum^7_{i=1}\hat{y}^C_{T+i,m}$ is first week's prediction, and $\hat{y}^C_{T+8:T+14,m}= \sum^{14}_{i=8}\hat{y}^C_{T+i,m}$ is the second week's prediction and etc, similarly for incremental weekly deaths predictions. We denote this method as ``ARGO Inspired Prediction''.

\subsubsection*{\textit{\%ILI}}
We obtain an accurate estimate $\hat{p}^*_{\tau,m}$ for the \%ILI for week $\tau=\{1,\ldots,\mathcal{T}\}$ of area $m$ using the ARGO method \cite{ARGO}, additionally incorporating area $m$ consecutive 1 week lagged COVID-19 incremental cases as exogenous variables.

\clearpage
\subsection*{Modifications of Previously-Proposed COVID and Influenza Methods}
\subsubsection*{ARGOX-2Step}
Here, ARGOX-2Step operates slightly differently for cases and deaths predictions. For COVID-19 deaths forecasts, similar to previous study \cite{ma2021covid}, the second step takes a dichotomous approach for the joint states and alone states, identified through geographical separations and multiple correlations on COVID-19 death growth trends, where the joint states are forecast jointly utilizing cross-state, cross-source information from each other, while alone states are forecast separately each with its own state and national information. On the other hand, we unify all the states and forecast jointly for cases predictions. The reason behind different second step approaches in the ARGOX framework lies in the sparsity of state-level COVID-19 cases and death data. COVID-19 incremental deaths are sparse and not well correlated with other states and regions for those isolated (alone) states, which weakens the impact from cross-state, cross-region information in the joint framework. Meanwhile, COVID-19 incremental cases are dense and well-correlated cross state and regions, which indicate advantages of using joint estimations. 

Lastly, we estimate 1 week ahead state-level \%ILI using ARGOX \cite{ARGOX}

\subsubsection*{ARGOX-NatConstraint}
Similar to prior study, we incorporate a constrained second step to ARGOX inspired state level prediction above \cite{ma2021covid} for both COVID-19 cases and death forecasts, while separating out HI and VT during death forecasting, and no such separation in cases forecasting, due to data sparsity.

\clearpage
\subsection*{Newly Proposed ARGOX-Idv}
\subsection*{ARGOX State-by-State Forecast (ARGOX-Idv)}
Since \%ILI and COVID-19 cases and death have strong connections between each state and its neighbours (figure \ref{fig:GA_COVID_ILI}), this section introduces a modified ARGOX framework \cite{ARGOX, ma2021covid}, by incorporating spatial temporal flu information for COVID-19 cases/death state level estimates, and considering COVID-19 cases information for \%ILI estimates, while treating each state separately to produce individual state-level forecasts. 

For the state $m$, our raw estimates for its weekly COVID-19 incremental cases/deaths $y_{\tau,m}$ are $\hat{y}^{GT}_{\tau,m}$, $\hat{y}_{\tau, r_m}^{reg}$,  $\hat{y}^{nat}_{\tau}$ and $p^{(n)}_{\tau-1,m}$, where $r_m$ is the region number for state $m$, and $p^{(n)}_{\tau-1,m}$ is the weekly aggregated daily imputed ILI of week $\tau-1$ in state $m$ for $n^\text{th}$ imputation iteration (since ILI lag one week behind COVID-19 published data). Here, we overload the notations without superscript specifying COVID-19 cases ($C$) and death ($D$), as their predictions' frameworks are the same. Similar to the second step in ARGOX \cite{ARGOX}, we denote the death increment at week $\tau$ of state $m$ as ${Z}_{\tau,m}=\Delta{y}_{\tau,m}={y}_{\tau,m}-{y}_{\tau-1,m}$ (target scalar) and it has the following predictors: (i) ${Z}_{\tau-1,m}=\Delta{y}_{\tau-1,m}$, (ii) $\{\hat{{y}}_{\tau,m}^{GT} - {y}_{\tau-1,m}\}_{m\in\mathcal{M}}$, (iii) $\hat{{y}}_{\tau,r_m}^{reg} - {y}_{\tau-1,r_m}$, (iv) $\hat{{y}}_{\tau}^{nat} - {y}_{\tau-1}$, and (v) $\{p^{(n)}_{\tau-1,m} - p^{(n)}_{\tau-2,m}\}_{m\in\mathcal{M}}$, where $\mathcal{M}$ is a set containing all states in the same region as the target state $m$. In other words, there are $(2|\mathcal{M}|+3)$ predictors for state $m$, and let's denote the predictor vector for $n^\text{th}$ imputation iteration as $\bm{W}^{(n)}_{\tau,m}=({Z}_{\tau-1}, \{\hat{{y}}_{\tau,m}^{GT} - {y}_{\tau-1,m}\}_{m\in\mathcal{M}}, (\hat{{y}}_{\tau}^{reg} - {y}_{\tau-1}), (\hat{{y}}_{\tau}^{nat} - {y}_{\tau-1}), \{{p}^{(n)}_{\tau-1,m} - {p}^{(n)}_{\tau-2,m}\}_{m\in\mathcal{M}})^\intercal$. 

We propose a modified structured variance-covariance matrix for reliable estimation and numerical stability (see subsection below). Lastly, the best linear predictor with ridge-regression inspired shrinkage \cite{ARGOX} is used to get the $n^\text{th}$ imputation iteration's estimate for each state, and we take the median across these $100$ estimations as our final predictions, considering 30 weeks training period.

Future \%ILI predictions also follow the similar framework as above, incorporating neighbouring state's COVID-19 cases information as the additional predictor in the second step of ARGOX \cite{ARGOX}. Particularly, our raw estimates for $p^*_{\tau,m}$, the \%ILI of state $m$ week $\tau$, are $\hat{p}^{GT}_{\tau,m}$ (from ARGOX first step \cite{ARGOX}), $\hat{p}^{reg}_{\tau,r_m}$ (from ARGO2 first step \cite{ARGO2_Regional}), $\hat{p}^{nat}_{\tau,m}$ (from ARGO \cite{ARGO}), and $y^{C}_{\tau,m}$ (weekly aggregated COVID-19 incremental cases of week $\tau$ in state $m$ from NYT), where $r_m$ is the region number for state $m$. Similarly, we denote the predictor vector for our target $Z^F_{\tau,m}=\Delta p^*_{\tau,m}= p^*_{\tau,m}-p^*_{\tau-1,m}$, as $\bm{W}^F_{\tau,m}=({Z}^F_{\tau-1}, \{\hat{{p}}_{\tau,m}^{GT} - {p}_{\tau-1,m}\}_{m\in\mathcal{M}}, (\hat{{p}}_{\tau}^{reg} - {p}^*_{\tau-1}), (\hat{{p}}_{\tau}^{nat} - {p}^*_{\tau-1}), \{y^{C}_{\tau,m} - y^C_{\tau-1,m}\}_{m\in\mathcal{M}})^\intercal$, where $F$ stands for ``Flu''. Since NYT publishes daily COVID-19 reported cases, we are able to use week $\tau$'s COVID-19 cases when prediction week $\tau$'s \%ILI. Lastly, the best linear predictor with ridge-regression inspired shrinkage \cite{ARGOX} is used to get the final estimate. 

\subsubsection*{Structured Variance-Covariance Matrix}
Here, we illustrate the proposed structured variance-covariance matrix and its assumption used in ARGOX-Idv, through COVID-19 cases/death prediction. \%ILI prediction also follows the same framework.

The predictor for COVID-19 incremental cases/death's increment $Z_{\tau,m}=y_{\tau,m}-y_{\tau-1,m}$ at week $\tau$ in state $m$ is   $\bm{W}^{(n)}_{\tau,m}=({Z}_{\tau-1,m}, \{\hat{{y}}_{\tau,m}^{GT} - {y}_{\tau-1,m}\}_{m\in\mathcal{M}}, (\hat{{y}}_{\tau}^{reg} - {y}_{\tau-1}), (\hat{{y}}_{\tau}^{nat} - {y}_{\tau-1}), \{{p}^{(n)}_{\tau-1,m} - {p}^{(n)}_{\tau-2,m}\}_{m\in\mathcal{M}})^\intercal$ in $n^{\text{th}}$ imputation iteration. Let's denote the $n^{\text{th}}$ imputed week $\tau$ ILI increment in state $m$ as $Z^{F,(n)}_{\tau,m} = \{{p}^{(n)}_{\tau,m} - {p}^{(n)}_{\tau-1,m}\}_{m\in\mathcal{M}}$. We will drop the super-script $n$ in equations below for simplicity, as the procedure will be the same across imputation iterations. 

From ARGOX \cite{ARGOX}, the final best linear predictor with ridge-regression-inspired shrinkage for $Z_{\tau,m}$ is 
\begin{equation}
    \hat{{Z}}_{\tau,m} = {\mu}^{(m)}_{Z} + \frac{1}{2}\Sigma_{ZW}^{(m)}(\frac{1}{2}\Sigma^{(m)}_{WW}+\frac{1}{2}D^{(m)}_{WW})^{-1}(\bm{W}_{\tau,m}-\bm{\mu}^{(m)}_{W}).\label{eq:2ndstep}
\end{equation}
where $\mu_Z^{(m)}$ and $\bm\mu_W^{(m)}$ are the mean scalar and vector of ${Z}_{\tau,m}$ and $\bm{W}_{\tau,m}$ respectively. $\sigma^{2^{(m)}}_{ZZ}, \Sigma^{(m)}_{ZW}, \Sigma^{(m)}_{WW}$ are covariance matrix of and between ${Z}_{\tau,m}$ and $\bm{W}_{\tau,m}$.

We structure the covariance matrices for reliable estimation, following the same assumptions in ARGOX \cite{ARGOX}:
\begin{enumerate}
    \item 
     The time series increments of COVID-19 cases/deaths and ILI are stationary and have a stable auto-correlation and cross-correlation across time and between COVID-19 and flu. Therefore, the covariances between the time series increments for state $m$ satisfy 
    \begin{itemize}
        \item $\mathrm{var}({Z}_{\tau,m}) = \mathrm{var}({Z}_{\tau-1,m})=\sigma^{2^{(m)}}_{ZZ} $
        \item $\mathrm{cov}(\bm{Z}_{\tau}, \bm{Z}_{\tau-1})=\rho^{(m)} \sigma^{2^{(m)}}_{ZZ}$
        \item $var({Z}^F_{\tau,m})=var({Z}^F_{\tau-1,m})=\sigma^{2^{(m)}}_{FF}$
        \item $cov({Z}_{\tau},{Z}^F_{\tau})=\sigma^{2^{(m)}}_{ZF}$ 
        \item $cov({Z}_{\tau},{Z}^F_{\tau-1})=\rho^{(m)}_F\sigma^{2^{(m)}}_{ZF}$.
    \end{itemize} 

    \item Different sources of information are independent for COVID and Flu for a given state $m$. In other words,  ${Z}^F_{\tau,m}, \hat{{y}}_{\tau,m}^{GT} - {y}_{\tau,m}, \hat{{y}}_{\tau,m}^{reg} - {y}_{\tau,m}, \hat{{y}}_{\tau,m}^{nat} - {y}_{\tau,m}, {Z}^F_{\tau}$ are mutually independent (where the boldface represents full vectors containing all the states).
\end{enumerate}

The covariance matrices are thereby simplified as:
\begin{equation}
    \Sigma^{(m)}_{ZW} = \begin{pmatrix}
\rho^{(m)} \sigma^{2^{(m)}}_{ZZ} & \sigma^{2^{(m)}}_{ZZ} & \sigma^{2^{(m)}}_{ZZ}& \sigma^{2^{(m)}}_{ZZ}& \rho^{(m)}_F\sigma^{2^{(m)}}_{ZF}
\end{pmatrix} 
\label{eqn:ARGOX_Idv_SigmaZW}
\end{equation}

\begin{equation}
\Sigma_{WW} = 
\begin{pmatrix}
\sigma^{2^{(m)}}_{ZZ} & \{\rho^{(m)}\sigma^{2^{(m)}}_{ZZ}\}_{m\in\mathcal{M}} & \rho^{(m)}\sigma^{2^{(m)}}_{ZZ}& \rho^{(m)}\sigma^{2^{(m)}}_{ZZ} & \{\sigma^{2^{(m)}}_{ZF}\}_{m\in\mathcal{M}}\\
\{\rho^{(m)}\sigma^{2^{(m)}}_{ZZ}\}_{m\in\mathcal{M}}^\intercal & \Sigma^\mathcal{M}_{ZZ} + \Sigma^{GT}_\mathcal{M} & \sigma^{2^{(m)}}_{ZZ}& \sigma^{2^{(m)}}_{ZZ} & \{\rho^{(m)}_F\sigma^{2^{(m)}}_{ZF}\}_{m\in\mathcal{M}}\\
\rho^{(m)}\sigma^{2^{(m)}}_{ZZ}& \sigma^{2^{(m)}}_{ZZ}& \sigma^{2^{(m)}}_{ZZ} +\sigma^{2^{reg}}& \sigma^{2^{(m)}}_{ZZ} & \{\rho^{(m)}_F\sigma^{2^{(m)}}_{ZF}\}_{m\in\mathcal{M}}\\
\rho^{(m)}\sigma^{2^{(m)}}_{ZZ}& \sigma^{2^{(m)}}_{ZZ}& \sigma^{2^{(m)}}_{ZZ}& \sigma^{2^{(m)}}_{ZZ} +\sigma^{2^{nat}} & \{\rho^{(m)}_F\sigma^{2^{(m)}}_{ZF}\}_{m\in\mathcal{M}}\\
 \{\sigma^{2^{(m)}}_{ZF}\}_{m\in\mathcal{M}}^\intercal &  \{\rho^{(m)}_F\sigma^{2^{(m)}}_{ZF}\}_{m\in\mathcal{M}}^\intercal & \{\rho^{(m)}_F\sigma^{2^{(m)}}_{ZF}\}_{m\in\mathcal{M}}^\intercal &  \{\rho^{(m)}_F\sigma^{2^{(m)}}_{ZF}\}_{m\in\mathcal{M}}^\intercal & \Sigma^\mathcal{M}_{FF}
\end{pmatrix}
\label{eqn:ARGOX_Idv_SigmaWW}
\end{equation}
where $\Sigma^\mathcal{M}_{ZZ} = \mathrm{var}(Z_{\tau,m\in\mathcal{M}})$, $\Sigma^\mathcal{M}_{FF} = \mathrm{var}(Z^F_{\tau,m\in\mathcal{M}})$ $\Sigma^{reg}_\mathcal{M} = \mathrm{var}(\hat{y}^{reg}_{\tau,m\in\mathcal{M}}-y_{\tau,m\in\mathcal{M}})$, $\Sigma^{nat}_\mathcal{M} = \mathrm{var}(\hat{y}^{nat}_{\tau,m\in\mathcal{M}}-y_{\tau,m\in\mathcal{M}} )$, and $\Sigma^{GT}_\mathcal{M} = \mathrm{var}(\hat{y}^{GT}_{\tau,m\in\mathcal{M}}- y_{\tau,m\in\mathcal{M}})$.

\clearpage
\subsection*{Google searches optimal lag}
Table \ref{tab:23query_optimallag} shows the selected 27 Google search queries' optimal lag (delay), selected through fitting regression of lagged terms against COVID-19 cases and death trends (separately) and select the lag with minimal mean-squared error. The optimal delays are derived during the period from April 1st 2020 to June 30th 2020.
\begin{table}[ht]
\footnotesize
\centering
\caption{selected 23 important terms' optimal lag}\label{tab:23query_optimallag}
\begin{tabular}{lll}
Google Search Term & Optimal Lag for COVID-19 Cases & Optimal Lag for COVID-19 Death \\
  \hline
coronavirus vaccine & 4 & 5\\
cough & 4 &24\\
covid 19 vaccine & 34 & 7\\
coronavirus exposure & 5 & 13\\
coronavirus cases & 35 & 30\\
coronavirus test & 35 & 30\\
covid 19 cases & 35 & 30\\
covid 19 & 34 & 21\\
exposed to coronavirus & 6 & 24\\
fever & 4 & 24\\
headache & 5 & 29\\
how long covid 19 & 6  & 21\\
how long contagious & 6 & 25\\
loss of smell & 10 &  25\\
loss of taste & 11 & 24\\
nausea & 10 &23\\
pneumonia &34 & 28 \\
rapid covid 19 &8 & 28\\
rapid coronavirus &8 & 27\\
robitussin & 35 & 15\\
sore throat & 8 & 30\\
sinus & 8&15 \\
symptoms of the covid 19 & 16 & 21\\
\hline
\end{tabular}
\end{table}

\clearpage
\subsection*{Google searches optimal lag visualization}
\begin{figure}[htbp]
\centering
\subfloat[National COVID-19 weekly incremental cases and query ``loss of taste'' search frequencies.]
{\includegraphics[width=0.6\textwidth]{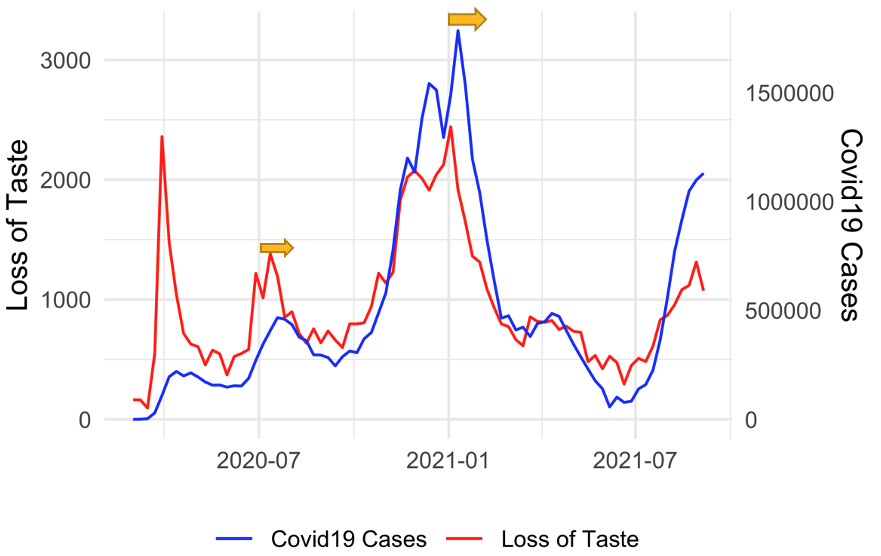}\label{fig:Nat_cases_lossoftaste}}
\hfill
\subfloat[National COVID-19 weekly incremental death and query ``loss of taste'' search frequencies.]
{\includegraphics[width=0.6\textwidth]{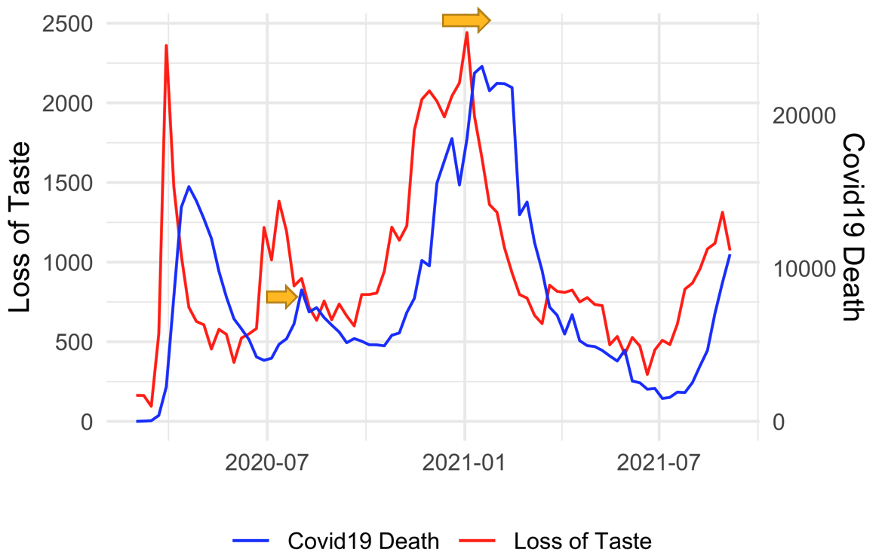}\label{fig:Nat_death_lossoftaste}}
\caption{Google search query ``loss of taste'' and COVID-19 weekly incremental cases (top) and death (bottom). Illustration of delay in peak between Google search query search frequencies (Loss of Taste in red) and COVID-19 national
level weekly incremental cases and death (blue). Y-axis are adjusted accordingly.}
\label{fig:Nat_CasesDeath_lossoftaste}
\end{figure}

Fig \ref{fig:Nat_cases_lossoftaste} and \ref{fig:Nat_death_lossoftaste} show the delay in peak between Google search query ``loss of taste'' and national level weekly incremental cases and death. 

\clearpage
\subsection*{ARGO Joint model parameter heatmaps when predicting COVID-19 Deaths}
\begin{figure}[htbp]
\centering
\subfloat[1 Week Ahead Coefficients]
{\includegraphics[width=0.5\textwidth]{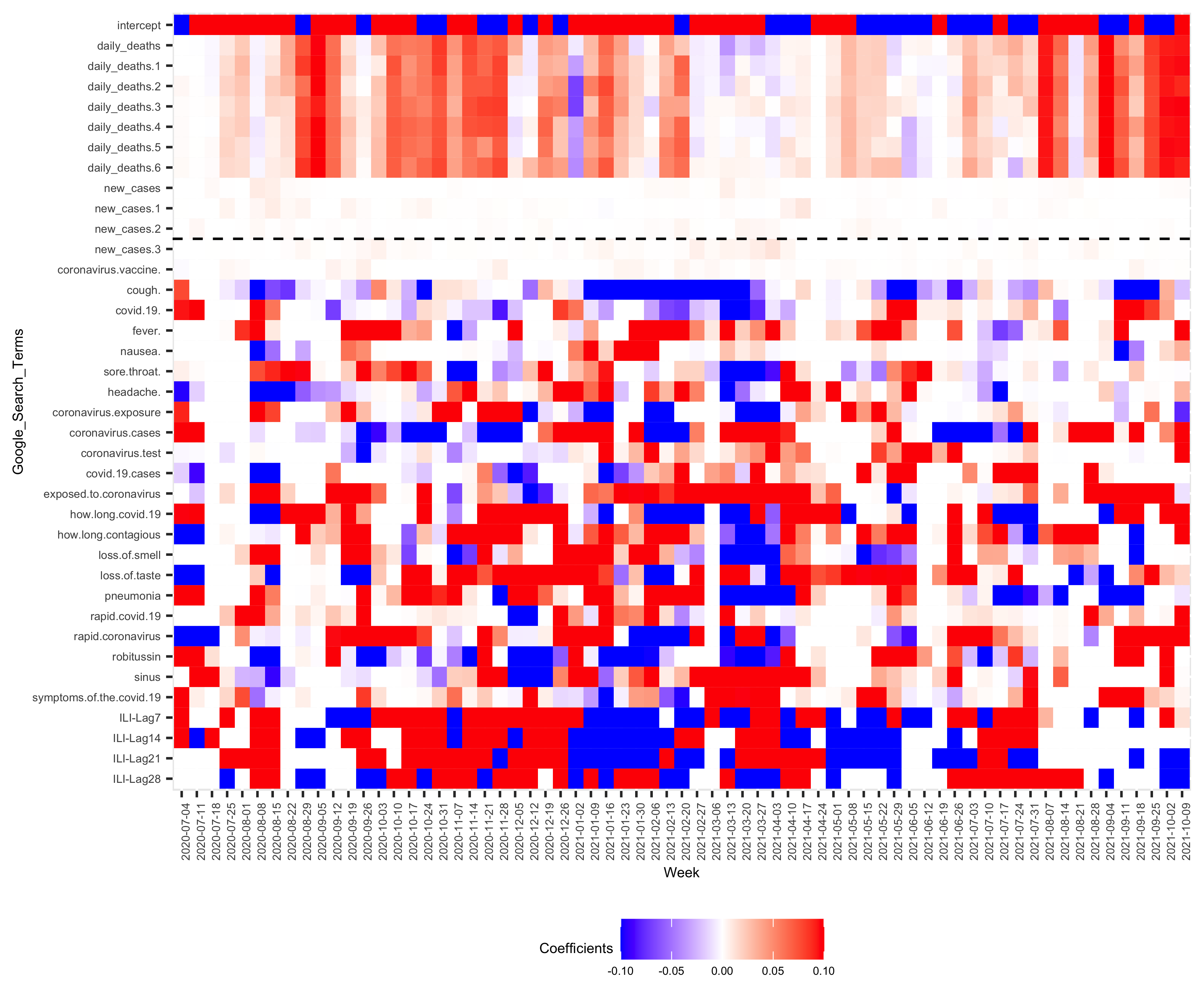}\label{fig:Nat_Coef_Death_1week}}
\subfloat[2 Week Ahead Coefficients]
{\includegraphics[width=0.5\textwidth]{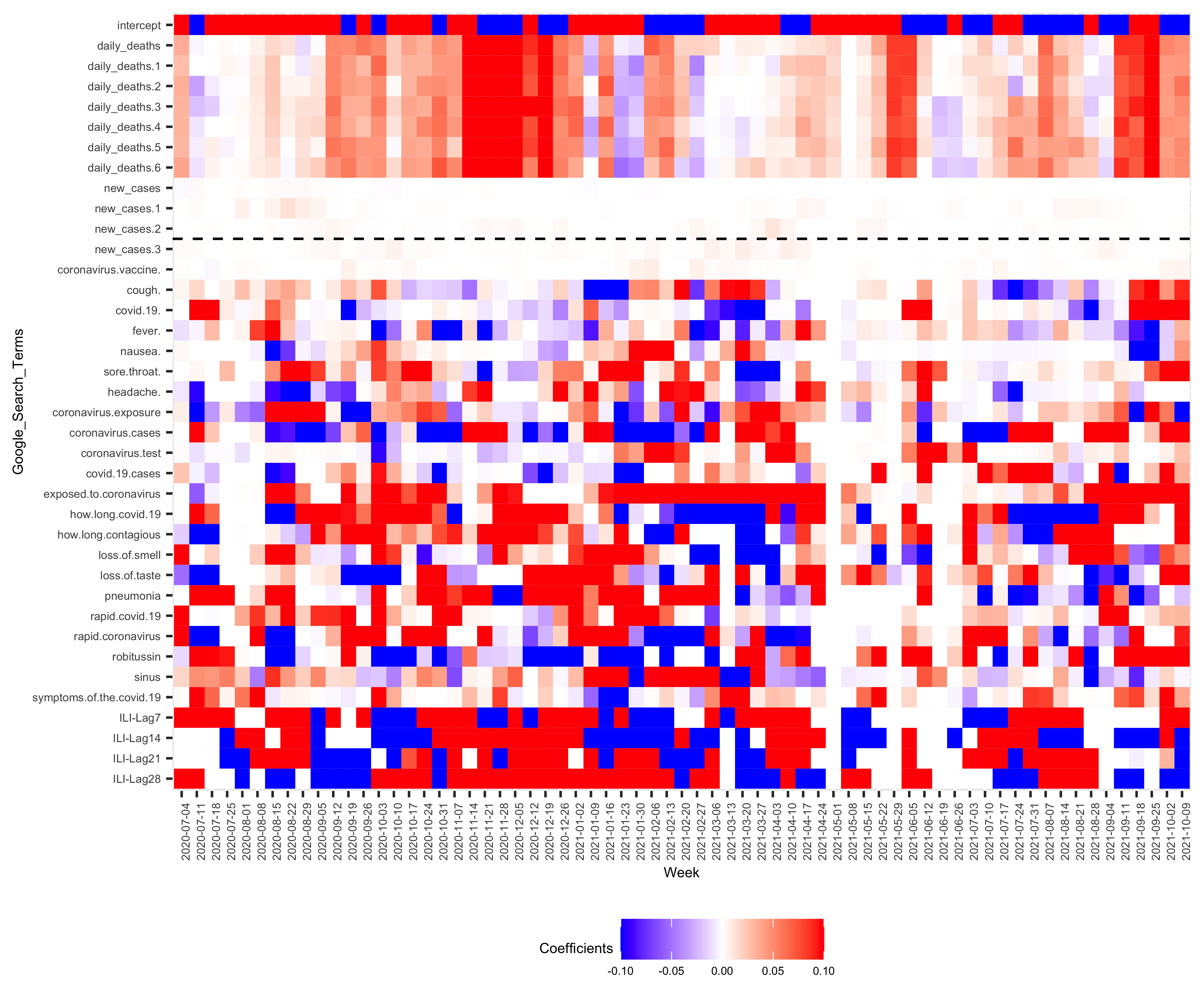}\label{fig:Nat_Coef_Death_2weeks}}
\hfill
\subfloat[3 Week Ahead Coefficients]
{\includegraphics[width=0.5\textwidth]{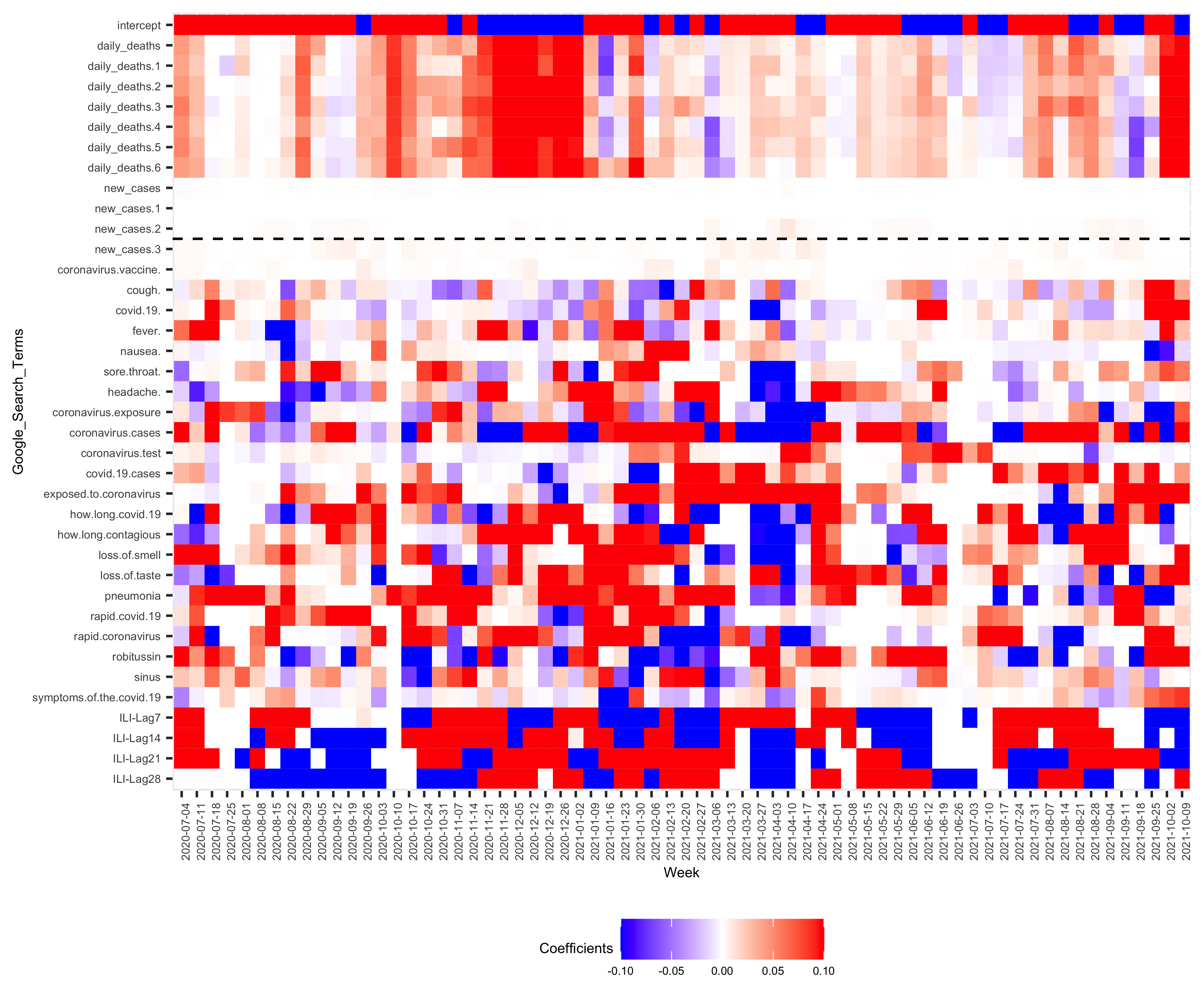}\label{fig:Nat_Coef_Death_3weeks}}
\subfloat[4 Week Ahead Coefficients]
{\includegraphics[width=0.5\textwidth]{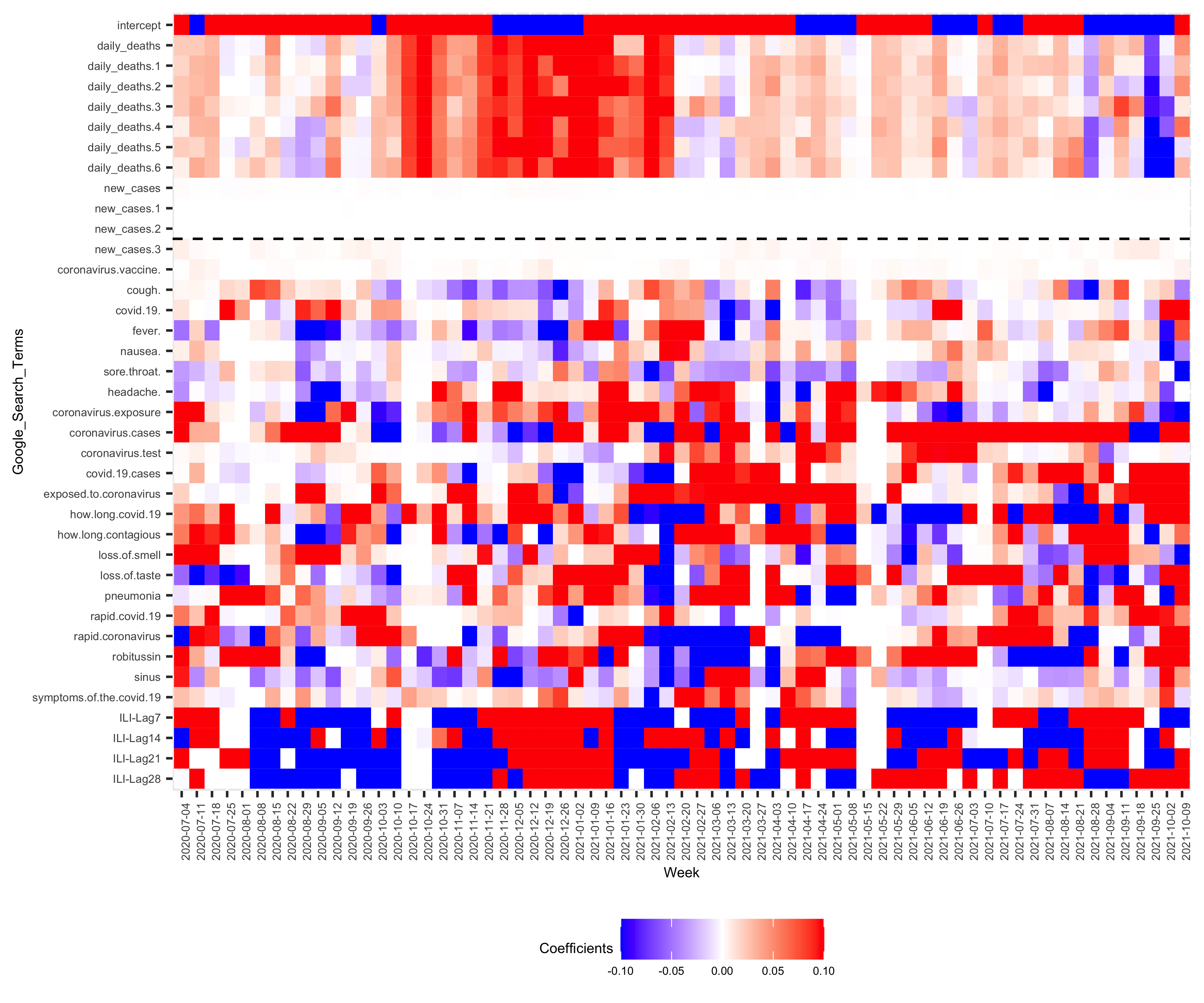}\label{fig:Nat_Coef_Death_4weeks}}
\caption{Smoothed coefficients for ARGO national level 1-4 weeks ahead predictions for COVID-19 Death. Coefficients larger than 0.1 are scaled to 0.1 and lower than -0.1 are scaled to -0.1, for simplicity. Red and blue represent positive and negative coefficients. Black horizontal dashed line separates Google query queries from autoregressive lags. The past 4 weeks \%ILI information as exogenous variables are at the bottom 4 rows.}
\label{fig:Nat_Coef_Death}
\end{figure}

\clearpage
\subsection*{ARGO Joint model parameter heatmaps when predicting COVID-19 Cases}
\begin{figure}[htbp]
\centering
\subfloat[1 Week Ahead Coefficients]
{\includegraphics[width=0.5\textwidth]{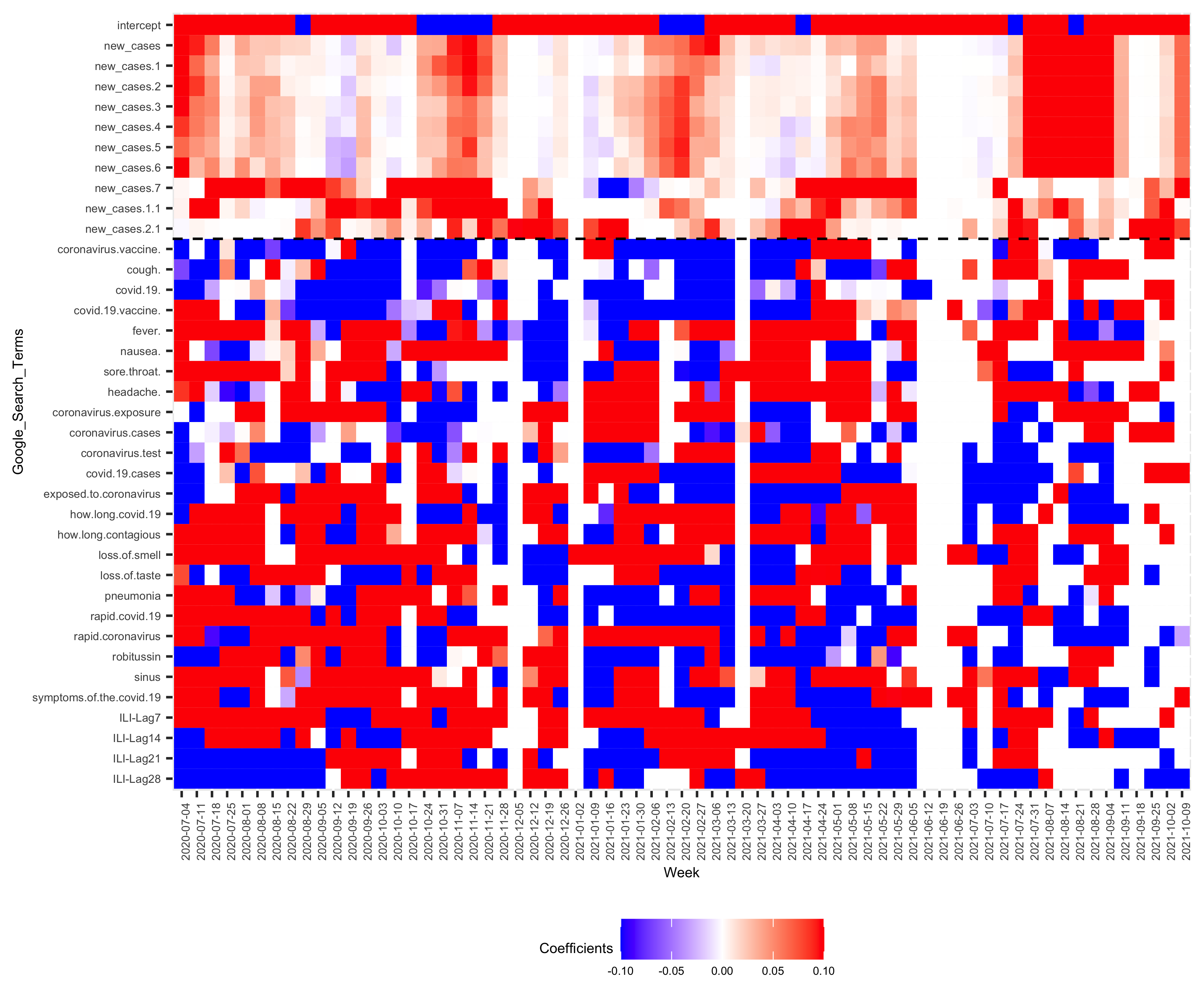}\label{fig:Nat_Coef_Cases_1week}}
\subfloat[2 Weeks Ahead Coefficients]
{\includegraphics[width=0.5\textwidth]{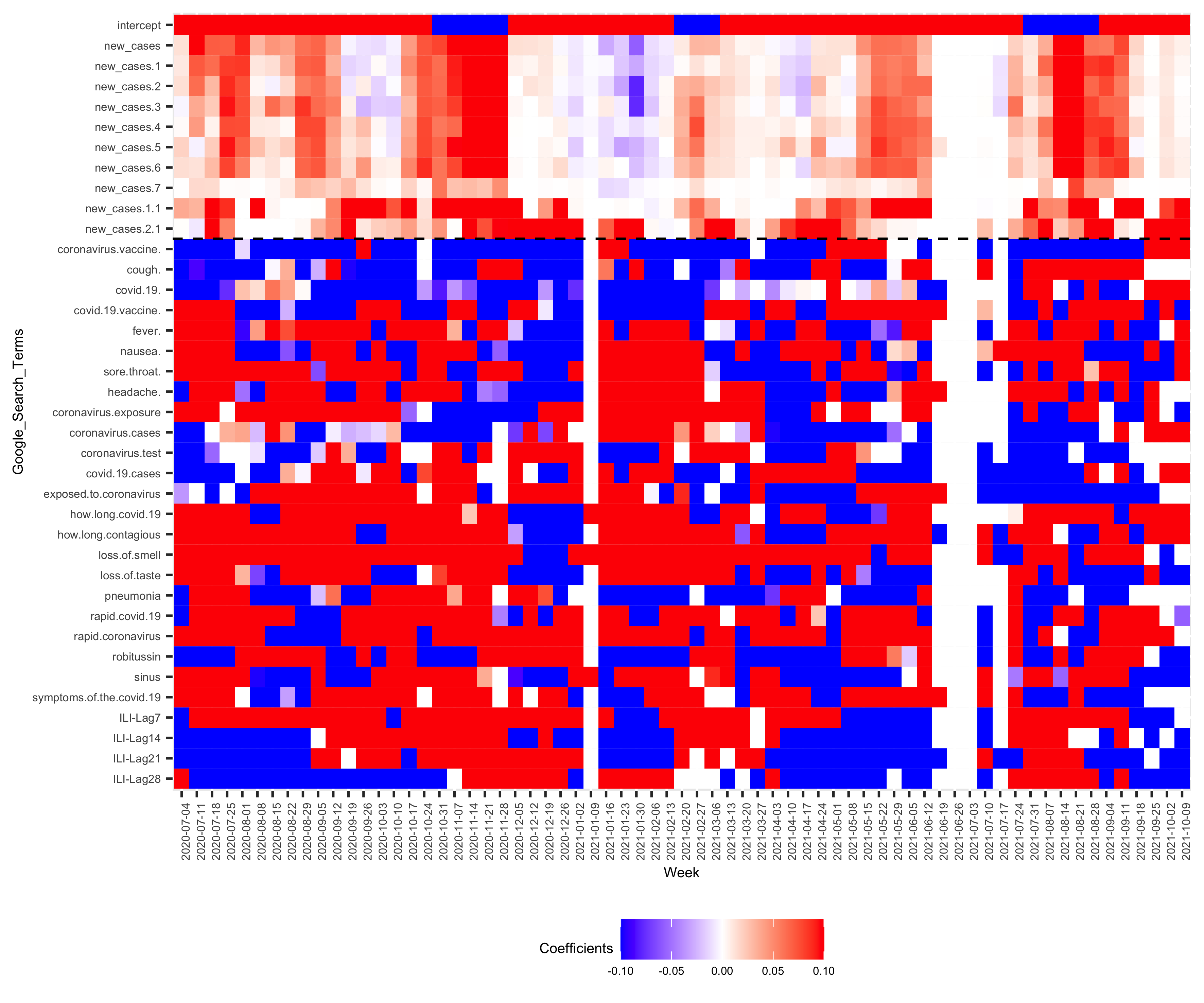}\label{fig:Nat_Coef_Cases_2weeks}}
\hfill
\subfloat[3 Weeks Ahead Coefficients]
{\includegraphics[width=0.5\textwidth]{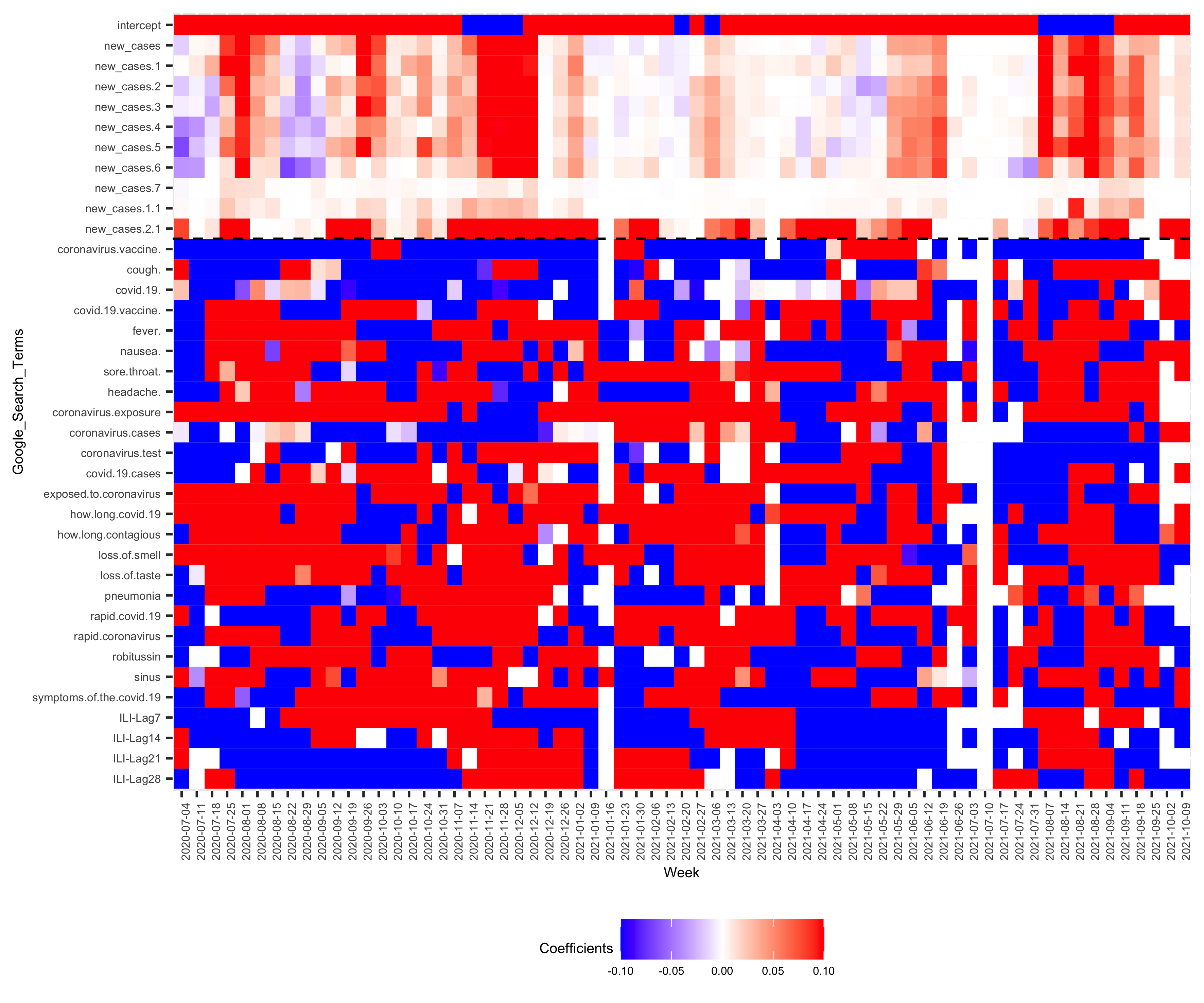}\label{fig:Nat_Coef_Cases_3weeks}}
\subfloat[4 Weeks Ahead Coefficients]
{\includegraphics[width=0.5\textwidth]{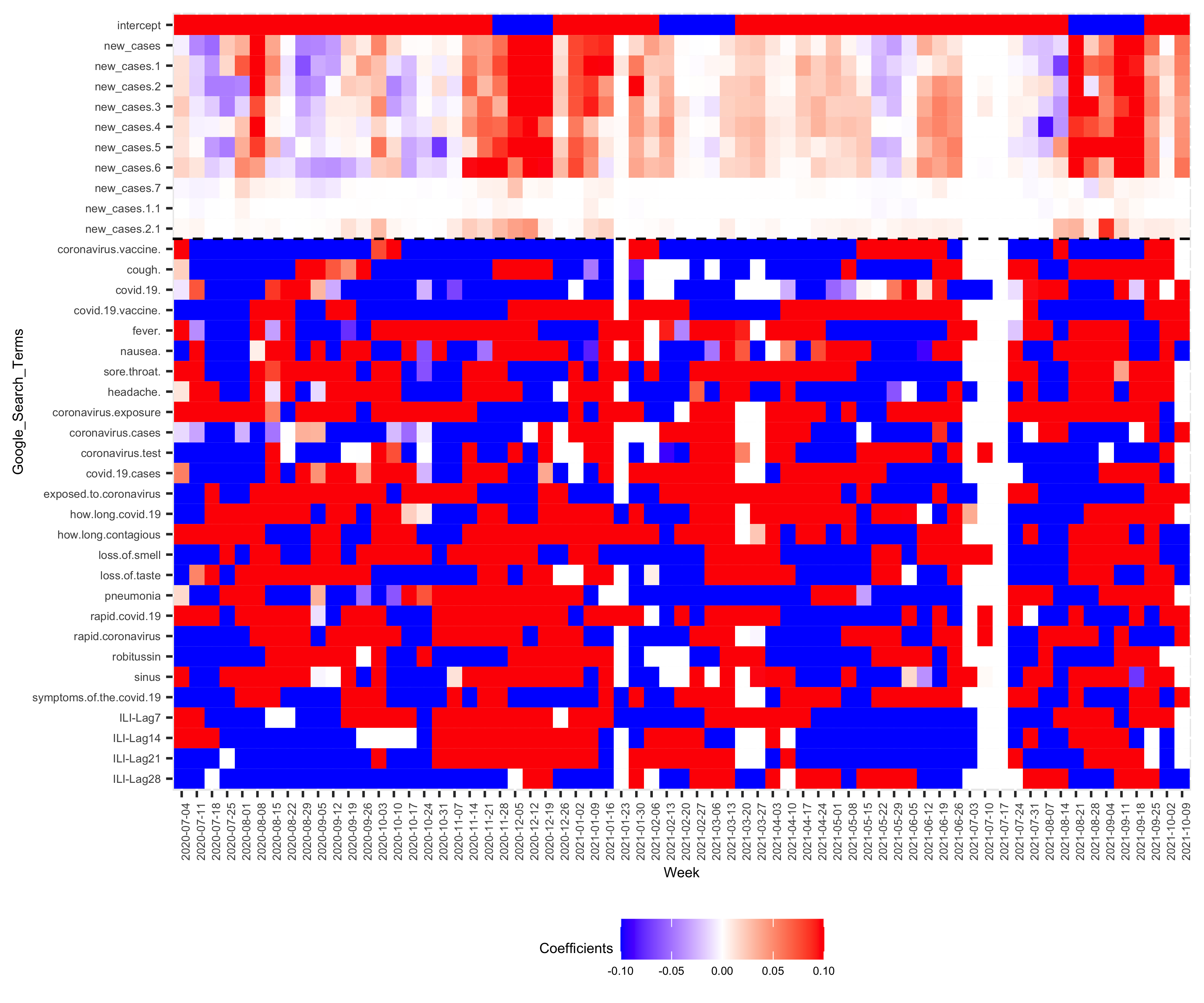}\label{fig:Nat_Coef_Cases_4weeks}}
\caption{Smoothed coefficients for ARGO national level 1-4 weeks ahead predictions for COVID-19 Cases. Coefficients larger than 0.1 are scaled to 0.1 and lower than -0.1 are scaled to -0.1, for simplicity. Red and blue color represent positive and negative coefficients. Black horizontal dashed line separates Google query queries from autoregressive lags. The past 4 weeks \%ILI information as exogenous variables are at the bottom.}
\label{fig:Nat_Coef_Case}
\end{figure}

\clearpage
\subsection*{ARGOX Joint Ensemble Selections}
\begin{table}[ht]
\sisetup{detect-weight,mode=text}
\renewrobustcmd{\bfseries}{\fontseries{b}\selectfont}
\renewrobustcmd{\boldmath}{}
\newrobustcmd{\B}{\bfseries}
\addtolength{\tabcolsep}{-4.1pt}
\footnotesize
\centering
\centering
\begin{tabular}{lrrrr}
\hline
& 1 Week Ahead & 2 Weeks Ahead & 3 Weeks Ahead & 4 Weeks Ahead\\
\hline \multicolumn{1}{l}{COVID-19 Cases} \\
 \hspace{1em} ARGO & 32.4\% & 31.2\% & 28.5\% & 26.1\% \\
 \hspace{1em} ARGOX-2Step & 22.2\% & 20.2\% & 19.8\% & 20.1\% \\
 \hspace{1em} ARGOX-NatConstraint & 18.2\% & 15.5\% & 15.1\% & 14.8\% \\
  \hspace{1em} ARGOX-Idv & 27.2\% & 33.1\% & 36.6\% & 39\% \\
\hline \multicolumn{1}{l}{COVID-19 Death} \\
  \hspace{1em} ARGO & 30.4\% & 29.8\% & 25.5\% & 22.9\% \\
 \hspace{1em} ARGOX-2Step & 20.2\% & 16.1\% & 11.2\% & 11.1\% \\
 \hspace{1em} ARGOX-NatConstraint & 15.1\% & 15.5\% & 14.1\% & 13.8\% \\
  \hspace{1em} ARGOX-Idv & 34.3\% & 38.6\% & 49.2\% & 52.2\% \\
\hline \multicolumn{1}{l}{\%ILI} \\
  \hspace{1em} ARGOX & 53.9\% & & & \\
 \hspace{1em} ARGOX-Idv & 46.1\% & & & \\
 \hline
\end{tabular}
\caption{Ensemble approach total selection across all U.S. states/area for 1 to 4 weeks ahead predictions for COVID-19 cases and death, and 1 week ahead prediction for \%ILI.}
\label{tab:Ensemble_Selection}
\end{table}

\clearpage
\subsection*{COVID-19 Death model comparisons}
\subsubsection*{Among our methods (national and state level)}
\begin{figure}[htbp]
\centering
\subfloat[1 Week Ahead National Level Predictions]
{\includegraphics[width=0.49\textwidth]{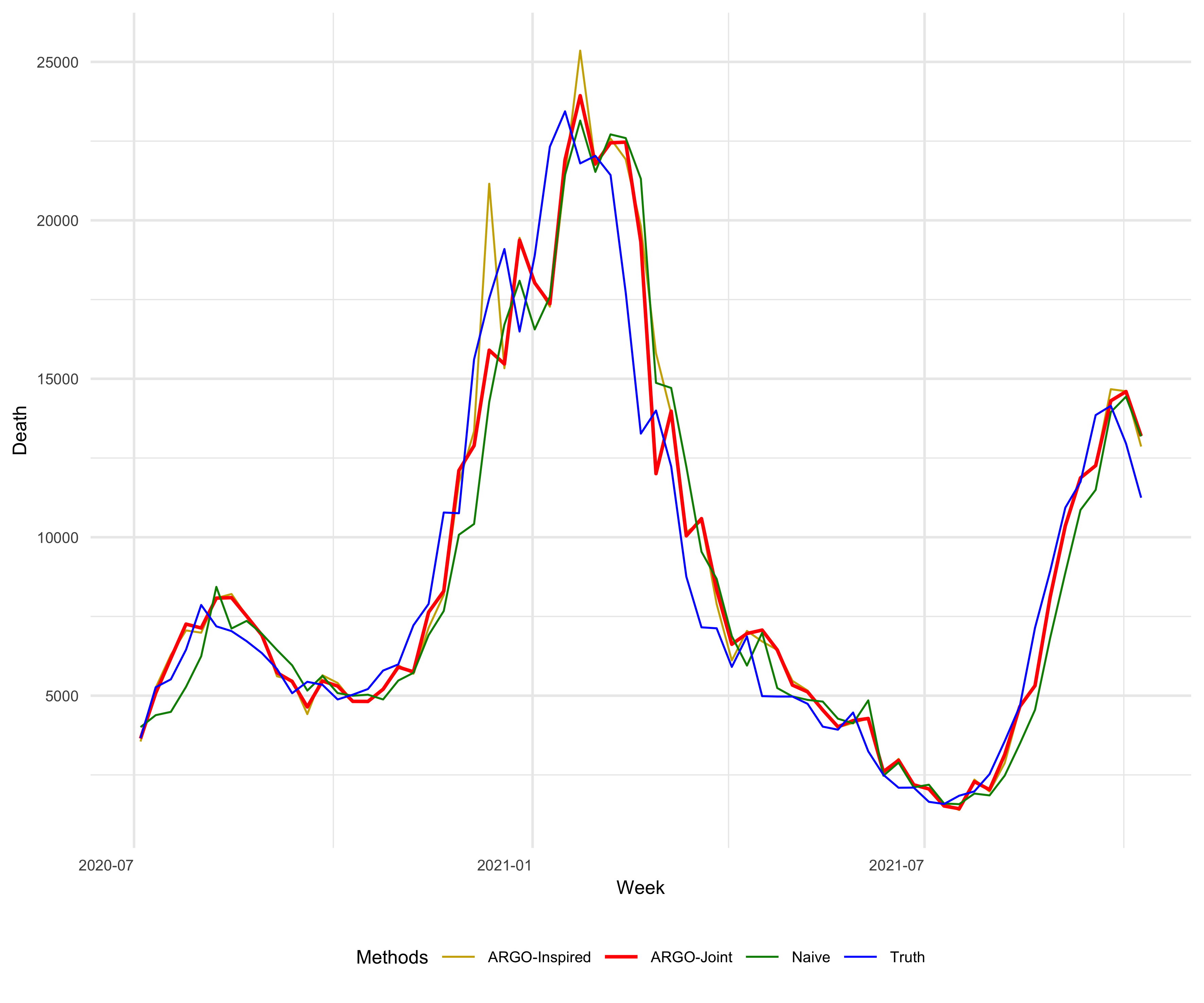}\label{fig:Nat_Compare_Death_Our_1Week}}
\hfill
\subfloat[2 Weeks Ahead National Level Predictions]
{\includegraphics[width=0.49\textwidth]{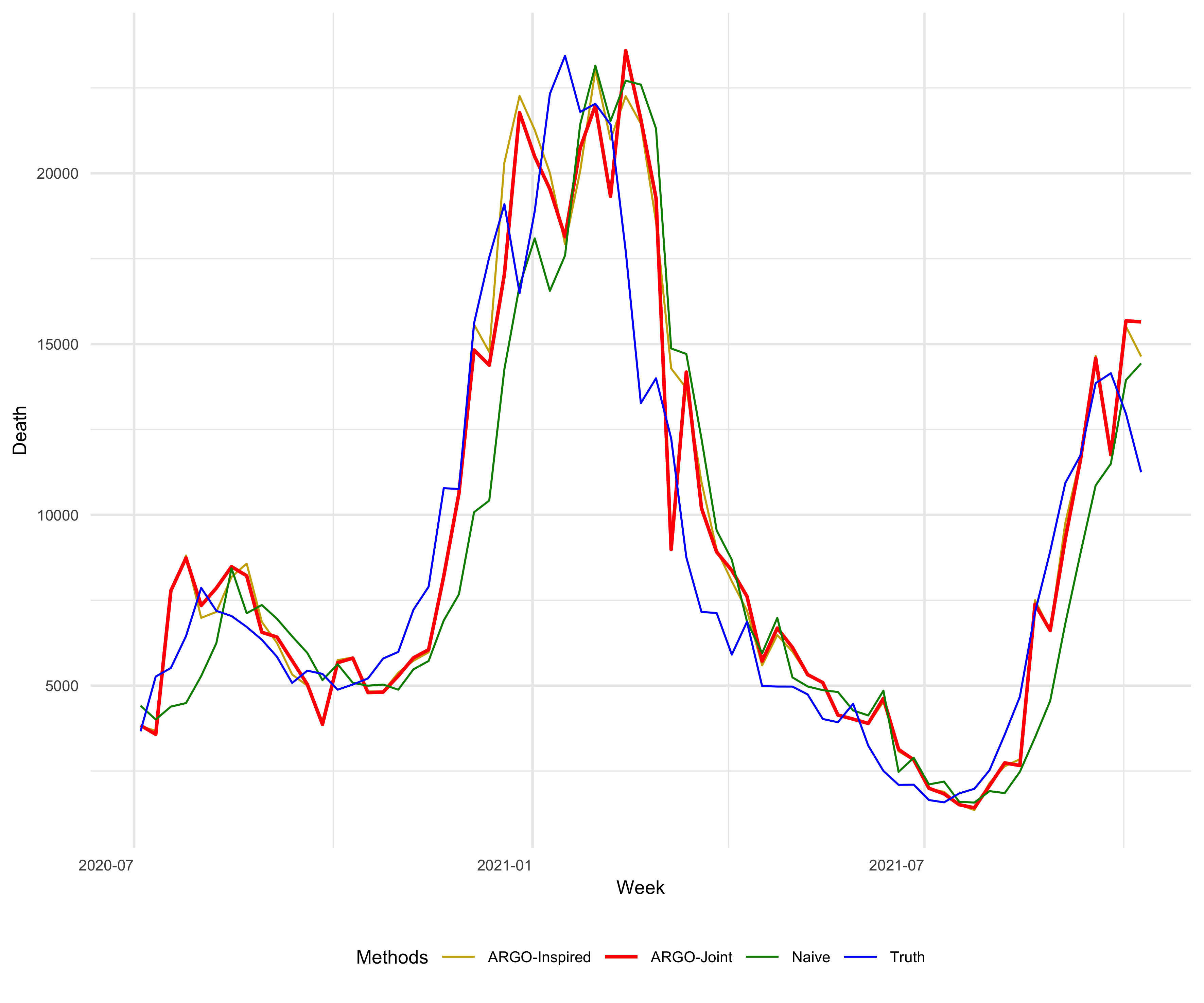}\label{fig:Nat_Compare_Death_Our_2Weeks}}
\hfill
\subfloat[3 Weeks Ahead National Level Predictions]
{\includegraphics[width=0.49\textwidth]{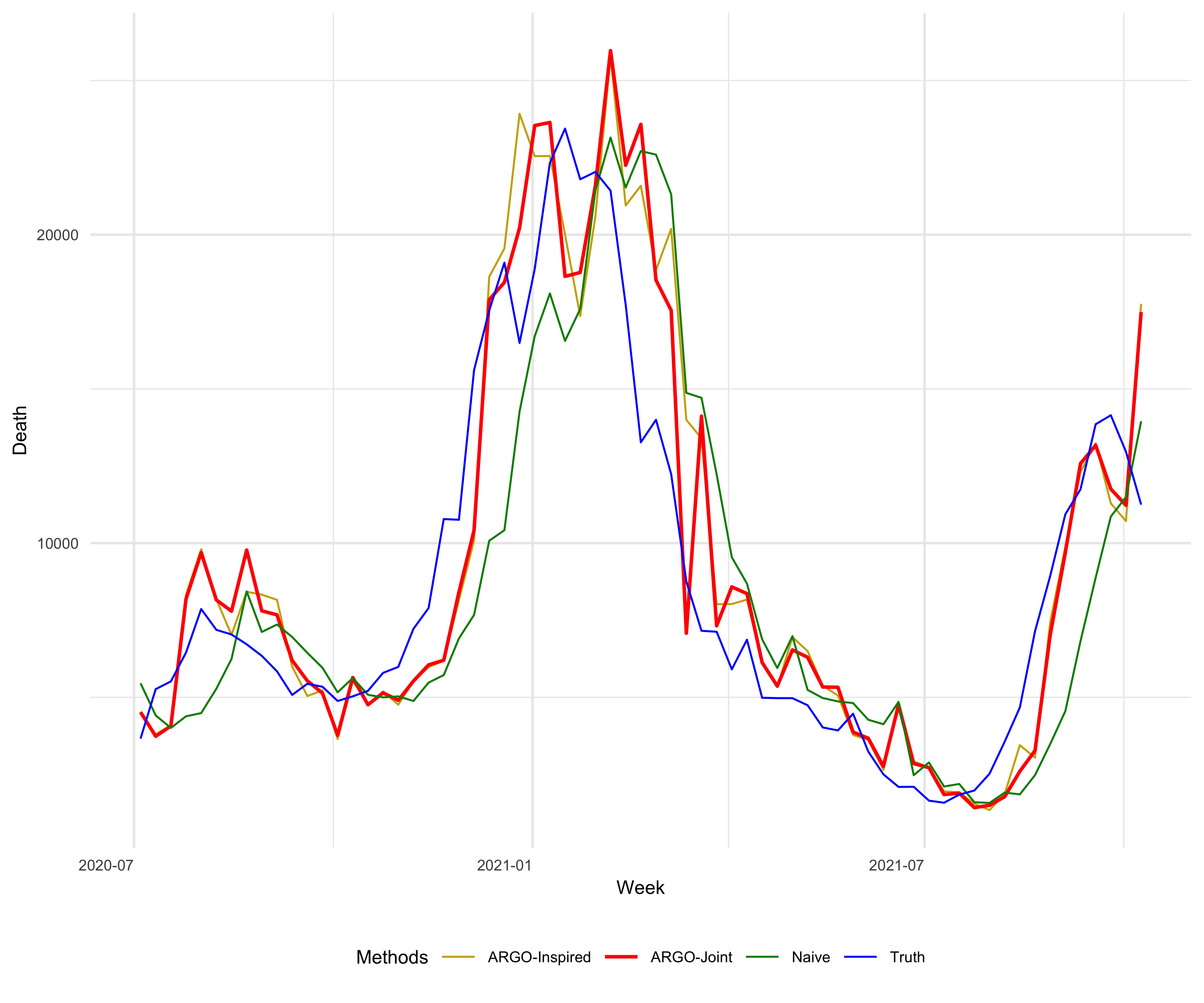}\label{fig:Nat_Compare_Death_Our_3Weeks}}
\hfill
\subfloat[4 Weeks Ahead National Level Predictions]
{\includegraphics[width=0.49\textwidth]{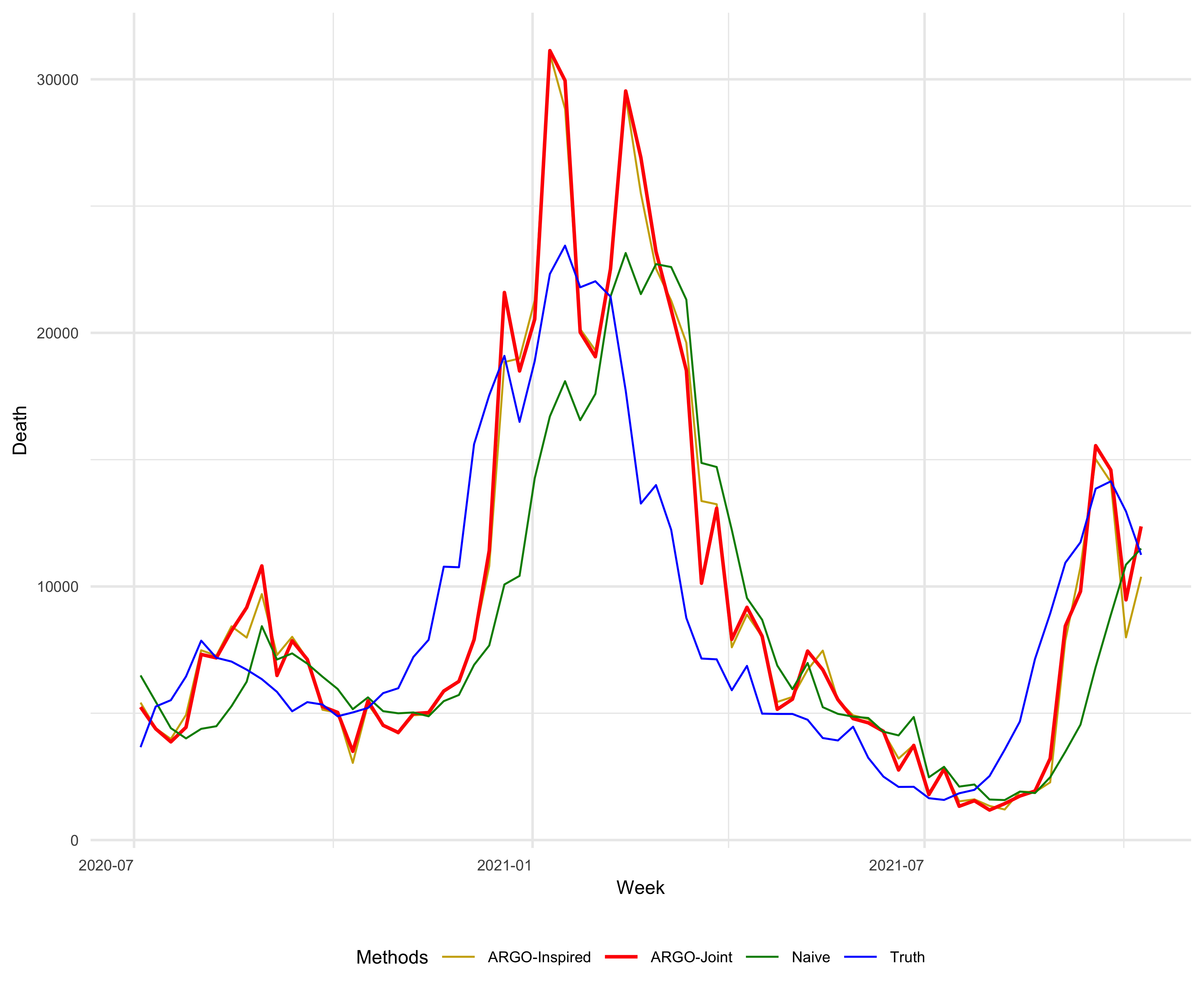}\label{fig:Nat_Compare_Death_Our_4Weeks}}
\caption{\textbf{National level 1 to 4 weeks ahead COVID-19 weekly incremental death predictions.} The method included are ARGO-Inspired (Ref \cite{ma2021covid}), ARGO-Joint, Naive (persistence), truth, weekly from 2020-07-04 to 2021-10-09. Estimation results for 1 (top left), 2 (top right), 3 (bottom left), and 4 (bottom right) weeks ahead weekly incremental death. ARGO-Joint estimations (thick red), contrasting with the true COVID-19 death from JHU dataset (blue) as well as the estimates from Ref \cite{ma2021covid} (gold) and Naive (green).}
\label{fig:Death_Nat_Our}
\end{figure}

\begin{figure}[htbp]
\centering
\includegraphics[width=0.6\textwidth]{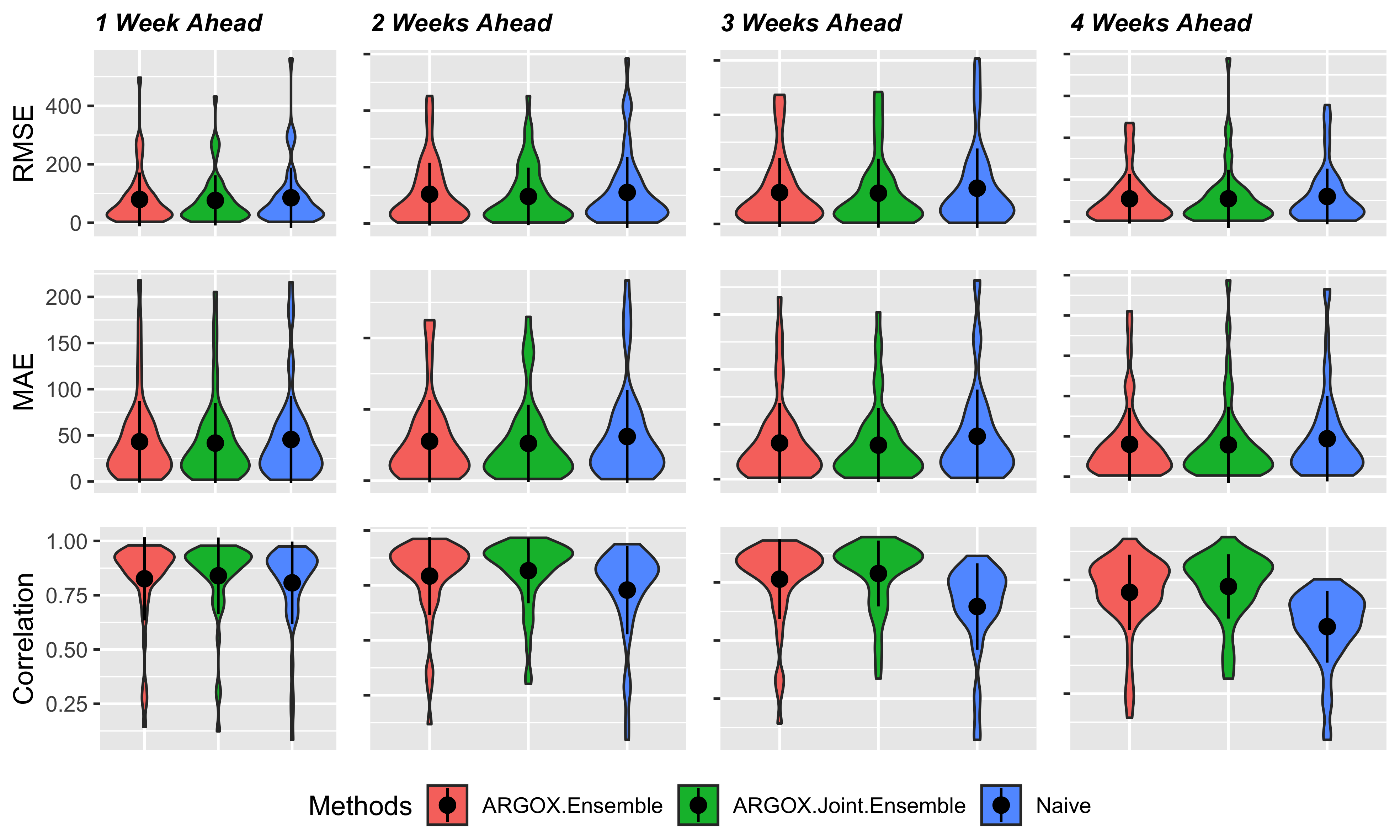}
\caption{\textbf{State Level Death Prediction Comparison in 3 Error Metrics.} Comparison among different versions of our models' 1 to 4 weeks (from left to right) ahead U.S. states level weekly incremental death predictions (from 2020-07-04 to 2021-10-10). The RMSE, MAE and Pearson correlation for each method across all states are reported in the violin plot. The methods (x-axis) are sorted based on their RMSE. Here, ARGOX-Ensemble denotes the previously proposed method \cite{ma2021covid}.}
\label{fig:Death_Our_Violine}
\end{figure}

Figure \ref{fig:Death_Nat_Our} shows 1 to 4 weeks ahead national level death prediction visualizations of our own method and Ref \cite{ma2021covid} against naive and truth, whereas table \ref{tab:Nat_Death_Our} shows the methods error in the 3 error metrics. Figure \ref{fig:Death_Our_Violine} shows 1 to 4 weeks ahead state level forecast result comparison among our methods through 3 error metrics in violin charts.

\begin{table}[htbp]
\sisetup{detect-weight,mode=text}
\renewrobustcmd{\bfseries}{\fontseries{b}\selectfont}
\renewrobustcmd{\boldmath}{}
\newrobustcmd{\B}{\bfseries}
\addtolength{\tabcolsep}{-4.1pt}
\footnotesize
\centering
\begin{tabular}{lrrrr}
  \hline
& 1 Week Ahead & 2 Weeks Ahead & 3 Weeks Ahead & 4 Weeks Ahead\\ 
    \hline \multicolumn{1}{l}{RMSE} \\
\hspace{1em} Naive &  1977.973  & 2954.151 & 3814.300 & 4671.885 \\
\hspace{1em} Ref \cite{ma2021covid} & 1749.212 & \B2224.436 &  2893.182  & 4123.993 \\
\hspace{1em} ARGO Joint & \B1665.172 & 2259.311 & \B2776.378  &  \B4106.152 \\
\multicolumn{1}{l}{MAE} \\
\hspace{1em} Naive & 1369.522   & 2110.806   & 2875.328   & 3575.806   \\
\hspace{1em} Ref \cite{ma2021covid}& 1183.567 &\B 1582.224   & 2103.164 & 2906.567  \\
\hspace{1em} ARGO Joint & \B1123.851 &   1595.164    & \B2020.970   & \B2901.448   \\
\multicolumn{1}{l}{Correlation} \\
\hspace{1em} Naive &  0.944 & 0.875 & 0.791 &0.684\\
\hspace{1em} Ref \cite{ma2021covid}  & 0.959 & \B0.938   & 0.903 &  0.843\\
\hspace{1em} ARGO Joint & \B0.960 & 0.924  & \B0.909 &  \B0.854 \\
\hline
\end{tabular}
\caption{\textbf{National Level Death Prediction Comparison in 3 Error Metrics.} Boldface highlights the best performance for each metric in each study period, for 1-4 weeks ahead predictions. All comparisons are based on the original scale of COVID-19 national incremental death. On average, ARGO Joint is able to achieve around 2.5\% RMSE, 2\% MAE reduction, and around 0.15\% correlation improvement from previously proposed ARGO Inspired method \cite{ma2021covid}.}\label{tab:Nat_Death_Our}
\end{table}

\subsubsection*{Among other publicly available methods (national and state level)}
\begin{figure}[htbp]
\centering
\includegraphics[width=0.8\textwidth]{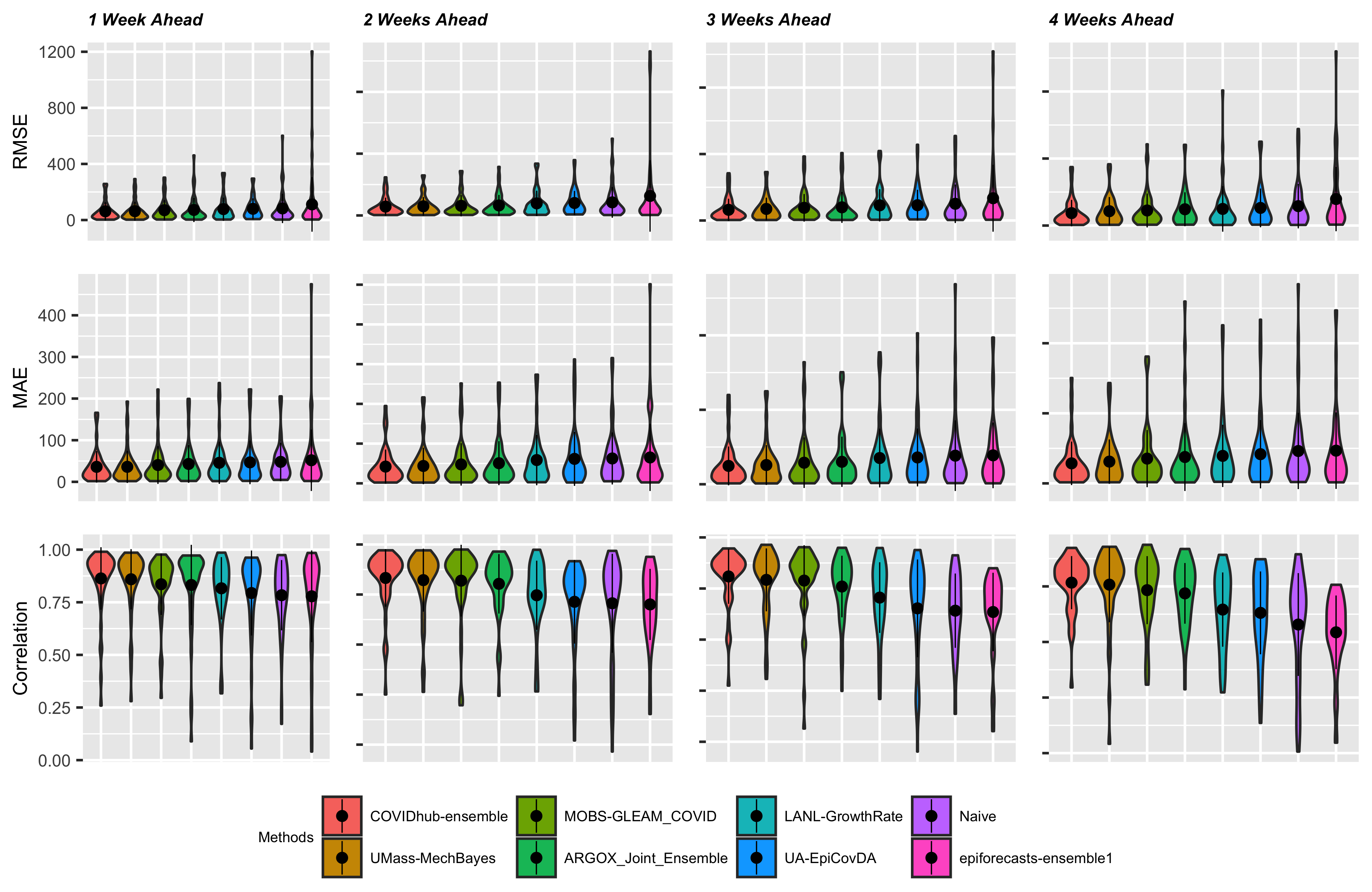}
\caption{\textbf{State Level Death Prediction Comparisons} among different CDC published teams' 1 to 4 weeks (from left to right) ahead weekly incremental death (from 2020-07-04 to 2021-10-10). The RMSE, MAE and Pearson correlation for each method across all states are reported in the violin plot. The methods (x-axis) are sorted based on their RMSE.}
\label{fig:Death_Other_Violine}
\end{figure}

Table \ref{tab:Nat_Death_Other} shows the comparison against other CDC published teams for 1-4 weeks ahead national level incremental death predictions. Figure \ref{fig:Death_Other_Violine} shows the state level prediction comparisons in violin charts, where the methods' error mean and standard deviations are displayed.

\begin{table}[htbp]
\sisetup{detect-weight,mode=text}
\renewrobustcmd{\bfseries}{\fontseries{b}\selectfont}
\renewrobustcmd{\boldmath}{}
\newrobustcmd{\B}{\bfseries}
\addtolength{\tabcolsep}{-4.1pt}
\footnotesize
\centering
\begin{tabular}{lrrrrr}
  \hline
 & 1 Week Ahead & 2 Weeks Ahead & 3 Weeks Ahead & 4 Weeks Ahead & Average \\ 
   \hline \multicolumn{1}{l}{RMSE} \\
\hspace{1em} UMass-MechBayes\cite{UMASS} & 1320.17 & 1741.42 & 2127.58 & 2242.40 & 1857.89 \\ 
 \hspace{1em} COVIDhub-ensemble\cite{CDC_Ensemble} & 1389.41 & 1706.14 & 2015.08 & 2431.96 & 1885.65 \\ 
\hspace{1em}  MOBS-GLEAM\_COVID\cite{MOBS_GLEAM} & 1403.75 & 1800.84 & 2241.63 & 2684.33 & 2032.63 \\ 
\hspace{1em}  ARGOX\_Joint\_Ensemble & (\#5)1665.17 & (\#4)2259.31 & (\#4)2776.38 & (\#5)4106.15 & (\#4)2701.75 \\ 
\hspace{1em}  LANL-GrowthRate\cite{LANL_GrowthRate} & 1614.51 & 2279.68 & 2893.12 & 4234.14 & 2755.36 \\ 
 \hspace{1em} UA-EpiCovDA\cite{UA_EpiGro} & 2051.75 & 2794.84 & 3346.70 & 3348.33 & 2885.40 \\ 
 \hspace{1em} epiforecasts-ensemble1\cite{epiforecasts_MIT} & 2236.67 & 2545.07 & 3169.34 & 4203.65 & 3038.68 \\ 
\hspace{1em}  Naive & 1977.97 & 2954.15 & 3814.30 & 4671.89 & 3354.58 \\ 
 \hline \multicolumn{1}{l}{MAE} \\
\hspace{1em} UMass-MechBayes\cite{UMASS} & 925.67 & 1115.24 & 1408.36 & 1630.82 & 1270.02 \\ 
 \hspace{1em} COVIDhub-ensemble\cite{CDC_Ensemble} & 965.55 & 1168.32 & 1436.95 & 1779.86 & 1337.67 \\ 
\hspace{1em}  MOBS-GLEAM\_COVID\cite{MOBS_GLEAM} & 1064.64 & 1362.83 & 1752.92 & 2033.36 & 1553.44 \\ 
 \hspace{1em} ARGOX\_Joint\_Ensemble & (\#4)1123.85 & (\#5)1695.16 & (\#4)2020.97 & (\#6)2901.45 & (\#4)1935.36 \\ 
\hspace{1em}  UA-EpiCovDA\cite{UA_EpiGro} & 1383.43 & 1770.43 & 2132.72 & 2601.15 & 1971.93 \\ 
 \hspace{1em} epiforecasts-ensemble1\cite{epiforecasts_MIT} & 1390.14 & 1656.43 & 2124.52 & 2832.15 & 2000.81 \\ 
\hspace{1em}  LANL-GrowthRate\cite{LANL_GrowthRate} & 1245.88 & 1808.40 & 2275.33 & 3096.98 & 2106.65 \\ 
 \hspace{1em} Naive & 1369.52 & 2110.81 & 2875.33 & 3575.81 & 2482.87 \\ 
 \hline \multicolumn{1}{l}{Correlation} \\
\hspace{1em} UMass-MechBayes\cite{UMASS} & 0.98 & 0.96 & 0.94 & 0.94 & 0.95 \\ 
\hspace{1em}  COVIDhub-ensemble\cite{CDC_Ensemble} & 0.97 & 0.96 & 0.94 & 0.92 & 0.95 \\ 
\hspace{1em}  MOBS-GLEAM\_COVID\cite{MOBS_GLEAM} & 0.97 & 0.96 & 0.94 & 0.91 & 0.94 \\ 
 \hspace{1em} LANL-GrowthRate\cite{LANL_GrowthRate} & 0.97 & 0.94 & 0.92 & 0.86 & 0.92 \\ 
 \hspace{1em} ARGOX\_Joint\_Ensemble & (\#5)0.96 & (\#5)0.92 & (\#5)0.91 & (\#6)0.85 & (\#5)0.91 \\ 
 \hspace{1em} UA-EpiCovDA\cite{UA_EpiGro} & 0.96 & 0.91 & 0.87 & 0.88 & 0.91 \\ 
 \hspace{1em} epiforecasts-ensemble1\cite{epiforecasts_MIT} & 0.93 & 0.91 & 0.86 & 0.80 & 0.87 \\ 
\hspace{1em}  Naive & 0.94 & 0.87 & 0.79 & 0.68 & 0.82 \\ 
   \hline
\end{tabular}
\caption{\textbf{National Level Death Prediction Comparisons} among different models' 1 to 4 weeks ahead weekly incremental death (from 2020-07-04 to 2021-10-10). The RMSE, MAE, Pearson correlation and their averages are reported. Methods are sorted based on their average. Our ARGOX-Joint-Ensemble's ranking for each error metric are included in parenthesis.}
\label{tab:Nat_Death_Other}
\end{table}

\clearpage
\subsection*{COVID-19 Cases model comparisons}
\subsubsection*{Among our own methods}
\begin{figure}[htbp]
\centering
\subfloat[1 Week Ahead National Level Predictions]
{\includegraphics[width=0.49\textwidth]{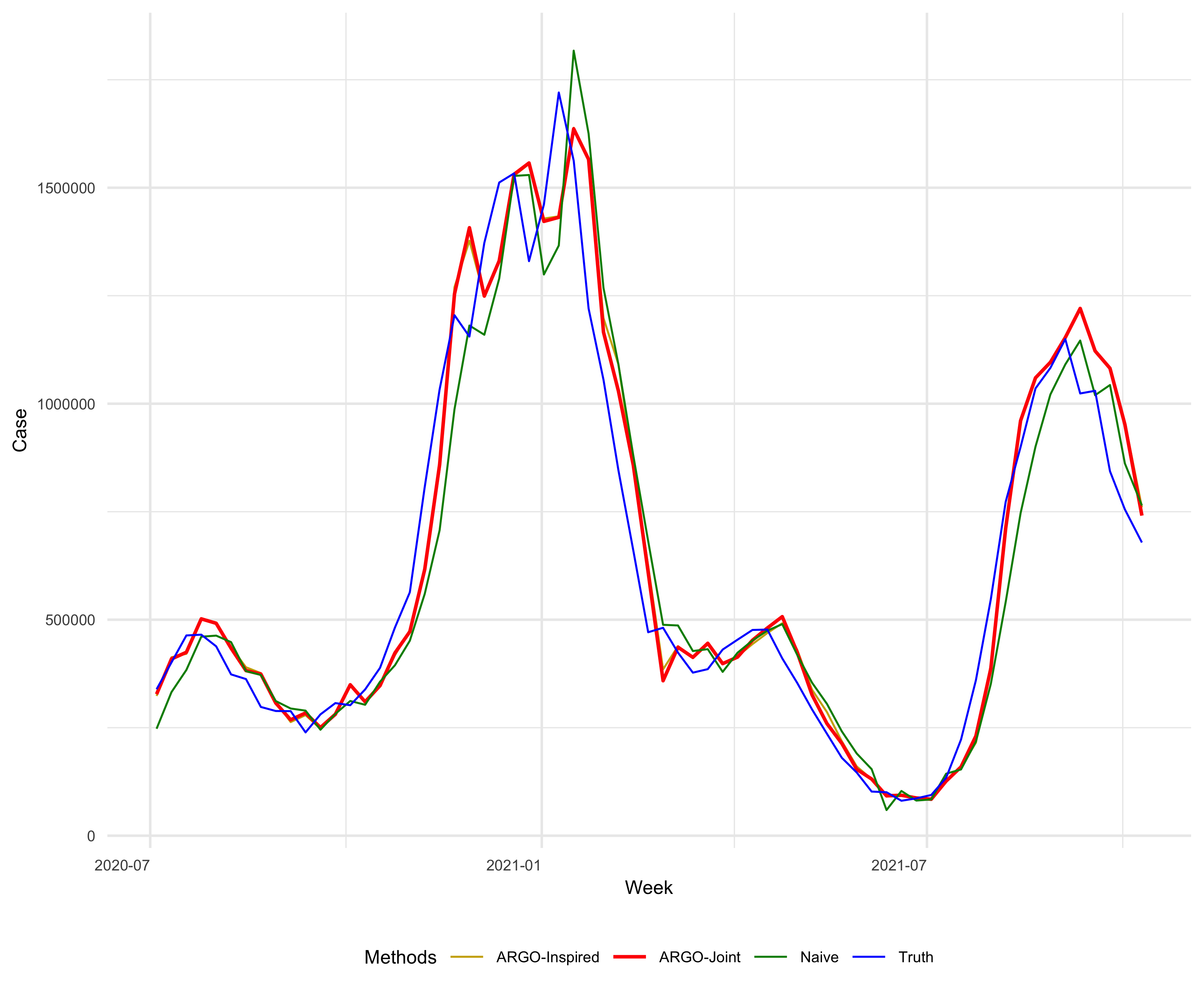}\label{fig:Nat_Compare_Our_Case_1Week}}
\hfill
\subfloat[2 Weeks Ahead National Level Predictions]
{\includegraphics[width=0.49\textwidth]{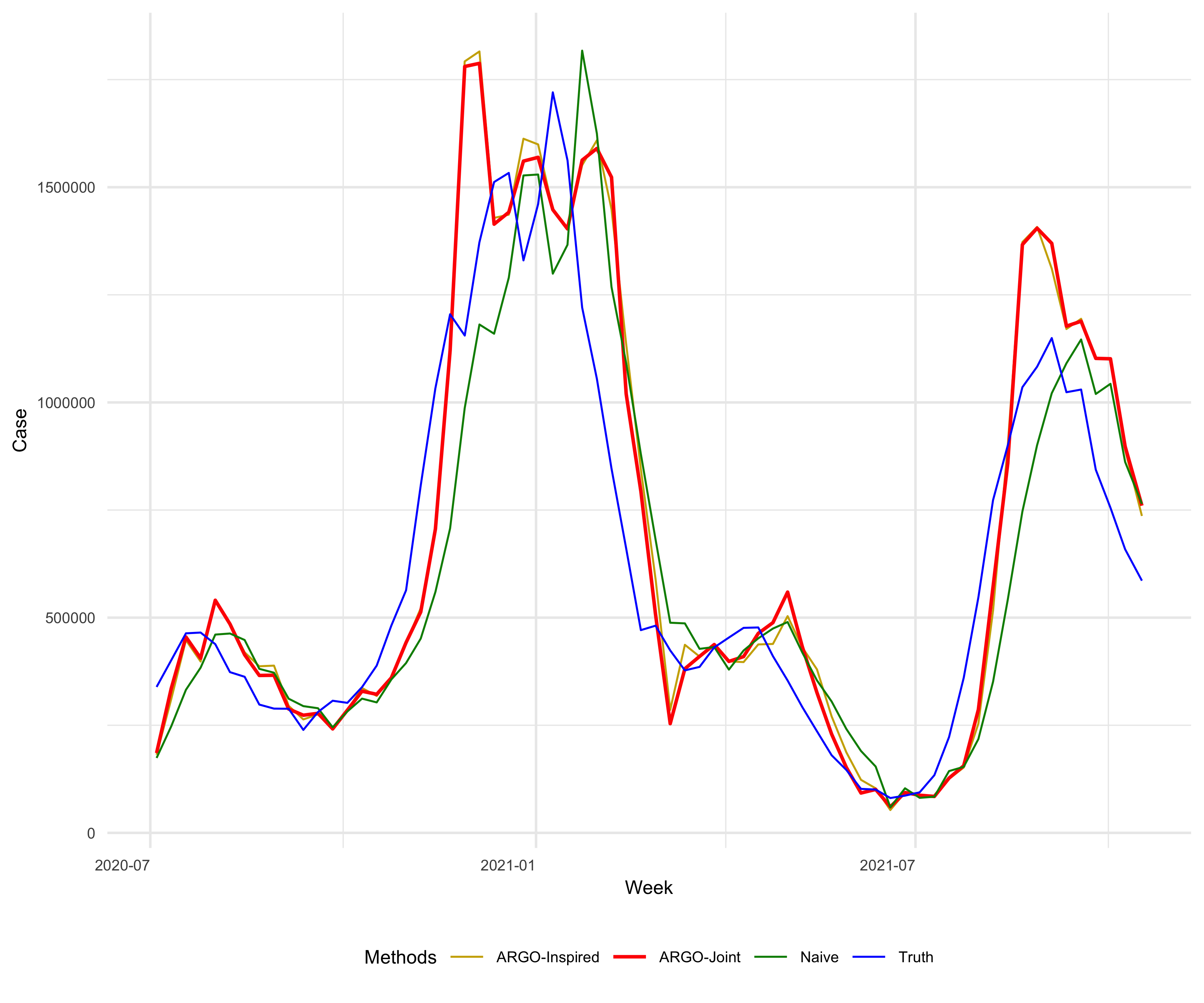}\label{fig:Nat_Compare_Our_Case_2Weeks}}
\hfill
\subfloat[3 Weeks Ahead National Level Predictions]
{\includegraphics[width=0.49\textwidth]{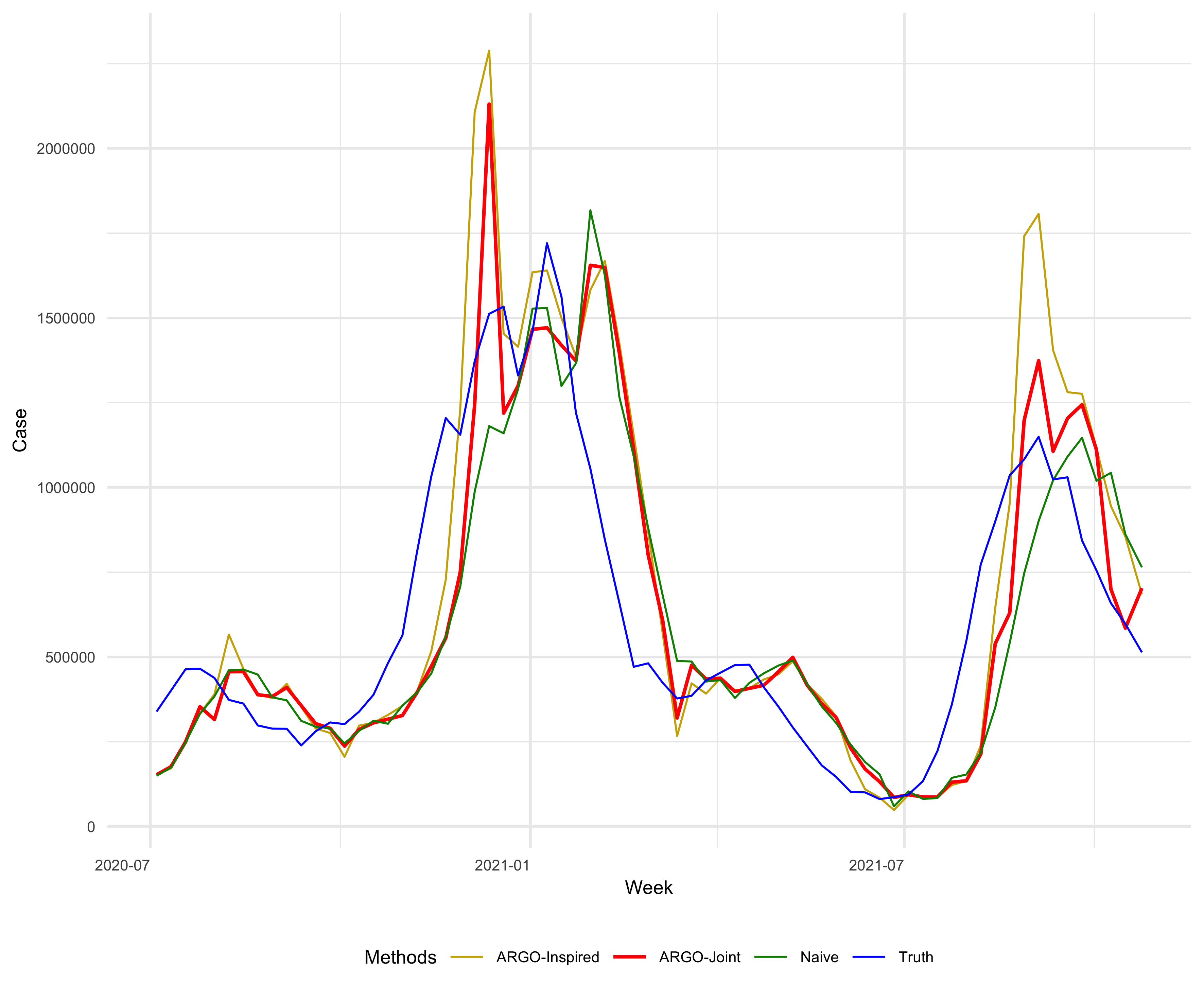}\label{fig:Nat_Compare_Our_Case_3Weeks}}
\hfill
\subfloat[4 Weeks Ahead National Level Predictions]
{\includegraphics[width=0.49\textwidth]{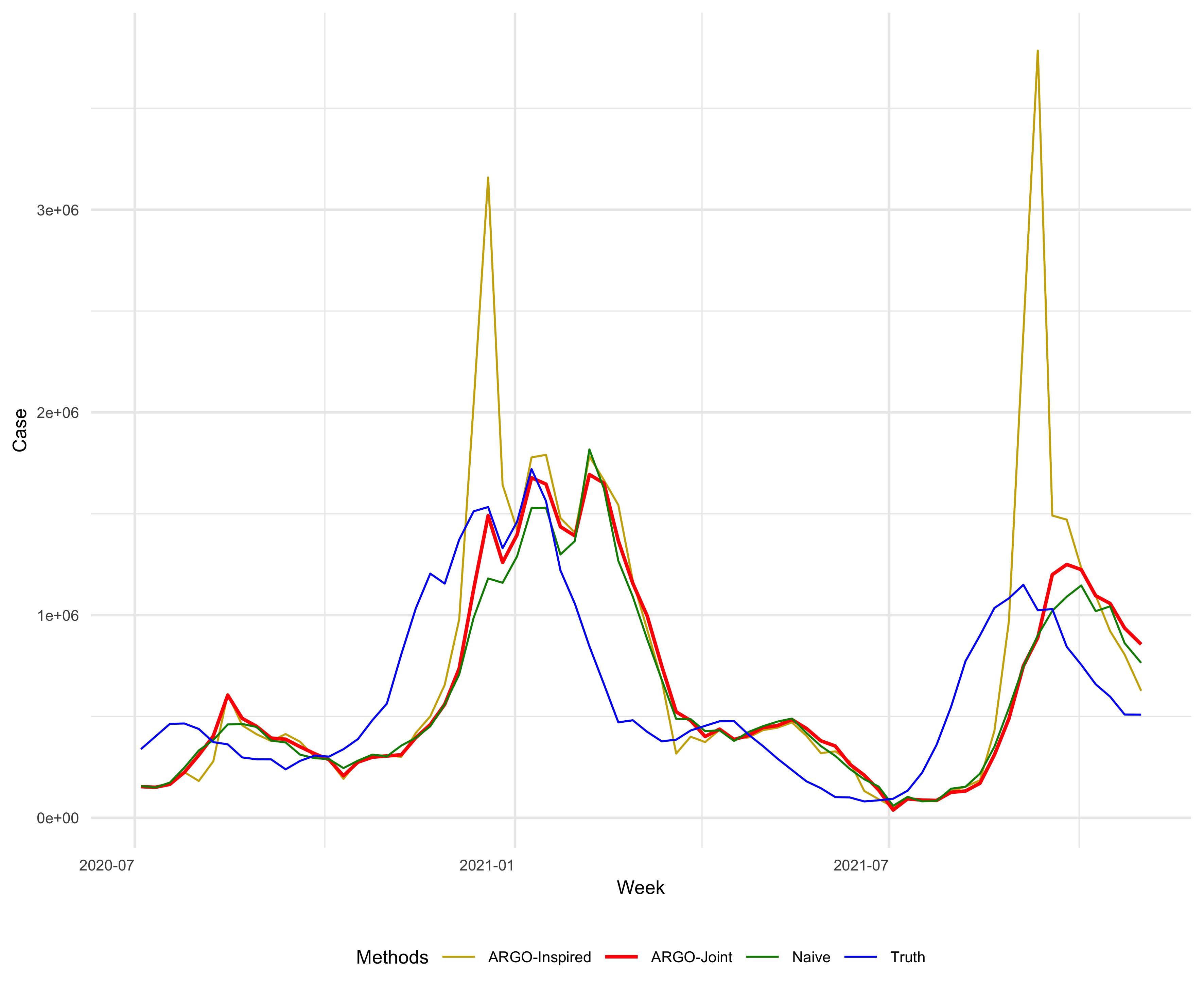}\label{fig:Nat_Compare_Our_Case_4Weeks}}
\caption{\textbf{National level 1 to 4 weeks ahead COVID-19 weekly incremental cases predictions} from 2020-07-04 to 2021-10-09. The method included are ARGO-Inspired (Ref \cite{ma2021covid}), ARGO-Joint, Naive (persistence), truth. Estimation results for 1 (top left), 2 (top right), 3 (bottom left), and 4 (bottom right) weeks ahead weekly incremental cases. ARGO-Joint estimations (thick red), contrasting with the true COVID-19 cases from JHU dataset (blue) as well as the estimates from Ref \cite{ma2021covid} (gold), and Naive (green).}
\label{fig:Case_Nat_Our}
\end{figure}

\begin{figure}[htbp]
\centering
\includegraphics[width=0.6\textwidth]{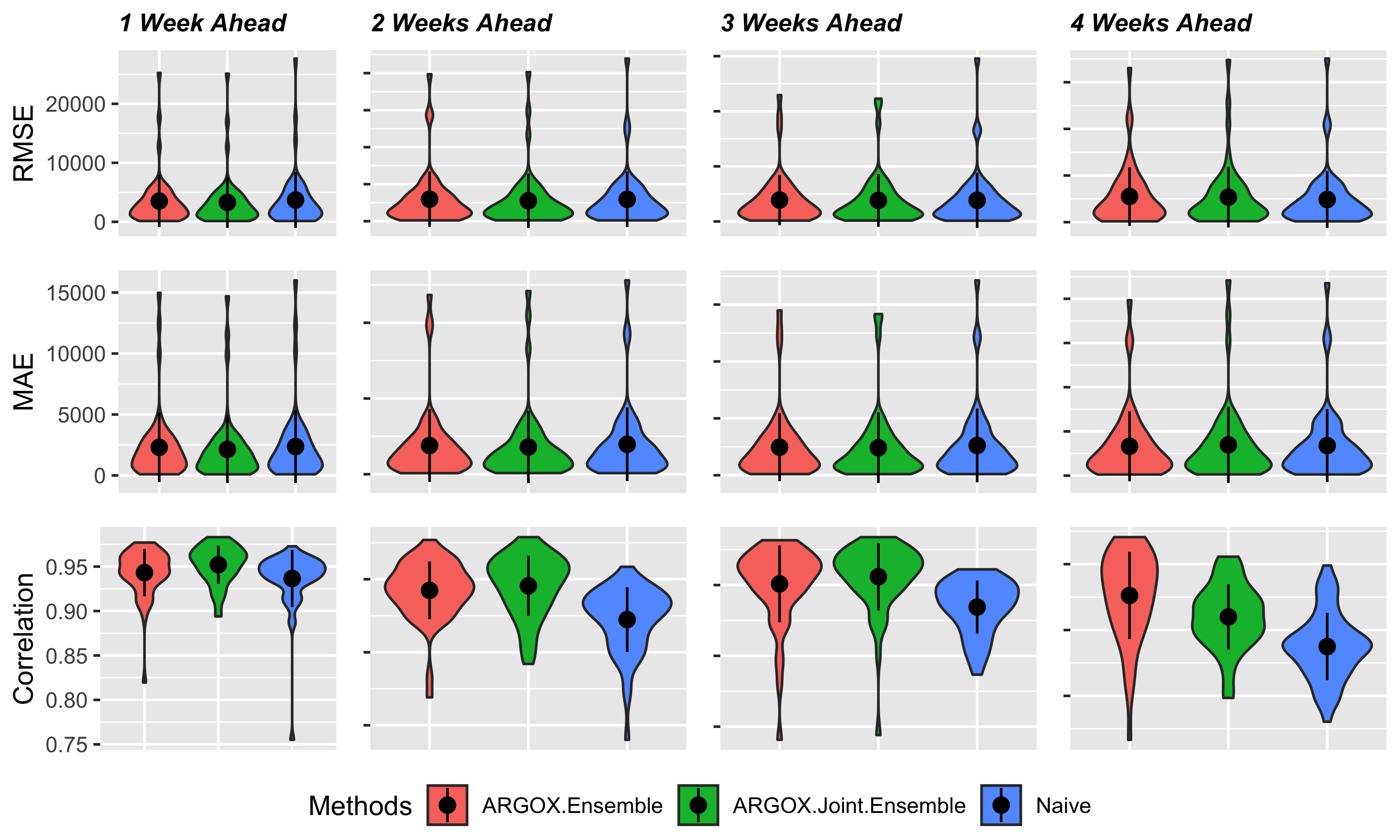}
\caption{\textbf{State Level Case Prediction Comparison in 3 Error Metrics.} Comparison among different versions of our models' 1 to 4 weeks (from left to right) ahead U.S. states level weekly incremental case predictions (from 2020-07-04 to 2021-10-10). The RMSE, MAE and Pearson correlation for each method across all states are reported in the violin plot. The methods (x-axis) are sorted based on their RMSE. Here, ARGOX-Ensemble denotes the previously proposed method \cite{ma2021covid}.}
\label{fig:Case_Our_Violine}
\end{figure}

Figure \ref{fig:Case_Nat_Our} shows 1 to 4 weeks ahead national level cases prediction visualizations of our own method and Ref \cite{ma2021covid} against naive and truth, whereas table \ref{tab:Nat_Case_Our} shows the methods error in the 3 error metrics. Figure \ref{fig:Case_Our_Violine} shows 1 to 4 weeks ahead state level forecast result comparison among our methods through 3 error metrics in violin charts.

\begin{table}[htbp]
\sisetup{detect-weight,mode=text}
\renewrobustcmd{\bfseries}{\fontseries{b}\selectfont}
\renewrobustcmd{\boldmath}{}
\newrobustcmd{\B}{\bfseries}
\addtolength{\tabcolsep}{-4.1pt}
\footnotesize
\centering
\begin{tabular}{lrrrr}
  \hline
& 1 Week Ahead & 2 Weeks Ahead & 3 Weeks Ahead & 4 Weeks Ahead\\ 
    \hline \multicolumn{1}{l}{RMSE} \\
\hspace{1em} Naive &  137656.8  &  225013.8 & 295799.2  & \B364383.0 \\
\hspace{1em} Ref \cite{ma2021covid} & 113181.5 & 215609.7 & 313043.2 &      453890.2      \\
\hspace{1em} ARGO Joint & \B110998.8 & \B210543.2 & \B283680.5     &  364460.7 \\
\multicolumn{1}{l}{MAE} \\
\hspace{1em} Naive & 100615.54   & 170079.9     & 223747.7 &  \B275350.5\\
\hspace{1em} Ref \cite{ma2021covid}& 81377.57 & 155505.8   & 226736.2 &305721.2  \\
\hspace{1em} ARGO Joint & \B78426.58  &   \B148767.5    & \B206003.7    & 275915.6   \\
\multicolumn{1}{l}{Correlation} \\
\hspace{1em} Naive &  0.949 &  0.863 &  0.761 & 0.636\\
\hspace{1em} Ref \cite{ma2021covid}  & 0.969 & 0.917   & 0.853 & \B0.689 \\
\hspace{1em} ARGO Joint & \B0.970 & \B0.920  & \B 0.811&  0.644 \\
\hline
\end{tabular}
\caption{\textbf{National Level Case Prediction Comparison in 3 Error Metrics.} Boldface highlights the best performance for each metric in each study period. All comparisons are based on the original scale of COVID-19 national incremental death. On average, ARGO Joint is able to achieve around 8.3\% RMSE, 6.7\% MAE reduction from previously proposed ARGO Inspired method \cite{ma2021covid}}\label{tab:Nat_Case_Our}
\end{table}

\subsubsection*{Among other publicly available methods (national and state level)}
\begin{figure}[htbp]
\centering
\includegraphics[width=0.8\textwidth]{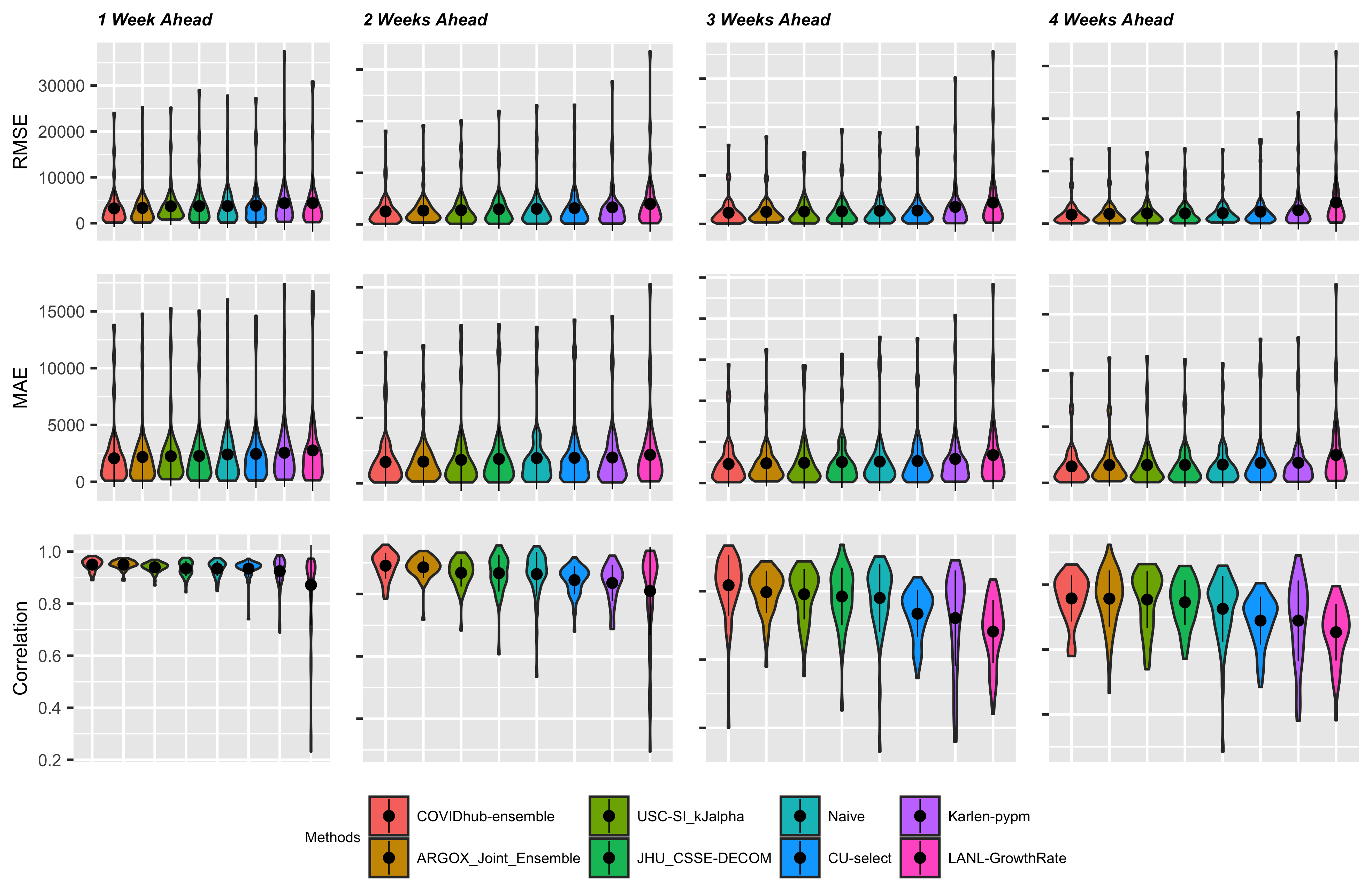}
\caption{\textbf{State Level Case Prediction Comparisons} among different CDC published teams' 1 to 4 weeks (from left to right) ahead weekly incremental death (from 2020-07-04 to 2021-10-10). The RMSE, MAE and Pearson correlation for each method across all states are reported in the violin plot. The methods (x-axis) are sorted based on their RMSE.}
\label{fig:Case_Other_Violine}
\end{figure}

Table \ref{tab:Nat_Case_Other} shows the comparison against other CDC published teams for 1-4 weeks ahead national level incremental case predictions. Figure \ref{fig:Case_Other_Violine} shows the state level prediction comparisons in violin charts, where the methods' error mean and standard deviations are displayed.

\begin{table}[htbp]
\sisetup{detect-weight,mode=text}
\renewrobustcmd{\bfseries}{\fontseries{b}\selectfont}
\renewrobustcmd{\boldmath}{}
\newrobustcmd{\B}{\bfseries}
\addtolength{\tabcolsep}{-4.1pt}
\footnotesize
\centering
\begin{tabular}{lrrrrr}
  \hline
 & 1 Week Ahead & 2 Weeks Ahead & 3 Weeks Ahead & 4 Weeks Ahead & Average \\ 
   \hline \multicolumn{1}{l}{RMSE} \\
\hspace{1em} COVIDhub-ensemble\cite{CDC_Ensemble} & 105307.53 & 176064.32 & 246069.78 & 304113.30 & 207888.73 \\ 
 \hspace{1em} USC-SI\_kJalpha\cite{USC-SI_kJalpha} & 110116.84 & 170671.92 & 239945.46 & 324840.32 & 211393.64 \\ 
\hspace{1em} LANL-GrowthRate\cite{LANL_GrowthRate} & 124698.57 & 179159.81 & 252036.56 & 321872.78 & 219441.93 \\ 
  \hspace{1em} ARGOX\_Joint\_Ensemble & (\#3)110998.63 & (\#5)210543.27 & (\#5)283680.51 & (\#5)364460.75 & (\#4)242420.83 \\ 
  \hspace{1em} JHU\_CSSE-DECOM\cite{JHU_CSSE-DECOM} & 122162.02 & 203493.86 & 280228.17 & 404584.75 & 252617.20 \\ 
  \hspace{1em} Naive & 137863.89 & 225635.06 & 296864.70 & 364598.59 & 256240.56 \\ 
  \hspace{1em} CU-select\cite{CU-Select} & 129680.42 & 253492.11 & 427770.12 & 523080.65 & 333505.82 \\ 
  \hspace{1em} Karlen-pypm\cite{karlen2020characterizing} & 131338.16 & 242952.74 & 423380.95 & 676409.41 & 368520.32 \\ 
 \hline \multicolumn{1}{l}{MAE} \\
\hspace{1em} USC-SI\_kJalpha & 72931.44 & 117312.58 & 176480.24 & 239699.35 & 151605.90 \\ 
  \hspace{1em} COVIDhub-ensemble & 74280.34 & 129640.57 & 182402.10 & 224256.88 & 152644.97 \\ 
  \hspace{1em} LANL-GrowthRate & 97880.20 & 136068.49 & 183986.27 & 228824.46 & 161689.85 \\ 
  \hspace{1em} ARGOX\_Joint\_Ensemble & (\#3)78426.58 & (\#5)148767.48 & (\#5)206003.73 & (\#5)275915.64 & (\#4)177278.43 \\ 
  \hspace{1em} JHU\_CSSE-DECOM & 84597.87 & 147923.86 & 204157.49 & 285166.74 & 180461.49 \\ 
  \hspace{1em} Naive & 100913.67 & 169953.72 & 222740.52 & 271961.01 & 191392.23 \\ 
  \hspace{1em} CU-select & 92742.52 & 151468.57 & 239511.63 & 287925.94 & 192912.16 \\ 
  \hspace{1em} Karlen-pypm\cite{karlen2020characterizing} & 88797.77 & 168000.82 & 281505.90 & 425809.86 & 241028.59 \\ 
 \hline \multicolumn{1}{l}{Correlation} \\
\hspace{1em} Karlen-pypm\cite{karlen2020characterizing} & 0.97 & 0.94 & 0.90 & 0.83 & 0.91 \\ 
  \hspace{1em} LANL-GrowthRate\cite{LANL_GrowthRate} & 0.98 & 0.94 & 0.87 & 0.79 & 0.89 \\ 
  \hspace{1em} COVIDhub-ensemble\cite{CDC_Ensemble} & 0.97 & 0.93 & 0.87 & 0.81 & 0.89 \\ 
  \hspace{1em} USC-SI\_kJalpha\cite{USC-SI_kJalpha} & 0.97 & 0.93 & 0.87 & 0.79 & 0.89 \\ 
  \hspace{1em} JHU\_CSSE-DECOM\cite{JHU_CSSE-DECOM} & 0.96 & 0.90 & 0.85 & 0.76 & 0.87 \\ 
  \hspace{1em} ARGOX\_Joint\_Ensemble & (\#5)0.97 & (\#5)0.92 & (\#6)0.81 & (\#7)0.64 & (\#6)0.83 \\ 
  \hspace{1em} Naive & 0.95 & 0.86 & 0.77 & 0.65 & 0.81 \\ 
  \hspace{1em} CU-select\cite{CU-Select} & 0.96 & 0.86 & 0.70 & 0.58 & 0.78 \\ 
   \hline
\end{tabular}
\caption{\textbf{National Level Case Prediction Comparisons} among different models' 1 to 4 weeks ahead weekly incremental cases (from 2020-07-04 to 2021-10-10). The RMSE, MAE, Pearson correlation and their averages are reported. Methods are sorted based on their average. Our ARGOX-Joint-Ensemble's ranking for each error metric are included in parenthesis.}
\label{tab:Nat_Case_Other}
\end{table}

\clearpage
\subsection*{\%ILI model comparisons}
\subsubsection*{Among our own methods}
\begin{figure}[htbp]
\centering
\includegraphics[width=0.6\textwidth]{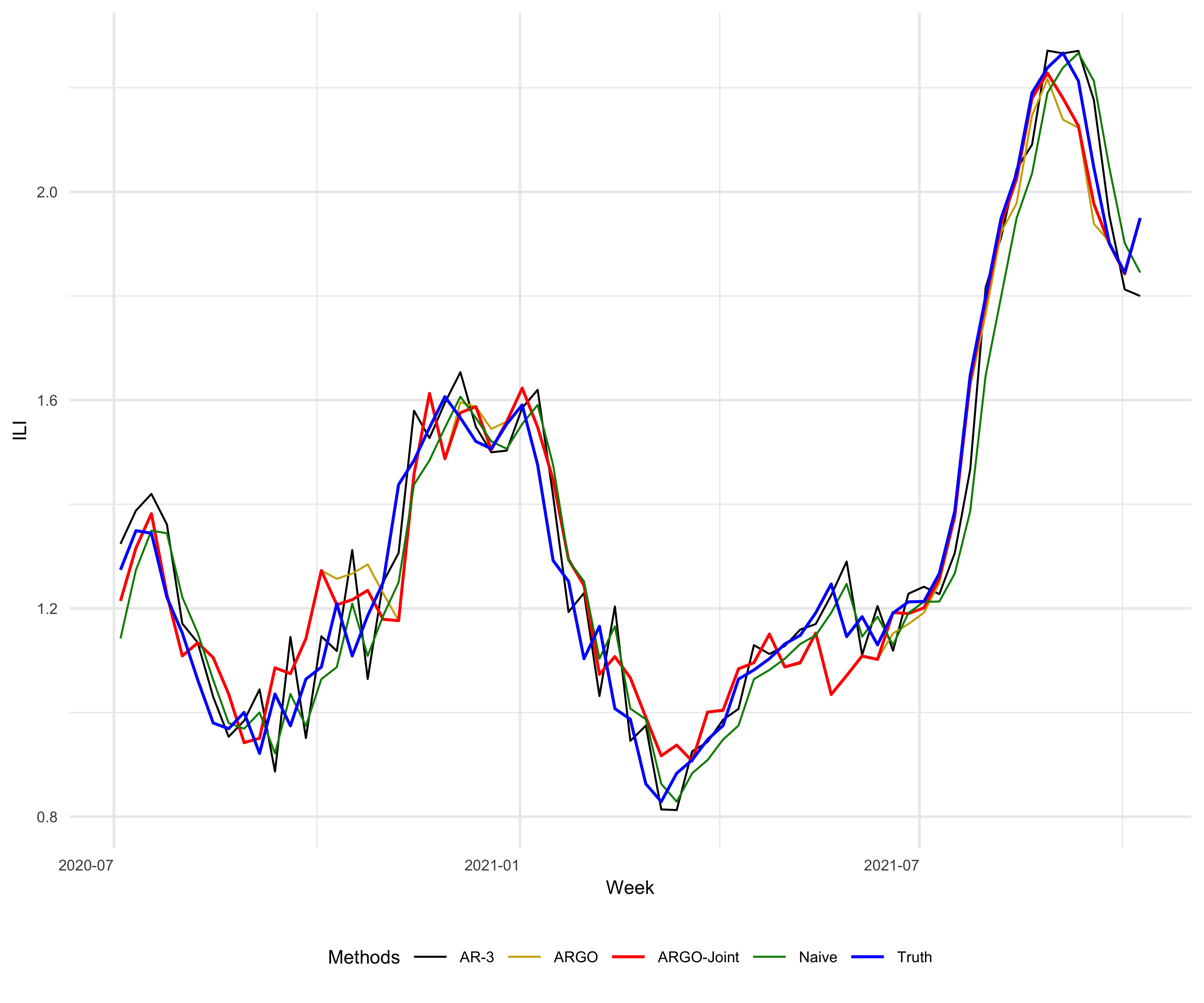}
\caption{\textbf{National level 1 week ahead \%ILI predictions} from 2020-07-04 to 2021-10-09. The method included are ARGO (Ref \cite{ARGO}), ARGO-Joint, AR-3, Naive (persistence), truth. ARGO-Joint estimations (thick red), contrasting with the true \%ILI (blue) as well as the estimates from Ref \cite{ARGO} (gold), Naive (green), and AR-3 (black).}
\label{fig:ILI_Nat_Our}
\end{figure}

Figure \ref{fig:ILI_Nat_Our} shows 1 week ahead national level \%ILI prediction visualizations of our own method and ARGO \cite{ARGO} against AR-3, naive and truth. 

\begin{table}[htbp]
\sisetup{detect-weight,mode=text}
\renewrobustcmd{\bfseries}{\fontseries{b}\selectfont}
\renewrobustcmd{\boldmath}{}
\newrobustcmd{\B}{\bfseries}
\addtolength{\tabcolsep}{-4.1pt}
\centering
\begin{tabular}{|c|c|c|c|}
  \hline
Method & RMSE & MAE & Correlation \\ \hline\hline
Naive &  0.092 & 0.075 & 0.970\\\hline
AR-3 & 0.086 & 0.067 & 0.973\\\hline
Ref \cite{ARGO} & 0.083 & 0.063 & 0.977\\\hline
ARGO Joint & \B0.077 & \B0.056 & \B0.979\\
\hline
\end{tabular}
\caption{\textbf{National Level \%ILI 1 week ahead Prediction Comparison in 3 Error Metrics.} Boldface highlights the best performance for each metric in each study period. All comparisons are based on the original scale of CDC published \%ILI. ARGO Joint is able to achieve 6.1\% RMSE, 11.1\% MAE reduction and 0.2\% increase in correlation from ARGO method \cite{ARGO}}\label{tab:Nat_ILI_Our}
\end{table}

\begin{figure}[htbp]
\centering
\includegraphics[width=0.8\textwidth]{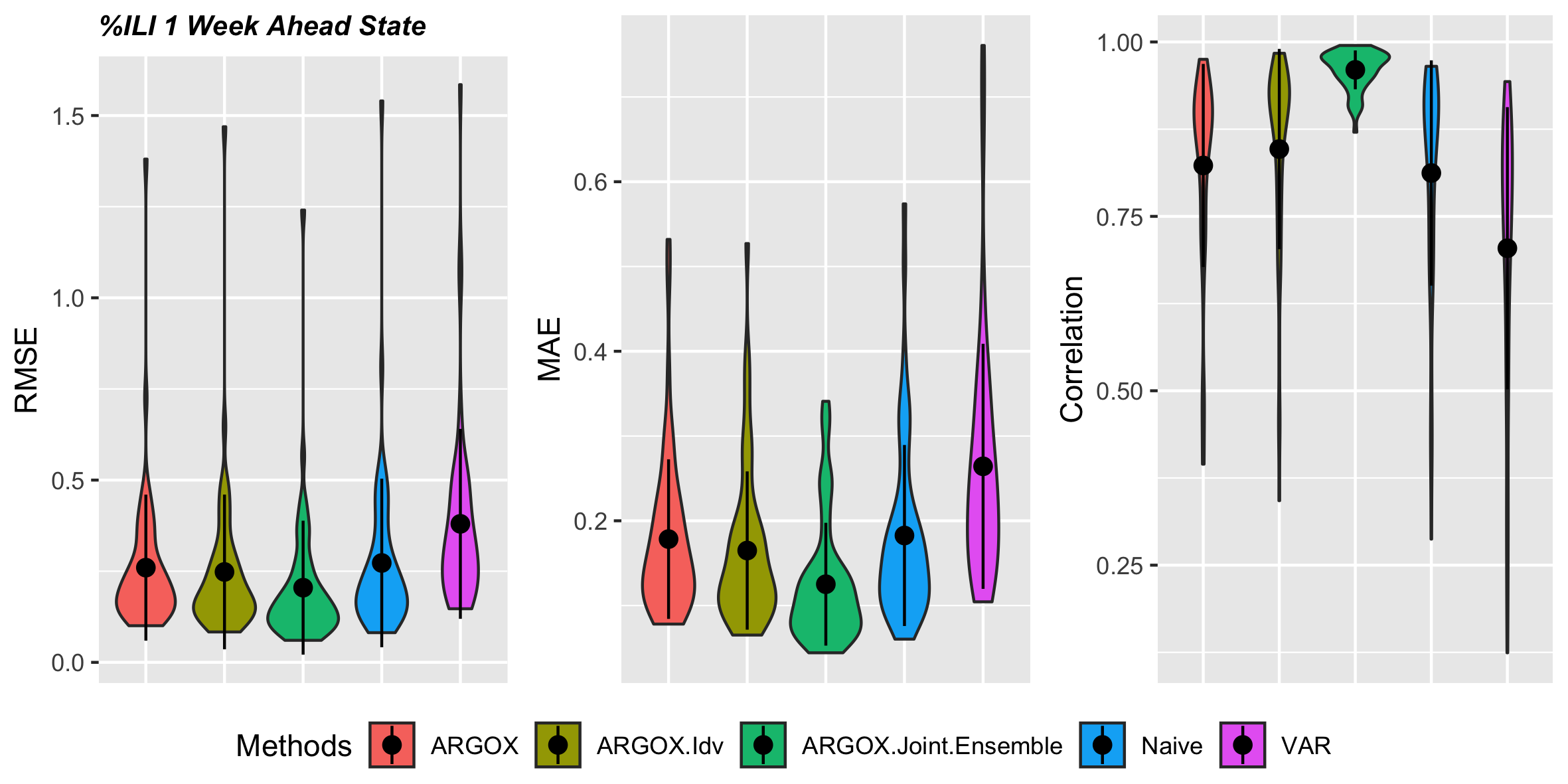}
\caption{\textbf{State Level \%ILI Prediction Comparison in 3 Error Metrics.} Comparison among different versions of our models' and benchmark methods for 1 week ahead U.S. states level weekly \%ILI predictions (from 2020-07-04 to 2021-10-10). The RMSE, MAE and Pearson correlation for each method across all states are reported in the violin plot. Here, ARGOX denotes Ref \cite{ARGOX}, ARGOX-Idv is in section \ref{sec:state_model}, and ARGOX-Joint-Ensemble is in section \ref{sec:state_model}.}
\label{fig:ILI_Our_Violine}
\end{figure}

\clearpage
\newgeometry{left=1.5cm,bottom=2cm}
\subsection*{Detailed COVID-19 incremental death estimation results for each state}
\begin{table}[ht]
\centering
\begin{tabular}{lrrrr}
  \hline
& 1 Week Ahead & 2 Weeks Ahead & 3 Weeks Ahead & 4 Weeks Ahead \\ 
    \hline \multicolumn{1}{l}{RMSE} \\
 \hspace{1em} Ref \cite{ma2021covid} &  10.96 & 8.82 & 10.59 & 8.08 \\ 
   \hspace{1em} ARGOX Joint Ensemble & 10.31 & 7.86 & 8.99 & 7.95 \\ 
   \hspace{1em} Naive & 12.16 & 12.64 & 12.28 & 10.93 \\ 
   \multicolumn{1}{l}{MAE} \\
  \hspace{1em}  Ref \cite{ma2021covid} &6.98 & 6.23 & 6.59 & 5.55 \\ 
   \hspace{1em} ARGOX Joint Ensemble & 5.74 & 5.27 & 5.19 & 5.73 \\ 
   \hspace{1em} Naive & 7.91 & 7.38 & 7.91 & 7.06 \\ 
   \multicolumn{1}{l}{Correlation} \\
 \hspace{1em} Ref \cite{ma2021covid} &0.50 & 0.65 & 0.46 & 0.74 \\ 
   \hspace{1em} ARGOX Joint Ensemble  & 0.51 & 0.71 & 0.65 & 0.74 \\ 
   \hspace{1em} Naive& 0.39 & 0.29 & 0.19 & 0.39 \\ 
   \hline
\end{tabular}
\caption{Comparison of different methods for state-level COVID-19 1 to 4 weeks ahead incremental death in Alaska (AK). The MSE, MAE, and correlation are reported and best performed method is highlighted in boldface.} 
\label{tab:State_Ours_Death_AK}
\end{table}

\begin{figure}[!h] 
  \centering 
\includegraphics[width=0.6\linewidth, page=1]{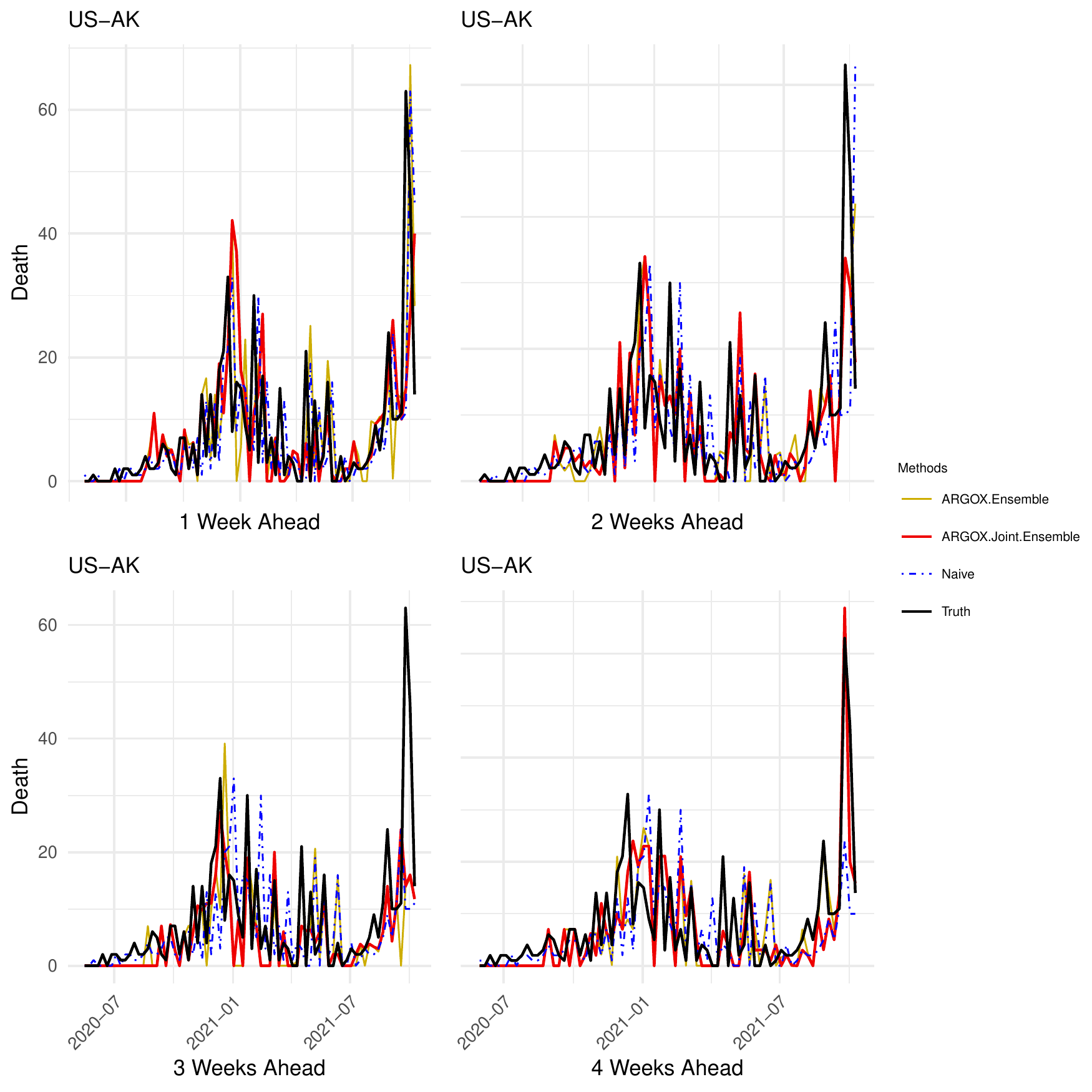} 
\caption{Plots of the COVID-19 1 week (top left), 2 weeks (top right), 3 weeks (bottom left), and 4 weeks (bottom right) ahead estimates for Alaska (AK). ARGOX-Ensemble is Ref \cite{ma2021covid}.}
\label{fig:State_Ours_Death_AK}
\end{figure}
\newpage

\begin{table}[ht]
\centering
\begin{tabular}{lrrrr}
  \hline
& 1 Week Ahead & 2 Weeks Ahead & 3 Weeks Ahead & 4 Weeks Ahead \\ 
    \hline \multicolumn{1}{l}{RMSE} \\
 \hspace{1em} Ref \cite{ma2021covid} & 
137.53 & 169.30 & 190.91 & 174.17 \\ 
   \hspace{1em} ARGOX Joint Ensemble & 146.72 & 144.77 & 171.29 & 175.85 \\ 
   \hspace{1em} Naive & 126.00 & 173.22 & 195.93 & 228.77 \\ 
   \multicolumn{1}{l}{MAE} \\
  \hspace{1em}  Ref \cite{ma2021covid} &75.58 & 97.31 & 113.67 & 102.71 \\ 
   \hspace{1em} ARGOX Joint Ensemble & 79.45 & 81.89 & 105.37 & 106.51 \\ 
   \hspace{1em} Naive & 72.71 & 97.58 & 113.59 & 139.46 \\ 
   \multicolumn{1}{l}{Correlation} \\
 \hspace{1em} Ref \cite{ma2021covid} &0.81 & 0.73 & 0.63 & 0.65 \\ 
   \hspace{1em} ARGOX Joint Ensemble  & 0.78 & 0.77 & 0.68 & 0.64 \\ 
   \hspace{1em} Naive& 0.83 & 0.68 & 0.58 & 0.43 \\ 
   \hline
\end{tabular}
\caption{Comparison of different methods for state-level COVID-19 1 to 4 weeks ahead incremental death in Alabama (AL). The MSE, MAE, and correlation are reported and best performed method is highlighted in boldface.} 
\end{table}

\begin{figure}[!h] 
  \centering 
\includegraphics[width=0.6\linewidth, page=2]{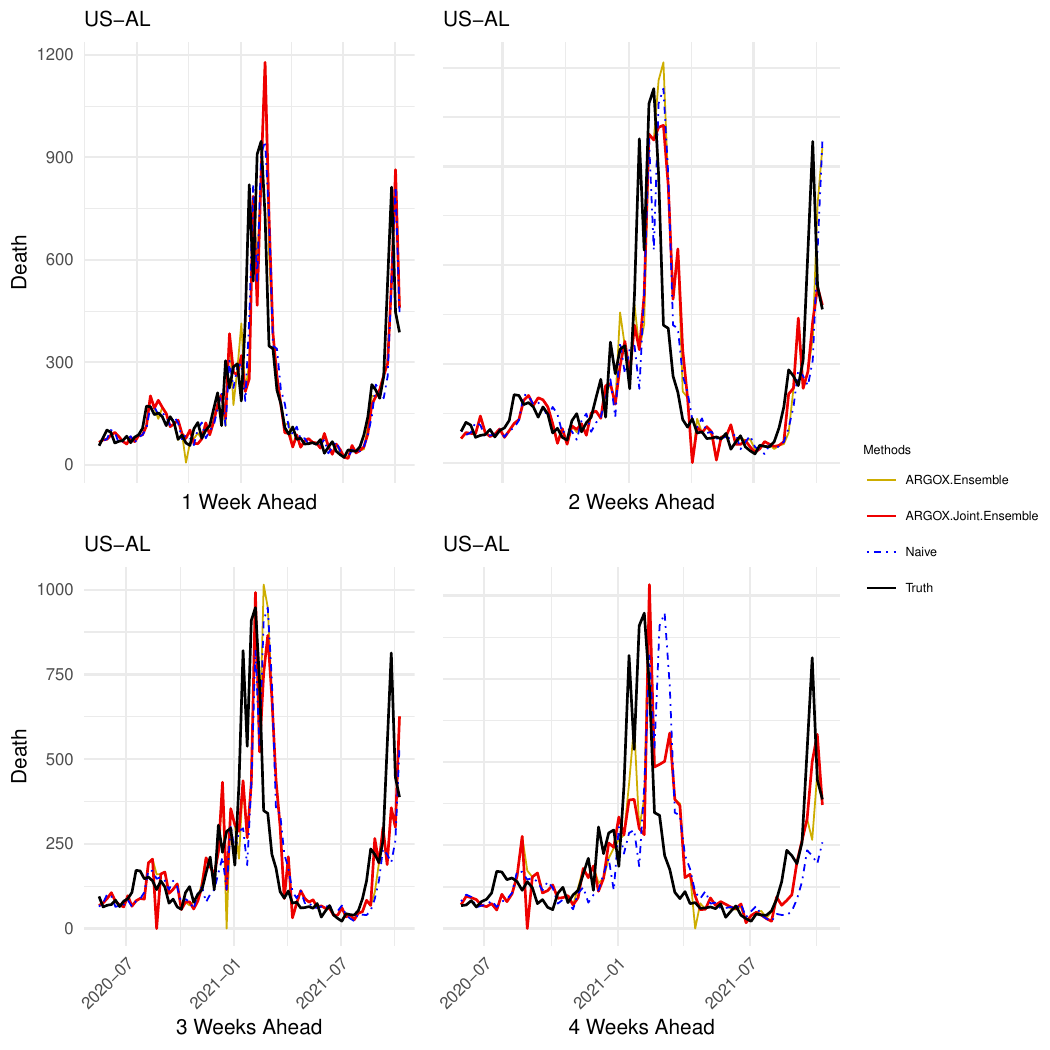} 
\caption{Plots of the COVID-19 1 week (top left), 2 weeks (top right), 3 weeks (bottom left), and 4 weeks (bottom right) ahead estimates for Alabama (AL). ARGOX-Ensemble is Ref \cite{ma2021covid}.}
\end{figure}
\newpage

\begin{table}[ht]
\centering
\begin{tabular}{lrrrr}
  \hline
& 1 Week Ahead & 2 Weeks Ahead & 3 Weeks Ahead & 4 Weeks Ahead \\ 
    \hline \multicolumn{1}{l}{RMSE} \\
 \hspace{1em} Ref \cite{ma2021covid} & 
36.22 & 42.84 & 48.89 & 60.99 \\ 
   \hspace{1em} ARGOX Joint Ensemble & 31.32 & 39.51 & 38.72 & 52.44 \\ 
   \hspace{1em} Naive & 42.20 & 52.24 & 64.14 & 76.68 \\ 
   \multicolumn{1}{l}{MAE} \\
  \hspace{1em}  Ref \cite{ma2021covid} &24.95 & 29.59 & 33.24 & 43.03 \\ 
   \hspace{1em} ARGOX Joint Ensemble & 21.49 & 29.01 & 28.15 & 37.89 \\ 
   \hspace{1em} Naive & 28.52 & 36.62 & 44.41 & 52.68 \\ 
   \multicolumn{1}{l}{Correlation} \\
 \hspace{1em} Ref \cite{ma2021covid} &0.91 & 0.88 & 0.84 & 0.74 \\ 
   \hspace{1em} ARGOX Joint Ensemble  & 0.93 & 0.89 & 0.90 & 0.82 \\ 
   \hspace{1em} Naive& 0.88 & 0.81 & 0.72 & 0.60 \\ 
   \hline
\end{tabular}
\caption{Comparison of different methods for state-level COVID-19 1 to 4 weeks ahead incremental death in Arkansas (AR). The MSE, MAE, and correlation are reported and best performed method is highlighted in boldface.} 
\end{table}

\begin{figure}[!h] 
  \centering 
\includegraphics[width=0.6\linewidth, page=3]{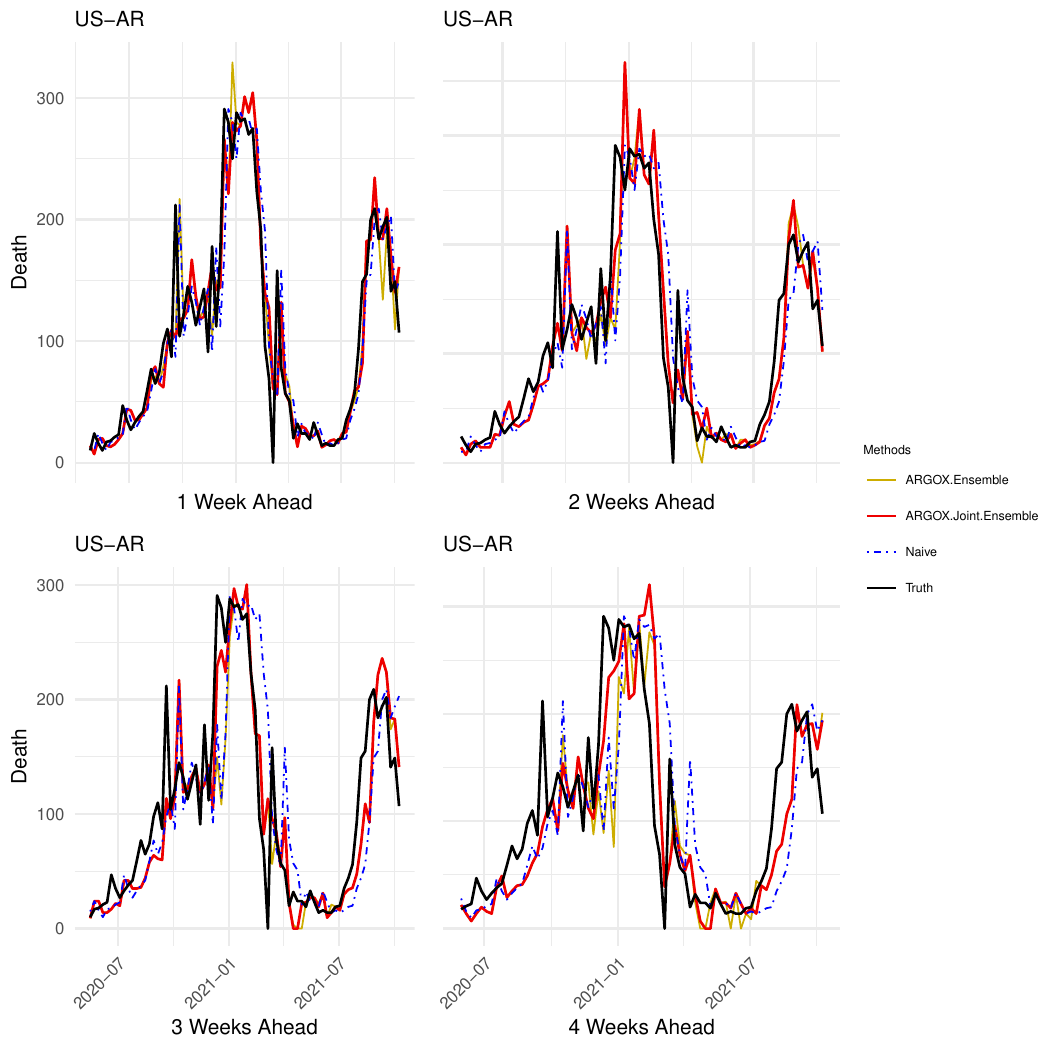} 
\caption{Plots of the COVID-19 1 week (top left), 2 weeks (top right), 3 weeks (bottom left), and 4 weeks (bottom right) ahead estimates for Arkansas (AR). ARGOX-Ensemble is Ref \cite{ma2021covid}.}
\end{figure}
\newpage

\begin{table}[ht]
\centering
\begin{tabular}{lrrrr}
  \hline
& 1 Week Ahead & 2 Weeks Ahead & 3 Weeks Ahead & 4 Weeks Ahead \\ 
    \hline \multicolumn{1}{l}{RMSE} \\
 \hspace{1em} Ref \cite{ma2021covid} & 
99.72 & 151.28 & 155.70 & 159.08 \\ 
   \hspace{1em} ARGOX Joint Ensemble & 95.21 & 144.17 & 153.11 & 185.31 \\ 
   \hspace{1em} Naive & 102.27 & 160.96 & 200.87 & 237.42 \\ 
   \multicolumn{1}{l}{MAE} \\
  \hspace{1em}  Ref \cite{ma2021covid} &65.88 & 98.15 & 102.44 & 103.35 \\ 
   \hspace{1em} ARGOX Joint Ensemble & 61.44 & 92.38 & 101.93 & 111.99 \\ 
   \hspace{1em} Naive & 65.65 & 102.66 & 137.42 & 167.35 \\ 
   \multicolumn{1}{l}{Correlation} \\
 \hspace{1em} Ref \cite{ma2021covid} &0.94 & 0.87 & 0.87 & 0.92 \\ 
   \hspace{1em} ARGOX Joint Ensemble  & 0.94 & 0.88 & 0.87 & 0.89 \\ 
   \hspace{1em} Naive& 0.93 & 0.83 & 0.73 & 0.63 \\ 
   \hline
\end{tabular}
\caption{Comparison of different methods for state-level COVID-19 1 to 4 weeks ahead incremental death in Arizona (AZ). The MSE, MAE, and correlation are reported and best performed method is highlighted in boldface.} 
\end{table}

\begin{figure}[!h] 
  \centering 
\includegraphics[width=0.6\linewidth, page=4]{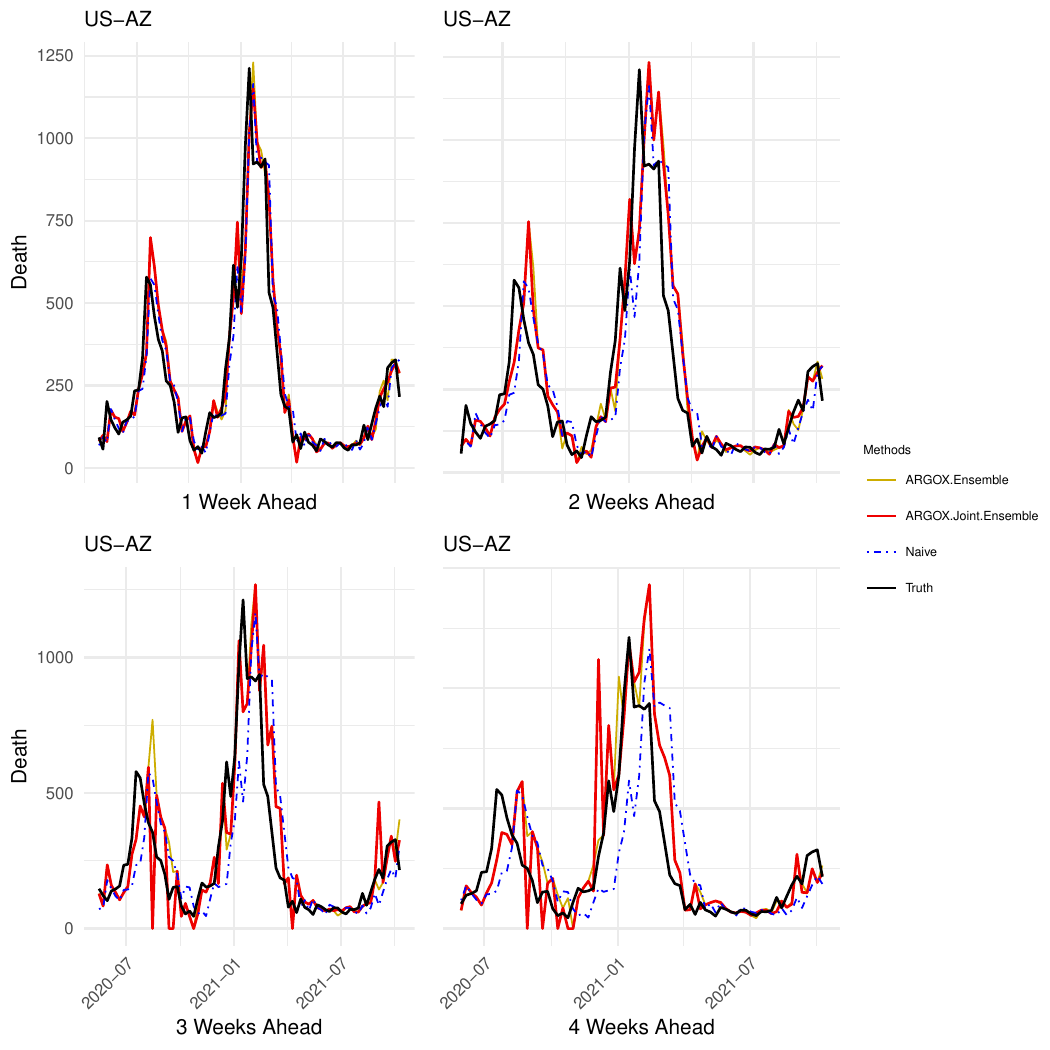} 
\caption{Plots of the COVID-19 1 week (top left), 2 weeks (top right), 3 weeks (bottom left), and 4 weeks (bottom right) ahead estimates for Arizona (AZ). ARGOX-Ensemble is Ref \cite{ma2021covid}.}
\end{figure}
\newpage

\begin{table}[ht]
\centering
\begin{tabular}{lrrrr}
  \hline
& 1 Week Ahead & 2 Weeks Ahead & 3 Weeks Ahead & 4 Weeks Ahead \\ 
    \hline \multicolumn{1}{l}{RMSE} \\
 \hspace{1em} Ref \cite{ma2021covid} & 
299.36 & 401.09 & 473.19 & 452.42 \\ 
   \hspace{1em} ARGOX Joint Ensemble & 294.66 & 362.84 & 376.47 & 425.91 \\ 
   \hspace{1em} Naive & 329.00 & 434.75 & 597.72 & 740.57 \\ 
   \multicolumn{1}{l}{MAE} \\
  \hspace{1em}  Ref \cite{ma2021covid} &193.59 & 241.34 & 290.10 & 336.28 \\ 
   \hspace{1em} ARGOX Joint Ensemble & 191.99 & 210.76 & 251.55 & 308.29 \\ 
   \hspace{1em} Naive & 214.15 & 272.45 & 395.09 & 511.56 \\ 
   \multicolumn{1}{l}{Correlation} \\
 \hspace{1em} Ref \cite{ma2021covid} &0.95 & 0.93 & 0.90 & 0.93 \\ 
   \hspace{1em} ARGOX Joint Ensemble  & 0.96 & 0.94 & 0.95 & 0.94 \\ 
   \hspace{1em} Naive& 0.94 & 0.90 & 0.81 & 0.71 \\ 
   \hline
\end{tabular}
\caption{Comparison of different methods for state-level COVID-19 1 to 4 weeks ahead incremental death in California (CA). The MSE, MAE, and correlation are reported and best performed method is highlighted in boldface.} 
\end{table}

\begin{figure}[!h] 
  \centering 
\includegraphics[width=0.6\linewidth, page=5]{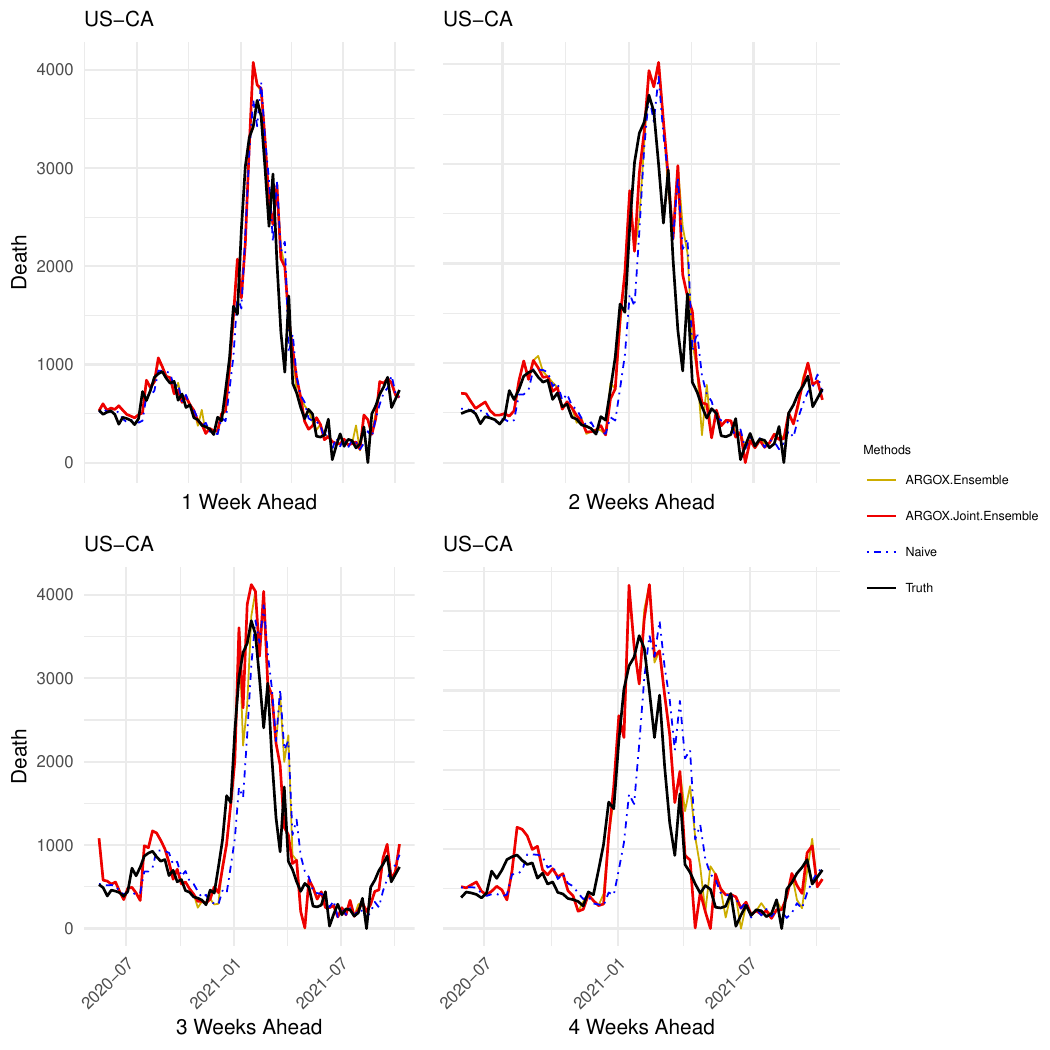} 
\caption{Plots of the COVID-19 1 week (top left), 2 weeks (top right), 3 weeks (bottom left), and 4 weeks (bottom right) ahead estimates for California (CA). ARGOX-Ensemble is Ref \cite{ma2021covid}.}
\end{figure}
\newpage

\begin{table}[ht]
\centering
\begin{tabular}{lrrrr}
  \hline
& 1 Week Ahead & 2 Weeks Ahead & 3 Weeks Ahead & 4 Weeks Ahead \\ 
    \hline \multicolumn{1}{l}{RMSE} \\
 \hspace{1em} Ref \cite{ma2021covid} & 
45.51 & 75.48 & 77.89 & 160.49 \\ 
   \hspace{1em} ARGOX Joint Ensemble & 43.75 & 52.66 & 64.59 & 150.72 \\ 
   \hspace{1em} Naive & 48.21 & 69.43 & 85.06 & 100.41 \\ 
   \multicolumn{1}{l}{MAE} \\
  \hspace{1em}  Ref \cite{ma2021covid} &24.63 & 38.16 & 39.98 & 69.93 \\ 
   \hspace{1em} ARGOX Joint Ensemble & 22.60 & 32.35 & 34.62 & 63.79 \\ 
   \hspace{1em} Naive & 28.44 & 39.14 & 52.05 & 60.68 \\ 
   \multicolumn{1}{l}{Correlation} \\
 \hspace{1em} Ref \cite{ma2021covid} &0.92 & 0.85 & 0.89 & 0.67 \\ 
   \hspace{1em} ARGOX Joint Ensemble  & 0.92 & 0.91 & 0.90 & 0.66 \\ 
   \hspace{1em} Naive& 0.89 & 0.78 & 0.67 & 0.54 \\ 
   \hline
\end{tabular}
\caption{Comparison of different methods for state-level COVID-19 1 to 4 weeks ahead incremental death in Colorado (CO). The MSE, MAE, and correlation are reported and best performed method is highlighted in boldface.} 
\end{table}

\begin{figure}[!h] 
  \centering 
\includegraphics[width=0.6\linewidth, page=6]{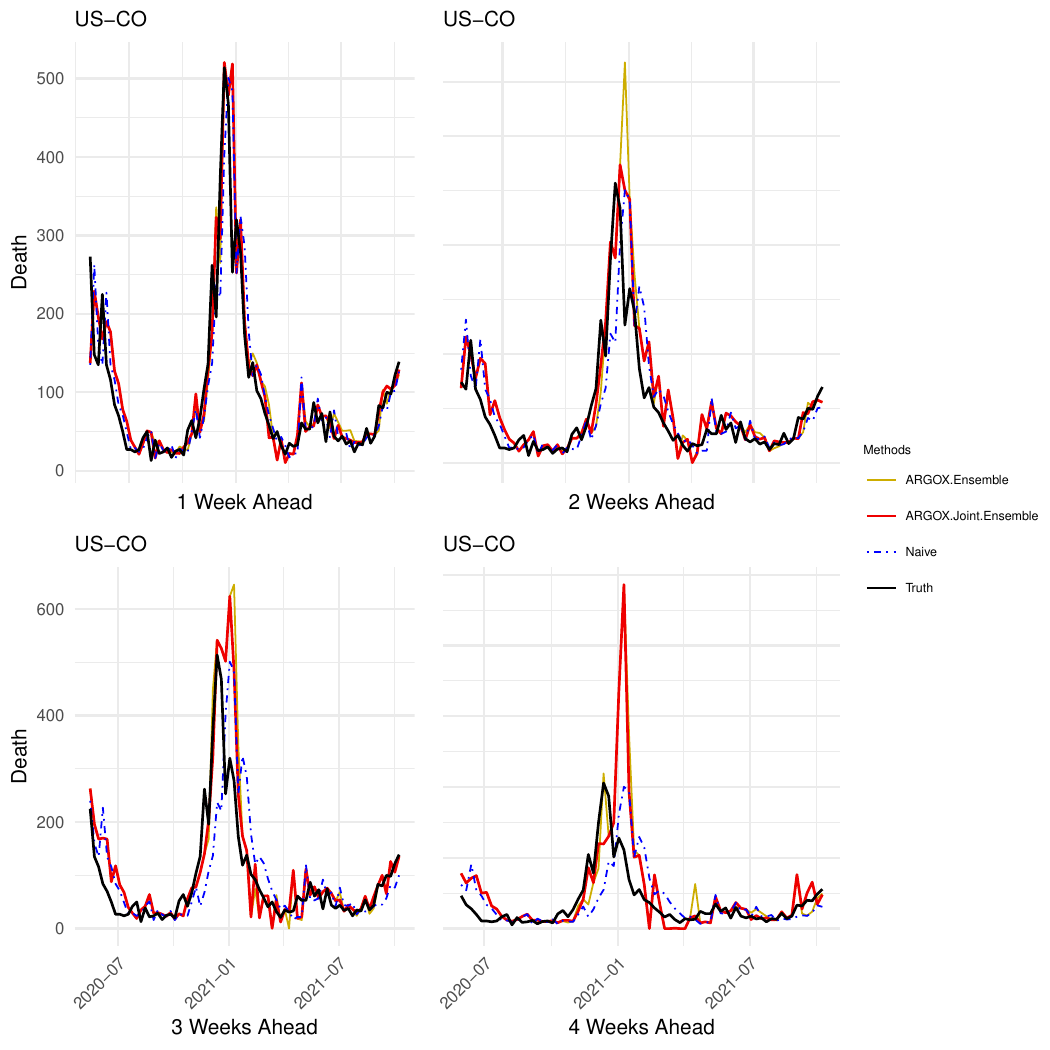} 
\caption{Plots of the COVID-19 1 week (top left), 2 weeks (top right), 3 weeks (bottom left), and 4 weeks (bottom right) ahead estimates for Colorado (CO). ARGOX-Ensemble is Ref \cite{ma2021covid}.}
\end{figure}
\newpage

\begin{table}[ht]
\centering
\begin{tabular}{lrrrr}
  \hline
& 1 Week Ahead & 2 Weeks Ahead & 3 Weeks Ahead & 4 Weeks Ahead \\ 
    \hline \multicolumn{1}{l}{RMSE} \\
 \hspace{1em} Ref \cite{ma2021covid} & 
27.43 & 30.69 & 28.12 & 39.26 \\ 
   \hspace{1em} ARGOX Joint Ensemble & 25.15 & 30.14 & 27.03 & 40.63 \\ 
   \hspace{1em} Naive & 32.38 & 33.52 & 45.95 & 55.55 \\ 
   \multicolumn{1}{l}{MAE} \\
  \hspace{1em}  Ref \cite{ma2021covid} &15.85 & 19.04 & 18.09 & 24.20 \\ 
   \hspace{1em} ARGOX Joint Ensemble & 14.16 & 18.29 & 16.78 & 24.76 \\ 
   \hspace{1em} Naive & 18.59 & 21.95 & 31.17 & 38.21 \\ 
   \multicolumn{1}{l}{Correlation} \\
 \hspace{1em} Ref \cite{ma2021covid} &0.94 & 0.92 & 0.94 & 0.88 \\ 
   \hspace{1em} ARGOX Joint Ensemble  & 0.95 & 0.93 & 0.95 & 0.92 \\ 
   \hspace{1em} Naive& 0.91 & 0.91 & 0.83 & 0.75 \\ 
   \hline
\end{tabular}
\caption{Comparison of different methods for state-level COVID-19 1 to 4 weeks ahead incremental death in Connecticut (CT). The MSE, MAE, and correlation are reported and best performed method is highlighted in boldface.} 
\end{table}

\begin{figure}[!h] 
  \centering 
\includegraphics[width=0.6\linewidth, page=7]{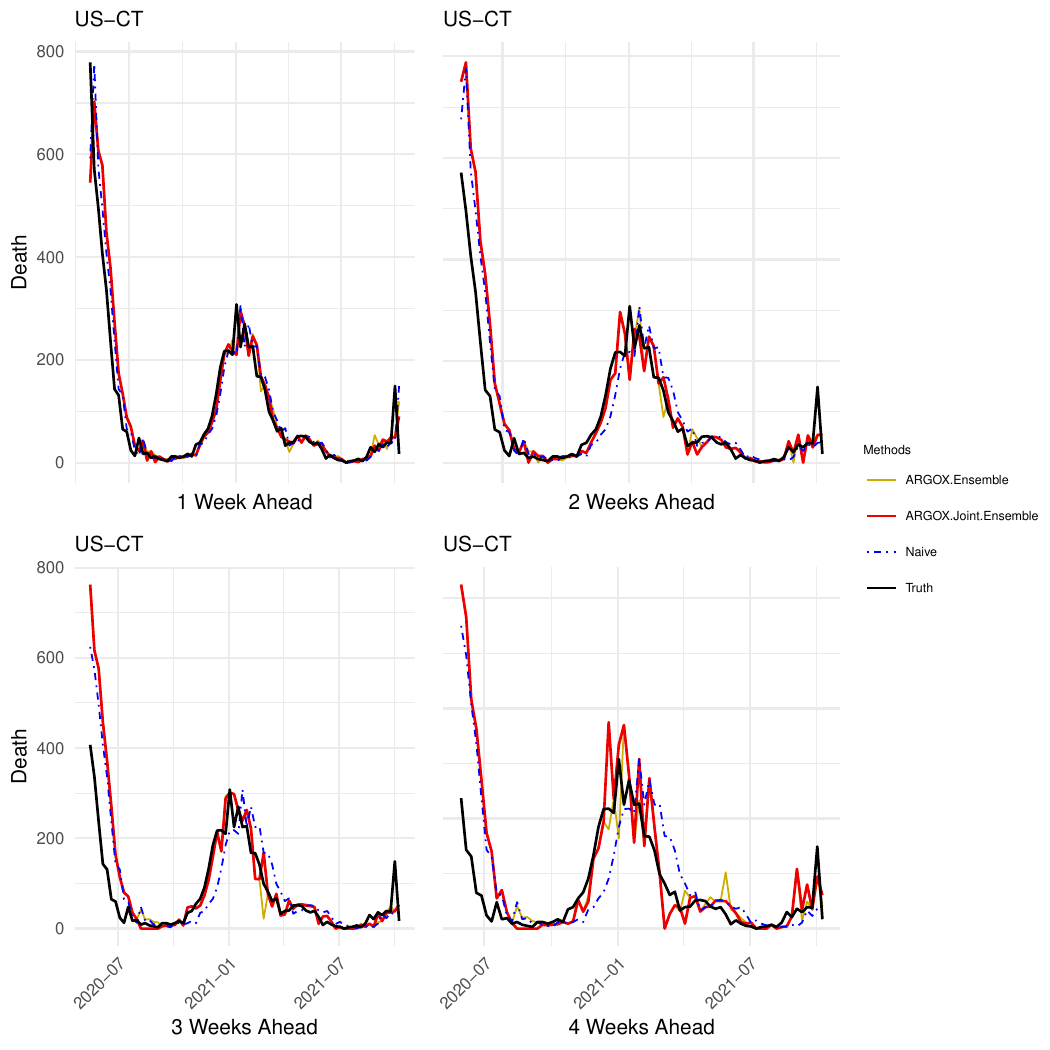} 
\caption{Plots of the COVID-19 1 week (top left), 2 weeks (top right), 3 weeks (bottom left), and 4 weeks (bottom right) ahead estimates for Connecticut (CT). ARGOX-Ensemble is Ref \cite{ma2021covid}.}
\end{figure}
\newpage

\begin{table}[ht]
\centering
\begin{tabular}{lrrrr}
  \hline
& 1 Week Ahead & 2 Weeks Ahead & 3 Weeks Ahead & 4 Weeks Ahead \\ 
    \hline \multicolumn{1}{l}{RMSE} \\
 \hspace{1em} Ref \cite{ma2021covid} & 
5.18 & 5.73 & 6.65 & 8.05 \\ 
   \hspace{1em} ARGOX Joint Ensemble & 4.98 & 5.35 & 6.16 & 8.68 \\ 
   \hspace{1em} Naive & 5.44 & 6.20 & 7.21 & 7.99 \\ 
   \multicolumn{1}{l}{MAE} \\
  \hspace{1em}  Ref \cite{ma2021covid} &3.61 & 3.90 & 4.64 & 5.50 \\ 
   \hspace{1em} ARGOX Joint Ensemble & 3.49 & 3.91 & 4.21 & 6.16 \\ 
   \hspace{1em} Naive & 3.77 & 4.34 & 5.20 & 5.62 \\ 
   \multicolumn{1}{l}{Correlation} \\
 \hspace{1em} Ref \cite{ma2021covid} &0.86 & 0.84 & 0.80 & 0.73 \\ 
   \hspace{1em} ARGOX Joint Ensemble  & 0.87 & 0.86 & 0.82 & 0.71 \\ 
   \hspace{1em} Naive& 0.84 & 0.79 & 0.72 & 0.66 \\ 
   \hline
\end{tabular}
\caption{Comparison of different methods for state-level COVID-19 1 to 4 weeks ahead incremental death in District of Columbia (DC). The MSE, MAE, and correlation are reported and best performed method is highlighted in boldface.} 
\end{table}

\begin{figure}[!h] 
  \centering 
\includegraphics[width=0.6\linewidth, page=8]{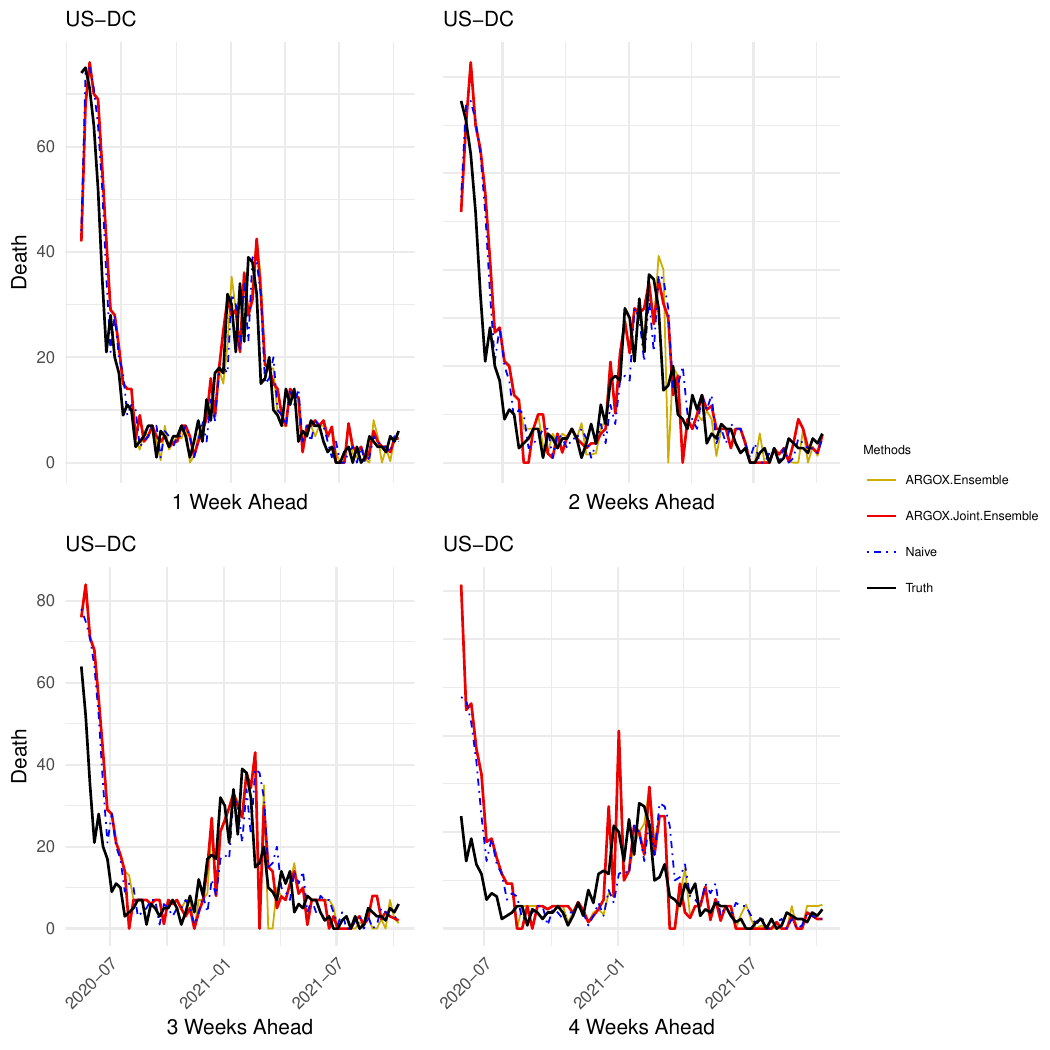} 
\caption{Plots of the COVID-19 1 week (top left), 2 weeks (top right), 3 weeks (bottom left), and 4 weeks (bottom right) ahead estimates for District of Columbia (DC). ARGOX-Ensemble is Ref \cite{ma2021covid}.}
\end{figure}
\newpage

\begin{table}[ht]
\centering
\begin{tabular}{lrrrr}
  \hline
& 1 Week Ahead & 2 Weeks Ahead & 3 Weeks Ahead & 4 Weeks Ahead \\ 
    \hline \multicolumn{1}{l}{RMSE} \\
 \hspace{1em} Ref \cite{ma2021covid} & 
32.07 & 29.25 & 28.36 & 26.47 \\ 
   \hspace{1em} ARGOX Joint Ensemble & 30.01 & 23.64 & 24.69 & 25.92 \\ 
   \hspace{1em} Naive & 30.33 & 31.11 & 30.03 & 30.75 \\ 
   \multicolumn{1}{l}{MAE} \\
  \hspace{1em}  Ref \cite{ma2021covid} &16.50 & 15.09 & 16.22 & 17.51 \\ 
   \hspace{1em} ARGOX Joint Ensemble & 15.45 & 13.73 & 13.78 & 15.86 \\ 
   \hspace{1em} Naive & 15.58 & 16.29 & 17.00 & 18.17 \\ 
   \multicolumn{1}{l}{Correlation} \\
 \hspace{1em} Ref \cite{ma2021covid} &0.21 & 0.28 & 0.29 & 0.26 \\ 
   \hspace{1em} ARGOX Joint Ensemble  & 0.25 & 0.42 & 0.42 & 0.35 \\ 
   \hspace{1em} Naive& 0.22 & 0.19 & 0.25 & 0.23 \\ 
   \hline
\end{tabular}
\caption{Comparison of different methods for state-level COVID-19 1 to 4 weeks ahead incremental death in Delaware (DE). The MSE, MAE, and correlation are reported and best performed method is highlighted in boldface.} 
\end{table}

\begin{figure}[!h] 
  \centering 
\includegraphics[width=0.6\linewidth, page=9]{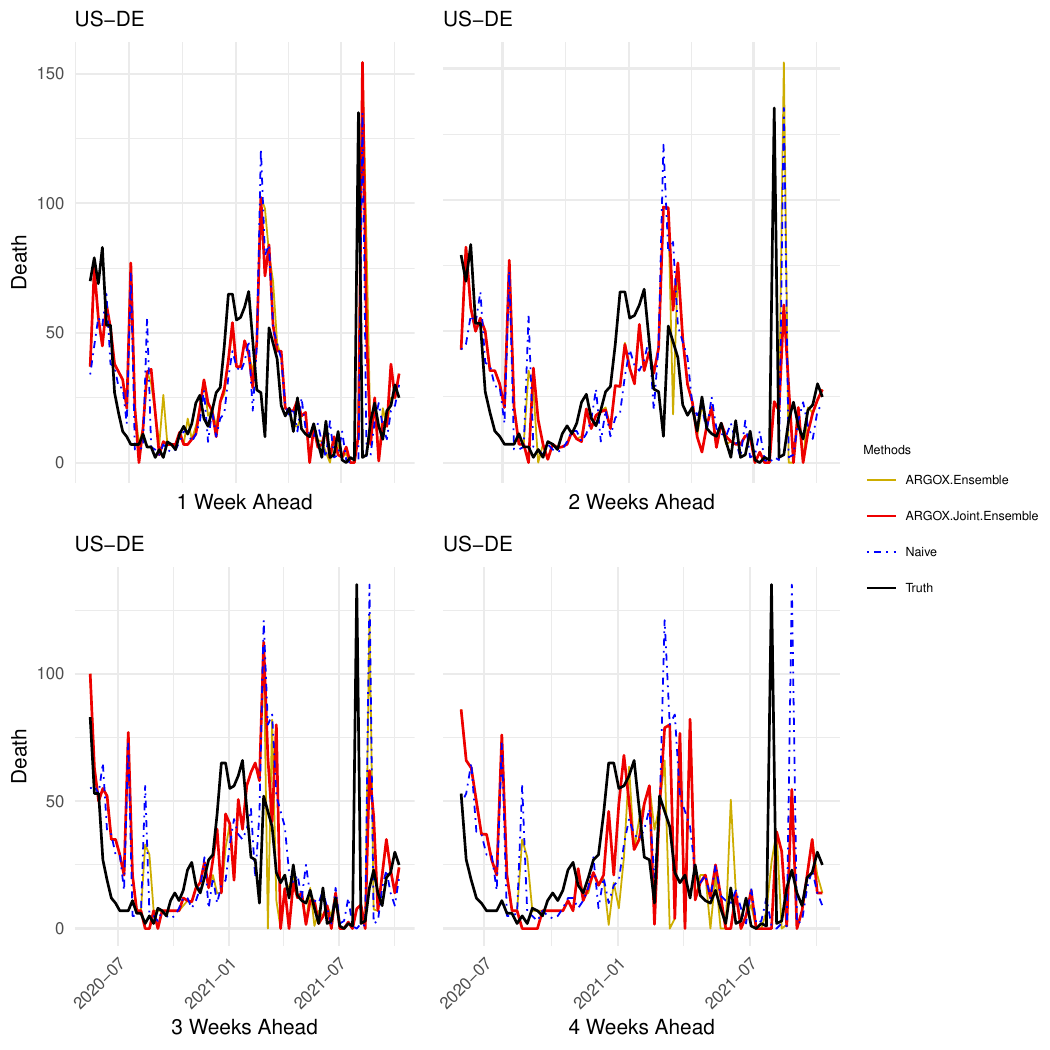} 
\caption{Plots of the COVID-19 1 week (top left), 2 weeks (top right), 3 weeks (bottom left), and 4 weeks (bottom right) ahead estimates for Delaware (DE). ARGOX-Ensemble is Ref \cite{ma2021covid}.}
\end{figure}
\newpage

\begin{table}[ht]
\centering
\begin{tabular}{lrrrr}
  \hline
& 1 Week Ahead & 2 Weeks Ahead & 3 Weeks Ahead & 4 Weeks Ahead \\ 
    \hline \multicolumn{1}{l}{RMSE} \\
 \hspace{1em} Ref \cite{ma2021covid} & 
162.03 & 240.46 & 407.64 & 589.42 \\ 
   \hspace{1em} ARGOX Joint Ensemble & 162.03 & 240.46 & 407.64 & 589.42 \\ 
   \hspace{1em} Naive & 189.48 & 319.21 & 428.23 & 519.32 \\ 
   \multicolumn{1}{l}{MAE} \\
  \hspace{1em}  Ref \cite{ma2021covid} &116.12 & 172.67 & 270.87 & 393.99 \\ 
   \hspace{1em} ARGOX Joint Ensemble & 116.12 & 172.67 & 270.87 & 393.99 \\ 
   \hspace{1em} Naive & 136.02 & 217.62 & 295.45 & 366.68 \\ 
   \multicolumn{1}{l}{Correlation} \\
 \hspace{1em} Ref \cite{ma2021covid} &0.97 & 0.94 & 0.82 & 0.53 \\ 
   \hspace{1em} ARGOX Joint Ensemble  & 0.97 & 0.94 & 0.82 & 0.53 \\ 
   \hspace{1em} Naive& 0.94 & 0.83 & 0.68 & 0.51 \\ 
   \hline
\end{tabular}
\caption{Comparison of different methods for state-level COVID-19 1 to 4 weeks ahead incremental death in Florida (FL). The MSE, MAE, and correlation are reported and best performed method is highlighted in boldface.} 
\end{table}

\begin{figure}[!h] 
  \centering 
\includegraphics[width=0.6\linewidth, page=10]{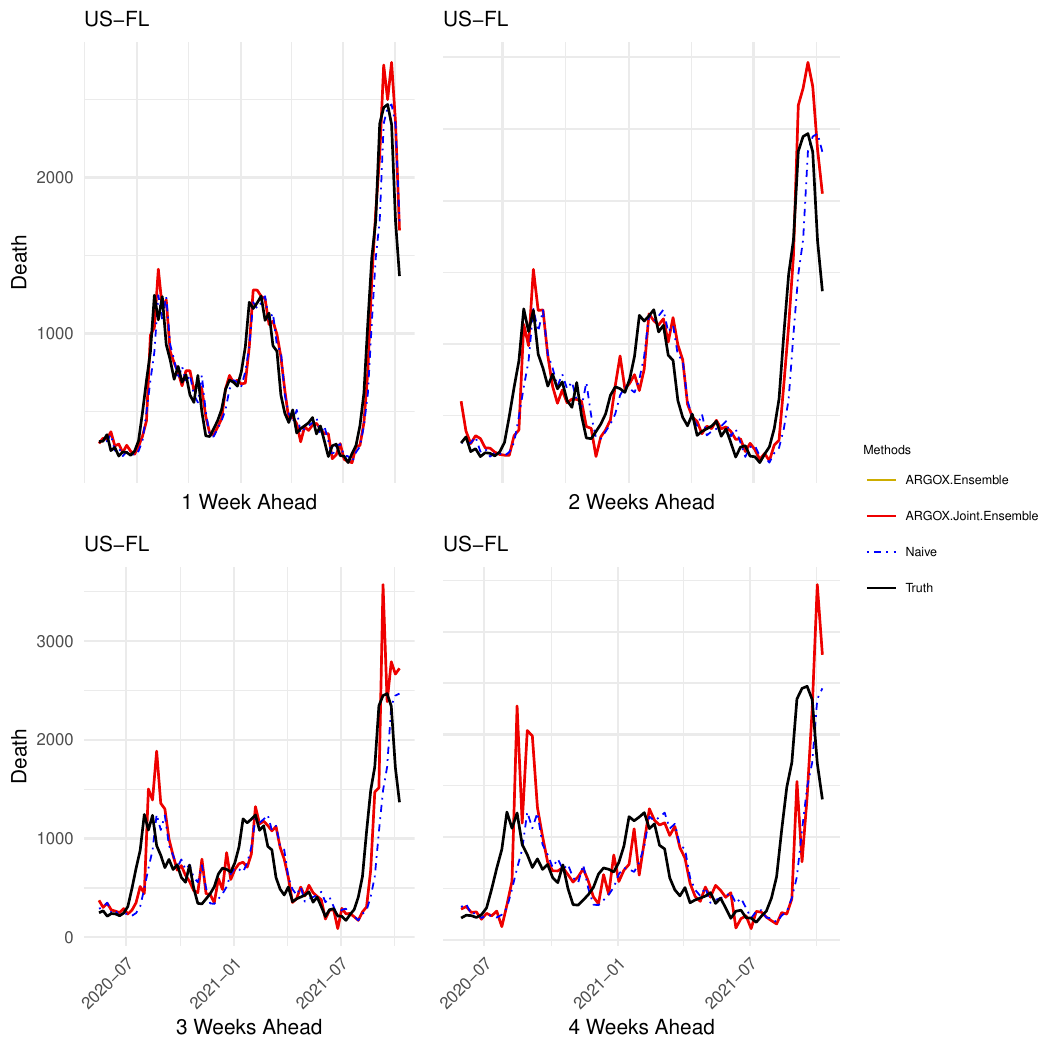} 
\caption{Plots of the COVID-19 1 week (top left), 2 weeks (top right), 3 weeks (bottom left), and 4 weeks (bottom right) ahead estimates for Florida (FL). ARGOX-Ensemble is Ref \cite{ma2021covid}.}
\end{figure}
\newpage

\begin{table}[ht]
\centering
\begin{tabular}{lrrrr}
  \hline
& 1 Week Ahead & 2 Weeks Ahead & 3 Weeks Ahead & 4 Weeks Ahead \\ 
    \hline \multicolumn{1}{l}{RMSE} \\
 \hspace{1em} Ref \cite{ma2021covid} & 
108.95 & 145.21 & 161.90 & 218.59 \\ 
   \hspace{1em} ARGOX Joint Ensemble & 99.23 & 133.47 & 154.34 & 199.83 \\ 
   \hspace{1em} Naive & 130.19 & 173.58 & 221.22 & 263.49 \\ 
   \multicolumn{1}{l}{MAE} \\
  \hspace{1em}  Ref \cite{ma2021covid} &69.09 & 88.70 & 108.38 & 158.29 \\ 
   \hspace{1em} ARGOX Joint Ensemble & 65.29 & 81.93 & 102.30 & 139.06 \\ 
   \hspace{1em} Naive & 86.83 & 117.89 & 156.95 & 192.62 \\ 
   \multicolumn{1}{l}{Correlation} \\
 \hspace{1em} Ref \cite{ma2021covid} &0.91 & 0.84 & 0.81 & 0.66 \\ 
   \hspace{1em} ARGOX Joint Ensemble  & 0.92 & 0.86 & 0.82 & 0.68 \\ 
   \hspace{1em} Naive& 0.87 & 0.76 & 0.60 & 0.43 \\ 
   \hline
\end{tabular}
\caption{Comparison of different methods for state-level COVID-19 1 to 4 weeks ahead incremental death in Georgia (GA). The MSE, MAE, and correlation are reported and best performed method is highlighted in boldface.} 
\end{table}

\begin{figure}[!h] 
  \centering 
\includegraphics[width=0.6\linewidth, page=11]{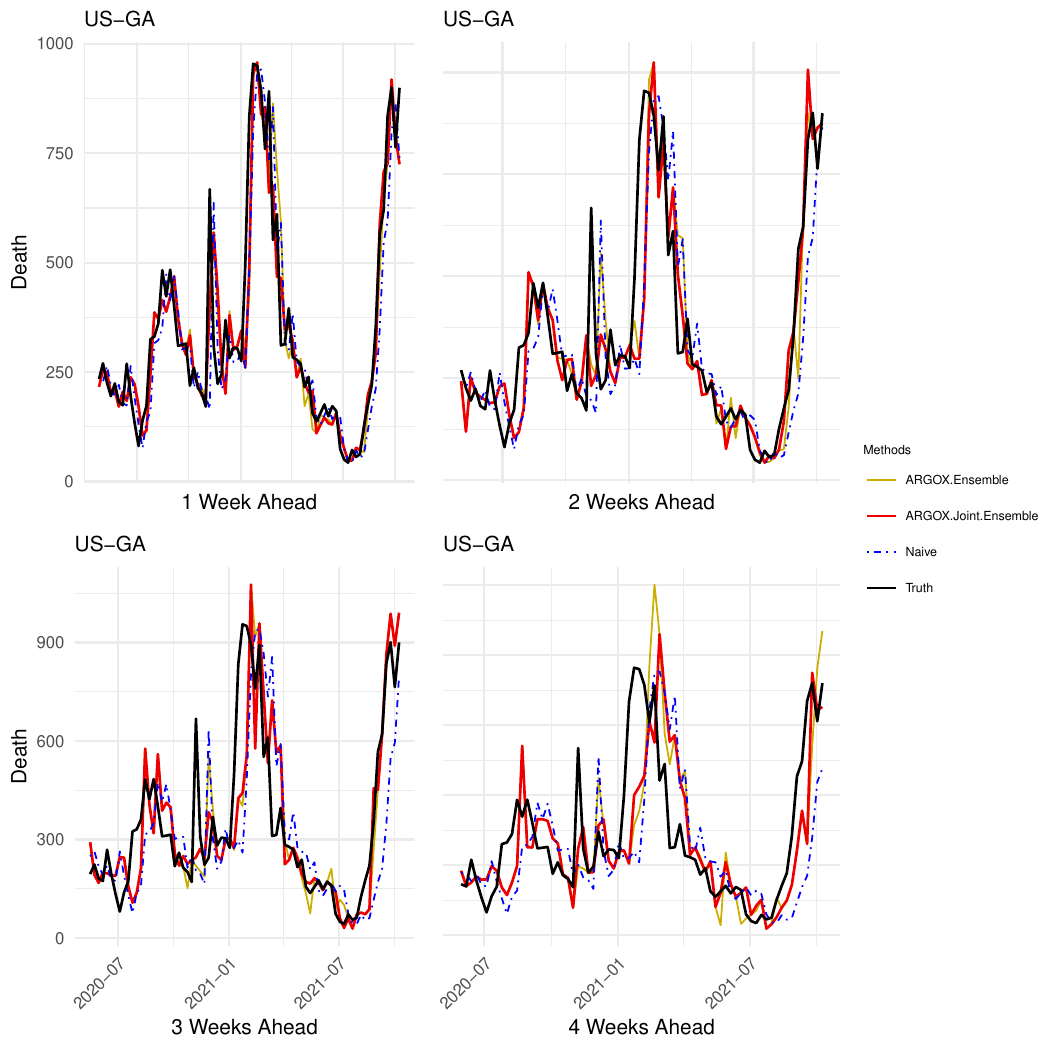} 
\caption{Plots of the COVID-19 1 week (top left), 2 weeks (top right), 3 weeks (bottom left), and 4 weeks (bottom right) ahead estimates for Georgia (GA). ARGOX-Ensemble is Ref \cite{ma2021covid}.}
\end{figure}
\newpage

\begin{table}[ht]
\centering
\begin{tabular}{lrrrr}
  \hline
& 1 Week Ahead & 2 Weeks Ahead & 3 Weeks Ahead & 4 Weeks Ahead \\ 
    \hline \multicolumn{1}{l}{RMSE} \\
 \hspace{1em} Ref \cite{ma2021covid} & 
12.03 & 12.25 & 12.68 & 14.94 \\ 
   \hspace{1em} ARGOX Joint Ensemble & 12.03 & 12.25 & 12.68 & 14.94 \\ 
   \hspace{1em} Naive & 12.14 & 13.06 & 14.22 & 15.96 \\ 
   \multicolumn{1}{l}{MAE} \\
  \hspace{1em}  Ref \cite{ma2021covid} &6.75 & 6.76 & 7.56 & 9.58 \\ 
   \hspace{1em} ARGOX Joint Ensemble & 6.75 & 6.76 & 7.56 & 9.58 \\ 
   \hspace{1em} Naive & 6.58 & 7.14 & 8.53 & 9.35 \\ 
   \multicolumn{1}{l}{Correlation} \\
 \hspace{1em} Ref \cite{ma2021covid} &0.63 & 0.64 & 0.56 & 0.38 \\ 
   \hspace{1em} ARGOX Joint Ensemble  & 0.63 & 0.64 & 0.56 & 0.38 \\ 
   \hspace{1em} Naive& 0.61 & 0.52 & 0.41 & 0.22 \\ 
   \hline
\end{tabular}
\caption{Comparison of different methods for state-level COVID-19 1 to 4 weeks ahead incremental death in Hawaii (HI). The MSE, MAE, and correlation are reported and best performed method is highlighted in boldface.} 
\end{table}

\begin{figure}[!h] 
  \centering 
\includegraphics[width=0.6\linewidth, page=12]{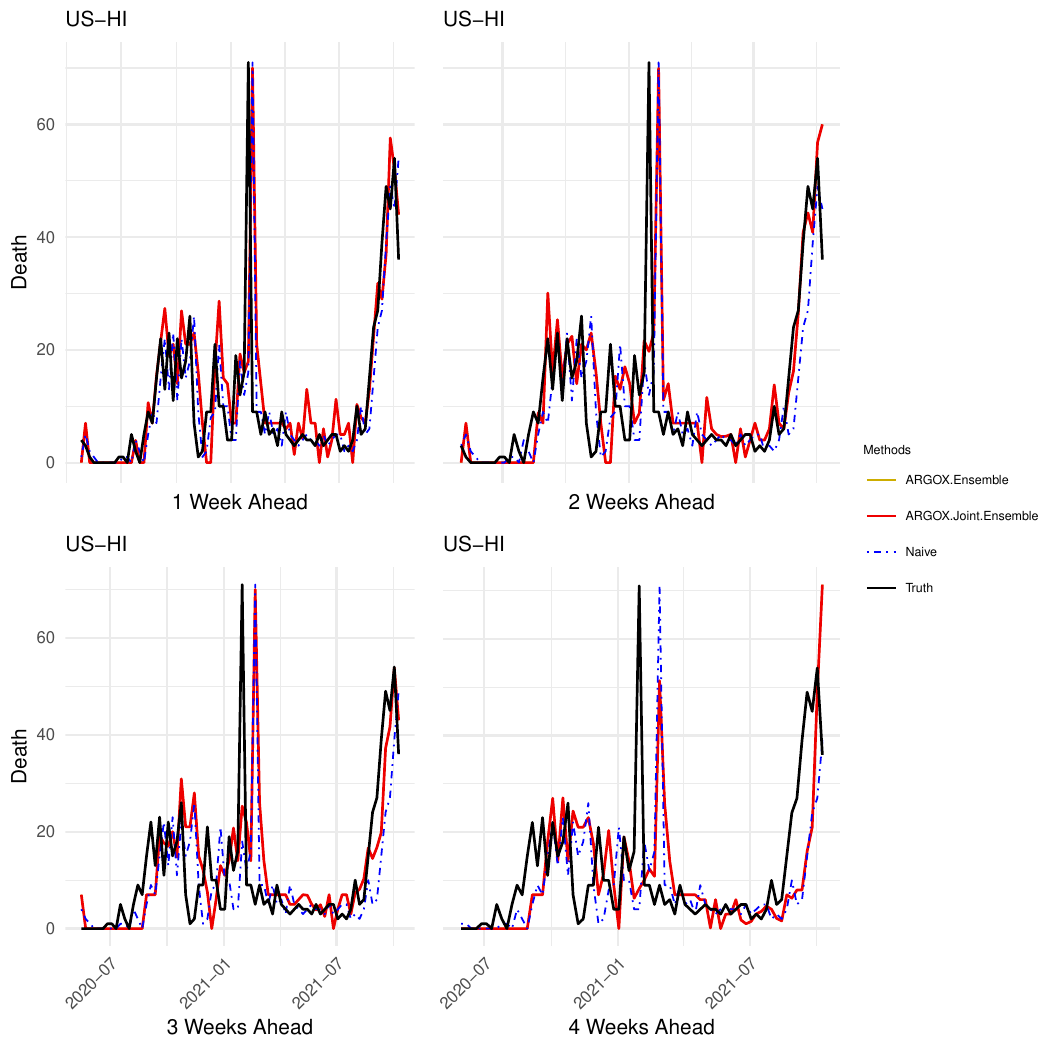} 
\caption{Plots of the COVID-19 1 week (top left), 2 weeks (top right), 3 weeks (bottom left), and 4 weeks (bottom right) ahead estimates for Hawaii (HI). ARGOX-Ensemble is Ref \cite{ma2021covid}.}
\end{figure}
\newpage

\begin{table}[ht]
\centering
\begin{tabular}{lrrrr}
  \hline
& 1 Week Ahead & 2 Weeks Ahead & 3 Weeks Ahead & 4 Weeks Ahead \\ 
    \hline \multicolumn{1}{l}{RMSE} \\
 \hspace{1em} Ref \cite{ma2021covid} & 
63.12 & 68.59 & 72.98 & 102.53 \\ 
   \hspace{1em} ARGOX Joint Ensemble & 53.10 & 63.38 & 64.48 & 97.42 \\ 
   \hspace{1em} Naive & 71.97 & 87.47 & 90.10 & 101.08 \\ 
   \multicolumn{1}{l}{MAE} \\
  \hspace{1em}  Ref \cite{ma2021covid} &33.87 & 35.58 & 40.22 & 53.23 \\ 
   \hspace{1em} ARGOX Joint Ensemble & 30.89 & 32.66 & 37.55 & 49.98 \\ 
   \hspace{1em} Naive & 34.41 & 41.28 & 48.94 & 56.62 \\ 
   \multicolumn{1}{l}{Correlation} \\
 \hspace{1em} Ref \cite{ma2021covid} &0.81 & 0.76 & 0.77 & 0.65 \\ 
   \hspace{1em} ARGOX Joint Ensemble  & 0.86 & 0.79 & 0.82 & 0.74 \\ 
   \hspace{1em} Naive& 0.74 & 0.62 & 0.60 & 0.50 \\ 
   \hline
\end{tabular}
\caption{Comparison of different methods for state-level COVID-19 1 to 4 weeks ahead incremental death in Iowa (IA). The MSE, MAE, and correlation are reported and best performed method is highlighted in boldface.} 
\end{table}

\begin{figure}[!h] 
  \centering 
\includegraphics[width=0.6\linewidth, page=13]{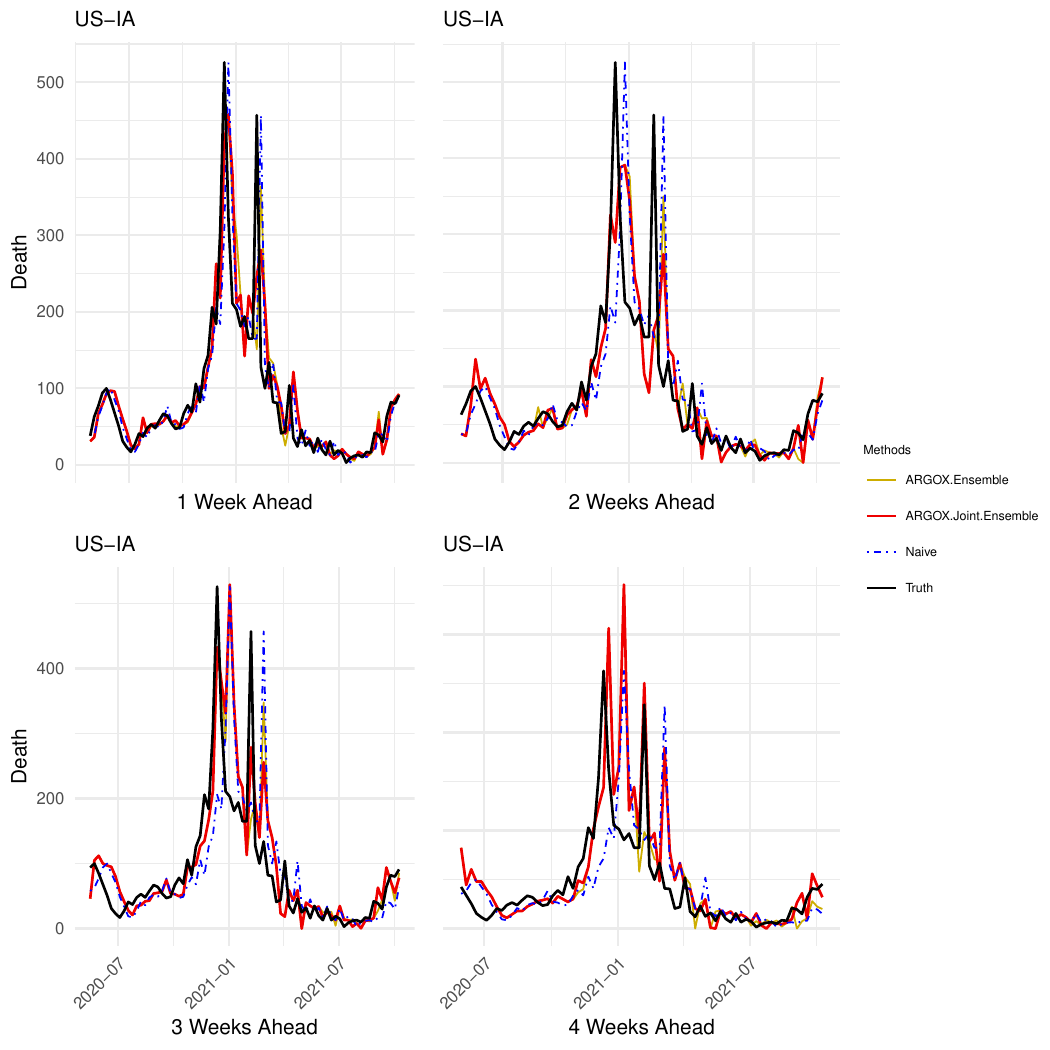} 
\caption{Plots of the COVID-19 1 week (top left), 2 weeks (top right), 3 weeks (bottom left), and 4 weeks (bottom right) ahead estimates for Iowa (IA). ARGOX-Ensemble is Ref \cite{ma2021covid}.}
\end{figure}
\newpage

\begin{table}[ht]
\centering
\begin{tabular}{lrrrr}
  \hline
& 1 Week Ahead & 2 Weeks Ahead & 3 Weeks Ahead & 4 Weeks Ahead \\ 
    \hline \multicolumn{1}{l}{RMSE} \\
 \hspace{1em} Ref \cite{ma2021covid} & 
21.42 & 20.97 & 27.20 & 32.79 \\ 
   \hspace{1em} ARGOX Joint Ensemble & 21.15 & 20.16 & 23.61 & 31.20 \\ 
   \hspace{1em} Naive & 22.60 & 24.78 & 31.70 & 38.55 \\ 
   \multicolumn{1}{l}{MAE} \\
  \hspace{1em}  Ref \cite{ma2021covid} &13.75 & 13.84 & 17.14 & 23.14 \\ 
   \hspace{1em} ARGOX Joint Ensemble & 13.57 & 13.93 & 15.21 & 21.95 \\ 
   \hspace{1em} Naive & 14.32 & 17.08 & 21.38 & 25.68 \\ 
   \multicolumn{1}{l}{Correlation} \\
 \hspace{1em} Ref \cite{ma2021covid} &0.87 & 0.87 & 0.79 & 0.69 \\ 
   \hspace{1em} ARGOX Joint Ensemble  & 0.88 & 0.88 & 0.84 & 0.71 \\ 
   \hspace{1em} Naive& 0.85 & 0.82 & 0.69 & 0.51 \\ 
   \hline
\end{tabular}
\caption{Comparison of different methods for state-level COVID-19 1 to 4 weeks ahead incremental death in Idaho (ID). The MSE, MAE, and correlation are reported and best performed method is highlighted in boldface.} 
\end{table}

\begin{figure}[!h] 
  \centering 
\includegraphics[width=0.6\linewidth, page=14]{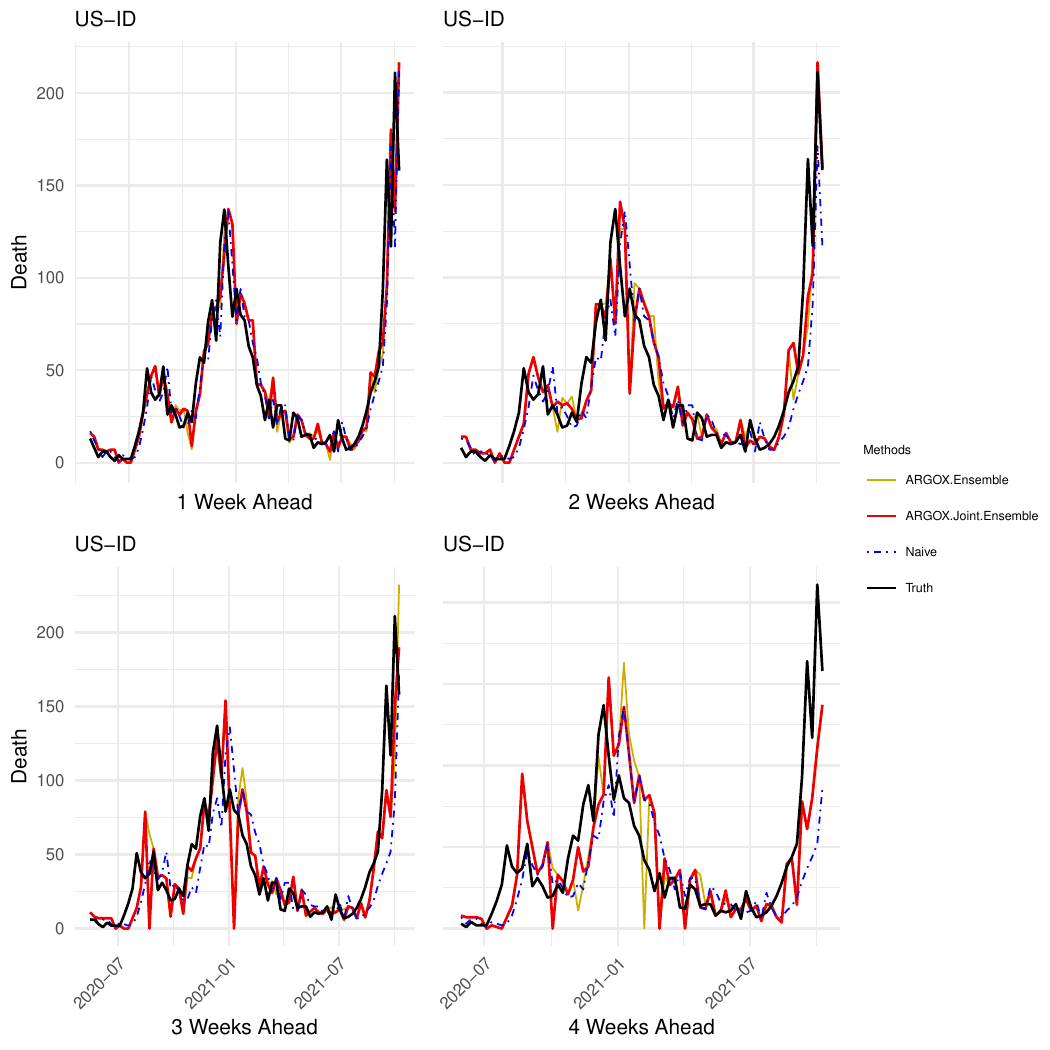} 
\caption{Plots of the COVID-19 1 week (top left), 2 weeks (top right), 3 weeks (bottom left), and 4 weeks (bottom right) ahead estimates for Idaho (ID). ARGOX-Ensemble is Ref \cite{ma2021covid}.}
\end{figure}
\newpage

\begin{table}[ht]
\centering
\begin{tabular}{lrrrr}
  \hline
& 1 Week Ahead & 2 Weeks Ahead & 3 Weeks Ahead & 4 Weeks Ahead \\ 
    \hline \multicolumn{1}{l}{RMSE} \\
 \hspace{1em} Ref \cite{ma2021covid} & 
82.54 & 111.98 & 137.87 & 163.03 \\ 
   \hspace{1em} ARGOX Joint Ensemble & 84.08 & 106.61 & 118.51 & 160.16 \\ 
   \hspace{1em} Naive & 83.65 & 131.78 & 180.93 & 225.68 \\ 
   \multicolumn{1}{l}{MAE} \\
  \hspace{1em}  Ref \cite{ma2021covid} &51.75 & 70.25 & 87.57 & 111.31 \\ 
   \hspace{1em} ARGOX Joint Ensemble & 52.28 & 66.81 & 77.44 & 110.11 \\ 
   \hspace{1em} Naive & 53.32 & 86.77 & 119.86 & 149.32 \\ 
   \multicolumn{1}{l}{Correlation} \\
 \hspace{1em} Ref \cite{ma2021covid} &0.97 & 0.95 & 0.93 & 0.89 \\ 
   \hspace{1em} ARGOX Joint Ensemble  & 0.96 & 0.95 & 0.94 & 0.88 \\ 
   \hspace{1em} Naive& 0.96 & 0.91 & 0.82 & 0.73 \\ 
   \hline
\end{tabular}
\caption{Comparison of different methods for state-level COVID-19 1 to 4 weeks ahead incremental death in Illinois (IL). The MSE, MAE, and correlation are reported and best performed method is highlighted in boldface.} 
\end{table}

\begin{figure}[!h] 
  \centering 
\includegraphics[width=0.6\linewidth, page=15]{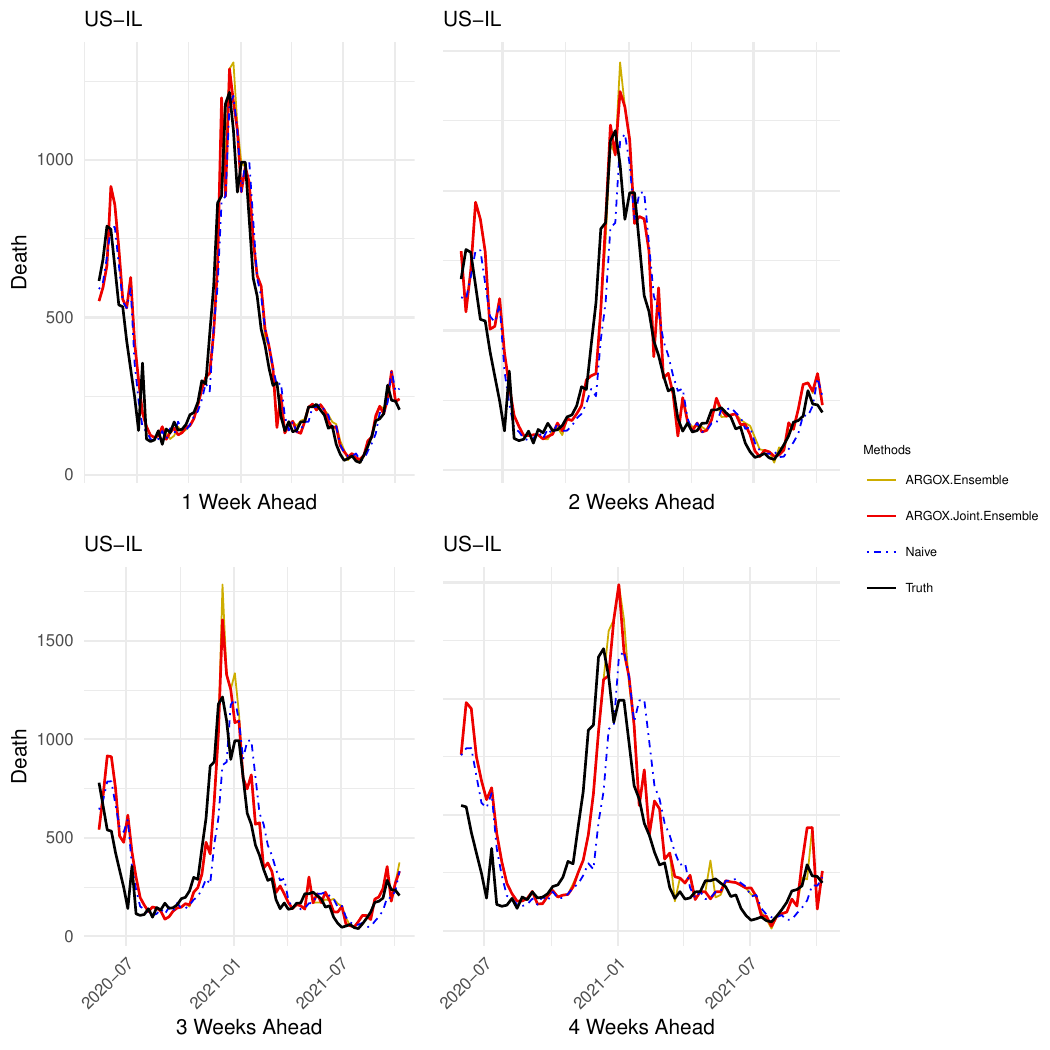} 
\caption{Plots of the COVID-19 1 week (top left), 2 weeks (top right), 3 weeks (bottom left), and 4 weeks (bottom right) ahead estimates for Illinois (IL). ARGOX-Ensemble is Ref \cite{ma2021covid}.}
\end{figure}
\newpage

\begin{table}[ht]
\centering
\begin{tabular}{lrrrr}
  \hline
& 1 Week Ahead & 2 Weeks Ahead & 3 Weeks Ahead & 4 Weeks Ahead \\ 
    \hline \multicolumn{1}{l}{RMSE} \\
 \hspace{1em} Ref \cite{ma2021covid} & 
155.55 & 170.55 & 159.16 & 191.62 \\ 
   \hspace{1em} ARGOX Joint Ensemble & 141.12 & 103.62 & 161.30 & 181.26 \\ 
   \hspace{1em} Naive & 188.50 & 219.46 & 230.13 & 248.88 \\ 
   \multicolumn{1}{l}{MAE} \\
  \hspace{1em}  Ref \cite{ma2021covid} &67.68 & 71.53 & 70.47 & 92.37 \\ 
   \hspace{1em} ARGOX Joint Ensemble & 58.35 & 55.18 & 68.20 & 87.04 \\ 
   \hspace{1em} Naive & 71.15 & 86.72 & 104.19 & 121.73 \\ 
   \multicolumn{1}{l}{Correlation} \\
 \hspace{1em} Ref \cite{ma2021covid} &0.74 & 0.70 & 0.72 & 0.65 \\ 
   \hspace{1em} ARGOX Joint Ensemble  & 0.75 & 0.85 & 0.74 & 0.68 \\ 
   \hspace{1em} Naive& 0.67 & 0.54 & 0.50 & 0.42 \\ 
   \hline
\end{tabular}
\caption{Comparison of different methods for state-level COVID-19 1 to 4 weeks ahead incremental death in Indiana (IN). The MSE, MAE, and correlation are reported and best performed method is highlighted in boldface.} 
\end{table}

\begin{figure}[!h] 
  \centering 
\includegraphics[width=0.6\linewidth, page=16]{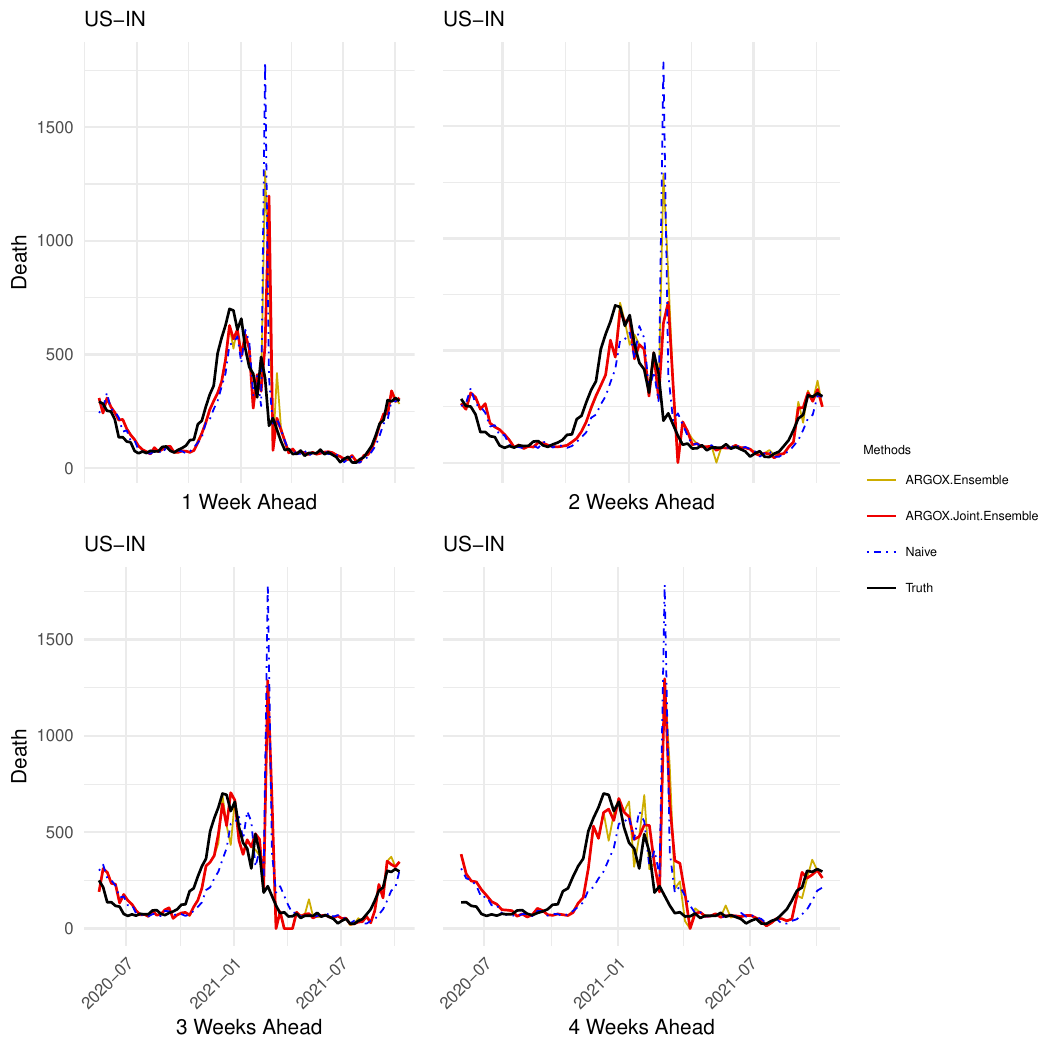} 
\caption{Plots of the COVID-19 1 week (top left), 2 weeks (top right), 3 weeks (bottom left), and 4 weeks (bottom right) ahead estimates for Indiana (IN). ARGOX-Ensemble is Ref \cite{ma2021covid}.}
\end{figure}
\newpage

\begin{table}[ht]
\centering
\begin{tabular}{lrrrr}
  \hline
& 1 Week Ahead & 2 Weeks Ahead & 3 Weeks Ahead & 4 Weeks Ahead \\ 
    \hline \multicolumn{1}{l}{RMSE} \\
 \hspace{1em} Ref \cite{ma2021covid} & 
47.29 & 58.79 & 62.15 & 58.67 \\ 
   \hspace{1em} ARGOX Joint Ensemble & 38.36 & 48.76 & 51.00 & 52.62 \\ 
   \hspace{1em} Naive & 62.10 & 67.10 & 81.06 & 79.74 \\ 
   \multicolumn{1}{l}{MAE} \\
  \hspace{1em}  Ref \cite{ma2021covid} &29.11 & 35.22 & 38.15 & 39.64 \\ 
   \hspace{1em} ARGOX Joint Ensemble & 24.10 & 32.43 & 33.58 & 39.34 \\ 
   \hspace{1em} Naive & 39.02 & 42.72 & 50.84 & 51.76 \\ 
   \multicolumn{1}{l}{Correlation} \\
 \hspace{1em} Ref \cite{ma2021covid} &0.87 & 0.81 & 0.78 & 0.81 \\ 
   \hspace{1em} ARGOX Joint Ensemble  & 0.91 & 0.87 & 0.84 & 0.84 \\ 
   \hspace{1em} Naive& 0.78 & 0.74 & 0.63 & 0.64 \\ 
   \hline
\end{tabular}
\caption{Comparison of different methods for state-level COVID-19 1 to 4 weeks ahead incremental death in Kansas (KS). The MSE, MAE, and correlation are reported and best performed method is highlighted in boldface.} 
\end{table}

\begin{figure}[!h] 
  \centering 
\includegraphics[width=0.6\linewidth, page=17]{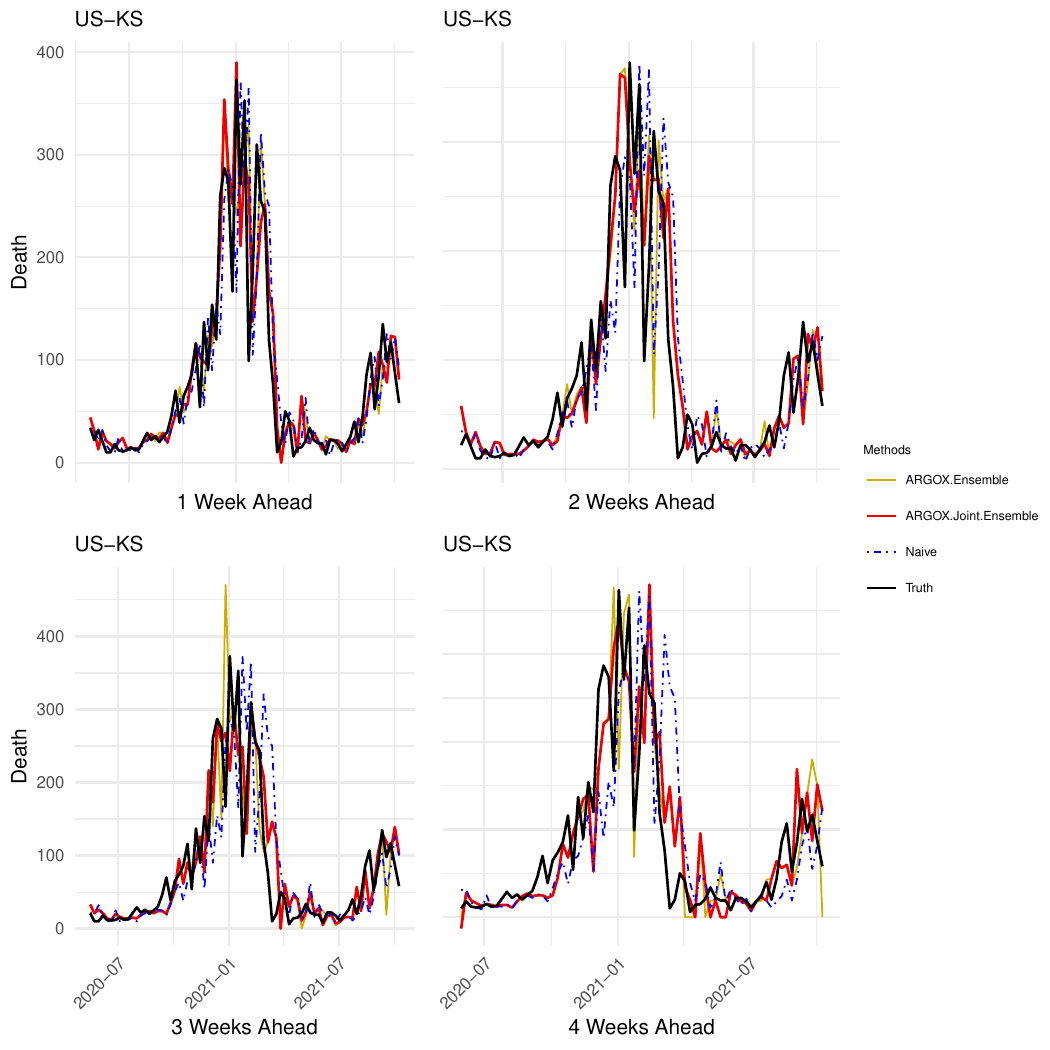} 
\caption{Plots of the COVID-19 1 week (top left), 2 weeks (top right), 3 weeks (bottom left), and 4 weeks (bottom right) ahead estimates for Kansas (KS). ARGOX-Ensemble is Ref \cite{ma2021covid}.}
\end{figure}
\newpage

\begin{table}[ht]
\centering
\begin{tabular}{lrrrr}
  \hline
& 1 Week Ahead & 2 Weeks Ahead & 3 Weeks Ahead & 4 Weeks Ahead \\ 
    \hline \multicolumn{1}{l}{RMSE} \\
 \hspace{1em} Ref \cite{ma2021covid} & 
98.73 & 107.30 & 108.61 & 98.52 \\ 
   \hspace{1em} ARGOX Joint Ensemble & 90.24 & 100.98 & 98.46 & 93.11 \\ 
   \hspace{1em} Naive & 100.35 & 116.41 & 120.34 & 130.72 \\ 
   \multicolumn{1}{l}{MAE} \\
  \hspace{1em}  Ref \cite{ma2021covid} &49.52 & 54.25 & 58.88 & 56.32 \\ 
   \hspace{1em} ARGOX Joint Ensemble & 41.82 & 50.44 & 53.77 & 52.65 \\ 
   \hspace{1em} Naive & 47.14 & 56.42 & 63.59 & 75.37 \\ 
   \multicolumn{1}{l}{Correlation} \\
 \hspace{1em} Ref \cite{ma2021covid} &0.61 & 0.51 & 0.52 & 0.58 \\ 
   \hspace{1em} ARGOX Joint Ensemble  & 0.66 & 0.55 & 0.59 & 0.63 \\ 
   \hspace{1em} Naive& 0.59 & 0.44 & 0.41 & 0.30 \\ 
   \hline
\end{tabular}
\caption{Comparison of different methods for state-level COVID-19 1 to 4 weeks ahead incremental death in Kentucky (KY). The MSE, MAE, and correlation are reported and best performed method is highlighted in boldface.} 
\end{table}

\begin{figure}[!h] 
  \centering 
\includegraphics[width=0.6\linewidth, page=18]{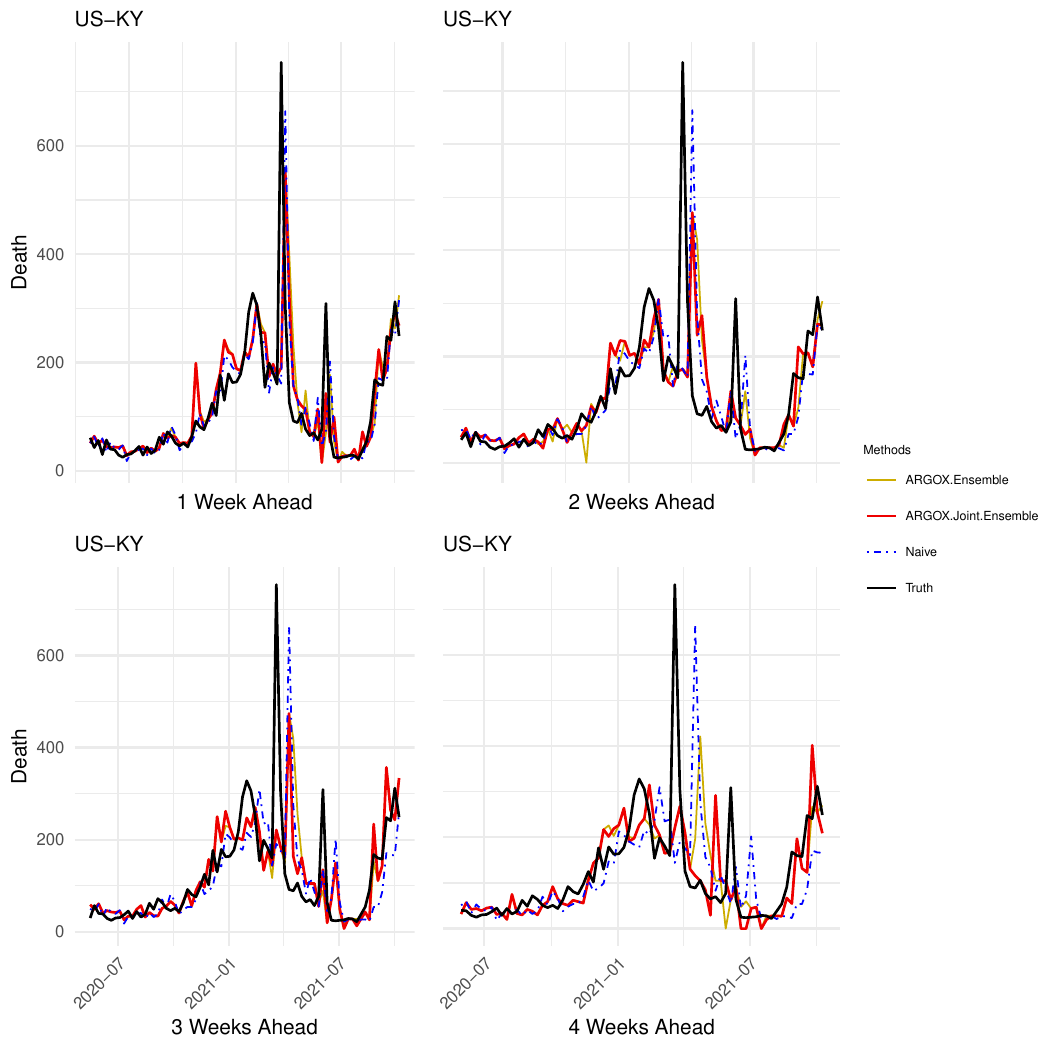} 
\caption{Plots of the COVID-19 1 week (top left), 2 weeks (top right), 3 weeks (bottom left), and 4 weeks (bottom right) ahead estimates for Kentucky (KY). ARGOX-Ensemble is Ref \cite{ma2021covid}.}
\end{figure}
\newpage

\begin{table}[ht]
\centering
\begin{tabular}{lrrrr}
  \hline
& 1 Week Ahead & 2 Weeks Ahead & 3 Weeks Ahead & 4 Weeks Ahead \\ 
    \hline \multicolumn{1}{l}{RMSE} \\
 \hspace{1em} Ref \cite{ma2021covid} & 
40.99 & 57.98 & 115.95 & 117.03 \\ 
   \hspace{1em} ARGOX Joint Ensemble & 31.98 & 53.40 & 519.37 & 109.64 \\ 
   \hspace{1em} Naive & 49.98 & 68.17 & 93.31 & 109.21 \\ 
   \multicolumn{1}{l}{MAE} \\
  \hspace{1em}  Ref \cite{ma2021covid} &27.61 & 37.22 & 64.61 & 76.66 \\ 
   \hspace{1em} ARGOX Joint Ensemble & 21.39 & 34.18 & 116.07 & 75.51 \\ 
   \hspace{1em} Naive & 35.35 & 47.28 & 65.06 & 79.43 \\ 
   \multicolumn{1}{l}{Correlation} \\
 \hspace{1em} Ref \cite{ma2021covid} &0.94 & 0.91 & 0.75 & 0.59 \\ 
   \hspace{1em} ARGOX Joint Ensemble  & 0.96 & 0.91 & 0.40 & 0.59 \\ 
   \hspace{1em} Naive& 0.89 & 0.80 & 0.63 & 0.49 \\ 
   \hline
\end{tabular}
\caption{Comparison of different methods for state-level COVID-19 1 to 4 weeks ahead incremental death in Louisiana (LA). The MSE, MAE, and correlation are reported and best performed method is highlighted in boldface.} 
\end{table}

\begin{figure}[!h] 
  \centering 
\includegraphics[width=0.6\linewidth, page=19]{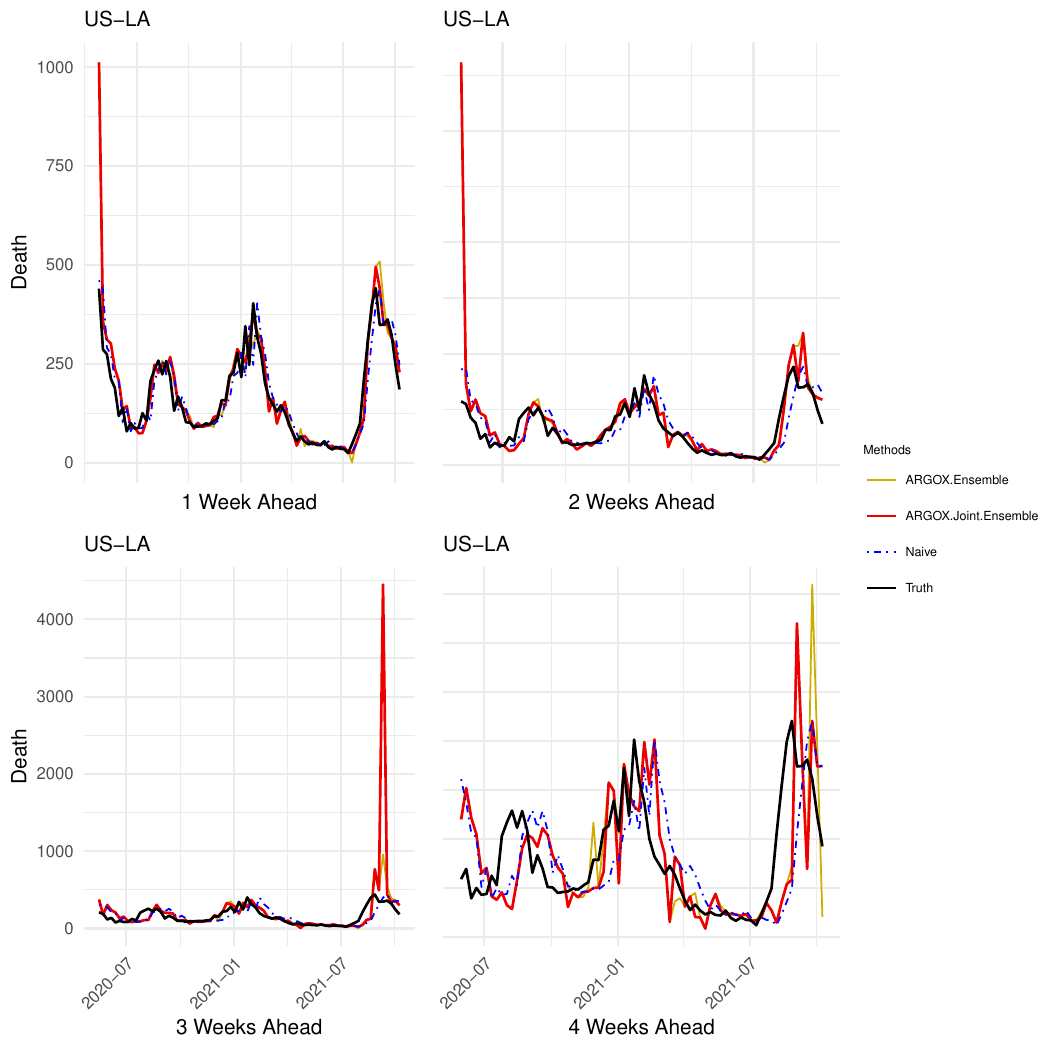} 
\caption{Plots of the COVID-19 1 week (top left), 2 weeks (top right), 3 weeks (bottom left), and 4 weeks (bottom right) ahead estimates for Louisiana (LA). ARGOX-Ensemble is Ref \cite{ma2021covid}.}
\end{figure}
\newpage

\begin{table}[ht]
\centering
\begin{tabular}{lrrrr}
  \hline
& 1 Week Ahead & 2 Weeks Ahead & 3 Weeks Ahead & 4 Weeks Ahead \\ 
    \hline \multicolumn{1}{l}{RMSE} \\
 \hspace{1em} Ref \cite{ma2021covid} & 
36.68 & 44.13 & 43.74 & 51.34 \\ 
   \hspace{1em} ARGOX Joint Ensemble & 34.68 & 38.24 & 38.36 & 52.31 \\ 
   \hspace{1em} Naive & 44.43 & 62.07 & 76.26 & 94.51 \\ 
   \multicolumn{1}{l}{MAE} \\
  \hspace{1em}  Ref \cite{ma2021covid} &25.83 & 26.72 & 31.14 & 36.67 \\ 
   \hspace{1em} ARGOX Joint Ensemble & 25.40 & 23.23 & 28.21 & 38.17 \\ 
   \hspace{1em} Naive & 30.52 & 39.08 & 53.77 & 66.81 \\ 
   \multicolumn{1}{l}{Correlation} \\
 \hspace{1em} Ref \cite{ma2021covid} &0.97 & 0.95 & 0.96 & 0.95 \\ 
   \hspace{1em} ARGOX Joint Ensemble  & 0.97 & 0.97 & 0.97 & 0.95 \\ 
   \hspace{1em} Naive& 0.95 & 0.91 & 0.86 & 0.79 \\ 
   \hline
\end{tabular}
\caption{Comparison of different methods for state-level COVID-19 1 to 4 weeks ahead incremental death in Massachusetts (MA). The MSE, MAE, and correlation are reported and best performed method is highlighted in boldface.} 
\end{table}

\begin{figure}[!h] 
  \centering 
\includegraphics[width=0.6\linewidth, page=20]{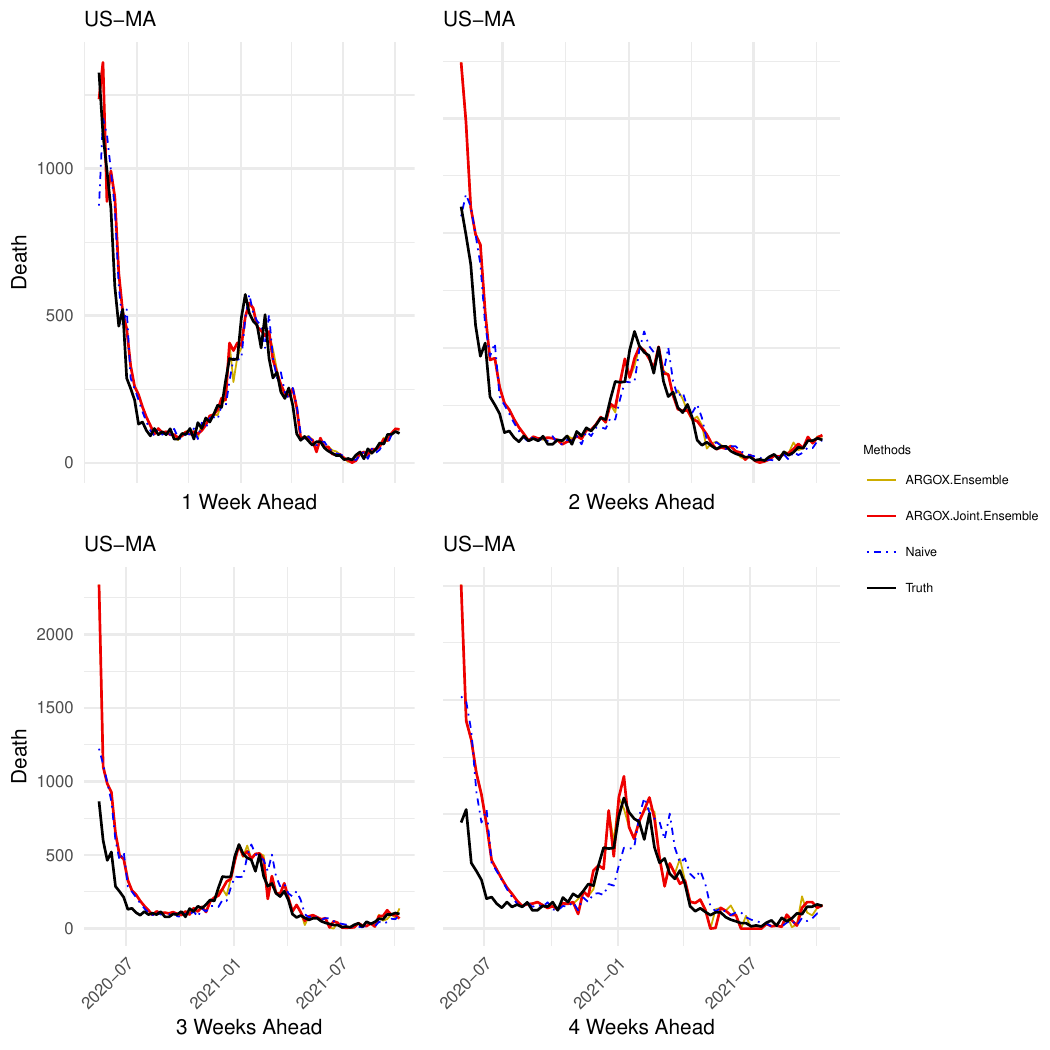} 
\caption{Plots of the COVID-19 1 week (top left), 2 weeks (top right), 3 weeks (bottom left), and 4 weeks (bottom right) ahead estimates for Massachusetts (MA). ARGOX-Ensemble is Ref \cite{ma2021covid}.}
\end{figure}
\newpage

\begin{table}[ht]
\centering
\begin{tabular}{lrrrr}
  \hline
& 1 Week Ahead & 2 Weeks Ahead & 3 Weeks Ahead & 4 Weeks Ahead \\ 
    \hline \multicolumn{1}{l}{RMSE} \\
 \hspace{1em} Ref \cite{ma2021covid} & 
88.07 & 85.55 & 85.69 & 96.05 \\ 
   \hspace{1em} ARGOX Joint Ensemble & 90.96 & 85.96 & 84.52 & 74.94 \\ 
   \hspace{1em} Naive & 99.48 & 102.23 & 110.19 & 114.35 \\ 
   \multicolumn{1}{l}{MAE} \\
  \hspace{1em}  Ref \cite{ma2021covid} &32.78 & 35.13 & 40.18 & 53.63 \\ 
   \hspace{1em} ARGOX Joint Ensemble & 33.48 & 36.25 & 40.10 & 39.19 \\ 
   \hspace{1em} Naive & 32.26 & 41.35 & 49.66 & 58.30 \\ 
   \multicolumn{1}{l}{Correlation} \\
 \hspace{1em} Ref \cite{ma2021covid} &0.61 & 0.63 & 0.64 & 0.63 \\ 
   \hspace{1em} ARGOX Joint Ensemble  & 0.59 & 0.64 & 0.66 & 0.74 \\ 
   \hspace{1em} Naive& 0.53 & 0.51 & 0.44 & 0.41 \\ 
   \hline
\end{tabular}
\caption{Comparison of different methods for state-level COVID-19 1 to 4 weeks ahead incremental death in Maryland (MD). The MSE, MAE, and correlation are reported and best performed method is highlighted in boldface.} 
\end{table}

\begin{figure}[!h] 
  \centering 
\includegraphics[width=0.6\linewidth, page=21]{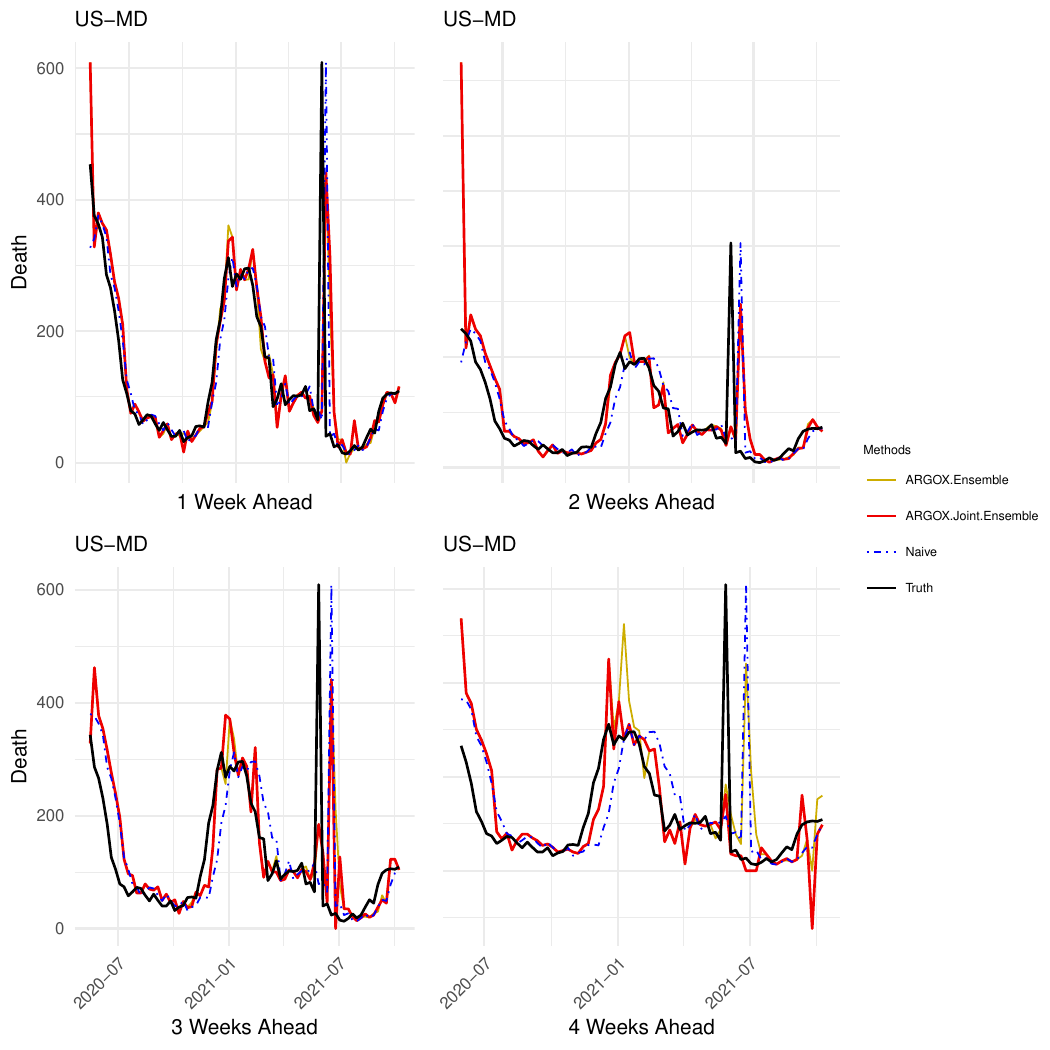} 
\caption{Plots of the COVID-19 1 week (top left), 2 weeks (top right), 3 weeks (bottom left), and 4 weeks (bottom right) ahead estimates for Maryland (MD). ARGOX-Ensemble is Ref \cite{ma2021covid}.}
\end{figure}
\newpage

\begin{table}[ht]
\centering
\begin{tabular}{lrrrr}
  \hline
& 1 Week Ahead & 2 Weeks Ahead & 3 Weeks Ahead & 4 Weeks Ahead \\ 
    \hline \multicolumn{1}{l}{RMSE} \\
 \hspace{1em} Ref \cite{ma2021covid} & 
12.04 & 10.91 & 10.20 & 16.31 \\ 
   \hspace{1em} ARGOX Joint Ensemble & 11.91 & 11.53 & 10.80 & 15.59 \\ 
   \hspace{1em} Naive & 13.23 & 14.41 & 15.14 & 16.99 \\ 
   \multicolumn{1}{l}{MAE} \\
  \hspace{1em}  Ref \cite{ma2021covid} &7.78 & 7.40 & 7.51 & 10.51 \\ 
   \hspace{1em} ARGOX Joint Ensemble & 7.67 & 7.48 & 7.46 & 10.04 \\ 
   \hspace{1em} Naive & 8.52 & 9.38 & 10.25 & 11.00 \\ 
   \multicolumn{1}{l}{Correlation} \\
 \hspace{1em} Ref \cite{ma2021covid} &0.74 & 0.78 & 0.81 & 0.63 \\ 
   \hspace{1em} ARGOX Joint Ensemble  & 0.74 & 0.78 & 0.80 & 0.65 \\ 
   \hspace{1em} Naive& 0.68 & 0.62 & 0.59 & 0.48 \\ 
   \hline
\end{tabular}
\caption{Comparison of different methods for state-level COVID-19 1 to 4 weeks ahead incremental death in Maine (ME). The MSE, MAE, and correlation are reported and best performed method is highlighted in boldface.} 
\end{table}

\begin{figure}[!h] 
  \centering 
\includegraphics[width=0.6\linewidth, page=22]{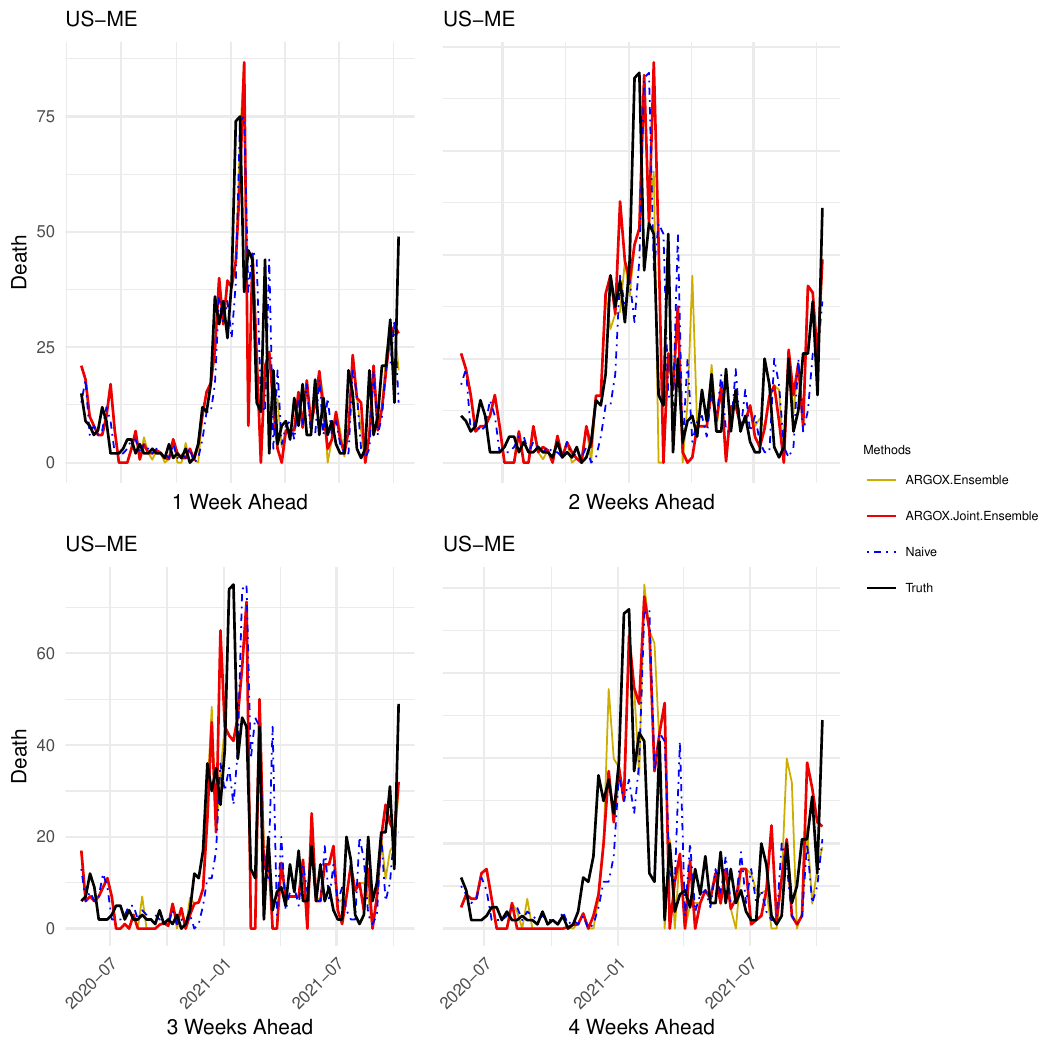} 
\caption{Plots of the COVID-19 1 week (top left), 2 weeks (top right), 3 weeks (bottom left), and 4 weeks (bottom right) ahead estimates for Maine (ME). ARGOX-Ensemble is Ref \cite{ma2021covid}.}
\end{figure}
\newpage

\begin{table}[ht]
\centering
\begin{tabular}{lrrrr}
  \hline
& 1 Week Ahead & 2 Weeks Ahead & 3 Weeks Ahead & 4 Weeks Ahead \\ 
    \hline \multicolumn{1}{l}{RMSE} \\
 \hspace{1em} Ref \cite{ma2021covid} & 
73.37 & 84.83 & 112.44 & 157.83 \\ 
   \hspace{1em} ARGOX Joint Ensemble & 69.08 & 83.70 & 101.94 & 144.50 \\ 
   \hspace{1em} Naive & 84.54 & 125.09 & 164.92 & 204.31 \\ 
   \multicolumn{1}{l}{MAE} \\
  \hspace{1em}  Ref \cite{ma2021covid} &44.87 & 61.70 & 81.72 & 112.77 \\ 
   \hspace{1em} ARGOX Joint Ensemble & 43.02 & 60.80 & 76.38 & 106.76 \\ 
   \hspace{1em} Naive & 51.65 & 89.28 & 122.72 & 151.48 \\ 
   \multicolumn{1}{l}{Correlation} \\
 \hspace{1em} Ref \cite{ma2021covid} &0.95 & 0.95 & 0.92 & 0.83 \\ 
   \hspace{1em} ARGOX Joint Ensemble  & 0.96 & 0.95 & 0.92 & 0.85 \\ 
   \hspace{1em} Naive& 0.93 & 0.85 & 0.74 & 0.60 \\ 
   \hline
\end{tabular}
\caption{Comparison of different methods for state-level COVID-19 1 to 4 weeks ahead incremental death in Michigan (MI). The MSE, MAE, and correlation are reported and best performed method is highlighted in boldface.} 
\end{table}

\begin{figure}[!h] 
  \centering 
\includegraphics[width=0.6\linewidth, page=23]{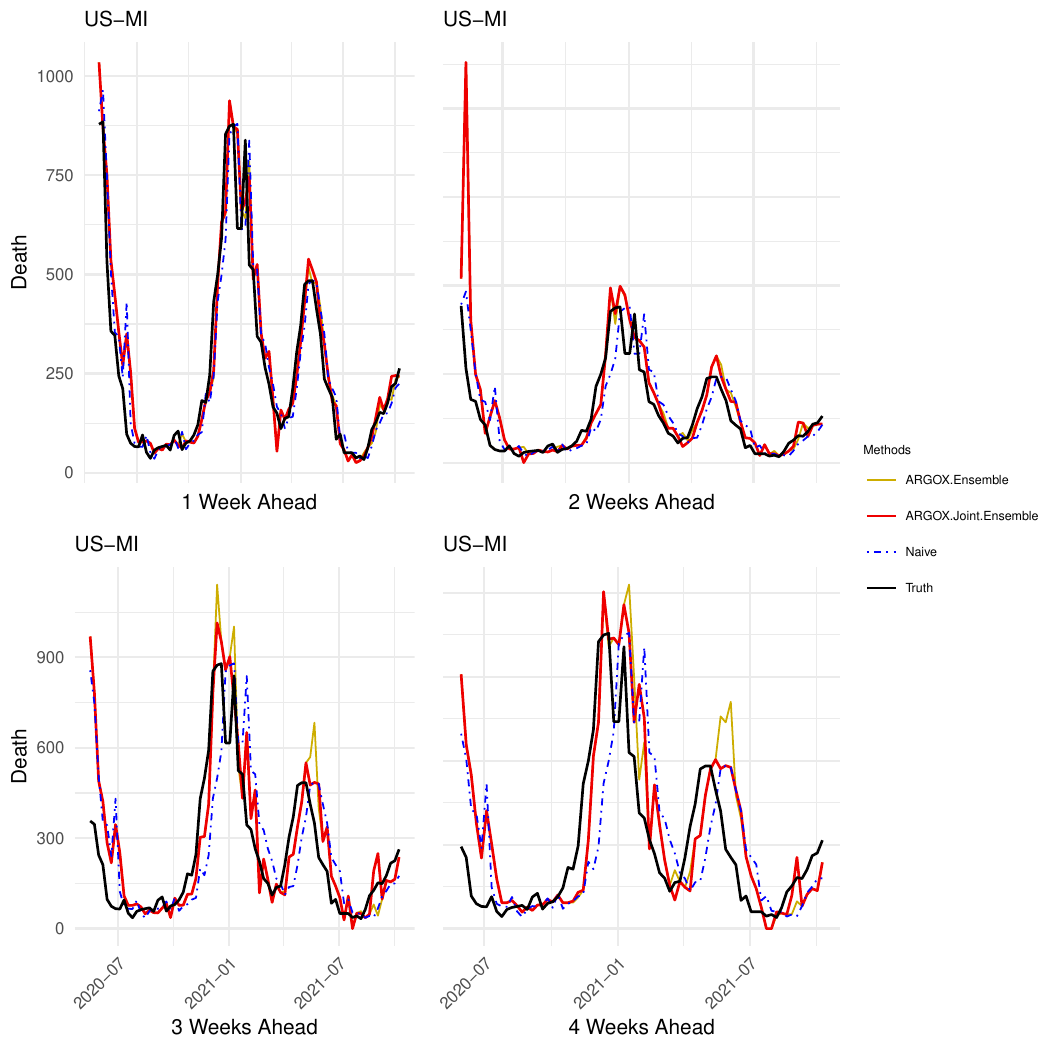} 
\caption{Plots of the COVID-19 1 week (top left), 2 weeks (top right), 3 weeks (bottom left), and 4 weeks (bottom right) ahead estimates for Michigan (MI). ARGOX-Ensemble is Ref \cite{ma2021covid}.}
\end{figure}
\newpage

\begin{table}[ht]
\centering
\begin{tabular}{lrrrr}
  \hline
& 1 Week Ahead & 2 Weeks Ahead & 3 Weeks Ahead & 4 Weeks Ahead \\ 
    \hline \multicolumn{1}{l}{RMSE} \\
 \hspace{1em} Ref \cite{ma2021covid} & 
36.44 & 39.92 & 48.81 & 93.10 \\ 
   \hspace{1em} ARGOX Joint Ensemble & 35.46 & 40.82 & 47.15 & 70.22 \\ 
   \hspace{1em} Naive & 39.22 & 56.08 & 72.58 & 88.78 \\ 
   \multicolumn{1}{l}{MAE} \\
  \hspace{1em}  Ref \cite{ma2021covid} &22.90 & 25.37 & 32.06 & 49.23 \\ 
   \hspace{1em} ARGOX Joint Ensemble & 21.89 & 26.47 & 30.92 & 45.25 \\ 
   \hspace{1em} Naive & 25.70 & 37.75 & 50.34 & 61.56 \\ 
   \multicolumn{1}{l}{Correlation} \\
 \hspace{1em} Ref \cite{ma2021covid} &0.95 & 0.93 & 0.92 & 0.84 \\ 
   \hspace{1em} ARGOX Joint Ensemble  & 0.95 & 0.93 & 0.93 & 0.87 \\ 
   \hspace{1em} Naive& 0.93 & 0.86 & 0.77 & 0.65 \\ 
   \hline
\end{tabular}
\caption{Comparison of different methods for state-level COVID-19 1 to 4 weeks ahead incremental death in Minnesota (MN). The MSE, MAE, and correlation are reported and best performed method is highlighted in boldface.} 
\end{table}

\begin{figure}[!h] 
  \centering 
\includegraphics[width=0.6\linewidth, page=24]{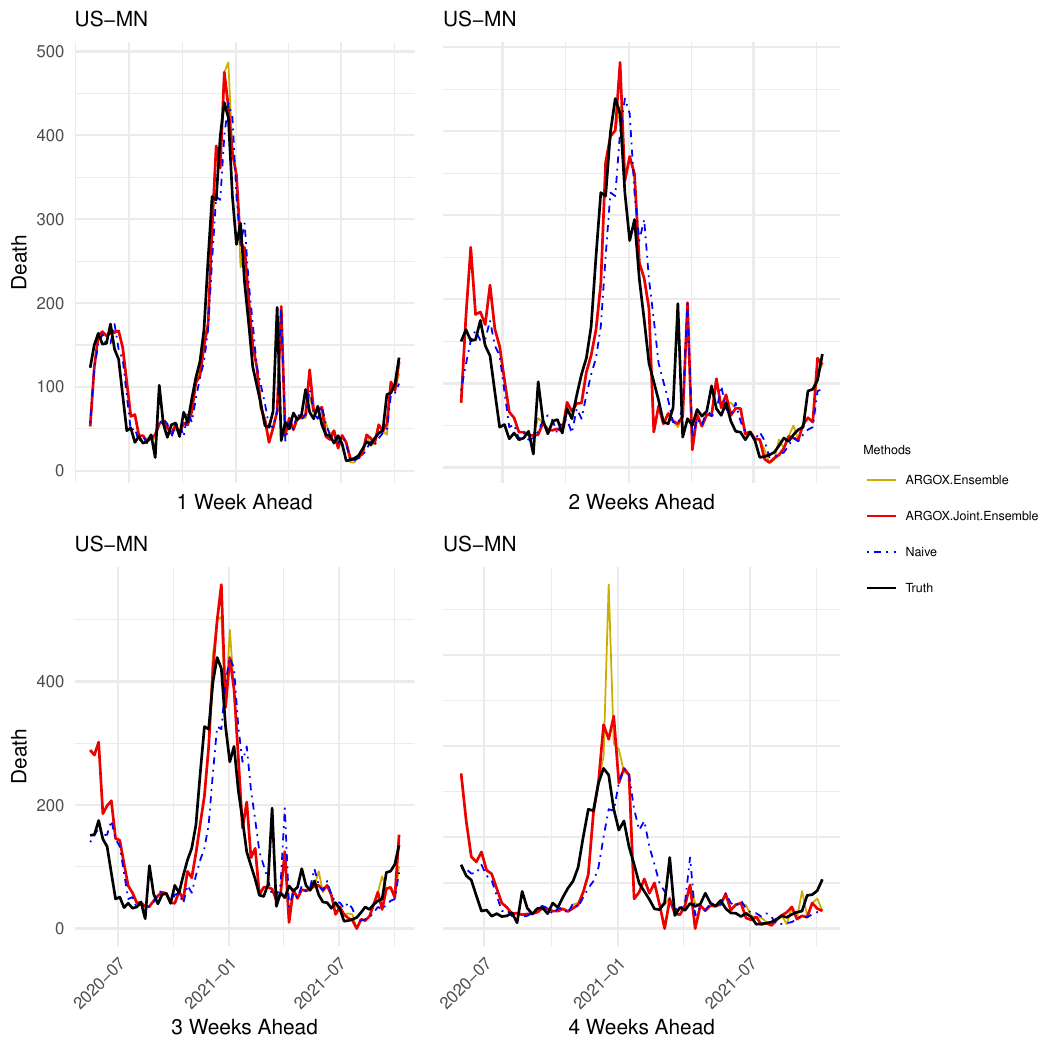} 
\caption{Plots of the COVID-19 1 week (top left), 2 weeks (top right), 3 weeks (bottom left), and 4 weeks (bottom right) ahead estimates for Minnesota (MN). ARGOX-Ensemble is Ref \cite{ma2021covid}.}
\end{figure}
\newpage

\begin{table}[ht]
\centering
\begin{tabular}{lrrrr}
  \hline
& 1 Week Ahead & 2 Weeks Ahead & 3 Weeks Ahead & 4 Weeks Ahead \\ 
    \hline \multicolumn{1}{l}{RMSE} \\
 \hspace{1em} Ref \cite{ma2021covid} & 
64.59 & 63.21 & 75.17 & 87.23 \\ 
   \hspace{1em} ARGOX Joint Ensemble & 64.92 & 53.90 & 63.99 & 83.27 \\ 
   \hspace{1em} Naive & 71.14 & 71.80 & 81.23 & 99.58 \\ 
   \multicolumn{1}{l}{MAE} \\
  \hspace{1em}  Ref \cite{ma2021covid} &48.53 & 45.78 & 56.49 & 66.42 \\ 
   \hspace{1em} ARGOX Joint Ensemble & 46.70 & 40.24 & 48.27 & 63.09 \\ 
   \hspace{1em} Naive & 55.08 & 53.98 & 62.95 & 73.38 \\ 
   \multicolumn{1}{l}{Correlation} \\
 \hspace{1em} Ref \cite{ma2021covid} &0.85 & 0.86 & 0.77 & 0.69 \\ 
   \hspace{1em} ARGOX Joint Ensemble  & 0.86 & 0.90 & 0.85 & 0.77 \\ 
   \hspace{1em} Naive& 0.80 & 0.80 & 0.74 & 0.62 \\ 
   \hline
\end{tabular}
\caption{Comparison of different methods for state-level COVID-19 1 to 4 weeks ahead incremental death in Missouri (MO). The MSE, MAE, and correlation are reported and best performed method is highlighted in boldface.} 
\end{table}

\begin{figure}[!h] 
  \centering 
\includegraphics[width=0.6\linewidth, page=25]{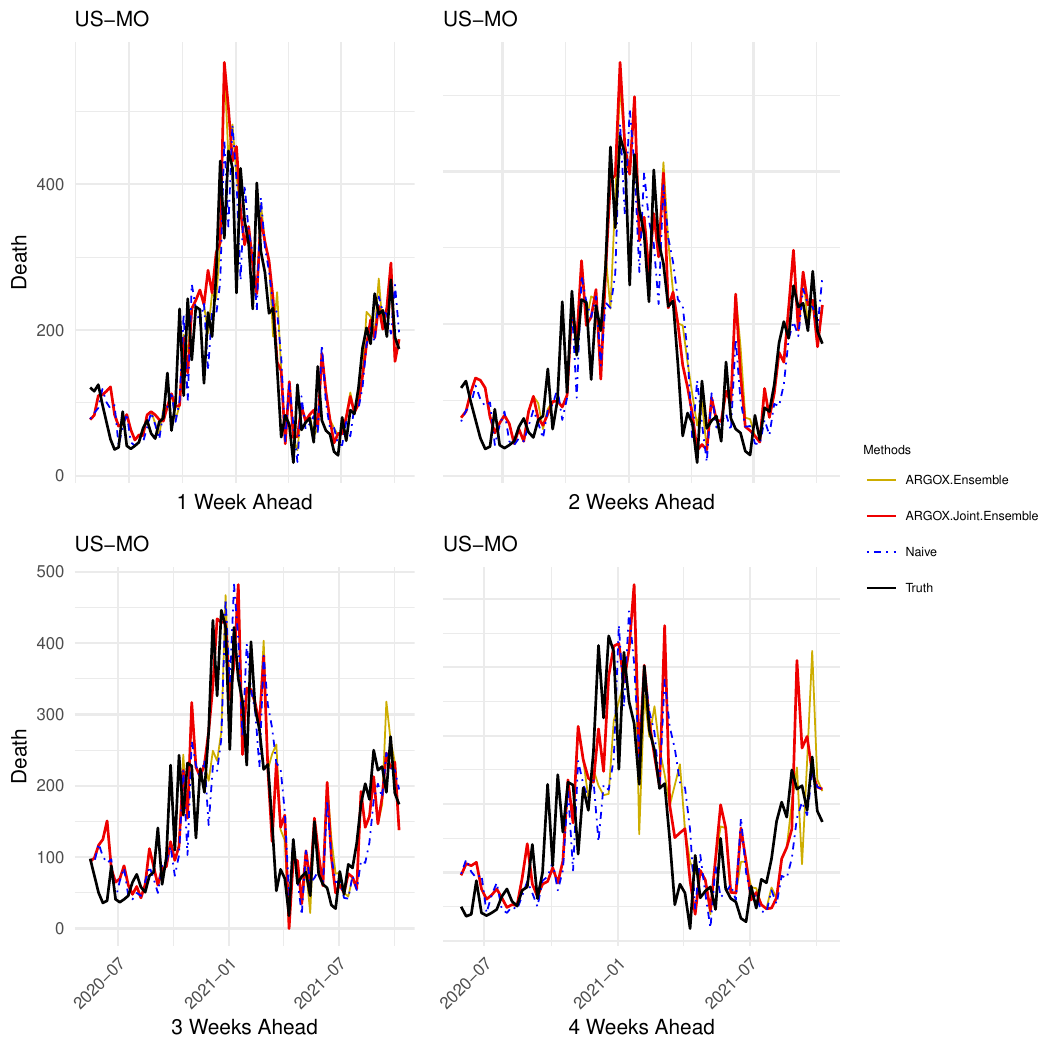} 
\caption{Plots of the COVID-19 1 week (top left), 2 weeks (top right), 3 weeks (bottom left), and 4 weeks (bottom right) ahead estimates for Missouri (MO). ARGOX-Ensemble is Ref \cite{ma2021covid}.}
\end{figure}
\newpage

\begin{table}[ht]
\centering
\begin{tabular}{lrrrr}
  \hline
& 1 Week Ahead & 2 Weeks Ahead & 3 Weeks Ahead & 4 Weeks Ahead \\ 
    \hline \multicolumn{1}{l}{RMSE} \\
 \hspace{1em} Ref \cite{ma2021covid} & 
35.00 & 53.11 & 72.28 & 108.11 \\ 
   \hspace{1em} ARGOX Joint Ensemble & 35.37 & 51.04 & 59.97 & 86.30 \\ 
   \hspace{1em} Naive & 40.40 & 58.56 & 74.68 & 89.73 \\ 
   \multicolumn{1}{l}{MAE} \\
  \hspace{1em}  Ref \cite{ma2021covid} &25.96 & 36.21 & 45.86 & 60.95 \\ 
   \hspace{1em} ARGOX Joint Ensemble & 23.79 & 36.02 & 40.50 & 56.13 \\ 
   \hspace{1em} Naive & 29.53 & 41.28 & 54.19 & 66.81 \\ 
   \multicolumn{1}{l}{Correlation} \\
 \hspace{1em} Ref \cite{ma2021covid} &0.94 & 0.88 & 0.81 & 0.68 \\ 
   \hspace{1em} ARGOX Joint Ensemble  & 0.94 & 0.88 & 0.83 & 0.73 \\ 
   \hspace{1em} Naive& 0.90 & 0.80 & 0.68 & 0.53 \\ 
   \hline
\end{tabular}
\caption{Comparison of different methods for state-level COVID-19 1 to 4 weeks ahead incremental death in Mississippi (MS). The MSE, MAE, and correlation are reported and best performed method is highlighted in boldface.} 
\end{table}

\begin{figure}[!h] 
  \centering 
\includegraphics[width=0.6\linewidth, page=26]{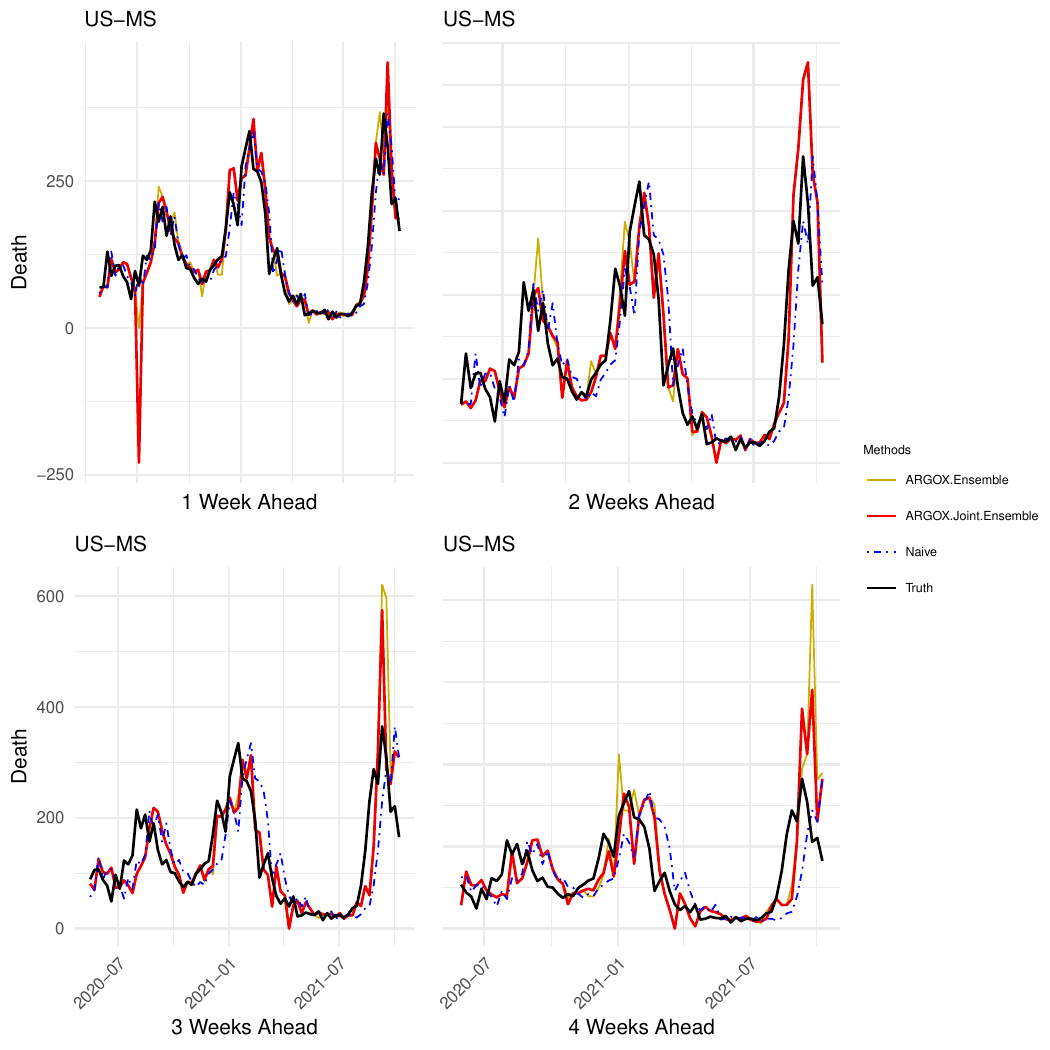} 
\caption{Plots of the COVID-19 1 week (top left), 2 weeks (top right), 3 weeks (bottom left), and 4 weeks (bottom right) ahead estimates for Mississippi (MS). ARGOX-Ensemble is Ref \cite{ma2021covid}.}
\end{figure}
\newpage

\begin{table}[ht]
\centering
\begin{tabular}{lrrrr}
  \hline
& 1 Week Ahead & 2 Weeks Ahead & 3 Weeks Ahead & 4 Weeks Ahead \\ 
    \hline \multicolumn{1}{l}{RMSE} \\
 \hspace{1em} Ref \cite{ma2021covid} & 
15.60 & 17.69 & 17.62 & 20.93 \\ 
   \hspace{1em} ARGOX Joint Ensemble & 14.14 & 17.92 & 15.61 & 18.72 \\ 
   \hspace{1em} Naive & 16.56 & 21.78 & 23.16 & 24.47 \\ 
   \multicolumn{1}{l}{MAE} \\
  \hspace{1em}  Ref \cite{ma2021covid} &11.59 & 12.85 & 12.63 & 14.80 \\ 
   \hspace{1em} ARGOX Joint Ensemble & 10.08 & 12.76 & 10.80 & 13.34 \\ 
   \hspace{1em} Naive & 11.02 & 14.92 & 15.20 & 16.78 \\ 
   \multicolumn{1}{l}{Correlation} \\
 \hspace{1em} Ref \cite{ma2021covid} &0.81 & 0.77 & 0.80 & 0.70 \\ 
   \hspace{1em} ARGOX Joint Ensemble  & 0.84 & 0.75 & 0.84 & 0.76 \\ 
   \hspace{1em} Naive& 0.79 & 0.64 & 0.59 & 0.55 \\ 
   \hline
\end{tabular}
\caption{Comparison of different methods for state-level COVID-19 1 to 4 weeks ahead incremental death in Montana (MT). The MSE, MAE, and correlation are reported and best performed method is highlighted in boldface.} 
\end{table}

\begin{figure}[!h] 
  \centering 
\includegraphics[width=0.6\linewidth, page=27]{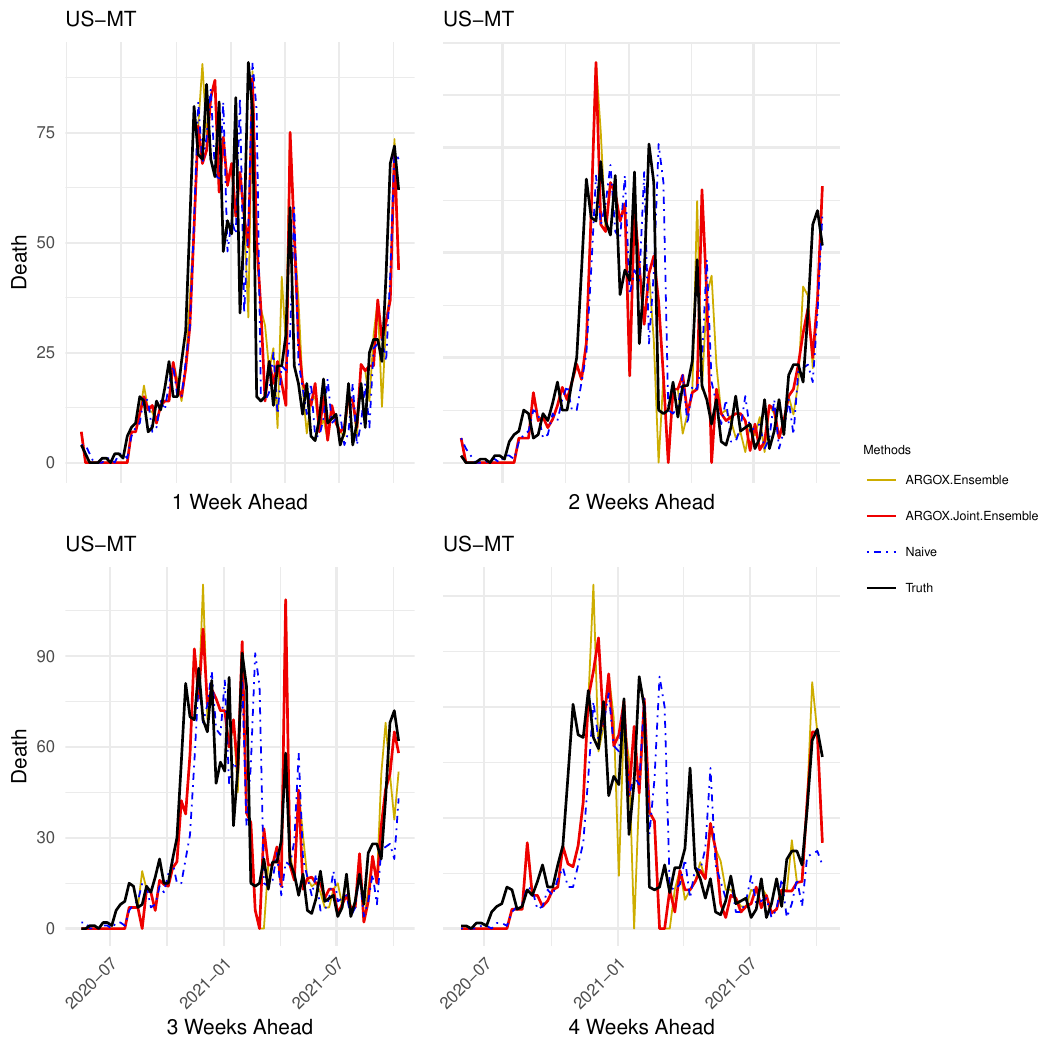} 
\caption{Plots of the COVID-19 1 week (top left), 2 weeks (top right), 3 weeks (bottom left), and 4 weeks (bottom right) ahead estimates for Montana (MT). ARGOX-Ensemble is Ref \cite{ma2021covid}.}
\end{figure}
\newpage

\begin{table}[ht]
\centering
\begin{tabular}{lrrrr}
  \hline
& 1 Week Ahead & 2 Weeks Ahead & 3 Weeks Ahead & 4 Weeks Ahead \\ 
    \hline \multicolumn{1}{l}{RMSE} \\
 \hspace{1em} Ref \cite{ma2021covid} & 
58.62 & 71.71 & 84.85 & 107.76 \\ 
   \hspace{1em} ARGOX Joint Ensemble & 55.37 & 66.77 & 76.89 & 108.80 \\ 
   \hspace{1em} Naive & 66.62 & 88.81 & 120.20 & 147.81 \\ 
   \multicolumn{1}{l}{MAE} \\
  \hspace{1em}  Ref \cite{ma2021covid} &44.22 & 48.86 & 58.63 & 71.13 \\ 
   \hspace{1em} ARGOX Joint Ensemble & 40.78 & 45.01 & 51.94 & 71.52 \\ 
   \hspace{1em} Naive & 47.38 & 64.29 & 83.75 & 105.03 \\ 
   \multicolumn{1}{l}{Correlation} \\
 \hspace{1em} Ref \cite{ma2021covid} &0.94 & 0.91 & 0.87 & 0.79 \\ 
   \hspace{1em} ARGOX Joint Ensemble  & 0.94 & 0.92 & 0.89 & 0.82 \\ 
   \hspace{1em} Naive& 0.92 & 0.85 & 0.73 & 0.59 \\ 
   \hline
\end{tabular}
\caption{Comparison of different methods for state-level COVID-19 1 to 4 weeks ahead incremental death in North Carolina (NC). The MSE, MAE, and correlation are reported and best performed method is highlighted in boldface.} 
\end{table}

\begin{figure}[!h] 
  \centering 
\includegraphics[width=0.6\linewidth, page=28]{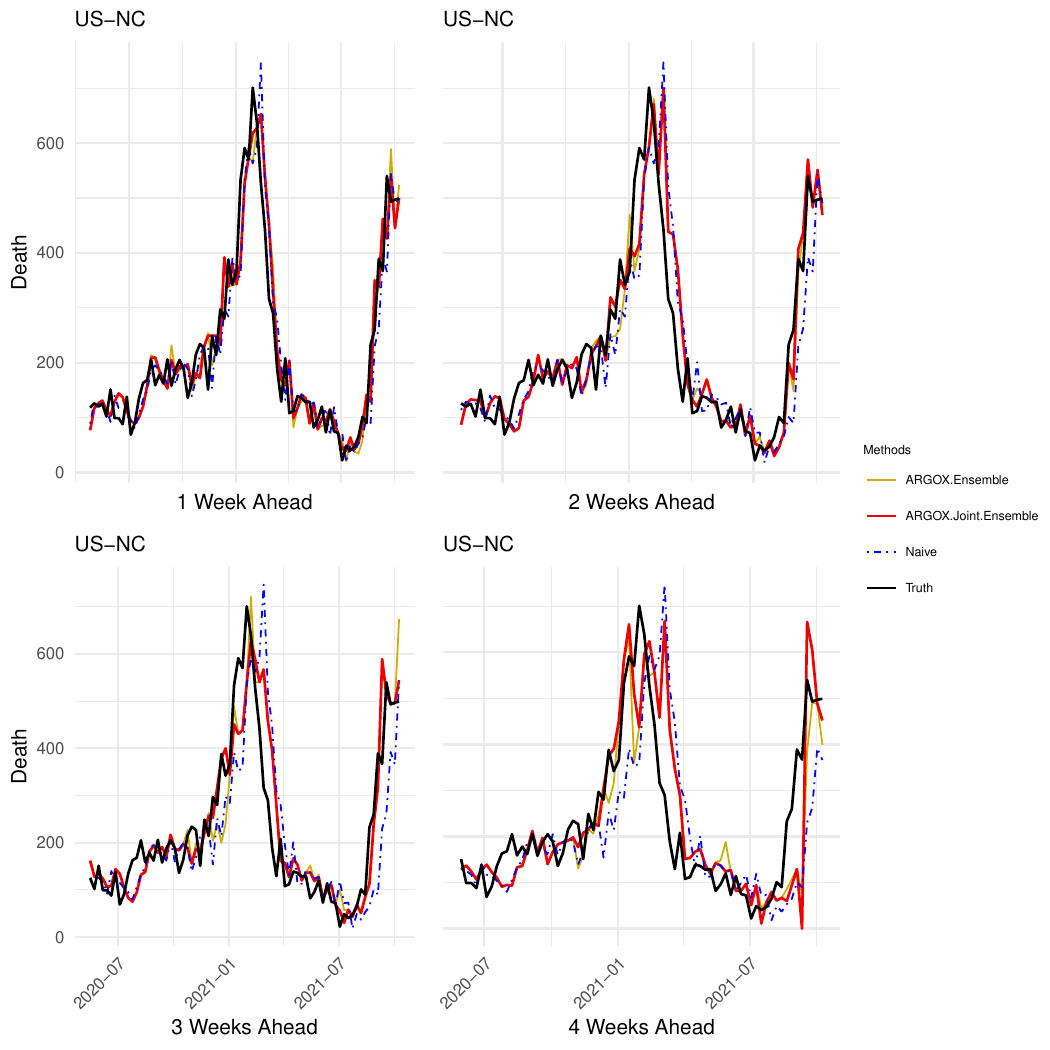} 
\caption{Plots of the COVID-19 1 week (top left), 2 weeks (top right), 3 weeks (bottom left), and 4 weeks (bottom right) ahead estimates for North Carolina (NC). ARGOX-Ensemble is Ref \cite{ma2021covid}.}
\end{figure}
\newpage

\begin{table}[ht]
\centering
\begin{tabular}{lrrrr}
  \hline
& 1 Week Ahead & 2 Weeks Ahead & 3 Weeks Ahead & 4 Weeks Ahead \\ 
    \hline \multicolumn{1}{l}{RMSE} \\
 \hspace{1em} Ref \cite{ma2021covid} & 
12.77 & 16.36 & 17.37 & 22.83 \\ 
   \hspace{1em} ARGOX Joint Ensemble & 13.05 & 15.95 & 16.72 & 26.25 \\ 
   \hspace{1em} Naive & 14.58 & 20.11 & 21.25 & 24.54 \\ 
   \multicolumn{1}{l}{MAE} \\
  \hspace{1em}  Ref \cite{ma2021covid} &7.42 & 8.50 & 11.08 & 12.80 \\ 
   \hspace{1em} ARGOX Joint Ensemble & 7.54 & 7.83 & 10.43 & 14.75 \\ 
   \hspace{1em} Naive & 8.36 & 10.28 & 12.61 & 14.97 \\ 
   \multicolumn{1}{l}{Correlation} \\
 \hspace{1em} Ref \cite{ma2021covid} &0.93 & 0.89 & 0.90 & 0.80 \\ 
   \hspace{1em} ARGOX Joint Ensemble  & 0.92 & 0.89 & 0.88 & 0.74 \\ 
   \hspace{1em} Naive& 0.90 & 0.81 & 0.79 & 0.72 \\ 
   \hline
\end{tabular}
\caption{Comparison of different methods for state-level COVID-19 1 to 4 weeks ahead incremental death in North Dakota (ND). The MSE, MAE, and correlation are reported and best performed method is highlighted in boldface.} 
\end{table}

\begin{figure}[!h] 
  \centering 
\includegraphics[width=0.6\linewidth, page=29]{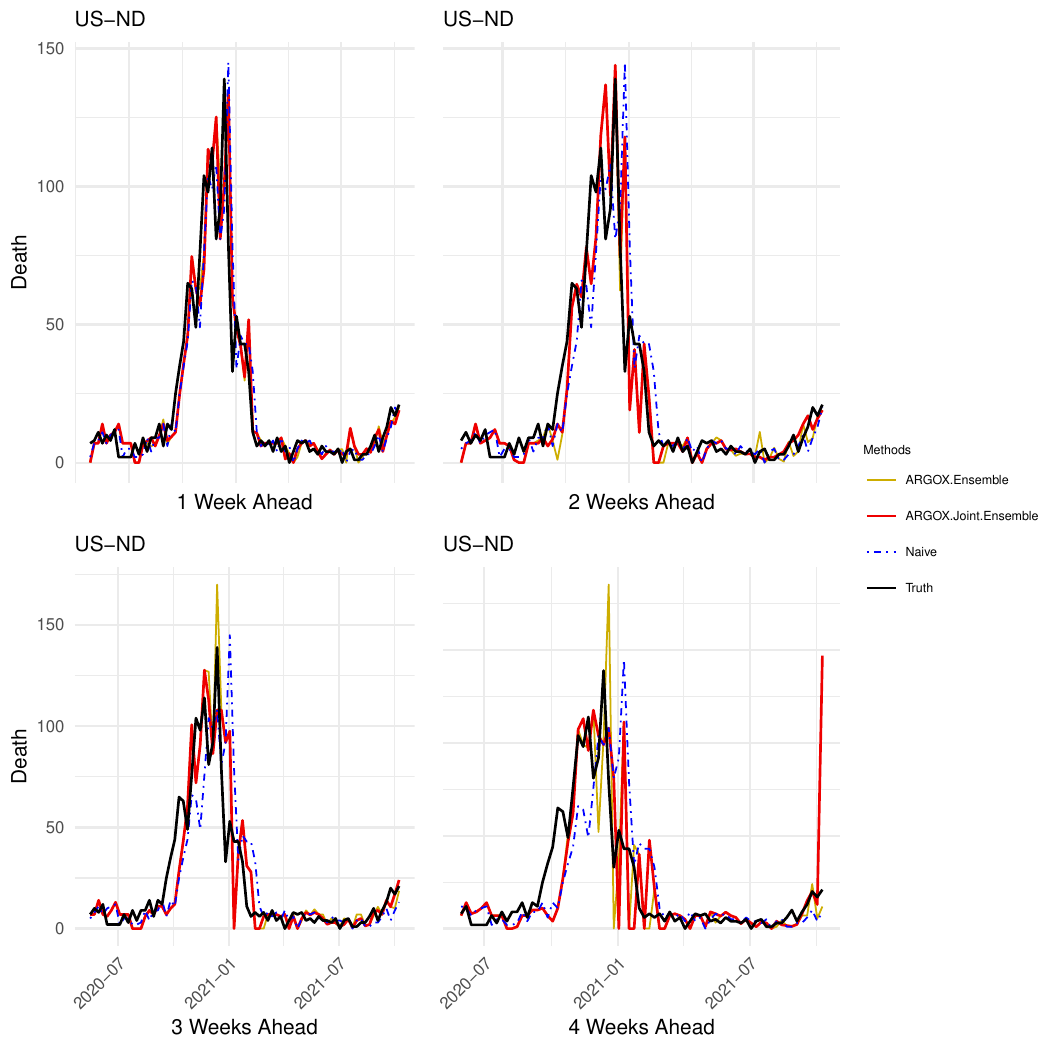} 
\caption{Plots of the COVID-19 1 week (top left), 2 weeks (top right), 3 weeks (bottom left), and 4 weeks (bottom right) ahead estimates for North Dakota (ND). ARGOX-Ensemble is Ref \cite{ma2021covid}.}
\end{figure}
\newpage

\begin{table}[ht]
\centering
\begin{tabular}{lrrrr}
  \hline
& 1 Week Ahead & 2 Weeks Ahead & 3 Weeks Ahead & 4 Weeks Ahead \\ 
    \hline \multicolumn{1}{l}{RMSE} \\
 \hspace{1em} Ref \cite{ma2021covid} & 
25.64 & 25.71 & 33.19 & 30.49 \\ 
   \hspace{1em} ARGOX Joint Ensemble & 25.93 & 27.23 & 33.20 & 28.90 \\ 
   \hspace{1em} Naive & 30.86 & 34.22 & 38.72 & 42.74 \\ 
   \multicolumn{1}{l}{MAE} \\
  \hspace{1em}  Ref \cite{ma2021covid} &16.73 & 17.33 & 21.83 & 21.18 \\ 
   \hspace{1em} ARGOX Joint Ensemble & 17.01 & 18.34 & 21.98 & 20.25 \\ 
   \hspace{1em} Naive & 19.03 & 20.09 & 24.42 & 27.33 \\ 
   \multicolumn{1}{l}{Correlation} \\
 \hspace{1em} Ref \cite{ma2021covid} &0.84 & 0.85 & 0.81 & 0.80 \\ 
   \hspace{1em} ARGOX Joint Ensemble  & 0.83 & 0.81 & 0.82 & 0.83 \\ 
   \hspace{1em} Naive& 0.75 & 0.69 & 0.61 & 0.52 \\ 
   \hline
\end{tabular}
\caption{Comparison of different methods for state-level COVID-19 1 to 4 weeks ahead incremental death in Nebraska (NE). The MSE, MAE, and correlation are reported and best performed method is highlighted in boldface.} 
\end{table}

\begin{figure}[!h] 
  \centering 
\includegraphics[width=0.6\linewidth, page=30]{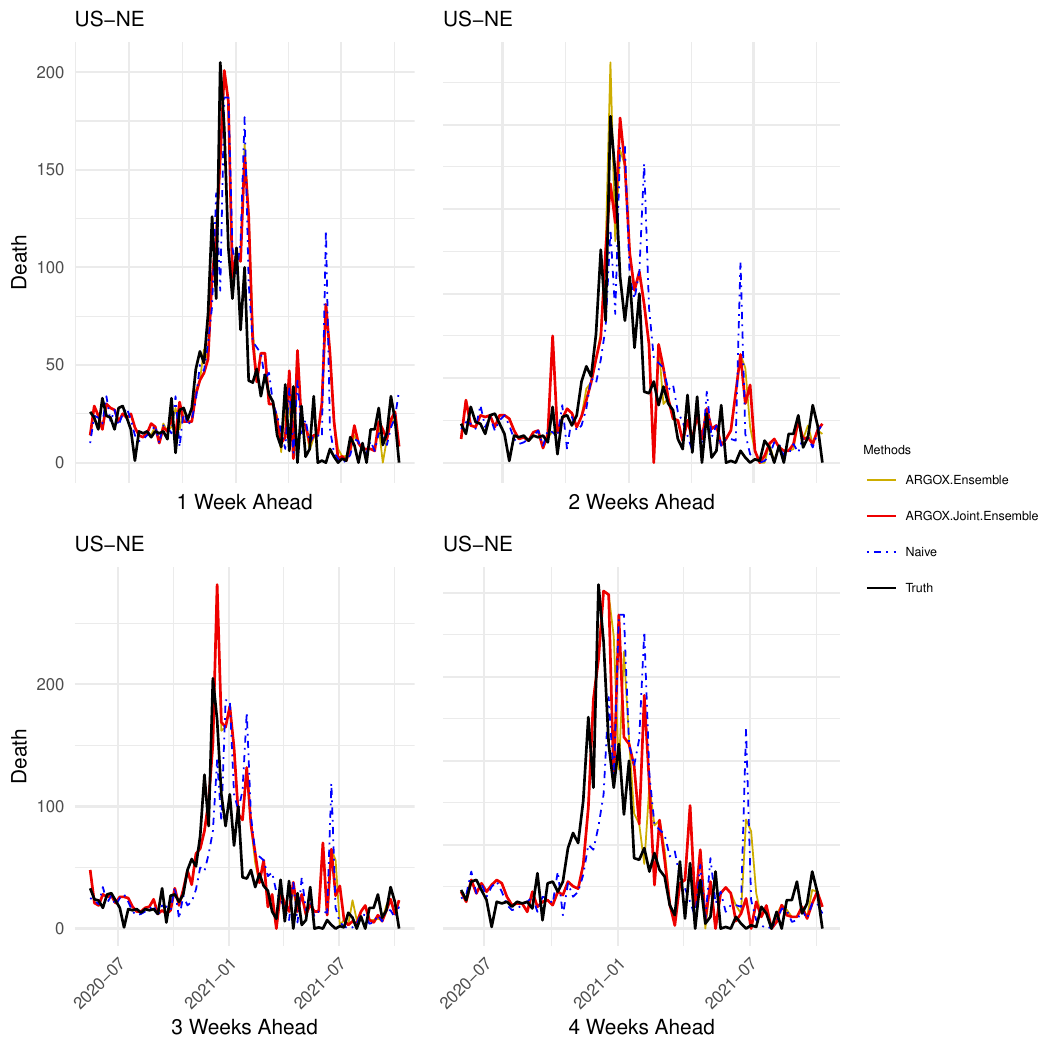} 
\caption{Plots of the COVID-19 1 week (top left), 2 weeks (top right), 3 weeks (bottom left), and 4 weeks (bottom right) ahead estimates for Nebraska (NE). ARGOX-Ensemble is Ref \cite{ma2021covid}.}
\end{figure}
\newpage

\begin{table}[ht]
\centering
\begin{tabular}{lrrrr}
  \hline
& 1 Week Ahead & 2 Weeks Ahead & 3 Weeks Ahead & 4 Weeks Ahead \\ 
    \hline \multicolumn{1}{l}{RMSE} \\
 \hspace{1em} Ref \cite{ma2021covid} & 
8.39 & 11.36 & 11.08 & 16.15 \\ 
   \hspace{1em} ARGOX Joint Ensemble & 6.65 & 10.60 & 9.11 & 13.37 \\ 
   \hspace{1em} Naive & 8.23 & 11.90 & 14.22 & 17.08 \\ 
   \multicolumn{1}{l}{MAE} \\
  \hspace{1em}  Ref \cite{ma2021covid} &5.83 & 7.47 & 7.82 & 9.98 \\ 
   \hspace{1em} ARGOX Joint Ensemble & 5.04 & 6.99 & 6.22 & 9.17 \\ 
   \hspace{1em} Naive & 5.76 & 7.62 & 9.23 & 10.92 \\ 
   \multicolumn{1}{l}{Correlation} \\
 \hspace{1em} Ref \cite{ma2021covid} &0.92 & 0.86 & 0.86 & 0.79 \\ 
   \hspace{1em} ARGOX Joint Ensemble  & 0.94 & 0.87 & 0.92 & 0.86 \\ 
   \hspace{1em} Naive& 0.91 & 0.81 & 0.73 & 0.62 \\ 
   \hline
\end{tabular}
\caption{Comparison of different methods for state-level COVID-19 1 to 4 weeks ahead incremental death in New Hampshire (NH). The MSE, MAE, and correlation are reported and best performed method is highlighted in boldface.} 
\end{table}

\begin{figure}[!h] 
  \centering 
\includegraphics[width=0.6\linewidth, page=31]{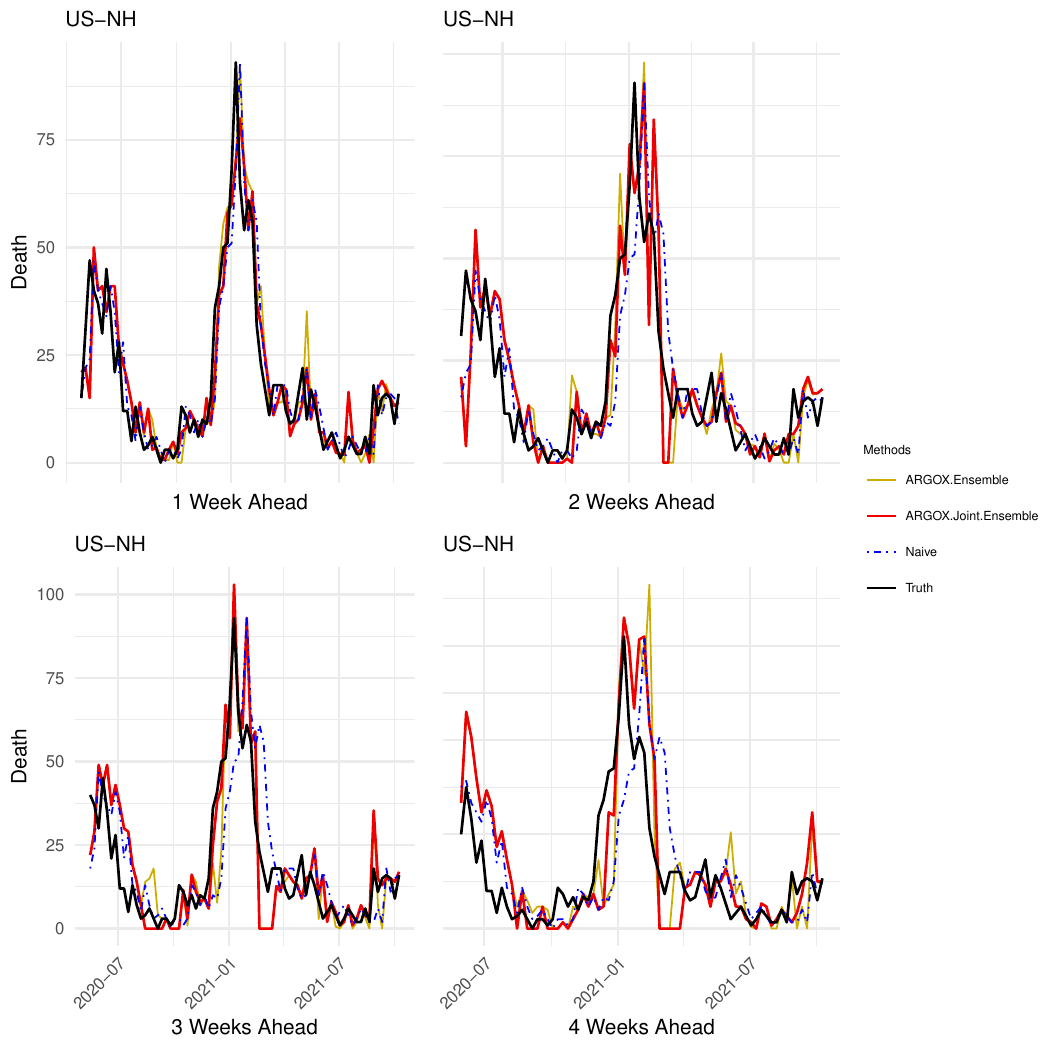} 
\caption{Plots of the COVID-19 1 week (top left), 2 weeks (top right), 3 weeks (bottom left), and 4 weeks (bottom right) ahead estimates for New Hampshire (NH). ARGOX-Ensemble is Ref \cite{ma2021covid}.}
\end{figure}
\newpage

\begin{table}[ht]
\centering
\begin{tabular}{lrrrr}
  \hline
& 1 Week Ahead & 2 Weeks Ahead & 3 Weeks Ahead & 4 Weeks Ahead \\ 
    \hline \multicolumn{1}{l}{RMSE} \\
 \hspace{1em} Ref \cite{ma2021covid} & 
44.11 & 58.12 & 79.77 & 91.35 \\ 
   \hspace{1em} ARGOX Joint Ensemble & 45.85 & 40.48 & 59.45 & 93.63 \\ 
   \hspace{1em} Naive & 49.47 & 75.27 & 97.37 & 118.80 \\ 
   \multicolumn{1}{l}{MAE} \\
  \hspace{1em}  Ref \cite{ma2021covid} &29.53 & 36.00 & 52.57 & 60.78 \\ 
   \hspace{1em} ARGOX Joint Ensemble & 30.16 & 29.77 & 41.02 & 58.32 \\ 
   \hspace{1em} Naive & 33.08 & 51.80 & 71.41 & 89.83 \\ 
   \multicolumn{1}{l}{Correlation} \\
 \hspace{1em} Ref \cite{ma2021covid} &0.97 & 0.94 & 0.92 & 0.91 \\ 
   \hspace{1em} ARGOX Joint Ensemble  & 0.96 & 0.97 & 0.96 & 0.91 \\ 
   \hspace{1em} Naive& 0.96 & 0.90 & 0.84 & 0.77 \\ 
   \hline
\end{tabular}
\caption{Comparison of different methods for state-level COVID-19 1 to 4 weeks ahead incremental death in New Jersey (NJ). The MSE, MAE, and correlation are reported and best performed method is highlighted in boldface.} 
\end{table}

\begin{figure}[!h] 
  \centering 
\includegraphics[width=0.6\linewidth, page=32]{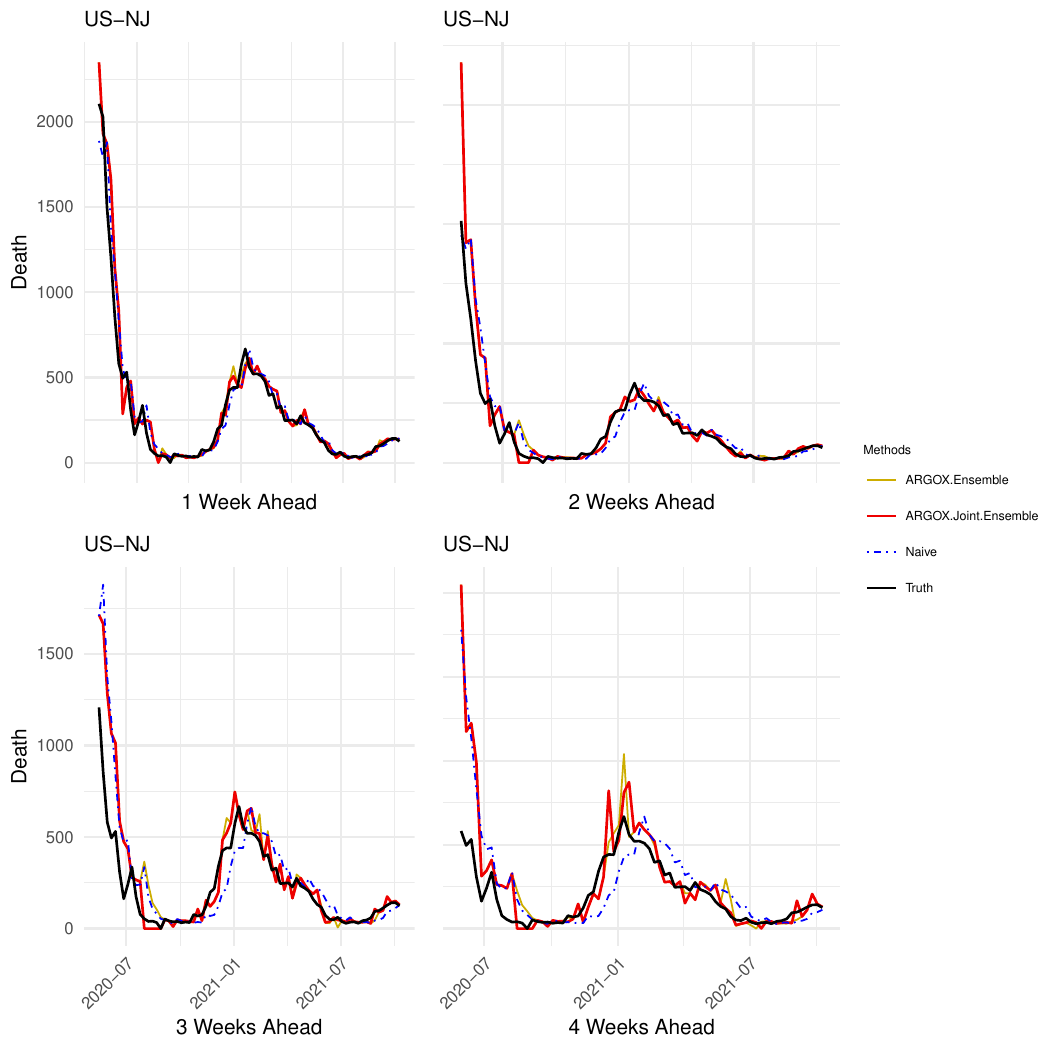} 
\caption{Plots of the COVID-19 1 week (top left), 2 weeks (top right), 3 weeks (bottom left), and 4 weeks (bottom right) ahead estimates for New Jersey (NJ). ARGOX-Ensemble is Ref \cite{ma2021covid}.}
\end{figure}
\newpage

\begin{table}[ht]
\centering
\begin{tabular}{lrrrr}
  \hline
& 1 Week Ahead & 2 Weeks Ahead & 3 Weeks Ahead & 4 Weeks Ahead \\ 
    \hline \multicolumn{1}{l}{RMSE} \\
 \hspace{1em} Ref \cite{ma2021covid} & 
24.58 & 25.95 & 28.98 & 35.66 \\ 
   \hspace{1em} ARGOX Joint Ensemble & 23.25 & 24.48 & 28.08 & 39.41 \\ 
   \hspace{1em} Naive & 29.90 & 33.27 & 40.51 & 49.25 \\ 
   \multicolumn{1}{l}{MAE} \\
  \hspace{1em}  Ref \cite{ma2021covid} &15.13 & 15.84 & 19.68 & 23.92 \\ 
   \hspace{1em} ARGOX Joint Ensemble & 15.02 & 15.76 & 18.61 & 26.47 \\ 
   \hspace{1em} Naive & 18.14 & 21.94 & 28.47 & 34.10 \\ 
   \multicolumn{1}{l}{Correlation} \\
 \hspace{1em} Ref \cite{ma2021covid} &0.92 & 0.91 & 0.90 & 0.86 \\ 
   \hspace{1em} ARGOX Joint Ensemble  & 0.93 & 0.92 & 0.91 & 0.87 \\ 
   \hspace{1em} Naive& 0.88 & 0.86 & 0.79 & 0.70 \\ 
   \hline
\end{tabular}
\caption{Comparison of different methods for state-level COVID-19 1 to 4 weeks ahead incremental death in New Mexico (NM). The MSE, MAE, and correlation are reported and best performed method is highlighted in boldface.} 
\end{table}

\begin{figure}[!h] 
  \centering 
\includegraphics[width=0.6\linewidth, page=33]{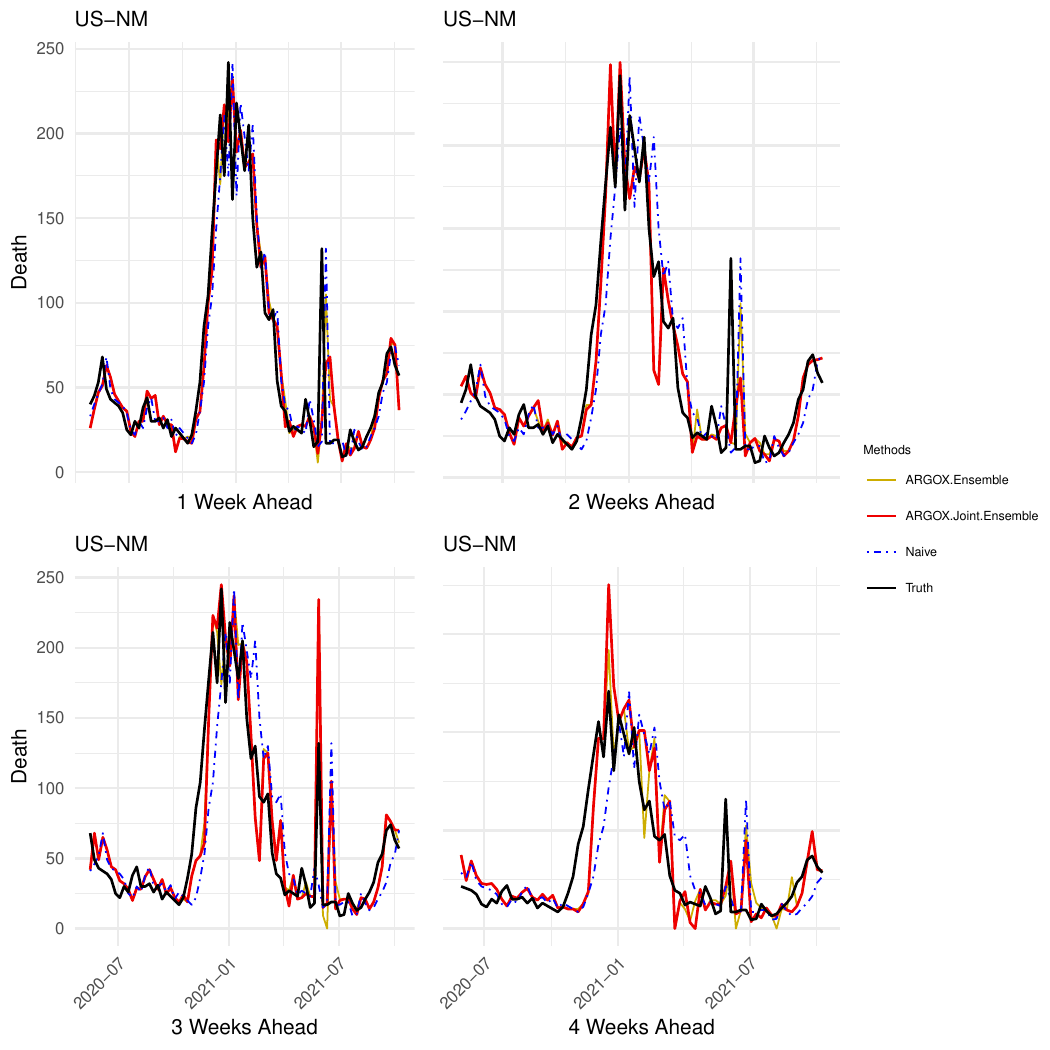} 
\caption{Plots of the COVID-19 1 week (top left), 2 weeks (top right), 3 weeks (bottom left), and 4 weeks (bottom right) ahead estimates for New Mexico (NM). ARGOX-Ensemble is Ref \cite{ma2021covid}.}
\end{figure}
\newpage

\begin{table}[ht]
\centering
\begin{tabular}{lrrrr}
  \hline
& 1 Week Ahead & 2 Weeks Ahead & 3 Weeks Ahead & 4 Weeks Ahead \\ 
    \hline \multicolumn{1}{l}{RMSE} \\
 \hspace{1em} Ref \cite{ma2021covid} & 
26.83 & 29.08 & 34.11 & 51.80 \\ 
   \hspace{1em} ARGOX Joint Ensemble & 27.38 & 29.49 & 39.64 & 52.96 \\ 
   \hspace{1em} Naive & 29.04 & 41.48 & 53.32 & 64.43 \\ 
   \multicolumn{1}{l}{MAE} \\
  \hspace{1em}  Ref \cite{ma2021covid} &18.55 & 20.74 & 24.67 & 34.43 \\ 
   \hspace{1em} ARGOX Joint Ensemble & 19.35 & 20.96 & 25.74 & 35.32 \\ 
   \hspace{1em} Naive & 21.85 & 30.72 & 39.64 & 49.56 \\ 
   \multicolumn{1}{l}{Correlation} \\
 \hspace{1em} Ref \cite{ma2021covid} &0.93 & 0.92 & 0.90 & 0.82 \\ 
   \hspace{1em} ARGOX Joint Ensemble  & 0.93 & 0.92 & 0.87 & 0.82 \\ 
   \hspace{1em} Naive& 0.92 & 0.84 & 0.74 & 0.62 \\ 
   \hline
\end{tabular}
\caption{Comparison of different methods for state-level COVID-19 1 to 4 weeks ahead incremental death in Nevada (NV). The MSE, MAE, and correlation are reported and best performed method is highlighted in boldface.} 
\end{table}

\begin{figure}[!h] 
  \centering 
\includegraphics[width=0.6\linewidth, page=34]{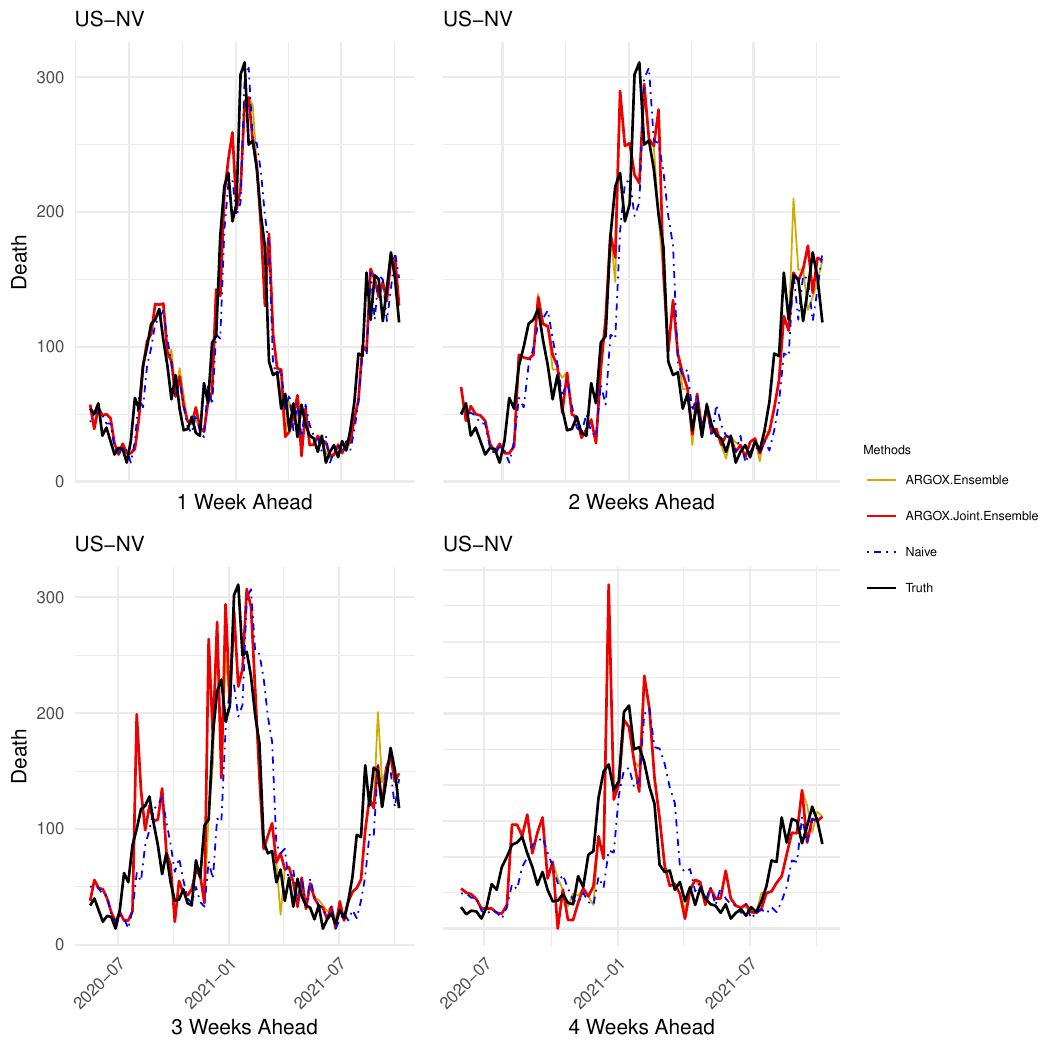} 
\caption{Plots of the COVID-19 1 week (top left), 2 weeks (top right), 3 weeks (bottom left), and 4 weeks (bottom right) ahead estimates for Nevada (NV). ARGOX-Ensemble is Ref \cite{ma2021covid}.}
\end{figure}
\newpage

\begin{table}[ht]
\centering
\begin{tabular}{lrrrr}
  \hline
& 1 Week Ahead & 2 Weeks Ahead & 3 Weeks Ahead & 4 Weeks Ahead \\ 
    \hline \multicolumn{1}{l}{RMSE} \\
 \hspace{1em} Ref \cite{ma2021covid} & 
93.22 & 98.62 & 110.09 & 178.53 \\ 
   \hspace{1em} ARGOX Joint Ensemble & 90.76 & 96.70 & 103.04 & 176.03 \\ 
   \hspace{1em} Naive & 103.01 & 153.84 & 205.53 & 261.13 \\ 
   \multicolumn{1}{l}{MAE} \\
  \hspace{1em}  Ref \cite{ma2021covid} &51.56 & 70.01 & 78.49 & 119.93 \\ 
   \hspace{1em} ARGOX Joint Ensemble & 49.07 & 67.26 & 70.29 & 117.44 \\ 
   \hspace{1em} Naive & 60.48 & 103.77 & 146.08 & 189.30 \\ 
   \multicolumn{1}{l}{Correlation} \\
 \hspace{1em} Ref \cite{ma2021covid} &0.97 & 0.97 & 0.97 & 0.94 \\ 
   \hspace{1em} ARGOX Joint Ensemble  & 0.97 & 0.97 & 0.97 & 0.95 \\ 
   \hspace{1em} Naive& 0.96 & 0.92 & 0.85 & 0.77 \\ 
   \hline
\end{tabular}
\caption{Comparison of different methods for state-level COVID-19 1 to 4 weeks ahead incremental death in New York (NY). The MSE, MAE, and correlation are reported and best performed method is highlighted in boldface.} 
\end{table}

\begin{figure}[!h] 
  \centering 
\includegraphics[width=0.6\linewidth, page=35]{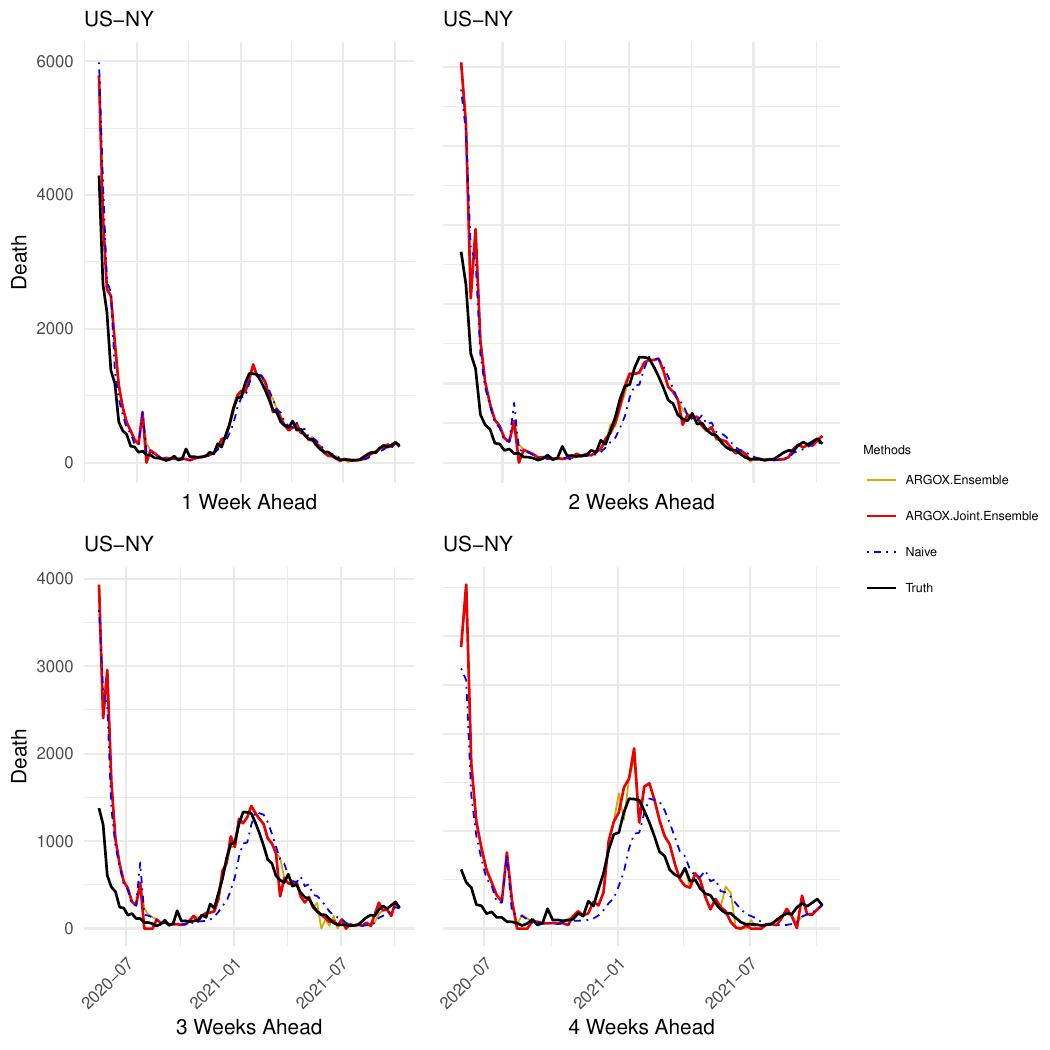} 
\caption{Plots of the COVID-19 1 week (top left), 2 weeks (top right), 3 weeks (bottom left), and 4 weeks (bottom right) ahead estimates for New York (NY). ARGOX-Ensemble is Ref \cite{ma2021covid}.}
\end{figure}
\newpage

\begin{table}[ht]
\centering
\begin{tabular}{lrrrr}
  \hline
& 1 Week Ahead & 2 Weeks Ahead & 3 Weeks Ahead & 4 Weeks Ahead \\ 
    \hline \multicolumn{1}{l}{RMSE} \\
 \hspace{1em} Ref \cite{ma2021covid} & 
534.73 & 449.58 & 412.09 & 469.51 \\ 
   \hspace{1em} ARGOX Joint Ensemble & 464.35 & 294.77 & 290.89 & 418.84 \\ 
   \hspace{1em} Naive & 605.88 & 630.51 & 650.83 & 669.76 \\ 
   \multicolumn{1}{l}{MAE} \\
  \hspace{1em}  Ref \cite{ma2021covid} &239.07 & 234.32 & 220.29 & 241.85 \\ 
   \hspace{1em} ARGOX Joint Ensemble & 224.38 & 187.29 & 184.38 & 232.22 \\ 
   \hspace{1em} Naive & 240.71 & 257.48 & 279.28 & 297.57 \\ 
   \multicolumn{1}{l}{Correlation} \\
 \hspace{1em} Ref \cite{ma2021covid} &0.24 & 0.28 & 0.33 & 0.25 \\ 
   \hspace{1em} ARGOX Joint Ensemble  & 0.28 & 0.55 & 0.58 & 0.33 \\ 
   \hspace{1em} Naive& 0.20 & 0.14 & 0.09 & 0.04 \\ 
   \hline
\end{tabular}
\caption{Comparison of different methods for state-level COVID-19 1 to 4 weeks ahead incremental death in Ohio (OH). The MSE, MAE, and correlation are reported and best performed method is highlighted in boldface.} 
\end{table}

\begin{figure}[!h] 
  \centering 
\includegraphics[width=0.6\linewidth, page=36]{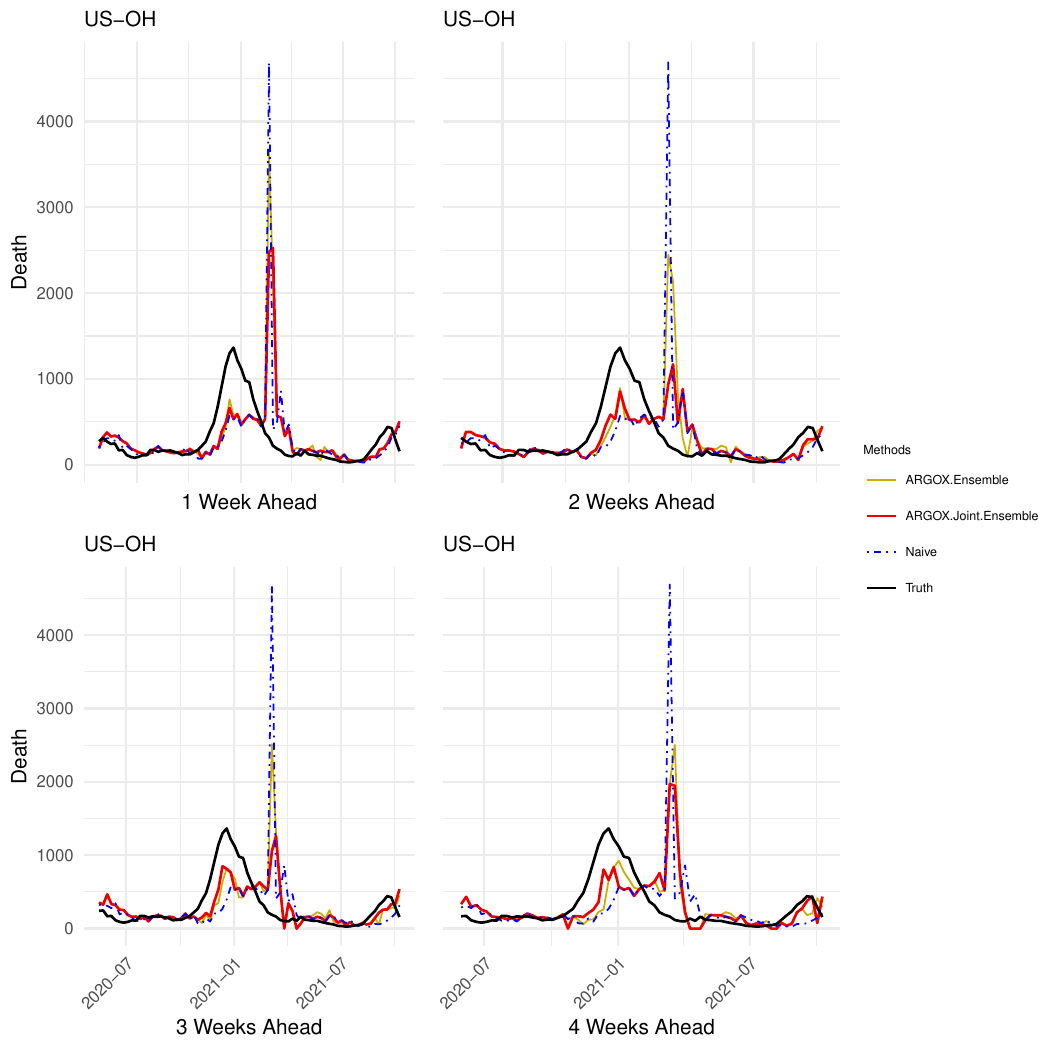} 
\caption{Plots of the COVID-19 1 week (top left), 2 weeks (top right), 3 weeks (bottom left), and 4 weeks (bottom right) ahead estimates for Ohio (OH). ARGOX-Ensemble is Ref \cite{ma2021covid}.}
\end{figure}
\newpage

\begin{table}[ht]
\centering
\begin{tabular}{lrrrr}
  \hline
& 1 Week Ahead & 2 Weeks Ahead & 3 Weeks Ahead & 4 Weeks Ahead \\ 
    \hline \multicolumn{1}{l}{RMSE} \\
 \hspace{1em} Ref \cite{ma2021covid} & 
268.43 & 273.02 & 274.80 & 276.93 \\ 
   \hspace{1em} ARGOX Joint Ensemble & 278.84 & 211.68 & 209.60 & 205.68 \\ 
   \hspace{1em} Naive & 294.89 & 302.65 & 299.94 & 299.35 \\ 
   \multicolumn{1}{l}{MAE} \\
  \hspace{1em}  Ref \cite{ma2021covid} &72.00 & 81.92 & 92.96 & 98.90 \\ 
   \hspace{1em} ARGOX Joint Ensemble & 81.00 & 69.60 & 71.52 & 73.05 \\ 
   \hspace{1em} Naive & 81.42 & 90.51 & 99.95 & 107.14 \\ 
   \multicolumn{1}{l}{Correlation} \\
 \hspace{1em} Ref \cite{ma2021covid} &0.11 & 0.08 & 0.11 & 0.11 \\ 
   \hspace{1em} ARGOX Joint Ensemble  & 0.08 & 0.25 & 0.29 & 0.35 \\ 
   \hspace{1em} Naive& 0.05 & 0.01 & 0.04 & 0.06 \\ 
   \hline
\end{tabular}
\caption{Comparison of different methods for state-level COVID-19 1 to 4 weeks ahead incremental death in Oklahoma (OK). The MSE, MAE, and correlation are reported and best performed method is highlighted in boldface.} 
\end{table}

\begin{figure}[!h] 
  \centering 
\includegraphics[width=0.6\linewidth, page=37]{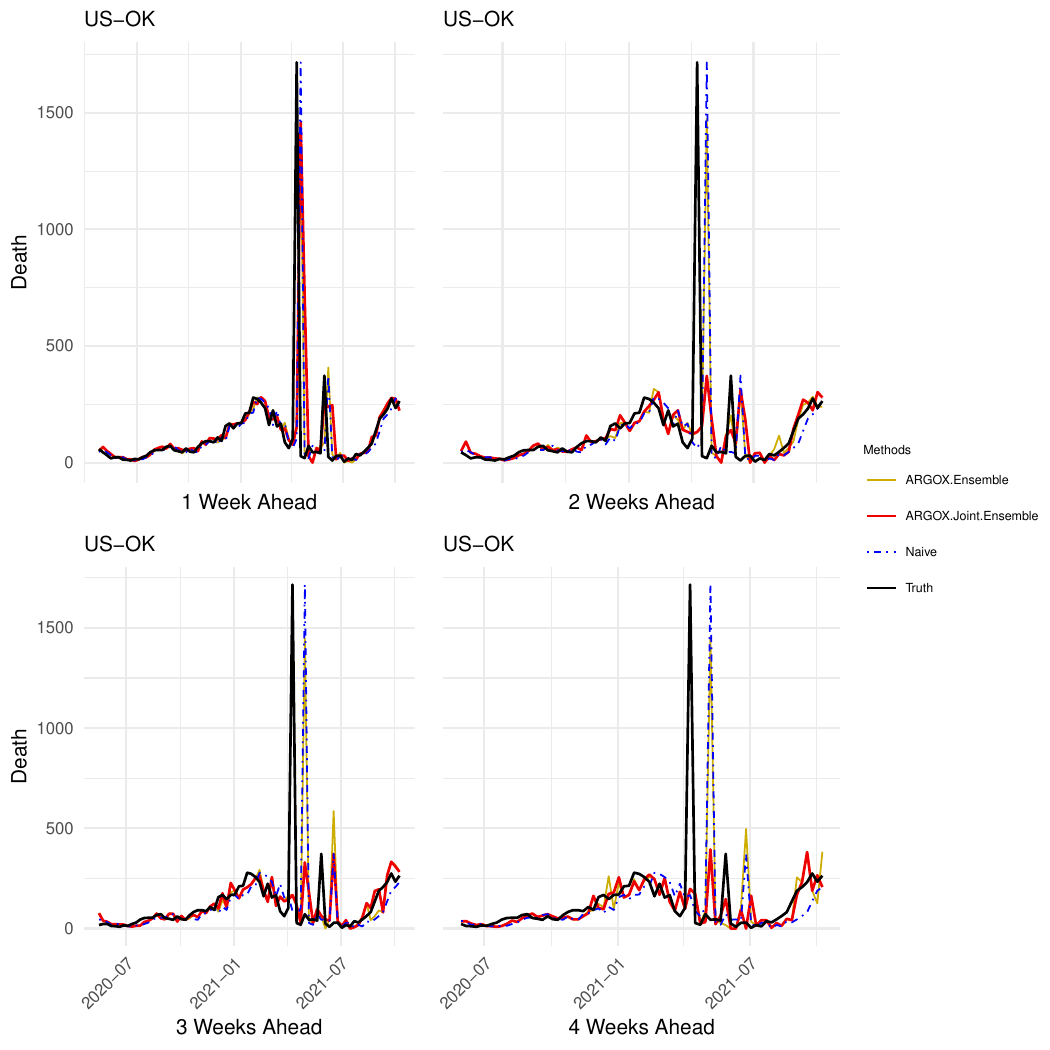} 
\caption{Plots of the COVID-19 1 week (top left), 2 weeks (top right), 3 weeks (bottom left), and 4 weeks (bottom right) ahead estimates for Oklahoma (OK). ARGOX-Ensemble is Ref \cite{ma2021covid}.}
\end{figure}
\newpage

\begin{table}[ht]
\centering
\begin{tabular}{lrrrr}
  \hline
& 1 Week Ahead & 2 Weeks Ahead & 3 Weeks Ahead & 4 Weeks Ahead \\ 
    \hline \multicolumn{1}{l}{RMSE} \\
 \hspace{1em} Ref \cite{ma2021covid} & 
25.89 & 28.05 & 35.28 & 43.84 \\ 
   \hspace{1em} ARGOX Joint Ensemble & 28.68 & 27.90 & 33.32 & 39.01 \\ 
   \hspace{1em} Naive & 33.71 & 38.77 & 43.43 & 43.87 \\ 
   \multicolumn{1}{l}{MAE} \\
  \hspace{1em}  Ref \cite{ma2021covid} &17.60 & 19.96 & 24.01 & 29.25 \\ 
   \hspace{1em} ARGOX Joint Ensemble & 19.16 & 19.61 & 21.79 & 27.12 \\ 
   \hspace{1em} Naive & 22.61 & 26.80 & 29.72 & 30.95 \\ 
   \multicolumn{1}{l}{Correlation} \\
 \hspace{1em} Ref \cite{ma2021covid} &0.84 & 0.82 & 0.71 & 0.70 \\ 
   \hspace{1em} ARGOX Joint Ensemble  & 0.82 & 0.83 & 0.76 & 0.77 \\ 
   \hspace{1em} Naive& 0.73 & 0.64 & 0.56 & 0.54 \\ 
   \hline
\end{tabular}
\caption{Comparison of different methods for state-level COVID-19 1 to 4 weeks ahead incremental death in Oregon (OR). The MSE, MAE, and correlation are reported and best performed method is highlighted in boldface.} 
\end{table}

\begin{figure}[!h] 
  \centering 
\includegraphics[width=0.6\linewidth, page=38]{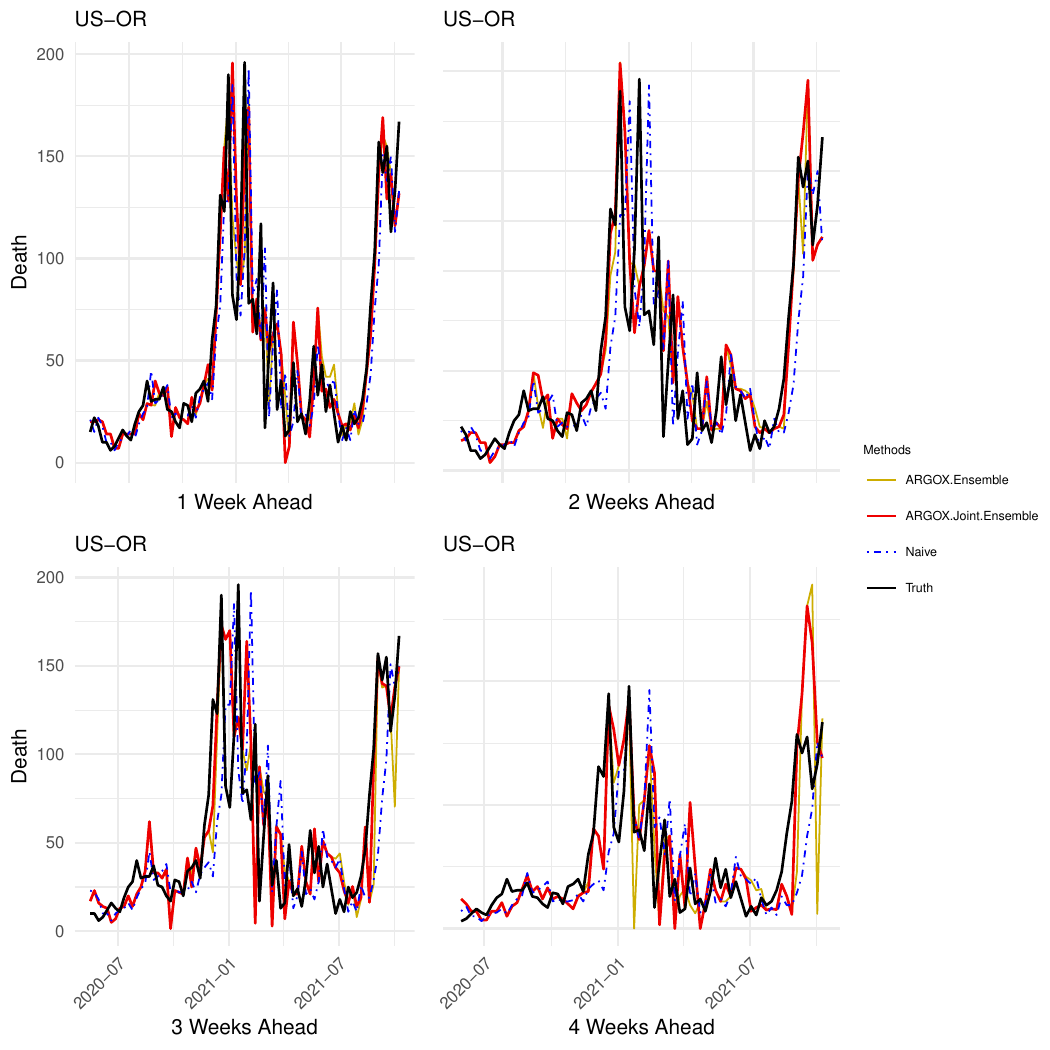} 
\caption{Plots of the COVID-19 1 week (top left), 2 weeks (top right), 3 weeks (bottom left), and 4 weeks (bottom right) ahead estimates for Oregon (OR). ARGOX-Ensemble is Ref \cite{ma2021covid}.}
\end{figure}
\newpage

\begin{table}[ht]
\centering
\begin{tabular}{lrrrr}
  \hline
& 1 Week Ahead & 2 Weeks Ahead & 3 Weeks Ahead & 4 Weeks Ahead \\ 
    \hline \multicolumn{1}{l}{RMSE} \\
 \hspace{1em} Ref \cite{ma2021covid} & 
121.83 & 138.44 & 222.49 & 339.33 \\ 
   \hspace{1em} ARGOX Joint Ensemble & 115.94 & 151.94 & 241.68 & 288.44 \\ 
   \hspace{1em} Naive & 121.63 & 186.77 & 261.13 & 319.53 \\ 
   \multicolumn{1}{l}{MAE} \\
  \hspace{1em}  Ref \cite{ma2021covid} &70.59 & 87.70 & 133.85 & 183.36 \\ 
   \hspace{1em} ARGOX Joint Ensemble & 69.13 & 92.83 & 123.21 & 165.91 \\ 
   \hspace{1em} Naive & 76.65 & 111.92 & 157.28 & 201.24 \\ 
   \multicolumn{1}{l}{Correlation} \\
 \hspace{1em} Ref \cite{ma2021covid} &0.96 & 0.95 & 0.91 & 0.87 \\ 
   \hspace{1em} ARGOX Joint Ensemble  & 0.96 & 0.93 & 0.82 & 0.86 \\ 
   \hspace{1em} Naive& 0.95 & 0.89 & 0.79 & 0.68 \\ 
   \hline
\end{tabular}
\caption{Comparison of different methods for state-level COVID-19 1 to 4 weeks ahead incremental death in Pennsylvania (PA). The MSE, MAE, and correlation are reported and best performed method is highlighted in boldface.} 
\end{table}

\begin{figure}[!h] 
  \centering 
\includegraphics[width=0.6\linewidth, page=39]{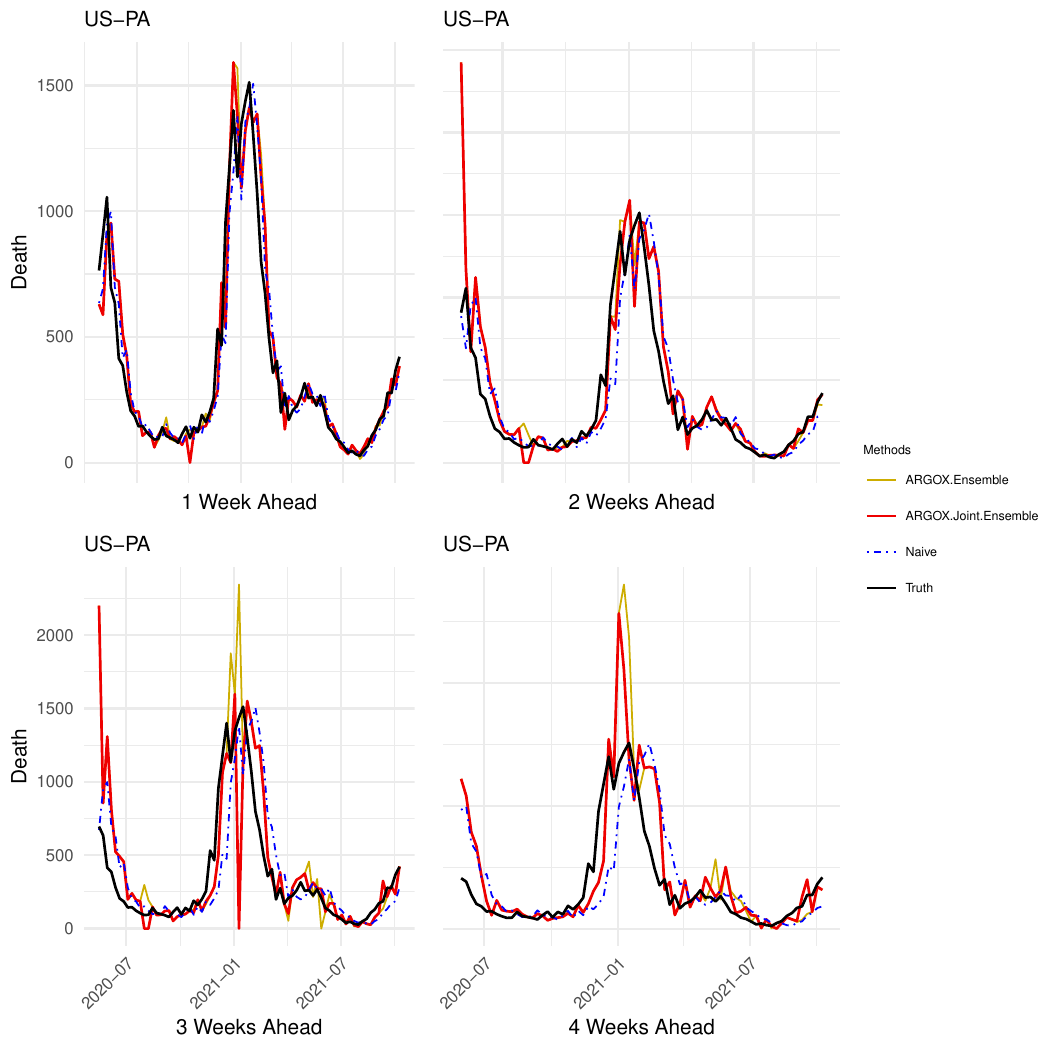} 
\caption{Plots of the COVID-19 1 week (top left), 2 weeks (top right), 3 weeks (bottom left), and 4 weeks (bottom right) ahead estimates for Pennsylvania (PA). ARGOX-Ensemble is Ref \cite{ma2021covid}.}
\end{figure}
\newpage

\begin{table}[ht]
\centering
\begin{tabular}{lrrrr}
  \hline
& 1 Week Ahead & 2 Weeks Ahead & 3 Weeks Ahead & 4 Weeks Ahead \\ 
    \hline \multicolumn{1}{l}{RMSE} \\
 \hspace{1em} Ref \cite{ma2021covid} & 
19.25 & 17.76 & 20.51 & 29.96 \\ 
   \hspace{1em} ARGOX Joint Ensemble & 19.03 & 18.22 & 32.68 & 24.71 \\ 
   \hspace{1em} Naive & 21.21 & 24.70 & 28.14 & 32.34 \\ 
   \multicolumn{1}{l}{MAE} \\
  \hspace{1em}  Ref \cite{ma2021covid} &11.10 & 11.22 & 12.61 & 18.33 \\ 
   \hspace{1em} ARGOX Joint Ensemble & 10.74 & 11.38 & 15.11 & 15.67 \\ 
   \hspace{1em} Naive & 11.82 & 14.51 & 16.72 & 19.62 \\ 
   \multicolumn{1}{l}{Correlation} \\
 \hspace{1em} Ref \cite{ma2021covid} &0.84 & 0.86 & 0.81 & 0.70 \\ 
   \hspace{1em} ARGOX Joint Ensemble  & 0.84 & 0.85 & 0.65 & 0.79 \\ 
   \hspace{1em} Naive& 0.80 & 0.73 & 0.65 & 0.55 \\ 
   \hline
\end{tabular}
\caption{Comparison of different methods for state-level COVID-19 1 to 4 weeks ahead incremental death in Rhode Island (RI). The MSE, MAE, and correlation are reported and best performed method is highlighted in boldface.} 
\end{table}

\begin{figure}[!h] 
  \centering 
\includegraphics[width=0.6\linewidth, page=40]{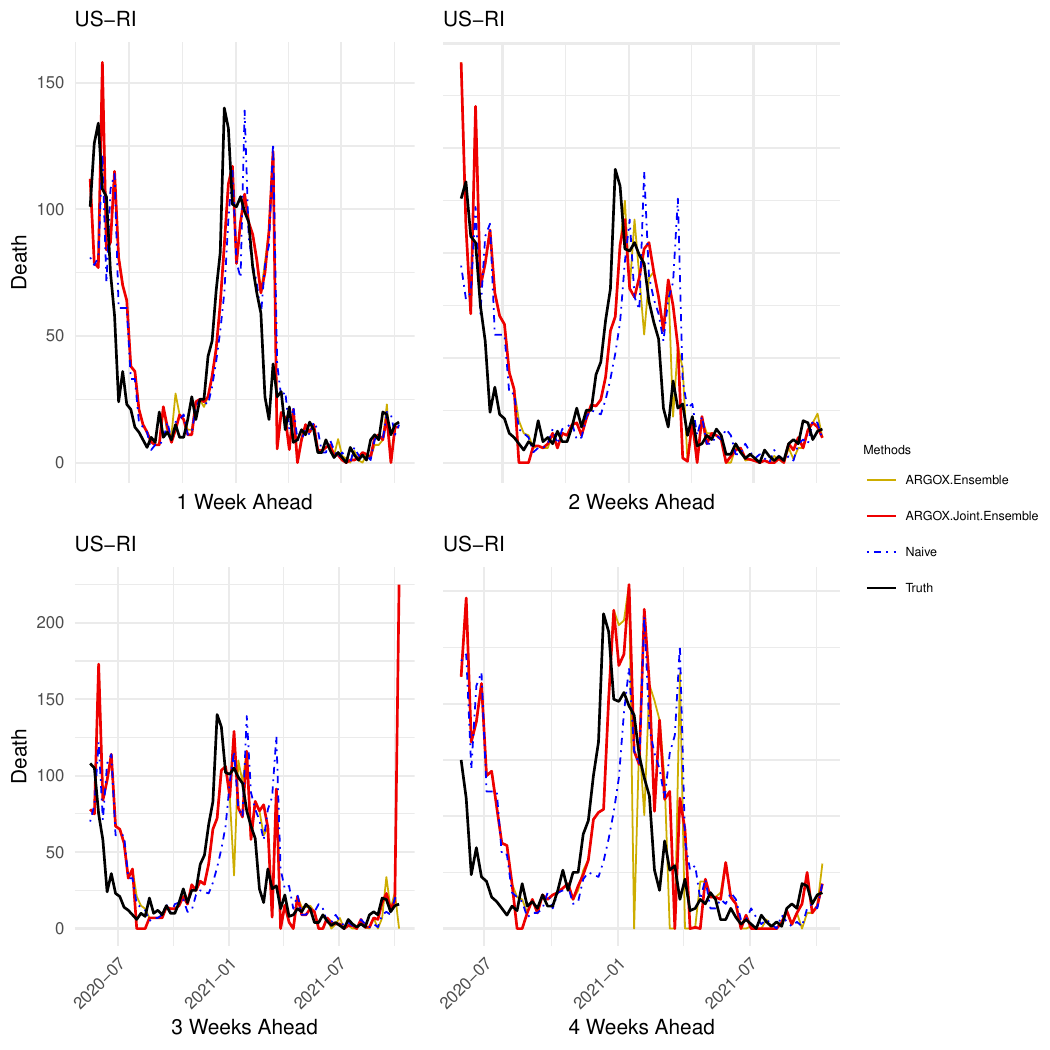} 
\caption{Plots of the COVID-19 1 week (top left), 2 weeks (top right), 3 weeks (bottom left), and 4 weeks (bottom right) ahead estimates for Rhode Island (RI). ARGOX-Ensemble is Ref \cite{ma2021covid}.}
\end{figure}
\newpage

\begin{table}[ht]
\centering
\begin{tabular}{lrrrr}
  \hline
& 1 Week Ahead & 2 Weeks Ahead & 3 Weeks Ahead & 4 Weeks Ahead \\ 
    \hline \multicolumn{1}{l}{RMSE} \\
 \hspace{1em} Ref \cite{ma2021covid} & 
60.01 & 70.73 & 89.95 & 103.81 \\ 
   \hspace{1em} ARGOX Joint Ensemble & 52.71 & 69.21 & 91.23 & 94.75 \\ 
   \hspace{1em} Naive & 64.75 & 89.44 & 112.50 & 133.43 \\ 
   \multicolumn{1}{l}{MAE} \\
  \hspace{1em}  Ref \cite{ma2021covid} &37.73 & 49.49 & 62.85 & 73.11 \\ 
   \hspace{1em} ARGOX Joint Ensemble & 33.20 & 46.16 & 62.33 & 66.25 \\ 
   \hspace{1em} Naive & 42.94 & 66.03 & 84.09 & 98.21 \\ 
   \multicolumn{1}{l}{Correlation} \\
 \hspace{1em} Ref \cite{ma2021covid} &0.92 & 0.88 & 0.83 & 0.77 \\ 
   \hspace{1em} ARGOX Joint Ensemble  & 0.93 & 0.89 & 0.82 & 0.80 \\ 
   \hspace{1em} Naive& 0.89 & 0.79 & 0.67 & 0.52 \\ 
   \hline
\end{tabular}
\caption{Comparison of different methods for state-level COVID-19 1 to 4 weeks ahead incremental death in South Carolina (SC). The MSE, MAE, and correlation are reported and best performed method is highlighted in boldface.} 
\end{table}

\begin{figure}[!h] 
  \centering 
\includegraphics[width=0.6\linewidth, page=41]{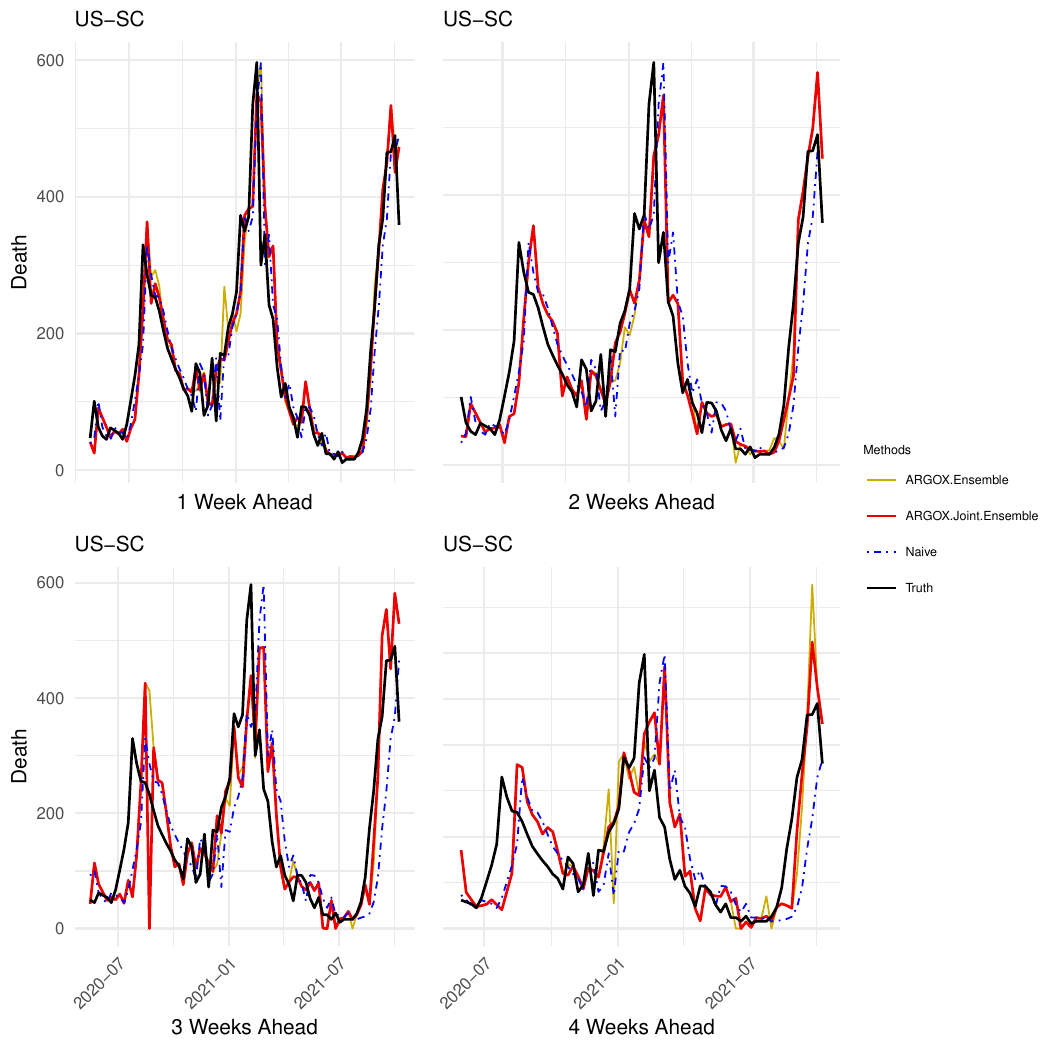} 
\caption{Plots of the COVID-19 1 week (top left), 2 weeks (top right), 3 weeks (bottom left), and 4 weeks (bottom right) ahead estimates for South Carolina (SC). ARGOX-Ensemble is Ref \cite{ma2021covid}.}
\end{figure}
\newpage

\begin{table}[ht]
\centering
\begin{tabular}{lrrrr}
  \hline
& 1 Week Ahead & 2 Weeks Ahead & 3 Weeks Ahead & 4 Weeks Ahead \\ 
    \hline \multicolumn{1}{l}{RMSE} \\
 \hspace{1em} Ref \cite{ma2021covid} & 
15.28 & 23.99 & 15.74 & 27.88 \\ 
   \hspace{1em} ARGOX Joint Ensemble & 15.09 & 14.21 & 17.64 & 26.82 \\ 
   \hspace{1em} Naive & 14.59 & 20.32 & 28.26 & 34.69 \\ 
   \multicolumn{1}{l}{MAE} \\
  \hspace{1em}  Ref \cite{ma2021covid} &8.70 & 11.24 & 9.87 & 16.00 \\ 
   \hspace{1em} ARGOX Joint Ensemble & 8.42 & 8.51 & 10.31 & 15.28 \\ 
   \hspace{1em} Naive & 8.33 & 12.34 & 16.75 & 22.16 \\ 
   \multicolumn{1}{l}{Correlation} \\
 \hspace{1em} Ref \cite{ma2021covid} &0.94 & 0.83 & 0.93 & 0.84 \\ 
   \hspace{1em} ARGOX Joint Ensemble  & 0.94 & 0.94 & 0.91 & 0.85 \\ 
   \hspace{1em} Naive& 0.94 & 0.88 & 0.77 & 0.66 \\ 
   \hline
\end{tabular}
\caption{Comparison of different methods for state-level COVID-19 1 to 4 weeks ahead incremental death in South Dakota (SD). The MSE, MAE, and correlation are reported and best performed method is highlighted in boldface.} 
\end{table}

\begin{figure}[!h] 
  \centering 
\includegraphics[width=0.6\linewidth, page=42]{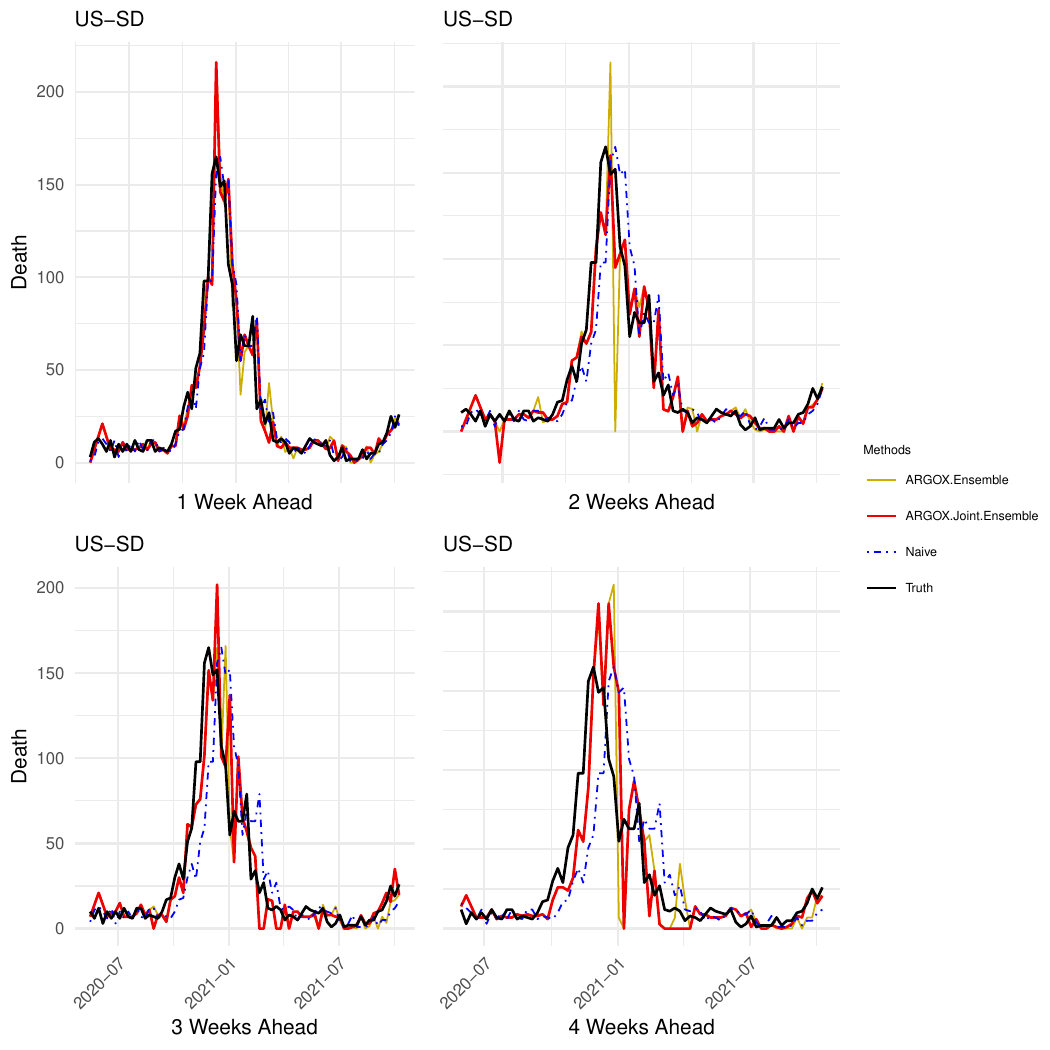} 
\caption{Plots of the COVID-19 1 week (top left), 2 weeks (top right), 3 weeks (bottom left), and 4 weeks (bottom right) ahead estimates for South Dakota (SD). ARGOX-Ensemble is Ref \cite{ma2021covid}.}
\end{figure}
\newpage

\begin{table}[ht]
\centering
\begin{tabular}{lrrrr}
  \hline
& 1 Week Ahead & 2 Weeks Ahead & 3 Weeks Ahead & 4 Weeks Ahead \\ 
    \hline \multicolumn{1}{l}{RMSE} \\
 \hspace{1em} Ref \cite{ma2021covid} & 
98.50 & 116.36 & 103.11 & 133.82 \\ 
   \hspace{1em} ARGOX Joint Ensemble & 94.53 & 110.41 & 99.35 & 135.19 \\ 
   \hspace{1em} Naive & 98.16 & 140.15 & 143.68 & 165.45 \\ 
   \multicolumn{1}{l}{MAE} \\
  \hspace{1em}  Ref \cite{ma2021covid} &58.55 & 66.19 & 64.27 & 84.18 \\ 
   \hspace{1em} ARGOX Joint Ensemble & 57.07 & 61.49 & 62.41 & 87.32 \\ 
   \hspace{1em} Naive & 62.29 & 85.65 & 89.30 & 110.03 \\ 
   \multicolumn{1}{l}{Correlation} \\
 \hspace{1em} Ref \cite{ma2021covid} &0.88 & 0.82 & 0.87 & 0.83 \\ 
   \hspace{1em} ARGOX Joint Ensemble  & 0.89 & 0.85 & 0.88 & 0.83 \\ 
   \hspace{1em} Naive& 0.88 & 0.75 & 0.74 & 0.66 \\ 
   \hline
\end{tabular}
\caption{Comparison of different methods for state-level COVID-19 1 to 4 weeks ahead incremental death in Tennessee (TN). The MSE, MAE, and correlation are reported and best performed method is highlighted in boldface.} 
\end{table}

\begin{figure}[!h] 
  \centering 
\includegraphics[width=0.6\linewidth, page=43]{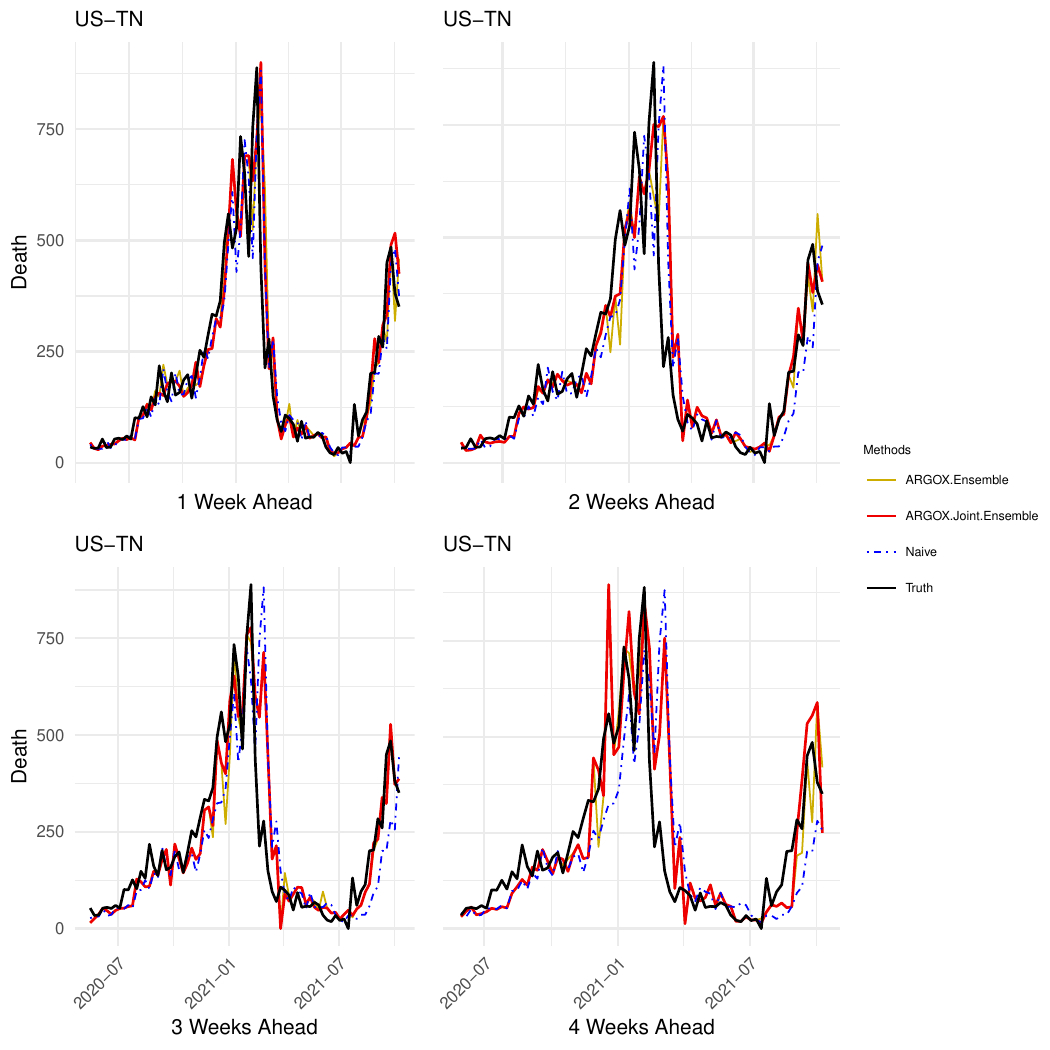} 
\caption{Plots of the COVID-19 1 week (top left), 2 weeks (top right), 3 weeks (bottom left), and 4 weeks (bottom right) ahead estimates for Tennessee (TN). ARGOX-Ensemble is Ref \cite{ma2021covid}.}
\end{figure}
\newpage

\begin{table}[ht]
\centering
\begin{tabular}{lrrrr}
  \hline
& 1 Week Ahead & 2 Weeks Ahead & 3 Weeks Ahead & 4 Weeks Ahead \\ 
    \hline \multicolumn{1}{l}{RMSE} \\
 \hspace{1em} Ref \cite{ma2021covid} & 
244.86 & 349.43 & 482.08 & 604.86 \\ 
   \hspace{1em} ARGOX Joint Ensemble & 247.67 & 353.23 & 443.73 & 1025.41 \\ 
   \hspace{1em} Naive & 303.12 & 439.14 & 519.86 & 618.45 \\ 
   \multicolumn{1}{l}{MAE} \\
  \hspace{1em}  Ref \cite{ma2021covid} &162.75 & 245.74 & 353.85 & 424.87 \\ 
   \hspace{1em} ARGOX Joint Ensemble & 165.08 & 251.07 & 322.27 & 512.44 \\ 
   \hspace{1em} Naive & 198.79 & 310.74 & 388.38 & 476.46 \\ 
   \multicolumn{1}{l}{Correlation} \\
 \hspace{1em} Ref \cite{ma2021covid} &0.93 & 0.87 & 0.83 & 0.72 \\ 
   \hspace{1em} ARGOX Joint Ensemble  & 0.93 & 0.87 & 0.83 & 0.62 \\ 
   \hspace{1em} Naive& 0.89 & 0.76 & 0.66 & 0.52 \\ 
   \hline
\end{tabular}
\caption{Comparison of different methods for state-level COVID-19 1 to 4 weeks ahead incremental death in Texas (TX). The MSE, MAE, and correlation are reported and best performed method is highlighted in boldface.} 
\end{table}

\begin{figure}[!h] 
  \centering 
\includegraphics[width=0.6\linewidth, page=44]{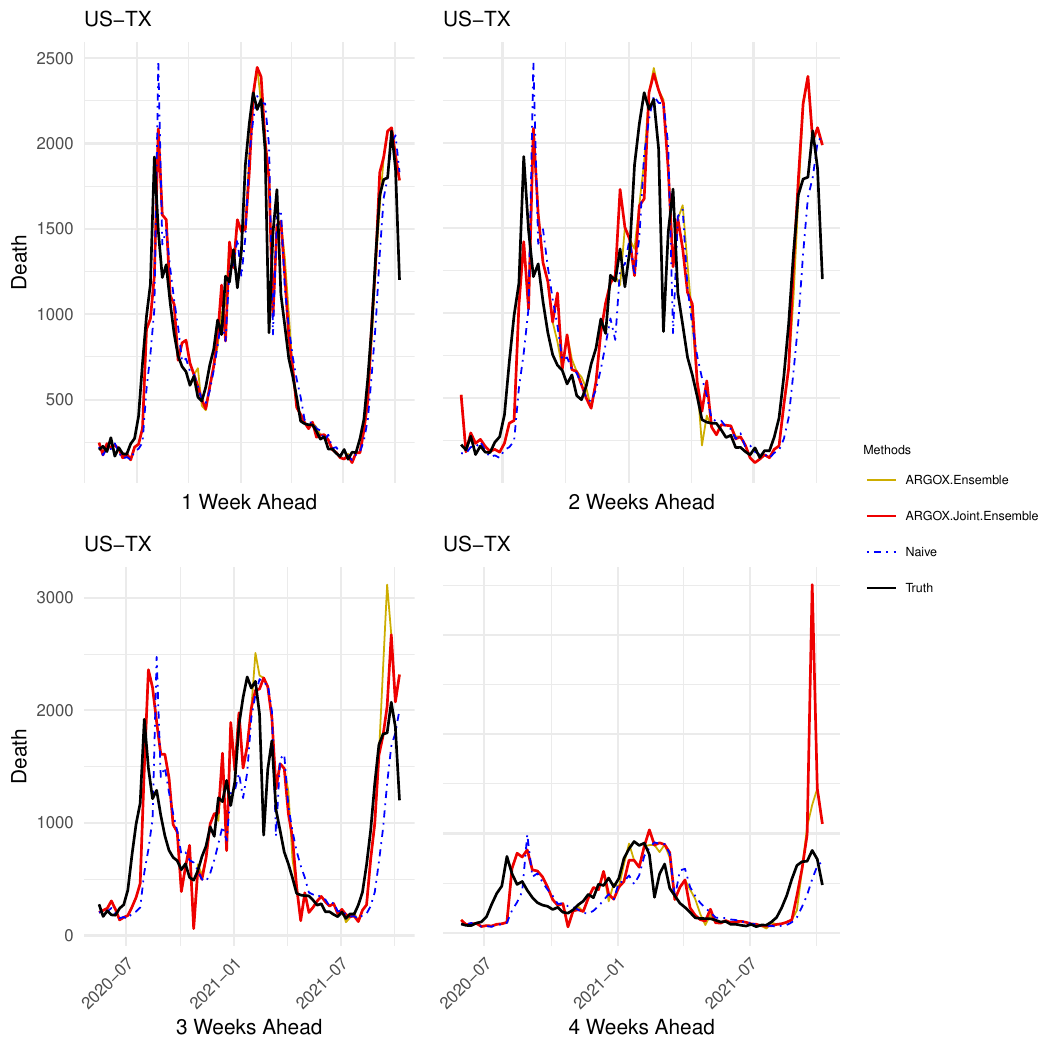} 
\caption{Plots of the COVID-19 1 week (top left), 2 weeks (top right), 3 weeks (bottom left), and 4 weeks (bottom right) ahead estimates for Texas (TX). ARGOX-Ensemble is Ref \cite{ma2021covid}.}
\end{figure}
\newpage

\begin{table}[ht]
\centering
\begin{tabular}{lrrrr}
  \hline
& 1 Week Ahead & 2 Weeks Ahead & 3 Weeks Ahead & 4 Weeks Ahead \\ 
    \hline \multicolumn{1}{l}{RMSE} \\
 \hspace{1em} Ref \cite{ma2021covid} & 
13.94 & 13.39 & 17.29 & 24.40 \\ 
   \hspace{1em} ARGOX Joint Ensemble & 13.41 & 14.82 & 16.64 & 23.30 \\ 
   \hspace{1em} Naive & 14.50 & 17.91 & 19.78 & 23.03 \\ 
   \multicolumn{1}{l}{MAE} \\
  \hspace{1em}  Ref \cite{ma2021covid} &10.92 & 11.14 & 14.08 & 17.59 \\ 
   \hspace{1em} ARGOX Joint Ensemble & 10.42 & 11.66 & 13.43 & 15.92 \\ 
   \hspace{1em} Naive & 10.64 & 14.69 & 16.62 & 19.68 \\ 
   \multicolumn{1}{l}{Correlation} \\
 \hspace{1em} Ref \cite{ma2021covid} &0.87 & 0.88 & 0.82 & 0.77 \\ 
   \hspace{1em} ARGOX Joint Ensemble  & 0.88 & 0.86 & 0.85 & 0.83 \\ 
   \hspace{1em} Naive& 0.86 & 0.78 & 0.74 & 0.65 \\ 
   \hline
\end{tabular}
\caption{Comparison of different methods for state-level COVID-19 1 to 4 weeks ahead incremental death in Utah (UT). The MSE, MAE, and correlation are reported and best performed method is highlighted in boldface.} 
\end{table}

\begin{figure}[!h] 
  \centering 
\includegraphics[width=0.6\linewidth, page=45]{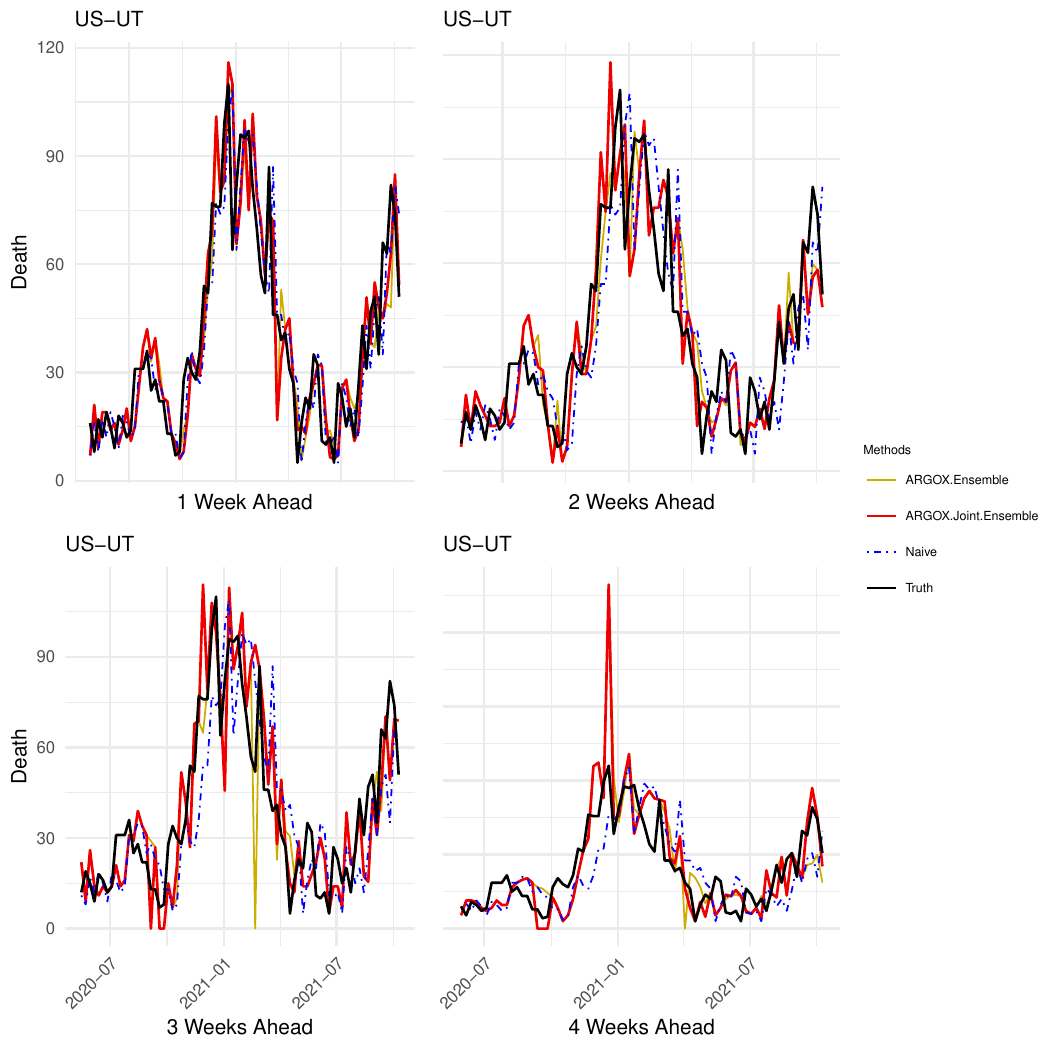} 
\caption{Plots of the COVID-19 1 week (top left), 2 weeks (top right), 3 weeks (bottom left), and 4 weeks (bottom right) ahead estimates for Utah (UT). ARGOX-Ensemble is Ref \cite{ma2021covid}.}
\end{figure}
\newpage

\begin{table}[ht]
\centering
\begin{tabular}{lrrrr}
  \hline
& 1 Week Ahead & 2 Weeks Ahead & 3 Weeks Ahead & 4 Weeks Ahead \\ 
    \hline \multicolumn{1}{l}{RMSE} \\
 \hspace{1em} Ref \cite{ma2021covid} & 
206.13 & 205.30 & 244.20 & 242.92 \\ 
   \hspace{1em} ARGOX Joint Ensemble & 153.02 & 214.26 & 178.19 & 200.31 \\ 
   \hspace{1em} Naive & 161.84 & 241.54 & 260.90 & 256.94 \\ 
   \multicolumn{1}{l}{MAE} \\
  \hspace{1em}  Ref \cite{ma2021covid} &82.34 & 94.19 & 109.25 & 113.59 \\ 
   \hspace{1em} ARGOX Joint Ensemble & 64.60 & 97.52 & 86.45 & 107.21 \\ 
   \hspace{1em} Naive & 67.74 & 111.98 & 122.42 & 125.00 \\ 
   \multicolumn{1}{l}{Correlation} \\
 \hspace{1em} Ref \cite{ma2021covid} &0.33 & 0.38 & 0.27 & 0.33 \\ 
   \hspace{1em} ARGOX Joint Ensemble  & 0.70 & 0.35 & 0.50 & 0.43 \\ 
   \hspace{1em} Naive& 0.67 & 0.28 & 0.17 & 0.21 \\ 
   \hline
\end{tabular}
\caption{Comparison of different methods for state-level COVID-19 1 to 4 weeks ahead incremental death in Virginia (VA). The MSE, MAE, and correlation are reported and best performed method is highlighted in boldface.} 
\end{table}

\begin{figure}[!h] 
  \centering 
\includegraphics[width=0.6\linewidth, page=46]{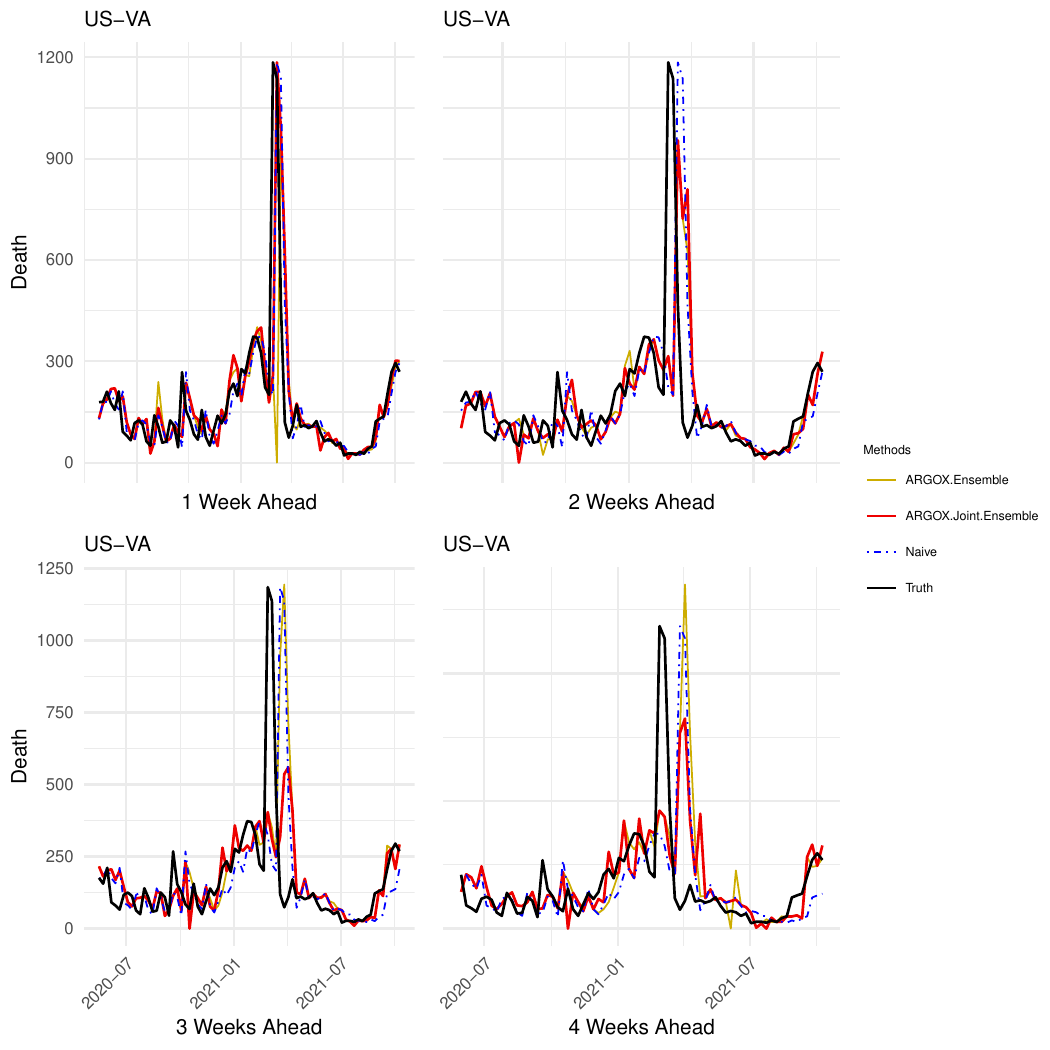} 
\caption{Plots of the COVID-19 1 week (top left), 2 weeks (top right), 3 weeks (bottom left), and 4 weeks (bottom right) ahead estimates for Virginia (VA). ARGOX-Ensemble is Ref \cite{ma2021covid}.}
\end{figure}
\newpage

\begin{table}[ht]
\centering
\begin{tabular}{lrrrr}
  \hline
& 1 Week Ahead & 2 Weeks Ahead & 3 Weeks Ahead & 4 Weeks Ahead \\ 
    \hline \multicolumn{1}{l}{RMSE} \\
 \hspace{1em} Ref \cite{ma2021covid} & 
3.23 & 4.04 & 4.33 & 4.64 \\ 
   \hspace{1em} ARGOX Joint Ensemble & 3.23 & 4.04 & 4.33 & 4.64 \\ 
   \hspace{1em} Naive & 3.05 & 3.57 & 4.27 & 4.84 \\ 
   \multicolumn{1}{l}{MAE} \\
  \hspace{1em}  Ref \cite{ma2021covid} &1.87 & 2.54 & 2.70 & 3.01 \\ 
   \hspace{1em} ARGOX Joint Ensemble & 1.87 & 2.54 & 2.70 & 3.01 \\ 
   \hspace{1em} Naive & 1.79 & 2.23 & 2.64 & 3.17 \\ 
   \multicolumn{1}{l}{Correlation} \\
 \hspace{1em} Ref \cite{ma2021covid} &0.84 & 0.73 & 0.66 & 0.63 \\ 
   \hspace{1em} ARGOX Joint Ensemble  & 0.84 & 0.73 & 0.66 & 0.63 \\ 
   \hspace{1em} Naive& 0.81 & 0.73 & 0.61 & 0.50 \\ 
   \hline
\end{tabular}
\caption{Comparison of different methods for state-level COVID-19 1 to 4 weeks ahead incremental death in Vermont (VT). The MSE, MAE, and correlation are reported and best performed method is highlighted in boldface.} 
\end{table}

\begin{figure}[!h] 
  \centering 
\includegraphics[width=0.6\linewidth, page=47]{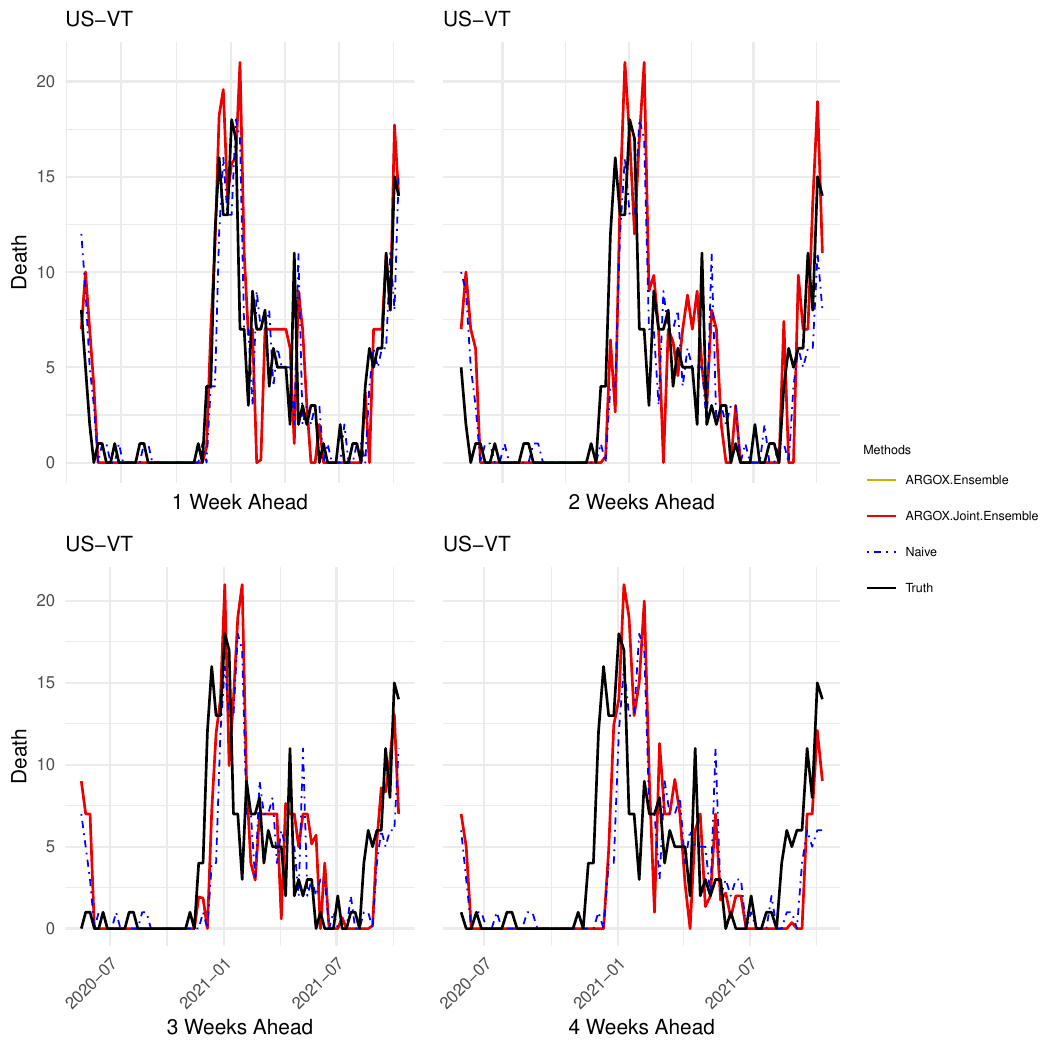} 
\caption{Plots of the COVID-19 1 week (top left), 2 weeks (top right), 3 weeks (bottom left), and 4 weeks (bottom right) ahead estimates for Vermont (VT). ARGOX-Ensemble is Ref \cite{ma2021covid}.}
\end{figure}
\newpage

\begin{table}[ht]
\centering
\begin{tabular}{lrrrr}
  \hline
& 1 Week Ahead & 2 Weeks Ahead & 3 Weeks Ahead & 4 Weeks Ahead \\ 
    \hline \multicolumn{1}{l}{RMSE} \\
 \hspace{1em} Ref \cite{ma2021covid} & 
42.00 & 43.60 & 52.21 & 52.12 \\ 
   \hspace{1em} ARGOX Joint Ensemble & 37.41 & 43.61 & 47.32 & 57.35 \\ 
   \hspace{1em} Naive & 41.27 & 44.50 & 59.82 & 64.14 \\ 
   \multicolumn{1}{l}{MAE} \\
  \hspace{1em}  Ref \cite{ma2021covid} &24.35 & 24.09 & 32.90 & 38.97 \\ 
   \hspace{1em} ARGOX Joint Ensemble & 23.05 & 24.02 & 32.64 & 41.54 \\ 
   \hspace{1em} Naive & 26.94 & 32.25 & 43.08 & 48.32 \\ 
   \multicolumn{1}{l}{Correlation} \\
 \hspace{1em} Ref \cite{ma2021covid} &0.84 & 0.83 & 0.78 & 0.81 \\ 
   \hspace{1em} ARGOX Joint Ensemble  & 0.87 & 0.83 & 0.84 & 0.81 \\ 
   \hspace{1em} Naive& 0.84 & 0.81 & 0.64 & 0.57 \\ 
   \hline
\end{tabular}
\caption{Comparison of different methods for state-level COVID-19 1 to 4 weeks ahead incremental death in Washington (WA). The MSE, MAE, and correlation are reported and best performed method is highlighted in boldface.} 
\end{table}

\begin{figure}[!h] 
  \centering 
\includegraphics[width=0.6\linewidth, page=48]{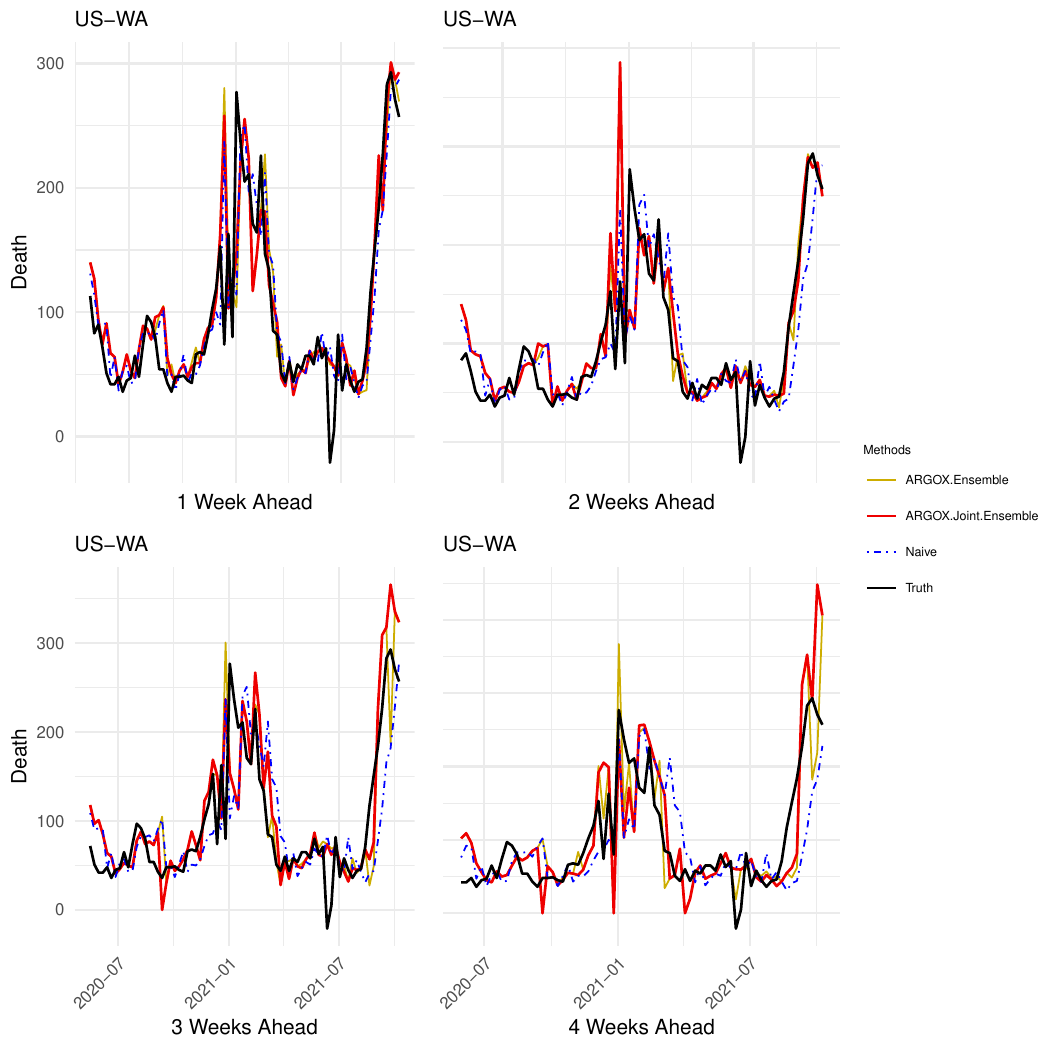} 
\caption{Plots of the COVID-19 1 week (top left), 2 weeks (top right), 3 weeks (bottom left), and 4 weeks (bottom right) ahead estimates for Washington (WA). ARGOX-Ensemble is Ref \cite{ma2021covid}.}
\end{figure}
\newpage

\begin{table}[ht]
\centering
\begin{tabular}{lrrrr}
  \hline
& 1 Week Ahead & 2 Weeks Ahead & 3 Weeks Ahead & 4 Weeks Ahead \\ 
    \hline \multicolumn{1}{l}{RMSE} \\
 \hspace{1em} Ref \cite{ma2021covid} & 
49.57 & 51.20 & 66.46 & 103.74 \\ 
   \hspace{1em} ARGOX Joint Ensemble & 50.88 & 51.36 & 66.58 & 81.28 \\ 
   \hspace{1em} Naive & 49.16 & 62.45 & 72.28 & 84.17 \\ 
   \multicolumn{1}{l}{MAE} \\
  \hspace{1em}  Ref \cite{ma2021covid} &32.36 & 37.39 & 45.58 & 65.57 \\ 
   \hspace{1em} ARGOX Joint Ensemble & 33.35 & 38.38 & 45.79 & 52.31 \\ 
   \hspace{1em} Naive & 34.30 & 44.35 & 51.31 & 57.40 \\ 
   \multicolumn{1}{l}{Correlation} \\
 \hspace{1em} Ref \cite{ma2021covid} &0.92 & 0.91 & 0.86 & 0.82 \\ 
   \hspace{1em} ARGOX Joint Ensemble  & 0.91 & 0.91 & 0.86 & 0.81 \\ 
   \hspace{1em} Naive& 0.91 & 0.86 & 0.81 & 0.74 \\ 
   \hline
\end{tabular}
\caption{Comparison of different methods for state-level COVID-19 1 to 4 weeks ahead incremental death in Wisconsin (WI). The MSE, MAE, and correlation are reported and best performed method is highlighted in boldface.} 
\end{table}

\begin{figure}[!h] 
  \centering 
\includegraphics[width=0.6\linewidth, page=49]{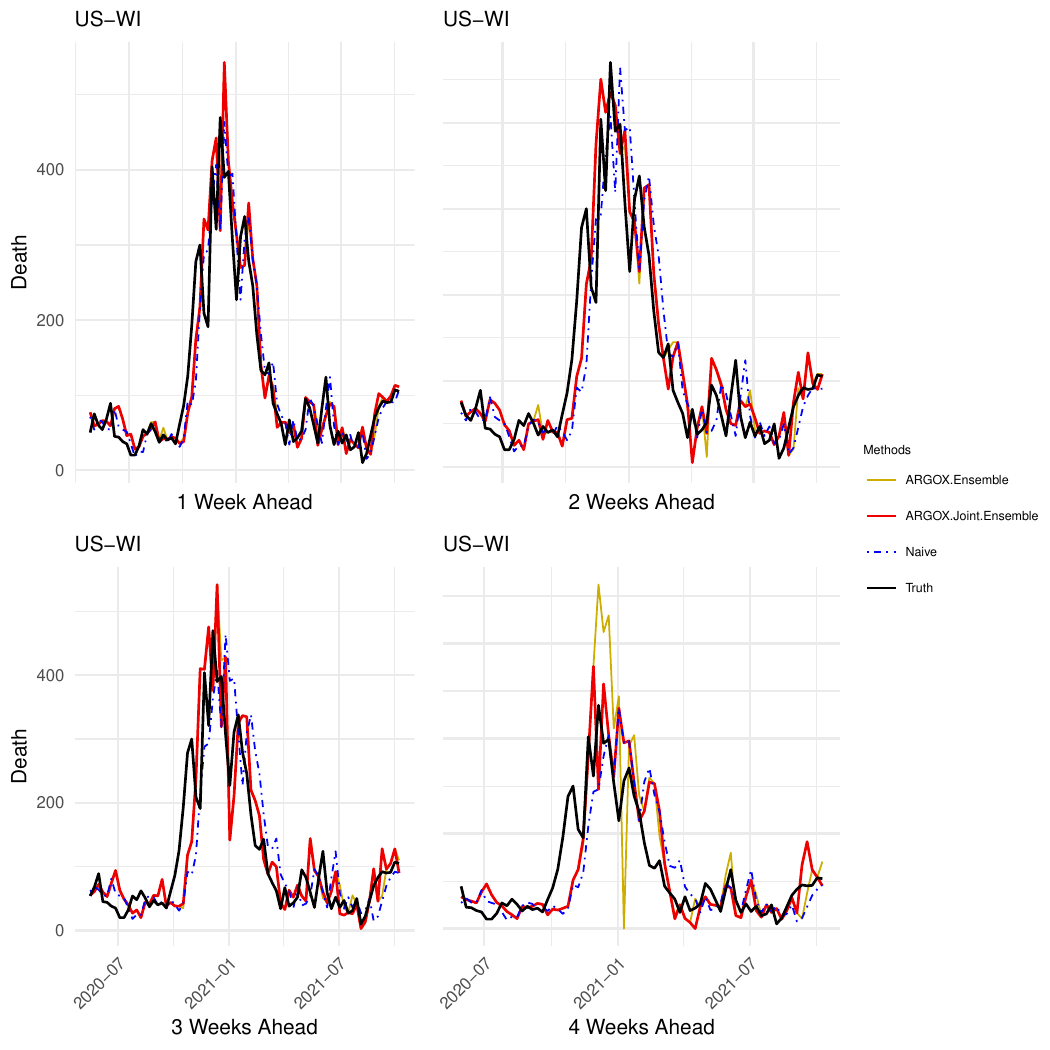} 
\caption{Plots of the COVID-19 1 week (top left), 2 weeks (top right), 3 weeks (bottom left), and 4 weeks (bottom right) ahead estimates for Wisconsin (WI). ARGOX-Ensemble is Ref \cite{ma2021covid}.}
\end{figure}
\newpage

\begin{table}[ht]
\centering
\begin{tabular}{lrrrr}
  \hline
& 1 Week Ahead & 2 Weeks Ahead & 3 Weeks Ahead & 4 Weeks Ahead \\ 
    \hline \multicolumn{1}{l}{RMSE} \\
 \hspace{1em} Ref \cite{ma2021covid} & 
32.30 & 35.62 & 38.12 & 54.69 \\ 
   \hspace{1em} ARGOX Joint Ensemble & 29.51 & 30.56 & 32.35 & 56.76 \\ 
   \hspace{1em} Naive & 35.79 & 43.84 & 47.25 & 55.26 \\ 
   \multicolumn{1}{l}{MAE} \\
  \hspace{1em}  Ref \cite{ma2021covid} &20.68 & 24.63 & 25.76 & 35.80 \\ 
   \hspace{1em} ARGOX Joint Ensemble & 18.98 & 19.77 & 21.56 & 35.80 \\ 
   \hspace{1em} Naive & 22.30 & 29.55 & 32.56 & 39.17 \\ 
   \multicolumn{1}{l}{Correlation} \\
 \hspace{1em} Ref \cite{ma2021covid} &0.84 & 0.81 & 0.79 & 0.61 \\ 
   \hspace{1em} ARGOX Joint Ensemble  & 0.86 & 0.86 & 0.85 & 0.60 \\ 
   \hspace{1em} Naive& 0.80 & 0.69 & 0.64 & 0.49 \\ 
   \hline
\end{tabular}
\caption{Comparison of different methods for state-level COVID-19 1 to 4 weeks ahead incremental death in West Virginia (WV). The MSE, MAE, and correlation are reported and best performed method is highlighted in boldface.} 
\end{table}

\begin{figure}[!h] 
  \centering 
\includegraphics[width=0.6\linewidth, page=50]{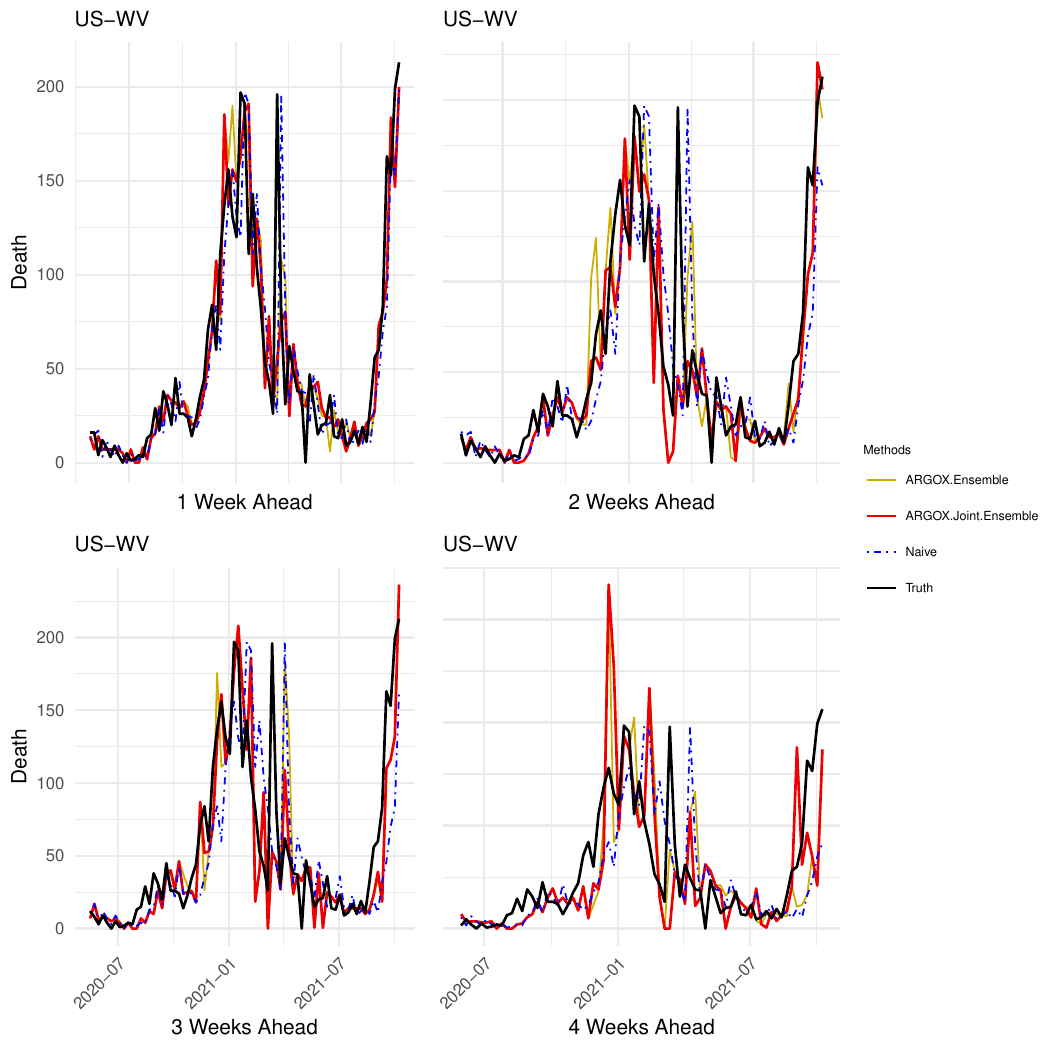} 
\caption{Plots of the COVID-19 1 week (top left), 2 weeks (top right), 3 weeks (bottom left), and 4 weeks (bottom right) ahead estimates for West Virginia (WV). ARGOX-Ensemble is Ref \cite{ma2021covid}.}
\end{figure}
\newpage

\begin{table}[ht]
\centering
\begin{tabular}{lrrrr}
  \hline
& 1 Week Ahead & 2 Weeks Ahead & 3 Weeks Ahead & 4 Weeks Ahead \\ 
    \hline \multicolumn{1}{l}{RMSE} \\
 \hspace{1em} Ref \cite{ma2021covid} & 
11.16 & 11.61 & 11.69 & 12.87 \\ 
   \hspace{1em} ARGOX Joint Ensemble & 9.44 & 10.74 & 9.00 & 13.09 \\ 
   \hspace{1em} Naive & 9.80 & 11.66 & 11.05 & 14.18 \\ 
   \multicolumn{1}{l}{MAE} \\
  \hspace{1em}  Ref \cite{ma2021covid} &6.74 & 7.80 & 6.88 & 8.09 \\ 
   \hspace{1em} ARGOX Joint Ensemble & 5.57 & 6.61 & 5.51 & 8.26 \\ 
   \hspace{1em} Naive & 6.29 & 7.31 & 8.14 & 10.08 \\ 
   \multicolumn{1}{l}{Correlation} \\
 \hspace{1em} Ref \cite{ma2021covid} &0.81 & 0.78 & 0.78 & 0.78 \\ 
   \hspace{1em} ARGOX Joint Ensemble  & 0.84 & 0.79 & 0.88 & 0.79 \\ 
   \hspace{1em} Naive& 0.82 & 0.75 & 0.78 & 0.63 \\ 
   \hline
\end{tabular}
\caption{Comparison of different methods for state-level COVID-19 1 to 4 weeks ahead incremental death in Wyoming (WY). The MSE, MAE, and correlation are reported and best performed method is highlighted in boldface.} 
\label{tab:State_Ours_Death_WY}
\end{table}

\begin{figure}[!h] 
  \centering 
\includegraphics[width=0.6\linewidth, page=51]{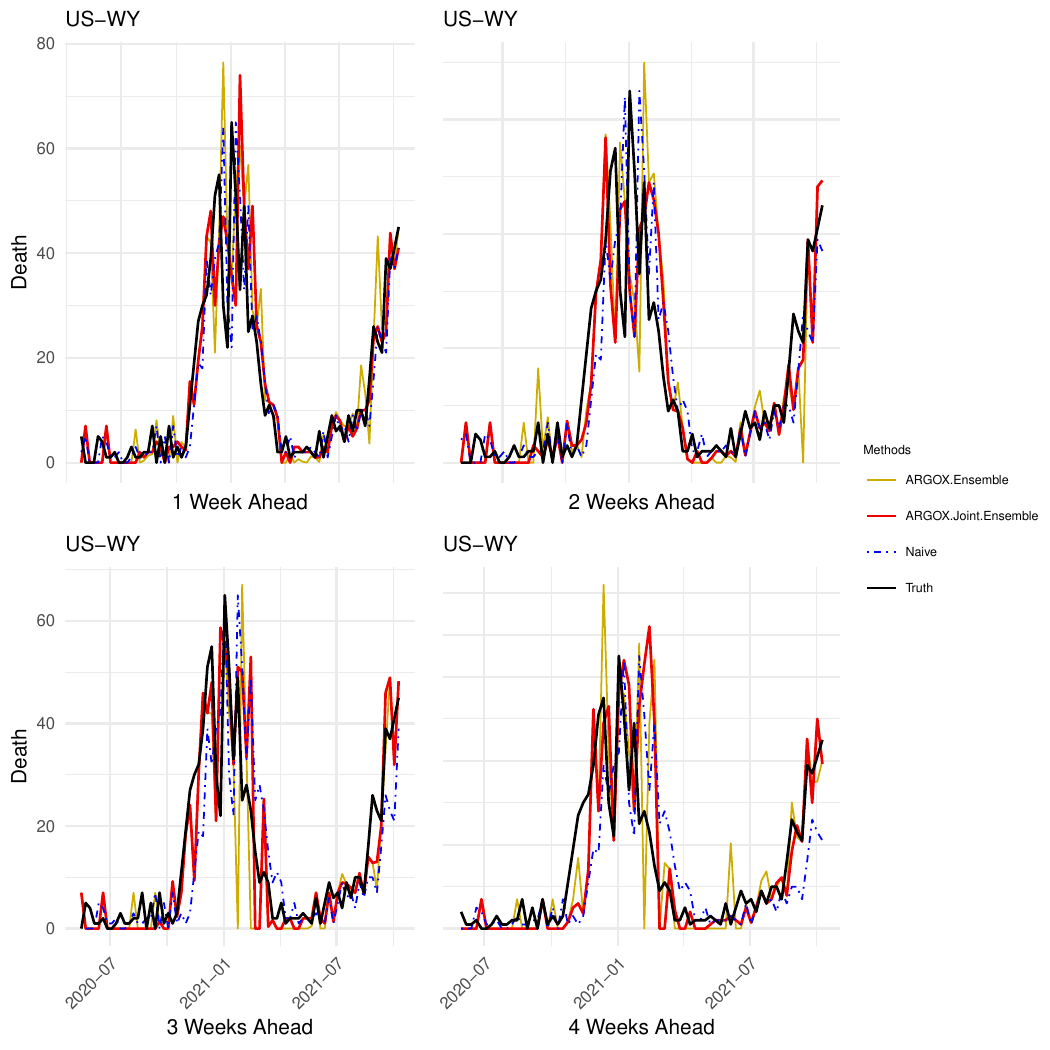} 
\caption{Plots of the COVID-19 1 week (top left), 2 weeks (top right), 3 weeks (bottom left), and 4 weeks (bottom right) ahead estimates for Wyoming (WY). ARGOX-Ensemble is Ref \cite{ma2021covid}.}
\label{fig:State_Ours_Death_WY}
\end{figure}
\newpage

\restoregeometry

\clearpage
\newgeometry{left=1.5cm,bottom=2cm}
\subsection*{Detailed COVID-19 incremental case estimation results for each state}
\begin{table}[ht]
\centering
\begin{tabular}{lrrrr}
  \hline
& 1 Week Ahead & 2 Weeks Ahead & 3 Weeks Ahead & 4 Weeks Ahead \\ 
    \hline \multicolumn{1}{l}{RMSE} \\
 \hspace{1em} Ref \cite{ma2021covid} & 
477.01 & 1124.92 & 2039.82 & 1242.97 \\ 
   \hspace{1em} ARGOX Joint Ensemble & 479.33 & 1002.90 & 831.77 & 1272.76 \\ 
   \hspace{1em} Naive & 484.09 & 848.96 & 1090.76 & 1308.85 \\ 
   \multicolumn{1}{l}{MAE} \\
  \hspace{1em}  Ref \cite{ma2021covid} &301.74 & 665.25 & 962.18 & 852.77 \\ 
   \hspace{1em} ARGOX Joint Ensemble & 296.25 & 632.83 & 561.17 & 912.34 \\ 
   \hspace{1em} Naive & 303.44 & 566.97 & 739.22 & 939.48 \\ 
   \multicolumn{1}{l}{Correlation} \\
 \hspace{1em} Ref \cite{ma2021covid} &0.96 & 0.90 & 0.60 & 0.88 \\ 
   \hspace{1em} ARGOX Joint Ensemble  & 0.96 & 0.88 & 0.90 & 0.80 \\ 
   \hspace{1em} Naive& 0.96 & 0.88 & 0.81 & 0.80 \\ 
   \hline
\end{tabular}
\caption{Comparison of different methods for state-level COVID-19 1 to 4 weeks ahead incremental case in Alaska (AK). The MSE, MAE, and correlation are reported and best performed method is highlighted in boldface.}
\label{tab:State_Ours_Case_AK}
\end{table}       

\begin{figure}[!h]    
\centering  
\includegraphics[width=0.6\linewidth, page=1]{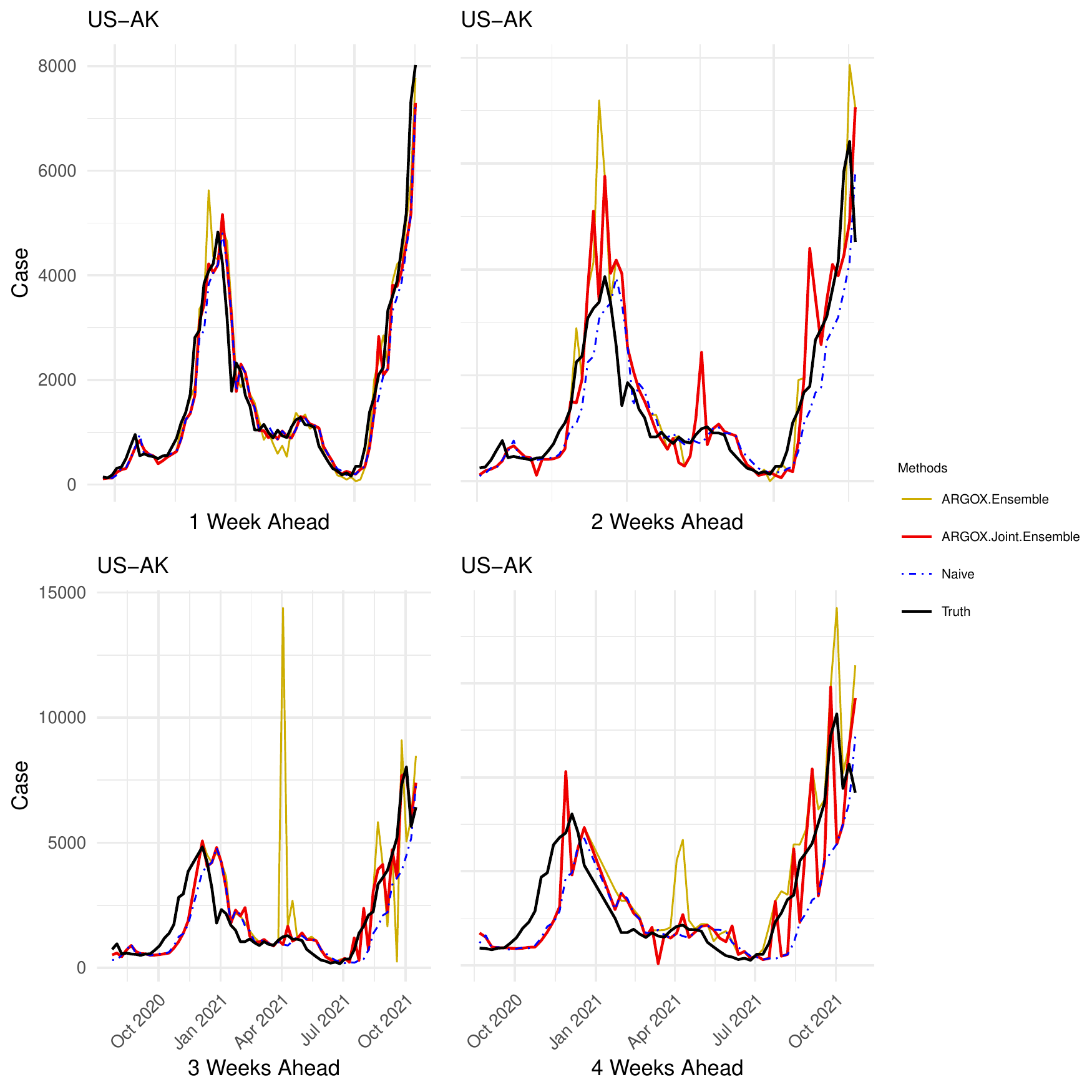}  \caption{Plots of the COVID-19 1 week (top left), 2 weeks (top right), 3 weeks (bottom left), and 4 weeks (bottom right) ahead estimates for Alaska (AK). ARGOX-Ensemble is Ref \cite{ma2021covid}.}   
\label{fig:State_Ours_Case_AK}
\end{figure} 
\newpage    

\begin{table}[ht]
\centering
\begin{tabular}{lrrrr}
  \hline
& 1 Week Ahead & 2 Weeks Ahead & 3 Weeks Ahead & 4 Weeks Ahead \\ 
    \hline \multicolumn{1}{l}{RMSE} \\
 \hspace{1em} Ref \cite{ma2021covid} & 
3697.48 & 4202.77 & 7043.91 & 7224.20 \\ 
   \hspace{1em} ARGOX Joint Ensemble & 4372.27 & 4172.79 & 6983.70 & 8361.37 \\ 
   \hspace{1em} Naive & 3520.71 & 5188.44 & 6974.31 & 8872.70 \\ 
   \multicolumn{1}{l}{MAE} \\
  \hspace{1em}  Ref \cite{ma2021covid} &2827.36 & 3190.90 & 5027.55 & 5188.65 \\ 
   \hspace{1em} ARGOX Joint Ensemble & 2996.11 & 3193.18 & 4987.95 & 6174.15 \\ 
   \hspace{1em} Naive & 2590.56 & 3915.55 & 5108.22 & 6689.60 \\ 
   \multicolumn{1}{l}{Correlation} \\
 \hspace{1em} Ref \cite{ma2021covid} &0.91 & 0.91 & 0.78 & 0.82 \\ 
   \hspace{1em} ARGOX Joint Ensemble  & 0.89 & 0.89 & 0.78 & 0.56 \\ 
   \hspace{1em} Naive& 0.92 & 0.82 & 0.69 & 0.45 \\ 
   \hline
\end{tabular}
\caption{Comparison of different methods for state-level COVID-19 1 to 4 weeks ahead incremental case in Alabama (AL). The MSE, MAE, and correlation are reported and best performed method is highlighted in boldface.} 
\end{table}       \begin{figure}[!h]    \centering  \includegraphics[width=0.6\linewidth, page=2]{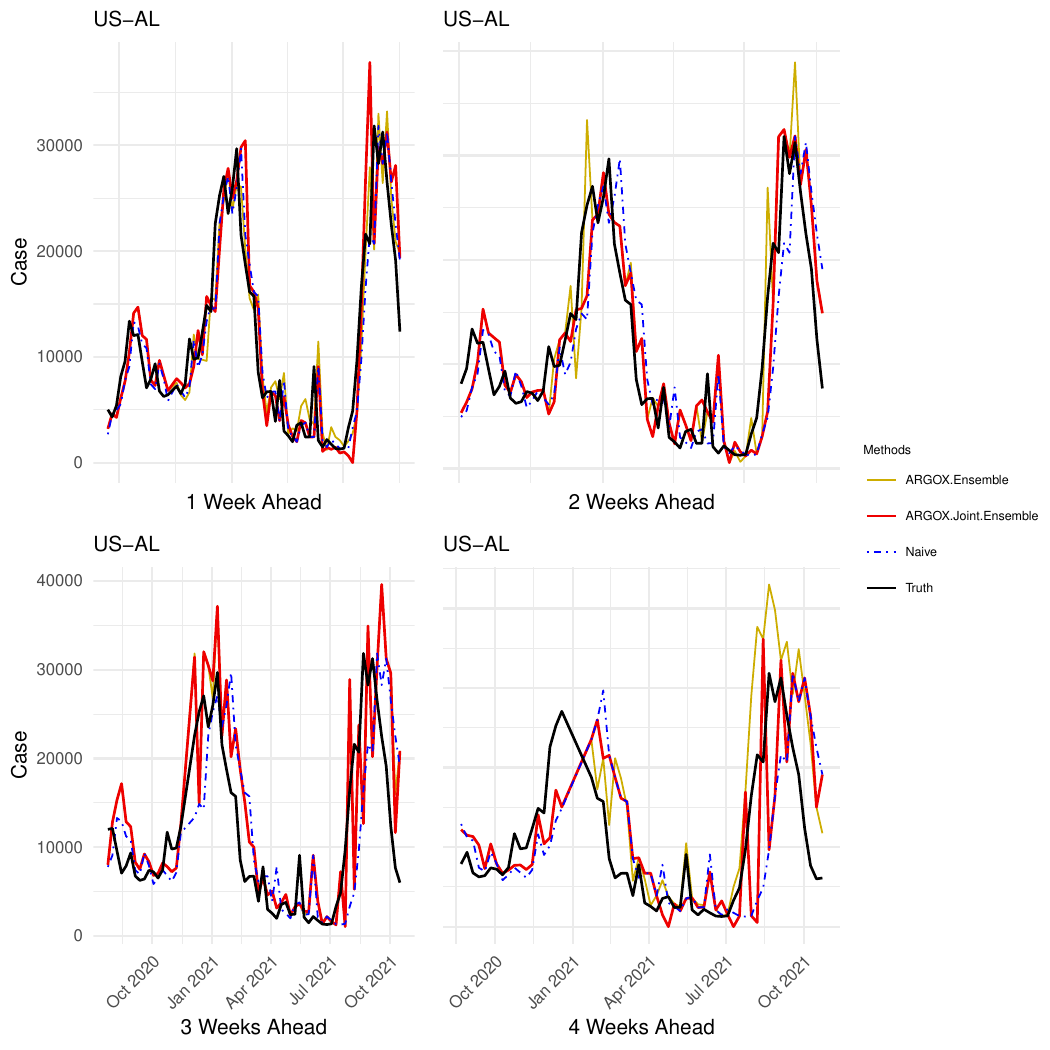}  \caption{Plots of the COVID-19 1 week (top left), 2 weeks (top right), 3 weeks (bottom left), and 4 weeks (bottom right) ahead estimates for Alabama (AL). ARGOX-Ensemble is Ref \cite{ma2021covid}.}  \end{figure} \newpage    
\begin{table}[ht]
\centering
\begin{tabular}{lrrrr}
  \hline
& 1 Week Ahead & 2 Weeks Ahead & 3 Weeks Ahead & 4 Weeks Ahead \\ 
    \hline \multicolumn{1}{l}{RMSE} \\
 \hspace{1em} Ref \cite{ma2021covid} & 
1895.64 & 2579.68 & 3102.44 & 4617.16 \\ 
   \hspace{1em} ARGOX Joint Ensemble & 1945.03 & 2814.12 & 3098.57 & 4764.24 \\ 
   \hspace{1em} Naive & 1827.87 & 2906.52 & 3825.71 & 4870.53 \\ 
   \multicolumn{1}{l}{MAE} \\
  \hspace{1em}  Ref \cite{ma2021covid} &1284.85 & 1893.90 & 2094.85 & 2963.80 \\ 
   \hspace{1em} ARGOX Joint Ensemble & 1253.89 & 2115.58 & 2084.06 & 3436.92 \\ 
   \hspace{1em} Naive & 1214.03 & 2010.85 & 2801.51 & 3657.17 \\ 
   \multicolumn{1}{l}{Correlation} \\
 \hspace{1em} Ref \cite{ma2021covid} &0.94 & 0.87 & 0.88 & 0.79 \\ 
   \hspace{1em} ARGOX Joint Ensemble  & 0.94 & 0.85 & 0.88 & 0.61 \\ 
   \hspace{1em} Naive& 0.94 & 0.85 & 0.75 & 0.54 \\ 
   \hline
\end{tabular}
\caption{Comparison of different methods for state-level COVID-19 1 to 4 weeks ahead incremental case in Arkansas (AR). The MSE, MAE, and correlation are reported and best performed method is highlighted in boldface.} 
\end{table}       \begin{figure}[!h]    \centering  \includegraphics[width=0.6\linewidth, page=3]{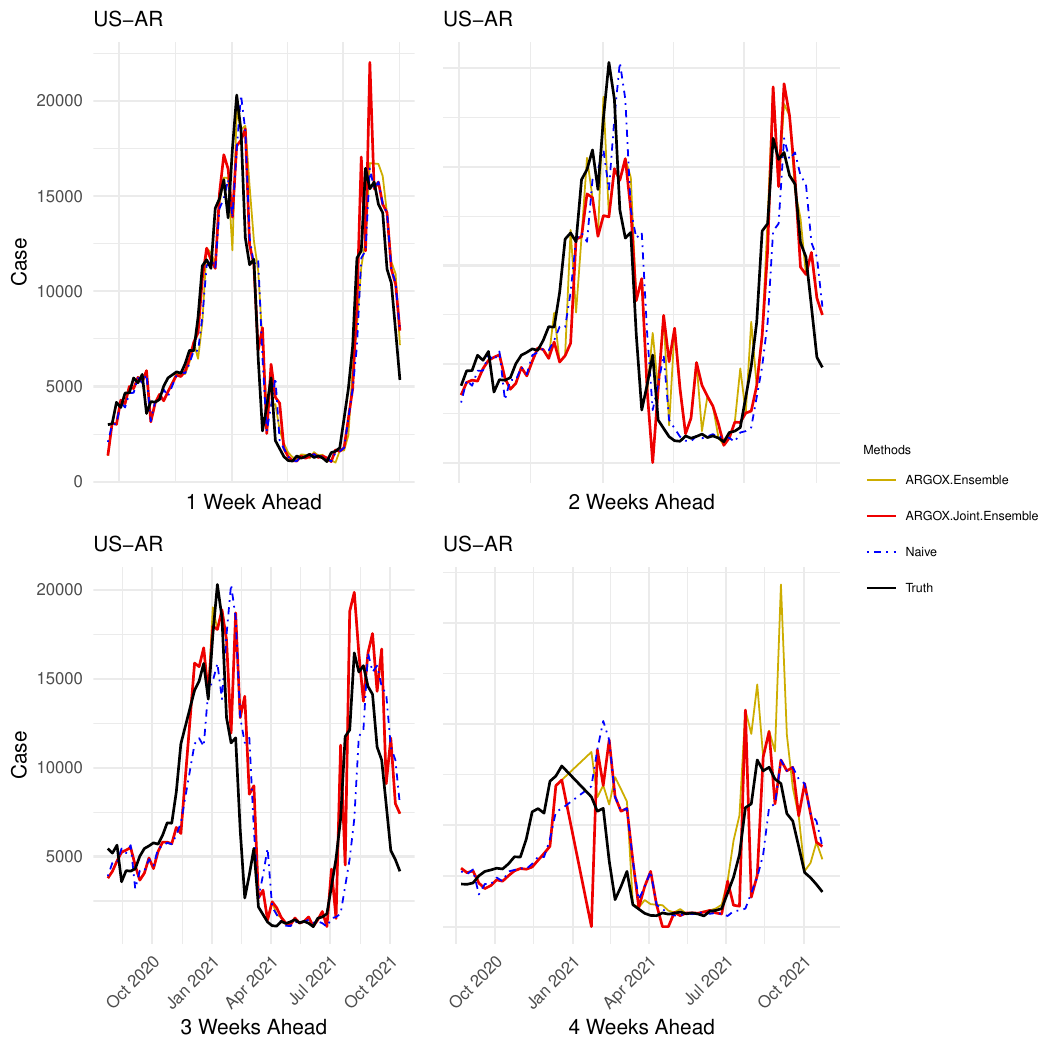}  \caption{Plots of the COVID-19 1 week (top left), 2 weeks (top right), 3 weeks (bottom left), and 4 weeks (bottom right) ahead estimates for Arkansas (AR). ARGOX-Ensemble is Ref \cite{ma2021covid}.}  \end{figure} \newpage   
\begin{table}[ht]
\centering
\begin{tabular}{lrrrr}
  \hline
& 1 Week Ahead & 2 Weeks Ahead & 3 Weeks Ahead & 4 Weeks Ahead \\ 
    \hline \multicolumn{1}{l}{RMSE} \\
 \hspace{1em} Ref \cite{ma2021covid} & 
4819.88 & 8063.63 & 10560.67 & 12050.02 \\ 
   \hspace{1em} ARGOX Joint Ensemble & 4744.19 & 8419.44 & 10491.41 & 13376.99 \\ 
   \hspace{1em} Naive & 4910.16 & 8298.27 & 11016.70 & 13446.76 \\ 
   \multicolumn{1}{l}{MAE} \\
  \hspace{1em}  Ref \cite{ma2021covid} &3035.24 & 5093.53 & 6956.50 & 7208.03 \\ 
   \hspace{1em} ARGOX Joint Ensemble & 2912.89 & 5420.28 & 6938.26 & 8346.38 \\ 
   \hspace{1em} Naive & 3104.72 & 5376.90 & 7087.40 & 8342.40 \\ 
   \multicolumn{1}{l}{Correlation} \\
 \hspace{1em} Ref \cite{ma2021covid} &0.95 & 0.89 & 0.84 & 0.63 \\ 
   \hspace{1em} ARGOX Joint Ensemble  & 0.95 & 0.87 & 0.85 & 0.56 \\ 
   \hspace{1em} Naive& 0.95 & 0.85 & 0.74 & 0.49 \\ 
   \hline
\end{tabular}
\caption{Comparison of different methods for state-level COVID-19 1 to 4 weeks ahead incremental case in Arizona (AZ). The MSE, MAE, and correlation are reported and best performed method is highlighted in boldface.} 
\end{table}       \begin{figure}[!h]    \centering  \includegraphics[width=0.6\linewidth, page=4]{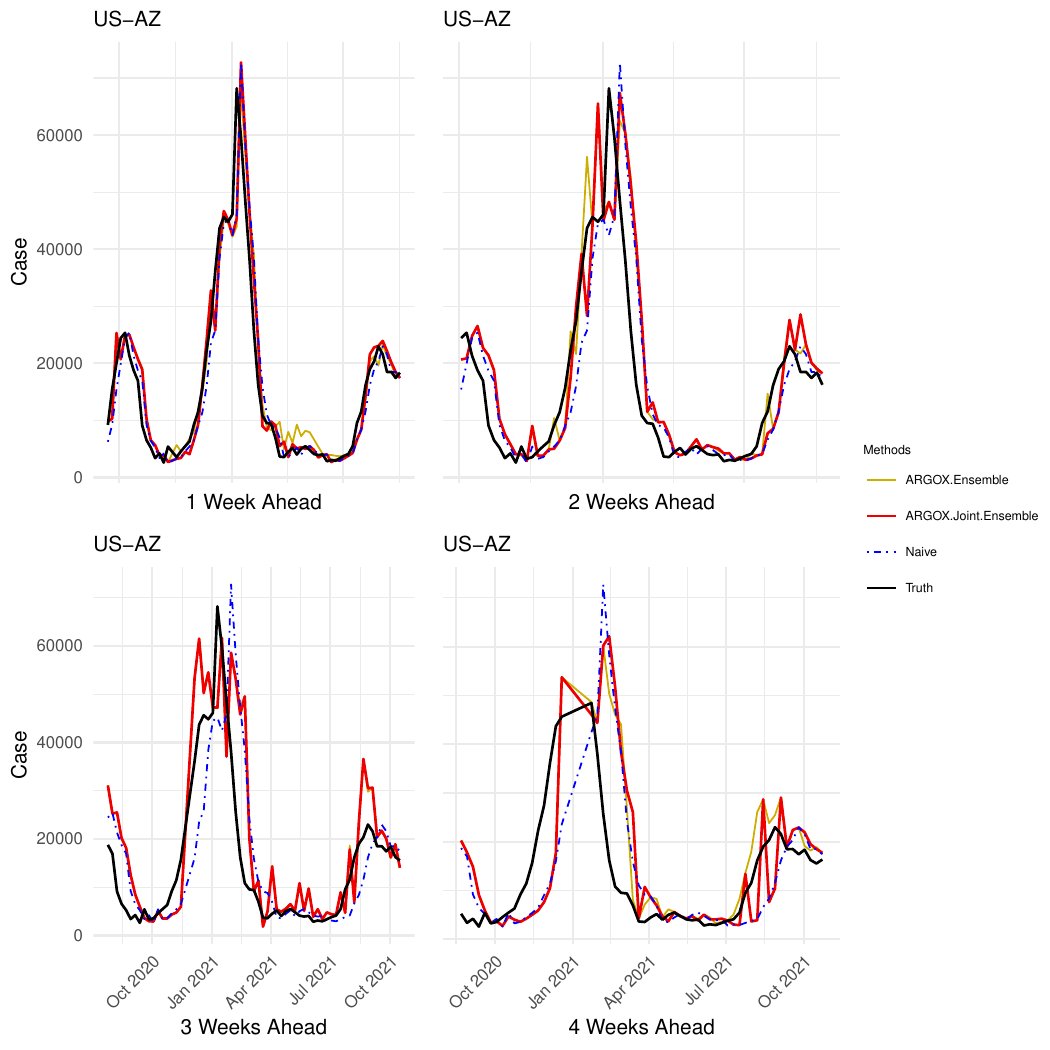}  \caption{Plots of the COVID-19 1 week (top left), 2 weeks (top right), 3 weeks (bottom left), and 4 weeks (bottom right) ahead estimates for Arizona (AZ). ARGOX-Ensemble is Ref \cite{ma2021covid}.}  \end{figure} \newpage   
\begin{table}[ht]
\centering
\begin{tabular}{lrrrr}
  \hline
& 1 Week Ahead & 2 Weeks Ahead & 3 Weeks Ahead & 4 Weeks Ahead \\ 
    \hline \multicolumn{1}{l}{RMSE} \\
 \hspace{1em} Ref \cite{ma2021covid} & 
25242.92 & 39728.81 & 45857.52 & 66118.03 \\ 
   \hspace{1em} ARGOX Joint Ensemble & 25124.43 & 40300.73 & 44564.72 & 69574.99 \\ 
   \hspace{1em} Naive & 27668.63 & 43948.75 & 59122.03 & 70159.79 \\ 
   \multicolumn{1}{l}{MAE} \\
  \hspace{1em}  Ref \cite{ma2021covid} &14974.97 & 23675.90 & 28970.25 & 39686.31 \\ 
   \hspace{1em} ARGOX Joint Ensemble & 14692.14 & 24189.96 & 28333.01 & 44165.27 \\ 
   \hspace{1em} Naive & 15999.16 & 25604.12 & 34255.35 & 43530.78 \\ 
   \multicolumn{1}{l}{Correlation} \\
 \hspace{1em} Ref \cite{ma2021covid} &0.94 & 0.89 & 0.91 & 0.46 \\ 
   \hspace{1em} ARGOX Joint Ensemble  & 0.94 & 0.87 & 0.92 & 0.39 \\ 
   \hspace{1em} Naive& 0.93 & 0.83 & 0.70 & 0.40 \\ 
   \hline
\end{tabular}
\caption{Comparison of different methods for state-level COVID-19 1 to 4 weeks ahead incremental case in California (CA). The MSE, MAE, and correlation are reported and best performed method is highlighted in boldface.} 
\end{table}       \begin{figure}[!h]    \centering  \includegraphics[width=0.6\linewidth, page=5]{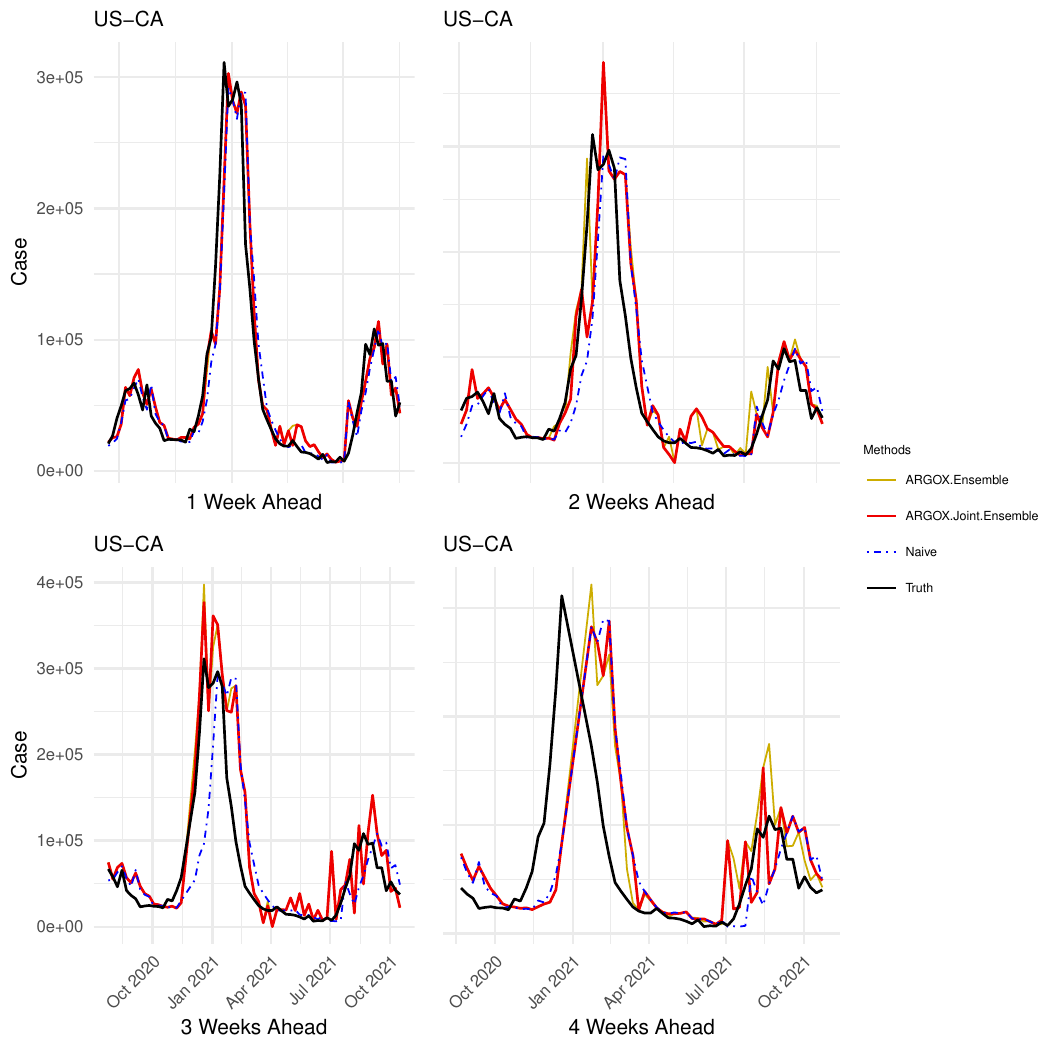}  \caption{Plots of the COVID-19 1 week (top left), 2 weeks (top right), 3 weeks (bottom left), and 4 weeks (bottom right) ahead estimates for California (CA). ARGOX-Ensemble is Ref \cite{ma2021covid}.}  \end{figure} \newpage   
\begin{table}[ht]
\centering
\begin{tabular}{lrrrr}
  \hline
& 1 Week Ahead & 2 Weeks Ahead & 3 Weeks Ahead & 4 Weeks Ahead \\ 
    \hline \multicolumn{1}{l}{RMSE} \\
 \hspace{1em} Ref \cite{ma2021covid} & 
2240.72 & 5136.79 & 5612.57 & 14617.59 \\ 
   \hspace{1em} ARGOX Joint Ensemble & 2226.81 & 3509.02 & 4781.78 & 14711.41 \\ 
   \hspace{1em} Naive & 2539.14 & 4429.68 & 5359.04 & 7152.79 \\ 
   \multicolumn{1}{l}{MAE} \\
  \hspace{1em}  Ref \cite{ma2021covid} &1526.76 & 2988.12 & 3528.98 & 6439.45 \\ 
   \hspace{1em} ARGOX Joint Ensemble & 1376.78 & 2219.28 & 3041.46 & 6707.70 \\ 
   \hspace{1em} Naive & 1602.03 & 2781.87 & 3494.86 & 4819.24 \\ 
   \multicolumn{1}{l}{Correlation} \\
 \hspace{1em} Ref \cite{ma2021covid} &0.96 & 0.90 & 0.77 & 0.54 \\ 
   \hspace{1em} ARGOX Joint Ensemble  & 0.97 & 0.92 & 0.78 & 0.53 \\ 
   \hspace{1em} Naive& 0.95 & 0.84 & 0.74 & 0.54 \\ 
   \hline
\end{tabular}
\caption{Comparison of different methods for state-level COVID-19 1 to 4 weeks ahead incremental case in Colorado (CO). The MSE, MAE, and correlation are reported and best performed method is highlighted in boldface.} 
\end{table}       \begin{figure}[!h]    \centering  \includegraphics[width=0.6\linewidth, page=6]{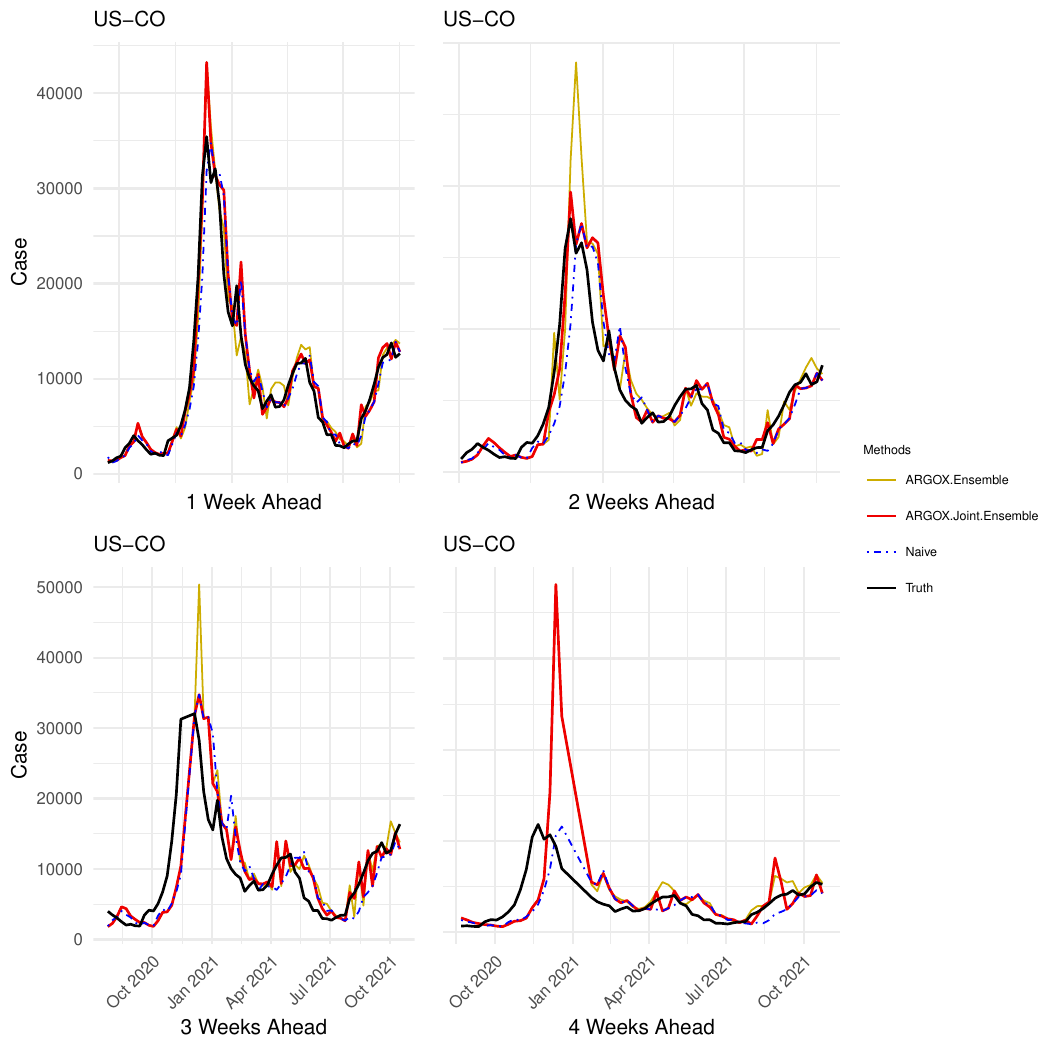}  \caption{Plots of the COVID-19 1 week (top left), 2 weeks (top right), 3 weeks (bottom left), and 4 weeks (bottom right) ahead estimates for Colorado (CO). ARGOX-Ensemble is Ref \cite{ma2021covid}.}  \end{figure} \newpage   
\begin{table}[ht]
\centering
\begin{tabular}{lrrrr}
  \hline
& 1 Week Ahead & 2 Weeks Ahead & 3 Weeks Ahead & 4 Weeks Ahead \\ 
    \hline \multicolumn{1}{l}{RMSE} \\
 \hspace{1em} Ref \cite{ma2021covid} & 
1988.61 & 2459.98 & 3341.58 & 4329.49 \\ 
   \hspace{1em} ARGOX Joint Ensemble & 1712.39 & 2129.20 & 3339.84 & 4801.53 \\ 
   \hspace{1em} Naive & 1830.87 & 2551.39 & 3000.41 & 3888.72 \\ 
   \multicolumn{1}{l}{MAE} \\
  \hspace{1em}  Ref \cite{ma2021covid} &1228.15 & 1716.27 & 2176.69 & 2341.18 \\ 
   \hspace{1em} ARGOX Joint Ensemble & 1039.10 & 1383.01 & 2168.80 & 2871.35 \\ 
   \hspace{1em} Naive & 1176.53 & 1666.55 & 2155.11 & 2896.72 \\ 
   \multicolumn{1}{l}{Correlation} \\
 \hspace{1em} Ref \cite{ma2021covid} &0.93 & 0.90 & 0.89 & 0.78 \\ 
   \hspace{1em} ARGOX Joint Ensemble  & 0.94 & 0.92 & 0.89 & 0.72 \\ 
   \hspace{1em} Naive& 0.93 & 0.87 & 0.82 & 0.62 \\ 
   \hline
\end{tabular}
\caption{Comparison of different methods for state-level COVID-19 1 to 4 weeks ahead incremental case in Connecticut (CT). The MSE, MAE, and correlation are reported and best performed method is highlighted in boldface.} 
\end{table}       \begin{figure}[!h]    \centering  \includegraphics[width=0.6\linewidth, page=7]{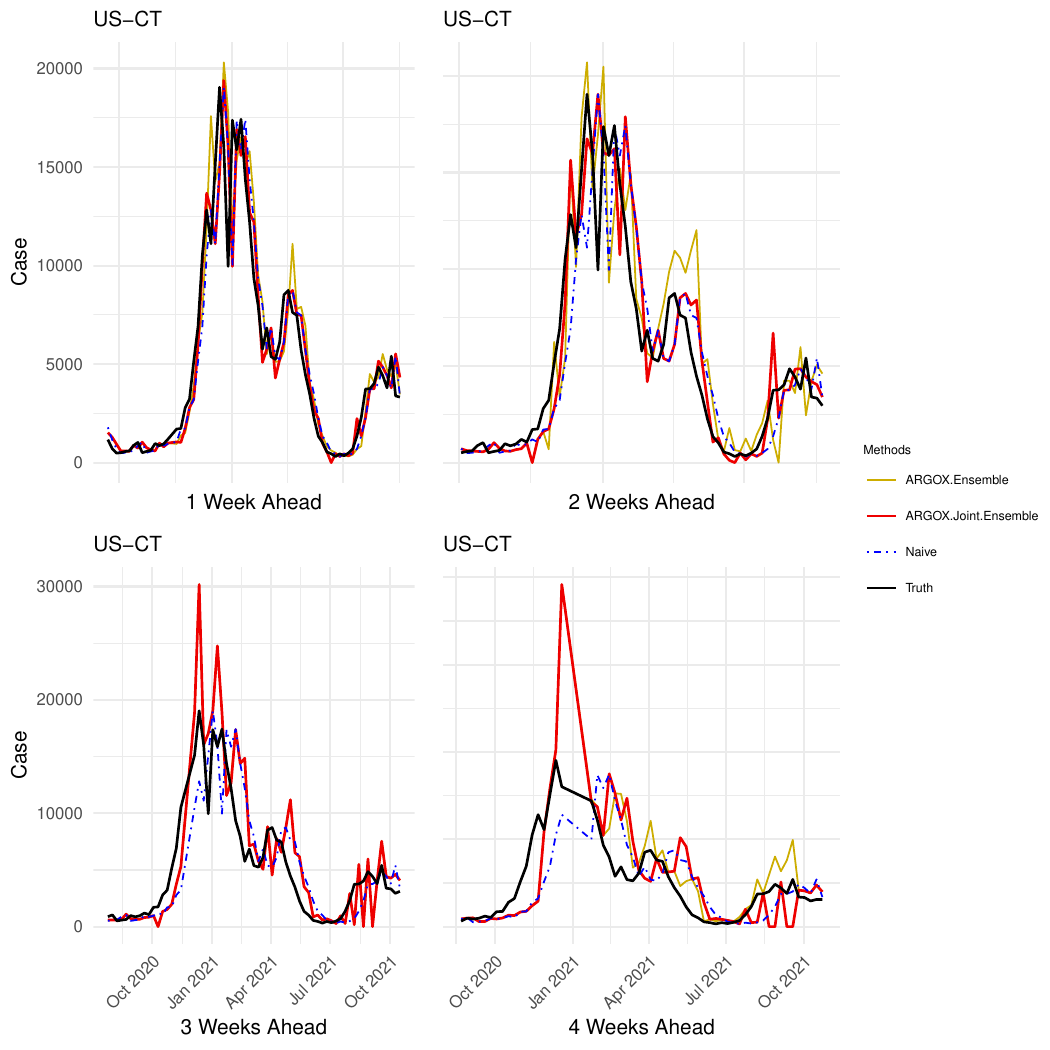}  \caption{Plots of the COVID-19 1 week (top left), 2 weeks (top right), 3 weeks (bottom left), and 4 weeks (bottom right) ahead estimates for Connecticut (CT). ARGOX-Ensemble is Ref \cite{ma2021covid}.}  \end{figure} \newpage   
\begin{table}[ht]
\centering
\begin{tabular}{lrrrr}
  \hline
& 1 Week Ahead & 2 Weeks Ahead & 3 Weeks Ahead & 4 Weeks Ahead \\ 
    \hline \multicolumn{1}{l}{RMSE} \\
 \hspace{1em} Ref \cite{ma2021covid} & 
229.81 & 267.16 & 380.20 & 485.26 \\ 
   \hspace{1em} ARGOX Joint Ensemble & 242.20 & 261.48 & 379.46 & 491.45 \\ 
   \hspace{1em} Naive & 219.04 & 284.33 & 376.78 & 510.21 \\ 
   \multicolumn{1}{l}{MAE} \\
  \hspace{1em}  Ref \cite{ma2021covid} &174.69 & 204.17 & 246.55 & 354.17 \\ 
   \hspace{1em} ARGOX Joint Ensemble & 166.44 & 183.14 & 242.15 & 362.96 \\ 
   \hspace{1em} Naive & 156.10 & 218.00 & 281.71 & 385.34 \\ 
   \multicolumn{1}{l}{Correlation} \\
 \hspace{1em} Ref \cite{ma2021covid} &0.91 & 0.89 & 0.85 & 0.67 \\ 
   \hspace{1em} ARGOX Joint Ensemble  & 0.91 & 0.90 & 0.85 & 0.61 \\ 
   \hspace{1em} Naive& 0.92 & 0.86 & 0.76 & 0.49 \\ 
   \hline
\end{tabular}
\caption{Comparison of different methods for state-level COVID-19 1 to 4 weeks ahead incremental case in District of Columbia (DC). The MSE, MAE, and correlation are reported and best performed method is highlighted in boldface.} 
\end{table}       \begin{figure}[!h]    \centering  \includegraphics[width=0.6\linewidth, page=8]{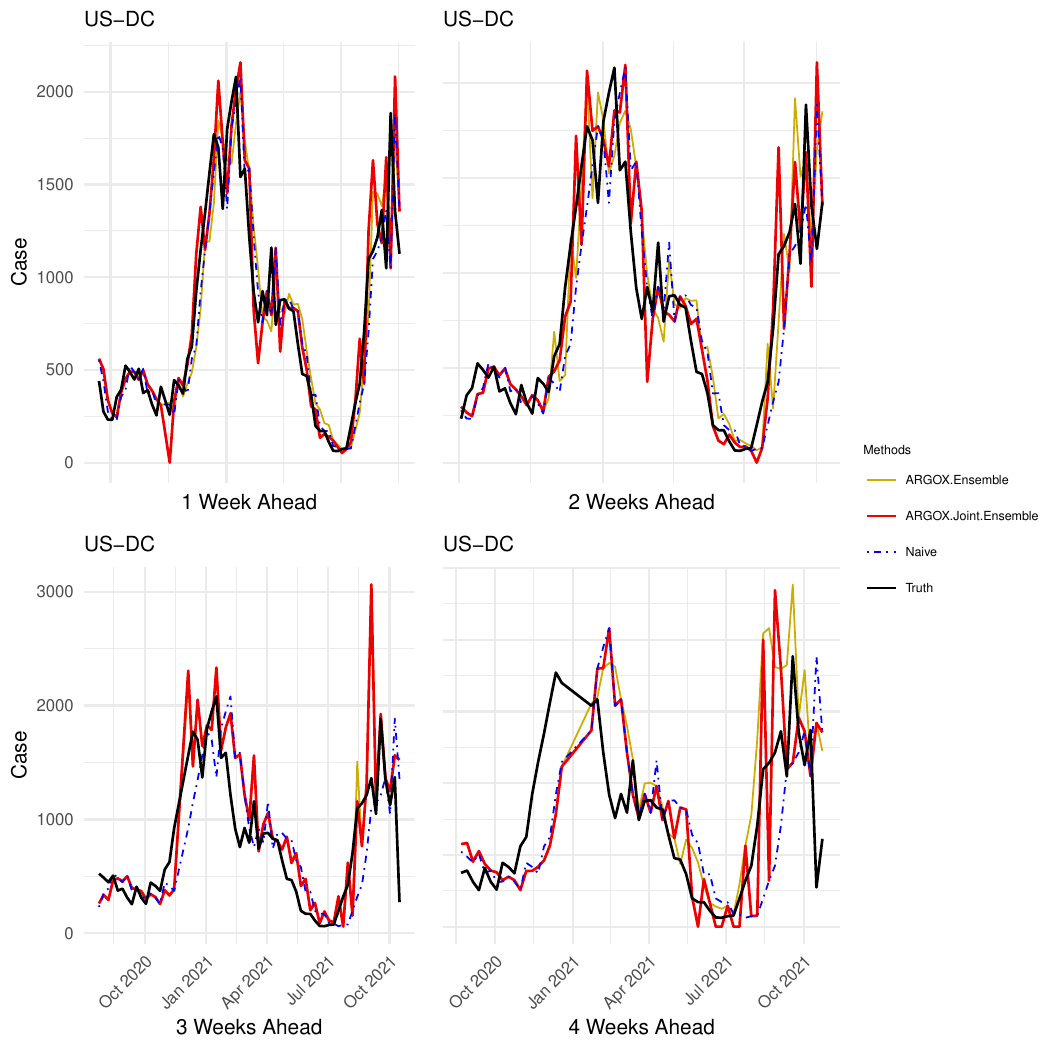}  \caption{Plots of the COVID-19 1 week (top left), 2 weeks (top right), 3 weeks (bottom left), and 4 weeks (bottom right) ahead estimates for District of Columbia (DC). ARGOX-Ensemble is Ref \cite{ma2021covid}.}  \end{figure} \newpage   
\begin{table}[ht]
\centering
\begin{tabular}{lrrrr}
  \hline
& 1 Week Ahead & 2 Weeks Ahead & 3 Weeks Ahead & 4 Weeks Ahead \\ 
    \hline \multicolumn{1}{l}{RMSE} \\
 \hspace{1em} Ref \cite{ma2021covid} & 
509.89 & 672.64 & 960.74 & 1224.22 \\ 
   \hspace{1em} ARGOX Joint Ensemble & 499.60 & 683.40 & 937.90 & 1251.42 \\ 
   \hspace{1em} Naive & 493.74 & 750.41 & 949.59 & 1274.14 \\ 
   \multicolumn{1}{l}{MAE} \\
  \hspace{1em}  Ref \cite{ma2021covid} &353.88 & 472.41 & 623.48 & 832.16 \\ 
   \hspace{1em} ARGOX Joint Ensemble & 360.16 & 491.77 & 608.52 & 885.55 \\ 
   \hspace{1em} Naive & 357.60 & 534.94 & 666.21 & 928.48 \\ 
   \multicolumn{1}{l}{Correlation} \\
 \hspace{1em} Ref \cite{ma2021covid} &0.95 & 0.93 & 0.86 & 0.62 \\ 
   \hspace{1em} ARGOX Joint Ensemble  & 0.94 & 0.91 & 0.86 & 0.57 \\ 
   \hspace{1em} Naive& 0.94 & 0.87 & 0.79 & 0.53 \\ 
   \hline
\end{tabular}
\caption{Comparison of different methods for state-level COVID-19 1 to 4 weeks ahead incremental case in Delaware (DE). The MSE, MAE, and correlation are reported and best performed method is highlighted in boldface.} 
\end{table}       \begin{figure}[!h]    \centering  \includegraphics[width=0.6\linewidth, page=9]{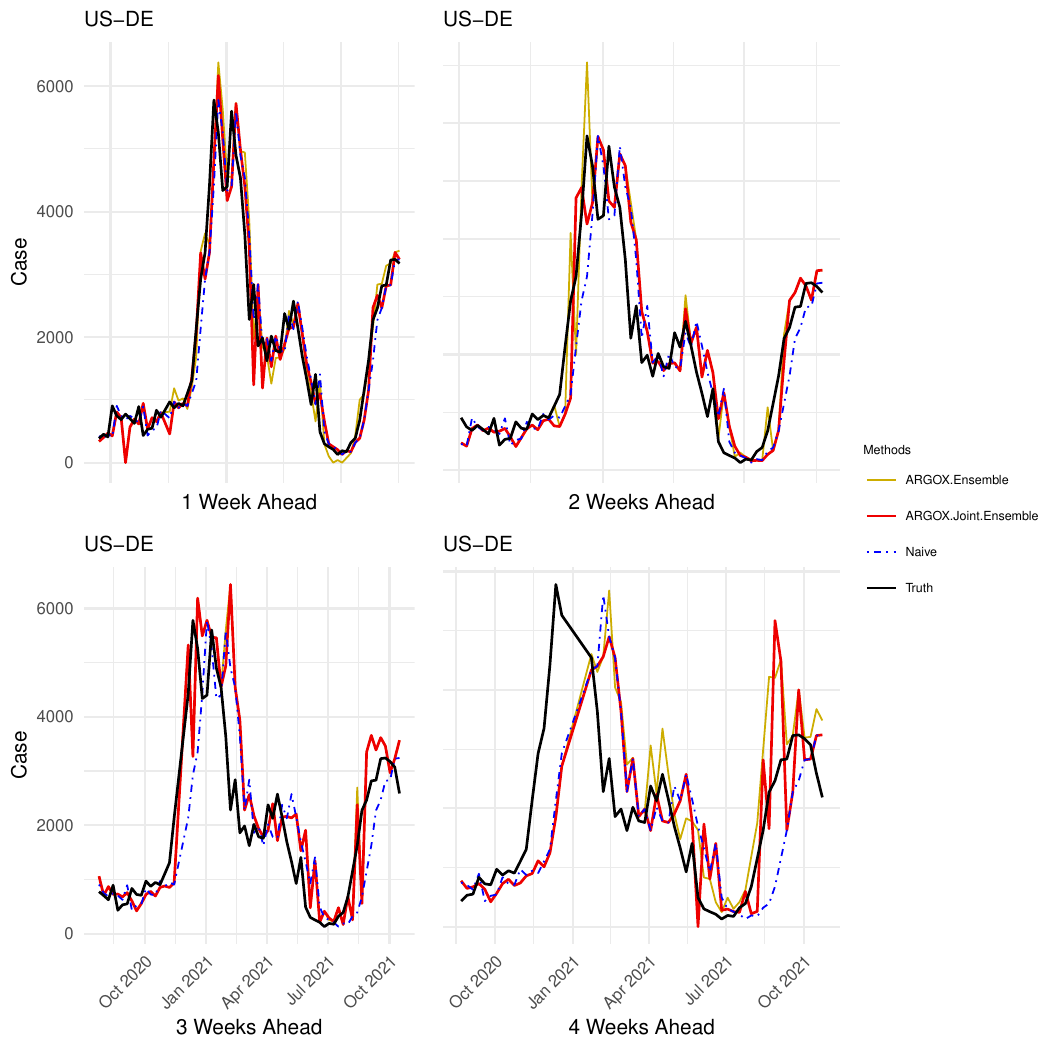}  \caption{Plots of the COVID-19 1 week (top left), 2 weeks (top right), 3 weeks (bottom left), and 4 weeks (bottom right) ahead estimates for Delaware (DE). ARGOX-Ensemble is Ref \cite{ma2021covid}.}  \end{figure} \newpage   
\begin{table}[ht]
\centering
\begin{tabular}{lrrrr}
  \hline
& 1 Week Ahead & 2 Weeks Ahead & 3 Weeks Ahead & 4 Weeks Ahead \\ 
    \hline \multicolumn{1}{l}{RMSE} \\
 \hspace{1em} Ref \cite{ma2021covid} & 
12739.27 & 28246.55 & 38513.29 & 44610.79 \\ 
   \hspace{1em} ARGOX Joint Ensemble & 12674.36 & 29736.92 & 42591.51 & 51232.58 \\ 
   \hspace{1em} Naive & 13856.70 & 23984.27 & 33301.60 & 42992.79 \\ 
   \multicolumn{1}{l}{MAE} \\
  \hspace{1em}  Ref \cite{ma2021covid} &9953.86 & 19201.88 & 25667.93 & 30360.96 \\ 
   \hspace{1em} ARGOX Joint Ensemble & 9870.77 & 20913.76 & 27761.80 & 36410.97 \\ 
   \hspace{1em} Naive & 10186.46 & 17912.46 & 24746.48 & 31780.38 \\ 
   \multicolumn{1}{l}{Correlation} \\
 \hspace{1em} Ref \cite{ma2021covid} &0.96 & 0.87 & 0.77 & 0.87 \\ 
   \hspace{1em} ARGOX Joint Ensemble  & 0.96 & 0.85 & 0.77 & 0.55 \\ 
   \hspace{1em} Naive& 0.93 & 0.79 & 0.62 & 0.39 \\ 
   \hline
\end{tabular}
\caption{Comparison of different methods for state-level COVID-19 1 to 4 weeks ahead incremental case in Florida (FL). The MSE, MAE, and correlation are reported and best performed method is highlighted in boldface.} 
\end{table}       \begin{figure}[!h]    \centering  \includegraphics[width=0.6\linewidth, page=10]{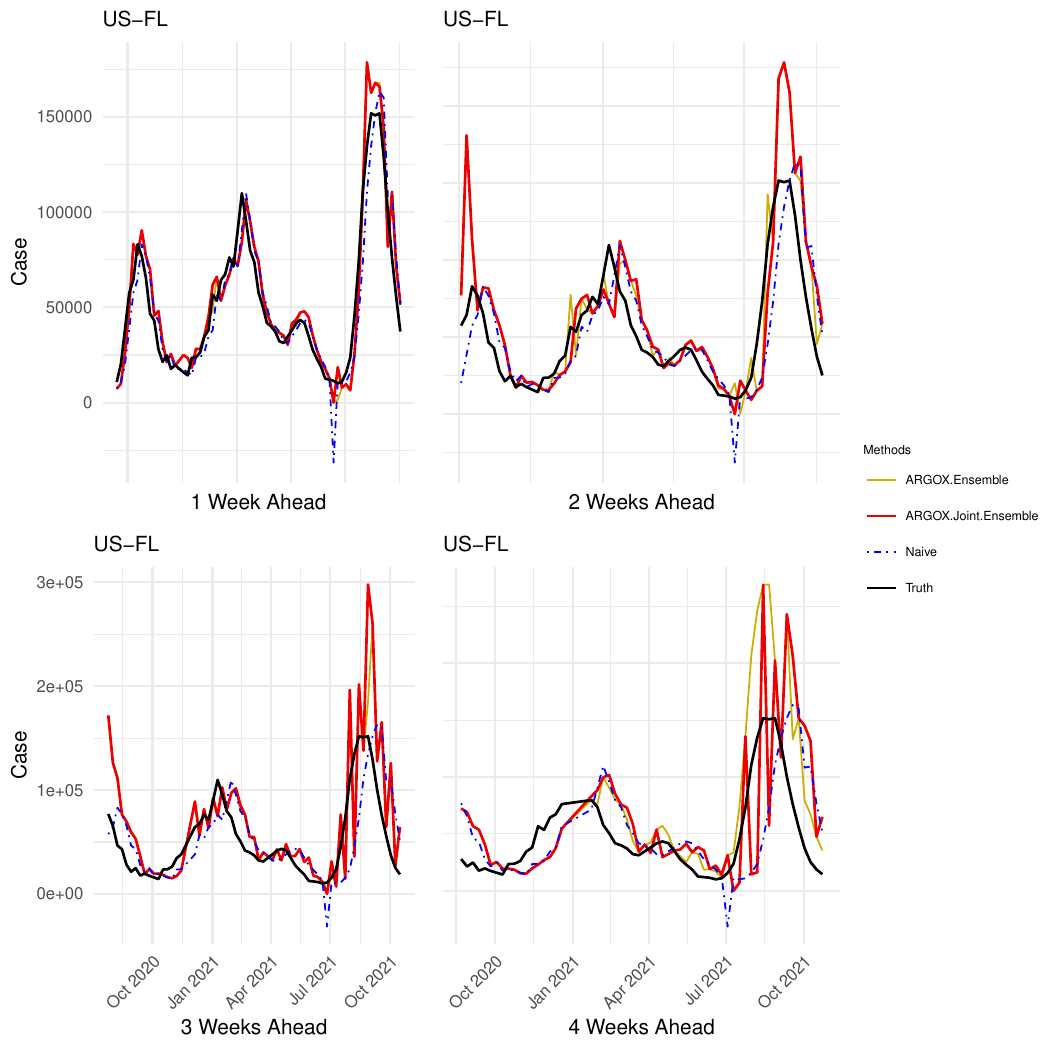}  \caption{Plots of the COVID-19 1 week (top left), 2 weeks (top right), 3 weeks (bottom left), and 4 weeks (bottom right) ahead estimates for Florida (FL). ARGOX-Ensemble is Ref \cite{ma2021covid}.}  \end{figure} \newpage   
\begin{table}[ht]
\centering
\begin{tabular}{lrrrr}
  \hline
& 1 Week Ahead & 2 Weeks Ahead & 3 Weeks Ahead & 4 Weeks Ahead \\ 
    \hline \multicolumn{1}{l}{RMSE} \\
 \hspace{1em} Ref \cite{ma2021covid} & 
6515.31 & 10735.52 & 15518.68 & 24139.44 \\ 
   \hspace{1em} ARGOX Joint Ensemble & 6506.05 & 11733.52 & 16289.14 & 16670.24 \\ 
   \hspace{1em} Naive & 7160.32 & 11161.17 & 15105.24 & 17835.22 \\ 
   \multicolumn{1}{l}{MAE} \\
  \hspace{1em}  Ref \cite{ma2021covid} &4562.84 & 7237.47 & 10240.10 & 14578.32 \\ 
   \hspace{1em} ARGOX Joint Ensemble & 4507.80 & 7322.46 & 10426.15 & 12706.10 \\ 
   \hspace{1em} Naive & 4764.54 & 8215.93 & 11232.21 & 13228.34 \\ 
   \multicolumn{1}{l}{Correlation} \\
 \hspace{1em} Ref \cite{ma2021covid} &0.93 & 0.85 & 0.82 & 0.75 \\ 
   \hspace{1em} ARGOX Joint Ensemble  & 0.93 & 0.81 & 0.81 & 0.69 \\ 
   \hspace{1em} Naive& 0.92 & 0.79 & 0.63 & 0.45 \\ 
   \hline
\end{tabular}
\caption{Comparison of different methods for state-level COVID-19 1 to 4 weeks ahead incremental case in Georgia (GA). The MSE, MAE, and correlation are reported and best performed method is highlighted in boldface.} 
\end{table}       \begin{figure}[!h]    \centering  \includegraphics[width=0.6\linewidth, page=11]{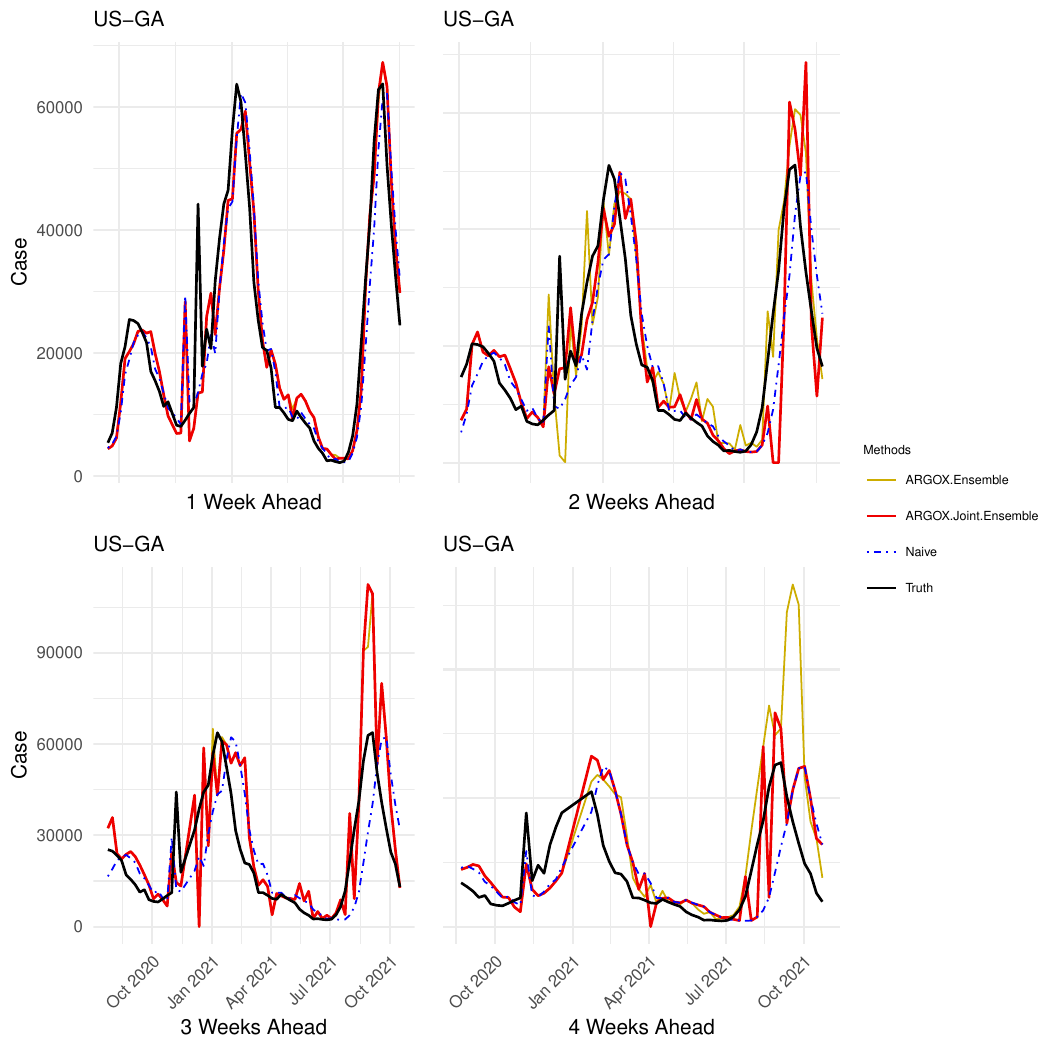}  \caption{Plots of the COVID-19 1 week (top left), 2 weeks (top right), 3 weeks (bottom left), and 4 weeks (bottom right) ahead estimates for Georgia (GA). ARGOX-Ensemble is Ref \cite{ma2021covid}.}  \end{figure} \newpage   
\begin{table}[ht]
\centering
\begin{tabular}{lrrrr}
  \hline
& 1 Week Ahead & 2 Weeks Ahead & 3 Weeks Ahead & 4 Weeks Ahead \\ 
    \hline \multicolumn{1}{l}{RMSE} \\
 \hspace{1em} Ref \cite{ma2021covid} & 
407.73 & 709.11 & 1189.88 & 1425.96 \\ 
   \hspace{1em} ARGOX Joint Ensemble & 374.64 & 746.60 & 1038.59 & 1533.96 \\ 
   \hspace{1em} Naive & 442.55 & 755.12 & 1086.93 & 1400.91 \\ 
   \multicolumn{1}{l}{MAE} \\
  \hspace{1em}  Ref \cite{ma2021covid} &249.88 & 437.43 & 671.82 & 668.96 \\ 
   \hspace{1em} ARGOX Joint Ensemble & 235.24 & 449.51 & 607.11 & 833.55 \\ 
   \hspace{1em} Naive & 269.79 & 463.40 & 666.98 & 837.72 \\ 
   \multicolumn{1}{l}{Correlation} \\
 \hspace{1em} Ref \cite{ma2021covid} &0.97 & 0.92 & 0.85 & 0.86 \\ 
   \hspace{1em} ARGOX Joint Ensemble  & 0.97 & 0.93 & 0.84 & 0.69 \\ 
   \hspace{1em} Naive& 0.94 & 0.83 & 0.66 & 0.47 \\ 
   \hline
\end{tabular}
\caption{Comparison of different methods for state-level COVID-19 1 to 4 weeks ahead incremental case in Hawaii (HI). The MSE, MAE, and correlation are reported and best performed method is highlighted in boldface.} 
\end{table}       \begin{figure}[!h]    \centering  \includegraphics[width=0.6\linewidth, page=12]{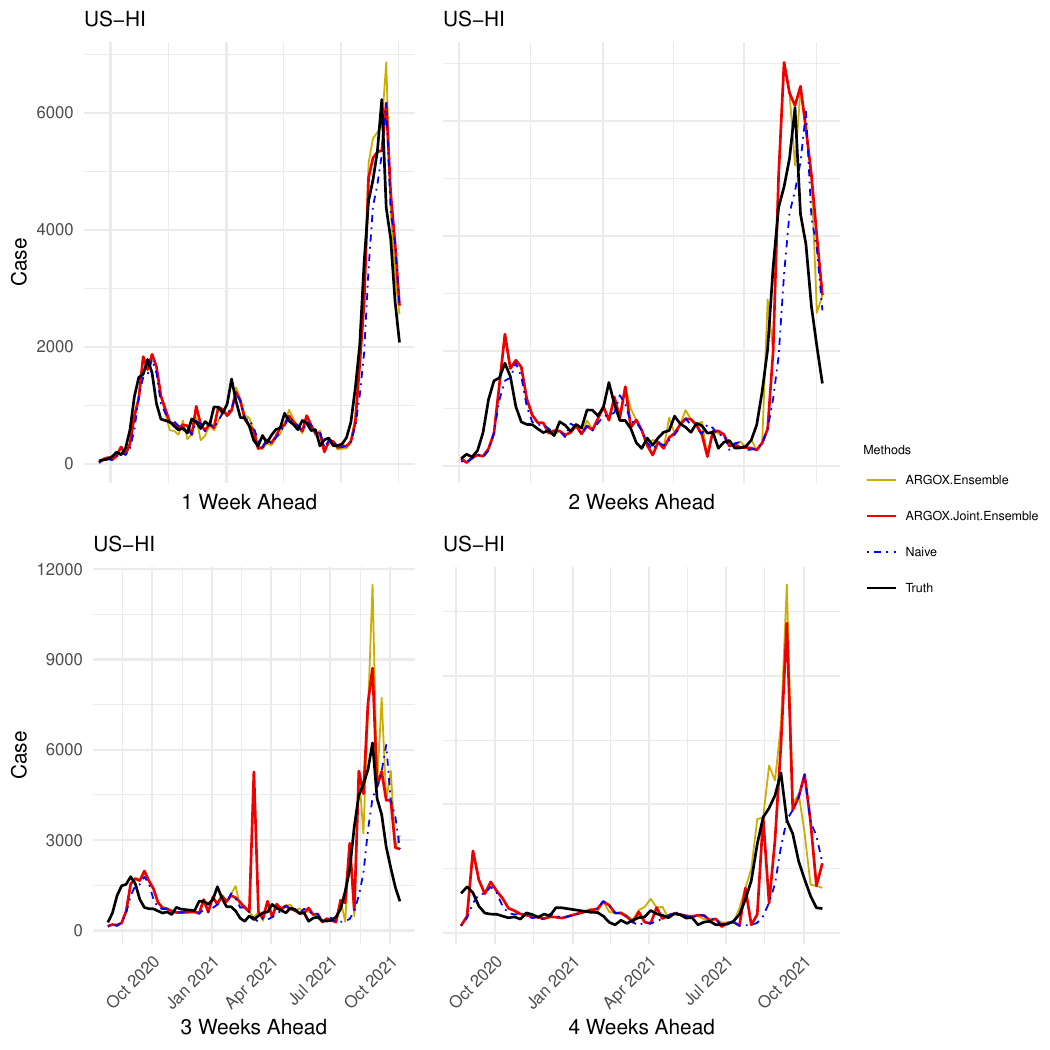}  \caption{Plots of the COVID-19 1 week (top left), 2 weeks (top right), 3 weeks (bottom left), and 4 weeks (bottom right) ahead estimates for Hawaii (HI). ARGOX-Ensemble is Ref \cite{ma2021covid}.}  \end{figure} \newpage   
\begin{table}[ht]
\centering
\begin{tabular}{lrrrr}
  \hline
& 1 Week Ahead & 2 Weeks Ahead & 3 Weeks Ahead & 4 Weeks Ahead \\ 
    \hline \multicolumn{1}{l}{RMSE} \\
 \hspace{1em} Ref \cite{ma2021covid} & 
2379.49 & 5478.46 & 5485.34 & 6700.35 \\ 
   \hspace{1em} ARGOX Joint Ensemble & 2360.10 & 4014.24 & 4924.91 & 6744.54 \\ 
   \hspace{1em} Naive & 2618.83 & 4270.16 & 5314.03 & 6374.89 \\ 
   \multicolumn{1}{l}{MAE} \\
  \hspace{1em}  Ref \cite{ma2021covid} &1411.44 & 2814.51 & 2904.52 & 3496.38 \\ 
   \hspace{1em} ARGOX Joint Ensemble & 1386.04 & 2130.49 & 2465.15 & 3648.86 \\ 
   \hspace{1em} Naive & 1479.63 & 2465.82 & 3001.79 & 3829.91 \\ 
   \multicolumn{1}{l}{Correlation} \\
 \hspace{1em} Ref \cite{ma2021covid} &0.94 & 0.74 & 0.58 & 0.46 \\ 
   \hspace{1em} ARGOX Joint Ensemble  & 0.94 & 0.78 & 0.60 & 0.45 \\ 
   \hspace{1em} Naive& 0.90 & 0.75 & 0.55 & 0.46 \\ 
   \hline
\end{tabular}
\caption{Comparison of different methods for state-level COVID-19 1 to 4 weeks ahead incremental case in Iowa (IA). The MSE, MAE, and correlation are reported and best performed method is highlighted in boldface.} 
\end{table}       \begin{figure}[!h]    \centering  \includegraphics[width=0.6\linewidth, page=13]{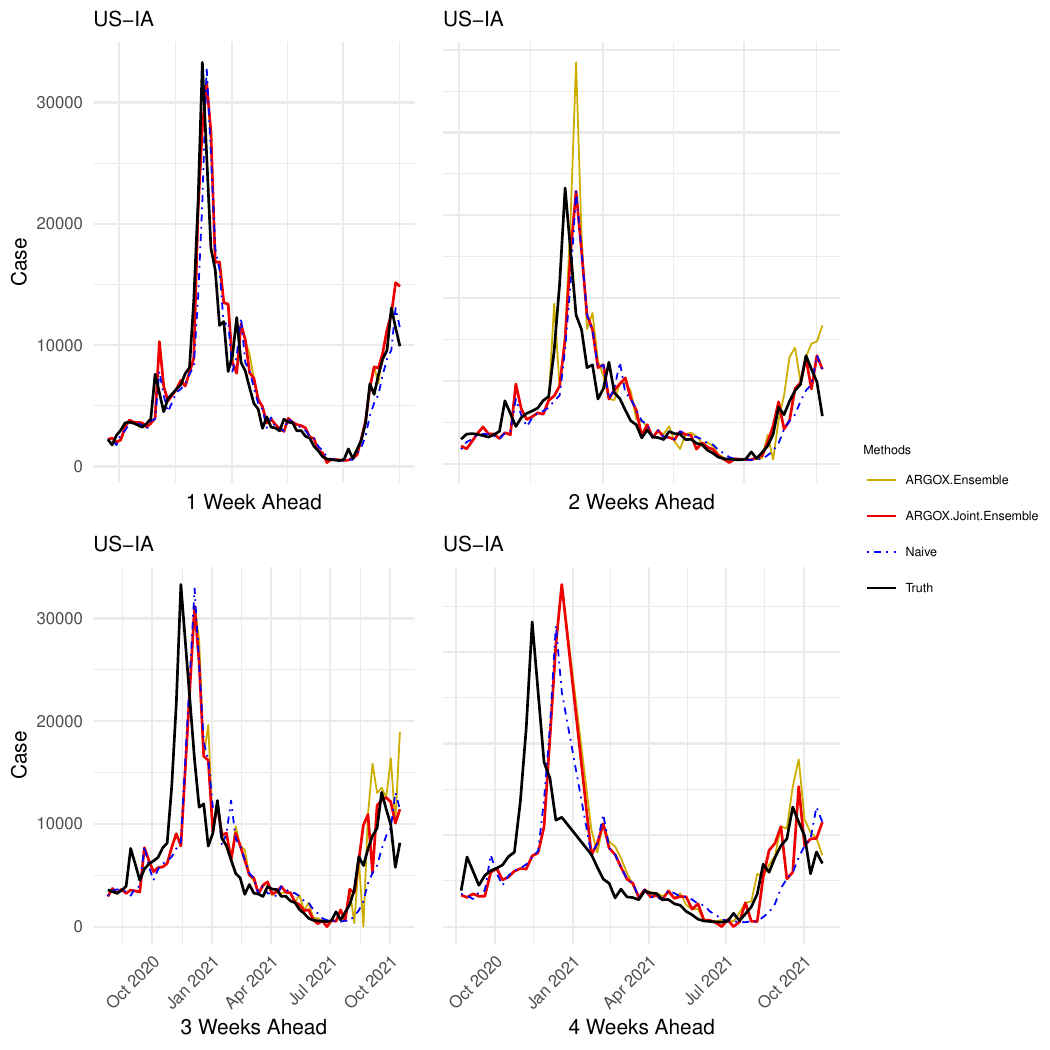}  \caption{Plots of the COVID-19 1 week (top left), 2 weeks (top right), 3 weeks (bottom left), and 4 weeks (bottom right) ahead estimates for Iowa (IA). ARGOX-Ensemble is Ref \cite{ma2021covid}.}  \end{figure} \newpage   
\begin{table}[ht]
\centering
\begin{tabular}{lrrrr}
  \hline
& 1 Week Ahead & 2 Weeks Ahead & 3 Weeks Ahead & 4 Weeks Ahead \\ 
    \hline \multicolumn{1}{l}{RMSE} \\
 \hspace{1em} Ref \cite{ma2021covid} & 
898.78 & 1309.16 & 1635.41 & 2116.53 \\ 
   \hspace{1em} ARGOX Joint Ensemble & 824.72 & 1200.15 & 1622.44 & 2071.22 \\ 
   \hspace{1em} Naive & 854.24 & 1360.61 & 1739.25 & 2152.20 \\ 
   \multicolumn{1}{l}{MAE} \\
  \hspace{1em}  Ref \cite{ma2021covid} &622.61 & 934.79 & 1177.53 & 1482.86 \\ 
   \hspace{1em} ARGOX Joint Ensemble & 571.24 & 852.32 & 1158.16 & 1581.39 \\ 
   \hspace{1em} Naive & 624.13 & 990.90 & 1310.22 & 1691.64 \\ 
   \multicolumn{1}{l}{Correlation} \\
 \hspace{1em} Ref \cite{ma2021covid} &0.96 & 0.93 & 0.86 & 0.80 \\ 
   \hspace{1em} ARGOX Joint Ensemble  & 0.96 & 0.92 & 0.86 & 0.77 \\ 
   \hspace{1em} Naive& 0.96 & 0.89 & 0.81 & 0.75 \\ 
   \hline
\end{tabular}
\caption{Comparison of different methods for state-level COVID-19 1 to 4 weeks ahead incremental case in Idaho (ID). The MSE, MAE, and correlation are reported and best performed method is highlighted in boldface.} 
\end{table}       \begin{figure}[!h]    \centering  \includegraphics[width=0.6\linewidth, page=14]{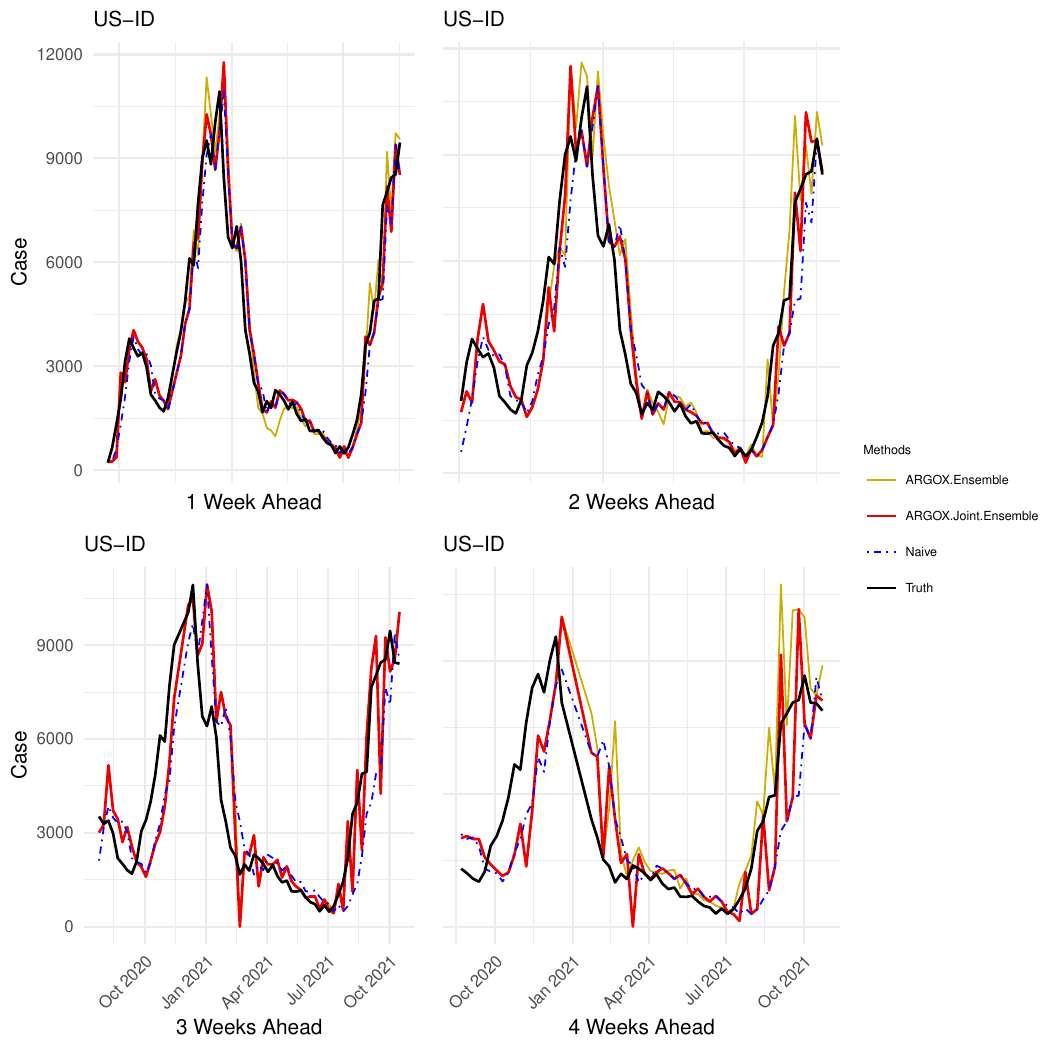}  \caption{Plots of the COVID-19 1 week (top left), 2 weeks (top right), 3 weeks (bottom left), and 4 weeks (bottom right) ahead estimates for Idaho (ID). ARGOX-Ensemble is Ref \cite{ma2021covid}.}  \end{figure} \newpage   
\begin{table}[ht]
\centering
\begin{tabular}{lrrrr}
  \hline
& 1 Week Ahead & 2 Weeks Ahead & 3 Weeks Ahead & 4 Weeks Ahead \\ 
    \hline \multicolumn{1}{l}{RMSE} \\
 \hspace{1em} Ref \cite{ma2021covid} & 
5005.50 & 11347.75 & 15608.17 & 28020.26 \\ 
   \hspace{1em} ARGOX Joint Ensemble & 4853.04 & 10372.31 & 15654.46 & 28055.18 \\ 
   \hspace{1em} Naive & 5885.99 & 10358.79 & 13145.72 & 17128.84 \\ 
   \multicolumn{1}{l}{MAE} \\
  \hspace{1em}  Ref \cite{ma2021covid} &3512.41 & 7210.79 & 10035.73 & 14601.50 \\ 
   \hspace{1em} ARGOX Joint Ensemble & 3404.77 & 7236.55 & 10039.67 & 14855.45 \\ 
   \hspace{1em} Naive & 3921.91 & 7072.78 & 9189.97 & 11991.71 \\ 
   \multicolumn{1}{l}{Correlation} \\
 \hspace{1em} Ref \cite{ma2021covid} &0.98 & 0.92 & 0.68 & 0.50 \\ 
   \hspace{1em} ARGOX Joint Ensemble  & 0.98 & 0.92 & 0.68 & 0.51 \\ 
   \hspace{1em} Naive& 0.95 & 0.85 & 0.73 & 0.56 \\ 
   \hline
\end{tabular}
\caption{Comparison of different methods for state-level COVID-19 1 to 4 weeks ahead incremental case in Illinois (IL). The MSE, MAE, and correlation are reported and best performed method is highlighted in boldface.} 
\end{table}       \begin{figure}[!h]    \centering  \includegraphics[width=0.6\linewidth, page=15]{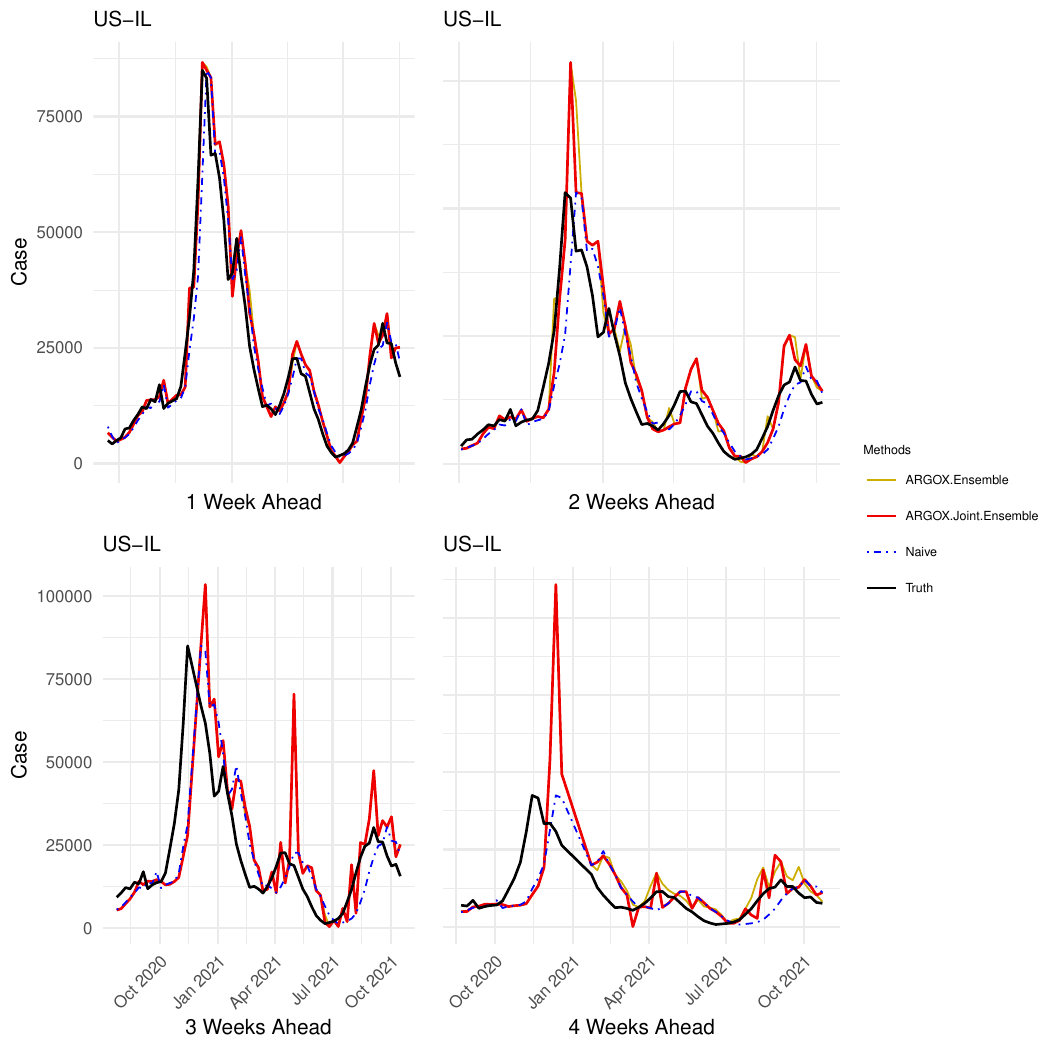}  \caption{Plots of the COVID-19 1 week (top left), 2 weeks (top right), 3 weeks (bottom left), and 4 weeks (bottom right) ahead estimates for Illinois (IL). ARGOX-Ensemble is Ref \cite{ma2021covid}.}  \end{figure} \newpage   
\begin{table}[ht]
\centering
\begin{tabular}{lrrrr}
  \hline
& 1 Week Ahead & 2 Weeks Ahead & 3 Weeks Ahead & 4 Weeks Ahead \\ 
    \hline \multicolumn{1}{l}{RMSE} \\
 \hspace{1em} Ref \cite{ma2021covid} & 
2981.63 & 6945.91 & 7516.61 & 16349.13 \\ 
   \hspace{1em} ARGOX Joint Ensemble & 2764.89 & 4780.42 & 7599.93 & 15409.21 \\ 
   \hspace{1em} Naive & 3486.17 & 6061.26 & 7779.50 & 10912.03 \\ 
   \multicolumn{1}{l}{MAE} \\
  \hspace{1em}  Ref \cite{ma2021covid} &2144.38 & 3724.54 & 5331.85 & 9173.61 \\ 
   \hspace{1em} ARGOX Joint Ensemble & 2005.22 & 3143.17 & 5380.83 & 8786.84 \\ 
   \hspace{1em} Naive & 2445.19 & 4046.70 & 5393.81 & 7870.17 \\ 
   \multicolumn{1}{l}{Correlation} \\
 \hspace{1em} Ref \cite{ma2021covid} &0.97 & 0.91 & 0.87 & 0.70 \\ 
   \hspace{1em} ARGOX Joint Ensemble  & 0.98 & 0.94 & 0.87 & 0.67 \\ 
   \hspace{1em} Naive& 0.96 & 0.88 & 0.80 & 0.55 \\ 
   \hline
\end{tabular}
\caption{Comparison of different methods for state-level COVID-19 1 to 4 weeks ahead incremental case in Indiana (IN). The MSE, MAE, and correlation are reported and best performed method is highlighted in boldface.} 
\end{table}       \begin{figure}[!h]    \centering  \includegraphics[width=0.6\linewidth, page=16]{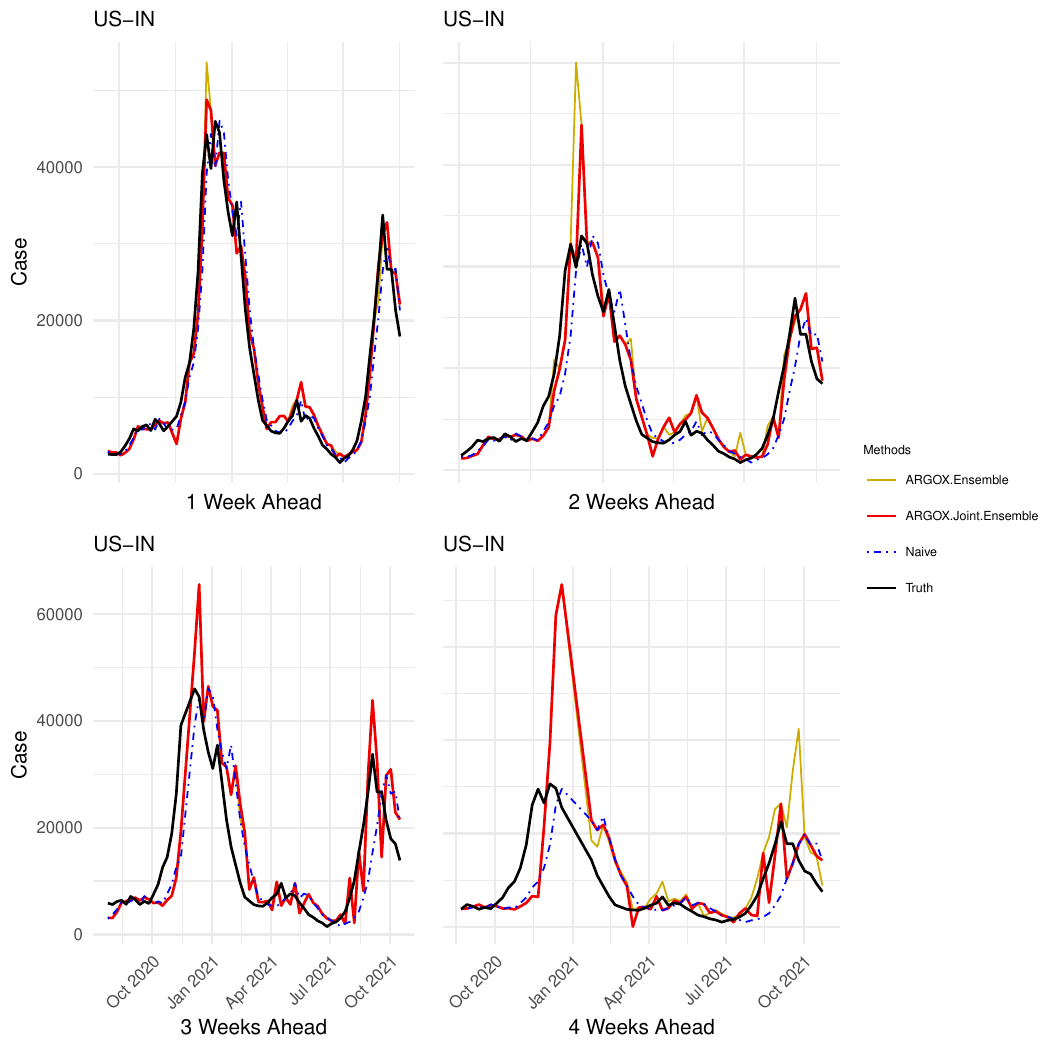}  \caption{Plots of the COVID-19 1 week (top left), 2 weeks (top right), 3 weeks (bottom left), and 4 weeks (bottom right) ahead estimates for Indiana (IN). ARGOX-Ensemble is Ref \cite{ma2021covid}.}  \end{figure} \newpage   
\begin{table}[ht]
\centering
\begin{tabular}{lrrrr}
  \hline
& 1 Week Ahead & 2 Weeks Ahead & 3 Weeks Ahead & 4 Weeks Ahead \\ 
    \hline \multicolumn{1}{l}{RMSE} \\
 \hspace{1em} Ref \cite{ma2021covid} & 
1970.24 & 3244.43 & 4450.01 & 8547.97 \\ 
   \hspace{1em} ARGOX Joint Ensemble & 1903.74 & 2305.29 & 3137.05 & 8617.44 \\ 
   \hspace{1em} Naive & 1810.00 & 2694.03 & 3260.63 & 4393.65 \\ 
   \multicolumn{1}{l}{MAE} \\
  \hspace{1em}  Ref \cite{ma2021covid} &1462.01 & 2124.59 & 2762.95 & 3870.69 \\ 
   \hspace{1em} ARGOX Joint Ensemble & 1385.23 & 1457.81 & 1984.88 & 4138.98 \\ 
   \hspace{1em} Naive & 1103.09 & 1816.66 & 2225.41 & 3029.38 \\ 
   \multicolumn{1}{l}{Correlation} \\
 \hspace{1em} Ref \cite{ma2021covid} &0.93 & 0.84 & 0.69 & 0.62 \\ 
   \hspace{1em} ARGOX Joint Ensemble  & 0.93 & 0.91 & 0.84 & 0.60 \\ 
   \hspace{1em} Naive& 0.94 & 0.86 & 0.79 & 0.59 \\ 
   \hline
\end{tabular}
\caption{Comparison of different methods for state-level COVID-19 1 to 4 weeks ahead incremental case in Kansas (KS). The MSE, MAE, and correlation are reported and best performed method is highlighted in boldface.} 
\end{table}       \begin{figure}[!h]    \centering  \includegraphics[width=0.6\linewidth, page=17]{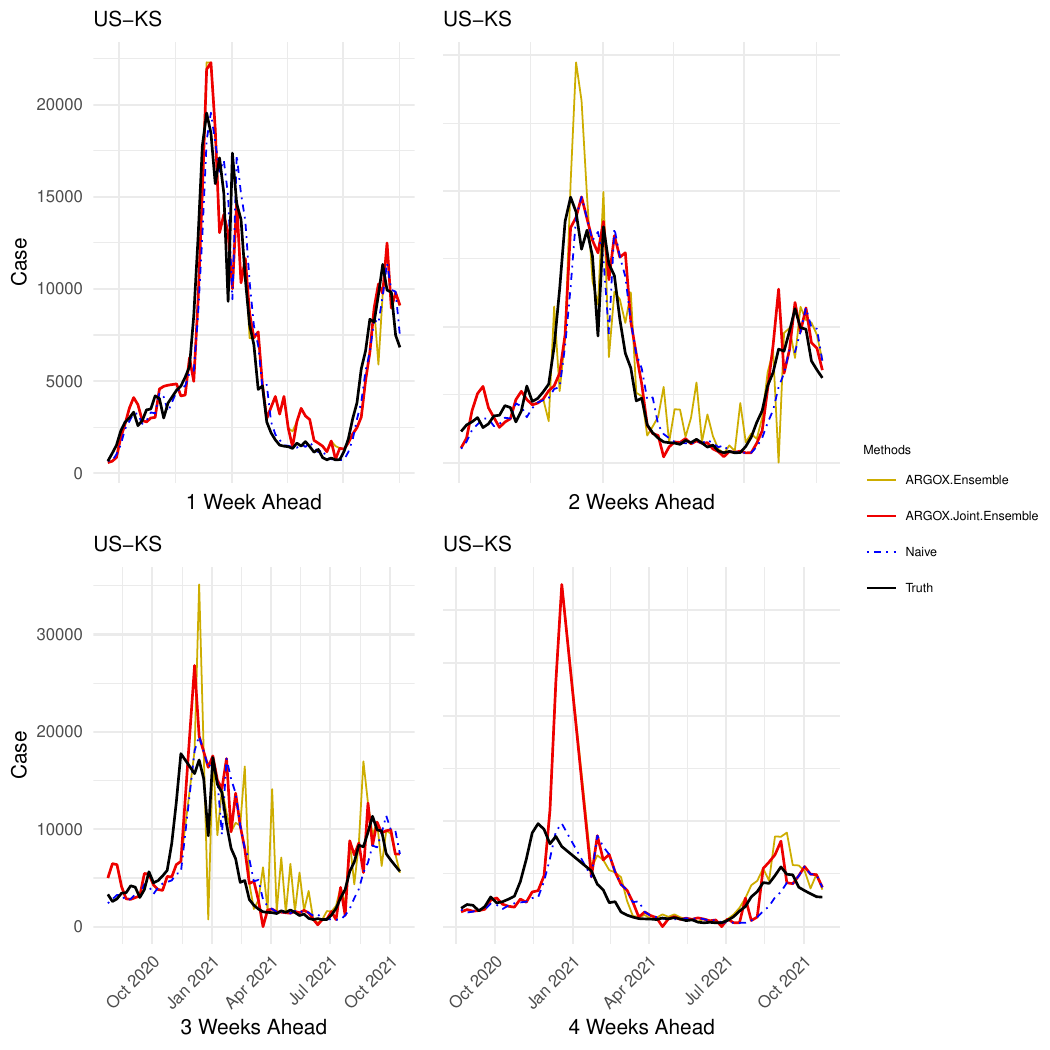}  \caption{Plots of the COVID-19 1 week (top left), 2 weeks (top right), 3 weeks (bottom left), and 4 weeks (bottom right) ahead estimates for Kansas (KS). ARGOX-Ensemble is Ref \cite{ma2021covid}.}  \end{figure} \newpage   
\begin{table}[ht]
\centering
\begin{tabular}{lrrrr}
  \hline
& 1 Week Ahead & 2 Weeks Ahead & 3 Weeks Ahead & 4 Weeks Ahead \\ 
    \hline \multicolumn{1}{l}{RMSE} \\
 \hspace{1em} Ref \cite{ma2021covid} & 
2549.54 & 3730.84 & 5714.65 & 7418.31 \\ 
   \hspace{1em} ARGOX Joint Ensemble & 2491.98 & 3519.34 & 5833.99 & 7145.59 \\ 
   \hspace{1em} Naive & 2532.03 & 4404.70 & 6081.72 & 7869.61 \\ 
   \multicolumn{1}{l}{MAE} \\
  \hspace{1em}  Ref \cite{ma2021covid} &1775.92 & 2620.65 & 3949.47 & 4942.55 \\ 
   \hspace{1em} ARGOX Joint Ensemble & 1680.27 & 2337.08 & 3613.54 & 5119.36 \\ 
   \hspace{1em} Naive & 1787.93 & 3079.64 & 4287.38 & 5643.90 \\ 
   \multicolumn{1}{l}{Correlation} \\
 \hspace{1em} Ref \cite{ma2021covid} &0.96 & 0.92 & 0.87 & 0.87 \\ 
   \hspace{1em} ARGOX Joint Ensemble  & 0.96 & 0.93 & 0.89 & 0.75 \\ 
   \hspace{1em} Naive& 0.96 & 0.87 & 0.76 & 0.57 \\ 
   \hline
\end{tabular}
\caption{Comparison of different methods for state-level COVID-19 1 to 4 weeks ahead incremental case in Kentucky (KY). The MSE, MAE, and correlation are reported and best performed method is highlighted in boldface.} 
\end{table}       \begin{figure}[!h]    \centering  \includegraphics[width=0.6\linewidth, page=18]{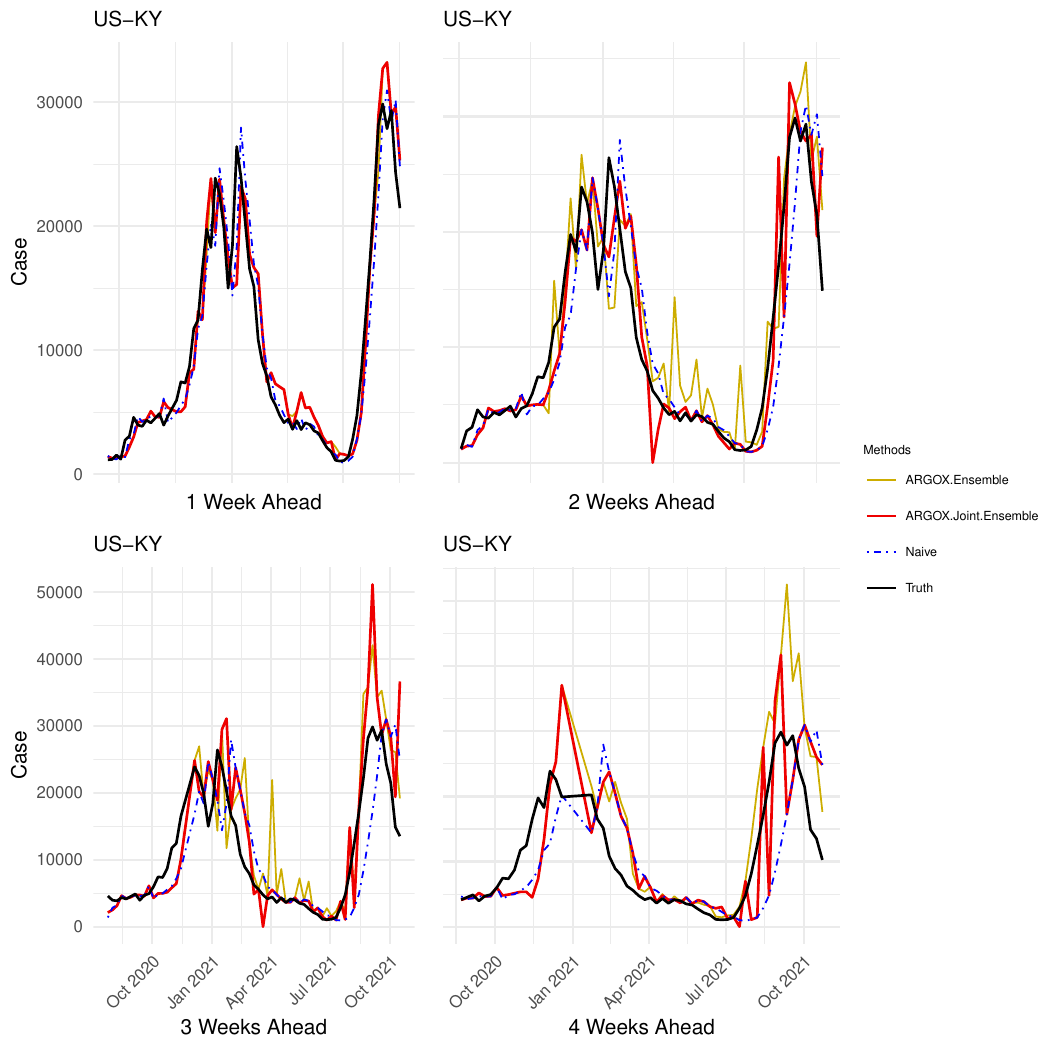}  \caption{Plots of the COVID-19 1 week (top left), 2 weeks (top right), 3 weeks (bottom left), and 4 weeks (bottom right) ahead estimates for Kentucky (KY). ARGOX-Ensemble is Ref \cite{ma2021covid}.}  \end{figure} \newpage   
\begin{table}[ht]
\centering
\begin{tabular}{lrrrr}
  \hline
& 1 Week Ahead & 2 Weeks Ahead & 3 Weeks Ahead & 4 Weeks Ahead \\ 
    \hline \multicolumn{1}{l}{RMSE} \\
 \hspace{1em} Ref \cite{ma2021covid} & 
3815.56 & 8590.95 & 12499.31 & 18338.72 \\ 
   \hspace{1em} ARGOX Joint Ensemble & 4088.57 & 5694.74 & 12493.66 & 9877.72 \\ 
   \hspace{1em} Naive & 3765.16 & 6246.78 & 8373.61 & 10638.62 \\ 
   \multicolumn{1}{l}{MAE} \\
  \hspace{1em}  Ref \cite{ma2021covid} &2798.96 & 4888.28 & 6112.51 & 8553.47 \\ 
   \hspace{1em} ARGOX Joint Ensemble & 2428.51 & 3604.79 & 6087.94 & 6348.23 \\ 
   \hspace{1em} Naive & 2530.18 & 4068.33 & 5348.65 & 6991.45 \\ 
   \multicolumn{1}{l}{Correlation} \\
 \hspace{1em} Ref \cite{ma2021covid} &0.91 & 0.85 & 0.71 & 0.76 \\ 
   \hspace{1em} ARGOX Joint Ensemble  & 0.92 & 0.81 & 0.71 & 0.64 \\ 
   \hspace{1em} Naive& 0.91 & 0.75 & 0.57 & 0.32 \\ 
   \hline
\end{tabular}
\caption{Comparison of different methods for state-level COVID-19 1 to 4 weeks ahead incremental case in Louisiana (LA). The MSE, MAE, and correlation are reported and best performed method is highlighted in boldface.} 
\end{table}       \begin{figure}[!h]    \centering  \includegraphics[width=0.6\linewidth, page=19]{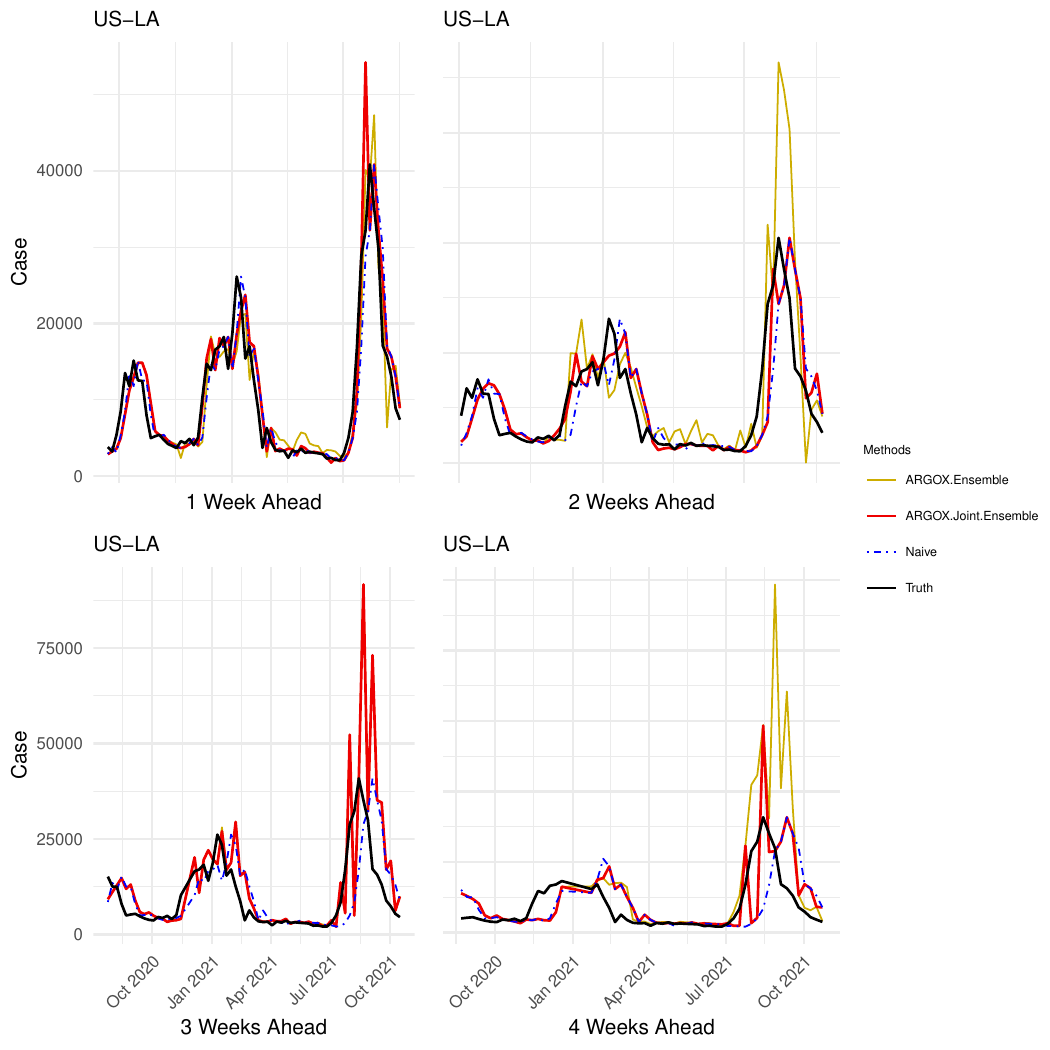}  \caption{Plots of the COVID-19 1 week (top left), 2 weeks (top right), 3 weeks (bottom left), and 4 weeks (bottom right) ahead estimates for Louisiana (LA). ARGOX-Ensemble is Ref \cite{ma2021covid}.}  \end{figure} \newpage   
\begin{table}[ht]
\centering
\begin{tabular}{lrrrr}
  \hline
& 1 Week Ahead & 2 Weeks Ahead & 3 Weeks Ahead & 4 Weeks Ahead \\ 
    \hline \multicolumn{1}{l}{RMSE} \\
 \hspace{1em} Ref \cite{ma2021covid} & 
2816.61 & 4162.13 & 5284.67 & 7042.37 \\ 
   \hspace{1em} ARGOX Joint Ensemble & 2767.27 & 4012.42 & 5644.61 & 8149.64 \\ 
   \hspace{1em} Naive & 3050.86 & 4959.35 & 6526.60 & 8225.02 \\ 
   \multicolumn{1}{l}{MAE} \\
  \hspace{1em}  Ref \cite{ma2021covid} &1917.18 & 2696.73 & 3503.88 & 4726.77 \\ 
   \hspace{1em} ARGOX Joint Ensemble & 1846.14 & 2667.30 & 3636.25 & 6048.57 \\ 
   \hspace{1em} Naive & 2001.74 & 3353.00 & 4521.33 & 5916.86 \\ 
   \multicolumn{1}{l}{Correlation} \\
 \hspace{1em} Ref \cite{ma2021covid} &0.96 & 0.93 & 0.92 & 0.69 \\ 
   \hspace{1em} ARGOX Joint Ensemble  & 0.96 & 0.93 & 0.91 & 0.64 \\ 
   \hspace{1em} Naive& 0.96 & 0.89 & 0.81 & 0.60 \\ 
   \hline
\end{tabular}
\caption{Comparison of different methods for state-level COVID-19 1 to 4 weeks ahead incremental case in Massachusetts (MA). The MSE, MAE, and correlation are reported and best performed method is highlighted in boldface.} 
\end{table}       \begin{figure}[!h]    \centering  \includegraphics[width=0.6\linewidth, page=20]{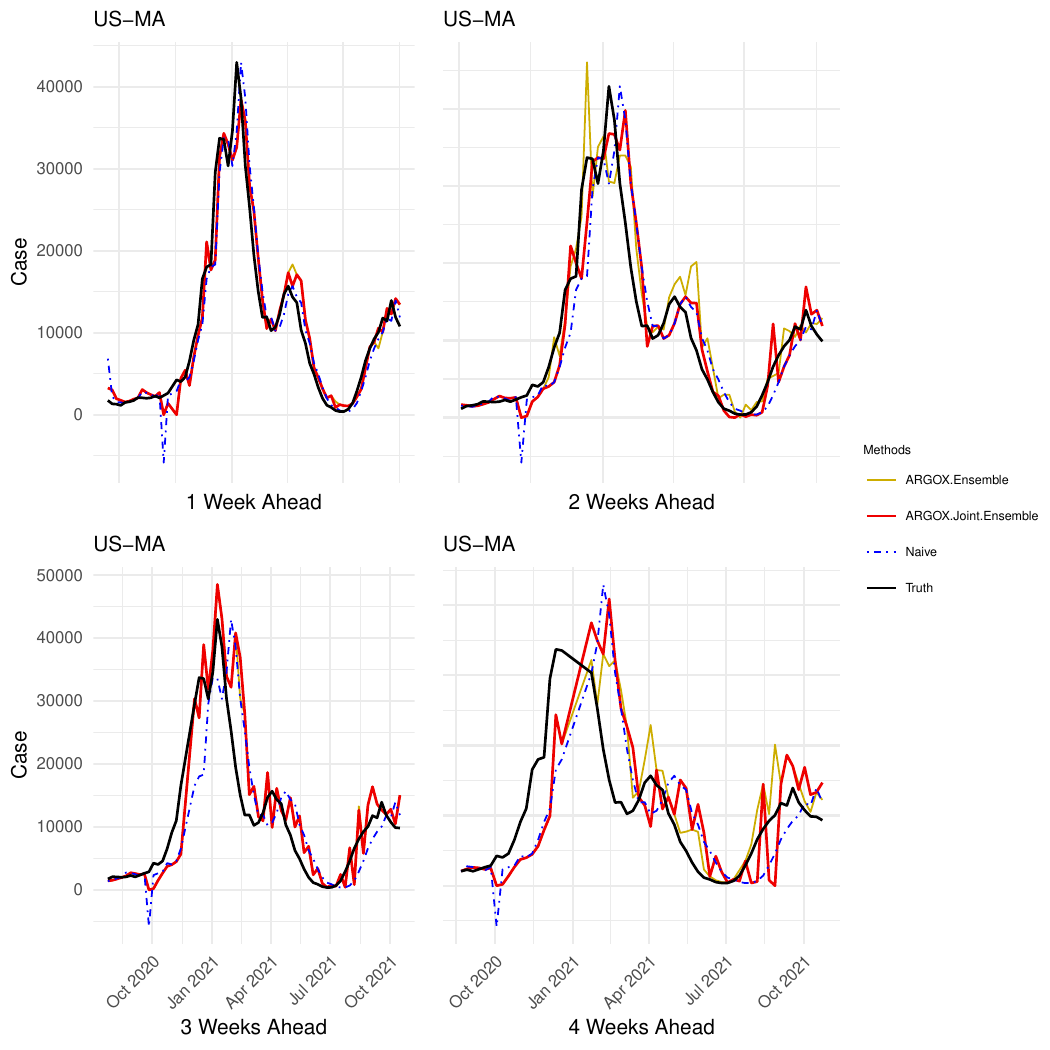}  \caption{Plots of the COVID-19 1 week (top left), 2 weeks (top right), 3 weeks (bottom left), and 4 weeks (bottom right) ahead estimates for Massachusetts (MA). ARGOX-Ensemble is Ref \cite{ma2021covid}.}  \end{figure} \newpage   
\begin{table}[ht]
\centering
\begin{tabular}{lrrrr}
  \hline
& 1 Week Ahead & 2 Weeks Ahead & 3 Weeks Ahead & 4 Weeks Ahead \\ 
    \hline \multicolumn{1}{l}{RMSE} \\
 \hspace{1em} Ref \cite{ma2021covid} & 
1988.47 & 2773.60 & 3968.08 & 4422.06 \\ 
   \hspace{1em} ARGOX Joint Ensemble & 1439.95 & 3056.32 & 3965.83 & 4518.41 \\ 
   \hspace{1em} Naive & 1593.75 & 2637.20 & 3350.24 & 4672.42 \\ 
   \multicolumn{1}{l}{MAE} \\
  \hspace{1em}  Ref \cite{ma2021covid} &1178.63 & 1824.85 & 2505.59 & 3094.78 \\ 
   \hspace{1em} ARGOX Joint Ensemble & 989.04 & 2110.68 & 2489.32 & 3184.11 \\ 
   \hspace{1em} Naive & 1127.75 & 1961.31 & 2522.46 & 3422.57 \\ 
   \multicolumn{1}{l}{Correlation} \\
 \hspace{1em} Ref \cite{ma2021covid} &0.93 & 0.88 & 0.82 & 0.58 \\ 
   \hspace{1em} ARGOX Joint Ensemble  & 0.96 & 0.83 & 0.82 & 0.55 \\ 
   \hspace{1em} Naive& 0.95 & 0.87 & 0.79 & 0.49 \\ 
   \hline
\end{tabular}
\caption{Comparison of different methods for state-level COVID-19 1 to 4 weeks ahead incremental case in Maryland (MD). The MSE, MAE, and correlation are reported and best performed method is highlighted in boldface.} 
\end{table}       \begin{figure}[!h]    \centering  \includegraphics[width=0.6\linewidth, page=21]{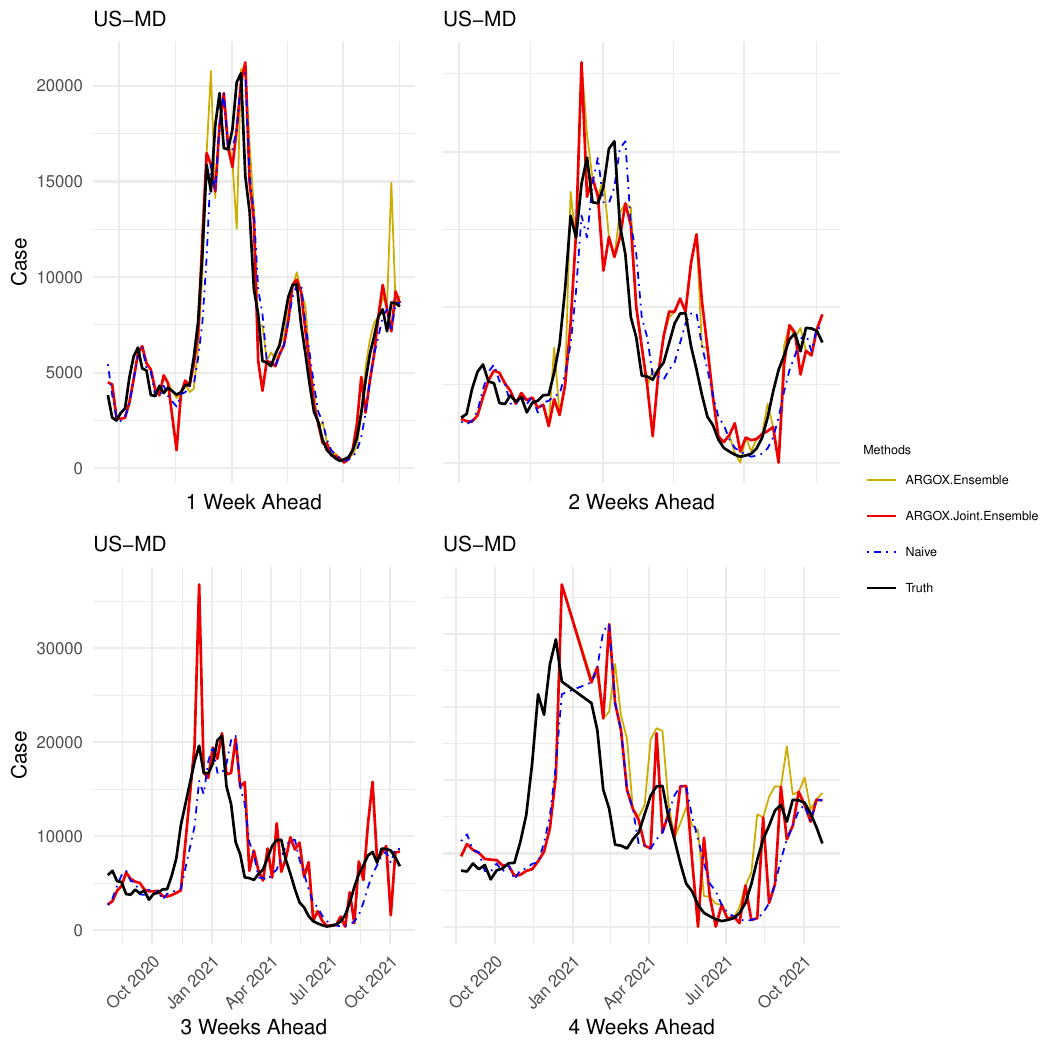}  \caption{Plots of the COVID-19 1 week (top left), 2 weeks (top right), 3 weeks (bottom left), and 4 weeks (bottom right) ahead estimates for Maryland (MD). ARGOX-Ensemble is Ref \cite{ma2021covid}.}  \end{figure} \newpage   
\begin{table}[ht]
\centering
\begin{tabular}{lrrrr}
  \hline
& 1 Week Ahead & 2 Weeks Ahead & 3 Weeks Ahead & 4 Weeks Ahead \\ 
    \hline \multicolumn{1}{l}{RMSE} \\
 \hspace{1em} Ref \cite{ma2021covid} & 
420.56 & 733.48 & 900.40 & 1170.92 \\ 
   \hspace{1em} ARGOX Joint Ensemble & 364.69 & 518.40 & 895.99 & 1131.20 \\ 
   \hspace{1em} Naive & 408.63 & 641.67 & 906.61 & 1091.08 \\ 
   \multicolumn{1}{l}{MAE} \\
  \hspace{1em}  Ref \cite{ma2021covid} &296.30 & 466.76 & 634.43 & 858.19 \\ 
   \hspace{1em} ARGOX Joint Ensemble & 252.75 & 328.25 & 621.16 & 859.49 \\ 
   \hspace{1em} Naive & 286.40 & 464.64 & 677.70 & 842.76 \\ 
   \multicolumn{1}{l}{Correlation} \\
 \hspace{1em} Ref \cite{ma2021covid} &0.94 & 0.85 & 0.80 & 0.80 \\ 
   \hspace{1em} ARGOX Joint Ensemble  & 0.95 & 0.92 & 0.80 & 0.60 \\ 
   \hspace{1em} Naive& 0.94 & 0.86 & 0.72 & 0.54 \\ 
   \hline
\end{tabular}
\caption{Comparison of different methods for state-level COVID-19 1 to 4 weeks ahead incremental case in Maine (ME). The MSE, MAE, and correlation are reported and best performed method is highlighted in boldface.} 
\end{table}       \begin{figure}[!h]    \centering  \includegraphics[width=0.6\linewidth, page=22]{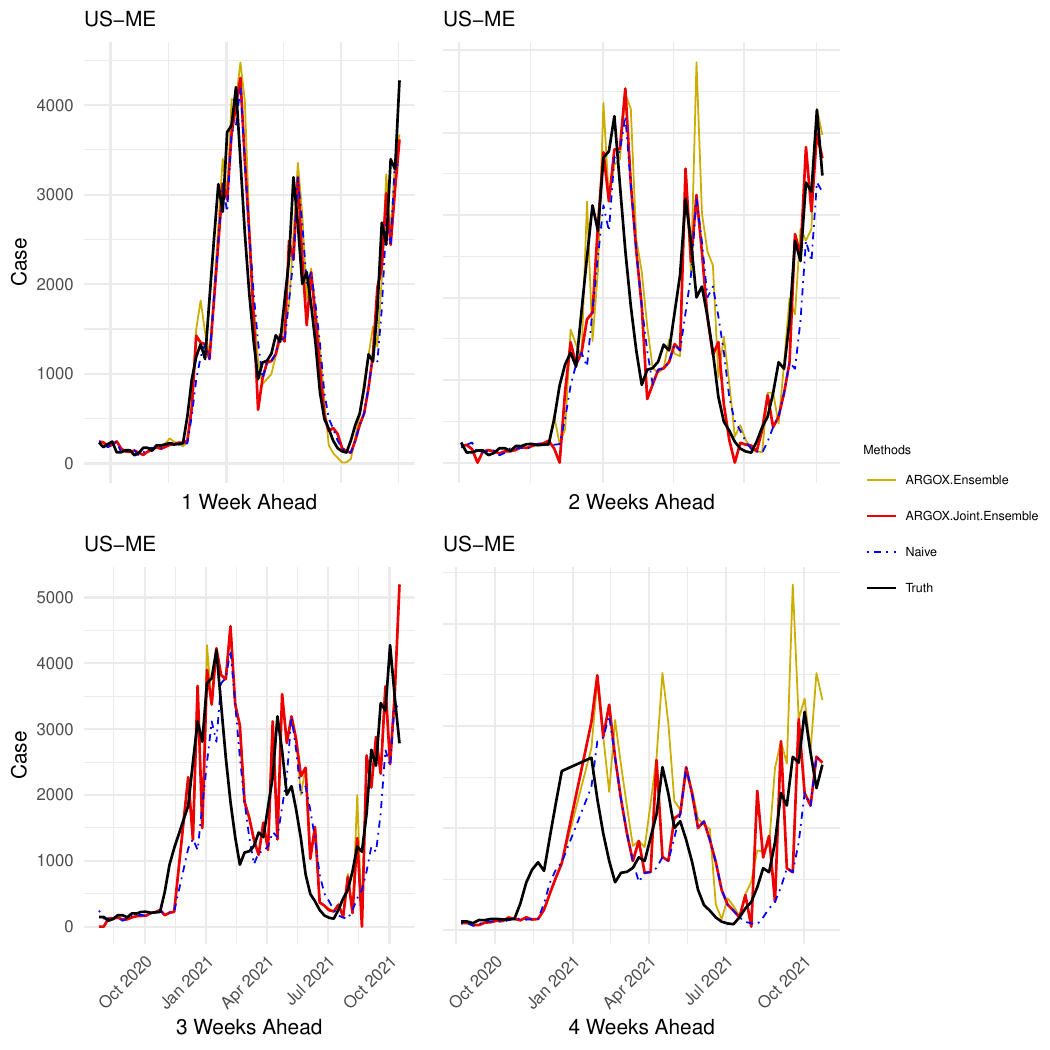}  \caption{Plots of the COVID-19 1 week (top left), 2 weeks (top right), 3 weeks (bottom left), and 4 weeks (bottom right) ahead estimates for Maine (ME). ARGOX-Ensemble is Ref \cite{ma2021covid}.}  \end{figure} \newpage   
\begin{table}[ht]
\centering
\begin{tabular}{lrrrr}
  \hline
& 1 Week Ahead & 2 Weeks Ahead & 3 Weeks Ahead & 4 Weeks Ahead \\ 
    \hline \multicolumn{1}{l}{RMSE} \\
 \hspace{1em} Ref \cite{ma2021covid} & 
4438.75 & 11797.07 & 15749.99 & 23816.72 \\ 
   \hspace{1em} ARGOX Joint Ensemble & 4431.16 & 9905.10 & 15755.84 & 22001.94 \\ 
   \hspace{1em} Naive & 5166.03 & 9417.61 & 12746.37 & 16453.87 \\ 
   \multicolumn{1}{l}{MAE} \\
  \hspace{1em}  Ref \cite{ma2021covid} &2938.56 & 6631.07 & 9220.09 & 13815.66 \\ 
   \hspace{1em} ARGOX Joint Ensemble & 2909.26 & 6410.90 & 9225.51 & 13604.57 \\ 
   \hspace{1em} Naive & 3634.18 & 6617.76 & 9150.43 & 12158.55 \\ 
   \multicolumn{1}{l}{Correlation} \\
 \hspace{1em} Ref \cite{ma2021covid} &0.96 & 0.85 & 0.70 & 0.58 \\ 
   \hspace{1em} ARGOX Joint Ensemble  & 0.96 & 0.84 & 0.70 & 0.48 \\ 
   \hspace{1em} Naive& 0.94 & 0.80 & 0.60 & 0.38 \\ 
   \hline
\end{tabular}
\caption{Comparison of different methods for state-level COVID-19 1 to 4 weeks ahead incremental case in Michigan (MI). The MSE, MAE, and correlation are reported and best performed method is highlighted in boldface.} 
\end{table}       \begin{figure}[!h]    \centering  \includegraphics[width=0.6\linewidth, page=23]{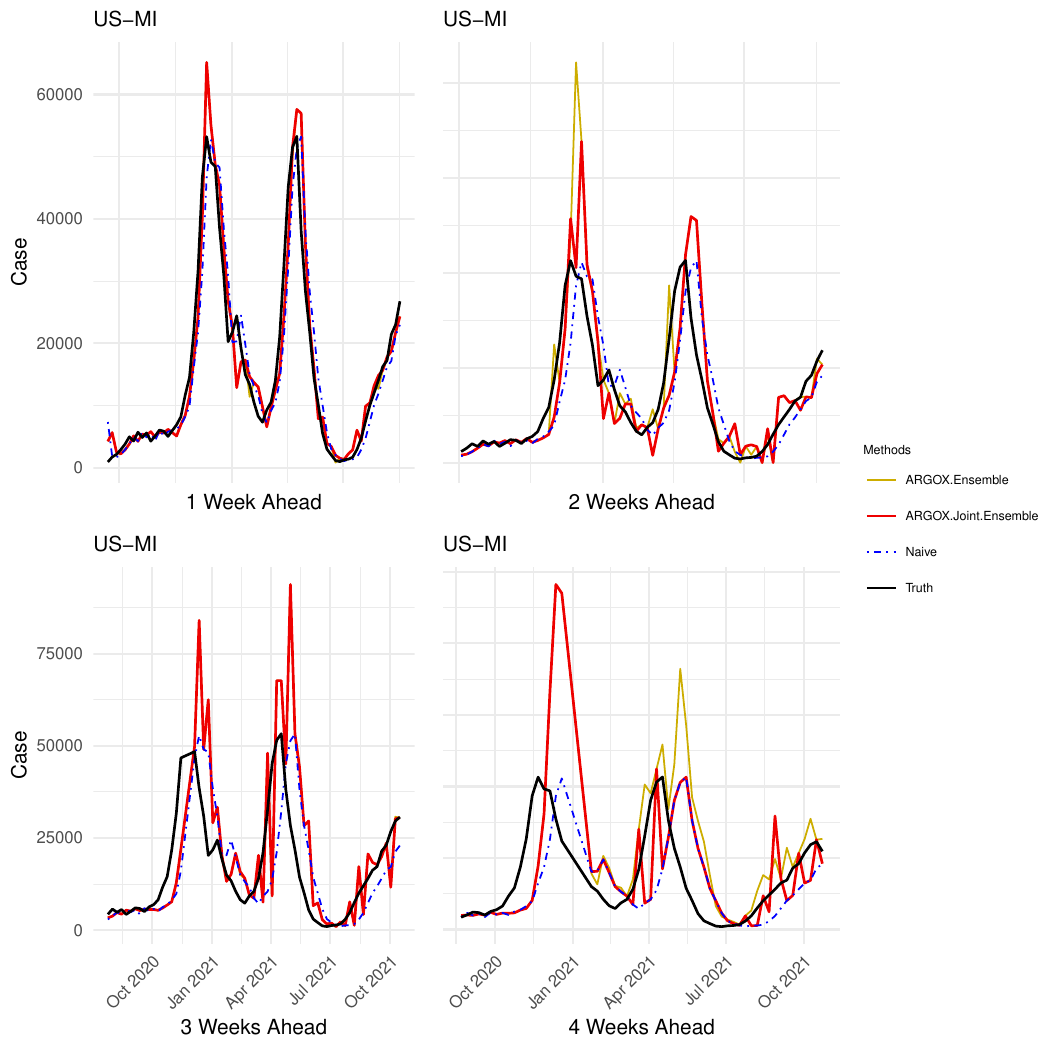}  \caption{Plots of the COVID-19 1 week (top left), 2 weeks (top right), 3 weeks (bottom left), and 4 weeks (bottom right) ahead estimates for Michigan (MI). ARGOX-Ensemble is Ref \cite{ma2021covid}.}  \end{figure} \newpage   
\begin{table}[ht]
\centering
\begin{tabular}{lrrrr}
  \hline
& 1 Week Ahead & 2 Weeks Ahead & 3 Weeks Ahead & 4 Weeks Ahead \\ 
    \hline \multicolumn{1}{l}{RMSE} \\
 \hspace{1em} Ref \cite{ma2021covid} & 
3029.80 & 6514.91 & 8482.79 & 20397.71 \\ 
   \hspace{1em} ARGOX Joint Ensemble & 3019.35 & 7789.02 & 8485.85 & 20350.84 \\ 
   \hspace{1em} Naive & 3592.95 & 6430.96 & 7944.41 & 9778.82 \\ 
   \multicolumn{1}{l}{MAE} \\
  \hspace{1em}  Ref \cite{ma2021covid} &1872.98 & 3566.97 & 4635.20 & 8667.50 \\ 
   \hspace{1em} ARGOX Joint Ensemble & 1844.13 & 3678.43 & 4621.32 & 8518.69 \\ 
   \hspace{1em} Naive & 2087.34 & 3844.97 & 4920.83 & 6438.79 \\ 
   \multicolumn{1}{l}{Correlation} \\
 \hspace{1em} Ref \cite{ma2021covid} &0.96 & 0.86 & 0.69 & 0.41 \\ 
   \hspace{1em} ARGOX Joint Ensemble  & 0.96 & 0.80 & 0.69 & 0.40 \\ 
   \hspace{1em} Naive& 0.94 & 0.79 & 0.64 & 0.51 \\ 
   \hline
\end{tabular}
\caption{Comparison of different methods for state-level COVID-19 1 to 4 weeks ahead incremental case in Minnesota (MN). The MSE, MAE, and correlation are reported and best performed method is highlighted in boldface.} 
\end{table}       \begin{figure}[!h]    \centering  \includegraphics[width=0.6\linewidth, page=24]{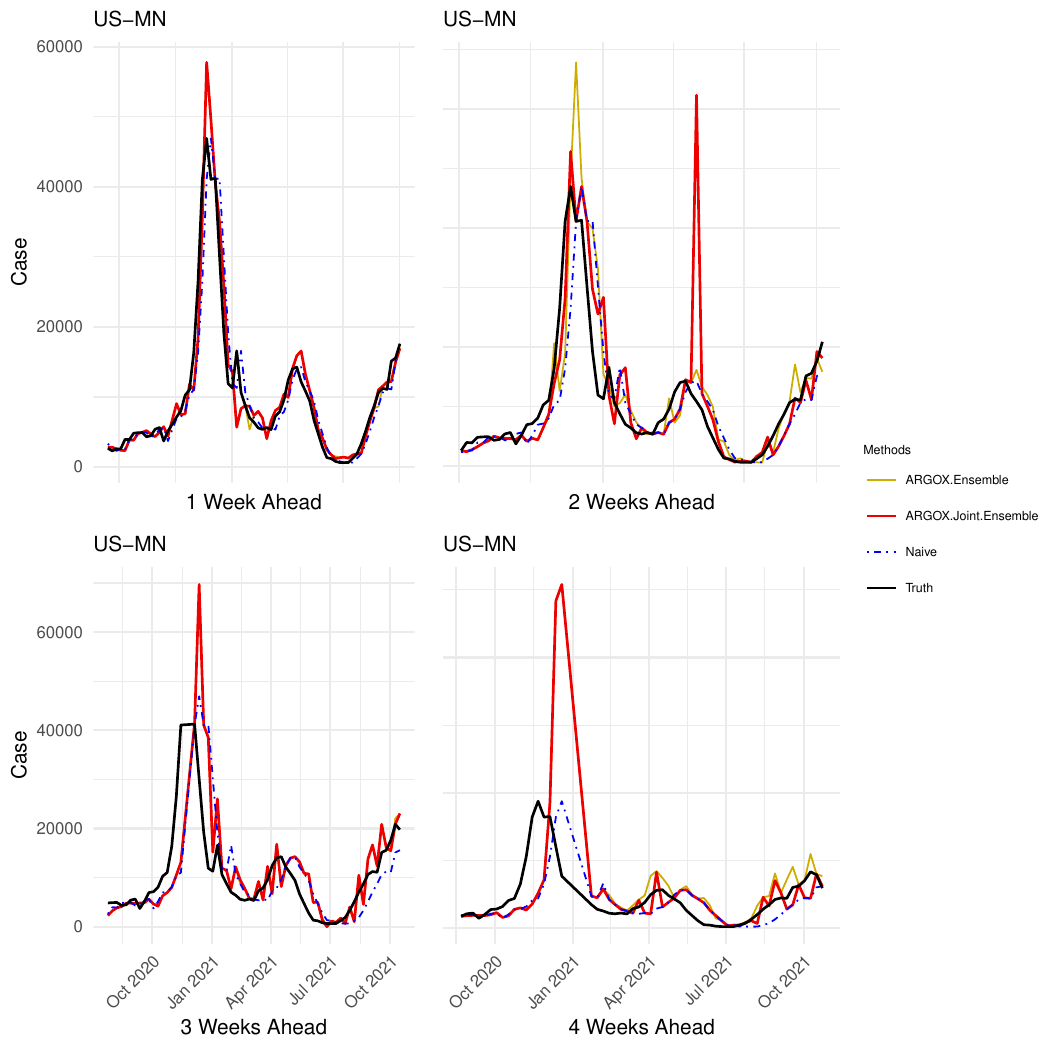}  \caption{Plots of the COVID-19 1 week (top left), 2 weeks (top right), 3 weeks (bottom left), and 4 weeks (bottom right) ahead estimates for Minnesota (MN). ARGOX-Ensemble is Ref \cite{ma2021covid}.}  \end{figure} \newpage   
\begin{table}[ht]
\centering
\begin{tabular}{lrrrr}
  \hline
& 1 Week Ahead & 2 Weeks Ahead & 3 Weeks Ahead & 4 Weeks Ahead \\ 
    \hline \multicolumn{1}{l}{RMSE} \\
 \hspace{1em} Ref \cite{ma2021covid} & 
5433.42 & 6293.27 & 13492.90 & 15008.47 \\ 
   \hspace{1em} ARGOX Joint Ensemble & 3502.34 & 4389.12 & 12620.30 & 7387.53 \\ 
   \hspace{1em} Naive & 6379.10 & 7171.35 & 7914.75 & 9183.48 \\ 
   \multicolumn{1}{l}{MAE} \\
  \hspace{1em}  Ref \cite{ma2021covid} &2799.19 & 3680.87 & 5706.57 & 7919.06 \\ 
   \hspace{1em} ARGOX Joint Ensemble & 2258.12 & 3067.61 & 4909.21 & 5540.40 \\ 
   \hspace{1em} Naive & 2757.82 & 3709.51 & 4465.19 & 5711.16 \\ 
   \multicolumn{1}{l}{Correlation} \\
 \hspace{1em} Ref \cite{ma2021covid} &0.82 & 0.77 & 0.36 & 0.27 \\ 
   \hspace{1em} ARGOX Joint Ensemble  & 0.92 & 0.86 & 0.38 & 0.60 \\ 
   \hspace{1em} Naive& 0.76 & 0.68 & 0.59 & 0.43 \\ 
   \hline
\end{tabular}
\caption{Comparison of different methods for state-level COVID-19 1 to 4 weeks ahead incremental case in Missouri (MO). The MSE, MAE, and correlation are reported and best performed method is highlighted in boldface.} 
\end{table}       \begin{figure}[!h]    \centering  \includegraphics[width=0.6\linewidth, page=25]{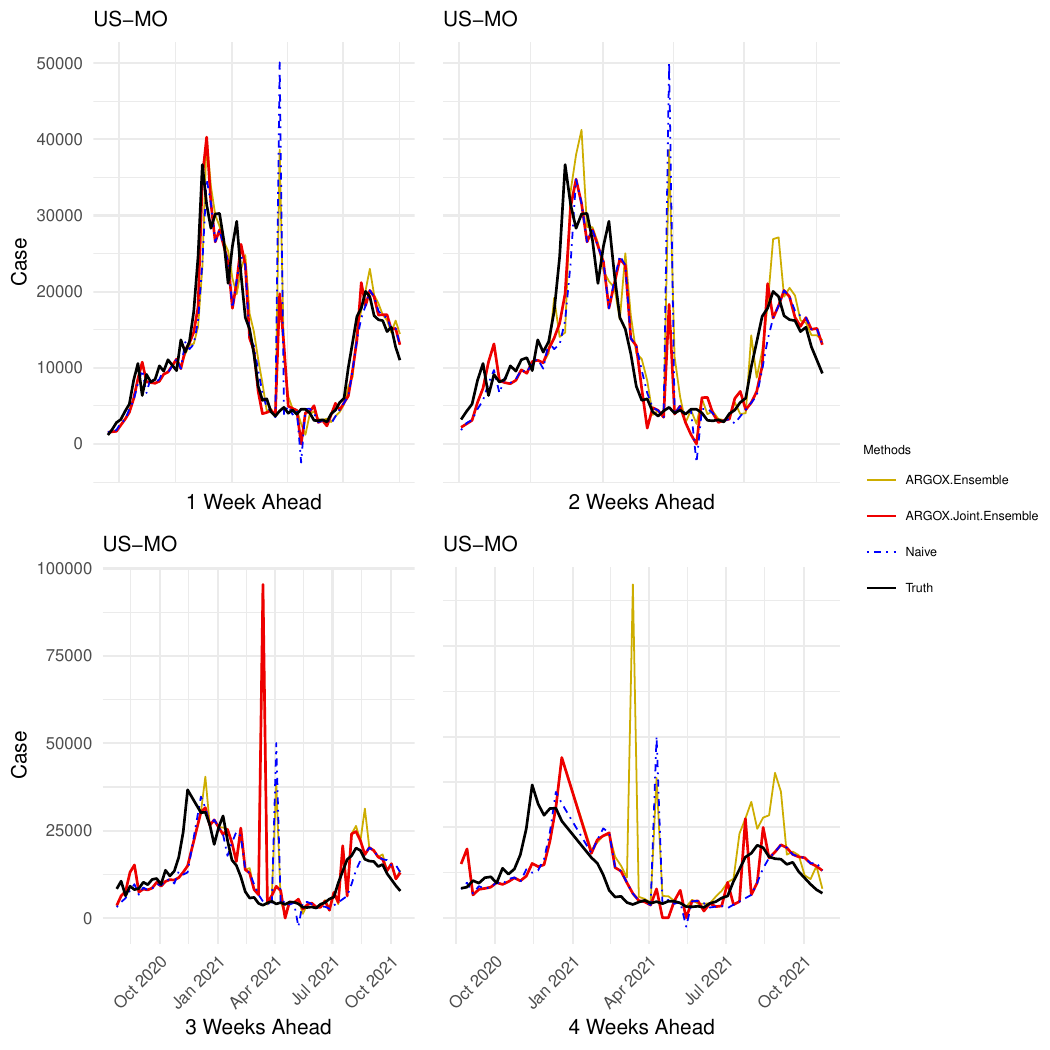}  \caption{Plots of the COVID-19 1 week (top left), 2 weeks (top right), 3 weeks (bottom left), and 4 weeks (bottom right) ahead estimates for Missouri (MO). ARGOX-Ensemble is Ref \cite{ma2021covid}.}  \end{figure} \newpage   
\begin{table}[ht]
\centering
\begin{tabular}{lrrrr}
  \hline
& 1 Week Ahead & 2 Weeks Ahead & 3 Weeks Ahead & 4 Weeks Ahead \\ 
    \hline \multicolumn{1}{l}{RMSE} \\
 \hspace{1em} Ref \cite{ma2021covid} & 
1979.07 & 3642.46 & 5630.05 & 9804.83 \\ 
   \hspace{1em} ARGOX Joint Ensemble & 1907.49 & 4796.81 & 4590.77 & 6379.00 \\ 
   \hspace{1em} Naive & 2183.95 & 3703.84 & 5143.08 & 6551.75 \\ 
   \multicolumn{1}{l}{MAE} \\
  \hspace{1em}  Ref \cite{ma2021covid} &1554.80 & 2335.68 & 3502.53 & 4890.27 \\ 
   \hspace{1em} ARGOX Joint Ensemble & 1473.88 & 2726.20 & 3187.71 & 4426.66 \\ 
   \hspace{1em} Naive & 1444.13 & 2484.15 & 3490.32 & 4549.24 \\ 
   \multicolumn{1}{l}{Correlation} \\
 \hspace{1em} Ref \cite{ma2021covid} &0.94 & 0.86 & 0.78 & 0.78 \\ 
   \hspace{1em} ARGOX Joint Ensemble  & 0.95 & 0.86 & 0.86 & 0.68 \\ 
   \hspace{1em} Naive& 0.93 & 0.80 & 0.63 & 0.39 \\ 
   \hline
\end{tabular}
\caption{Comparison of different methods for state-level COVID-19 1 to 4 weeks ahead incremental case in Mississippi (MS). The MSE, MAE, and correlation are reported and best performed method is highlighted in boldface.} 
\end{table}       \begin{figure}[!h]    \centering  \includegraphics[width=0.6\linewidth, page=26]{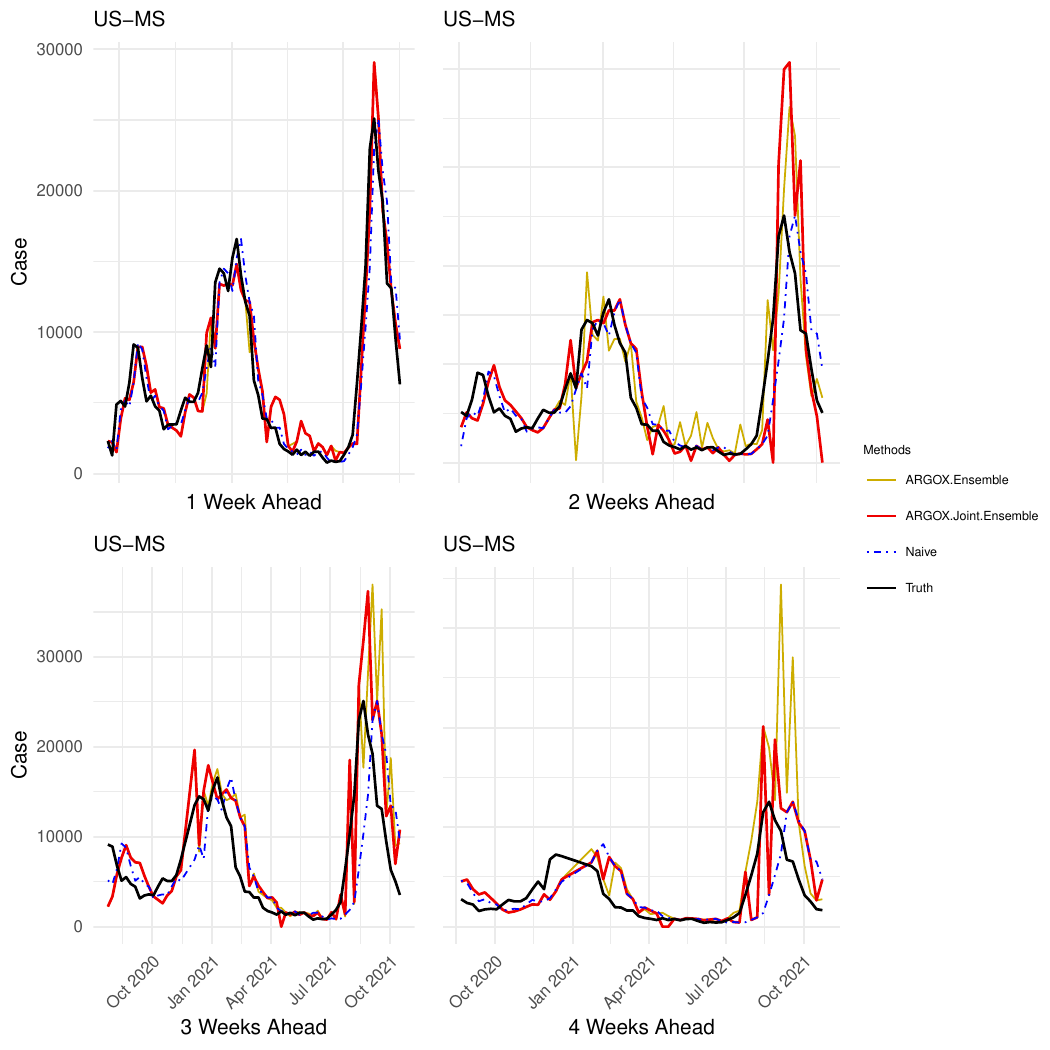}  \caption{Plots of the COVID-19 1 week (top left), 2 weeks (top right), 3 weeks (bottom left), and 4 weeks (bottom right) ahead estimates for Mississippi (MS). ARGOX-Ensemble is Ref \cite{ma2021covid}.}  \end{figure} \newpage   
\begin{table}[ht]
\centering
\begin{tabular}{lrrrr}
  \hline
& 1 Week Ahead & 2 Weeks Ahead & 3 Weeks Ahead & 4 Weeks Ahead \\ 
    \hline \multicolumn{1}{l}{RMSE} \\
 \hspace{1em} Ref \cite{ma2021covid} & 
738.68 & 904.98 & 1141.61 & 1472.87 \\ 
   \hspace{1em} ARGOX Joint Ensemble & 604.79 & 856.98 & 1148.13 & 1602.67 \\ 
   \hspace{1em} Naive & 646.39 & 1015.37 & 1322.34 & 1659.88 \\ 
   \multicolumn{1}{l}{MAE} \\
  \hspace{1em}  Ref \cite{ma2021covid} &468.87 & 596.76 & 720.92 & 974.98 \\ 
   \hspace{1em} ARGOX Joint Ensemble & 349.24 & 505.06 & 725.75 & 1107.08 \\ 
   \hspace{1em} Naive & 414.53 & 708.22 & 911.86 & 1146.88 \\ 
   \multicolumn{1}{l}{Correlation} \\
 \hspace{1em} Ref \cite{ma2021covid} &0.94 & 0.93 & 0.89 & 0.82 \\ 
   \hspace{1em} ARGOX Joint Ensemble  & 0.96 & 0.93 & 0.89 & 0.80 \\ 
   \hspace{1em} Naive& 0.95 & 0.89 & 0.79 & 0.73 \\ 
   \hline
\end{tabular}
\caption{Comparison of different methods for state-level COVID-19 1 to 4 weeks ahead incremental case in Montana (MT). The MSE, MAE, and correlation are reported and best performed method is highlighted in boldface.} 
\end{table}       \begin{figure}[!h]    \centering  \includegraphics[width=0.6\linewidth, page=27]{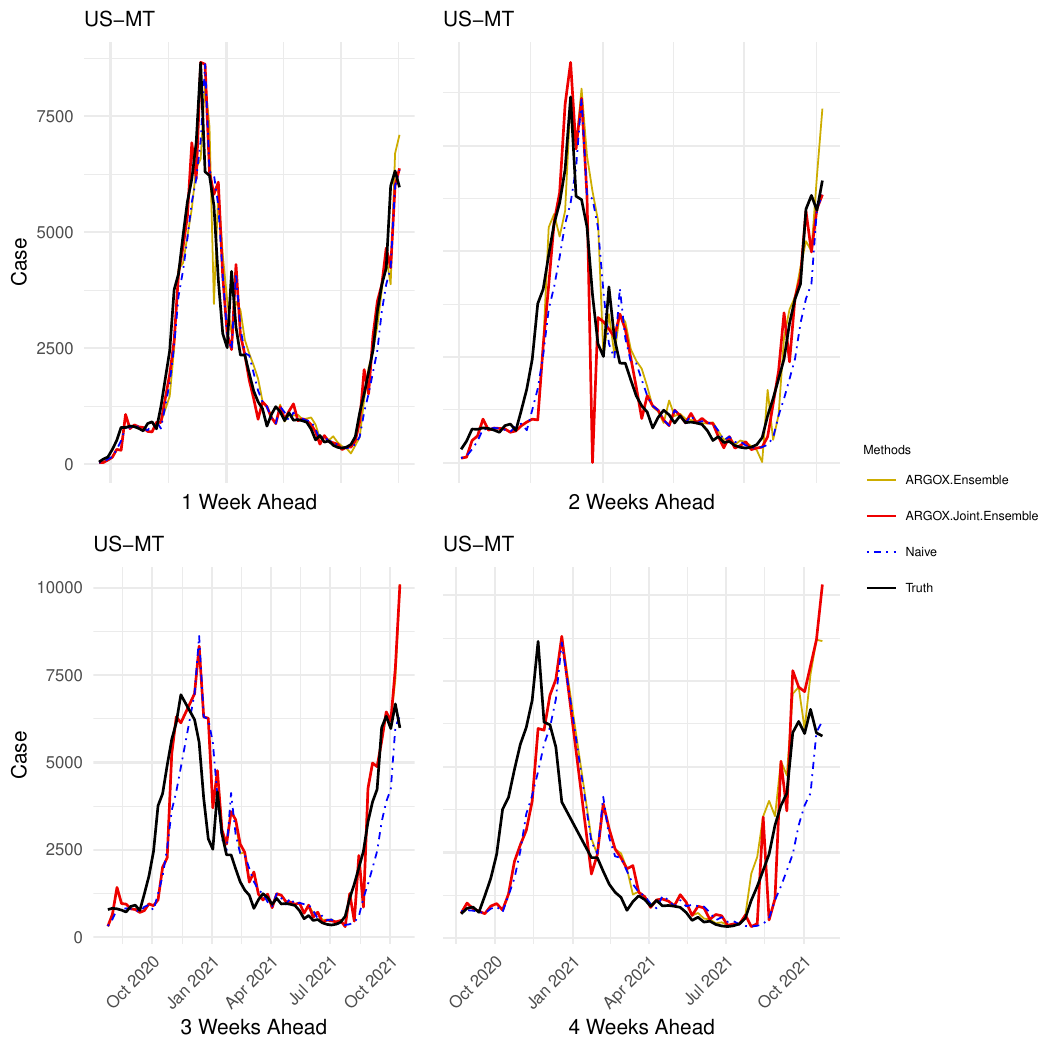}  \caption{Plots of the COVID-19 1 week (top left), 2 weeks (top right), 3 weeks (bottom left), and 4 weeks (bottom right) ahead estimates for Montana (MT). ARGOX-Ensemble is Ref \cite{ma2021covid}.}  \end{figure} \newpage   
\begin{table}[ht]
\centering
\begin{tabular}{lrrrr}
  \hline
& 1 Week Ahead & 2 Weeks Ahead & 3 Weeks Ahead & 4 Weeks Ahead \\ 
    \hline \multicolumn{1}{l}{RMSE} \\
 \hspace{1em} Ref \cite{ma2021covid} & 
4997.48 & 7273.55 & 9912.22 & 14553.55 \\ 
   \hspace{1em} ARGOX Joint Ensemble & 4808.66 & 6930.17 & 7609.99 & 13245.06 \\ 
   \hspace{1em} Naive & 4818.12 & 7612.88 & 10554.18 & 13674.22 \\ 
   \multicolumn{1}{l}{MAE} \\
  \hspace{1em}  Ref \cite{ma2021covid} &3500.79 & 5115.99 & 6729.07 & 9882.63 \\ 
   \hspace{1em} ARGOX Joint Ensemble & 3344.10 & 4579.16 & 5200.60 & 9765.17 \\ 
   \hspace{1em} Naive & 3305.74 & 5369.91 & 7655.29 & 10105.12 \\ 
   \multicolumn{1}{l}{Correlation} \\
 \hspace{1em} Ref \cite{ma2021covid} &0.94 & 0.90 & 0.87 & 0.83 \\ 
   \hspace{1em} ARGOX Joint Ensemble  & 0.95 & 0.91 & 0.91 & 0.72 \\ 
   \hspace{1em} Naive& 0.95 & 0.87 & 0.76 & 0.55 \\ 
   \hline
\end{tabular}
\caption{Comparison of different methods for state-level COVID-19 1 to 4 weeks ahead incremental case in North Carolina (NC). The MSE, MAE, and correlation are reported and best performed method is highlighted in boldface.} 
\end{table}       \begin{figure}[!h]    \centering  \includegraphics[width=0.6\linewidth, page=28]{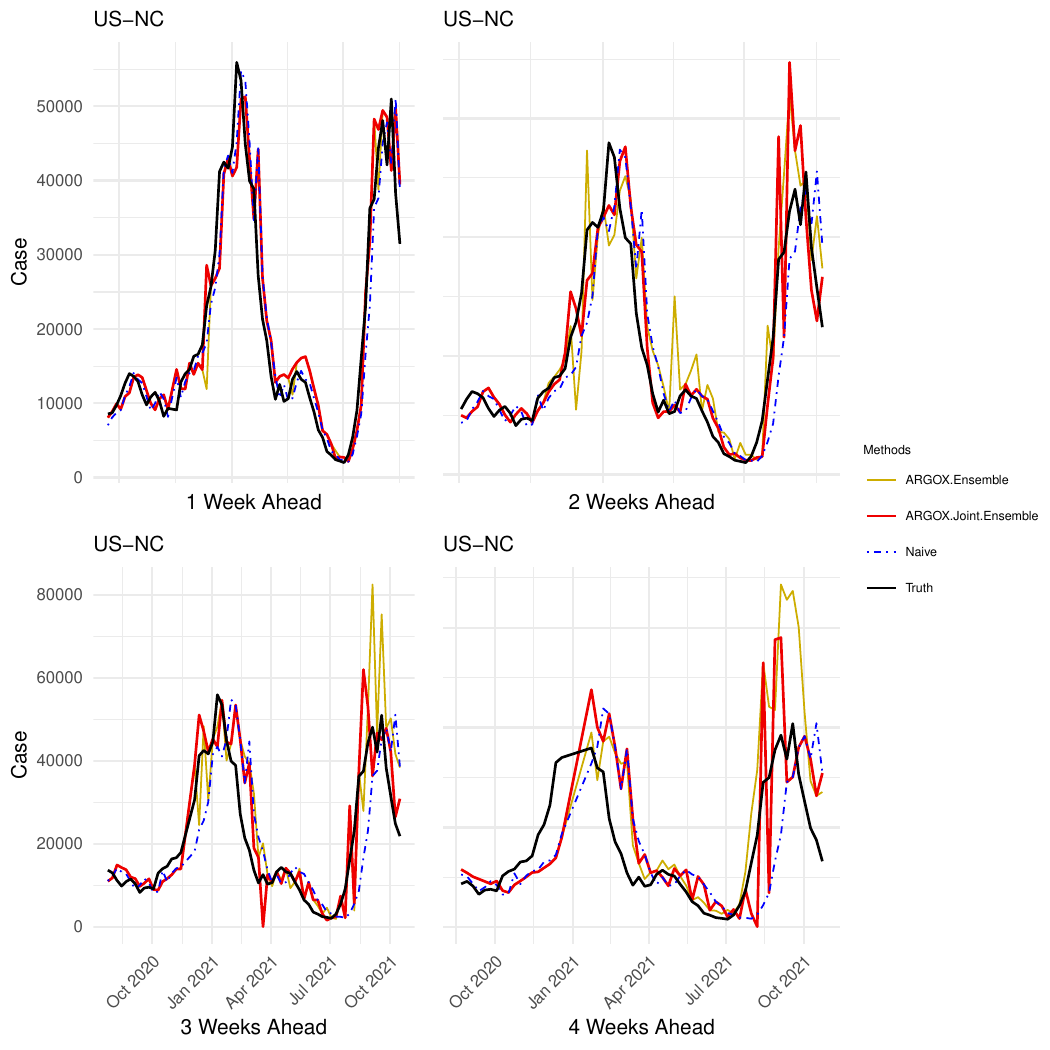}  \caption{Plots of the COVID-19 1 week (top left), 2 weeks (top right), 3 weeks (bottom left), and 4 weeks (bottom right) ahead estimates for North Carolina (NC). ARGOX-Ensemble is Ref \cite{ma2021covid}.}  \end{figure} \newpage   
\begin{table}[ht]
\centering
\begin{tabular}{lrrrr}
  \hline
& 1 Week Ahead & 2 Weeks Ahead & 3 Weeks Ahead & 4 Weeks Ahead \\ 
    \hline \multicolumn{1}{l}{RMSE} \\
 \hspace{1em} Ref \cite{ma2021covid} & 
747.90 & 1334.58 & 2467.90 & 1863.03 \\ 
   \hspace{1em} ARGOX Joint Ensemble & 686.92 & 1021.71 & 1495.11 & 1875.41 \\ 
   \hspace{1em} Naive & 688.74 & 1187.98 & 1613.64 & 1957.64 \\ 
   \multicolumn{1}{l}{MAE} \\
  \hspace{1em}  Ref \cite{ma2021covid} &410.64 & 788.02 & 1157.06 & 1105.73 \\ 
   \hspace{1em} ARGOX Joint Ensemble & 369.33 & 584.22 & 905.98 & 1132.52 \\ 
   \hspace{1em} Naive & 384.16 & 722.37 & 1015.29 & 1249.03 \\ 
   \multicolumn{1}{l}{Correlation} \\
 \hspace{1em} Ref \cite{ma2021covid} &0.96 & 0.84 & 0.52 & 0.66 \\ 
   \hspace{1em} ARGOX Joint Ensemble  & 0.96 & 0.91 & 0.73 & 0.67 \\ 
   \hspace{1em} Naive& 0.95 & 0.86 & 0.68 & 0.65 \\ 
   \hline
\end{tabular}
\caption{Comparison of different methods for state-level COVID-19 1 to 4 weeks ahead incremental case in North Dakota (ND). The MSE, MAE, and correlation are reported and best performed method is highlighted in boldface.} 
\end{table}       \begin{figure}[!h]    \centering  \includegraphics[width=0.6\linewidth, page=29]{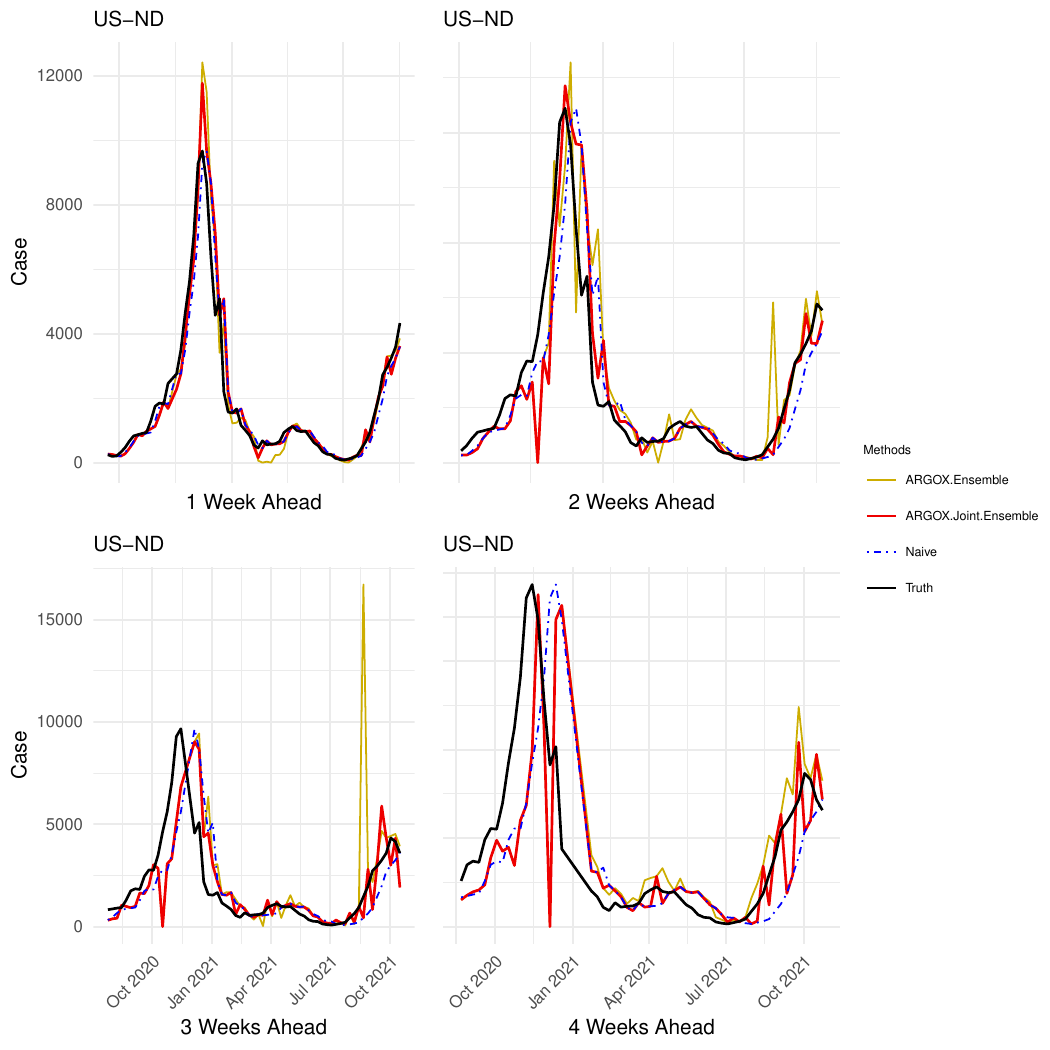}  \caption{Plots of the COVID-19 1 week (top left), 2 weeks (top right), 3 weeks (bottom left), and 4 weeks (bottom right) ahead estimates for North Dakota (ND). ARGOX-Ensemble is Ref \cite{ma2021covid}.}  \end{figure} \newpage   
\begin{table}[ht]
\centering
\begin{tabular}{lrrrr}
  \hline
& 1 Week Ahead & 2 Weeks Ahead & 3 Weeks Ahead & 4 Weeks Ahead \\ 
    \hline \multicolumn{1}{l}{RMSE} \\
 \hspace{1em} Ref \cite{ma2021covid} & 
1332.44 & 2123.25 & 2175.35 & 3799.94 \\ 
   \hspace{1em} ARGOX Joint Ensemble & 1160.15 & 1718.89 & 2210.36 & 3780.77 \\ 
   \hspace{1em} Naive & 1245.35 & 1888.76 & 2389.32 & 3049.57 \\ 
   \multicolumn{1}{l}{MAE} \\
  \hspace{1em}  Ref \cite{ma2021covid} &838.56 & 1352.34 & 1448.66 & 2132.51 \\ 
   \hspace{1em} ARGOX Joint Ensemble & 706.92 & 1043.13 & 1388.77 & 2152.39 \\ 
   \hspace{1em} Naive & 773.69 & 1220.87 & 1551.95 & 2036.52 \\ 
   \multicolumn{1}{l}{Correlation} \\
 \hspace{1em} Ref \cite{ma2021covid} &0.96 & 0.87 & 0.78 & 0.58 \\ 
   \hspace{1em} ARGOX Joint Ensemble  & 0.96 & 0.89 & 0.77 & 0.58 \\ 
   \hspace{1em} Naive& 0.94 & 0.86 & 0.74 & 0.62 \\ 
   \hline
\end{tabular}
\caption{Comparison of different methods for state-level COVID-19 1 to 4 weeks ahead incremental case in Nebraska (NE). The MSE, MAE, and correlation are reported and best performed method is highlighted in boldface.} 
\end{table}       \begin{figure}[!h]    \centering  \includegraphics[width=0.6\linewidth, page=30]{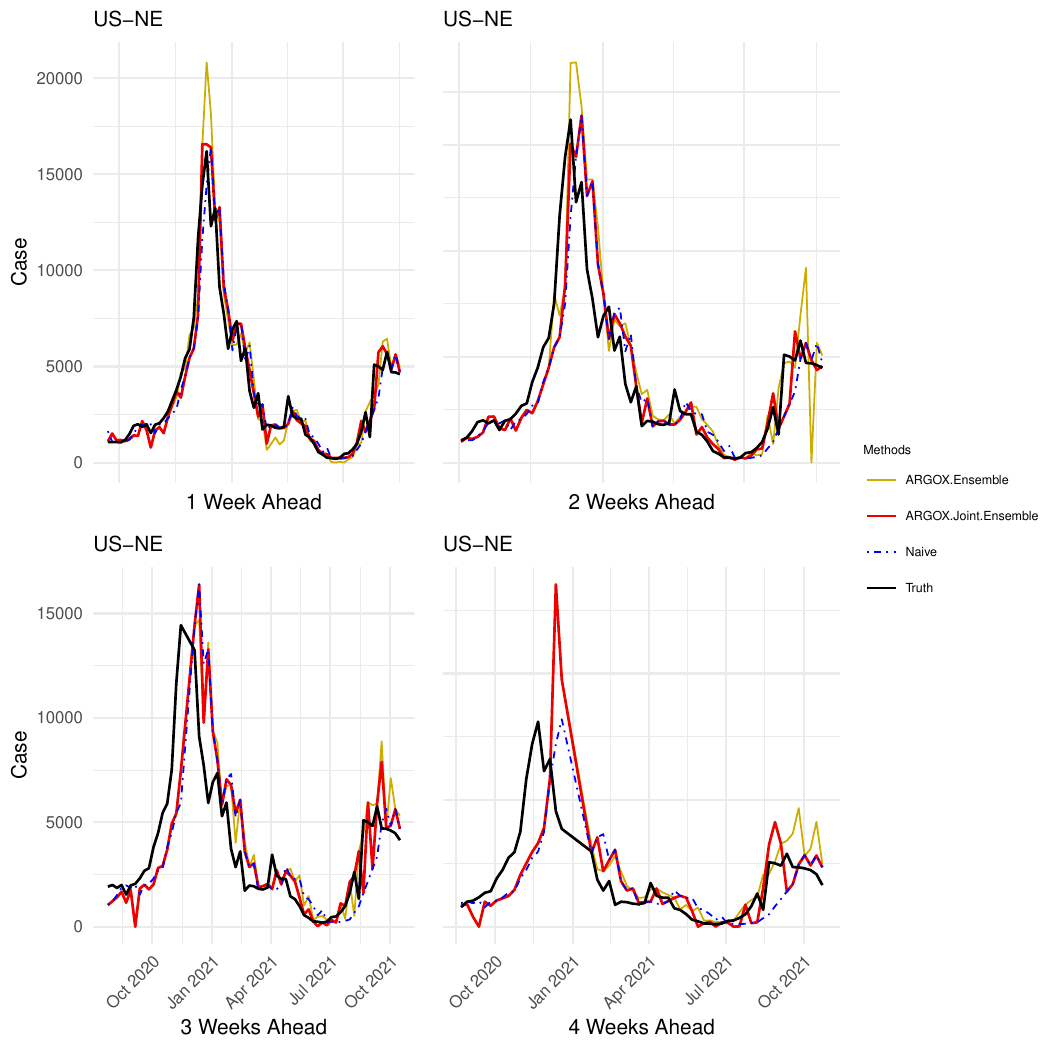}  \caption{Plots of the COVID-19 1 week (top left), 2 weeks (top right), 3 weeks (bottom left), and 4 weeks (bottom right) ahead estimates for Nebraska (NE). ARGOX-Ensemble is Ref \cite{ma2021covid}.}  \end{figure} \newpage   
\begin{table}[ht]
\centering
\begin{tabular}{lrrrr}
  \hline
& 1 Week Ahead & 2 Weeks Ahead & 3 Weeks Ahead & 4 Weeks Ahead \\ 
    \hline \multicolumn{1}{l}{RMSE} \\
 \hspace{1em} Ref \cite{ma2021covid} & 
571.09 & 848.81 & 909.69 & 1329.34 \\ 
   \hspace{1em} ARGOX Joint Ensemble & 548.32 & 831.55 & 888.13 & 1425.52 \\ 
   \hspace{1em} Naive & 581.70 & 878.47 & 1087.92 & 1416.51 \\ 
   \multicolumn{1}{l}{MAE} \\
  \hspace{1em}  Ref \cite{ma2021covid} &357.16 & 550.72 & 613.43 & 903.61 \\ 
   \hspace{1em} ARGOX Joint Ensemble & 303.86 & 555.08 & 597.27 & 999.12 \\ 
   \hspace{1em} Naive & 351.44 & 585.85 & 778.56 & 1059.67 \\ 
   \multicolumn{1}{l}{Correlation} \\
 \hspace{1em} Ref \cite{ma2021covid} &0.94 & 0.92 & 0.91 & 0.70 \\ 
   \hspace{1em} ARGOX Joint Ensemble  & 0.95 & 0.89 & 0.91 & 0.63 \\ 
   \hspace{1em} Naive& 0.94 & 0.86 & 0.79 & 0.56 \\ 
   \hline
\end{tabular}
\caption{Comparison of different methods for state-level COVID-19 1 to 4 weeks ahead incremental case in New Hampshire (NH). The MSE, MAE, and correlation are reported and best performed method is highlighted in boldface.} 
\end{table}       \begin{figure}[!h]    \centering  \includegraphics[width=0.6\linewidth, page=31]{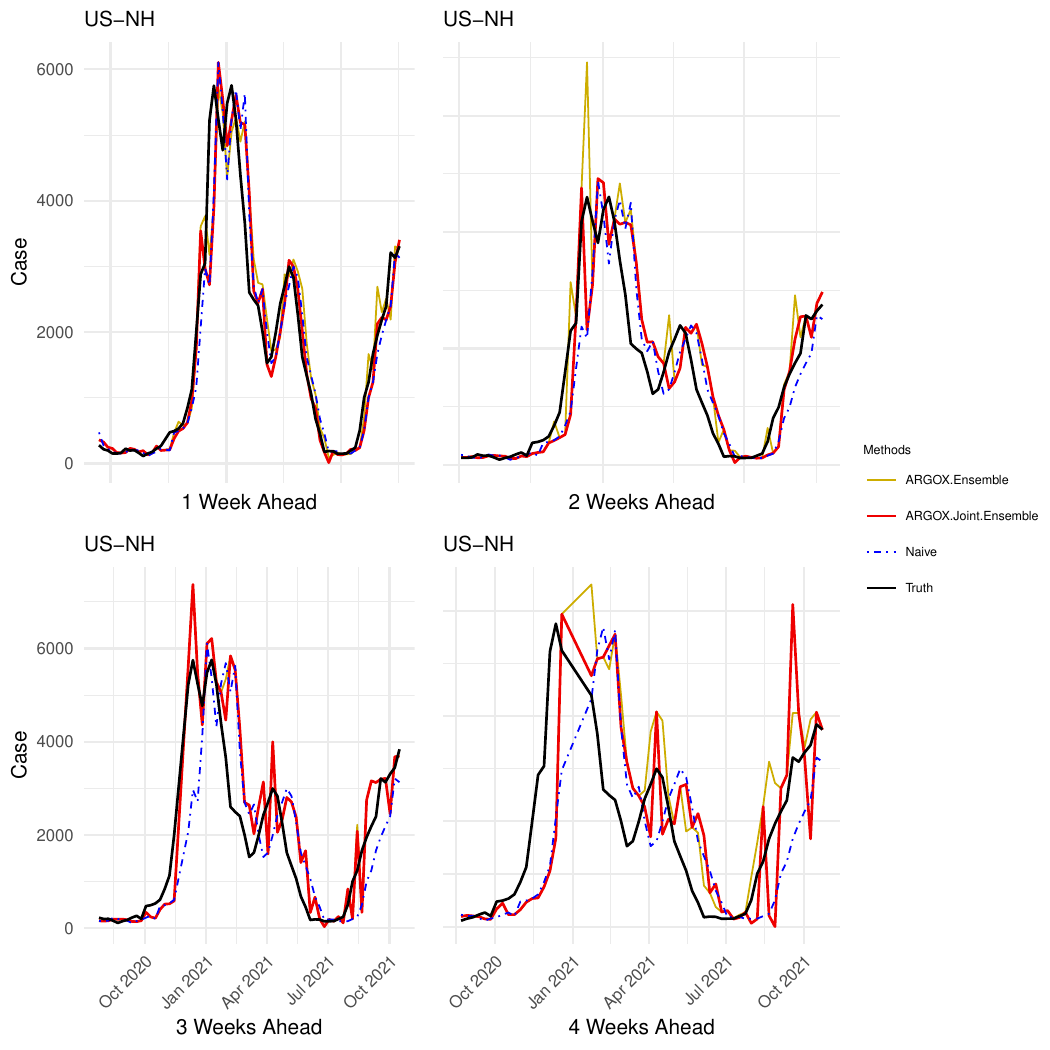}  \caption{Plots of the COVID-19 1 week (top left), 2 weeks (top right), 3 weeks (bottom left), and 4 weeks (bottom right) ahead estimates for New Hampshire (NH). ARGOX-Ensemble is Ref \cite{ma2021covid}.}  \end{figure} \newpage   
\begin{table}[ht]
\centering
\begin{tabular}{lrrrr}
  \hline
& 1 Week Ahead & 2 Weeks Ahead & 3 Weeks Ahead & 4 Weeks Ahead \\ 
    \hline \multicolumn{1}{l}{RMSE} \\
 \hspace{1em} Ref \cite{ma2021covid} & 
5705.50 & 7121.90 & 8434.91 & 10715.89 \\ 
   \hspace{1em} ARGOX Joint Ensemble & 3302.11 & 5179.48 & 8434.95 & 11544.17 \\ 
   \hspace{1em} Naive & 6837.96 & 8705.53 & 9895.80 & 12890.36 \\ 
   \multicolumn{1}{l}{MAE} \\
  \hspace{1em}  Ref \cite{ma2021covid} &3229.54 & 4203.79 & 4957.92 & 6809.82 \\ 
   \hspace{1em} ARGOX Joint Ensemble & 2275.10 & 3304.67 & 4997.31 & 7789.68 \\ 
   \hspace{1em} Naive & 3308.22 & 4794.91 & 5978.67 & 8153.84 \\ 
   \multicolumn{1}{l}{Correlation} \\
 \hspace{1em} Ref \cite{ma2021covid} &0.92 & 0.88 & 0.84 & 0.75 \\ 
   \hspace{1em} ARGOX Joint Ensemble  & 0.97 & 0.93 & 0.84 & 0.68 \\ 
   \hspace{1em} Naive& 0.89 & 0.81 & 0.76 & 0.55 \\ 
   \hline
\end{tabular}
\caption{Comparison of different methods for state-level COVID-19 1 to 4 weeks ahead incremental case in New Jersey (NJ). The MSE, MAE, and correlation are reported and best performed method is highlighted in boldface.} 
\end{table}       \begin{figure}[!h]    \centering  \includegraphics[width=0.6\linewidth, page=32]{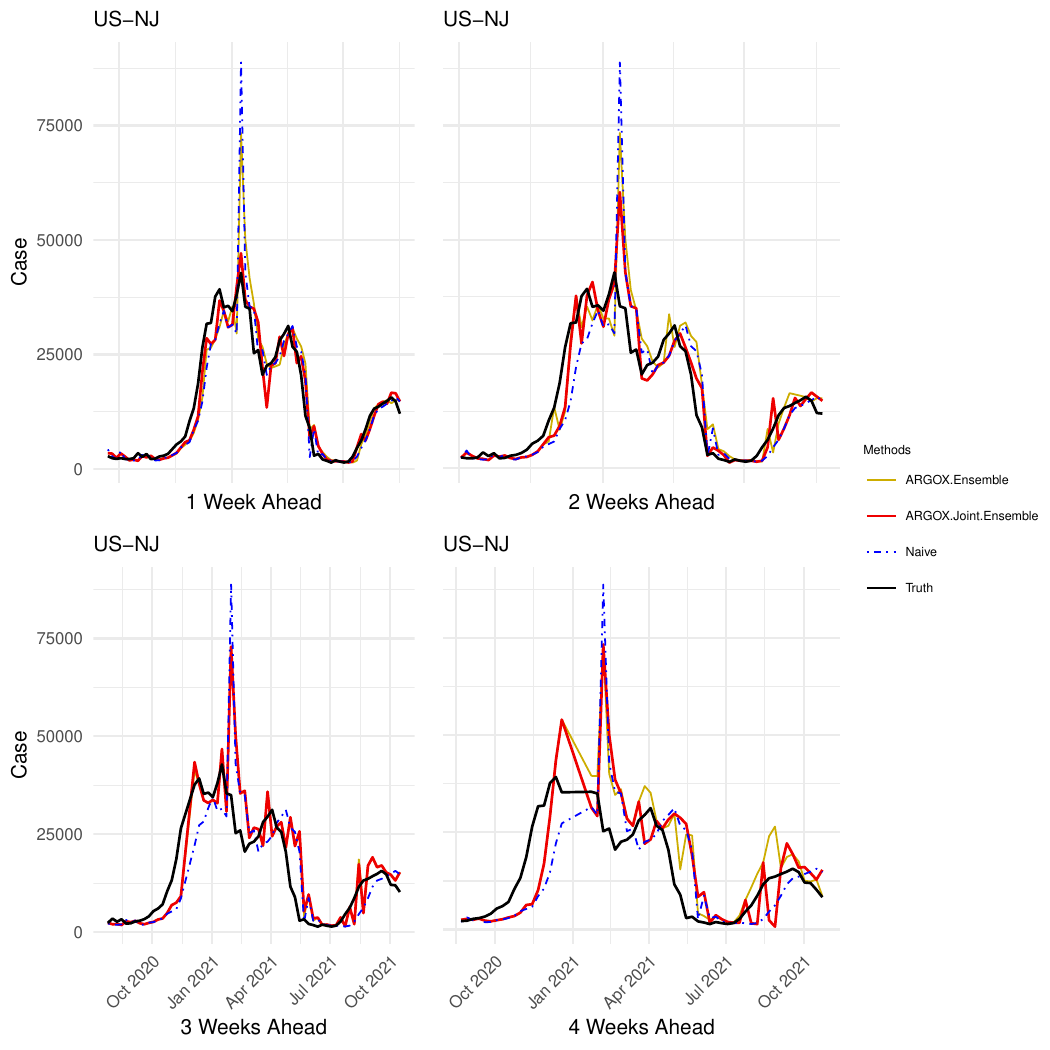}  \caption{Plots of the COVID-19 1 week (top left), 2 weeks (top right), 3 weeks (bottom left), and 4 weeks (bottom right) ahead estimates for New Jersey (NJ). ARGOX-Ensemble is Ref \cite{ma2021covid}.}  \end{figure} \newpage   
\begin{table}[ht]
\centering
\begin{tabular}{lrrrr}
  \hline
& 1 Week Ahead & 2 Weeks Ahead & 3 Weeks Ahead & 4 Weeks Ahead \\ 
    \hline \multicolumn{1}{l}{RMSE} \\
 \hspace{1em} Ref \cite{ma2021covid} & 
1086.06 & 2347.60 & 2621.09 & 3319.17 \\ 
   \hspace{1em} ARGOX Joint Ensemble & 1048.50 & 1707.07 & 2621.09 & 3268.07 \\ 
   \hspace{1em} Naive & 1187.67 & 1937.45 & 2035.02 & 3096.67 \\ 
   \multicolumn{1}{l}{MAE} \\
  \hspace{1em}  Ref \cite{ma2021covid} &672.79 & 1385.07 & 1529.47 & 1951.07 \\ 
   \hspace{1em} ARGOX Joint Ensemble & 672.17 & 1084.45 & 1529.47 & 1925.68 \\ 
   \hspace{1em} Naive & 700.74 & 1228.28 & 1451.90 & 2044.76 \\ 
   \multicolumn{1}{l}{Correlation} \\
 \hspace{1em} Ref \cite{ma2021covid} &0.97 & 0.87 & 0.87 & 0.69 \\ 
   \hspace{1em} ARGOX Joint Ensemble  & 0.98 & 0.90 & 0.87 & 0.68 \\ 
   \hspace{1em} Naive& 0.95 & 0.86 & 0.83 & 0.58 \\ 
   \hline
\end{tabular}
\caption{Comparison of different methods for state-level COVID-19 1 to 4 weeks ahead incremental case in New Mexico (NM). The MSE, MAE, and correlation are reported and best performed method is highlighted in boldface.} 
\end{table}       \begin{figure}[!h]    \centering  \includegraphics[width=0.6\linewidth, page=33]{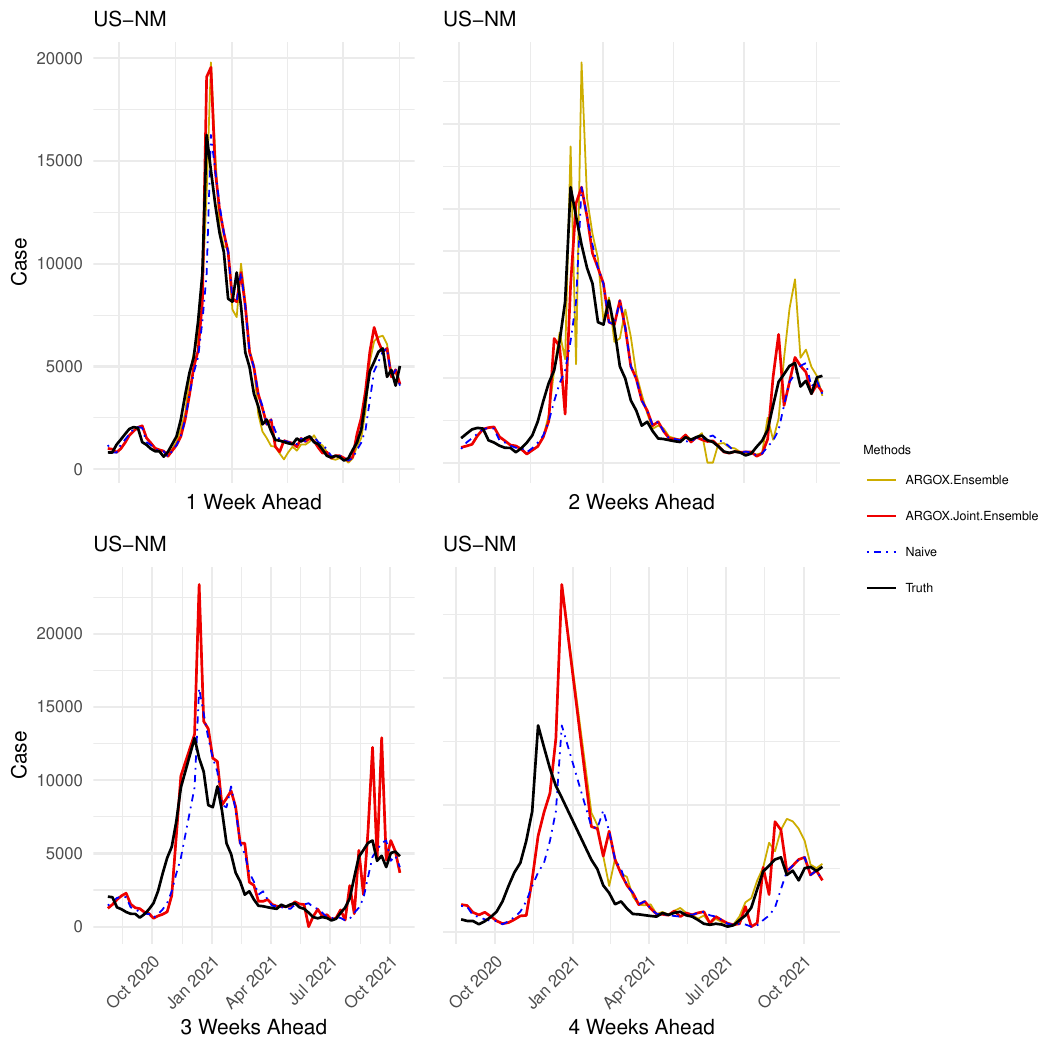}  \caption{Plots of the COVID-19 1 week (top left), 2 weeks (top right), 3 weeks (bottom left), and 4 weeks (bottom right) ahead estimates for New Mexico (NM). ARGOX-Ensemble is Ref \cite{ma2021covid}.}  \end{figure} \newpage   
\begin{table}[ht]
\centering
\begin{tabular}{lrrrr}
  \hline
& 1 Week Ahead & 2 Weeks Ahead & 3 Weeks Ahead & 4 Weeks Ahead \\ 
    \hline \multicolumn{1}{l}{RMSE} \\
 \hspace{1em} Ref \cite{ma2021covid} & 
1461.60 & 2079.76 & 3074.97 & 3913.64 \\ 
   \hspace{1em} ARGOX Joint Ensemble & 1364.14 & 2150.90 & 3074.23 & 3542.03 \\ 
   \hspace{1em} Naive & 1437.50 & 2382.50 & 2814.39 & 3995.81 \\ 
   \multicolumn{1}{l}{MAE} \\
  \hspace{1em}  Ref \cite{ma2021covid} &1069.69 & 1485.12 & 2001.13 & 2725.27 \\ 
   \hspace{1em} ARGOX Joint Ensemble & 1004.93 & 1626.15 & 1996.76 & 2618.17 \\ 
   \hspace{1em} Naive & 991.65 & 1714.48 & 2061.59 & 2921.21 \\ 
   \multicolumn{1}{l}{Correlation} \\
 \hspace{1em} Ref \cite{ma2021covid} &0.95 & 0.90 & 0.82 & 0.59 \\ 
   \hspace{1em} ARGOX Joint Ensemble  & 0.95 & 0.90 & 0.82 & 0.63 \\ 
   \hspace{1em} Naive& 0.95 & 0.86 & 0.80 & 0.47 \\ 
   \hline
\end{tabular}
\caption{Comparison of different methods for state-level COVID-19 1 to 4 weeks ahead incremental case in Nevada (NV). The MSE, MAE, and correlation are reported and best performed method is highlighted in boldface.} 
\end{table}       \begin{figure}[!h]    \centering  \includegraphics[width=0.6\linewidth, page=34]{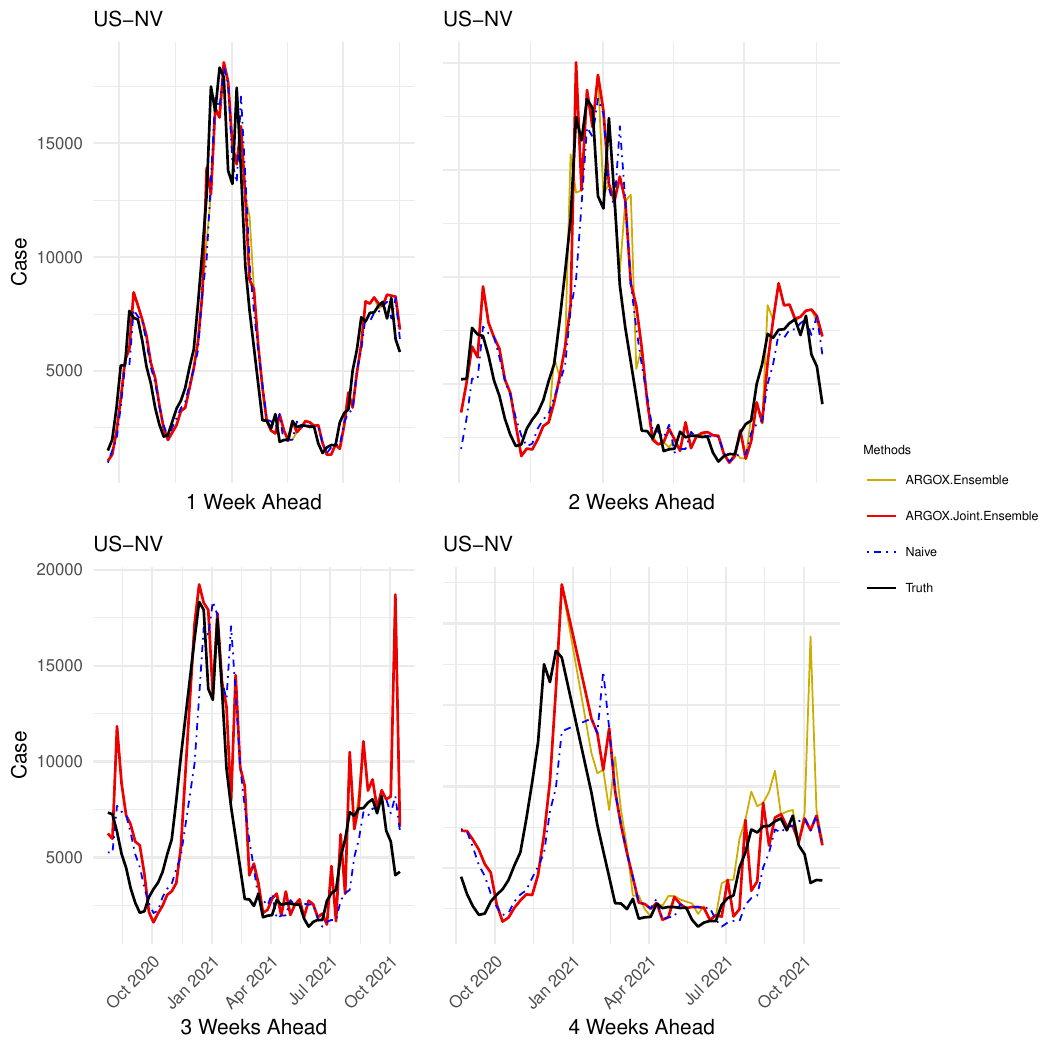}  \caption{Plots of the COVID-19 1 week (top left), 2 weeks (top right), 3 weeks (bottom left), and 4 weeks (bottom right) ahead estimates for Nevada (NV). ARGOX-Ensemble is Ref \cite{ma2021covid}.}  \end{figure} \newpage   
\begin{table}[ht]
\centering
\begin{tabular}{lrrrr}
  \hline
& 1 Week Ahead & 2 Weeks Ahead & 3 Weeks Ahead & 4 Weeks Ahead \\ 
    \hline \multicolumn{1}{l}{RMSE} \\
 \hspace{1em} Ref \cite{ma2021covid} & 
6569.15 & 11498.30 & 12577.83 & 16479.96 \\ 
   \hspace{1em} ARGOX Joint Ensemble & 5262.93 & 9850.97 & 12827.56 & 17216.52 \\ 
   \hspace{1em} Naive & 6657.64 & 11637.95 & 16082.69 & 19689.86 \\ 
   \multicolumn{1}{l}{MAE} \\
  \hspace{1em}  Ref \cite{ma2021covid} &4383.45 & 7526.77 & 7891.05 & 11329.30 \\ 
   \hspace{1em} ARGOX Joint Ensemble & 3543.95 & 6093.92 & 7893.67 & 12122.63 \\ 
   \hspace{1em} Naive & 4381.07 & 8184.76 & 11358.75 & 14260.07 \\ 
   \multicolumn{1}{l}{Correlation} \\
 \hspace{1em} Ref \cite{ma2021covid} &0.97 & 0.93 & 0.93 & 0.83 \\ 
   \hspace{1em} ARGOX Joint Ensemble  & 0.98 & 0.95 & 0.94 & 0.82 \\ 
   \hspace{1em} Naive& 0.97 & 0.92 & 0.84 & 0.73 \\ 
   \hline
\end{tabular}
\caption{Comparison of different methods for state-level COVID-19 1 to 4 weeks ahead incremental case in New York (NY). The MSE, MAE, and correlation are reported and best performed method is highlighted in boldface.} 
\end{table}       \begin{figure}[!h]    \centering  \includegraphics[width=0.6\linewidth, page=35]{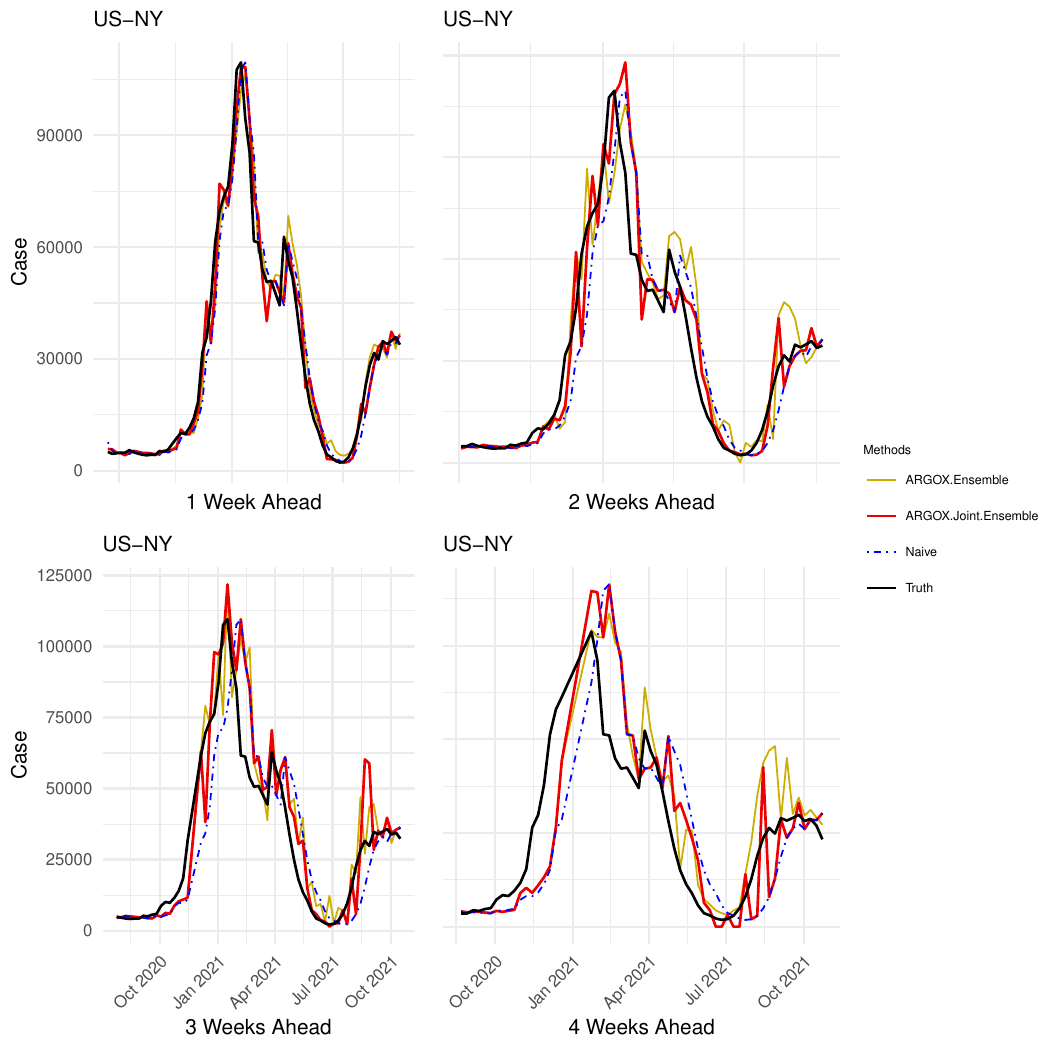}  \caption{Plots of the COVID-19 1 week (top left), 2 weeks (top right), 3 weeks (bottom left), and 4 weeks (bottom right) ahead estimates for New York (NY). ARGOX-Ensemble is Ref \cite{ma2021covid}.}  \end{figure} \newpage   
\begin{table}[ht]
\centering
\begin{tabular}{lrrrr}
  \hline
& 1 Week Ahead & 2 Weeks Ahead & 3 Weeks Ahead & 4 Weeks Ahead \\ 
    \hline \multicolumn{1}{l}{RMSE} \\
 \hspace{1em} Ref \cite{ma2021covid} & 
5624.09 & 9192.78 & 11126.25 & 19265.78 \\ 
   \hspace{1em} ARGOX Joint Ensemble & 5513.98 & 8588.03 & 13067.03 & 19800.68 \\ 
   \hspace{1em} Naive & 6283.06 & 9752.13 & 12077.30 & 16585.47 \\ 
   \multicolumn{1}{l}{MAE} \\
  \hspace{1em}  Ref \cite{ma2021covid} &3129.63 & 5576.53 & 7151.25 & 11330.17 \\ 
   \hspace{1em} ARGOX Joint Ensemble & 2926.86 & 5603.79 & 7897.45 & 12222.48 \\ 
   \hspace{1em} Naive & 3806.63 & 6488.12 & 8290.54 & 11807.00 \\ 
   \multicolumn{1}{l}{Correlation} \\
 \hspace{1em} Ref \cite{ma2021covid} &0.96 & 0.94 & 0.91 & 0.80 \\ 
   \hspace{1em} ARGOX Joint Ensemble  & 0.96 & 0.93 & 0.90 & 0.75 \\ 
   \hspace{1em} Naive& 0.95 & 0.88 & 0.81 & 0.57 \\ 
   \hline
\end{tabular}
\caption{Comparison of different methods for state-level COVID-19 1 to 4 weeks ahead incremental case in Ohio (OH). The MSE, MAE, and correlation are reported and best performed method is highlighted in boldface.} 
\end{table}       \begin{figure}[!h]    \centering  \includegraphics[width=0.6\linewidth, page=36]{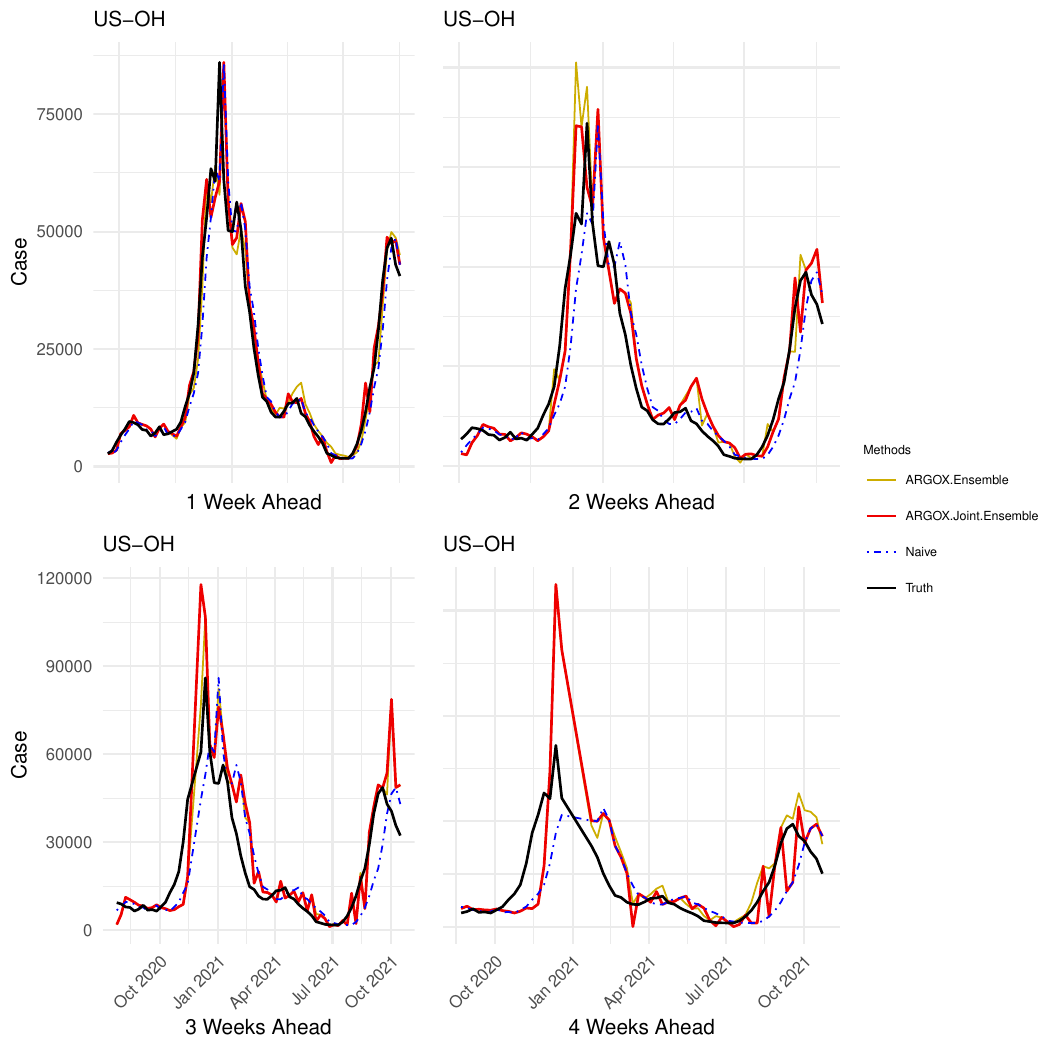}  \caption{Plots of the COVID-19 1 week (top left), 2 weeks (top right), 3 weeks (bottom left), and 4 weeks (bottom right) ahead estimates for Ohio (OH). ARGOX-Ensemble is Ref \cite{ma2021covid}.}  \end{figure} \newpage   
\begin{table}[ht]
\centering
\begin{tabular}{lrrrr}
  \hline
& 1 Week Ahead & 2 Weeks Ahead & 3 Weeks Ahead & 4 Weeks Ahead \\ 
    \hline \multicolumn{1}{l}{RMSE} \\
 \hspace{1em} Ref \cite{ma2021covid} & 
2870.32 & 3712.09 & 4466.10 & 6276.14 \\ 
   \hspace{1em} ARGOX Joint Ensemble & 2054.76 & 3008.33 & 4466.57 & 6132.35 \\ 
   \hspace{1em} Naive & 2424.84 & 3960.36 & 4556.24 & 6276.56 \\ 
   \multicolumn{1}{l}{MAE} \\
  \hspace{1em}  Ref \cite{ma2021covid} &1815.18 & 2391.56 & 2805.93 & 4015.48 \\ 
   \hspace{1em} ARGOX Joint Ensemble & 1372.54 & 2047.60 & 2805.51 & 4292.06 \\ 
   \hspace{1em} Naive & 1592.85 & 2726.21 & 3200.98 & 4470.00 \\ 
   \multicolumn{1}{l}{Correlation} \\
 \hspace{1em} Ref \cite{ma2021covid} &0.92 & 0.88 & 0.85 & 0.67 \\ 
   \hspace{1em} ARGOX Joint Ensemble  & 0.97 & 0.92 & 0.85 & 0.63 \\ 
   \hspace{1em} Naive& 0.95 & 0.85 & 0.80 & 0.57 \\ 
   \hline
\end{tabular}
\caption{Comparison of different methods for state-level COVID-19 1 to 4 weeks ahead incremental case in Oklahoma (OK). The MSE, MAE, and correlation are reported and best performed method is highlighted in boldface.} 
\end{table}       \begin{figure}[!h]    \centering  \includegraphics[width=0.6\linewidth, page=37]{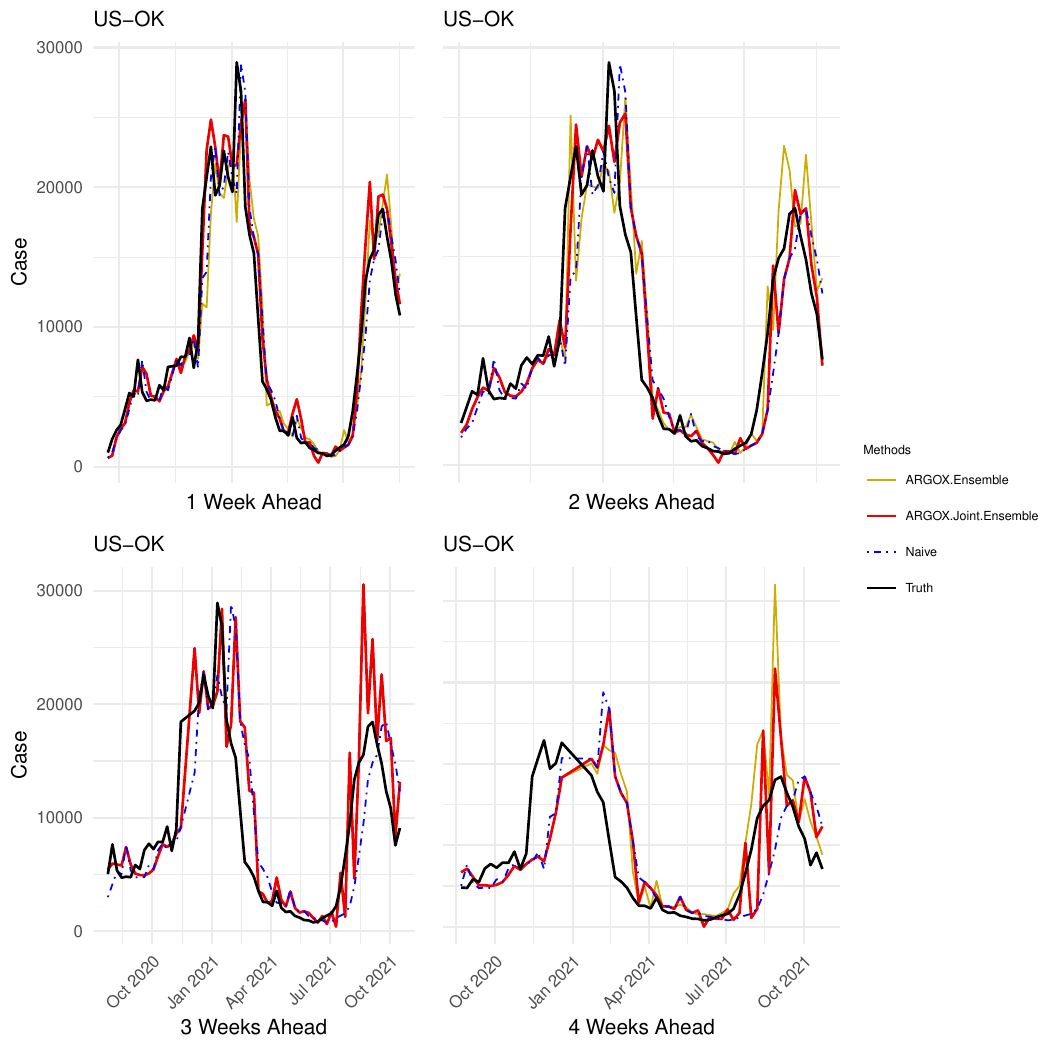}  \caption{Plots of the COVID-19 1 week (top left), 2 weeks (top right), 3 weeks (bottom left), and 4 weeks (bottom right) ahead estimates for Oklahoma (OK). ARGOX-Ensemble is Ref \cite{ma2021covid}.}  \end{figure} \newpage   
\begin{table}[ht]
\centering
\begin{tabular}{lrrrr}
  \hline
& 1 Week Ahead & 2 Weeks Ahead & 3 Weeks Ahead & 4 Weeks Ahead \\ 
    \hline \multicolumn{1}{l}{RMSE} \\
 \hspace{1em} Ref \cite{ma2021covid} & 
1104.02 & 2390.99 & 3589.45 & 4722.92 \\ 
   \hspace{1em} ARGOX Joint Ensemble & 1000.86 & 1474.11 & 2064.57 & 3078.08 \\ 
   \hspace{1em} Naive & 1152.21 & 2034.18 & 2861.51 & 3772.03 \\ 
   \multicolumn{1}{l}{MAE} \\
  \hspace{1em}  Ref \cite{ma2021covid} &767.31 & 1506.92 & 2364.84 & 2960.18 \\ 
   \hspace{1em} ARGOX Joint Ensemble & 639.12 & 1125.63 & 1590.76 & 2420.60 \\ 
   \hspace{1em} Naive & 779.03 & 1439.85 & 2038.14 & 2757.64 \\ 
   \multicolumn{1}{l}{Correlation} \\
 \hspace{1em} Ref \cite{ma2021covid} &0.97 & 0.84 & 0.81 & 0.84 \\ 
   \hspace{1em} ARGOX Joint Ensemble  & 0.98 & 0.94 & 0.90 & 0.78 \\ 
   \hspace{1em} Naive& 0.96 & 0.86 & 0.73 & 0.55 \\ 
   \hline
\end{tabular}
\caption{Comparison of different methods for state-level COVID-19 1 to 4 weeks ahead incremental case in Oregon (OR). The MSE, MAE, and correlation are reported and best performed method is highlighted in boldface.} 
\end{table}       \begin{figure}[!h]    \centering  \includegraphics[width=0.6\linewidth, page=38]{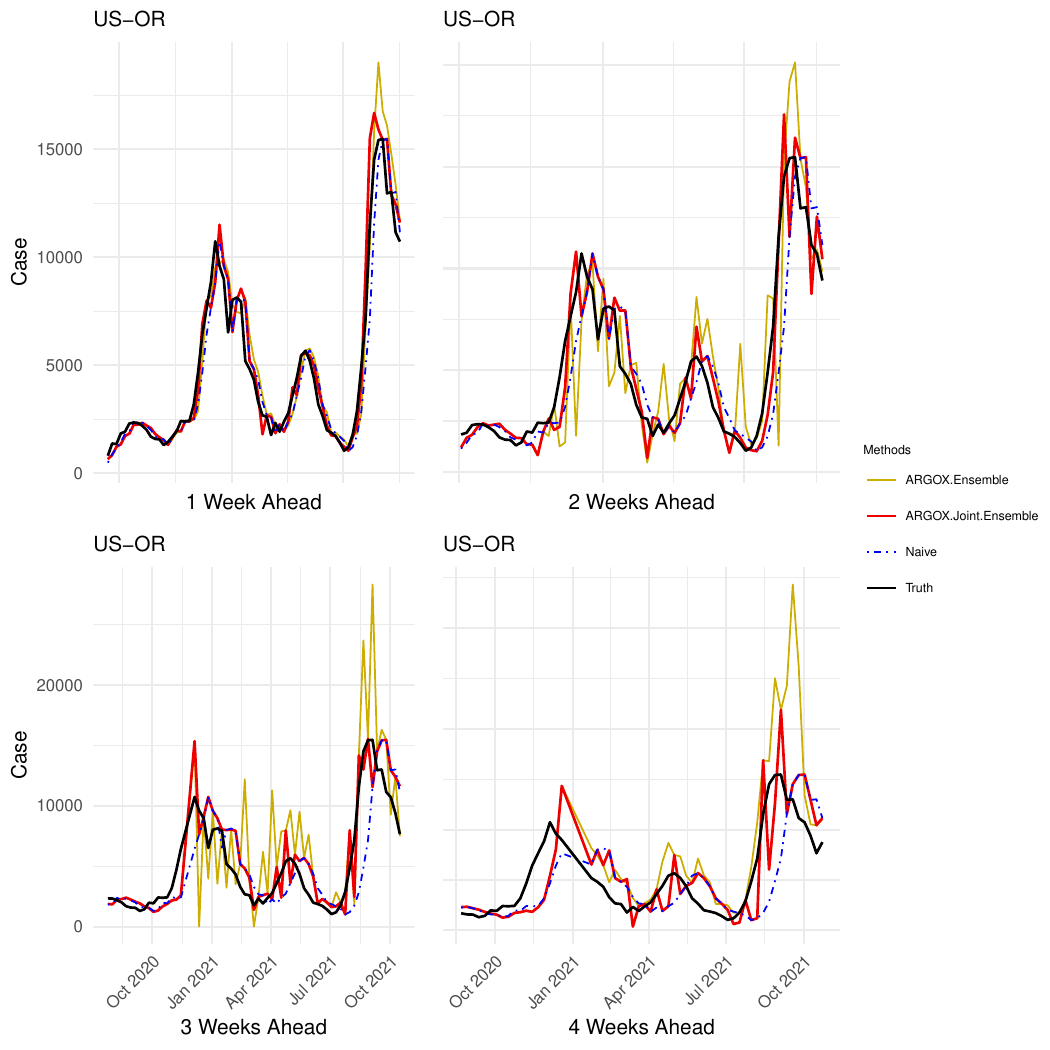}  \caption{Plots of the COVID-19 1 week (top left), 2 weeks (top right), 3 weeks (bottom left), and 4 weeks (bottom right) ahead estimates for Oregon (OR). ARGOX-Ensemble is Ref \cite{ma2021covid}.}  \end{figure} \newpage   
\begin{table}[ht]
\centering
\begin{tabular}{lrrrr}
  \hline
& 1 Week Ahead & 2 Weeks Ahead & 3 Weeks Ahead & 4 Weeks Ahead \\ 
    \hline \multicolumn{1}{l}{RMSE} \\
 \hspace{1em} Ref \cite{ma2021covid} & 
3898.21 & 6829.56 & 10729.37 & 13914.40 \\ 
   \hspace{1em} ARGOX Joint Ensemble & 3807.41 & 7041.95 & 10739.01 & 15279.40 \\ 
   \hspace{1em} Naive & 4823.39 & 8312.00 & 10620.64 & 15207.52 \\ 
   \multicolumn{1}{l}{MAE} \\
  \hspace{1em}  Ref \cite{ma2021covid} &2432.54 & 4529.05 & 7446.38 & 9218.69 \\ 
   \hspace{1em} ARGOX Joint Ensemble & 2267.23 & 4869.76 & 7413.61 & 10869.61 \\ 
   \hspace{1em} Naive & 3353.00 & 6108.70 & 8133.29 & 11667.84 \\ 
   \multicolumn{1}{l}{Correlation} \\
 \hspace{1em} Ref \cite{ma2021covid} &0.98 & 0.95 & 0.91 & 0.71 \\ 
   \hspace{1em} ARGOX Joint Ensemble  & 0.98 & 0.93 & 0.91 & 0.66 \\ 
   \hspace{1em} Naive& 0.96 & 0.89 & 0.82 & 0.53 \\ 
   \hline
\end{tabular}
\caption{Comparison of different methods for state-level COVID-19 1 to 4 weeks ahead incremental case in Pennsylvania (PA). The MSE, MAE, and correlation are reported and best performed method is highlighted in boldface.} 
\end{table}       \begin{figure}[!h]    \centering  \includegraphics[width=0.6\linewidth, page=39]{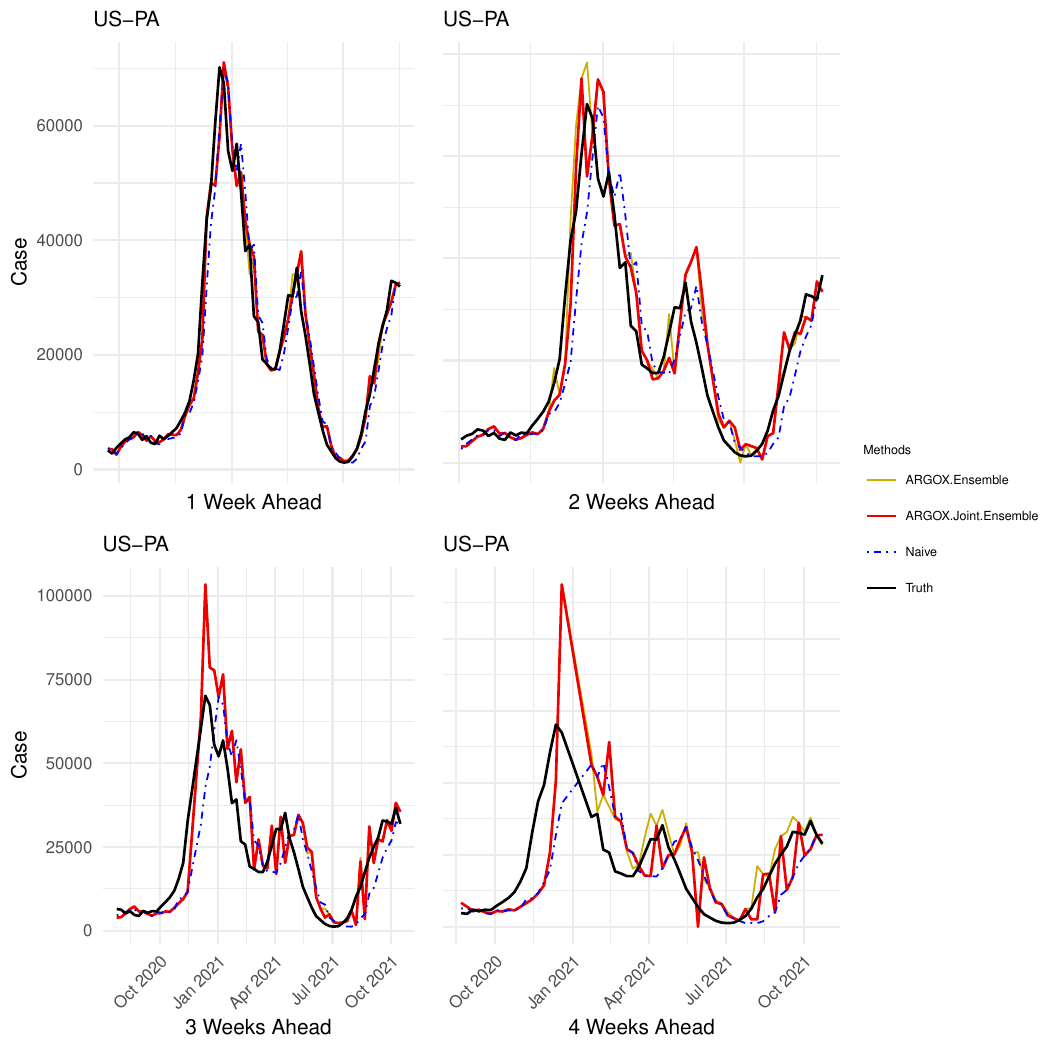}  \caption{Plots of the COVID-19 1 week (top left), 2 weeks (top right), 3 weeks (bottom left), and 4 weeks (bottom right) ahead estimates for Pennsylvania (PA). ARGOX-Ensemble is Ref \cite{ma2021covid}.}  \end{figure} \newpage   
\begin{table}[ht]
\centering
\begin{tabular}{lrrrr}
  \hline
& 1 Week Ahead & 2 Weeks Ahead & 3 Weeks Ahead & 4 Weeks Ahead \\ 
    \hline \multicolumn{1}{l}{RMSE} \\
 \hspace{1em} Ref \cite{ma2021covid} & 
1030.42 & 1531.51 & 1680.28 & 1509.22 \\ 
   \hspace{1em} ARGOX Joint Ensemble & 894.20 & 1027.01 & 1691.55 & 1653.89 \\ 
   \hspace{1em} Naive & 910.37 & 1369.32 & 1472.15 & 1794.96 \\ 
   \multicolumn{1}{l}{MAE} \\
  \hspace{1em}  Ref \cite{ma2021covid} &611.35 & 931.78 & 988.01 & 1009.69 \\ 
   \hspace{1em} ARGOX Joint Ensemble & 528.03 & 655.75 & 986.80 & 1187.54 \\ 
   \hspace{1em} Naive & 533.07 & 835.40 & 990.70 & 1215.10 \\ 
   \multicolumn{1}{l}{Correlation} \\
 \hspace{1em} Ref \cite{ma2021covid} &0.91 & 0.88 & 0.85 & 0.73 \\ 
   \hspace{1em} ARGOX Joint Ensemble  & 0.92 & 0.90 & 0.85 & 0.69 \\ 
   \hspace{1em} Naive& 0.92 & 0.82 & 0.79 & 0.58 \\ 
   \hline
\end{tabular}
\caption{Comparison of different methods for state-level COVID-19 1 to 4 weeks ahead incremental case in Rhode Island (RI). The MSE, MAE, and correlation are reported and best performed method is highlighted in boldface.} 
\end{table}       \begin{figure}[!h]    \centering  \includegraphics[width=0.6\linewidth, page=40]{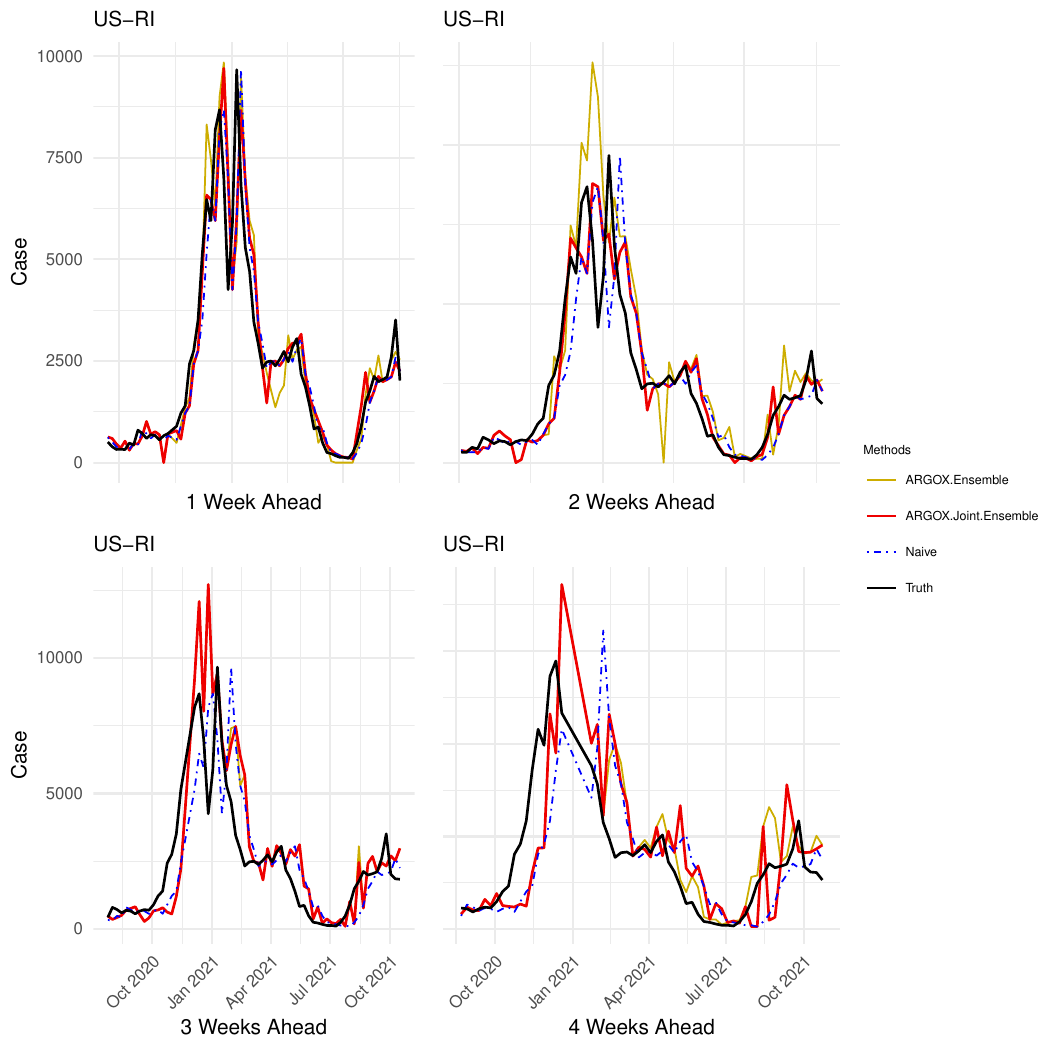}  \caption{Plots of the COVID-19 1 week (top left), 2 weeks (top right), 3 weeks (bottom left), and 4 weeks (bottom right) ahead estimates for Rhode Island (RI). ARGOX-Ensemble is Ref \cite{ma2021covid}.}  \end{figure} \newpage   
\begin{table}[ht]
\centering
\begin{tabular}{lrrrr}
  \hline
& 1 Week Ahead & 2 Weeks Ahead & 3 Weeks Ahead & 4 Weeks Ahead \\ 
    \hline \multicolumn{1}{l}{RMSE} \\
 \hspace{1em} Ref \cite{ma2021covid} & 
3451.22 & 5271.44 & 8056.01 & 10957.84 \\ 
   \hspace{1em} ARGOX Joint Ensemble & 3588.17 & 5405.27 & 8687.94 & 13927.57 \\ 
   \hspace{1em} Naive & 3611.79 & 5779.24 & 8024.31 & 9931.04 \\ 
   \multicolumn{1}{l}{MAE} \\
  \hspace{1em}  Ref \cite{ma2021covid} &2448.29 & 3802.96 & 5580.88 & 6656.14 \\ 
   \hspace{1em} ARGOX Joint Ensemble & 2120.98 & 3413.55 & 5416.07 & 7956.19 \\ 
   \hspace{1em} Naive & 2465.68 & 4037.24 & 5838.56 & 7327.48 \\ 
   \multicolumn{1}{l}{Correlation} \\
 \hspace{1em} Ref \cite{ma2021covid} &0.94 & 0.88 & 0.83 & 0.84 \\ 
   \hspace{1em} ARGOX Joint Ensemble  & 0.94 & 0.88 & 0.82 & 0.70 \\ 
   \hspace{1em} Naive& 0.93 & 0.83 & 0.68 & 0.51 \\ 
   \hline
\end{tabular}
\caption{Comparison of different methods for state-level COVID-19 1 to 4 weeks ahead incremental case in South Carolina (SC). The MSE, MAE, and correlation are reported and best performed method is highlighted in boldface.} 
\end{table}       \begin{figure}[!h]    \centering  \includegraphics[width=0.6\linewidth, page=41]{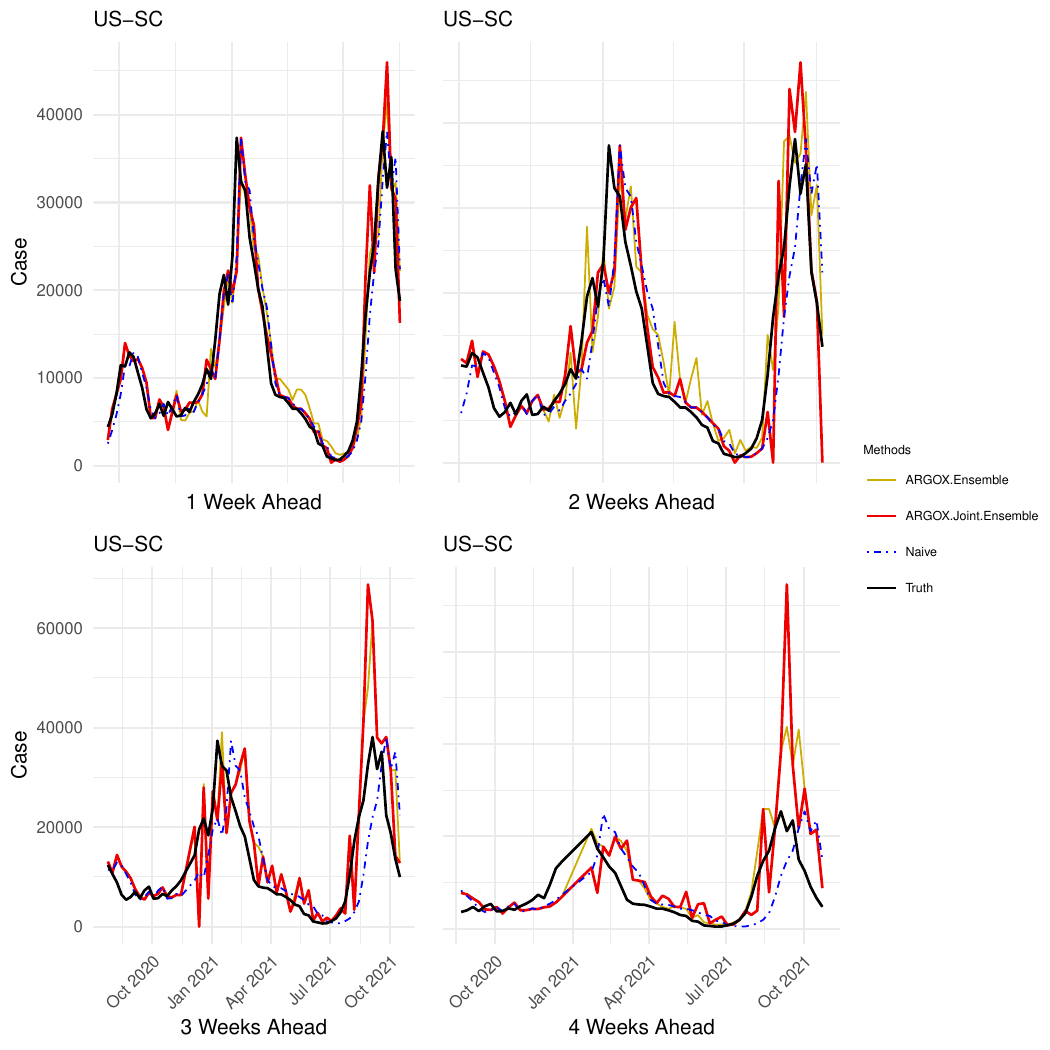}  \caption{Plots of the COVID-19 1 week (top left), 2 weeks (top right), 3 weeks (bottom left), and 4 weeks (bottom right) ahead estimates for South Carolina (SC). ARGOX-Ensemble is Ref \cite{ma2021covid}.}  \end{figure} \newpage   
\begin{table}[ht]
\centering
\begin{tabular}{lrrrr}
  \hline
& 1 Week Ahead & 2 Weeks Ahead & 3 Weeks Ahead & 4 Weeks Ahead \\ 
    \hline \multicolumn{1}{l}{RMSE} \\
 \hspace{1em} Ref \cite{ma2021covid} & 
797.16 & 1225.33 & 1455.94 & 2213.03 \\ 
   \hspace{1em} ARGOX Joint Ensemble & 603.41 & 982.95 & 1348.68 & 2260.94 \\ 
   \hspace{1em} Naive & 651.59 & 1065.98 & 1515.53 & 1808.07 \\ 
   \multicolumn{1}{l}{MAE} \\
  \hspace{1em}  Ref \cite{ma2021covid} &480.28 & 761.13 & 891.17 & 1313.22 \\ 
   \hspace{1em} ARGOX Joint Ensemble & 367.67 & 561.39 & 822.63 & 1415.68 \\ 
   \hspace{1em} Naive & 385.60 & 678.52 & 990.48 & 1222.24 \\ 
   \multicolumn{1}{l}{Correlation} \\
 \hspace{1em} Ref \cite{ma2021covid} &0.94 & 0.89 & 0.77 & 0.63 \\ 
   \hspace{1em} ARGOX Joint Ensemble  & 0.96 & 0.91 & 0.79 & 0.61 \\ 
   \hspace{1em} Naive& 0.96 & 0.89 & 0.73 & 0.69 \\ 
   \hline
\end{tabular}
\caption{Comparison of different methods for state-level COVID-19 1 to 4 weeks ahead incremental case in South Dakota (SD). The MSE, MAE, and correlation are reported and best performed method is highlighted in boldface.} 
\end{table}       \begin{figure}[!h]    \centering  \includegraphics[width=0.6\linewidth, page=42]{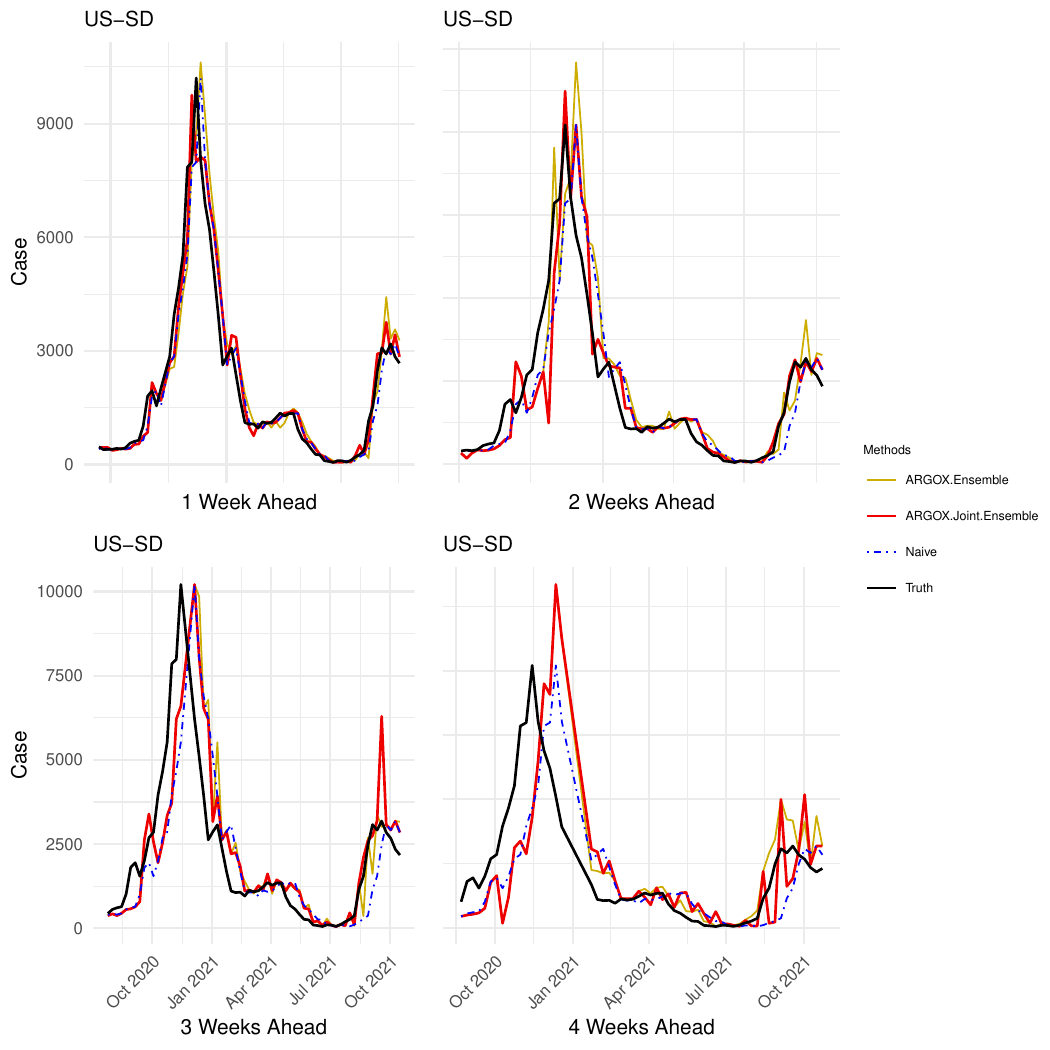}  \caption{Plots of the COVID-19 1 week (top left), 2 weeks (top right), 3 weeks (bottom left), and 4 weeks (bottom right) ahead estimates for South Dakota (SD). ARGOX-Ensemble is Ref \cite{ma2021covid}.}  \end{figure} \newpage   
\begin{table}[ht]
\centering
\begin{tabular}{lrrrr}
  \hline
& 1 Week Ahead & 2 Weeks Ahead & 3 Weeks Ahead & 4 Weeks Ahead \\ 
    \hline \multicolumn{1}{l}{RMSE} \\
 \hspace{1em} Ref \cite{ma2021covid} & 
5516.75 & 7515.26 & 10658.06 & 15893.80 \\ 
   \hspace{1em} ARGOX Joint Ensemble & 5461.38 & 8179.63 & 9861.83 & 13945.57 \\ 
   \hspace{1em} Naive & 5622.54 & 9510.47 & 12747.36 & 15502.30 \\ 
   \multicolumn{1}{l}{MAE} \\
  \hspace{1em}  Ref \cite{ma2021covid} &3442.21 & 5008.50 & 7611.33 & 10577.94 \\ 
   \hspace{1em} ARGOX Joint Ensemble & 3323.04 & 5317.33 & 7279.13 & 10169.47 \\ 
   \hspace{1em} Naive & 3828.07 & 6608.69 & 8950.32 & 11253.36 \\ 
   \multicolumn{1}{l}{Correlation} \\
 \hspace{1em} Ref \cite{ma2021covid} &0.94 & 0.90 & 0.88 & 0.75 \\ 
   \hspace{1em} ARGOX Joint Ensemble  & 0.94 & 0.87 & 0.89 & 0.60 \\ 
   \hspace{1em} Naive& 0.93 & 0.80 & 0.66 & 0.42 \\ 
   \hline
\end{tabular}
\caption{Comparison of different methods for state-level COVID-19 1 to 4 weeks ahead incremental case in Tennessee (TN). The MSE, MAE, and correlation are reported and best performed method is highlighted in boldface.} 
\end{table}       \begin{figure}[!h]    \centering  \includegraphics[width=0.6\linewidth, page=43]{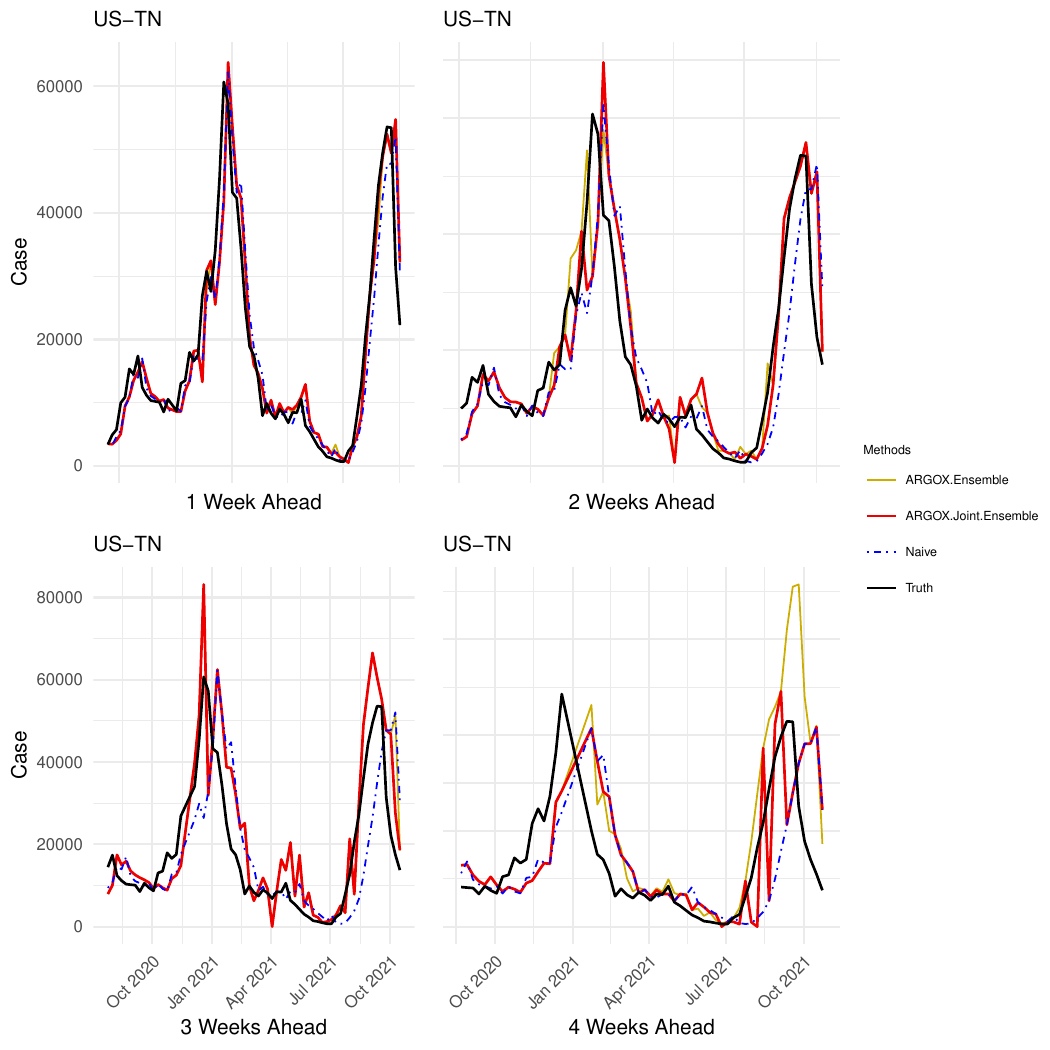}  \caption{Plots of the COVID-19 1 week (top left), 2 weeks (top right), 3 weeks (bottom left), and 4 weeks (bottom right) ahead estimates for Tennessee (TN). ARGOX-Ensemble is Ref \cite{ma2021covid}.}  \end{figure} \newpage   
\begin{table}[ht]
\centering
\begin{tabular}{lrrrr}
  \hline
& 1 Week Ahead & 2 Weeks Ahead & 3 Weeks Ahead & 4 Weeks Ahead \\ 
    \hline \multicolumn{1}{l}{RMSE} \\
 \hspace{1em} Ref \cite{ma2021covid} & 
17712.96 & 29055.77 & 33658.15 & 43817.95 \\ 
   \hspace{1em} ARGOX Joint Ensemble & 16941.41 & 23380.99 & 35500.70 & 41769.56 \\ 
   \hspace{1em} Naive & 17603.64 & 26439.11 & 32899.33 & 40461.44 \\ 
   \multicolumn{1}{l}{MAE} \\
  \hspace{1em}  Ref \cite{ma2021covid} &12433.07 & 20256.86 & 23137.43 & 30582.44 \\ 
   \hspace{1em} ARGOX Joint Ensemble & 11539.25 & 16592.33 & 24575.62 & 30460.49 \\ 
   \hspace{1em} Naive & 12324.09 & 19135.09 & 23853.40 & 30164.38 \\ 
   \multicolumn{1}{l}{Correlation} \\
 \hspace{1em} Ref \cite{ma2021covid} &0.91 & 0.84 & 0.81 & 0.77 \\ 
   \hspace{1em} ARGOX Joint Ensemble  & 0.92 & 0.86 & 0.79 & 0.59 \\ 
   \hspace{1em} Naive& 0.91 & 0.79 & 0.69 & 0.50 \\ 
   \hline
\end{tabular}
\caption{Comparison of different methods for state-level COVID-19 1 to 4 weeks ahead incremental case in Texas (TX). The MSE, MAE, and correlation are reported and best performed method is highlighted in boldface.} 
\end{table}       \begin{figure}[!h]    \centering  \includegraphics[width=0.6\linewidth, page=44]{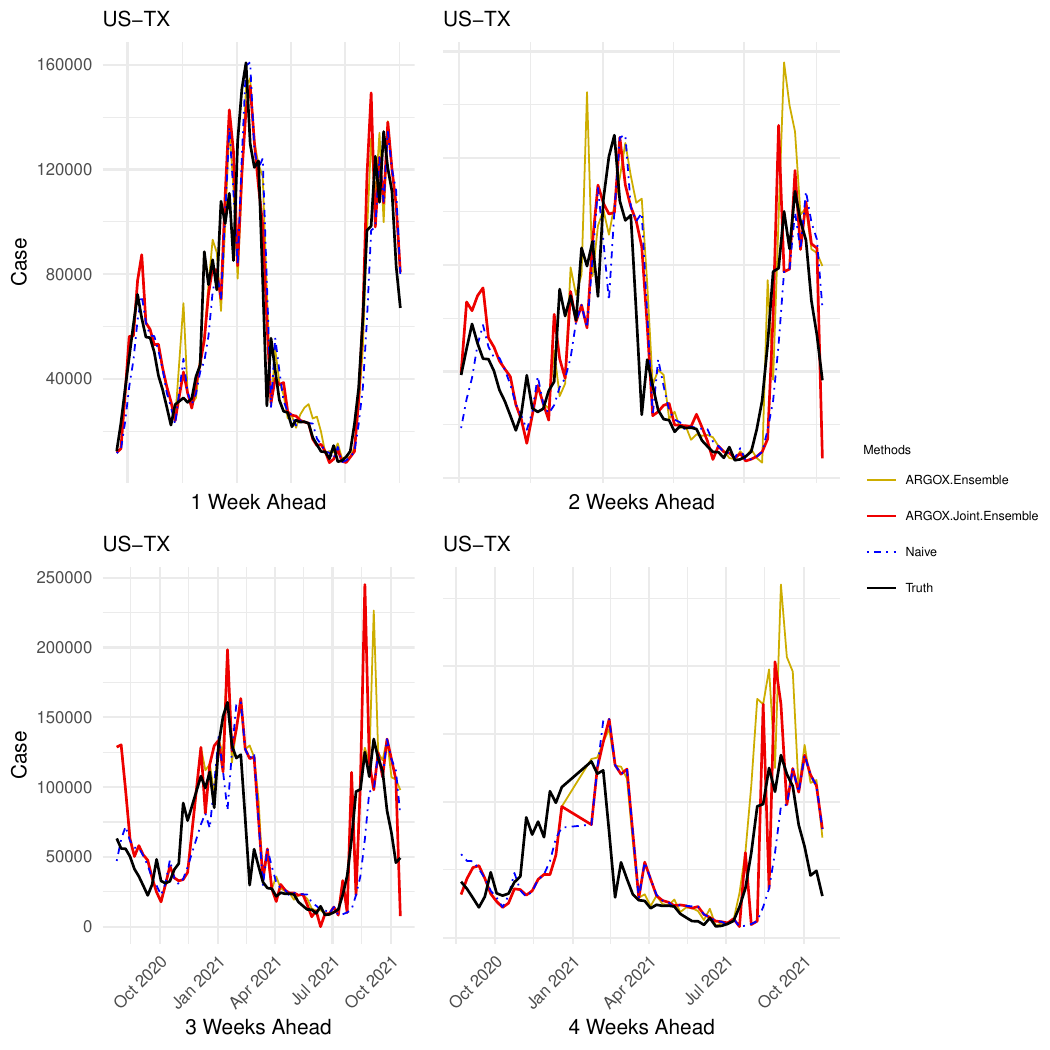}  \caption{Plots of the COVID-19 1 week (top left), 2 weeks (top right), 3 weeks (bottom left), and 4 weeks (bottom right) ahead estimates for Texas (TX). ARGOX-Ensemble is Ref \cite{ma2021covid}.}  \end{figure} \newpage   
\begin{table}[ht]
\centering
\begin{tabular}{lrrrr}
  \hline
& 1 Week Ahead & 2 Weeks Ahead & 3 Weeks Ahead & 4 Weeks Ahead \\ 
    \hline \multicolumn{1}{l}{RMSE} \\
 \hspace{1em} Ref \cite{ma2021covid} & 
1870.01 & 2795.10 & 3604.78 & 4883.49 \\ 
   \hspace{1em} ARGOX Joint Ensemble & 1497.39 & 2831.93 & 3184.06 & 4965.71 \\ 
   \hspace{1em} Naive & 1748.39 & 2862.24 & 3455.11 & 4572.21 \\ 
   \multicolumn{1}{l}{MAE} \\
  \hspace{1em}  Ref \cite{ma2021covid} &1163.85 & 1719.80 & 2204.21 & 3184.27 \\ 
   \hspace{1em} ARGOX Joint Ensemble & 924.73 & 1826.16 & 2060.12 & 3324.03 \\ 
   \hspace{1em} Naive & 1104.44 & 1859.21 & 2357.89 & 3189.67 \\ 
   \multicolumn{1}{l}{Correlation} \\
 \hspace{1em} Ref \cite{ma2021covid} &0.95 & 0.90 & 0.84 & 0.70 \\ 
   \hspace{1em} ARGOX Joint Ensemble  & 0.97 & 0.89 & 0.86 & 0.68 \\ 
   \hspace{1em} Naive& 0.96 & 0.88 & 0.82 & 0.65 \\ 
   \hline
\end{tabular}
\caption{Comparison of different methods for state-level COVID-19 1 to 4 weeks ahead incremental case in Utah (UT). The MSE, MAE, and correlation are reported and best performed method is highlighted in boldface.} 
\end{table}       \begin{figure}[!h]    \centering  \includegraphics[width=0.6\linewidth, page=45]{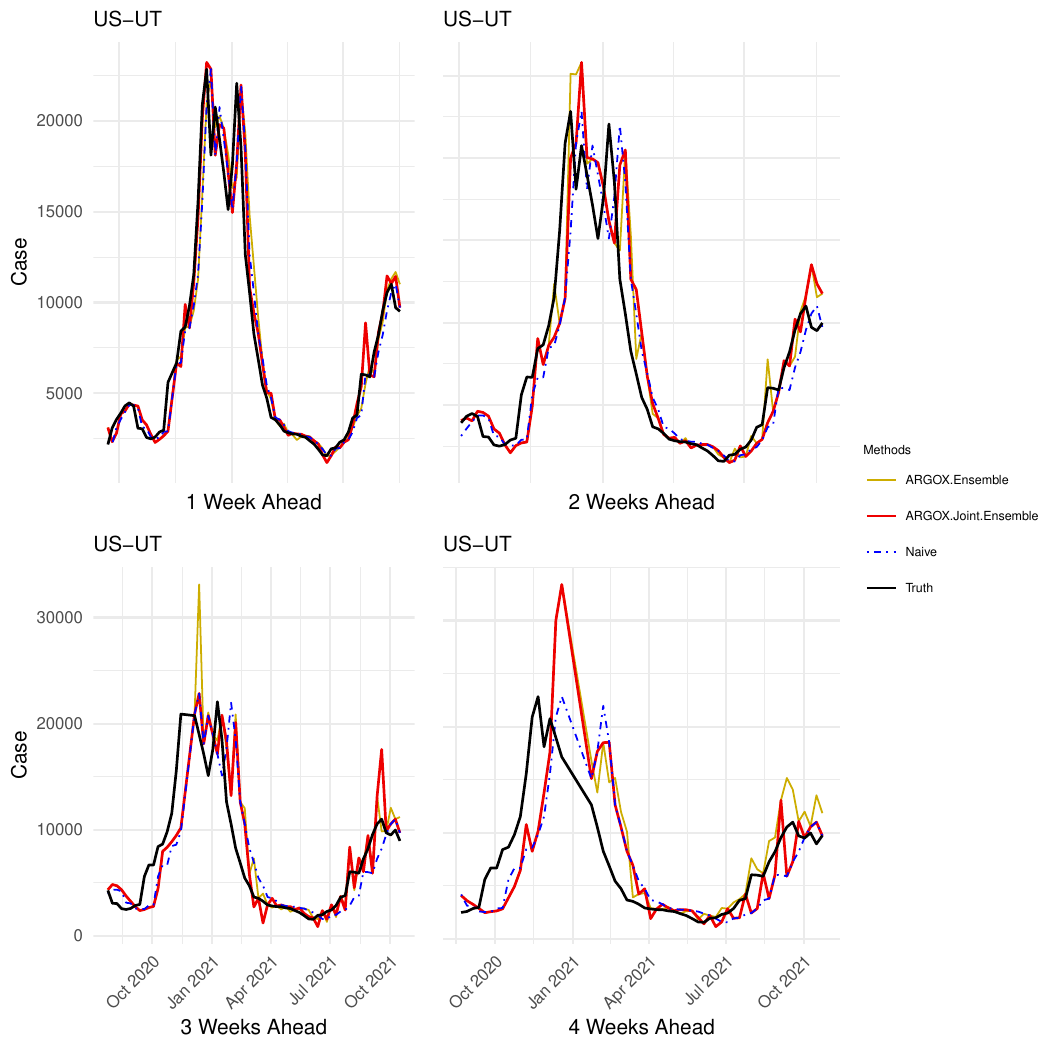}  \caption{Plots of the COVID-19 1 week (top left), 2 weeks (top right), 3 weeks (bottom left), and 4 weeks (bottom right) ahead estimates for Utah (UT). ARGOX-Ensemble is Ref \cite{ma2021covid}.}  \end{figure} \newpage   
\begin{table}[ht]
\centering
\begin{tabular}{lrrrr}
  \hline
& 1 Week Ahead & 2 Weeks Ahead & 3 Weeks Ahead & 4 Weeks Ahead \\ 
    \hline \multicolumn{1}{l}{RMSE} \\
 \hspace{1em} Ref \cite{ma2021covid} & 
2874.32 & 4578.45 & 6223.48 & 6905.69 \\ 
   \hspace{1em} ARGOX Joint Ensemble & 2360.04 & 4531.21 & 6196.60 & 7763.02 \\ 
   \hspace{1em} Naive & 2752.77 & 4554.38 & 6437.13 & 8277.92 \\ 
   \multicolumn{1}{l}{MAE} \\
  \hspace{1em}  Ref \cite{ma2021covid} &2033.78 & 3033.17 & 3991.24 & 4252.31 \\ 
   \hspace{1em} ARGOX Joint Ensemble & 1578.47 & 3212.22 & 3930.01 & 5385.62 \\ 
   \hspace{1em} Naive & 1828.66 & 3212.48 & 4615.81 & 6028.53 \\ 
   \multicolumn{1}{l}{Correlation} \\
 \hspace{1em} Ref \cite{ma2021covid} &0.95 & 0.89 & 0.85 & 0.78 \\ 
   \hspace{1em} ARGOX Joint Ensemble  & 0.97 & 0.88 & 0.85 & 0.67 \\ 
   \hspace{1em} Naive& 0.96 & 0.89 & 0.78 & 0.59 \\ 
   \hline
\end{tabular}
\caption{Comparison of different methods for state-level COVID-19 1 to 4 weeks ahead incremental case in Virginia (VA). The MSE, MAE, and correlation are reported and best performed method is highlighted in boldface.} 
\end{table}       \begin{figure}[!h]    \centering  \includegraphics[width=0.6\linewidth, page=46]{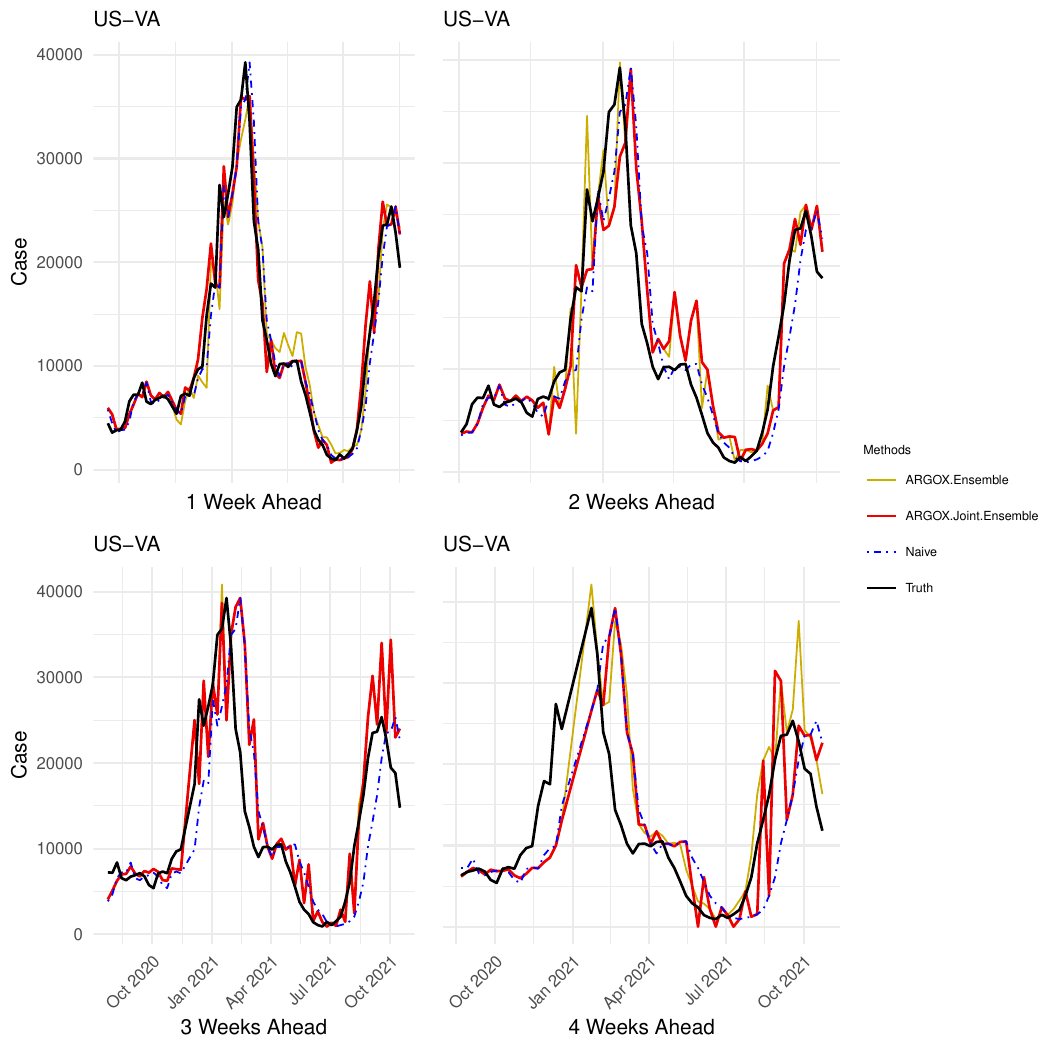}  \caption{Plots of the COVID-19 1 week (top left), 2 weeks (top right), 3 weeks (bottom left), and 4 weeks (bottom right) ahead estimates for Virginia (VA). ARGOX-Ensemble is Ref \cite{ma2021covid}.}  \end{figure} \newpage   
\begin{table}[ht]
\centering
\begin{tabular}{lrrrr}
  \hline
& 1 Week Ahead & 2 Weeks Ahead & 3 Weeks Ahead & 4 Weeks Ahead \\ 
    \hline \multicolumn{1}{l}{RMSE} \\
 \hspace{1em} Ref \cite{ma2021covid} & 
136.01 & 212.35 & 315.15 & 355.49 \\ 
   \hspace{1em} ARGOX Joint Ensemble & 133.04 & 225.42 & 313.55 & 403.88 \\ 
   \hspace{1em} Naive & 138.47 & 212.81 & 269.94 & 347.38 \\ 
   \multicolumn{1}{l}{MAE} \\
  \hspace{1em}  Ref \cite{ma2021covid} &93.99 & 134.50 & 197.14 & 248.42 \\ 
   \hspace{1em} ARGOX Joint Ensemble & 89.14 & 149.46 & 192.70 & 281.15 \\ 
   \hspace{1em} Naive & 93.07 & 144.69 & 199.73 & 264.71 \\ 
   \multicolumn{1}{l}{Correlation} \\
 \hspace{1em} Ref \cite{ma2021covid} &0.96 & 0.89 & 0.83 & 0.87 \\ 
   \hspace{1em} ARGOX Joint Ensemble  & 0.95 & 0.87 & 0.83 & 0.72 \\ 
   \hspace{1em} Naive& 0.95 & 0.89 & 0.83 & 0.73 \\ 
   \hline
\end{tabular}
\caption{Comparison of different methods for state-level COVID-19 1 to 4 weeks ahead incremental case in Vermont (VT). The MSE, MAE, and correlation are reported and best performed method is highlighted in boldface.} 
\end{table}       \begin{figure}[!h]    \centering  \includegraphics[width=0.6\linewidth, page=47]{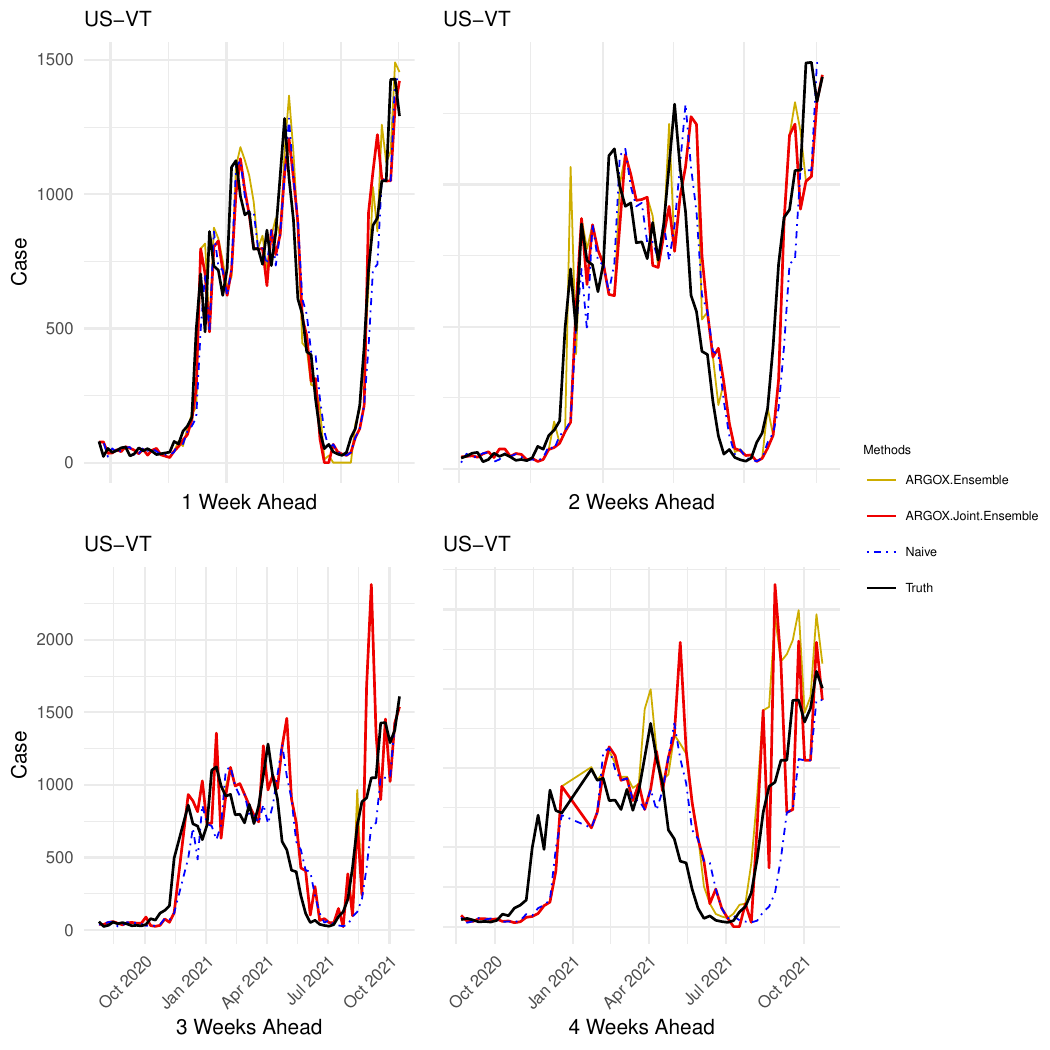}  \caption{Plots of the COVID-19 1 week (top left), 2 weeks (top right), 3 weeks (bottom left), and 4 weeks (bottom right) ahead estimates for Vermont (VT). ARGOX-Ensemble is Ref \cite{ma2021covid}.}  \end{figure} \newpage   
\begin{table}[ht]
\centering
\begin{tabular}{lrrrr}
  \hline
& 1 Week Ahead & 2 Weeks Ahead & 3 Weeks Ahead & 4 Weeks Ahead \\ 
    \hline \multicolumn{1}{l}{RMSE} \\
 \hspace{1em} Ref \cite{ma2021covid} & 
3087.13 & 3850.42 & 6219.64 & 7762.60 \\ 
   \hspace{1em} ARGOX Joint Ensemble & 3081.17 & 3272.04 & 3712.91 & 5080.54 \\ 
   \hspace{1em} Naive & 3153.05 & 3976.52 & 4904.10 & 6204.14 \\ 
   \multicolumn{1}{l}{MAE} \\
  \hspace{1em}  Ref \cite{ma2021covid} &1999.93 & 2524.59 & 3882.36 & 4872.31 \\ 
   \hspace{1em} ARGOX Joint Ensemble & 1994.32 & 2301.22 & 2729.54 & 3941.62 \\ 
   \hspace{1em} Naive & 1980.74 & 2845.04 & 3533.65 & 4596.79 \\ 
   \multicolumn{1}{l}{Correlation} \\
 \hspace{1em} Ref \cite{ma2021covid} &0.89 & 0.86 & 0.78 & 0.75 \\ 
   \hspace{1em} ARGOX Joint Ensemble  & 0.89 & 0.90 & 0.85 & 0.73 \\ 
   \hspace{1em} Naive& 0.89 & 0.82 & 0.73 & 0.55 \\ 
   \hline
\end{tabular}
\caption{Comparison of different methods for state-level COVID-19 1 to 4 weeks ahead incremental case in Washington (WA). The MSE, MAE, and correlation are reported and best performed method is highlighted in boldface.} 
\end{table}       \begin{figure}[!h]    \centering  \includegraphics[width=0.6\linewidth, page=48]{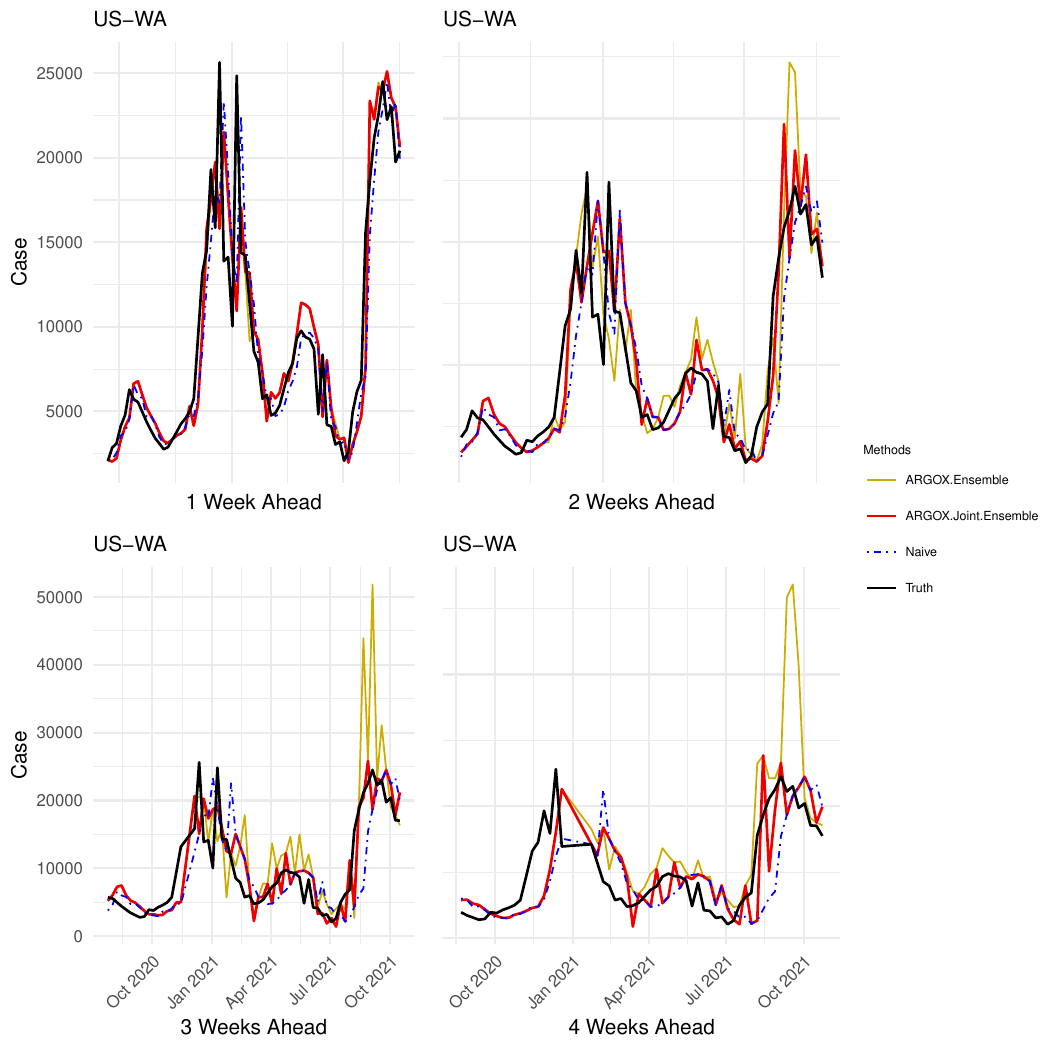}  \caption{Plots of the COVID-19 1 week (top left), 2 weeks (top right), 3 weeks (bottom left), and 4 weeks (bottom right) ahead estimates for Washington (WA). ARGOX-Ensemble is Ref \cite{ma2021covid}.}  \end{figure} \newpage   
\begin{table}[ht]
\centering
\begin{tabular}{lrrrr}
  \hline
& 1 Week Ahead & 2 Weeks Ahead & 3 Weeks Ahead & 4 Weeks Ahead \\ 
    \hline \multicolumn{1}{l}{RMSE} \\
 \hspace{1em} Ref \cite{ma2021covid} & 
3776.21 & 6658.41 & 7735.78 & 10518.60 \\ 
   \hspace{1em} ARGOX Joint Ensemble & 3729.16 & 7014.42 & 6833.72 & 12310.39 \\ 
   \hspace{1em} Naive & 4015.59 & 5946.46 & 7419.84 & 9247.15 \\ 
   \multicolumn{1}{l}{MAE} \\
  \hspace{1em}  Ref \cite{ma2021covid} &2278.68 & 4208.02 & 4992.58 & 6414.48 \\ 
   \hspace{1em} ARGOX Joint Ensemble & 2211.79 & 4587.86 & 4496.12 & 7843.64 \\ 
   \hspace{1em} Naive & 2269.62 & 3937.04 & 5093.87 & 6527.64 \\ 
   \multicolumn{1}{l}{Correlation} \\
 \hspace{1em} Ref \cite{ma2021covid} &0.95 & 0.87 & 0.75 & 0.62 \\ 
   \hspace{1em} ARGOX Joint Ensemble  & 0.95 & 0.83 & 0.79 & 0.56 \\ 
   \hspace{1em} Naive& 0.94 & 0.86 & 0.75 & 0.68 \\ 
   \hline
\end{tabular}
\caption{Comparison of different methods for state-level COVID-19 1 to 4 weeks ahead incremental case in Wisconsin (WI). The MSE, MAE, and correlation are reported and best performed method is highlighted in boldface.} 
\end{table}       \begin{figure}[!h]    \centering  \includegraphics[width=0.6\linewidth, page=49]{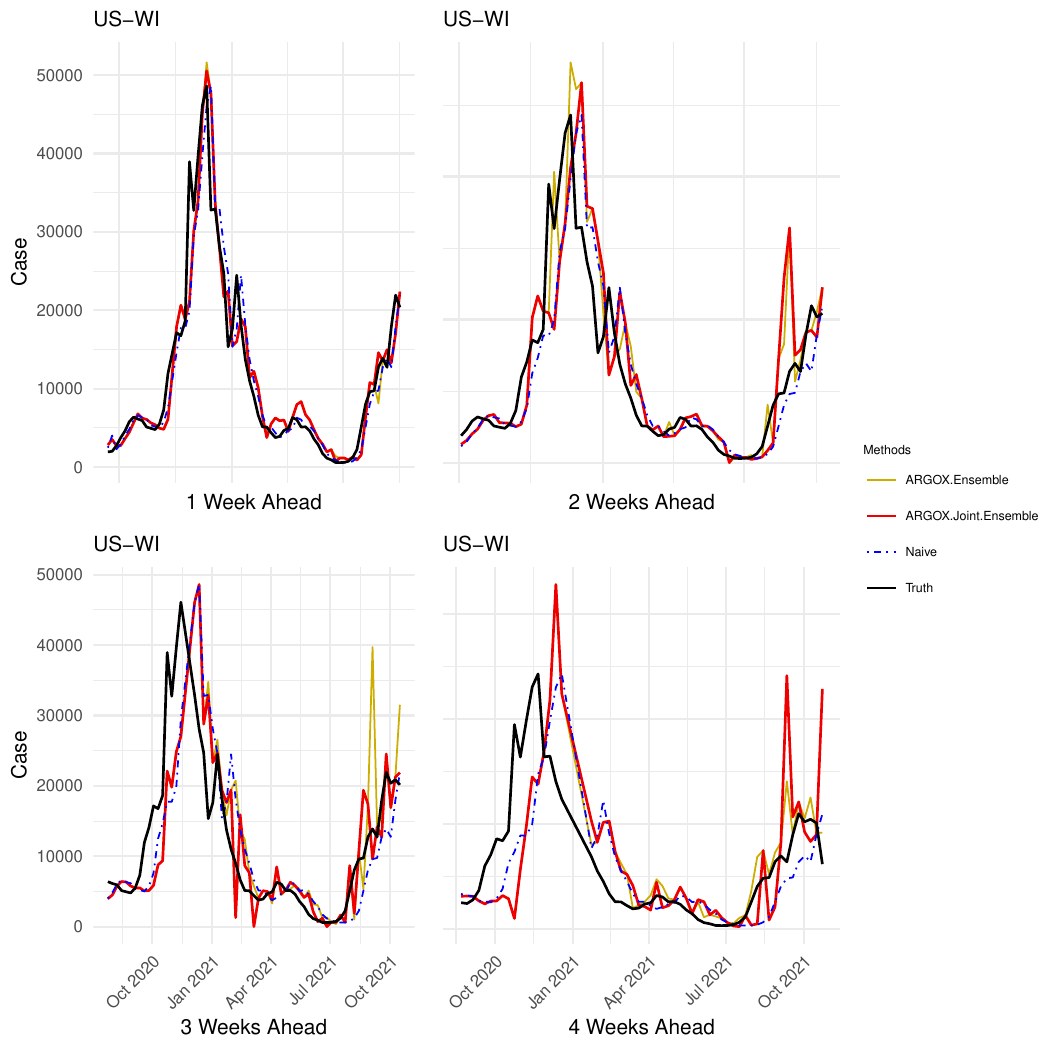}  \caption{Plots of the COVID-19 1 week (top left), 2 weeks (top right), 3 weeks (bottom left), and 4 weeks (bottom right) ahead estimates for Wisconsin (WI). ARGOX-Ensemble is Ref \cite{ma2021covid}.}  \end{figure} \newpage   
\begin{table}[ht]
\centering
\begin{tabular}{lrrrr}
  \hline
& 1 Week Ahead & 2 Weeks Ahead & 3 Weeks Ahead & 4 Weeks Ahead \\ 
    \hline \multicolumn{1}{l}{RMSE} \\
 \hspace{1em} Ref \cite{ma2021covid} & 
1015.81 & 1329.62 & 1886.12 & 2227.09 \\ 
   \hspace{1em} ARGOX Joint Ensemble & 747.18 & 1589.17 & 1885.89 & 3170.93 \\ 
   \hspace{1em} Naive & 986.31 & 1720.67 & 2433.17 & 3282.91 \\ 
   \multicolumn{1}{l}{MAE} \\
  \hspace{1em}  Ref \cite{ma2021covid} &650.57 & 868.54 & 1199.13 & 1513.65 \\ 
   \hspace{1em} ARGOX Joint Ensemble & 491.69 & 1070.64 & 1199.69 & 2034.03 \\ 
   \hspace{1em} Naive & 629.19 & 1110.67 & 1572.81 & 2309.29 \\ 
   \multicolumn{1}{l}{Correlation} \\
 \hspace{1em} Ref \cite{ma2021covid} &0.97 & 0.93 & 0.90 & 0.85 \\ 
   \hspace{1em} ARGOX Joint Ensemble  & 0.98 & 0.91 & 0.91 & 0.82 \\ 
   \hspace{1em} Naive& 0.96 & 0.88 & 0.76 & 0.53 \\ 
   \hline
\end{tabular}
\caption{Comparison of different methods for state-level COVID-19 1 to 4 weeks ahead incremental case in West Virginia (WV). The MSE, MAE, and correlation are reported and best performed method is highlighted in boldface.} 
\end{table}       \begin{figure}[!h]    \centering  \includegraphics[width=0.6\linewidth, page=50]{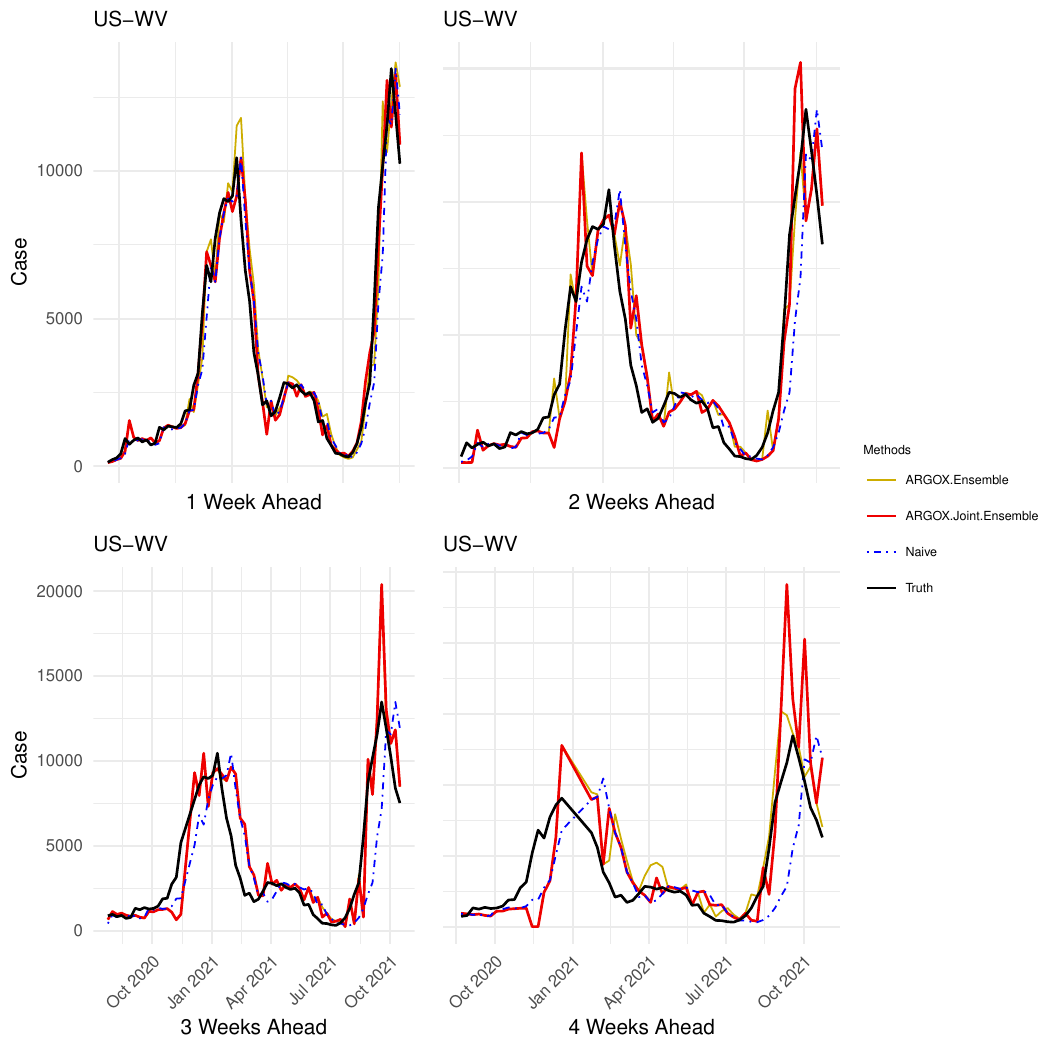}  \caption{Plots of the COVID-19 1 week (top left), 2 weeks (top right), 3 weeks (bottom left), and 4 weeks (bottom right) ahead estimates for West Virginia (WV). ARGOX-Ensemble is Ref \cite{ma2021covid}.}  \end{figure} \newpage   
\begin{table}[ht]
\centering
\begin{tabular}{lrrrr}
  \hline
& 1 Week Ahead & 2 Weeks Ahead & 3 Weeks Ahead & 4 Weeks Ahead \\ 
    \hline \multicolumn{1}{l}{RMSE} \\
 \hspace{1em} Ref \cite{ma2021covid} & 
598.83 & 889.38 & 917.80 & 1543.86 \\ 
   \hspace{1em} ARGOX Joint Ensemble & 326.11 & 432.05 & 1006.02 & 2079.97 \\ 
   \hspace{1em} Naive & 445.02 & 758.99 & 953.21 & 1212.06 \\ 
   \multicolumn{1}{l}{MAE} \\
  \hspace{1em}  Ref \cite{ma2021covid} &303.86 & 517.19 & 562.14 & 879.50 \\ 
   \hspace{1em} ARGOX Joint Ensemble & 197.93 & 284.35 & 565.85 & 1078.27 \\ 
   \hspace{1em} Naive & 275.63 & 492.07 & 599.81 & 776.67 \\ 
   \multicolumn{1}{l}{Correlation} \\
 \hspace{1em} Ref \cite{ma2021covid} &0.95 & 0.89 & 0.80 & 0.67 \\ 
   \hspace{1em} ARGOX Joint Ensemble  & 0.98 & 0.96 & 0.78 & 0.60 \\ 
   \hspace{1em} Naive& 0.95 & 0.85 & 0.73 & 0.63 \\ 
   \hline
\end{tabular}
\caption{Comparison of different methods for state-level COVID-19 1 to 4 weeks ahead incremental case in Wyoming (WY). The MSE, MAE, and correlation are reported and best performed method is highlighted in boldface.} \label{tab:State_Ours_Case_WY}
\end{table}       

\begin{figure}[!h]    \centering  \includegraphics[width=0.6\linewidth, page=51]{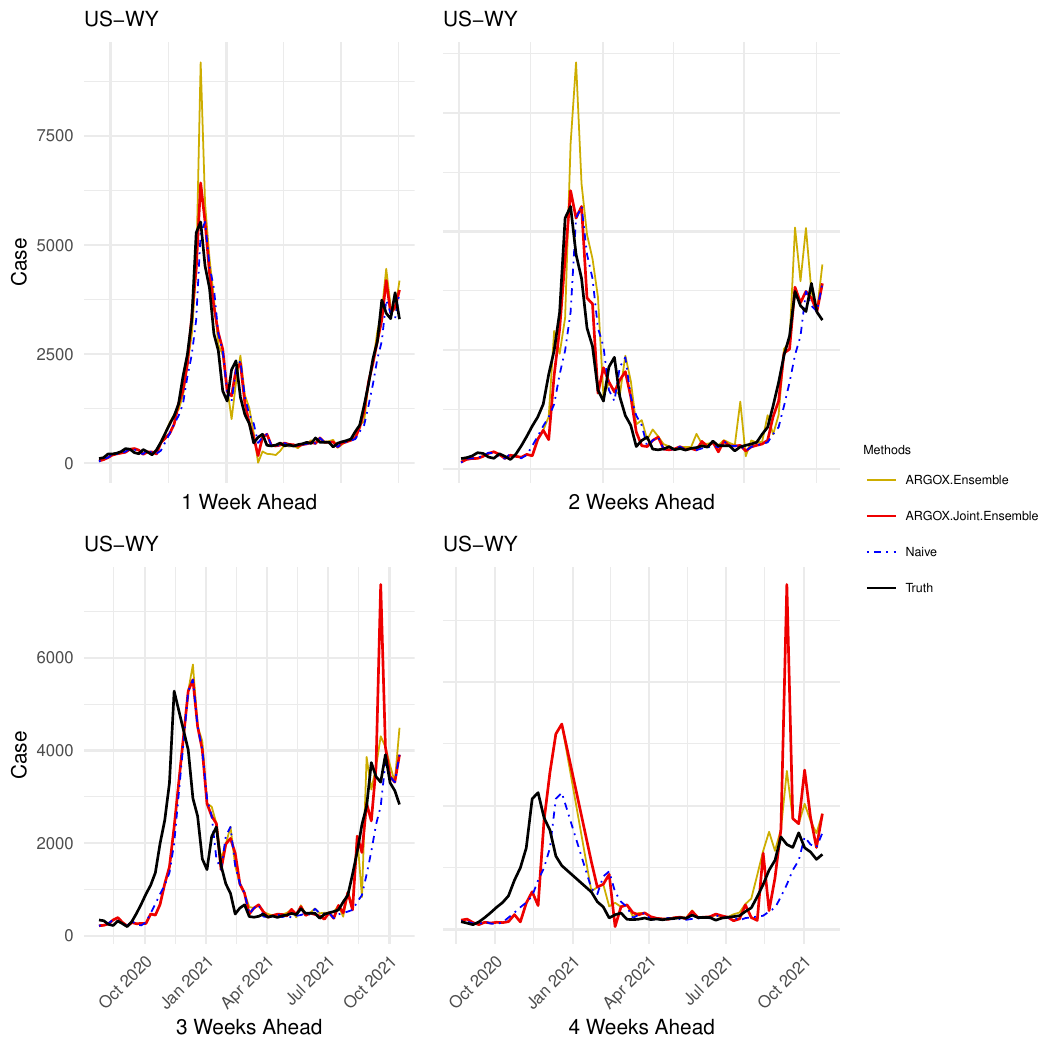}  \caption{Plots of the COVID-19 1 week (top left), 2 weeks (top right), 3 weeks (bottom left), and 4 weeks (bottom right) ahead estimates for Wyoming (WY). ARGOX-Ensemble is Ref \cite{ma2021covid}.} \label{fig:State_Ours_Case_WY} 
\end{figure} \newpage   

\restoregeometry

\clearpage
\newgeometry{left=1.5cm,bottom=2cm}
\subsection*{Detailed \%ILI estimation results for each state}
\begin{table}[ht]
\centering
\begin{tabular}{|c|c|c|c|}
  \hline
Methods & RMSE & MAE & Correlation \\ \hline
  Naive & 
0.17 & 0.13 & 0.89 \\ 
    \hline VAR &0.21 & 0.15 & 0.84 \\ 
    \hline Ref \cite{ARGOX} & 0.18 & 0.14 & 0.88 \\ 
    \hline ARGOX-Idv & 0.15 & 0.11 & 0.92 \\ 
    \hline ARGOX-Joint-Ensemble & 0.11 & 0.08 & 0.98 \\ 
   \hline
\end{tabular}
\caption{Comparison of different methods for state-level ILI 1 week ahead incremental death in Alabama (AL). The MSE, MAE, and correlation are reported and best performed method is highlighted in boldface.} 
\label{tab:State_Ours_ILI_AL}
\end{table}

\begin{figure}[!h] 
  \centering 
\includegraphics[width=0.6\linewidth, page=1]{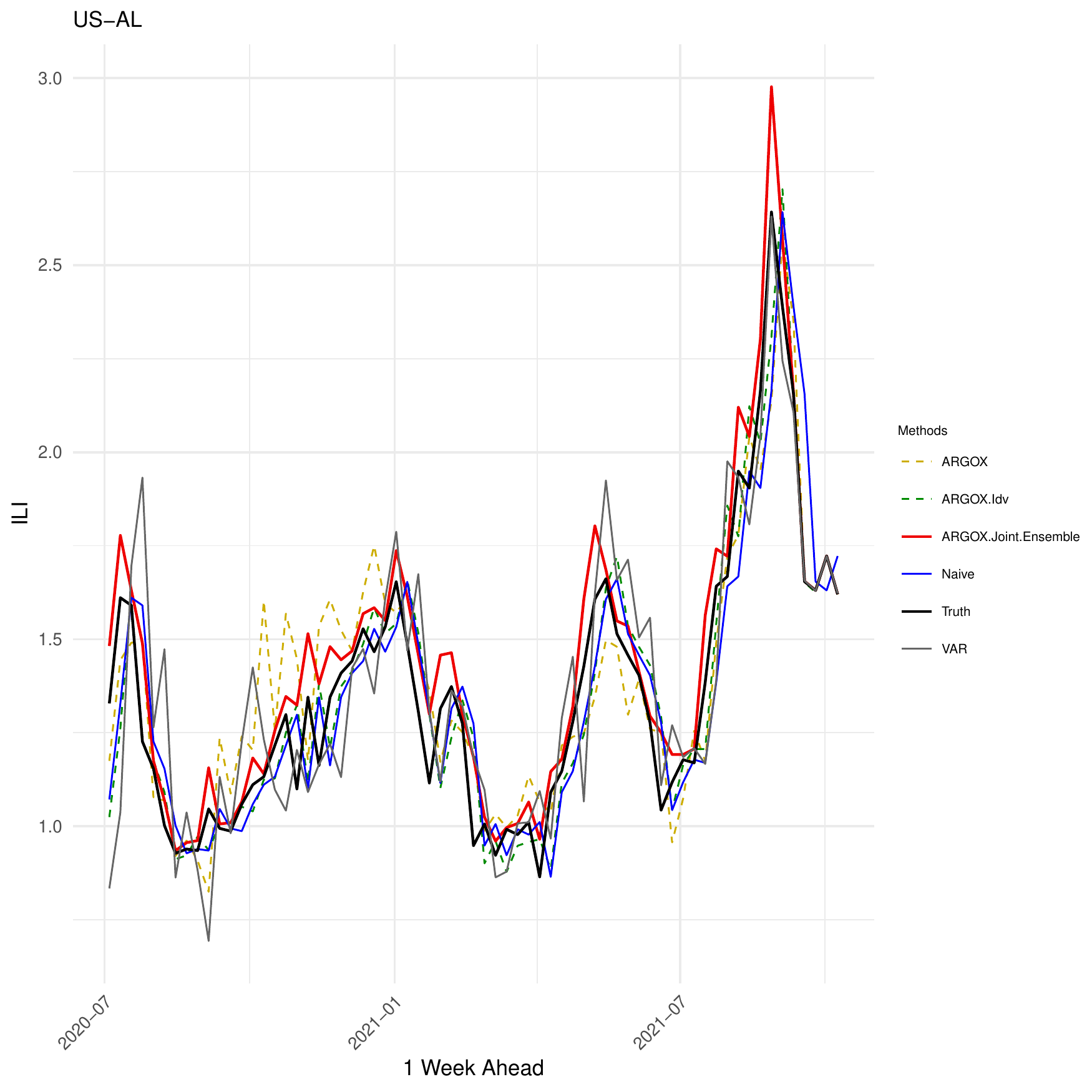} 
\caption{Plots of the \%ILI 1 week ahead estimates for Alabama (AL). ARGOX is Ref \cite{ARGOX}.}
\label{fig:State_Ours_ILI_AL}
\end{figure}
\newpage  

\begin{table}[ht]
\centering
\begin{tabular}{|c|c|c|c|}
  \hline
Methods & RMSE & MAE & Correlation \\ \hline
  Naive & 
0.42 & 0.28 & 0.40 \\ 
    \hline VAR &0.51 & 0.34 & 0.22 \\ 
    \hline Ref \cite{ARGOX} & 0.36 & 0.24 & 0.50 \\ 
    \hline ARGOX-Idv & 0.31 & 0.22 & 0.66 \\ 
    \hline ARGOX-Joint-Ensemble & 0.26 & 0.16 & 0.90 \\ 
   \hline
\end{tabular}
\caption{Comparison of different methods for state-level ILI 1 week ahead incremental death in Alaska (AK). The MSE, MAE, and correlation are reported and best performed method is highlighted in boldface.} 
\end{table}

\begin{figure}[!h] 
  \centering 
\includegraphics[width=0.6\linewidth, page=2]{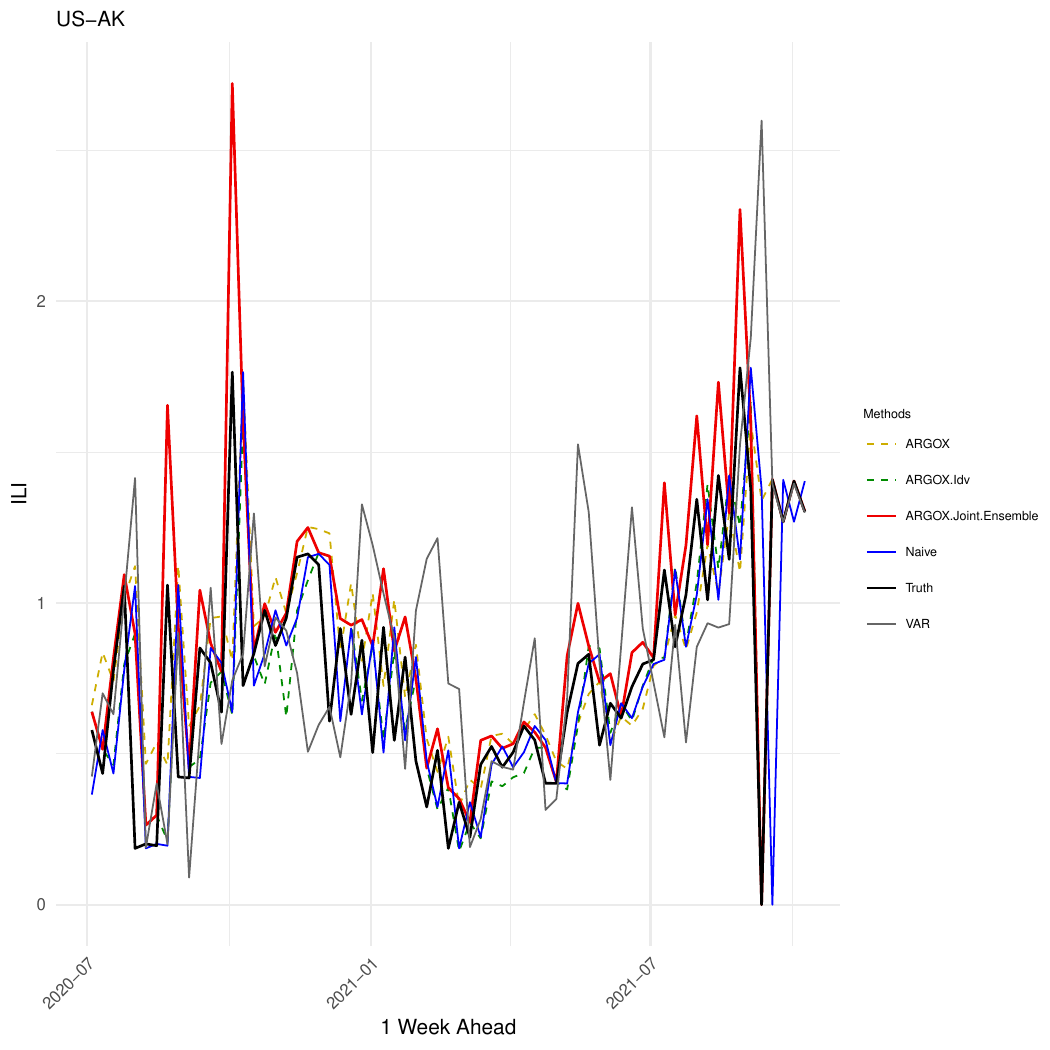} 
\caption{Plots of the \%ILI 1 week ahead estimates for Alaska (AK). ARGOX is Ref \cite{ARGOX}.}
\end{figure}
\newpage  

\begin{table}[ht]
\centering
\begin{tabular}{|c|c|c|c|}
  \hline
Methods & RMSE & MAE & Correlation \\ \hline
  Naive & 
0.27 & 0.20 & 0.85 \\ 
    \hline VAR &0.34 & 0.24 & 0.80 \\ 
    \hline Ref \cite{ARGOX} & 0.24 & 0.18 & 0.87 \\ 
    \hline ARGOX-Idv & 0.26 & 0.19 & 0.87 \\ 
    \hline ARGOX-Joint-Ensemble & 0.21 & 0.14 & 0.96 \\ 
   \hline
\end{tabular}
\caption{Comparison of different methods for state-level ILI 1 week ahead incremental death in Arizona (AZ). The MSE, MAE, and correlation are reported and best performed method is highlighted in boldface.} 
\end{table}

\begin{figure}[!h] 
  \centering 
\includegraphics[width=0.6\linewidth, page=3]{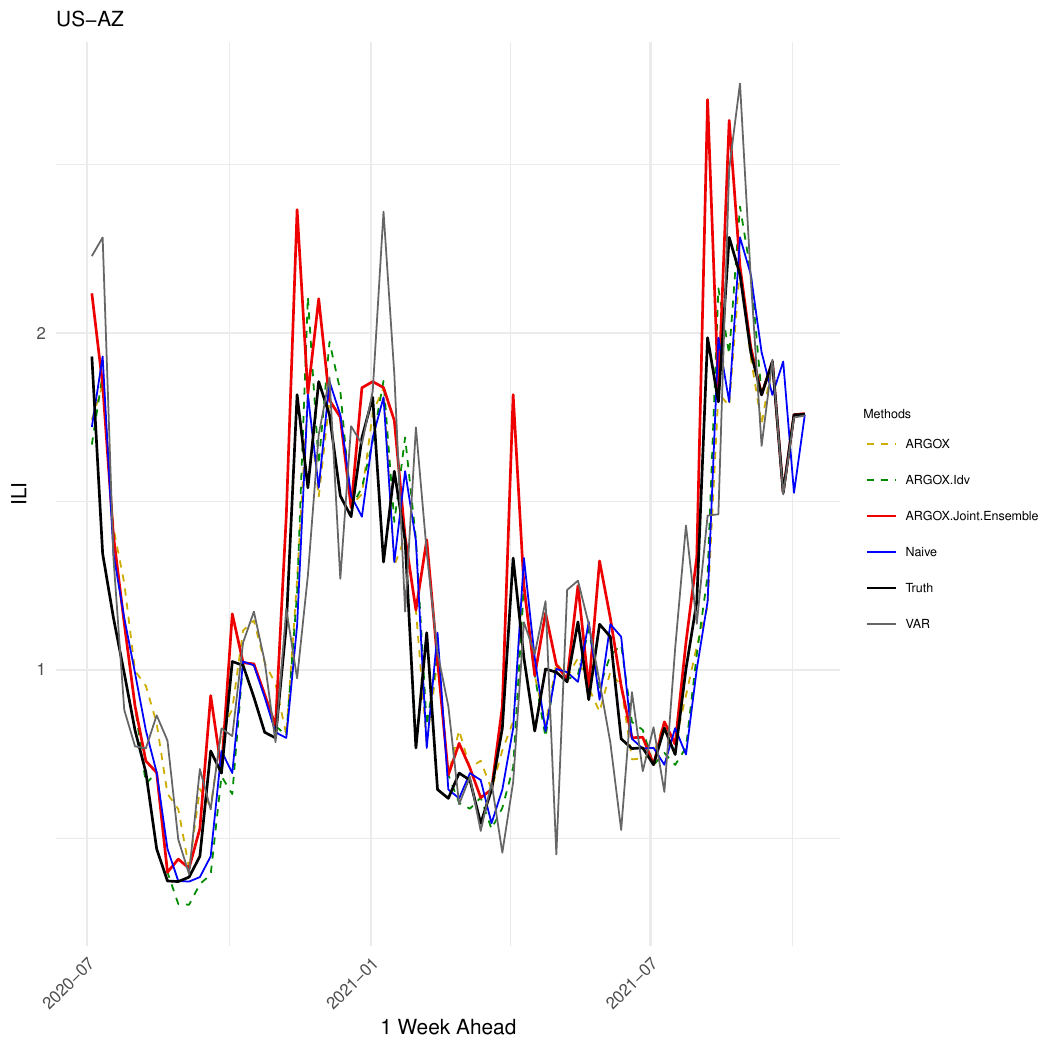} 
\caption{Plots of the \%ILI 1 week ahead estimates for Arizona (AZ). ARGOX is Ref \cite{ARGOX}.}
\end{figure}
\newpage

\begin{table}[ht]
\centering
\begin{tabular}{|c|c|c|c|}
  \hline
Methods & RMSE & MAE & Correlation \\ \hline
  Naive & 
0.20 & 0.16 & 0.83 \\ 
    \hline VAR &0.33 & 0.26 & 0.71 \\ 
    \hline Ref \cite{ARGOX} & 0.25 & 0.20 & 0.78 \\ 
    \hline ARGOX-Idv & 0.20 & 0.15 & 0.85 \\ 
    \hline ARGOX-Joint-Ensemble & 0.16 & 0.12 & 0.96 \\ 
   \hline
\end{tabular}
\caption{Comparison of different methods for state-level ILI 1 week ahead incremental death in Arkansas (AR). The MSE, MAE, and correlation are reported and best performed method is highlighted in boldface.} 
\end{table}

\begin{figure}[!h] 
  \centering 
\includegraphics[width=0.6\linewidth, page=4]{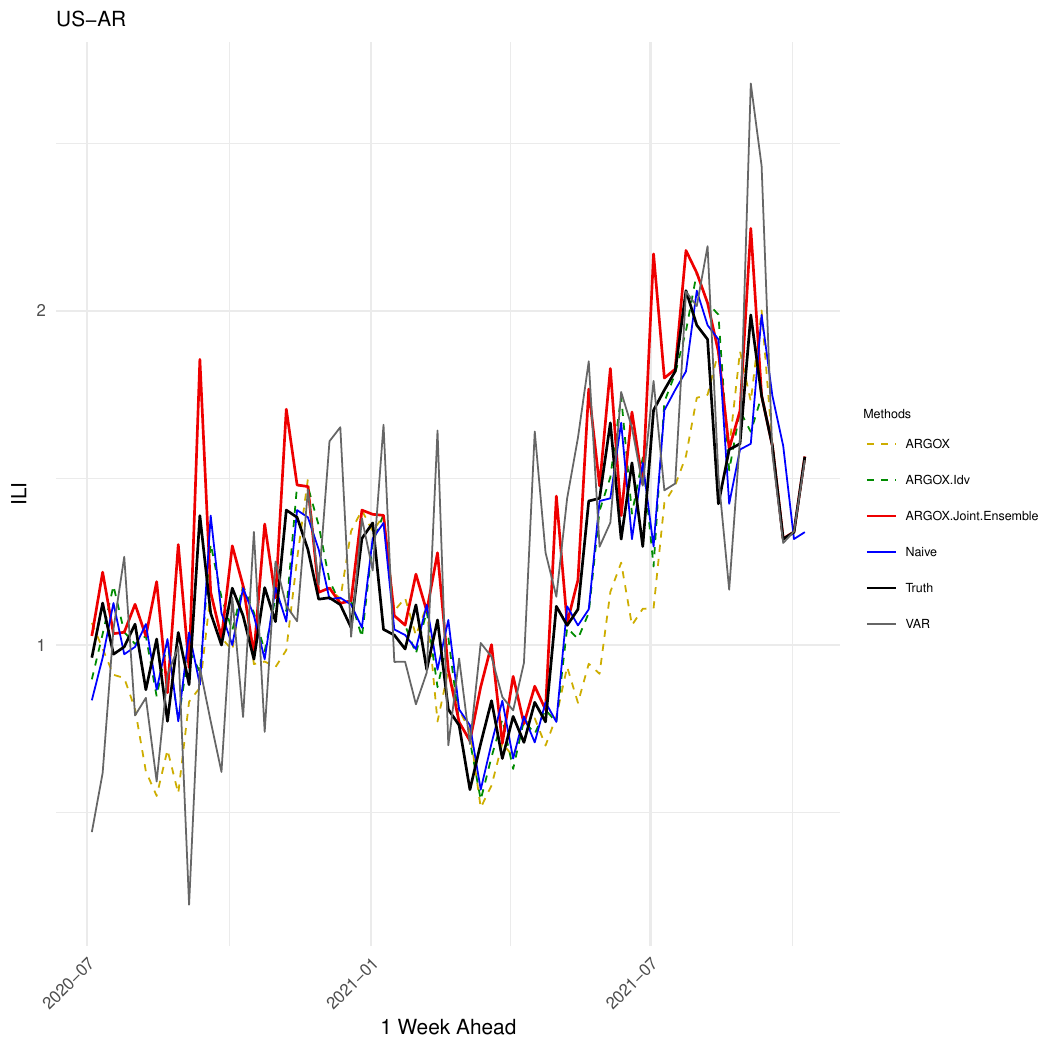} 
\caption{Plots of the \%ILI 1 week ahead estimates for Arkansas (AR). ARGOX is Ref \cite{ARGOX}.}
\end{figure}
\newpage

\begin{table}[ht]
\centering
\begin{tabular}{|c|c|c|c|}
  \hline
Methods & RMSE & MAE & Correlation \\ \hline
  Naive & 
0.16 & 0.13 & 0.95 \\ 
    \hline VAR &0.23 & 0.17 & 0.90 \\ 
    \hline Ref \cite{ARGOX} & 0.14 & 0.11 & 0.96 \\ 
    \hline ARGOX-Idv & 0.15 & 0.12 & 0.96 \\ 
    \hline ARGOX-Joint-Ensemble & 0.10 & 0.08 & 0.99 \\ 
   \hline
\end{tabular}
\caption{Comparison of different methods for state-level ILI 1 week ahead incremental death in California (CA). The MSE, MAE, and correlation are reported and best performed method is highlighted in boldface.} 
\end{table}

\begin{figure}[!h] 
  \centering 
\includegraphics[width=0.6\linewidth, page=5]{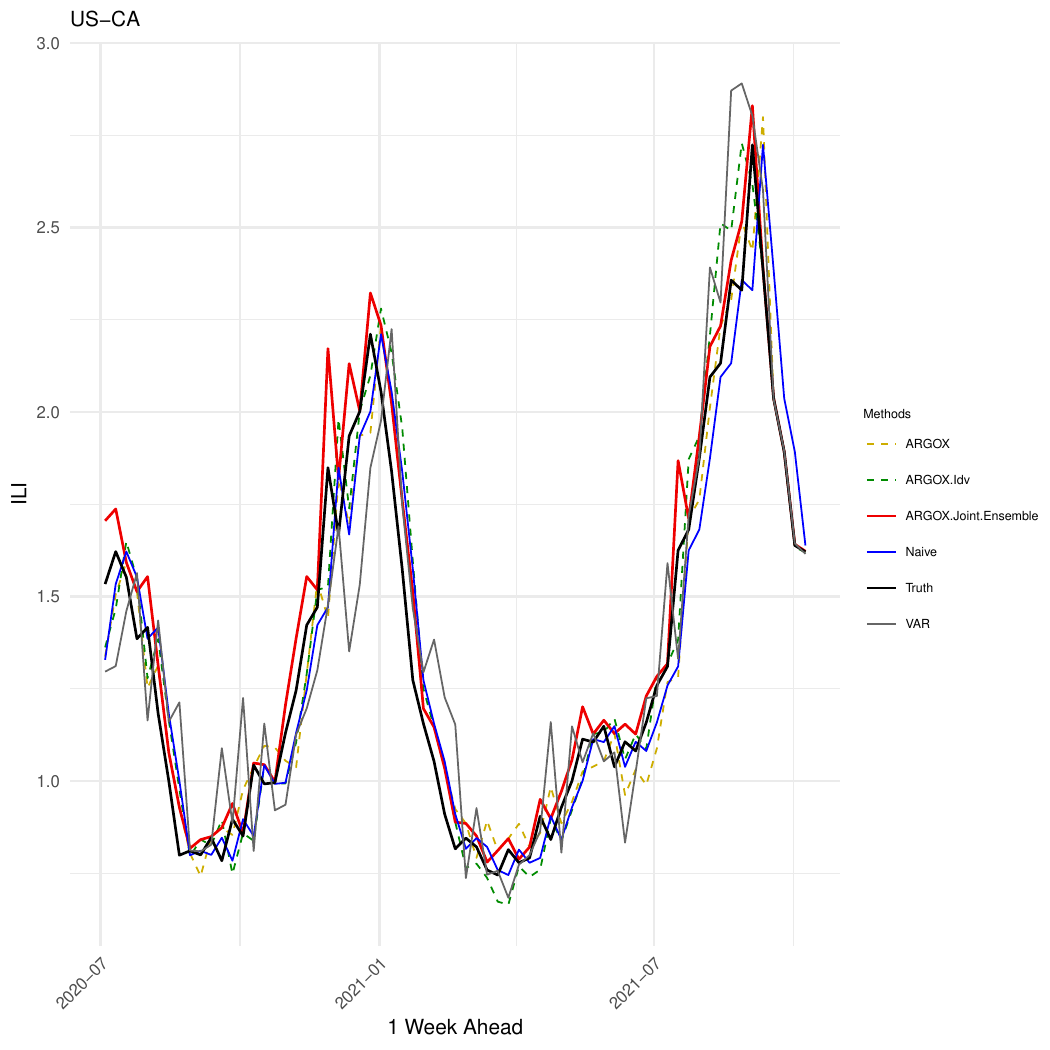} 
\caption{Plots of the \%ILI 1 week ahead estimates for California (CA). ARGOX is Ref \cite{ARGOX}.}
\end{figure}
\newpage 

\begin{table}[ht]
\centering
\begin{tabular}{|c|c|c|c|}
  \hline
Methods & RMSE & MAE & Correlation \\ \hline
  Naive & 0.14 & 0.10 & 0.95 \\ 
    \hline VAR &0.19 & 0.14 & 0.90 \\ 
    \hline Ref \cite{ARGOX} & 0.20 & 0.13 & 0.90 \\ 
    \hline ARGOX-Idv & 0.15 & 0.10 & 0.94 \\ 
    \hline ARGOX-Joint-Ensemble & 0.13 & 0.08 & 0.98 \\ 
   \hline
\end{tabular}
\caption{Comparison of different methods for state-level ILI 1 week ahead incremental death in Colorado (CO). The MSE, MAE, and correlation are reported and best performed method is highlighted in boldface.} 
\end{table}

\begin{figure}[!h] 
  \centering 
\includegraphics[width=0.6\linewidth, page=6]{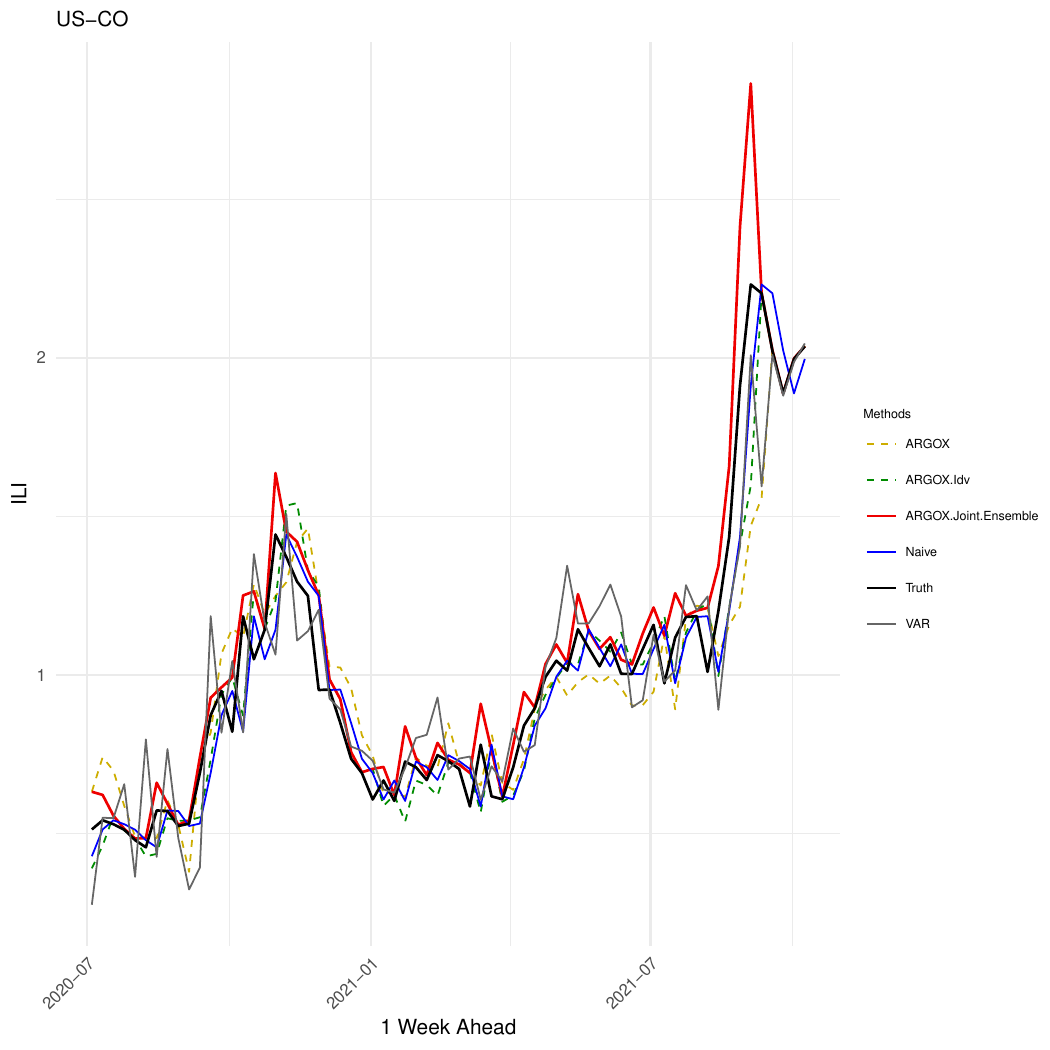} 
\caption{Plots of the \%ILI 1 week ahead estimates for Colorado (CO). ARGOX is Ref \cite{ARGOX}.}
\end{figure}
\newpage

\begin{table}[ht]
\centering
\begin{tabular}{|c|c|c|c|}
  \hline
Methods & RMSE & MAE & Correlation \\ \hline
  Naive & 
0.14 & 0.11 & 0.91 \\ 
    \hline VAR &0.28 & 0.22 & 0.76 \\ 
    \hline Ref \cite{ARGOX} & 0.14 & 0.10 & 0.91 \\ 
    \hline ARGOX-Idv & 0.15 & 0.11 & 0.91 \\ 
    \hline ARGOX-Joint-Ensemble & 0.11 & 0.07 & 0.97 \\ 
   \hline
\end{tabular}
\caption{Comparison of different methods for state-level ILI 1 week ahead incremental death in Connecticut (CT). The MSE, MAE, and correlation are reported and best performed method is highlighted in boldface.} 
\end{table}

\begin{figure}[!h] 
  \centering 
\includegraphics[width=0.6\linewidth, page=7]{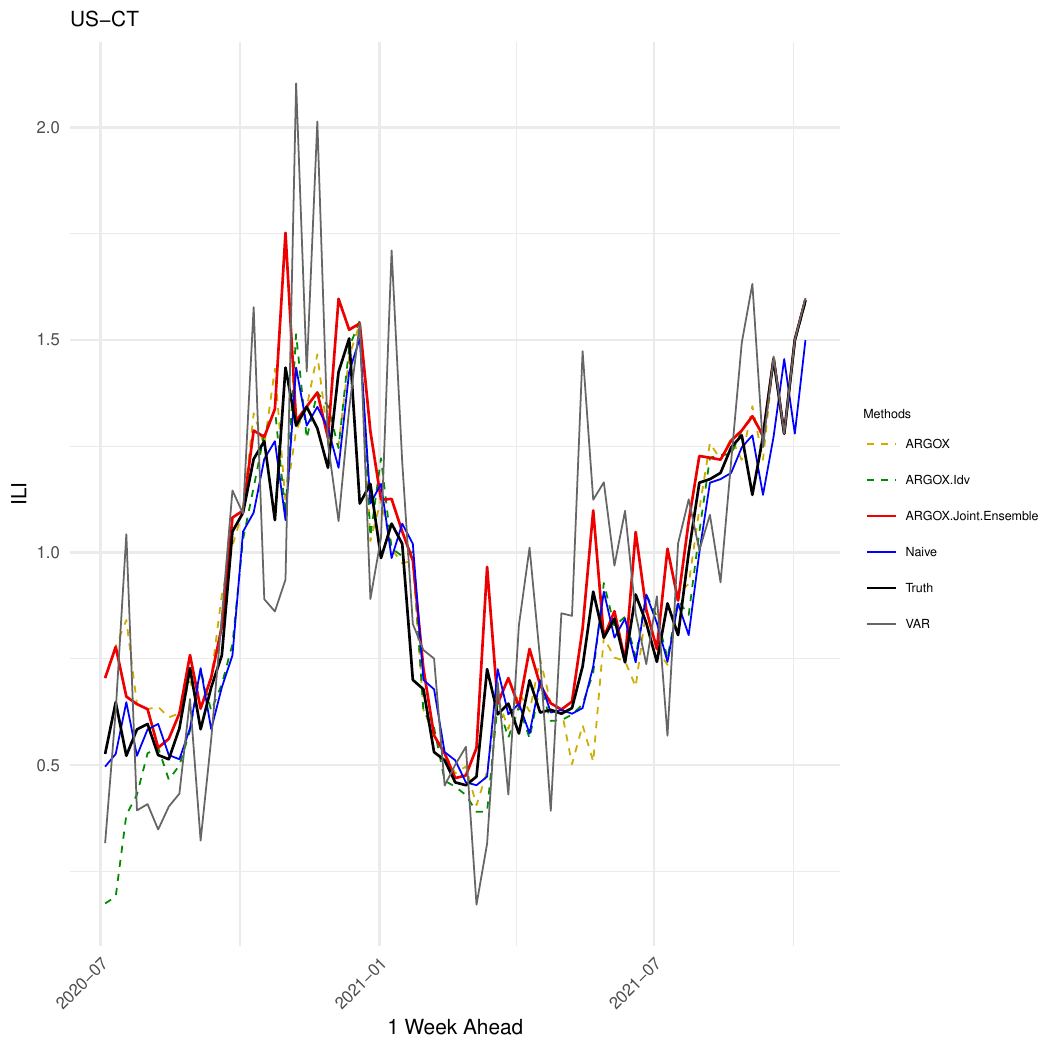} 
\caption{Plots of the \%ILI 1 week ahead estimates for Connecticut (CT). ARGOX is Ref \cite{ARGOX}.}
\end{figure}
\newpage 

\begin{table}[ht]
\centering
\begin{tabular}{|c|c|c|c|}
  \hline
Methods & RMSE & MAE & Correlation \\ \hline
  Naive & 
0.19 & 0.13 & 0.71 \\ 
    \hline VAR &0.38 & 0.27 & 0.52 \\ 
    \hline Ref \cite{ARGOX} & 0.19 & 0.15 & 0.70 \\ 
    \hline ARGOX-Idv & 0.18 & 0.12 & 0.75 \\ 
    \hline ARGOX-Joint-Ensemble & 0.15 & 0.09 & 0.91 \\ 
   \hline
\end{tabular}
\caption{Comparison of different methods for state-level ILI 1 week ahead incremental death in Delaware (DE). The MSE, MAE, and correlation are reported and best performed method is highlighted in boldface.} 
\end{table}

\begin{figure}[!h] 
  \centering 
\includegraphics[width=0.6\linewidth, page=8]{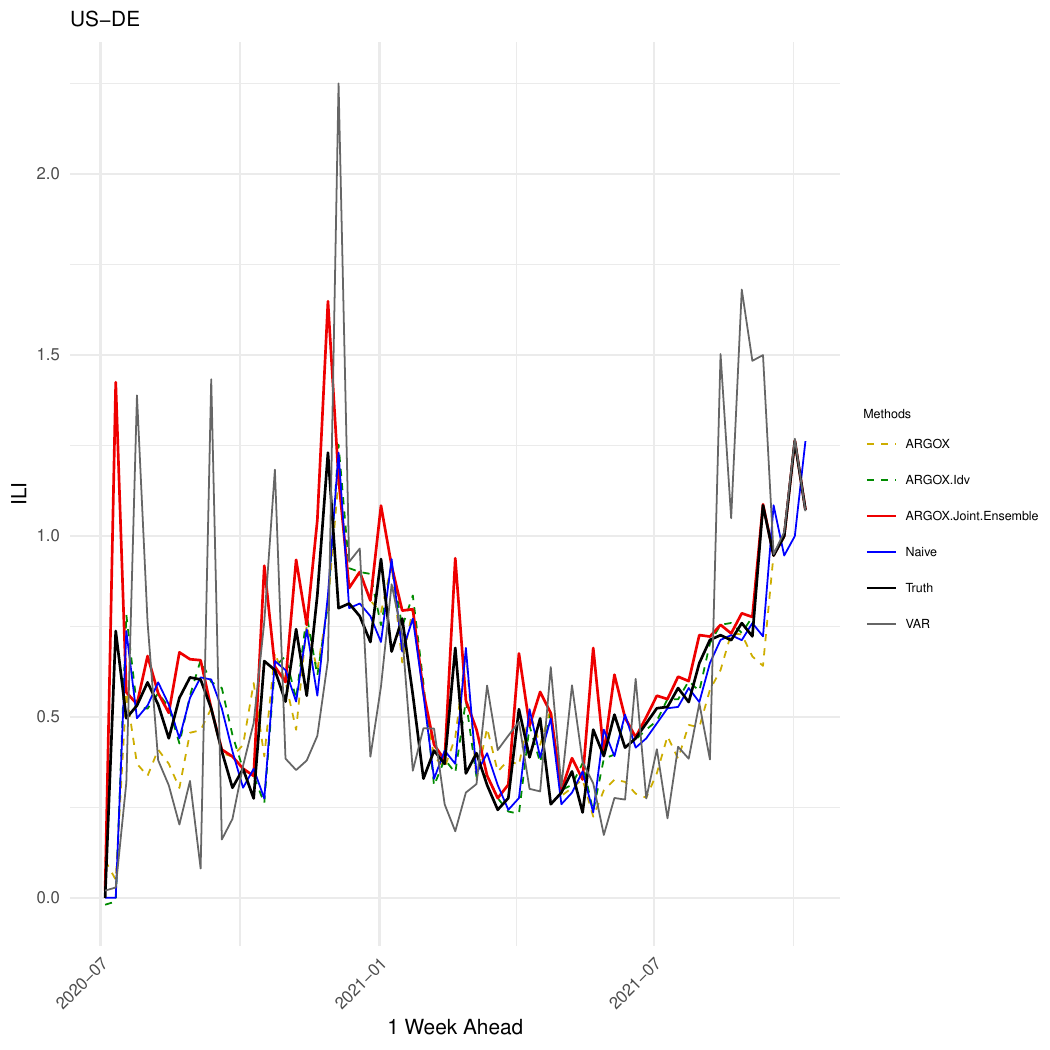} 
\caption{Plots of the \%ILI 1 week ahead estimates for Delaware (DE). ARGOX is Ref \cite{ARGOX}.}
\end{figure}
\newpage 

\begin{table}[ht]
\centering
\begin{tabular}{|c|c|c|c|}
  \hline
Methods & RMSE & MAE & Correlation \\ \hline
  Naive & 
0.81 & 0.50 & 0.83 \\ 
    \hline VAR &1.04 & 0.76 & 0.80 \\ 
    \hline Ref \cite{ARGOX} & 0.73 & 0.49 & 0.87 \\ 
    \hline ARGOX-Idv & 0.65 & 0.40 & 0.90 \\ 
    \hline ARGOX-Joint-Ensemble & 0.57 & 0.32 & 0.95 \\ 
   \hline
\end{tabular}
\caption{Comparison of different methods for state-level ILI 1 week ahead incremental death in District of Columbia (DC). The MSE, MAE, and correlation are reported and best performed method is highlighted in boldface.} 
\end{table}

\begin{figure}[!h] 
  \centering 
\includegraphics[width=0.6\linewidth, page=9]{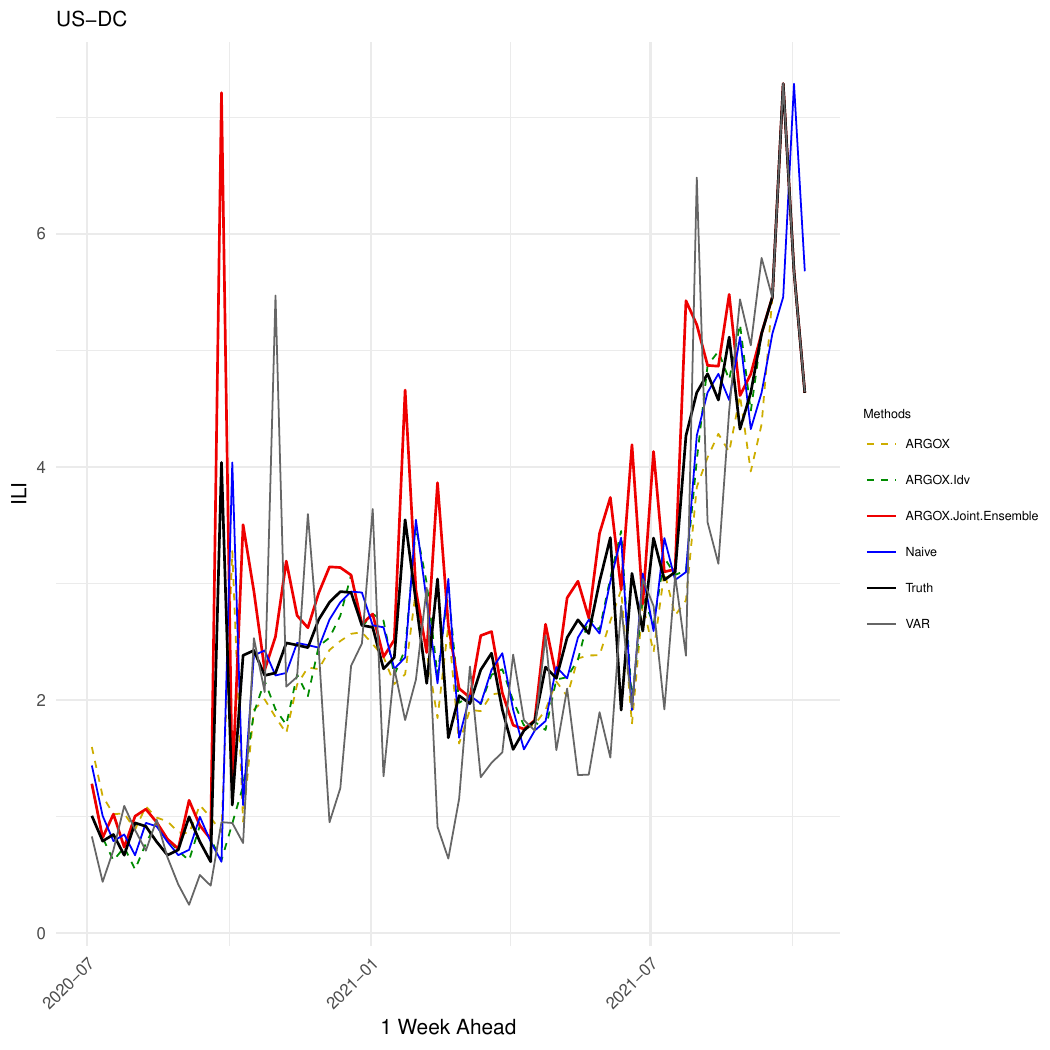} 
\caption{Plots of the \%ILI 1 week ahead estimates for District of Columbia (DC). ARGOX is Ref \cite{ARGOX}.}
\end{figure}
\newpage 

\begin{table}[ht]
\centering
\begin{tabular}{|c|c|c|c|}
  \hline
Methods & RMSE & MAE & Correlation \\ \hline
  Naive & 
0.33 & 0.23 & 0.90 \\ 
    \hline VAR &0.48 & 0.32 & 0.84 \\ 
    \hline Ref \cite{ARGOX} & 0.30 & 0.21 & 0.92 \\ 
    \hline ARGOX-Idv & 0.28 & 0.19 & 0.93 \\ 
    \hline ARGOX-Joint-Ensemble & 0.22 & 0.13 & 0.98 \\ 
   \hline
\end{tabular}
\caption{Comparison of different methods for state-level ILI 1 week ahead incremental death in Georgia (GA). The MSE, MAE, and correlation are reported and best performed method is highlighted in boldface.} 
\end{table}

\begin{figure}[!h] 
  \centering 
\includegraphics[width=0.6\linewidth, page=10]{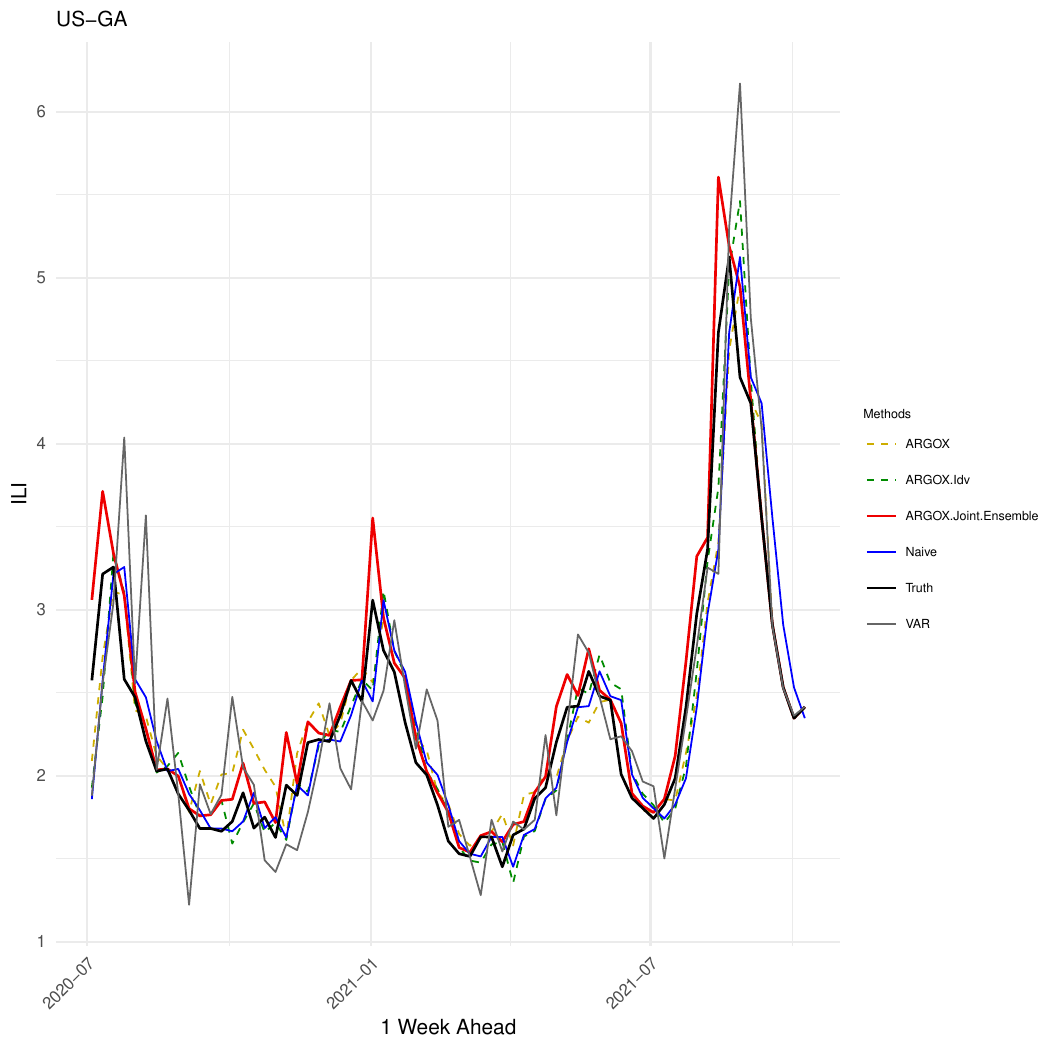} 
\caption{Plots of the \%ILI 1 week ahead estimates for Georgia (GA). ARGOX is Ref \cite{ARGOX}.}
\end{figure}
\newpage 

\begin{table}[ht]
\centering
\begin{tabular}{|c|c|c|c|}
  \hline
Methods & RMSE & MAE & Correlation \\ \hline
  Naive & 
0.40 & 0.31 & 0.45 \\ 
    \hline VAR &0.36 & 0.28 & 0.53 \\ 
    \hline Ref \cite{ARGOX} & 0.37 & 0.28 & 0.47 \\ 
    \hline ARGOX-Idv & 0.40 & 0.30 & 0.46 \\ 
    \hline ARGOX-Joint-Ensemble & 0.36 & 0.26 & 0.87 \\ 
   \hline
\end{tabular}
\caption{Comparison of different methods for state-level ILI 1 week ahead incremental death in Hawaii (HI). The MSE, MAE, and correlation are reported and best performed method is highlighted in boldface.} 
\end{table}

\begin{figure}[!h] 
  \centering 
\includegraphics[width=0.6\linewidth, page=11]{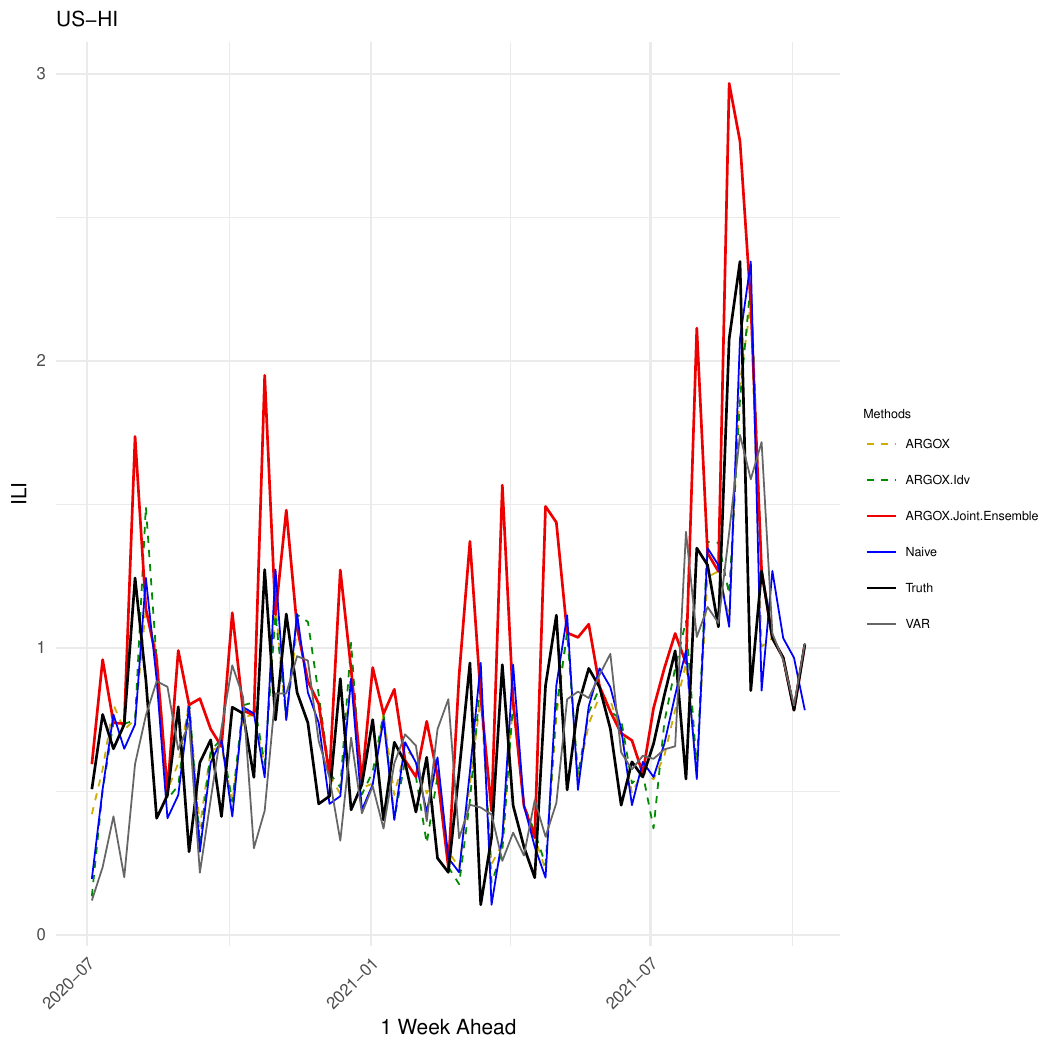} 
\caption{Plots of the \%ILI 1 week ahead estimates for Hawaii (HI). ARGOX is Ref \cite{ARGOX}.}
\end{figure}
\newpage

\begin{table}[ht]
\centering
\begin{tabular}{|c|c|c|c|}
  \hline
Methods & RMSE & MAE & Correlation \\ \hline
  Naive & 
0.26 & 0.19 & 0.86 \\ 
    \hline VAR &1.11 & 0.65 & 0.33 \\ 
    \hline Ref \cite{ARGOX} & 0.25 & 0.17 & 0.87 \\ 
    \hline ARGOX-Idv & 0.25 & 0.17 & 0.87 \\ 
    \hline ARGOX-Joint-Ensemble & 0.19 & 0.12 & 0.97 \\ 
   \hline
\end{tabular}
\caption{Comparison of different methods for state-level ILI 1 week ahead incremental death in Idaho (ID). The MSE, MAE, and correlation are reported and best performed method is highlighted in boldface.} 
\end{table}

\begin{figure}[!h] 
  \centering 
\includegraphics[width=0.6\linewidth, page=12]{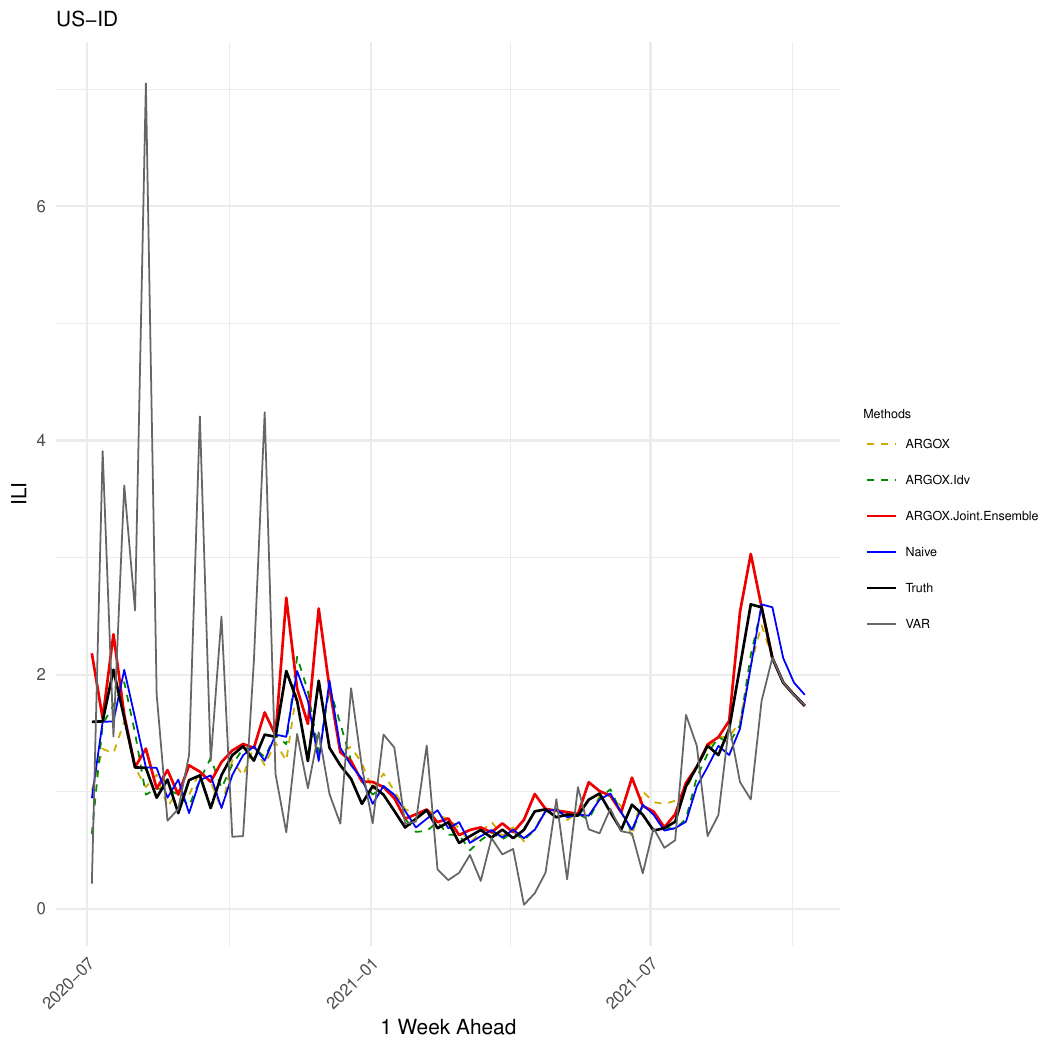} 
\caption{Plots of the \%ILI 1 week ahead estimates for Idaho (ID). ARGOX is Ref \cite{ARGOX}.}
\end{figure}
\newpage

\begin{table}[ht]
\centering
\begin{tabular}{|c|c|c|c|}
  \hline
Methods & RMSE & MAE & Correlation \\ \hline
  Naive & 
0.10 & 0.08 & 0.92 \\ 
    \hline VAR &0.15 & 0.12 & 0.85 \\ 
    \hline Ref \cite{ARGOX} & 0.12 & 0.09 & 0.90 \\ 
    \hline ARGOX-Idv & 0.10 & 0.08 & 0.94 \\ 
    \hline ARGOX-Joint-Ensemble & 0.08 & 0.06 & 0.98 \\ 
   \hline
\end{tabular}
\caption{Comparison of different methods for state-level ILI 1 week ahead incremental death in Illinois (IL). The MSE, MAE, and correlation are reported and best performed method is highlighted in boldface.} 
\end{table}

\begin{figure}[!h] 
  \centering 
\includegraphics[width=0.6\linewidth, page=13]{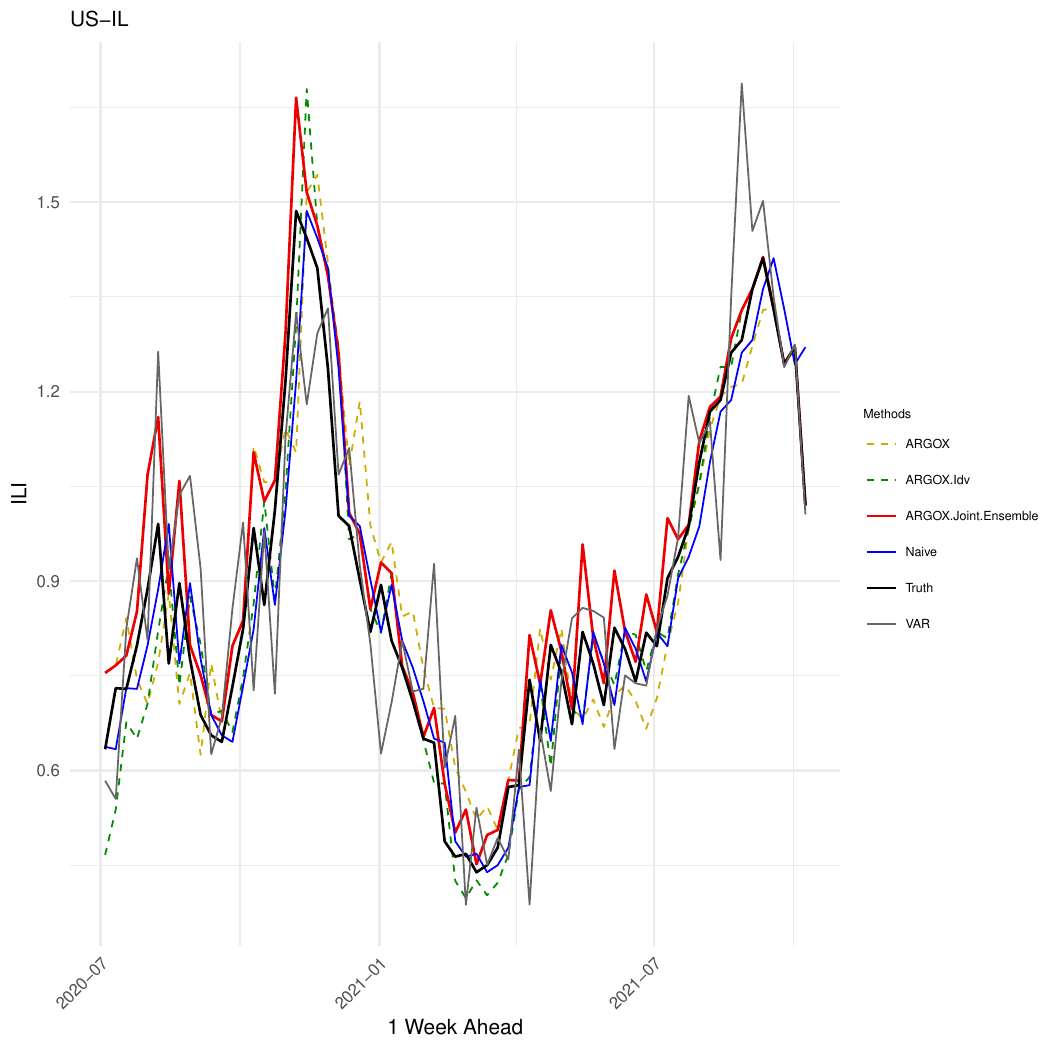} 
\caption{Plots of the \%ILI 1 week ahead estimates for Illinois (IL). ARGOX is Ref \cite{ARGOX}.}
\end{figure}
\newpage

\begin{table}[ht]
\centering
\begin{tabular}{|c|c|c|c|}
  \hline
Methods & RMSE & MAE & Correlation \\ \hline
  Naive & 
0.36 & 0.22 & 0.75 \\ 
    \hline VAR &0.50 & 0.35 & 0.49 \\ 
    \hline Ref \cite{ARGOX} & 0.35 & 0.22 & 0.73 \\ 
    \hline ARGOX-Idv & 0.34 & 0.20 & 0.78 \\ 
    \hline ARGOX-Joint-Ensemble & 0.27 & 0.14 & 0.95 \\ 
   \hline
\end{tabular}
\caption{Comparison of different methods for state-level ILI 1 week ahead incremental death in Indiana (IN). The MSE, MAE, and correlation are reported and best performed method is highlighted in boldface.} 
\end{table}

\begin{figure}[!h] 
  \centering 
\includegraphics[width=0.6\linewidth, page=14]{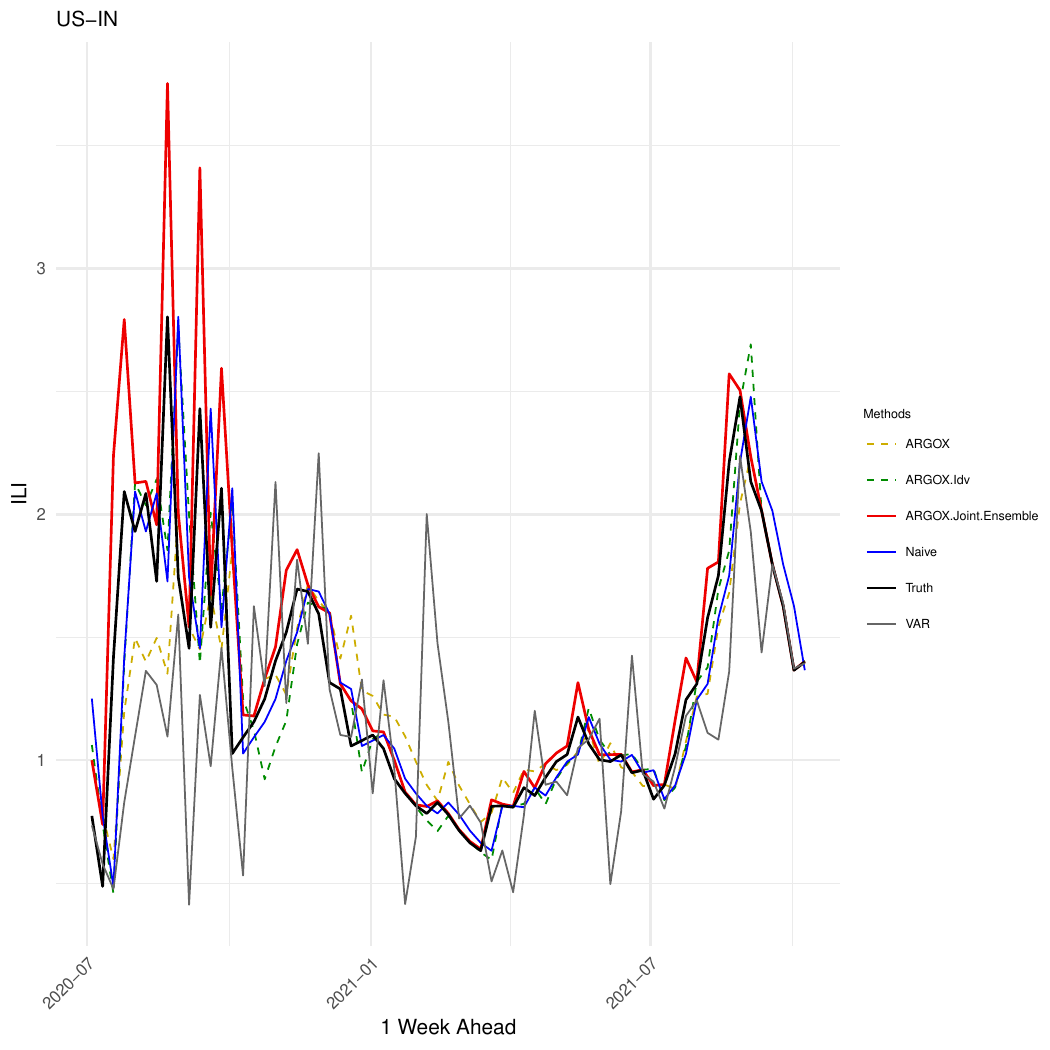} 
\caption{Plots of the \%ILI 1 week ahead estimates for Indiana (IN). ARGOX is Ref \cite{ARGOX}.}
\end{figure}
\newpage 

\begin{table}[ht]
\centering
\begin{tabular}{|c|c|c|c|}
  \hline
Methods & RMSE & MAE & Correlation \\ \hline
  Naive & 
1.54 & 0.57 & 0.52 \\ 
    \hline VAR &1.58 & 0.71 & 0.33 \\ 
    \hline ARGOX1.38 & 0.53 & 0.49 \\ 
    \hline ARGOX-Idv & 1.47 & 0.53 & 0.56 \\ 
    \hline ARGOX-Joint-Ensemble & 1.24 & 0.34 & 0.95 \\ 
   \hline
\end{tabular}
\caption{Comparison of different methods for state-level ILI 1 week ahead incremental death in Iowa (IA). The MSE, MAE, and correlation are reported and best performed method is highlighted in boldface.} 
\end{table}

\begin{figure}[!h] 
  \centering 
\includegraphics[width=0.6\linewidth, page=15]{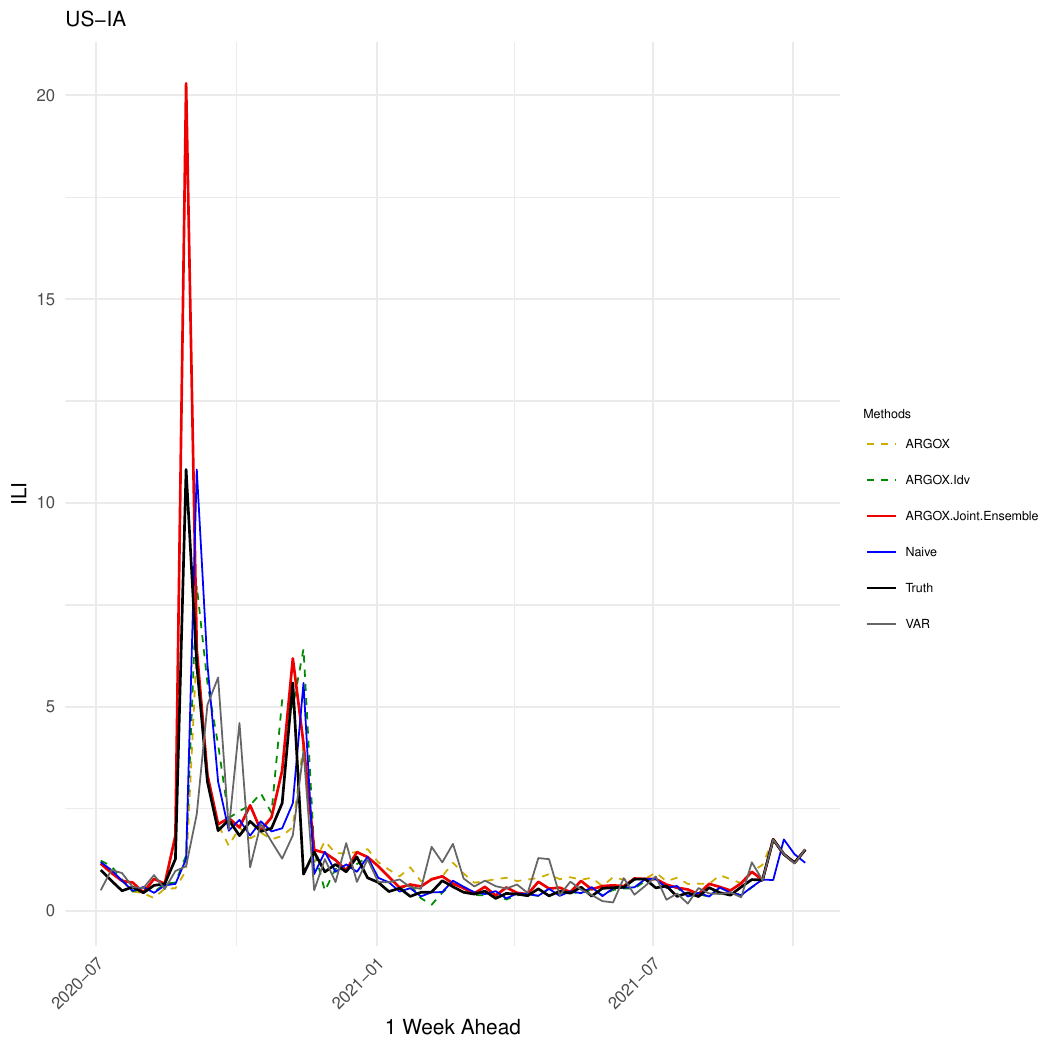} 
\caption{Plots of the \%ILI 1 week ahead estimates for Iowa (IA). ARGOX is Ref \cite{ARGOX}.}
\end{figure}
\newpage

\begin{table}[ht]
\centering
\begin{tabular}{|c|c|c|c|}
  \hline
Methods & RMSE & MAE & Correlation \\ \hline
  Naive & 
0.22 & 0.17 & 0.87 \\ 
    \hline VAR &0.31 & 0.24 & 0.77 \\ 
    \hline Ref \cite{ARGOX} & 0.22 & 0.16 & 0.87 \\ 
    \hline ARGOX-Idv & 0.21 & 0.15 & 0.90 \\ 
    \hline ARGOX-Joint-Ensemble & 0.17 & 0.11 & 0.97 \\ 
   \hline
\end{tabular}
\caption{Comparison of different methods for state-level ILI 1 week ahead incremental death in Kansas (KS). The MSE, MAE, and correlation are reported and best performed method is highlighted in boldface.} 
\end{table}

\begin{figure}[!h] 
  \centering 
\includegraphics[width=0.6\linewidth, page=16]{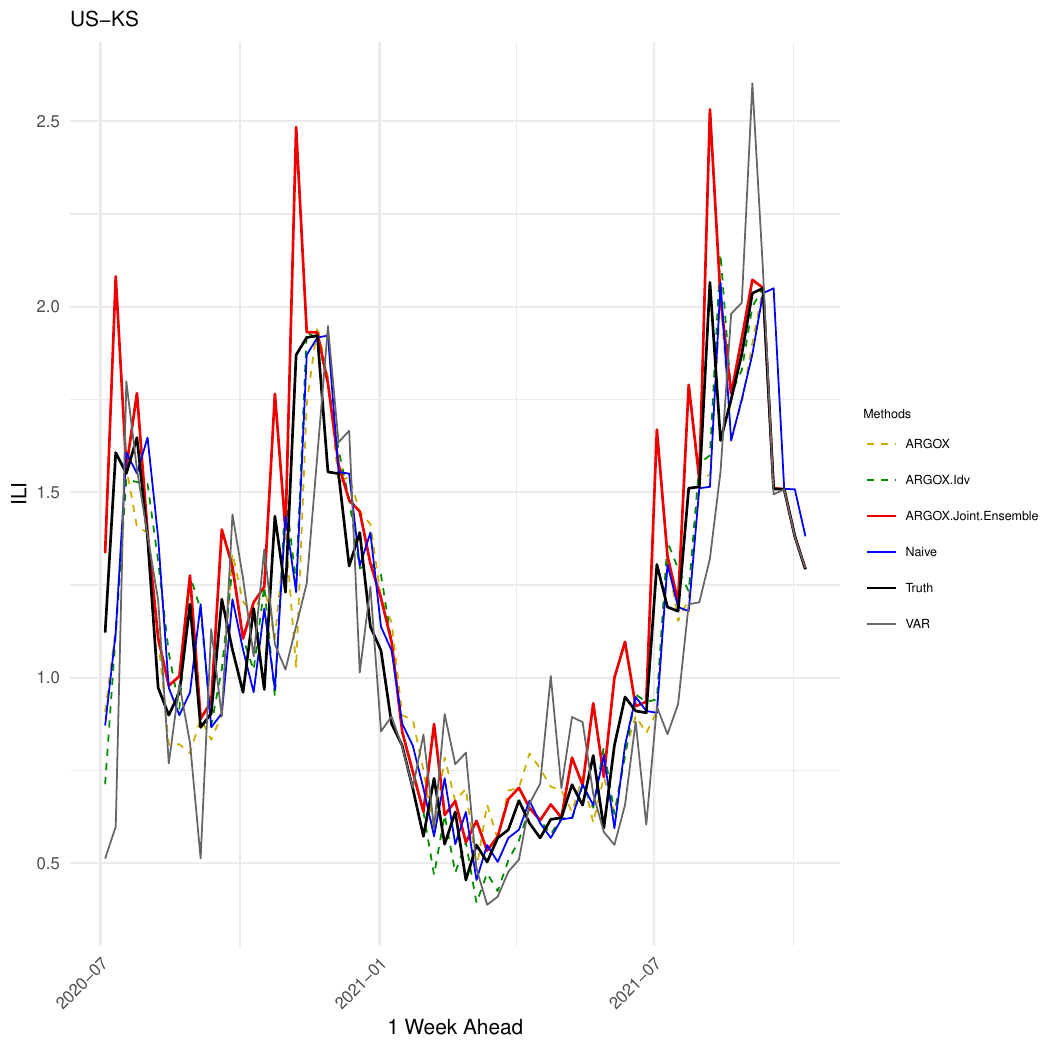} 
\caption{Plots of the \%ILI 1 week ahead estimates for Kansas (KS). ARGOX is Ref \cite{ARGOX}.}
\end{figure}
\newpage

\begin{table}[ht]
\centering
\begin{tabular}{|c|c|c|c|}
  \hline
Methods & RMSE & MAE & Correlation \\ \hline
  Naive & 
0.20 & 0.15 & 0.91 \\ 
    \hline VAR &0.44 & 0.33 & 0.68 \\ 
    \hline Ref \cite{ARGOX} & 0.21 & 0.15 & 0.91 \\ 
    \hline ARGOX-Idv & 0.19 & 0.14 & 0.93 \\ 
    \hline ARGOX-Joint-Ensemble & 0.15 & 0.10 & 0.98 \\ 
   \hline
\end{tabular}
\caption{Comparison of different methods for state-level ILI 1 week ahead incremental death in Kentucky (KY). The MSE, MAE, and correlation are reported and best performed method is highlighted in boldface.} 
\end{table}

\begin{figure}[!h] 
  \centering 
\includegraphics[width=0.6\linewidth, page=17]{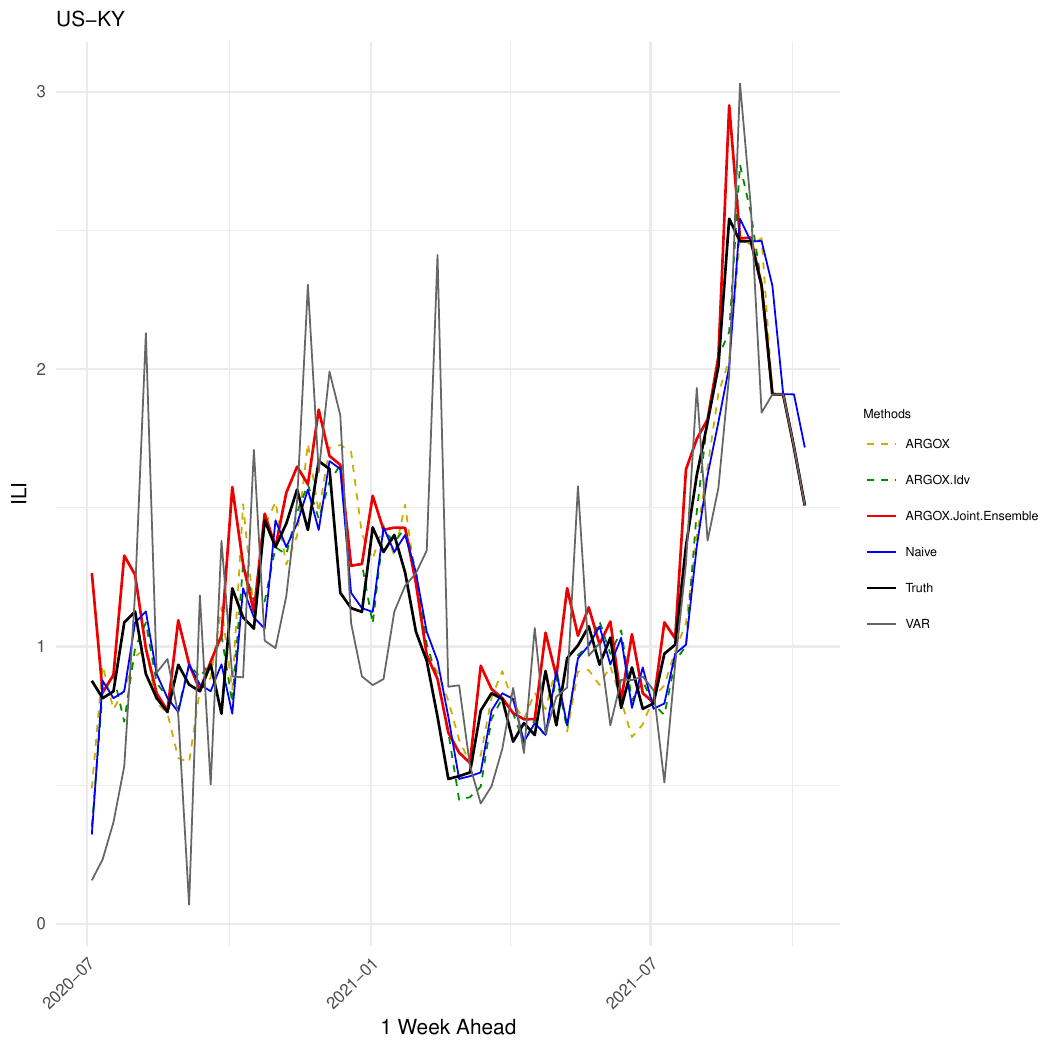} 
\caption{Plots of the \%ILI 1 week ahead estimates for Kentucky (KY). ARGOX is Ref \cite{ARGOX}.}
\end{figure}
\newpage

\begin{table}[ht]
\centering
\begin{tabular}{|c|c|c|c|}
  \hline
Methods & RMSE & MAE & Correlation \\ \hline
  Naive & 
0.30 & 0.17 & 0.94 \\ 
    \hline VAR &0.46 & 0.28 & 0.91 \\ 
    \hline Ref \cite{ARGOX} & 0.37 & 0.23 & 0.91 \\ 
    \hline ARGOX-Idv & 0.30 & 0.18 & 0.94 \\ 
    \hline ARGOX-Joint-Ensemble & 0.27 & 0.14 & 0.96 \\ 
   \hline
\end{tabular}
\caption{Comparison of different methods for state-level ILI 1 week ahead incremental death in Louisiana (LA). The MSE, MAE, and correlation are reported and best performed method is highlighted in boldface.} 
\end{table}

\begin{figure}[!h] 
  \centering 
\includegraphics[width=0.6\linewidth, page=18]{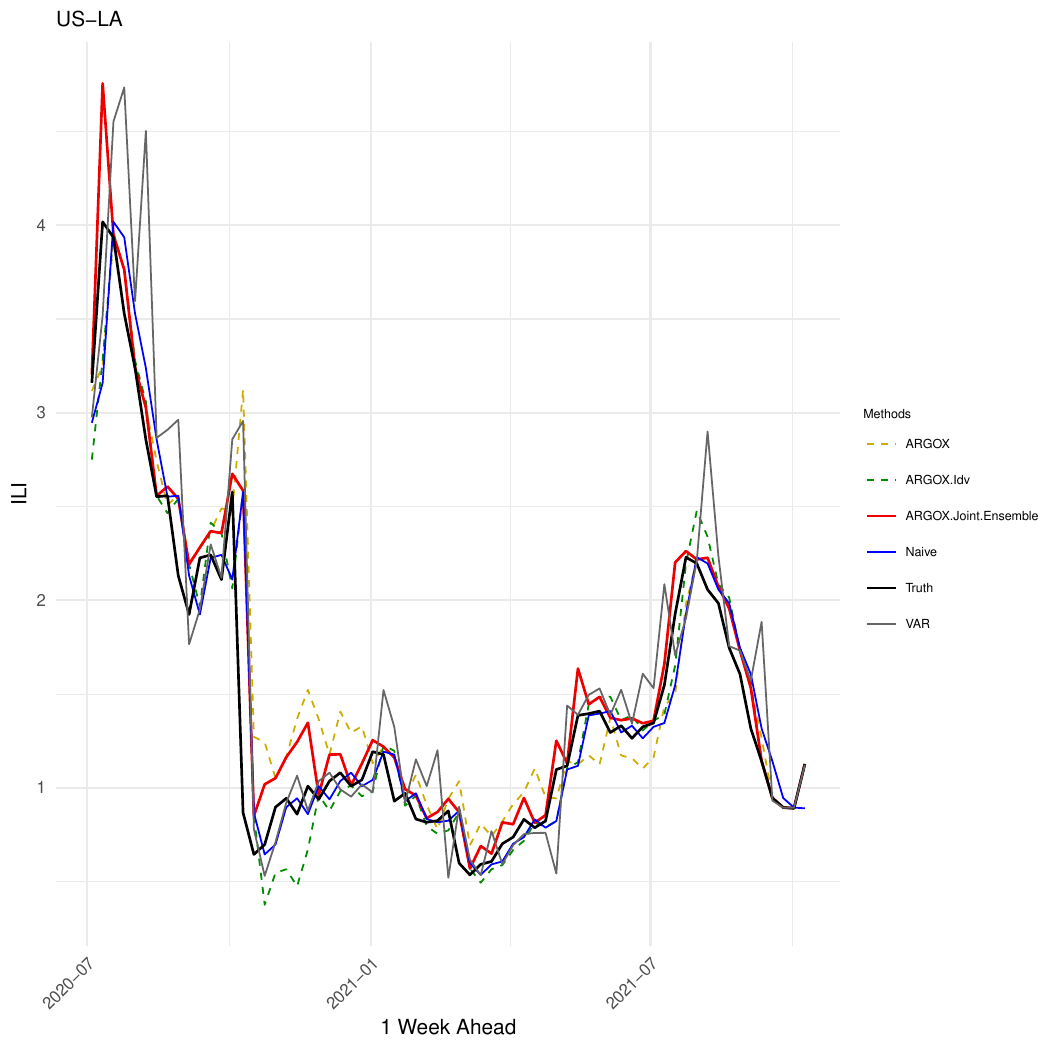} 
\caption{Plots of the \%ILI 1 week ahead estimates for Louisiana (LA). ARGOX is Ref \cite{ARGOX}.}
\end{figure}
\newpage

\begin{table}[ht]
\centering
\begin{tabular}{|c|c|c|c|}
  \hline
Methods & RMSE & MAE & Correlation \\ \hline
  Naive & 
0.12 & 0.10 & 0.88 \\ 
    \hline VAR &0.17 & 0.12 & 0.80 \\ 
    \hline Ref \cite{ARGOX} & 0.14 & 0.11 & 0.85 \\ 
    \hline ARGOX-Idv & 0.12 & 0.09 & 0.90 \\ 
    \hline ARGOX-Joint-Ensemble & 0.09 & 0.07 & 0.96 \\ 
   \hline
\end{tabular}
\caption{Comparison of different methods for state-level ILI 1 week ahead incremental death in Maine (ME). The MSE, MAE, and correlation are reported and best performed method is highlighted in boldface.} 
\end{table}

\begin{figure}[!h] 
  \centering 
\includegraphics[width=0.6\linewidth, page=19]{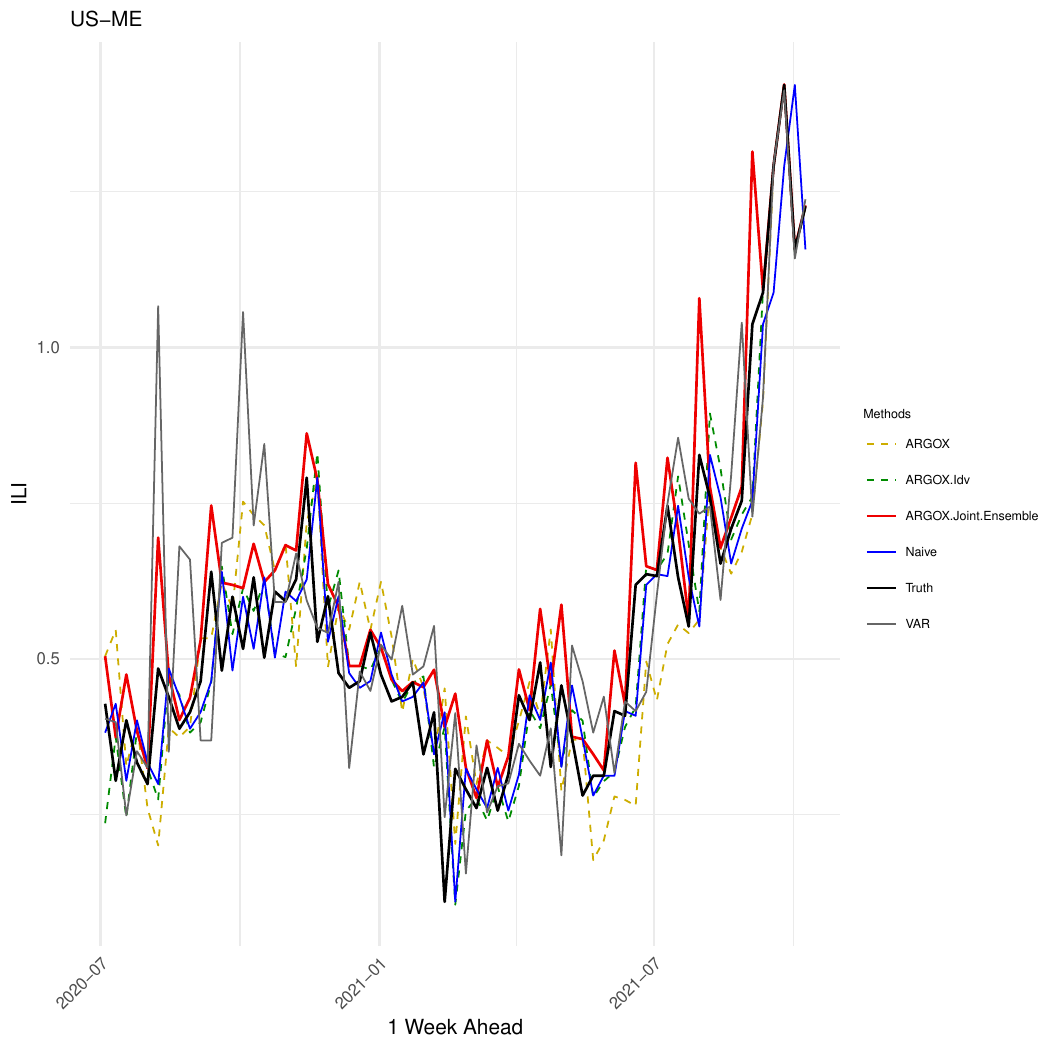} 
\caption{Plots of the \%ILI 1 week ahead estimates for Maine (ME). ARGOX is Ref \cite{ARGOX}.}
\end{figure}
\newpage

\begin{table}[ht]
\centering
\begin{tabular}{|c|c|c|c|}
  \hline
Methods & RMSE & MAE & Correlation \\ \hline
  Naive & 
0.14 & 0.11 & 0.95 \\ 
    \hline VAR &0.30 & 0.22 & 0.78 \\ 
    \hline Ref \cite{ARGOX} & 0.15 & 0.12 & 0.94 \\ 
    \hline ARGOX-Idv & 0.15 & 0.11 & 0.95 \\ 
    \hline ARGOX-Joint-Ensemble & 0.10 & 0.07 & 0.99 \\ 
   \hline
\end{tabular}
\caption{Comparison of different methods for state-level ILI 1 week ahead incremental death in Maryland (MD). The MSE, MAE, and correlation are reported and best performed method is highlighted in boldface.} 
\end{table}

\begin{figure}[!h] 
  \centering 
\includegraphics[width=0.6\linewidth, page=20]{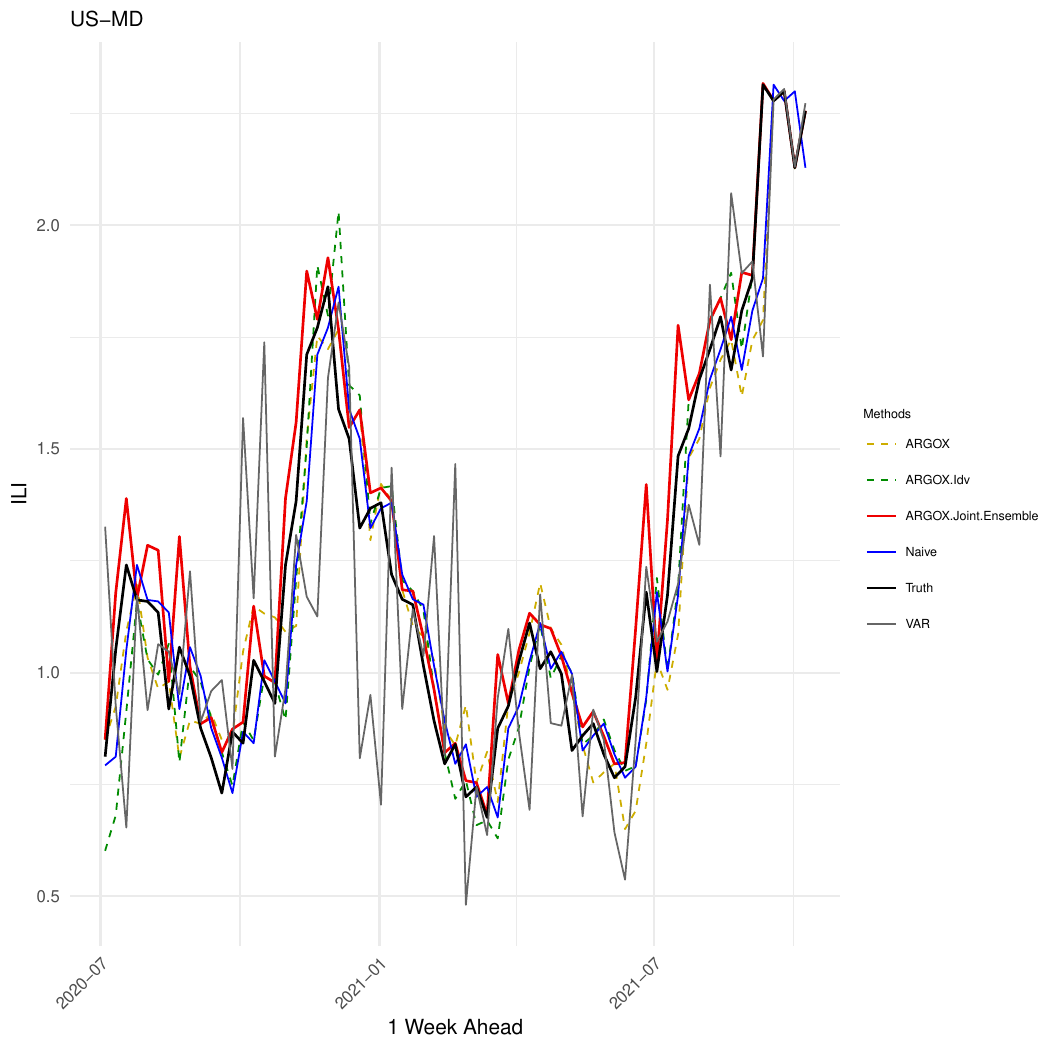} 
\caption{Plots of the \%ILI 1 week ahead estimates for Maryland (MD). ARGOX is Ref \cite{ARGOX}.}
\end{figure}
\newpage

\begin{table}[ht]
\centering
\begin{tabular}{|c|c|c|c|}
  \hline
Methods & RMSE & MAE & Correlation \\ \hline
  Naive & 
0.08 & 0.06 & 0.94 \\ 
    \hline VAR &0.16 & 0.10 & 0.77 \\ 
    \hline Ref \cite{ARGOX} & 0.12 & 0.09 & 0.87 \\ 
    \hline ARGOX-Idv & 0.09 & 0.07 & 0.94 \\ 
    \hline ARGOX-Joint-Ensemble & 0.07 & 0.05 & 0.98 \\ 
   \hline
\end{tabular}
\caption{Comparison of different methods for state-level ILI 1 week ahead incremental death in Massachusetts (MA). The MSE, MAE, and correlation are reported and best performed method is highlighted in boldface.} 
\end{table}

\begin{figure}[!h] 
  \centering 
\includegraphics[width=0.6\linewidth, page=21]{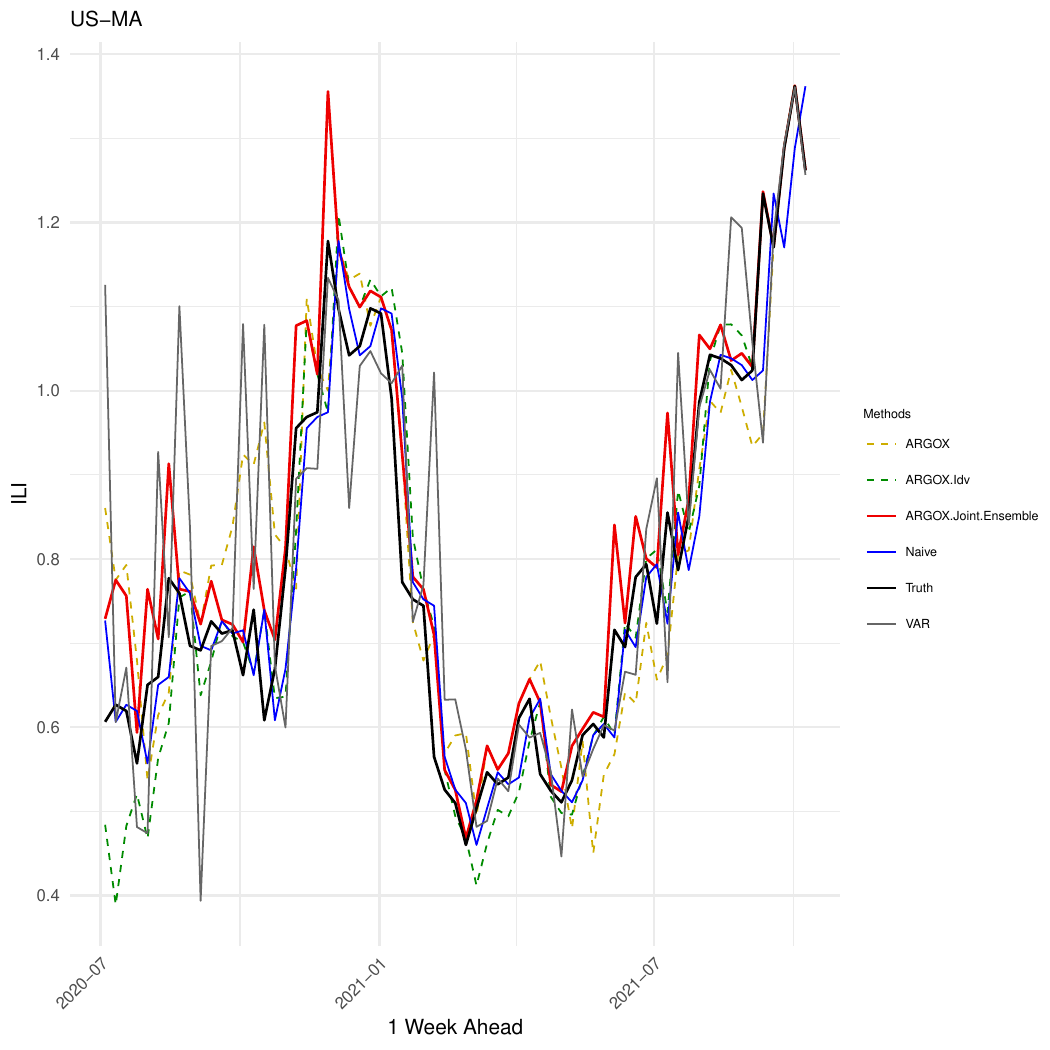} 
\caption{Plots of the \%ILI 1 week ahead estimates for Massachusetts (MA). ARGOX is Ref \cite{ARGOX}.}
\end{figure}
\newpage

\begin{table}[ht]
\centering
\begin{tabular}{|c|c|c|c|}
  \hline
Methods & RMSE & MAE & Correlation \\ \hline
  Naive & 
0.17 & 0.11 & 0.65 \\ 
    \hline VAR &0.15 & 0.11 & 0.76 \\ 
    \hline Ref \cite{ARGOX} & 0.13 & 0.10 & 0.82 \\ 
    \hline ARGOX-Idv & 0.12 & 0.09 & 0.84 \\ 
    \hline ARGOX-Joint-Ensemble & 0.11 & 0.08 & 0.94 \\ 
   \hline
\end{tabular}
\caption{Comparison of different methods for state-level ILI 1 week ahead incremental death in Michigan (MI). The MSE, MAE, and correlation are reported and best performed method is highlighted in boldface.} 
\end{table}

\begin{figure}[!h] 
  \centering 
\includegraphics[width=0.6\linewidth, page=22]{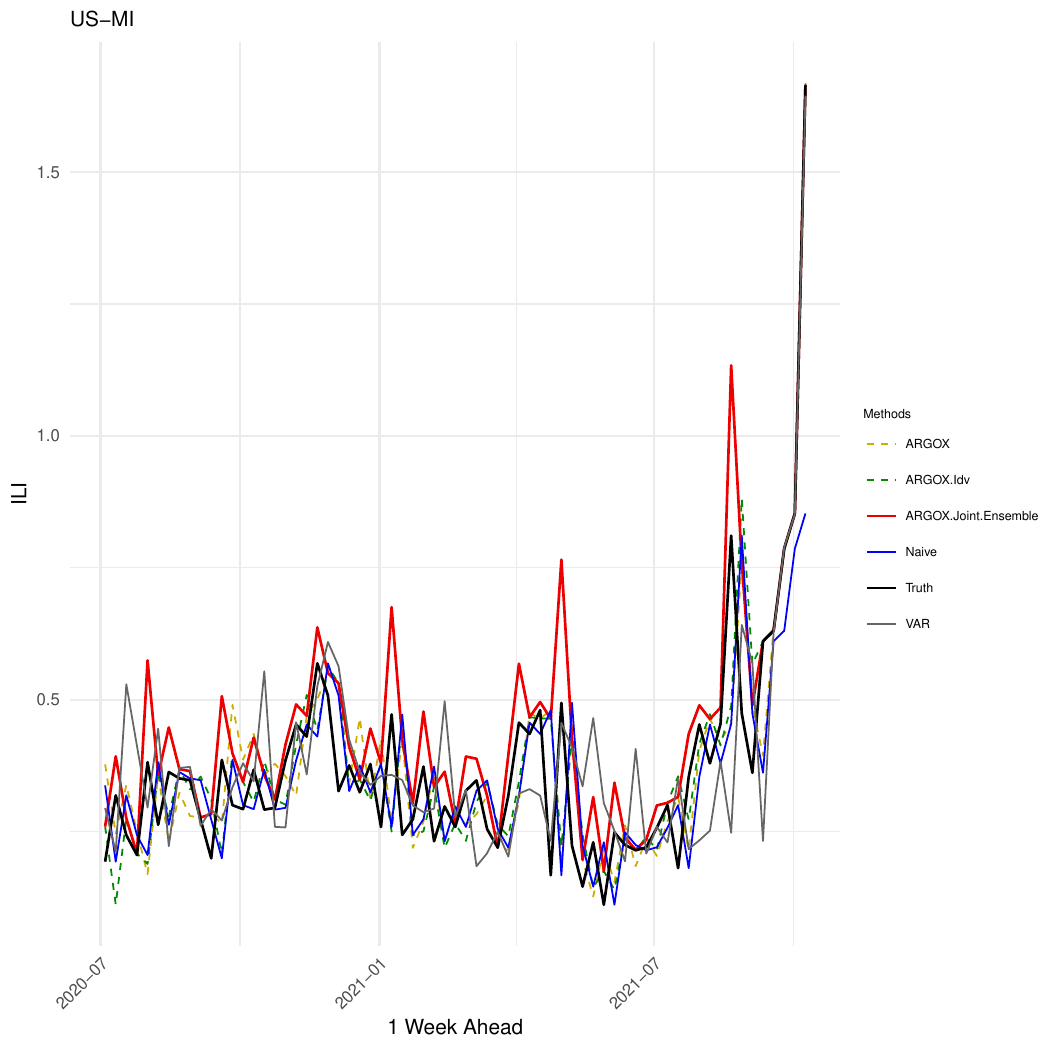} 
\caption{Plots of the \%ILI 1 week ahead estimates for Michigan (MI). ARGOX is Ref \cite{ARGOX}.}
\end{figure}
\newpage

\begin{table}[ht]
\centering
\begin{tabular}{|c|c|c|c|}
  \hline
Methods & RMSE & MAE & Correlation \\ \hline
  Naive & 
0.16 & 0.11 & 0.82 \\ 
    \hline VAR &0.26 & 0.18 & 0.68 \\ 
    \hline Ref \cite{ARGOX} & 0.20 & 0.15 & 0.81 \\ 
    \hline ARGOX-Idv & 0.14 & 0.10 & 0.87 \\ 
    \hline ARGOX-Joint-Ensemble & 0.11 & 0.07 & 0.96 \\ 
   \hline
\end{tabular}
\caption{Comparison of different methods for state-level ILI 1 week ahead incremental death in Minnesota (MN). The MSE, MAE, and correlation are reported and best performed method is highlighted in boldface.} 
\end{table}

\begin{figure}[!h] 
  \centering 
\includegraphics[width=0.6\linewidth, page=23]{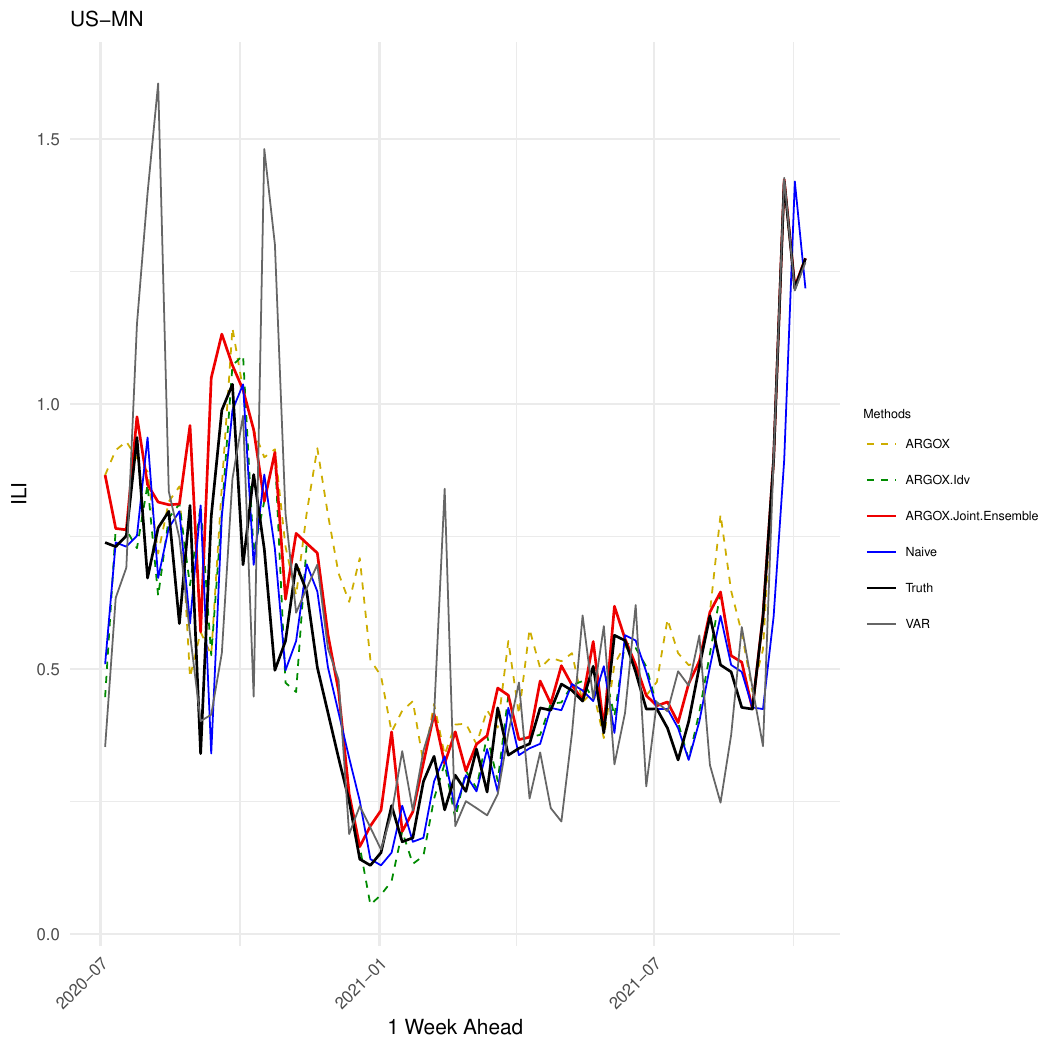} 
\caption{Plots of the \%ILI 1 week ahead estimates for Minnesota (MN). ARGOX is Ref \cite{ARGOX}.}
\end{figure}
\newpage

\begin{table}[ht]
\centering
\begin{tabular}{|c|c|c|c|}
  \hline
Methods & RMSE & MAE & Correlation \\ \hline
  Naive & 
0.44 & 0.33 & 0.65 \\ 
    \hline VAR &0.48 & 0.39 & 0.64 \\ 
    \hline Ref \cite{ARGOX} & 0.38 & 0.29 & 0.72 \\ 
    \hline ARGOX-Idv & 0.41 & 0.29 & 0.70 \\ 
    \hline ARGOX-Joint-Ensemble & 0.34 & 0.23 & 0.93 \\ 
   \hline
\end{tabular}
\caption{Comparison of different methods for state-level ILI 1 week ahead incremental death in Mississippi (MS). The MSE, MAE, and correlation are reported and best performed method is highlighted in boldface.} 
\end{table}

\begin{figure}[!h] 
  \centering 
\includegraphics[width=0.6\linewidth, page=24]{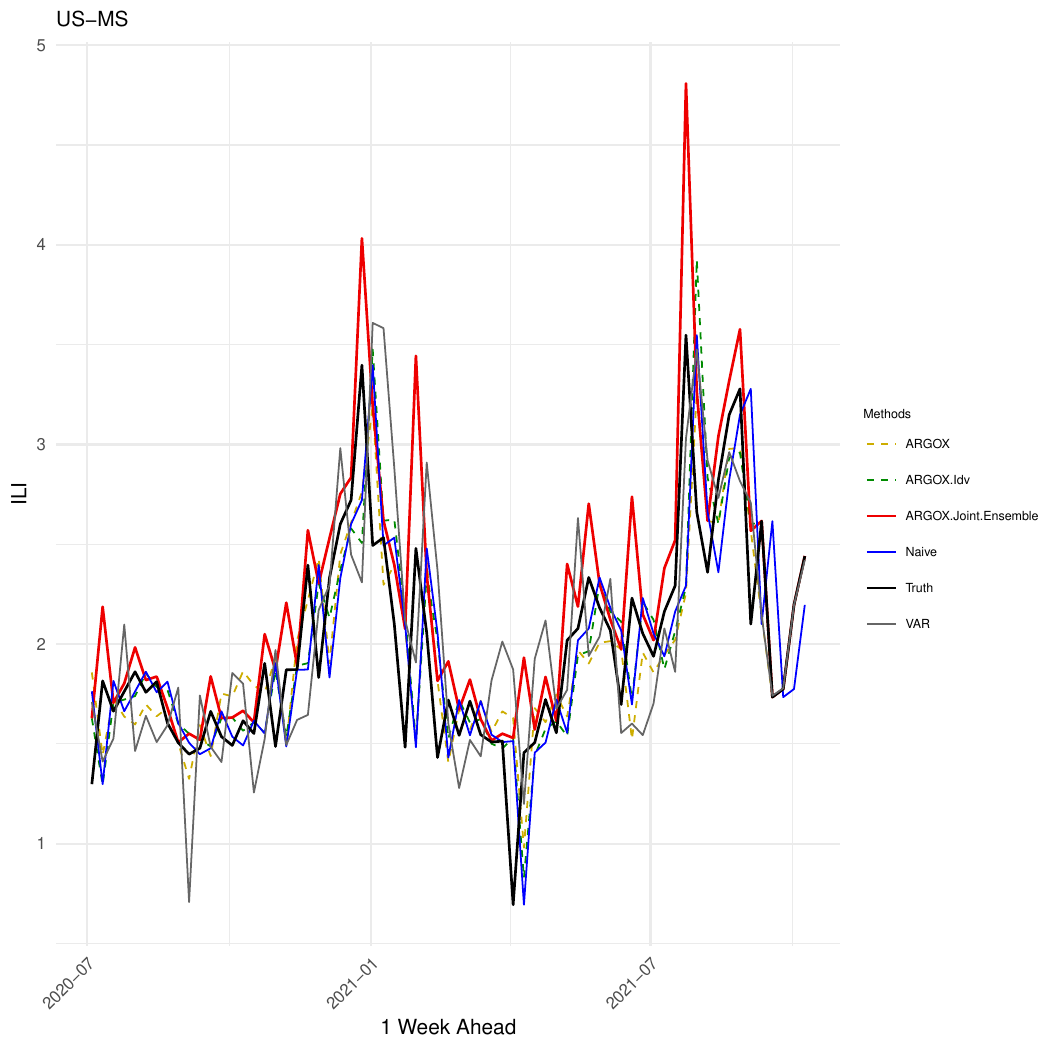} 
\caption{Plots of the \%ILI 1 week ahead estimates for Mississippi (MS). ARGOX is Ref \cite{ARGOX}.}
\end{figure}
\newpage

\begin{table}[ht]
\centering
\begin{tabular}{|c|c|c|c|}
  \hline
Methods & RMSE & MAE & Correlation \\ \hline
  Naive & 
0.15 & 0.12 & 0.93 \\ 
    \hline VAR &0.52 & 0.38 & 0.61 \\ 
    \hline Ref \cite{ARGOX} & 0.21 & 0.16 & 0.89 \\ 
    \hline ARGOX-Idv & 0.15 & 0.11 & 0.93 \\ 
    \hline ARGOX-Joint-Ensemble & 0.11 & 0.08 & 0.98 \\ 
   \hline
\end{tabular}
\caption{Comparison of different methods for state-level ILI 1 week ahead incremental death in Missouri (MO). The MSE, MAE, and correlation are reported and best performed method is highlighted in boldface.} 
\end{table}

\begin{figure}[!h] 
  \centering 
\includegraphics[width=0.6\linewidth, page=25]{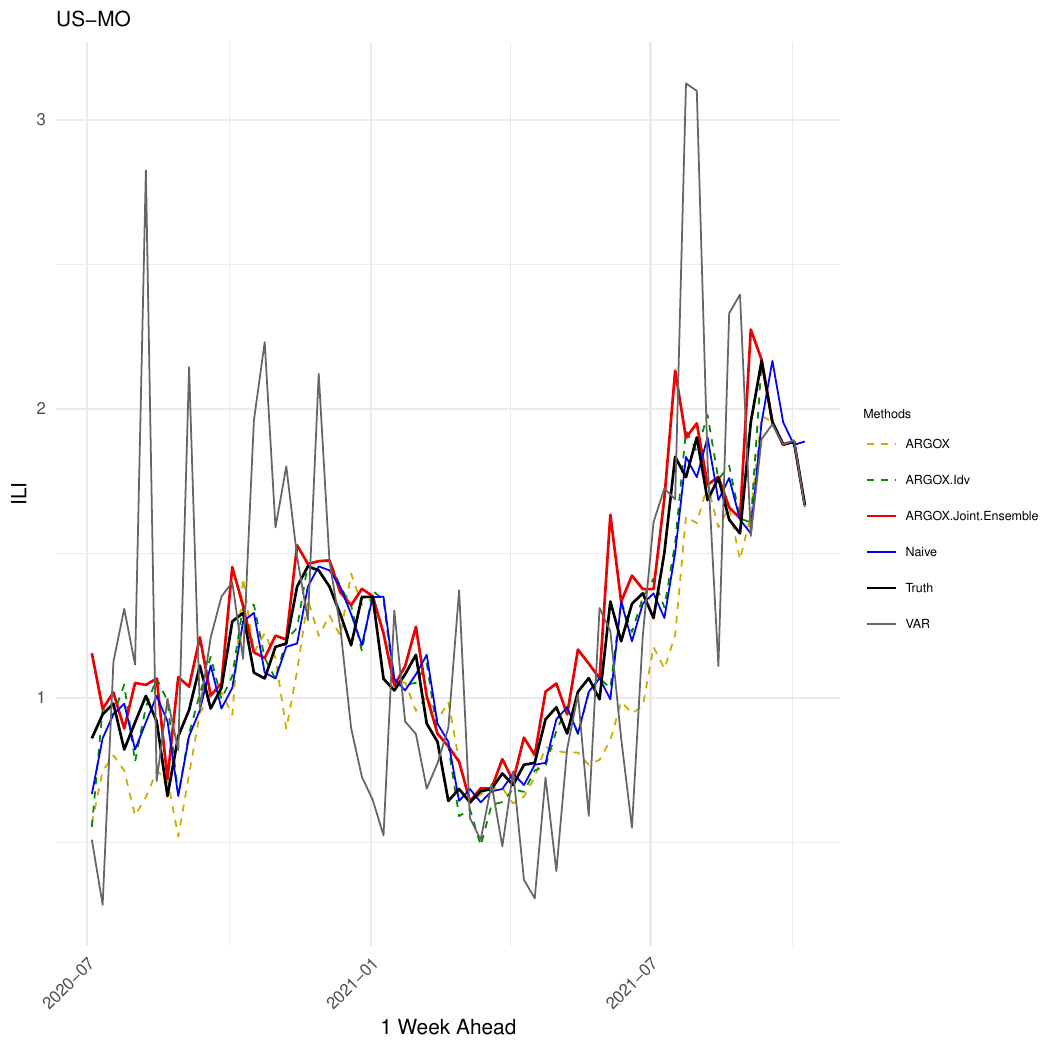} 
\caption{Plots of the \%ILI 1 week ahead estimates for Missouri (MO). ARGOX is Ref \cite{ARGOX}.}
\end{figure}
\newpage

\begin{table}[ht]
\centering
\begin{tabular}{|c|c|c|c|}
  \hline
Methods & RMSE & MAE & Correlation \\ \hline
  Naive & 
0.18 & 0.14 & 0.87 \\ 
    \hline VAR &0.23 & 0.18 & 0.81 \\ 
    \hline Ref \cite{ARGOX} & 0.17 & 0.13 & 0.89 \\ 
    \hline ARGOX-Idv & 0.18 & 0.13 & 0.89 \\ 
    \hline ARGOX-Joint-Ensemble & 0.14 & 0.10 & 0.96 \\ 
   \hline
\end{tabular}
\caption{Comparison of different methods for state-level ILI 1 week ahead incremental death in Montana (MT). The MSE, MAE, and correlation are reported and best performed method is highlighted in boldface.} 
\end{table}

\begin{figure}[!h] 
  \centering 
\includegraphics[width=0.6\linewidth, page=26]{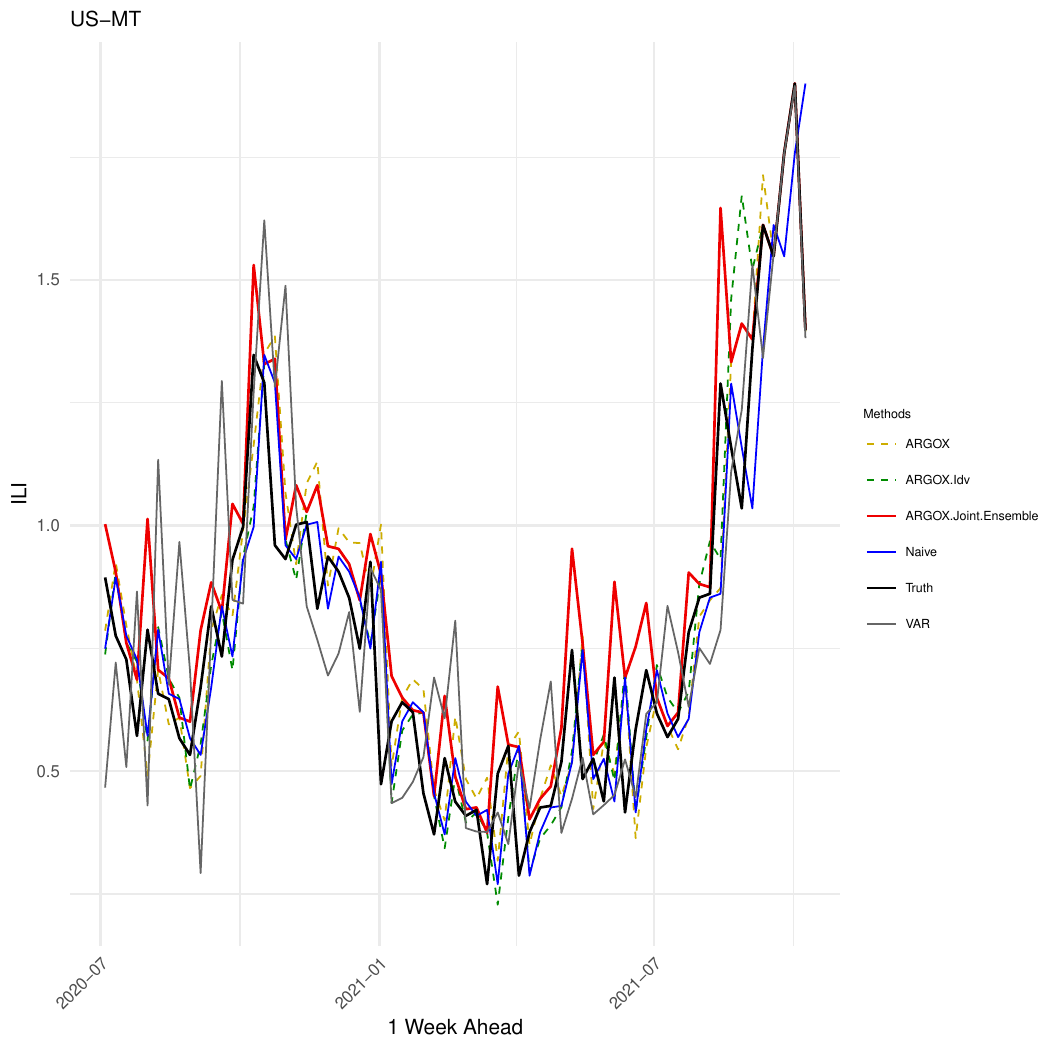} 
\caption{Plots of the \%ILI 1 week ahead estimates for Montana (MT). ARGOX is Ref \cite{ARGOX}.}
\end{figure}
\newpage

\begin{table}[ht]
\centering
\begin{tabular}{|c|c|c|c|}
  \hline
Methods & RMSE & MAE & Correlation \\ \hline
  Naive & 
0.24 & 0.19 & 0.91 \\ 
    \hline VAR &0.31 & 0.22 & 0.86 \\ 
    \hline Ref \cite{ARGOX} & 0.26 & 0.20 & 0.89 \\ 
    \hline ARGOX-Idv & 0.24 & 0.17 & 0.92 \\ 
    \hline ARGOX-Joint-Ensemble & 0.19 & 0.13 & 0.98 \\ 
   \hline
\end{tabular}
\caption{Comparison of different methods for state-level ILI 1 week ahead incremental death in Nebraska (NE). The MSE, MAE, and correlation are reported and best performed method is highlighted in boldface.} 
\end{table}

\begin{figure}[!h] 
  \centering 
\includegraphics[width=0.6\linewidth, page=27]{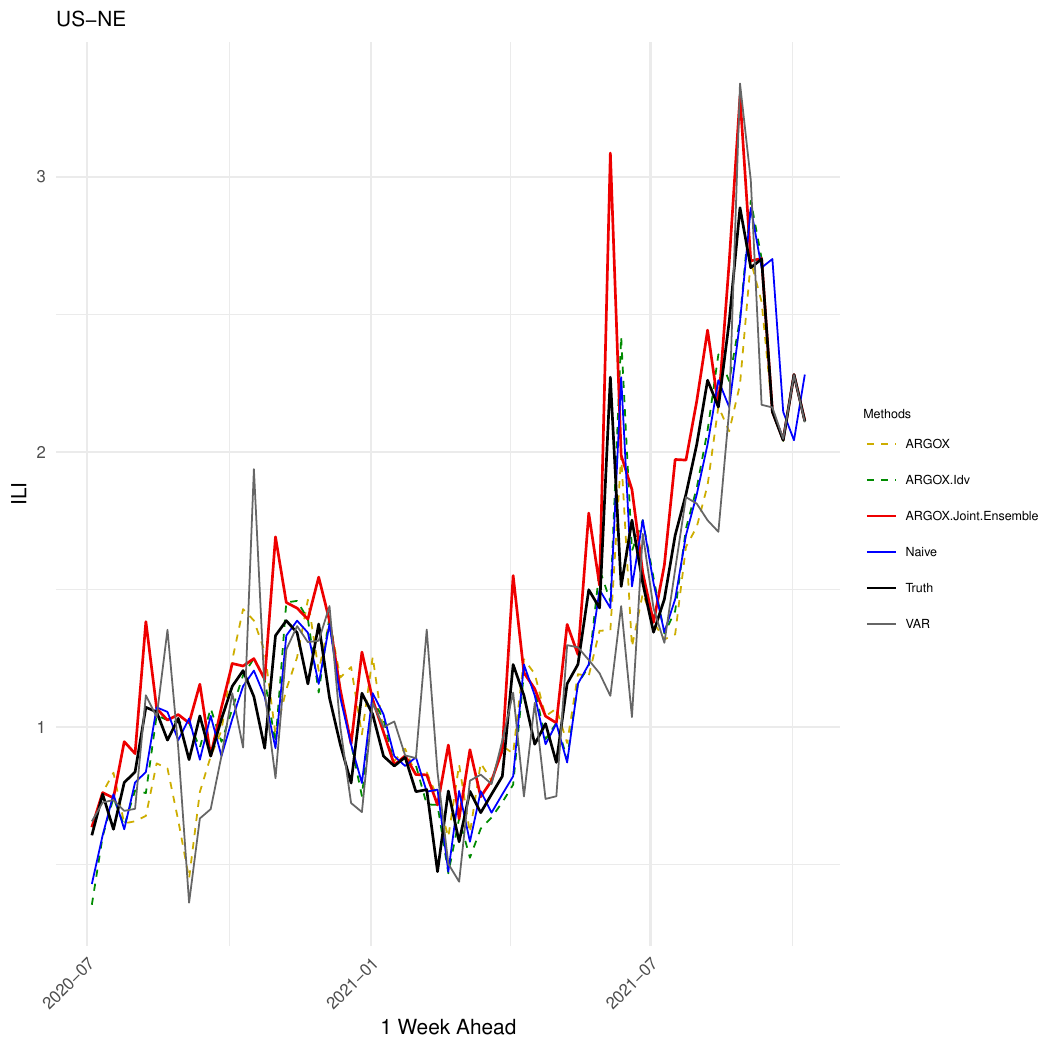} 
\caption{Plots of the \%ILI 1 week ahead estimates for Nebraska (NE). ARGOX is Ref \cite{ARGOX}.}
\end{figure}
\newpage

\begin{table}[ht]
\centering
\begin{tabular}{|c|c|c|c|}
  \hline
Methods & RMSE & MAE & Correlation \\ \hline
  Naive & 
0.44 & 0.33 & 0.76 \\ 
    \hline VAR &0.51 & 0.39 & 0.75 \\ 
    \hline Ref \cite{ARGOX} & 0.42 & 0.30 & 0.78 \\ 
    \hline ARGOX-Idv & 0.40 & 0.28 & 0.80 \\ 
    \hline ARGOX-Joint-Ensemble & 0.36 & 0.24 & 0.93 \\ 
   \hline
\end{tabular}
\caption{Comparison of different methods for state-level ILI 1 week ahead incremental death in Nevada (NV). The MSE, MAE, and correlation are reported and best performed method is highlighted in boldface.} 
\end{table}

\begin{figure}[!h] 
  \centering 
\includegraphics[width=0.6\linewidth, page=28]{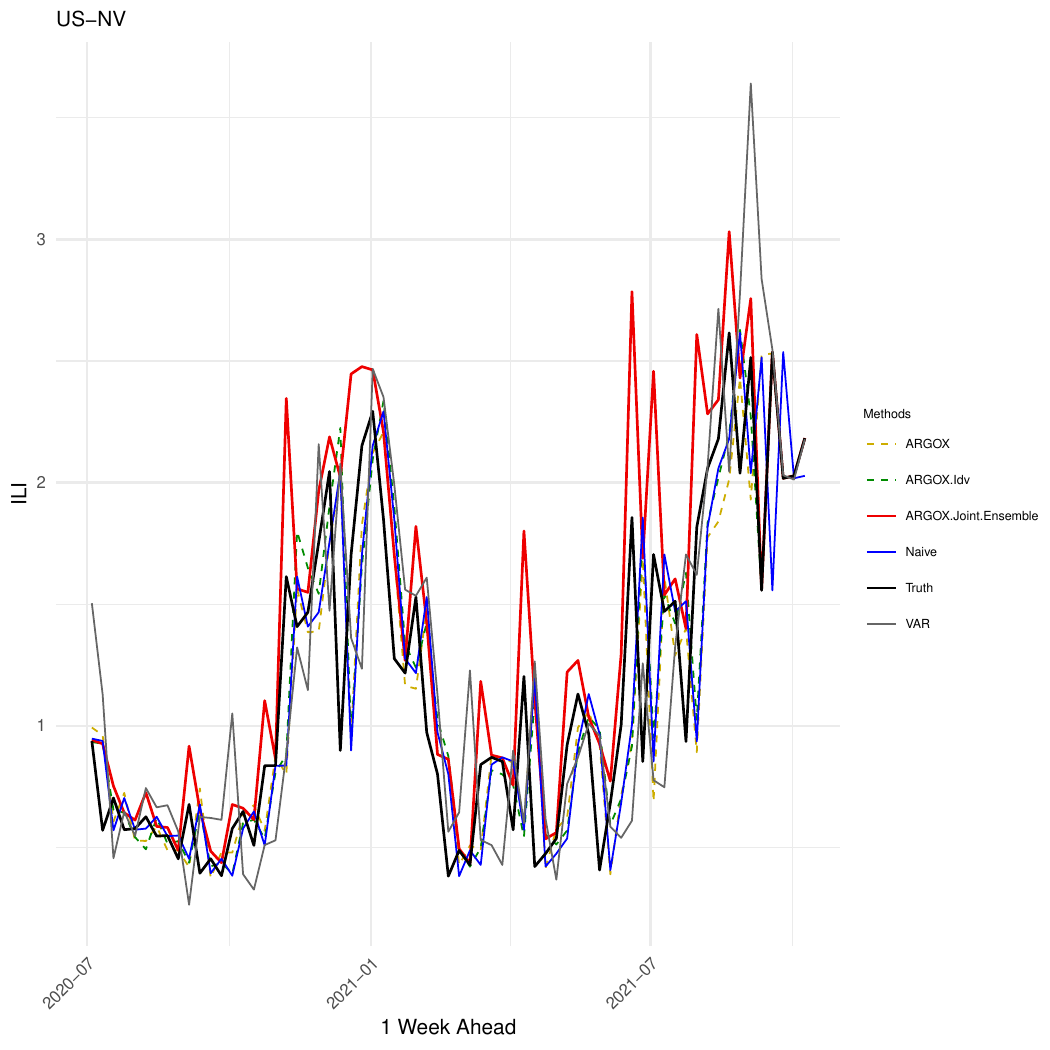} 
\caption{Plots of the \%ILI 1 week ahead estimates for Nevada (NV). ARGOX is Ref \cite{ARGOX}.}
\end{figure}
\newpage

\begin{table}[ht]
\centering
\begin{tabular}{|c|c|c|c|}
  \hline
Methods & RMSE & MAE & Correlation \\ \hline
  Naive & 
0.21 & 0.14 & 0.43 \\ 
    \hline VAR &0.28 & 0.20 & 0.22 \\ 
    \hline Ref \cite{ARGOX} & 0.20 & 0.14 & 0.39 \\ 
    \hline ARGOX-Idv & 0.20 & 0.13 & 0.45 \\ 
    \hline ARGOX-Joint-Ensemble & 0.17 & 0.11 & 0.90 \\ 
   \hline
\end{tabular}
\caption{Comparison of different methods for state-level ILI 1 week ahead incremental death in New Hampshire (NH). The MSE, MAE, and correlation are reported and best performed method is highlighted in boldface.} 
\end{table}

\begin{figure}[!h] 
  \centering 
\includegraphics[width=0.6\linewidth, page=29]{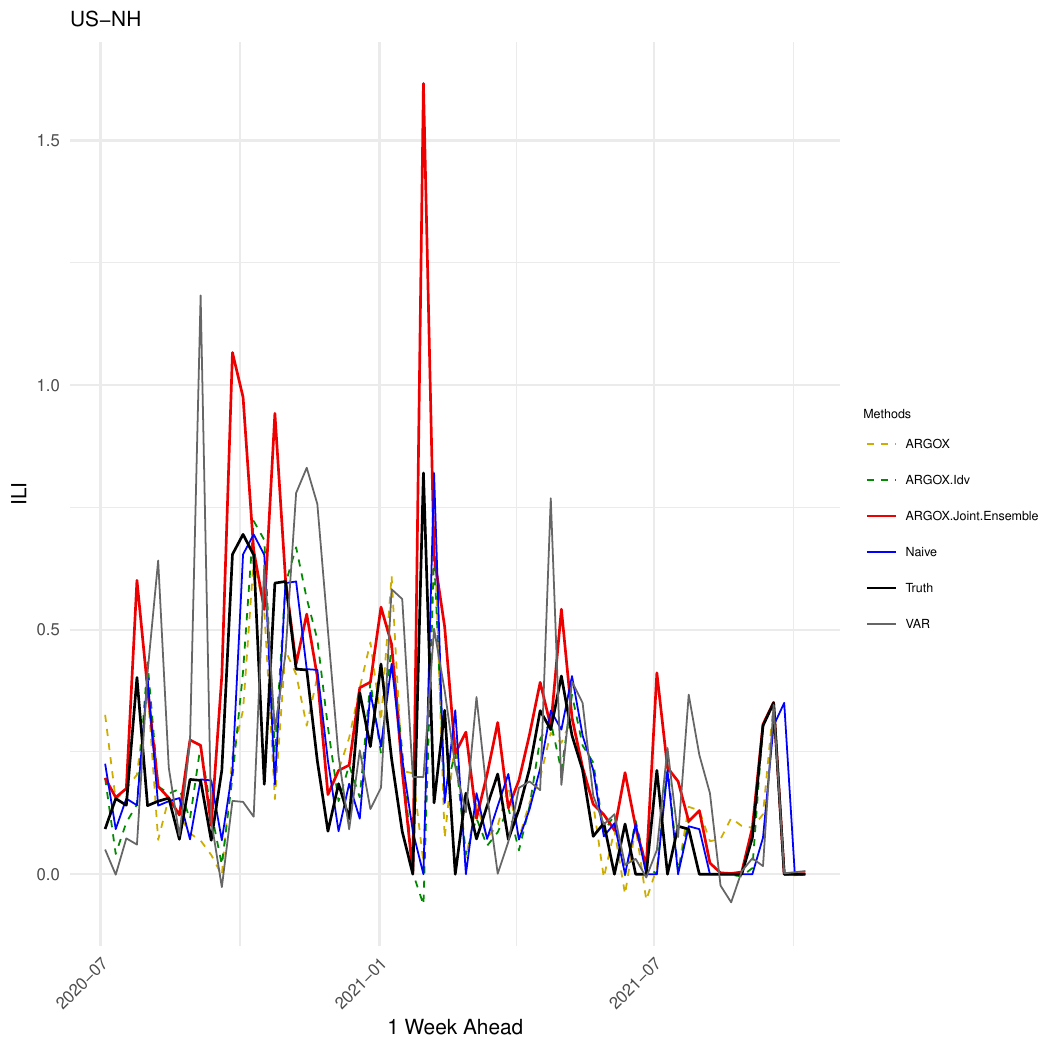} 
\caption{Plots of the \%ILI 1 week ahead estimates for New Hampshire (NH). ARGOX is Ref \cite{ARGOX}.}
\end{figure}
\newpage

\begin{table}[ht]
\centering
\begin{tabular}{|c|c|c|c|}
  \hline
Methods & RMSE & MAE & Correlation \\ \hline
  Naive & 
0.21 & 0.16 & 0.95 \\ 
    \hline VAR &0.33 & 0.26 & 0.87 \\ 
    \hline Ref \cite{ARGOX} & 0.22 & 0.17 & 0.94 \\ 
    \hline ARGOX-Idv & 0.23 & 0.17 & 0.95 \\ 
    \hline ARGOX-Joint-Ensemble & 0.16 & 0.11 & 0.98 \\ 
   \hline
\end{tabular}
\caption{Comparison of different methods for state-level ILI 1 week ahead incremental death in New Jersey (NJ). The MSE, MAE, and correlation are reported and best performed method is highlighted in boldface.} 
\end{table}

\begin{figure}[!h] 
  \centering 
\includegraphics[width=0.6\linewidth, page=30]{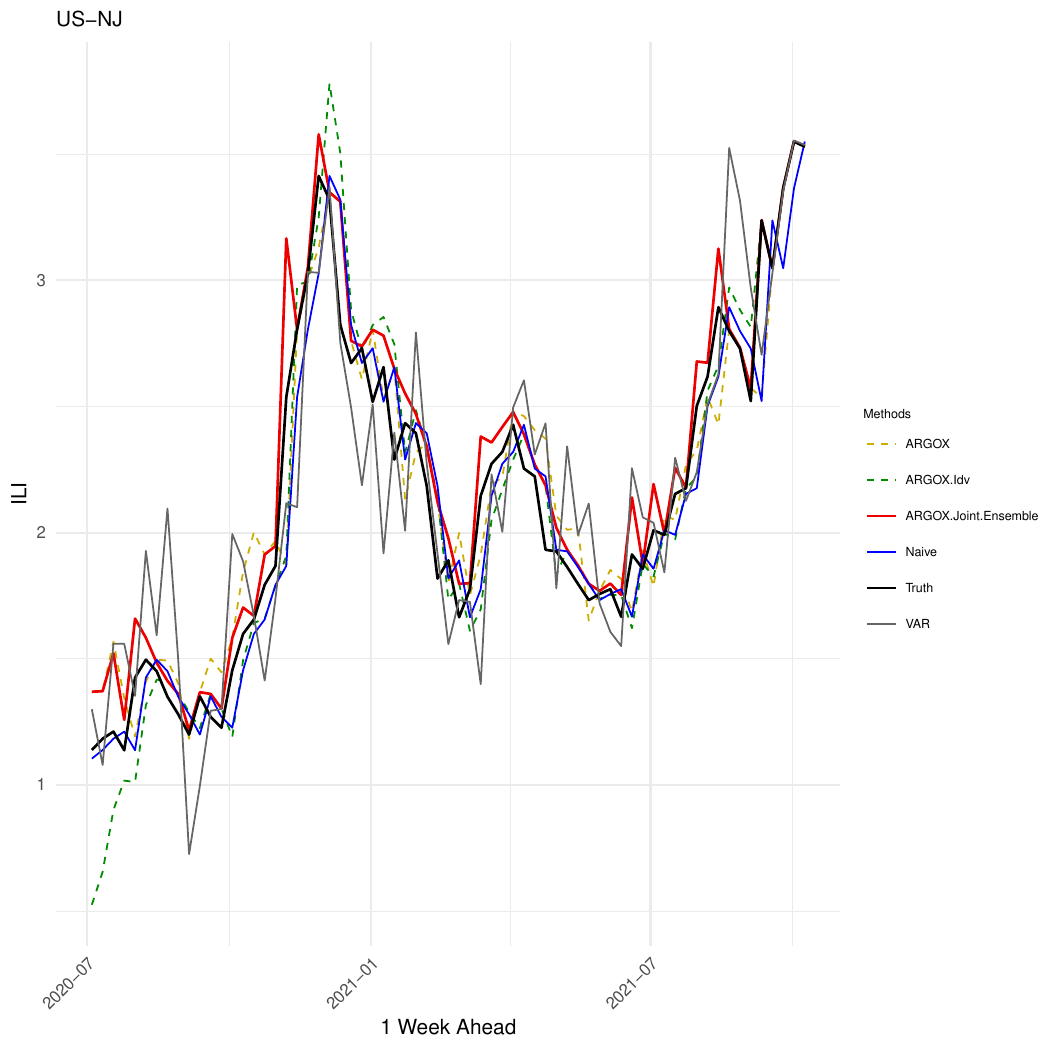} 
\caption{Plots of the \%ILI 1 week ahead estimates for New Jersey (NJ). ARGOX is Ref \cite{ARGOX}.}
\end{figure}
\newpage

\begin{table}[ht]
\centering
\begin{tabular}{|c|c|c|c|}
  \hline
Methods & RMSE & MAE & Correlation \\ \hline
  Naive & 
0.26 & 0.19 & 0.94 \\ 
    \hline VAR &0.35 & 0.26 & 0.89 \\ 
    \hline Ref \cite{ARGOX} & 0.25 & 0.19 & 0.94 \\ 
    \hline ARGOX-Idv & 0.24 & 0.17 & 0.95 \\ 
    \hline ARGOX-Joint-Ensemble & 0.19 & 0.13 & 0.99 \\ 
   \hline
\end{tabular}
\caption{Comparison of different methods for state-level ILI 1 week ahead incremental death in New Mexico (NM). The MSE, MAE, and correlation are reported and best performed method is highlighted in boldface.} 
\end{table}

\begin{figure}[!h] 
  \centering 
\includegraphics[width=0.6\linewidth, page=31]{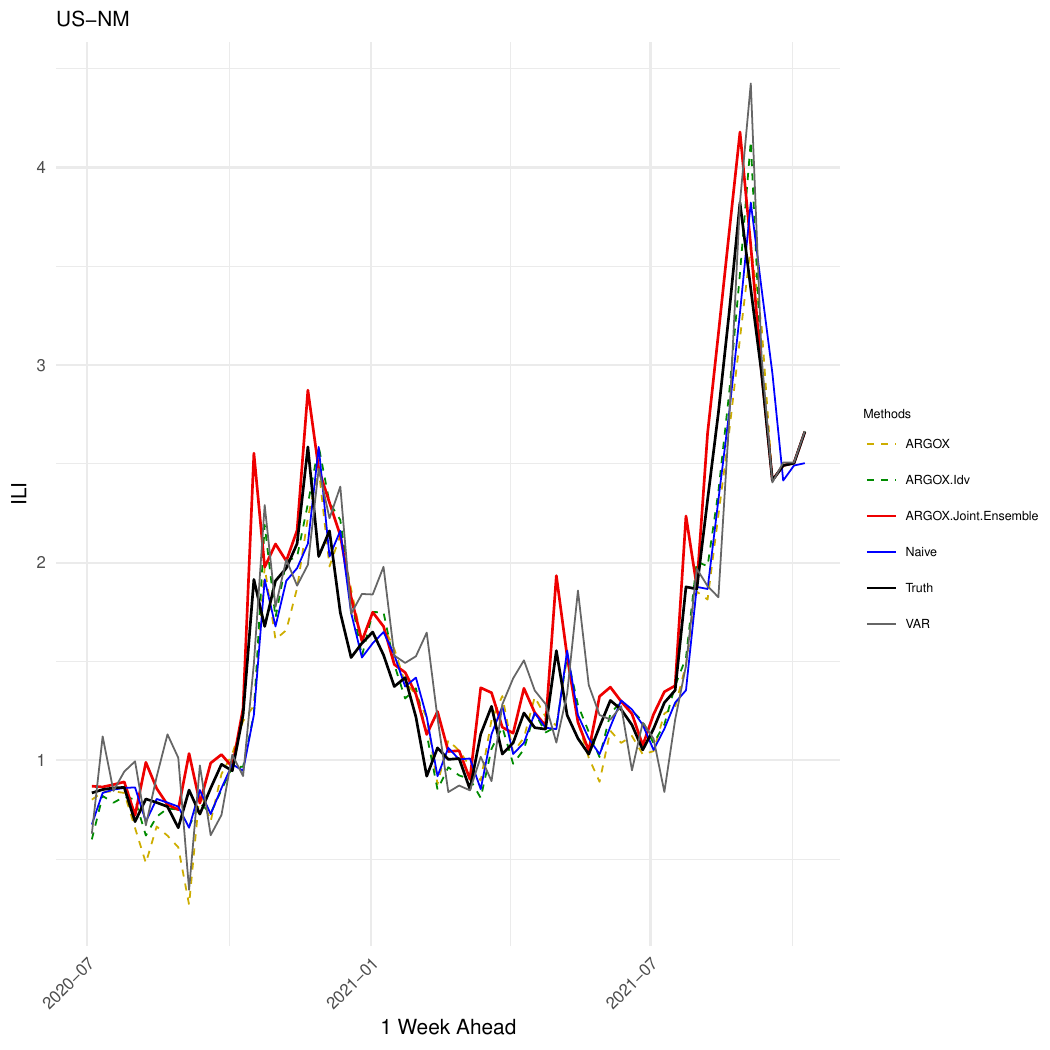} 
\caption{Plots of the \%ILI 1 week ahead estimates for New Mexico (NM). ARGOX is Ref \cite{ARGOX}.}
\end{figure}
\newpage

\begin{table}[ht]
\centering
\begin{tabular}{|c|c|c|c|}
  \hline
Methods & RMSE & MAE & Correlation \\ \hline
  Naive & 
0.13 & 0.10 & 0.95 \\ 
    \hline VAR &0.23 & 0.18 & 0.86 \\ 
    \hline Ref \cite{ARGOX} & 0.14 & 0.11 & 0.94 \\ 
    \hline ARGOX-Idv & 0.14 & 0.10 & 0.95 \\ 
    \hline ARGOX-Joint-Ensemble & 0.10 & 0.07 & 0.98 \\ 
   \hline
\end{tabular}
\caption{Comparison of different methods for state-level ILI 1 week ahead incremental death in New York (NY). The MSE, MAE, and correlation are reported and best performed method is highlighted in boldface.} 
\end{table}

\begin{figure}[!h] 
  \centering 
\includegraphics[width=0.6\linewidth, page=32]{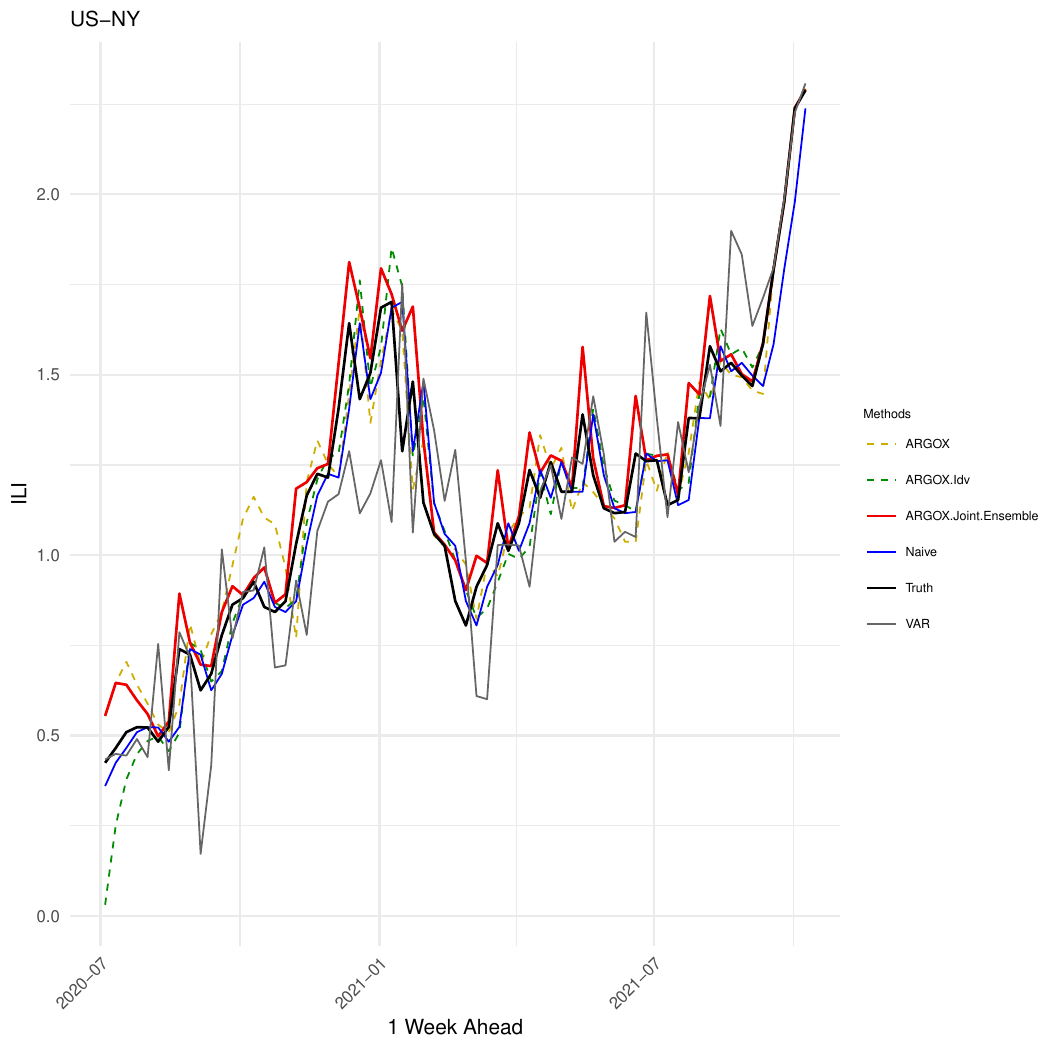} 
\caption{Plots of the \%ILI 1 week ahead estimates for New York (NY). ARGOX is Ref \cite{ARGOX}.}
\end{figure}
\newpage

\begin{table}[ht]
\centering
\begin{tabular}{|c|c|c|c|}
  \hline
Methods & RMSE & MAE & Correlation \\ \hline
  Naive & 
0.11 & 0.09 & 0.97 \\ 
    \hline VAR &0.17 & 0.13 & 0.91 \\ 
    \hline Ref \cite{ARGOX} & 0.10 & 0.08 & 0.97 \\ 
    \hline ARGOX-Idv & 0.08 & 0.07 & 0.98 \\ 
    \hline ARGOX-Joint-Ensemble & 0.06 & 0.04 & 0.99 \\ 
   \hline
\end{tabular}
\caption{Comparison of different methods for state-level ILI 1 week ahead incremental death in North Carolina (NC). The MSE, MAE, and correlation are reported and best performed method is highlighted in boldface.} 
\end{table}

\begin{figure}[!h] 
  \centering 
\includegraphics[width=0.6\linewidth, page=33]{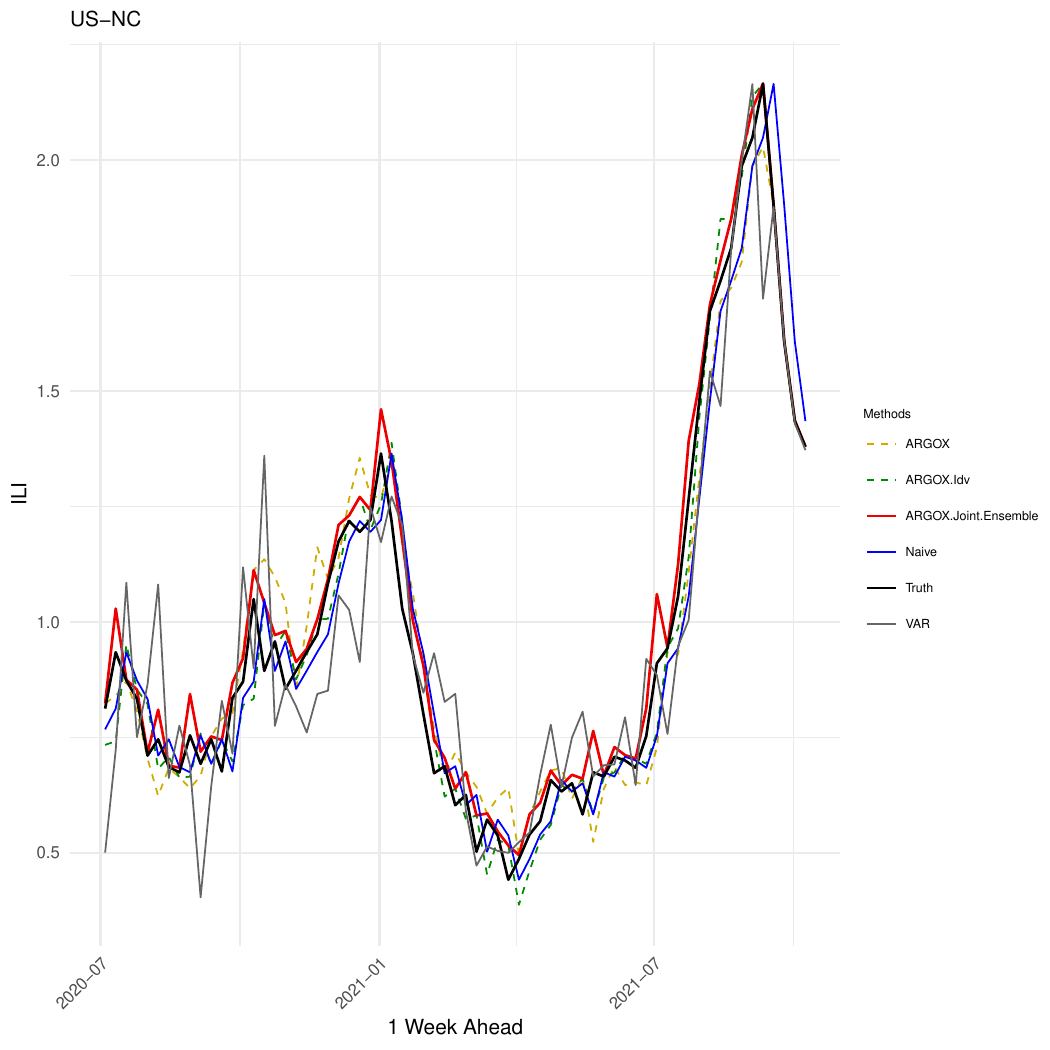} 
\caption{Plots of the \%ILI 1 week ahead estimates for North Carolina (NC). ARGOX is Ref \cite{ARGOX}.}
\end{figure}
\newpage

\begin{table}[ht]
\centering
\begin{tabular}{|c|c|c|c|}
  \hline
Methods & RMSE & MAE & Correlation \\ \hline
  Naive & 
0.48 & 0.35 & 0.69 \\ 
    \hline VAR &0.68 & 0.48 & 0.35 \\ 
    \hline Ref \cite{ARGOX} & 0.43 & 0.30 & 0.71 \\ 
    \hline ARGOX-Idv & 0.47 & 0.34 & 0.70 \\ 
    \hline ARGOX-Joint-Ensemble & 0.38 & 0.25 & 0.94 \\ 
   \hline
\end{tabular}
\caption{Comparison of different methods for state-level ILI 1 week ahead incremental death in North Dakota (ND). The MSE, MAE, and correlation are reported and best performed method is highlighted in boldface.} 
\end{table}

\begin{figure}[!h] 
  \centering 
\includegraphics[width=0.6\linewidth, page=34]{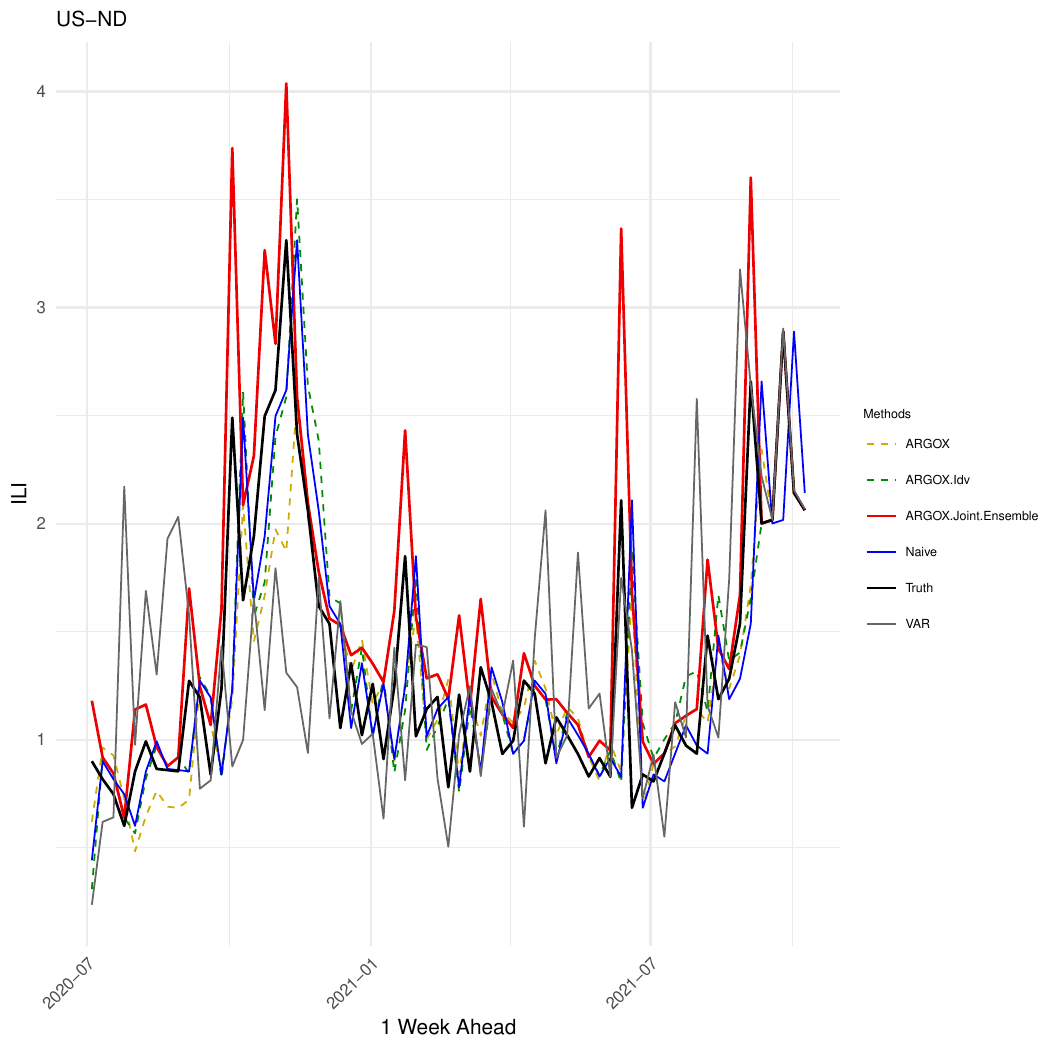} 
\caption{Plots of the \%ILI 1 week ahead estimates for North Dakota (ND). ARGOX is Ref \cite{ARGOX}.}
\end{figure}
\newpage

\begin{table}[ht]
\centering
\begin{tabular}{|c|c|c|c|}
  \hline
Methods & RMSE & MAE & Correlation \\ \hline
  Naive & 
0.27 & 0.18 & 0.83 \\ 
    \hline VAR &0.25 & 0.20 & 0.86 \\ 
    \hline Ref \cite{ARGOX} & 0.17 & 0.13 & 0.94 \\ 
    \hline ARGOX-Idv & 0.17 & 0.12 & 0.93 \\ 
    \hline ARGOX-Joint-Ensemble & 0.13 & 0.09 & 0.98 \\ 
   \hline
\end{tabular}
\caption{Comparison of different methods for state-level ILI 1 week ahead incremental death in Ohio (OH). The MSE, MAE, and correlation are reported and best performed method is highlighted in boldface.} 
\end{table}

\begin{figure}[!h] 
  \centering 
\includegraphics[width=0.6\linewidth, page=35]{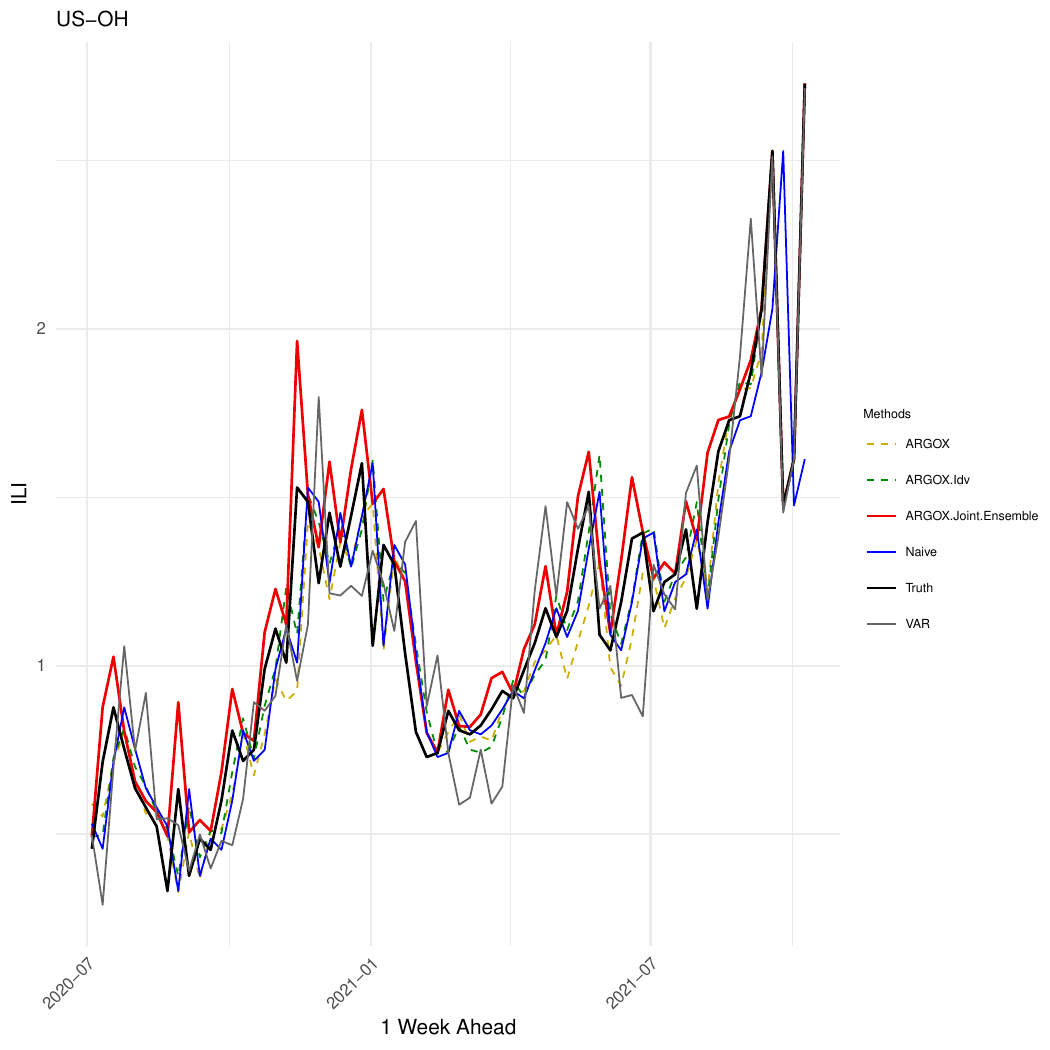} 
\caption{Plots of the \%ILI 1 week ahead estimates for Ohio (OH). ARGOX is Ref \cite{ARGOX}.}
\end{figure}
\newpage

\begin{table}[ht]
\centering
\begin{tabular}{|c|c|c|c|}
  \hline
Methods & RMSE & MAE & Correlation \\ \hline
  Naive & 
0.51 & 0.39 & 0.80 \\ 
    \hline VAR &0.61 & 0.44 & 0.71 \\ 
    \hline Ref \cite{ARGOX} & 0.49 & 0.37 & 0.81 \\ 
    \hline ARGOX-Idv & 0.50 & 0.37 & 0.82 \\ 
    \hline ARGOX-Joint-Ensemble & 0.44 & 0.31 & 0.95 \\ 
   \hline
\end{tabular}
\caption{Comparison of different methods for state-level ILI 1 week ahead incremental death in Oklahoma (OK). The MSE, MAE, and correlation are reported and best performed method is highlighted in boldface.} 
\end{table}

\begin{figure}[!h] 
  \centering 
\includegraphics[width=0.6\linewidth, page=36]{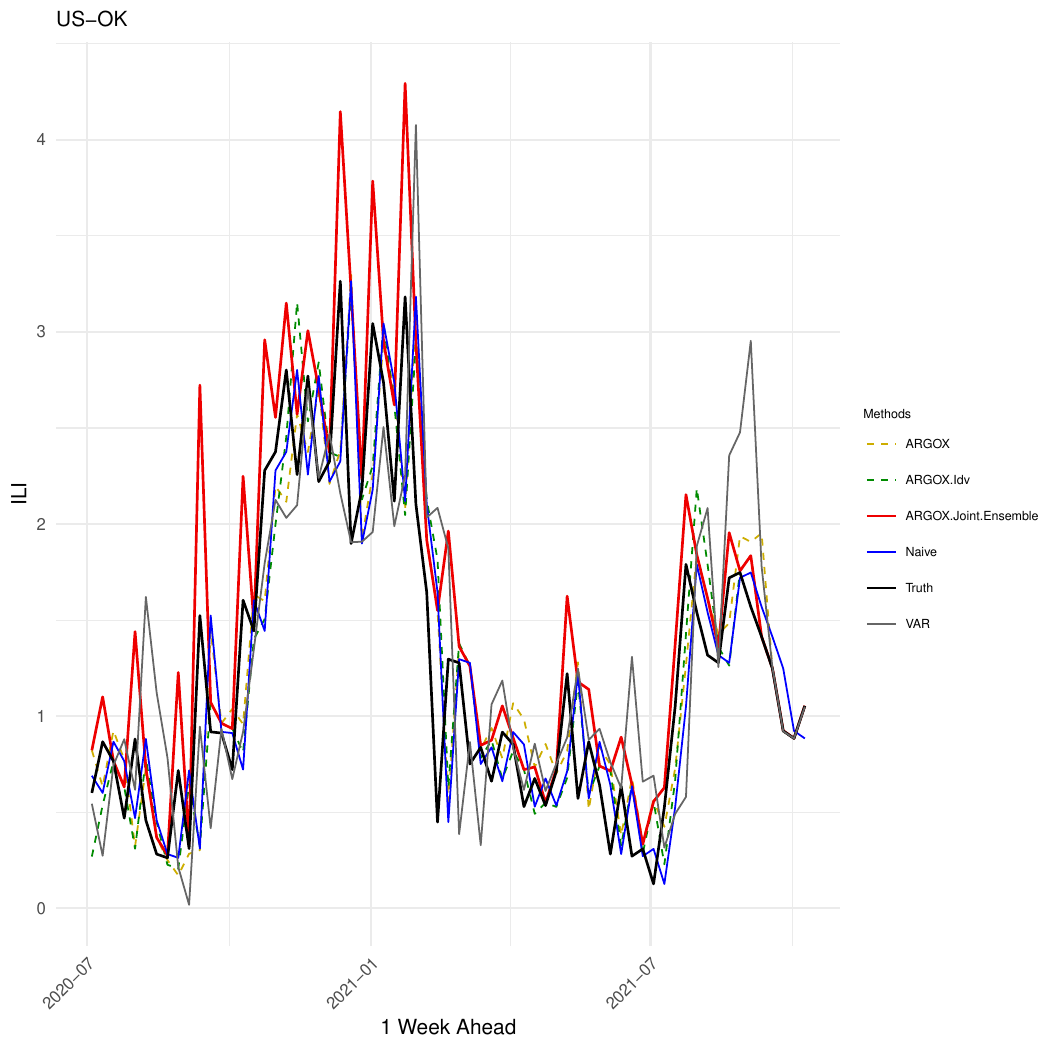} 
\caption{Plots of the \%ILI 1 week ahead estimates for Oklahoma (OK). ARGOX is Ref \cite{ARGOX}.}
\end{figure}
\newpage

\begin{table}[ht]
\centering
\begin{tabular}{|c|c|c|c|}
  \hline
Methods & RMSE & MAE & Correlation \\ \hline
  Naive & 
0.15 & 0.11 & 0.91 \\ 
    \hline VAR &0.19 & 0.14 & 0.86 \\ 
    \hline Ref \cite{ARGOX} & 0.13 & 0.10 & 0.93 \\ 
    \hline ARGOX-Idv & 0.13 & 0.09 & 0.94 \\ 
    \hline ARGOX-Joint-Ensemble & 0.09 & 0.06 & 0.99 \\ 
   \hline
\end{tabular}
\caption{Comparison of different methods for state-level ILI 1 week ahead incremental death in Oregon (OR). The MSE, MAE, and correlation are reported and best performed method is highlighted in boldface.} 
\end{table}

\begin{figure}[!h] 
  \centering 
\includegraphics[width=0.6\linewidth, page=37]{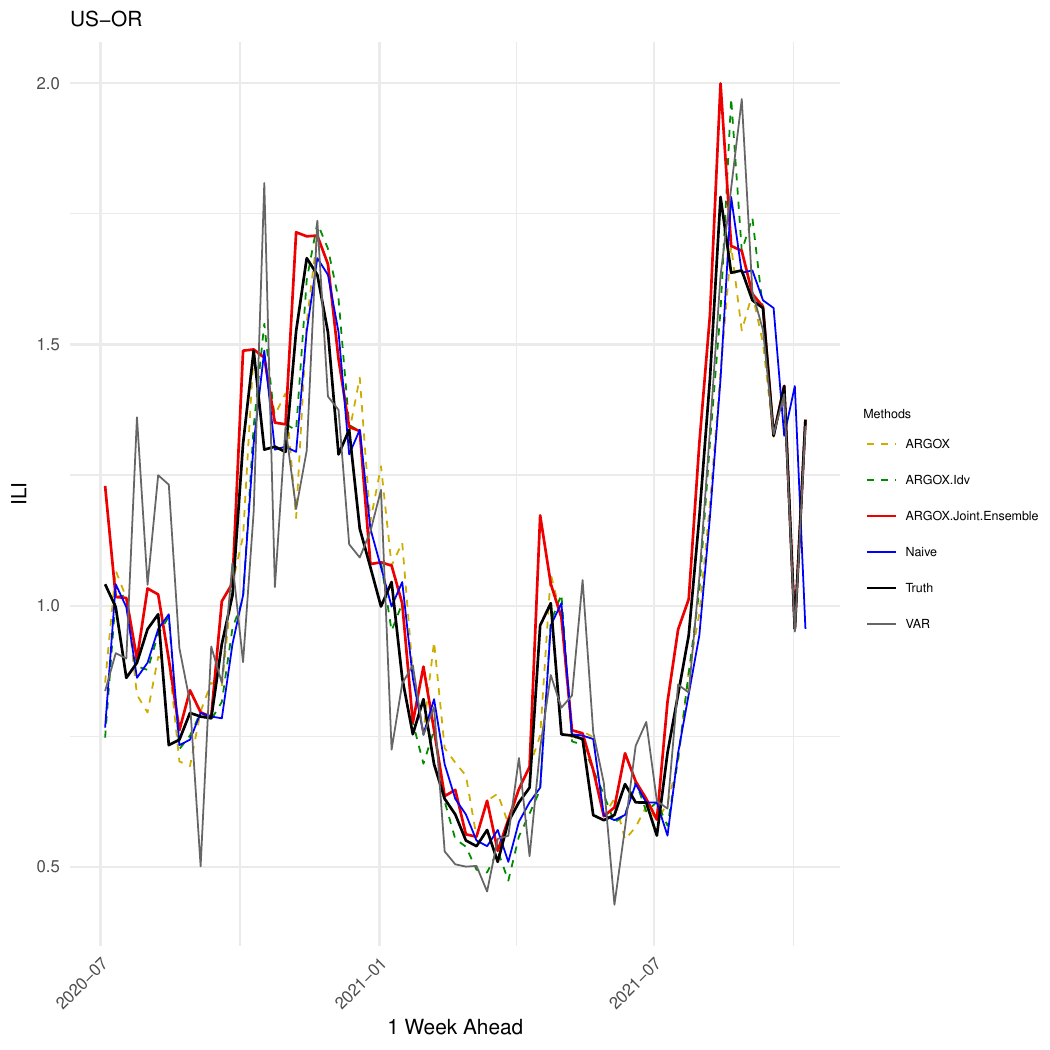} 
\caption{Plots of the \%ILI 1 week ahead estimates for Oregon (OR). ARGOX is Ref \cite{ARGOX}.}
\end{figure}
\newpage

\begin{table}[ht]
\centering
\begin{tabular}{|c|c|c|c|}
  \hline
Methods & RMSE & MAE & Correlation \\ \hline
  Naive & 
0.13 & 0.09 & 0.93 \\ 
    \hline VAR &0.16 & 0.12 & 0.88 \\ 
    \hline Ref \cite{ARGOX} & 0.13 & 0.10 & 0.92 \\ 
    \hline ARGOX-Idv & 0.12 & 0.09 & 0.94 \\ 
    \hline ARGOX-Joint-Ensemble & 0.10 & 0.07 & 0.98 \\ 
   \hline
\end{tabular}
\caption{Comparison of different methods for state-level ILI 1 week ahead incremental death in Pennsylvania (PA). The MSE, MAE, and correlation are reported and best performed method is highlighted in boldface.} 
\end{table}

\begin{figure}[!h] 
  \centering 
\includegraphics[width=0.6\linewidth, page=38]{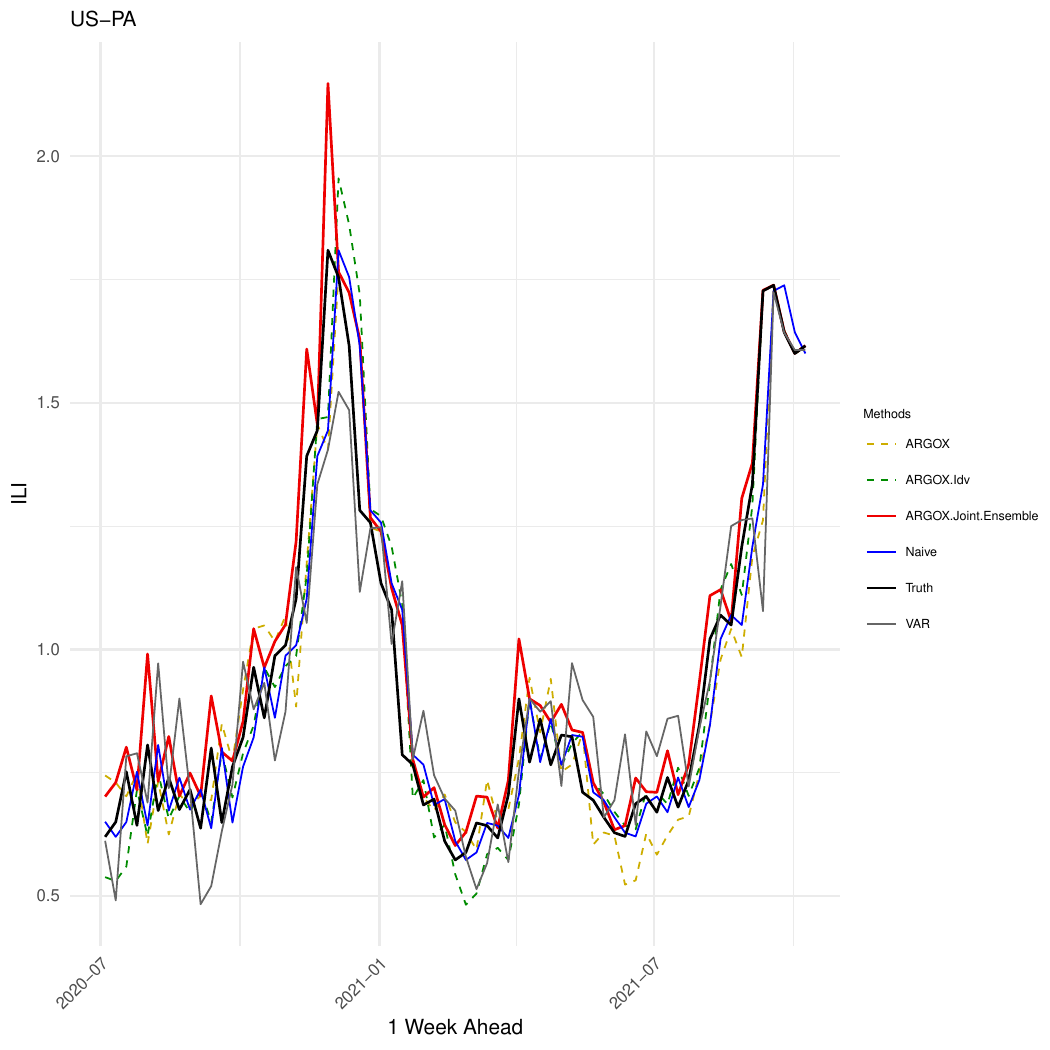} 
\caption{Plots of the \%ILI 1 week ahead estimates for Pennsylvania (PA). ARGOX is Ref \cite{ARGOX}.}
\end{figure}
\newpage

\begin{table}[ht]
\centering
\begin{tabular}{|c|c|c|c|}
  \hline
Methods & RMSE & MAE & Correlation \\ \hline
  Naive & 
0.12 & 0.09 & 0.73 \\ 
    \hline VAR &0.39 & 0.23 & 0.47 \\ 
    \hline Ref \cite{ARGOX} & 0.15 & 0.11 & 0.72 \\ 
    \hline ARGOX-Idv & 0.12 & 0.09 & 0.74 \\ 
    \hline ARGOX-Joint-Ensemble & 0.10 & 0.07 & 0.95 \\ 
   \hline
\end{tabular}
\caption{Comparison of different methods for state-level ILI 1 week ahead incremental death in Rhode Island (RI). The MSE, MAE, and correlation are reported and best performed method is highlighted in boldface.} 
\end{table}

\begin{figure}[!h] 
  \centering 
\includegraphics[width=0.6\linewidth, page=39]{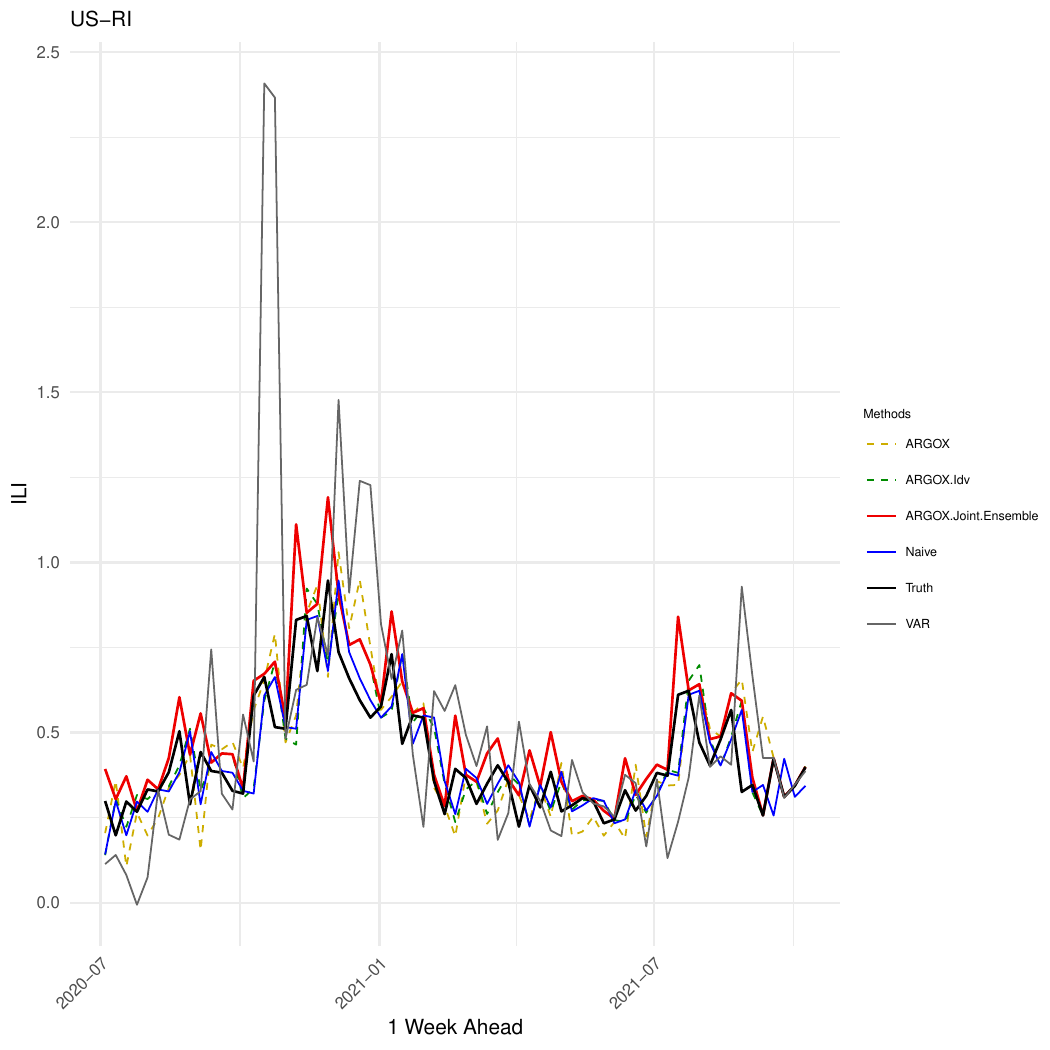} 
\caption{Plots of the \%ILI 1 week ahead estimates for Rhode Island (RI). ARGOX is Ref \cite{ARGOX}.}
\end{figure}
\newpage

\begin{table}[ht]
\centering
\begin{tabular}{|c|c|c|c|}
  \hline
Methods & RMSE & MAE & Correlation \\ \hline
  Naive & 
0.24 & 0.18 & 0.90 \\ 
    \hline VAR &0.44 & 0.34 & 0.68 \\ 
    \hline Ref \cite{ARGOX} & 0.27 & 0.22 & 0.89 \\ 
    \hline ARGOX-Idv & 0.24 & 0.18 & 0.91 \\ 
    \hline ARGOX-Joint-Ensemble & 0.18 & 0.13 & 0.97 \\ 
   \hline
\end{tabular}
\caption{Comparison of different methods for state-level ILI 1 week ahead incremental death in South Carolina (SC). The MSE, MAE, and correlation are reported and best performed method is highlighted in boldface.} 
\end{table}

\begin{figure}[!h] 
  \centering 
\includegraphics[width=0.6\linewidth, page=40]{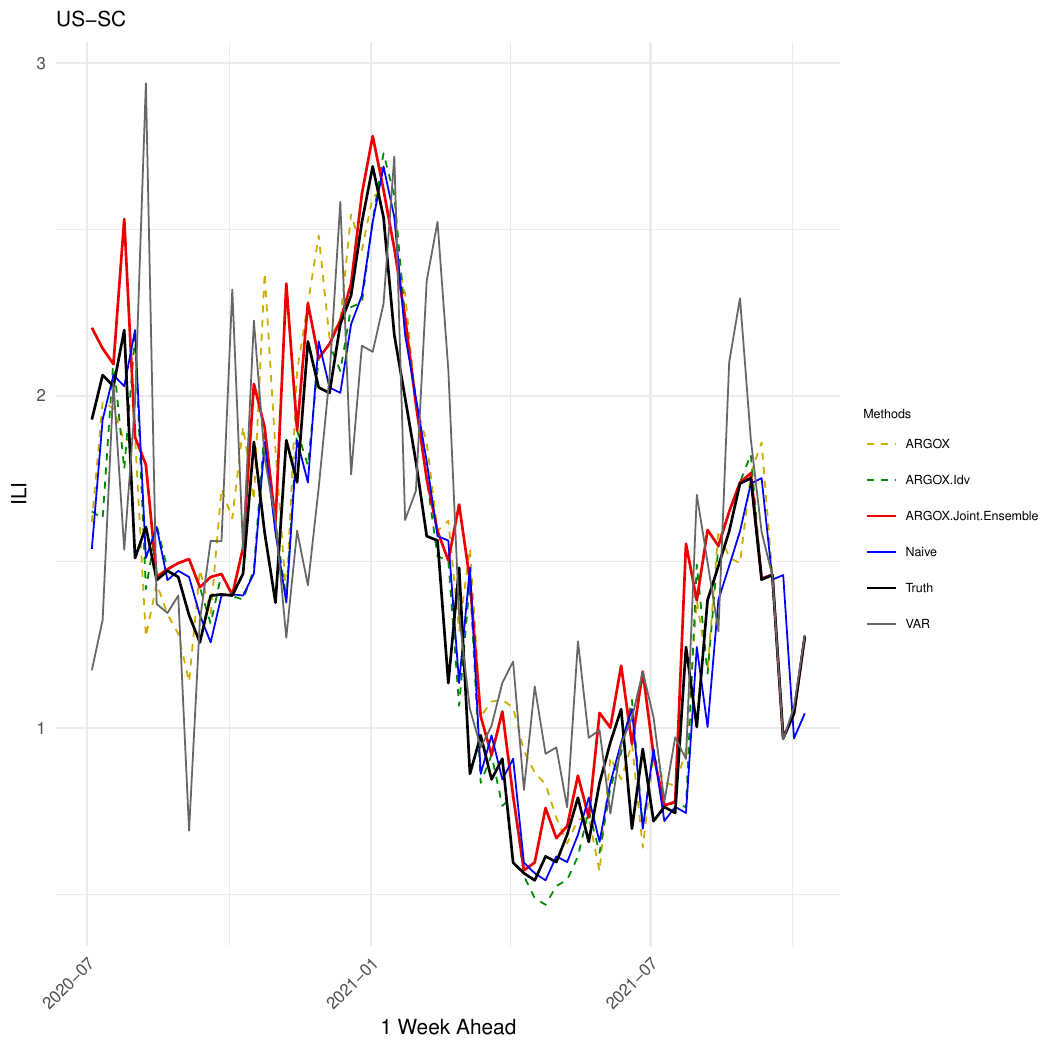} 
\caption{Plots of the \%ILI 1 week ahead estimates for South Carolina (SC). ARGOX is Ref \cite{ARGOX}.}
\end{figure}
\newpage

\begin{table}[ht]
\centering
\begin{tabular}{|c|c|c|c|}
  \hline
Methods & RMSE & MAE & Correlation \\ \hline
  Naive & 
0.17 & 0.14 & 0.71 \\ 
    \hline VAR &0.26 & 0.17 & 0.56 \\ 
    \hline Ref \cite{ARGOX} & 0.16 & 0.12 & 0.75 \\ 
    \hline ARGOX-Idv & 0.16 & 0.13 & 0.77 \\ 
    \hline ARGOX-Joint-Ensemble & 0.14 & 0.10 & 0.91 \\ 
   \hline
\end{tabular}
\caption{Comparison of different methods for state-level ILI 1 week ahead incremental death in South Dakota (SD). The MSE, MAE, and correlation are reported and best performed method is highlighted in boldface.} 
\end{table}

\begin{figure}[!h] 
  \centering 
\includegraphics[width=0.6\linewidth, page=41]{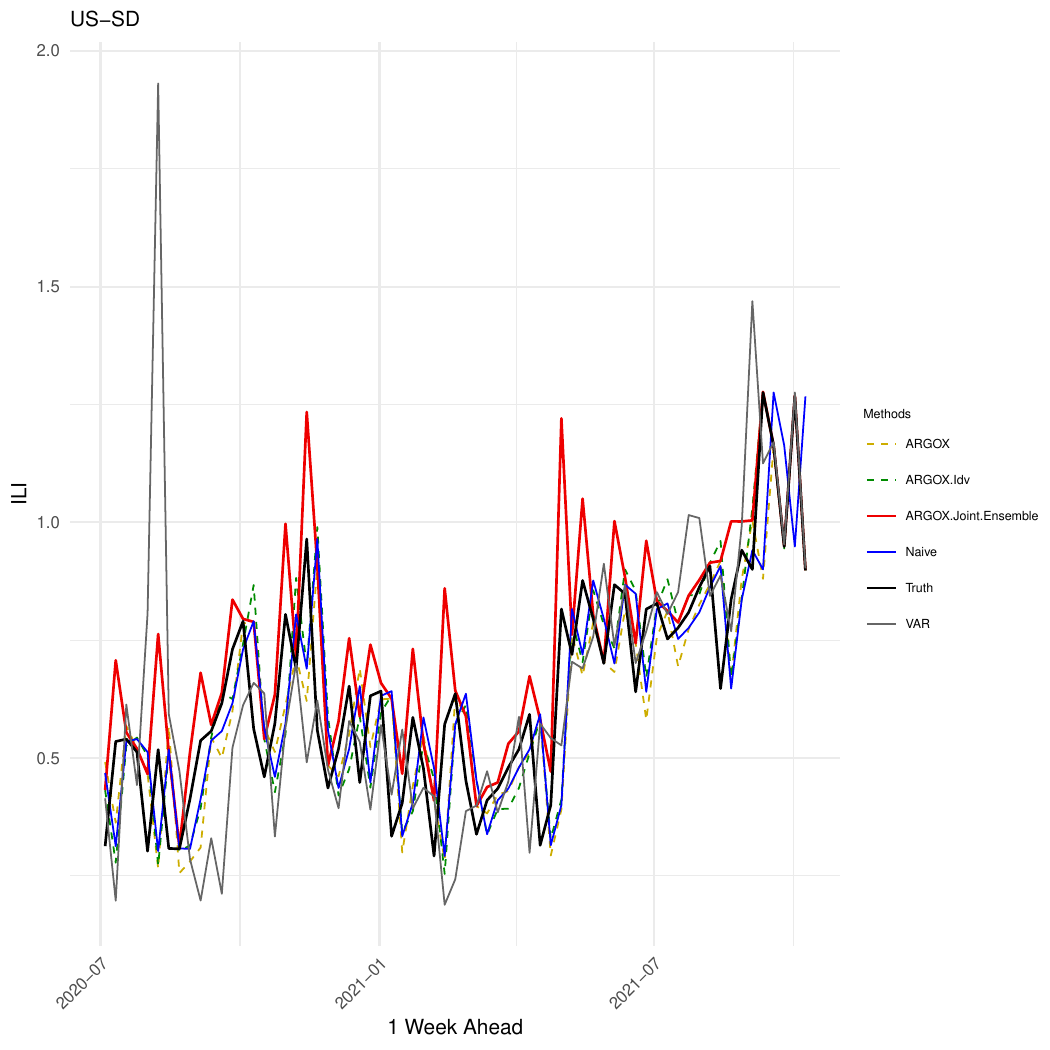} 
\caption{Plots of the \%ILI 1 week ahead estimates for South Dakota (SD). ARGOX is Ref \cite{ARGOX}.}
\end{figure}
\newpage

\begin{table}[ht]
\centering
\begin{tabular}{|c|c|c|c|}
  \hline
Methods & RMSE & MAE & Correlation \\ \hline
  Naive & 
0.21 & 0.14 & 0.91 \\ 
    \hline VAR &0.40 & 0.29 & 0.74 \\ 
    \hline Ref \cite{ARGOX} & 0.19 & 0.14 & 0.92 \\ 
    \hline ARGOX-Idv & 0.17 & 0.12 & 0.94 \\ 
    \hline ARGOX-Joint-Ensemble & 0.13 & 0.09 & 0.99 \\ 
   \hline
\end{tabular}
\caption{Comparison of different methods for state-level ILI 1 week ahead incremental death in Tennessee (TN). The MSE, MAE, and correlation are reported and best performed method is highlighted in boldface.} 
\end{table}

\begin{figure}[!h] 
  \centering 
\includegraphics[width=0.6\linewidth, page=42]{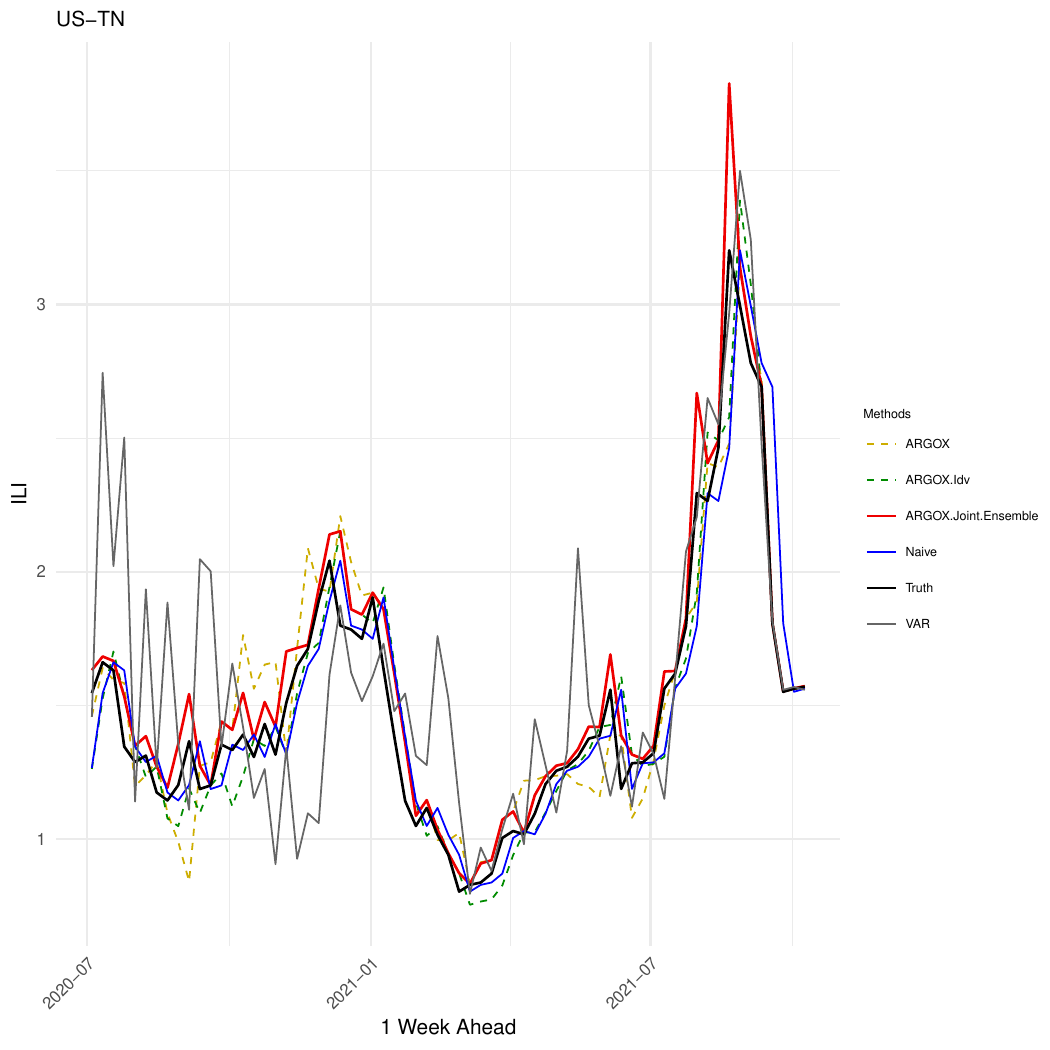} 
\caption{Plots of the \%ILI 1 week ahead estimates for Tennessee (TN). ARGOX is Ref \cite{ARGOX}.}
\end{figure}
\newpage

\begin{table}[ht]
\centering
\begin{tabular}{|c|c|c|c|}
  \hline
Methods & RMSE & MAE & Correlation \\ \hline
  Naive & 
0.29 & 0.20 & 0.87 \\ 
    \hline VAR &0.33 & 0.26 & 0.85 \\ 
    \hline Ref \cite{ARGOX} & 0.29 & 0.21 & 0.87 \\ 
    \hline ARGOX-Idv & 0.27 & 0.18 & 0.89 \\ 
    \hline ARGOX-Joint-Ensemble & 0.24 & 0.14 & 0.95 \\ 
   \hline
\end{tabular}
\caption{Comparison of different methods for state-level ILI 1 week ahead incremental death in Texas (TX). The MSE, MAE, and correlation are reported and best performed method is highlighted in boldface.} 
\end{table}

\begin{figure}[!h] 
  \centering 
\includegraphics[width=0.6\linewidth, page=43]{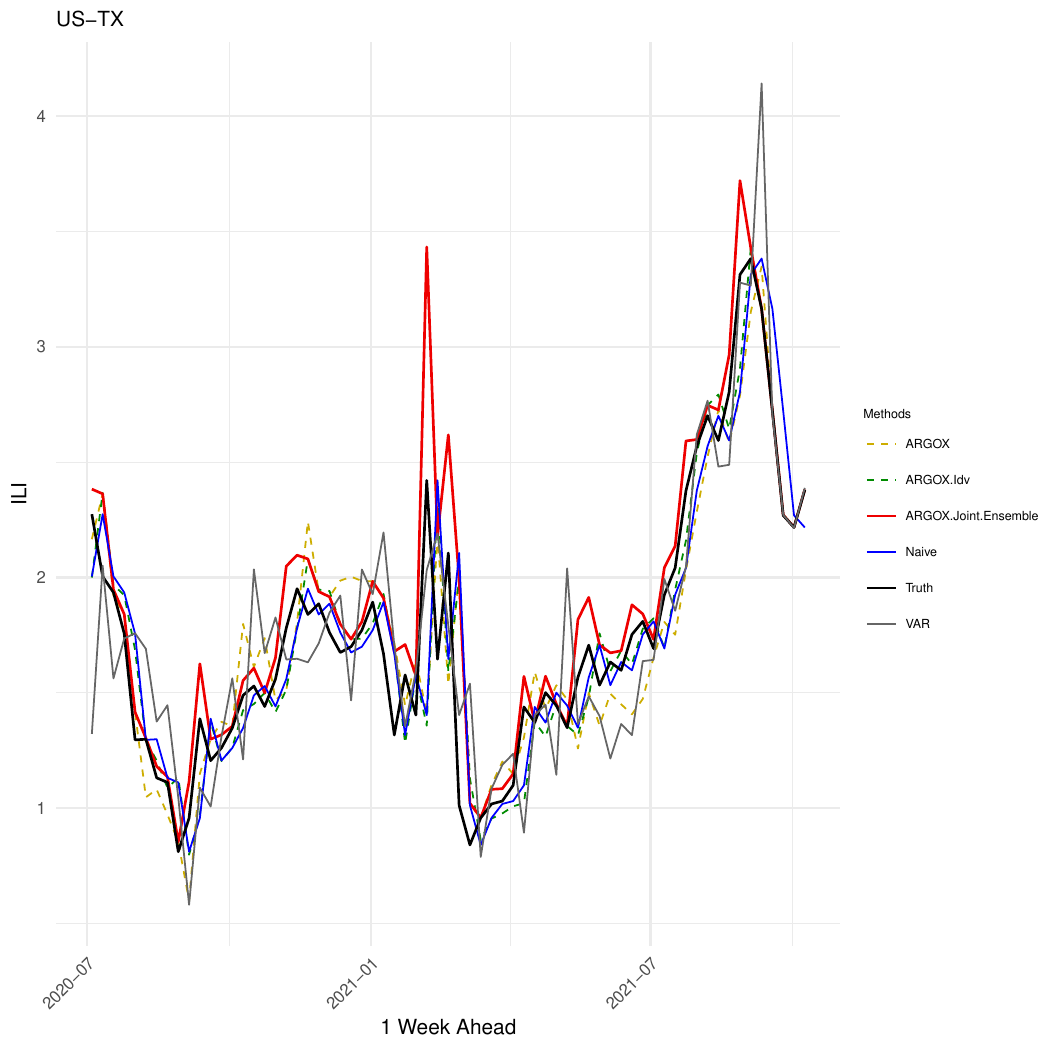} 
\caption{Plots of the \%ILI 1 week ahead estimates for Texas (TX). ARGOX is Ref \cite{ARGOX}.}
\end{figure}
\newpage

\begin{table}[ht]
\centering
\begin{tabular}{|c|c|c|c|}
  \hline
Methods & RMSE & MAE & Correlation \\ \hline
  Naive & 
0.18 & 0.14 & 0.92 \\ 
    \hline VAR &0.29 & 0.22 & 0.79 \\ 
    \hline Ref \cite{ARGOX} & 0.21 & 0.16 & 0.89 \\ 
    \hline ARGOX-Idv & 0.17 & 0.13 & 0.93 \\ 
    \hline ARGOX-Joint-Ensemble & 0.15 & 0.11 & 0.98 \\ 
   \hline
\end{tabular}
\caption{Comparison of different methods for state-level ILI 1 week ahead incremental death in Utah (UT). The MSE, MAE, and correlation are reported and best performed method is highlighted in boldface.} 
\end{table}

\begin{figure}[!h] 
  \centering 
\includegraphics[width=0.6\linewidth, page=44]{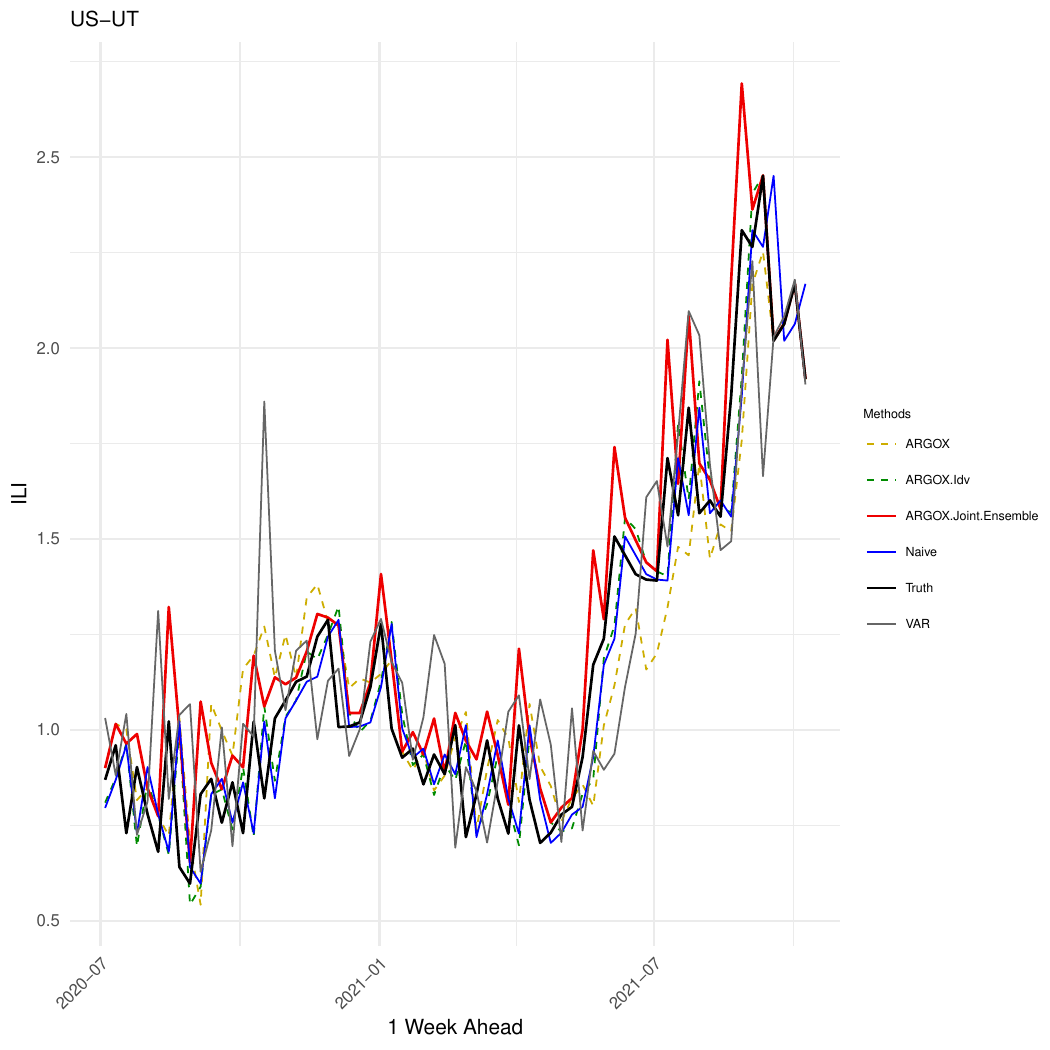} 
\caption{Plots of the \%ILI 1 week ahead estimates for Utah (UT). ARGOX is Ref \cite{ARGOX}.}
\end{figure}
\newpage

\begin{table}[ht]
\centering
\begin{tabular}{|c|c|c|c|}
  \hline
Methods & RMSE & MAE & Correlation \\ \hline
  Naive & 
0.23 & 0.17 & 0.64 \\ 
    \hline VAR &0.26 & 0.20 & 0.57 \\ 
    \hline Ref \cite{ARGOX} & 0.23 & 0.17 & 0.64 \\ 
    \hline ARGOX-Idv & 0.24 & 0.17 & 0.67 \\ 
    \hline ARGOX-Joint-Ensemble & 0.20 & 0.13 & 0.91 \\ 
   \hline
\end{tabular}
\caption{Comparison of different methods for state-level ILI 1 week ahead incremental death in Vermont (VT). The MSE, MAE, and correlation are reported and best performed method is highlighted in boldface.} 
\end{table}

\begin{figure}[!h] 
  \centering 
\includegraphics[width=0.6\linewidth, page=45]{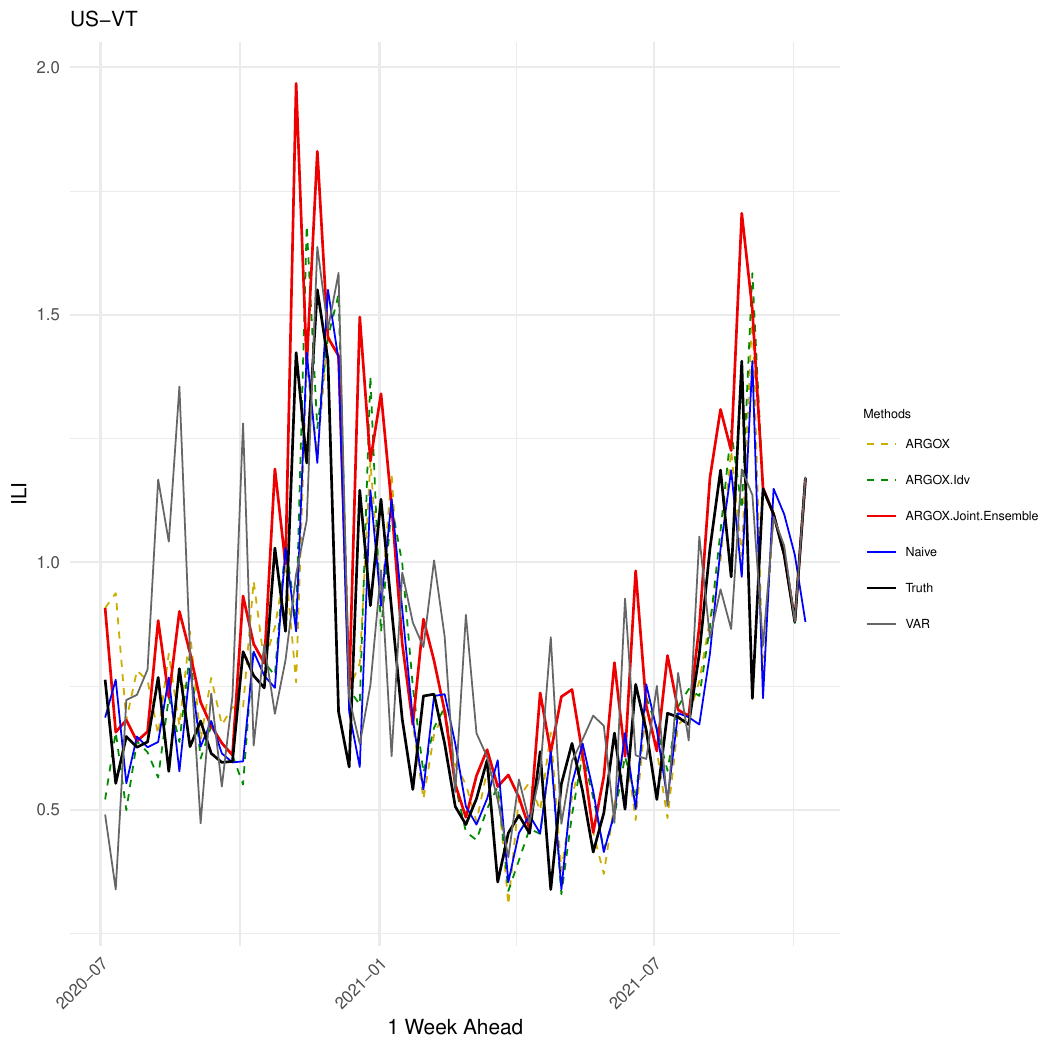} 
\caption{Plots of the \%ILI 1 week ahead estimates for Vermont (VT). ARGOX is Ref \cite{ARGOX}.}
\end{figure}
\newpage

\begin{table}[ht]
\centering
\begin{tabular}{|c|c|c|c|}
  \hline
Methods & RMSE & MAE & Correlation \\ \hline
  Naive & 
0.49 & 0.17 & 0.67 \\ 
    \hline VAR &0.20 & 0.15 & 0.94 \\ 
    \hline Ref \cite{ARGOX} & 0.14 & 0.11 & 0.98 \\ 
    \hline ARGOX-Idv & 0.11 & 0.08 & 0.98 \\ 
    \hline ARGOX-Joint-Ensemble & 0.09 & 0.06 & 0.99 \\ 
   \hline
\end{tabular}
\caption{Comparison of different methods for state-level ILI 1 week ahead incremental death in Virginia (VA). The MSE, MAE, and correlation are reported and best performed method is highlighted in boldface.} 
\end{table}

\begin{figure}[!h] 
  \centering 
\includegraphics[width=0.6\linewidth, page=46]{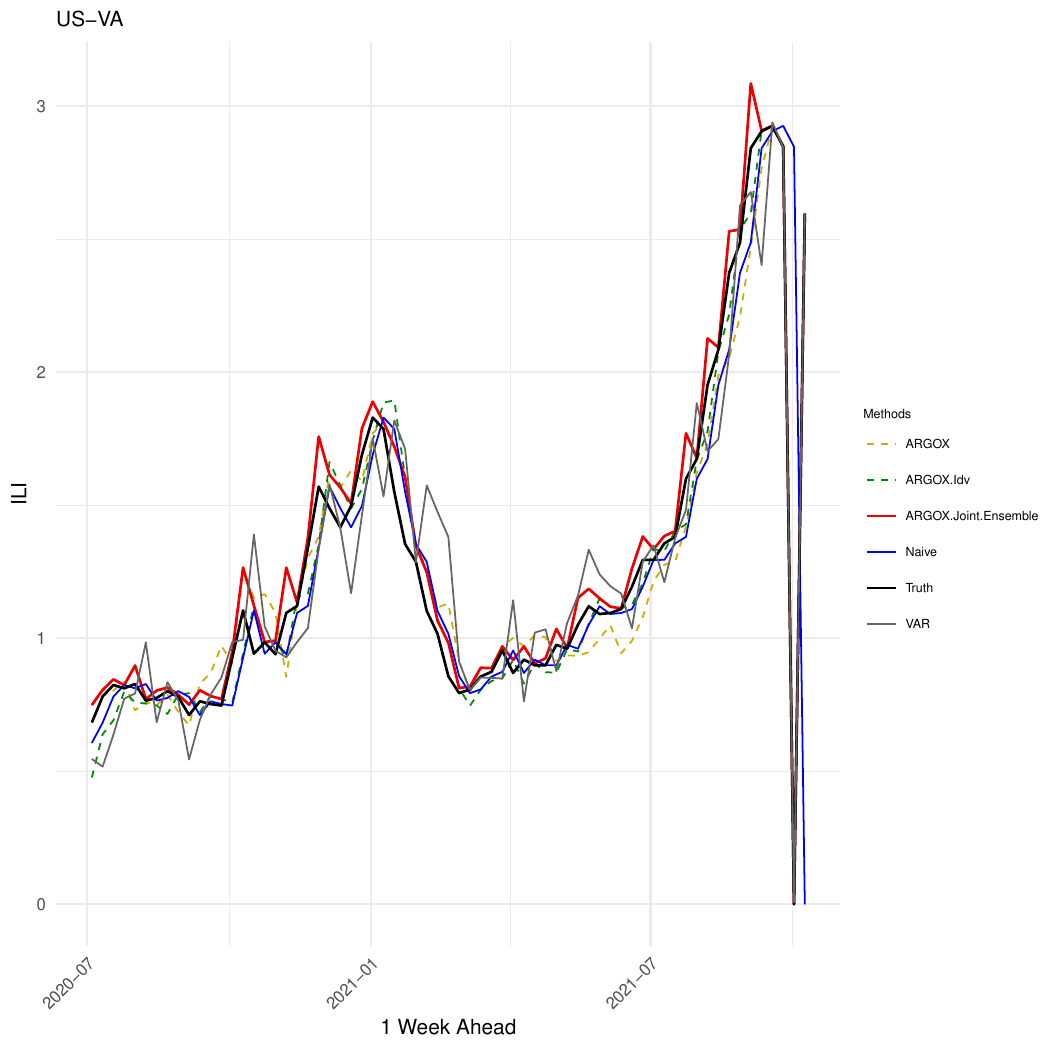} 
\caption{Plots of the \%ILI 1 week ahead estimates for Virginia (VA). ARGOX is Ref \cite{ARGOX}.}
\end{figure}
\newpage

\begin{table}[ht]
\centering
\begin{tabular}{|c|c|c|c|}
  \hline
Methods & RMSE & MAE & Correlation \\ \hline
  Naive & 
0.11 & 0.09 & 0.92 \\ 
    \hline VAR &0.24 & 0.18 & 0.69 \\ 
    \hline Ref \cite{ARGOX} & 0.12 & 0.10 & 0.89 \\ 
    \hline ARGOX-Idv & 0.11 & 0.08 & 0.92 \\ 
    \hline ARGOX-Joint-Ensemble & 0.08 & 0.06 & 0.98 \\ 
   \hline
\end{tabular}
\caption{Comparison of different methods for state-level ILI 1 week ahead incremental death in Washington (WA). The MSE, MAE, and correlation are reported and best performed method is highlighted in boldface.} 
\end{table}

\begin{figure}[!h] 
  \centering 
\includegraphics[width=0.6\linewidth, page=47]{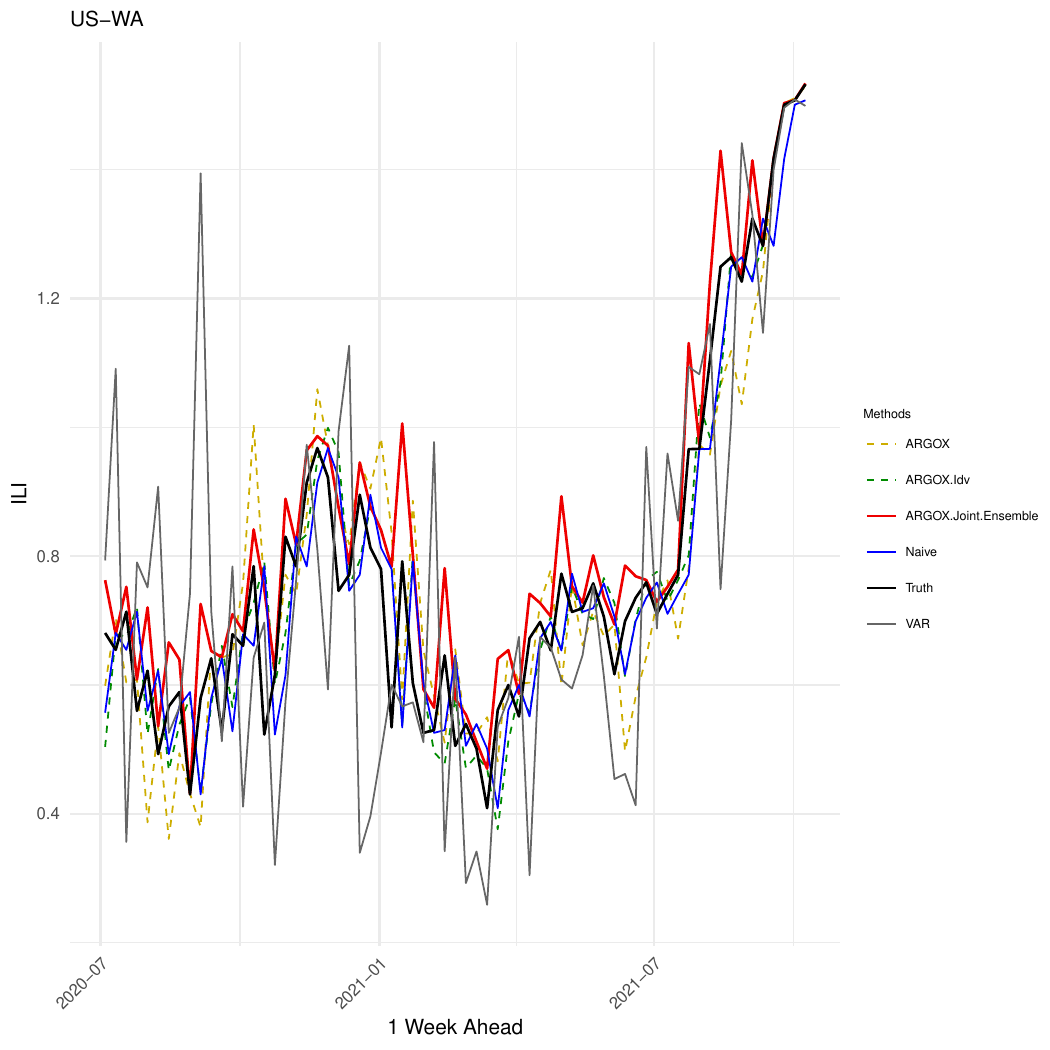} 
\caption{Plots of the \%ILI 1 week ahead estimates for Washington (WA). ARGOX is Ref \cite{ARGOX}.}
\end{figure}
\newpage

\begin{table}[ht]
\centering
\begin{tabular}{|c|c|c|c|}
  \hline
Methods & RMSE & MAE & Correlation \\ \hline
  Naive & 
0.54 & 0.30 & 0.29 \\ 
    \hline VAR &0.54 & 0.35 & 0.12 \\ 
    \hline Ref \cite{ARGOX} & 0.45 & 0.25 & 0.41 \\ 
    \hline ARGOX-Idv & 0.47 & 0.26 & 0.34 \\ 
    \hline ARGOX-Joint-Ensemble & 0.42 & 0.21 & 0.92 \\ 
   \hline
\end{tabular}
\caption{Comparison of different methods for state-level ILI 1 week ahead incremental death in West Virginia (WV). The MSE, MAE, and correlation are reported and best performed method is highlighted in boldface.} 
\end{table}

\begin{figure}[!h] 
  \centering 
\includegraphics[width=0.6\linewidth, page=48]{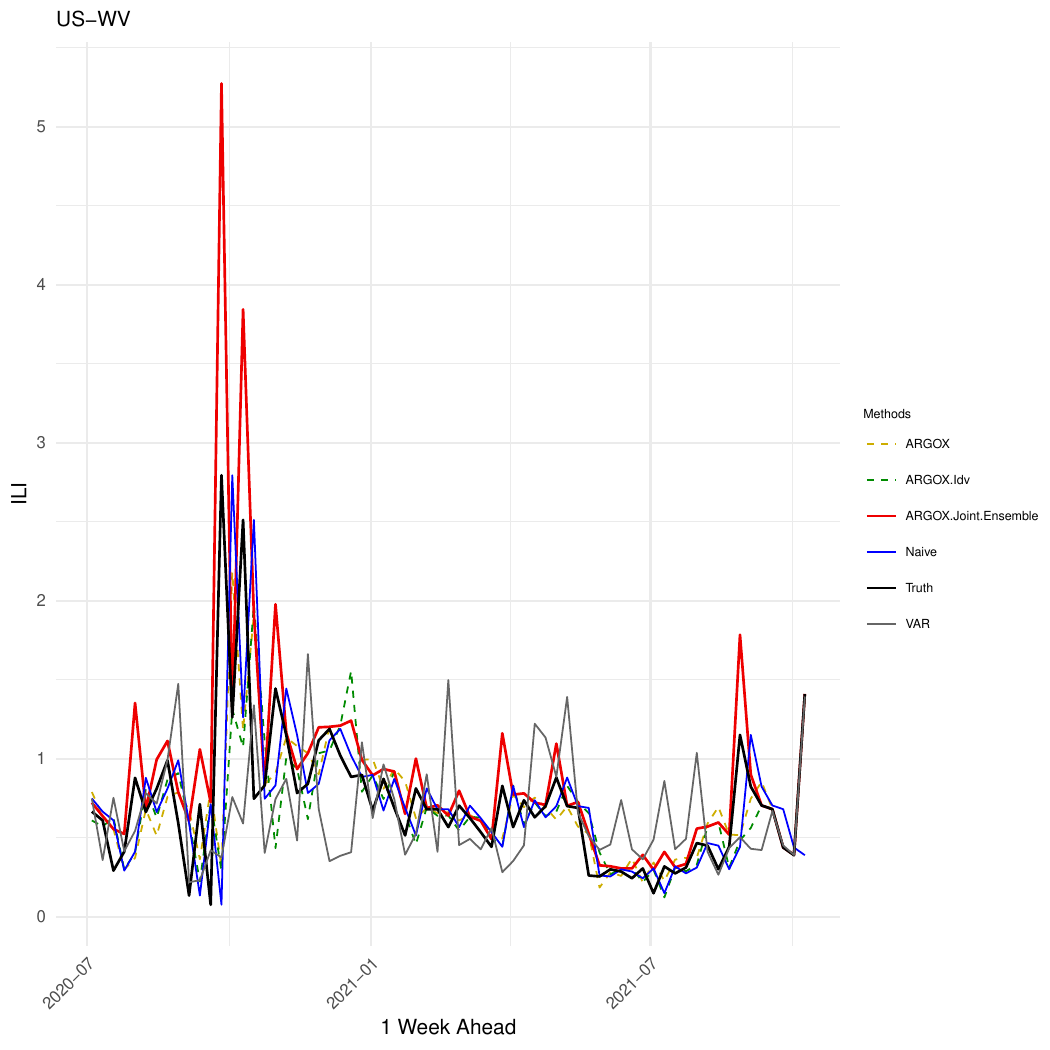} 
\caption{Plots of the \%ILI 1 week ahead estimates for West Virginia (WV). ARGOX is Ref \cite{ARGOX}.}
\end{figure}
\newpage

\begin{table}[ht]
\centering
\begin{tabular}{|c|c|c|c|}
  \hline
Methods & RMSE & MAE & Correlation \\ \hline
  Naive & 
0.11 & 0.08 & 0.91 \\ 
    \hline VAR &0.18 & 0.13 & 0.82 \\ 
    \hline Ref \cite{ARGOX} & 0.14 & 0.11 & 0.86 \\ 
    \hline ARGOX-Idv & 0.10 & 0.07 & 0.94 \\ 
    \hline ARGOX-Joint-Ensemble & 0.08 & 0.05 & 0.98 \\ 
   \hline
\end{tabular}
\caption{Comparison of different methods for state-level ILI 1 week ahead incremental death in Wisconsin (WI). The MSE, MAE, and correlation are reported and best performed method is highlighted in boldface.} 
\end{table}

\begin{figure}[!h] 
  \centering 
\includegraphics[width=0.6\linewidth, page=49]{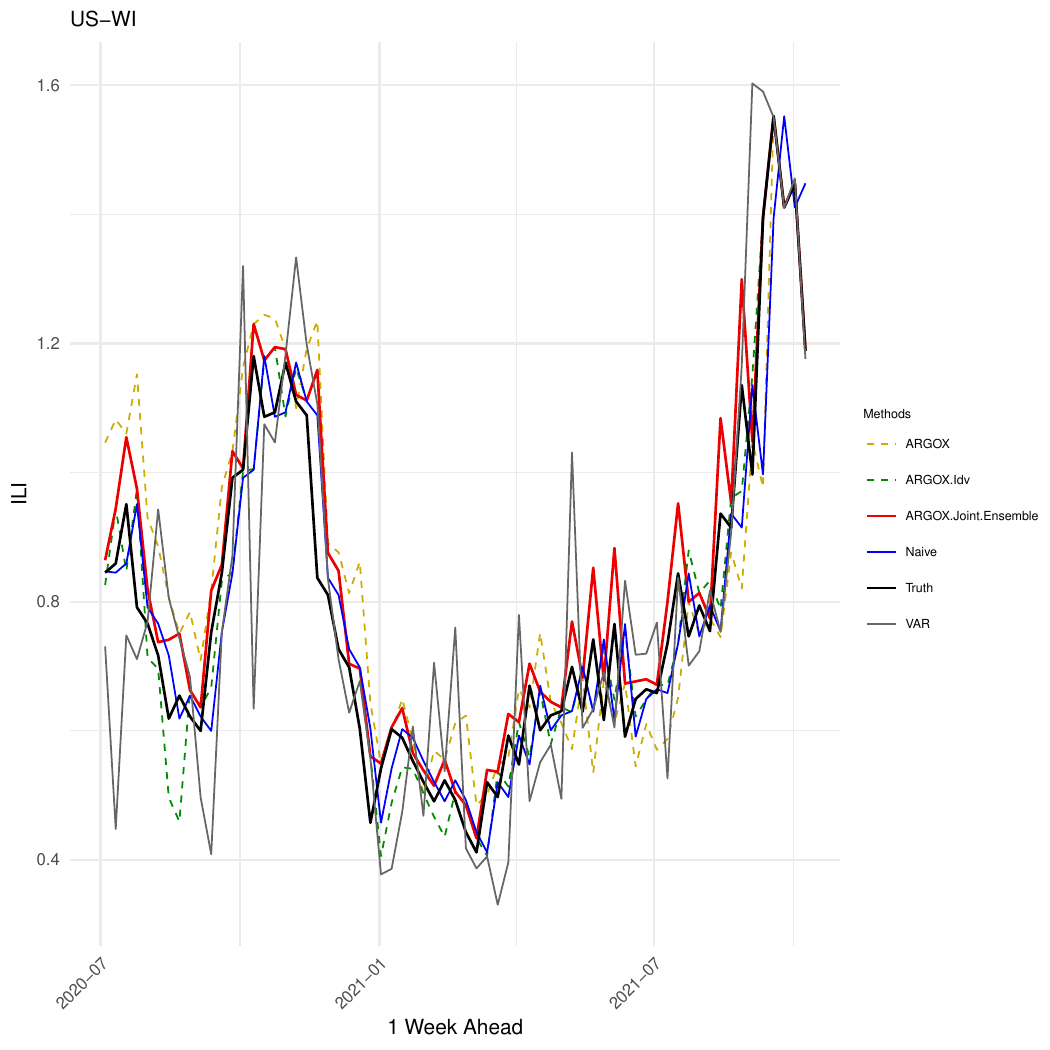} 
\caption{Plots of the \%ILI 1 week ahead estimates for Wisconsin (WI). ARGOX is Ref \cite{ARGOX}.}
\end{figure}
\newpage

\begin{table}[ht]
\centering
\begin{tabular}{|c|c|c|c|}
  \hline
Methods & RMSE & MAE & Correlation \\ \hline
  Naive & 
0.34 & 0.25 & 0.78 \\ 
    \hline VAR &0.41 & 0.29 & 0.71 \\ 
    \hline Ref \cite{ARGOX} & 0.32 & 0.23 & 0.82 \\ 
    \hline ARGOX-Idv & 0.32 & 0.23 & 0.81 \\ 
    \hline ARGOX-Joint-Ensemble & 0.28 & 0.19 & 0.94 \\ 
   \hline
\end{tabular}
\caption{Comparison of different methods for state-level ILI 1 week ahead incremental death in Wyoming (WY). The MSE, MAE, and correlation are reported and best performed method is highlighted in boldface.} 
\end{table}

\begin{figure}[!h] 
  \centering 
\includegraphics[width=0.6\linewidth, page=50]{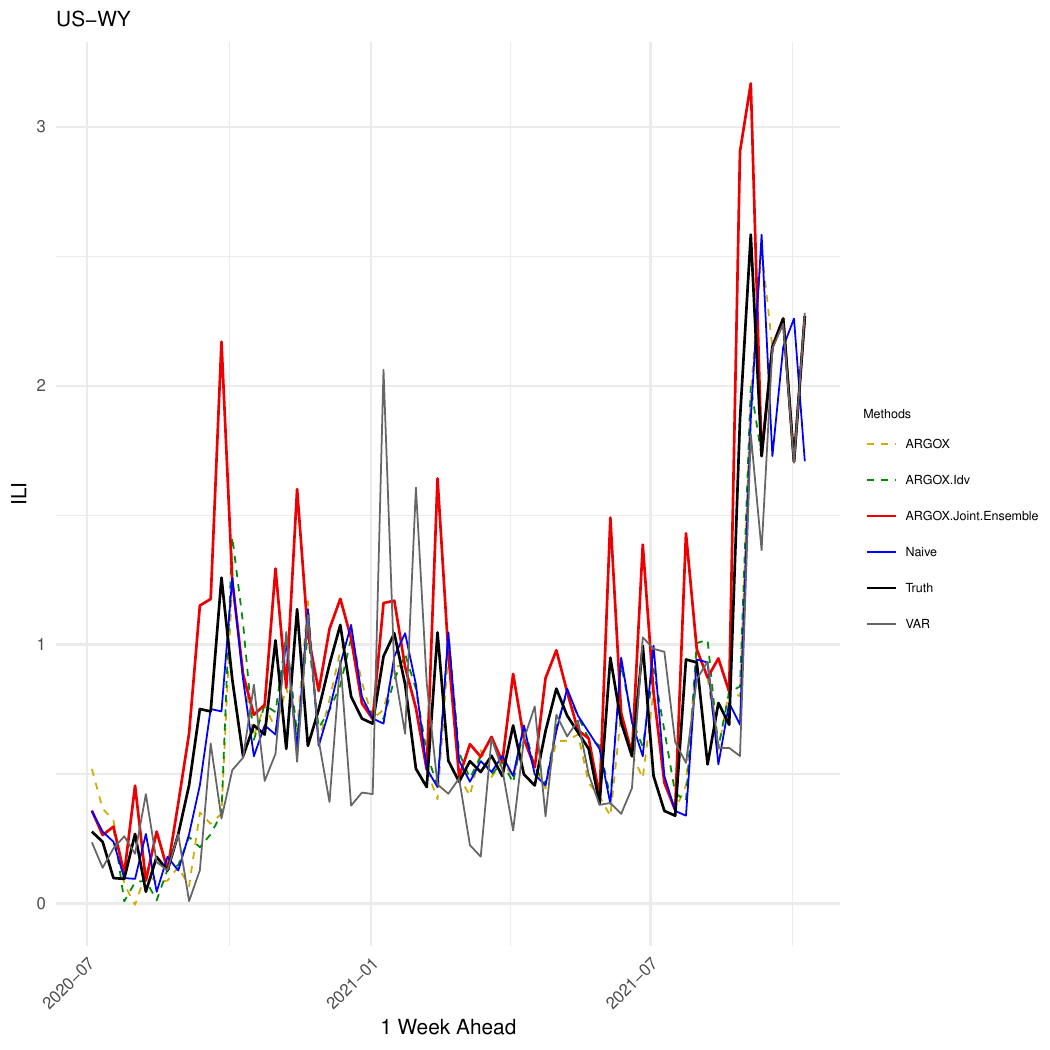} 
\caption{Plots of the \%ILI 1 week ahead estimates for Wyoming (WY). ARGOX is Ref \cite{ARGOX}.}
\end{figure}
\newpage

\begin{table}[ht]
\centering
\begin{tabular}{|c|c|c|c|}
  \hline
Methods & RMSE & MAE & Correlation \\ \hline
  Naive & 
0.11 & 0.08 & 0.96 \\ 
    \hline VAR &0.19 & 0.14 & 0.89 \\ 
    \hline Ref \cite{ARGOX} & 0.14 & 0.11 & 0.93 \\ 
    \hline ARGOX-Idv & 0.14 & 0.11 & 0.93 \\ 
    \hline ARGOX-Joint-Ensemble & 0.14 & 0.11 & 0.97 \\ 
   \hline
\end{tabular}
\caption{Comparison of different methods for state-level ILI 1 week ahead incremental death in US-NYC. The MSE, MAE, and correlation are reported and best performed method is highlighted in boldface.} 
\label{tab:State_Ours_ILI_NYC}
\end{table}

\begin{figure}[!h] 
  \centering 
\includegraphics[width=0.6\linewidth, page=51]{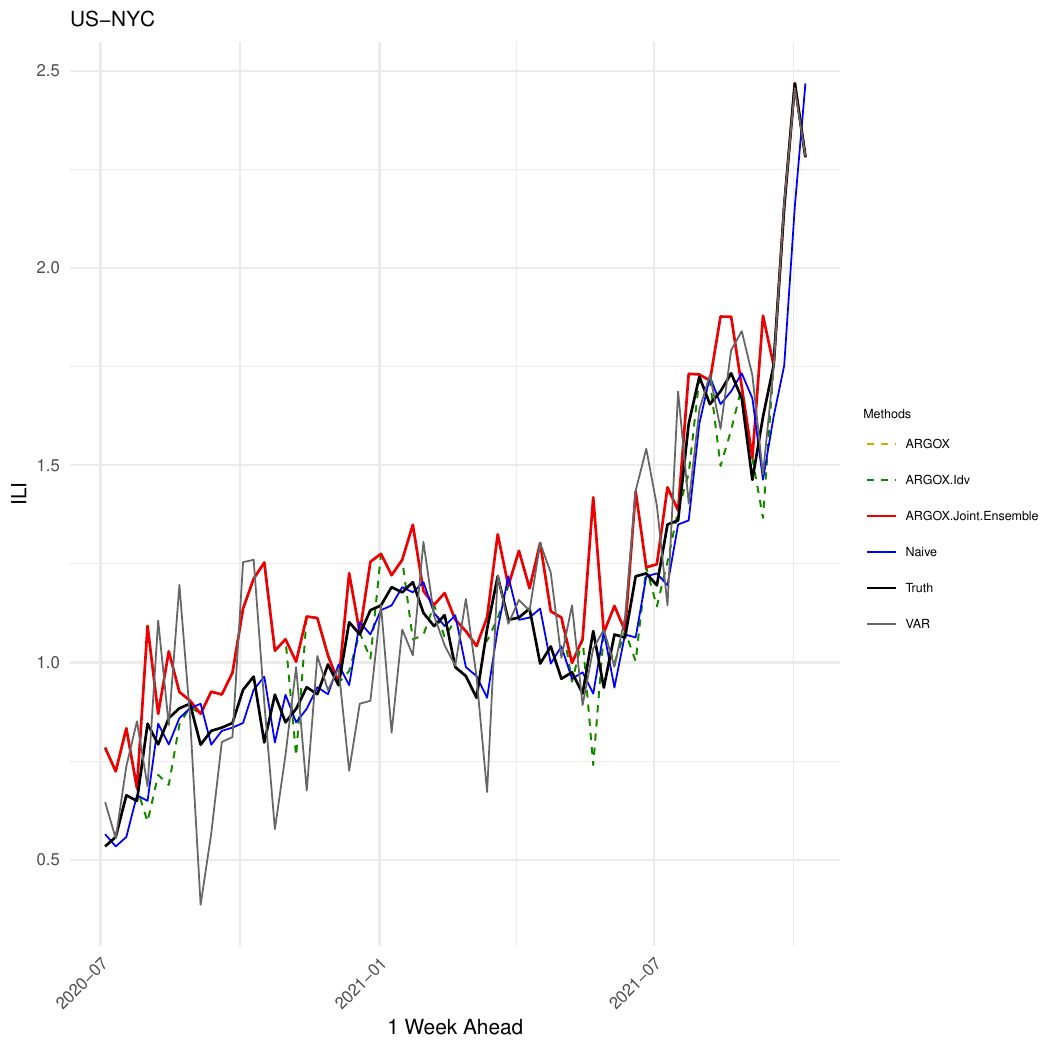} 
\caption{Plots of the \%ILI 1 week ahead estimates for New York City (NYC). ARGOX is Ref \cite{ARGOX}.}
\label{fig:State_Ours_ILI_NYC}
\end{figure}
\newpage

\restoregeometry


\begin{thebibliography}{10}
\bibitem{hassan_2022}
Jennifer Hassan.
\newblock What is `flurona'? coronavirus and influenza co-infections reported
  as omicron surges.
\newblock {\em The Washington Post}, Jan 2022.

\bibitem{berger2022twindemic}
Miriam Berger.
\newblock The world has avoided a `twindemic`, but as flu cases rise alongside
  covid, experts fear that could change.
\newblock {\em The Washington Post}, 2022.

\bibitem{rubin2020happens}
Rita Rubin.
\newblock What happens when covid-19 collides with flu season?
\newblock {\em Jama}, 324(10):923--925, 2020.

\bibitem{CDC_Flu_Stat}
{Center for Disease Control and Prevention}.
\newblock Disease burden of flus, 2021.
\newblock \url{https://www.cdc.gov/flu/about/burden/index.html}, Last accessed
  on 2021-12-30.

\bibitem{brooks2018nonmechanistic}
Logan~C Brooks, David~C Farrow, Sangwon Hyun, Ryan~J Tibshirani, and Roni
  Rosenfeld.
\newblock Nonmechanistic forecasts of seasonal influenza with iterative
  one-week-ahead distributions.
\newblock {\em PLoS computational biology}, 14(6):e1006134, 2018.

\bibitem{osthus2019dynamic}
Dave Osthus, James Gattiker, Reid Priedhorsky, and Sara~Y Del~Valle.
\newblock Dynamic bayesian influenza forecasting in the united states with
  hierarchical discrepancy (with discussion).
\newblock {\em Bayesian Analysis}, 14(1):261--312, 2019.

\bibitem{shaman2012forecasting}
Jeffrey Shaman and Alicia Karspeck.
\newblock Forecasting seasonal outbreaks of influenza.
\newblock {\em Proceedings of the National Academy of Sciences},
  109(50):20425--20430, 2012.

\bibitem{tizzoni2012real}
Michele Tizzoni, Paolo Bajardi, Chiara Poletto, Jos{\'e}~J Ramasco, Duygu
  Balcan, Bruno Gon{\c{c}}alves, Nicola Perra, Vittoria Colizza, and Alessandro
  Vespignani.
\newblock Real-time numerical forecast of global epidemic spreading: case study
  of 2009 a/h1n1pdm.
\newblock {\em BMC medicine}, 10(1):1--31, 2012.

\bibitem{yang2014comparison}
Wan Yang, Alicia Karspeck, and Jeffrey Shaman.
\newblock Comparison of filtering methods for the modeling and retrospective
  forecasting of influenza epidemics.
\newblock {\em PLoS computational biology}, 10(4):e1003583, 2014.

\bibitem{yang2015inference}
Wan Yang, Marc Lipsitch, and Jeffrey Shaman.
\newblock Inference of seasonal and pandemic influenza transmission dynamics.
\newblock {\em Proceedings of the National Academy of Sciences},
  112(9):2723--2728, 2015.

\bibitem{reich2019collaborative}
Nicholas~G Reich, Logan~C Brooks, Spencer~J Fox, Sasikiran Kandula, Craig~J
  McGowan, Evan Moore, Dave Osthus, Evan~L Ray, Abhinav Tushar, Teresa~K
  Yamana, et~al.
\newblock A collaborative multiyear, multimodel assessment of seasonal
  influenza forecasting in the united states.
\newblock {\em Proceedings of the National Academy of Sciences},
  116(8):3146--3154, 2019.

\bibitem{wang2019defsi}
Lijing Wang, Jiangzhuo Chen, and Madhav Marathe.
\newblock Defsi: Deep learning based epidemic forecasting with synthetic
  information.
\newblock In {\em Proceedings of the AAAI Conference on Artificial
  Intelligence}, volume~33, pages 9607--9612, 2019.

\bibitem{venna2018novel}
Siva~R Venna, Amirhossein Tavanaei, Raju~N Gottumukkala, Vijay~V Raghavan,
  Anthony~S Maida, and Stephen Nichols.
\newblock A novel data-driven model for real-time influenza forecasting.
\newblock {\em IEEE Access}, 7:7691--7701, 2018.

\bibitem{shaman2010absolute}
Jeffrey Shaman, Virginia~E Pitzer, C{\'e}cile Viboud, Bryan~T Grenfell, and
  Marc Lipsitch.
\newblock Absolute humidity and the seasonal onset of influenza in the
  continental united states.
\newblock {\em PLoS biology}, 8(2):e1000316, 2010.

\bibitem{tamerius2013environmental}
James~D Tamerius, Jeffrey Shaman, Wladmir~J Alonso, Kimberly Bloom-Feshbach,
  Christopher~K Uejio, Andrew Comrie, and C{\'e}cile Viboud.
\newblock Environmental predictors of seasonal influenza epidemics across
  temperate and tropical climates.
\newblock {\em PLoS pathogens}, 9(3):e1003194, 2013.

\bibitem{paul2014twitter}
Michael~J Paul, Mark Dredze, and David Broniatowski.
\newblock Twitter improves influenza forecasting.
\newblock {\em PLoS currents}, 6, 2014.

\bibitem{signorini2011use}
Alessio Signorini, Alberto~Maria Segre, and Philip~M Polgreen.
\newblock The use of twitter to track levels of disease activity and public
  concern in the us during the influenza a h1n1 pandemic.
\newblock {\em PloS one}, 6(5):e19467, 2011.

\bibitem{mciver2014wikipedia}
David~J McIver and John~S Brownstein.
\newblock Wikipedia usage estimates prevalence of influenza-like illness in the
  united states in near real-time.
\newblock {\em PLoS computational biology}, 10(4):e1003581, 2014.

\bibitem{generous2014global}
Nicholas Generous, Geoffrey Fairchild, Alina Deshpande, Sara~Y Del~Valle, and
  Reid Priedhorsky.
\newblock Global disease monitoring and forecasting with wikipedia.
\newblock {\em PLoS computational biology}, 10(11):e1003892, 2014.

\bibitem{GFT_2008}
Jeremy Ginsberg, Matthew Mohebbi, Rajan Patel, Lynnette Brammer, Mark
  Smolinski, and Larry Brilliant.
\newblock Detecting influenza epidemics using search engine query data.
\newblock {\em Nature}, 457:1012--4, 12 2008.

\bibitem{yang2015accurate}
Shihao Yang, Mauricio Santillana, and S.~C. Kou.
\newblock Accurate estimation of influenza epidemics using google search data
  via argo.
\newblock {\em Proceedings of the National Academy of Sciences},
  112(47):14473--14478, 2015.

\bibitem{ARGO2_Regional}
Shaoyang Ning and Shihao Yang.
\newblock Accurate regional influenza epidemics tracking using internet search
  data.
\newblock {\em Scientific Reports}, 9:5238, 03 2019.

\bibitem{ARGOX}
Shihao Yang, Shaoyang Ning, and S.~C. Kou.
\newblock Use internet search data to accurately track state level influenza
  epidemics.
\newblock {\em Sci Rep}, 11(4023), 2021.

\bibitem{polgreen2008using}
Philip~M Polgreen, Yiling Chen, David~M Pennock, Forrest~D Nelson, and Robert~A
  Weinstein.
\newblock Using internet searches for influenza surveillance.
\newblock {\em Clinical infectious diseases}, 47(11):1443--1448, 2008.

\bibitem{yuan2013monitoring}
Qingyu Yuan, Elaine~O Nsoesie, Benfu Lv, Geng Peng, Rumi Chunara, and John~S
  Brownstein.
\newblock Monitoring influenza epidemics in china with search query from baidu.
\newblock {\em PloS one}, 8(5):e64323, 2013.

\bibitem{Delphi_KF}
Maria Jahja, David Farrow, Roni Rosenfeld, and Ryan~J Tibshirani.
\newblock Kalman filter, sensor fusion, and constrained regression:
  Equivalences and insights.
\newblock In H.~Wallach, H.~Larochelle, A.~Beygelzimer, F.~d\textquotesingle
  Alch\'{e}-Buc, E.~Fox, and R.~Garnett, editors, {\em Advances in Neural
  Information Processing Systems}, volume~32. Curran Associates, Inc., 2019.

\bibitem{DeepCOVID_GT}
Alexander Rodriguez, Anika Tabassum, Jiaming Cui, Jiajia Xie, Javen Ho, Pulak
  Agarwal, Bijaya Adhikari, and B.~Aditya Prakash.
\newblock Deepcovid: An operational deep learning-driven framework for
  explainable real-time covid-19 forecasting.
\newblock {\em medRxiv}, 2020.

\bibitem{UCSB_attention}
Xiaoyong Jin, Yu-Xiang Wang, and Xifeng Yan.
\newblock Inter-series attention model for covid-19 forecasting, 2020.

\bibitem{COVID19Simulator}
J.~Chhatwal, O.~Dalgic, P.~Mueller, M.~Adee, Y.~Xiao, M.A. Ladd, B.P. Linas,
  and T.~Ayer.
\newblock Pin68 covid-19 simulator: An interactive tool to inform covid-19
  intervention policy decisions in the united states.
\newblock {\em Value in health : the journal of the International Society for
  Pharmacoeconomics and Outcomes Research}, 23(12):S556—S556, December 2020.

\bibitem{UCLA_sueir}
Difan Zou, Lingxiao Wang, Pan Xu, Jinghui Chen, Weitong Zhang, and Quanquan Gu.
\newblock Epidemic model guided machine learning for covid-19 forecasts in the
  united states.
\newblock {\em medRxiv}, 2020.

\bibitem{yang2021estimating}
Wan Yang, Sasikiran Kandula, Mary Huynh, Sharon~K Greene, Gretchen Van~Wye,
  Wenhui Li, Hiu~Tai Chan, Emily McGibbon, Alice Yeung, Don Olson, et~al.
\newblock Estimating the infection-fatality risk of sars-cov-2 in new york city
  during the spring 2020 pandemic wave: a model-based analysis.
\newblock {\em The Lancet Infectious Diseases}, 21(2):203--212, 2021.

\bibitem{CDC_Ensemble}
Evan~L Ray, Nutcha Wattanachit, Jarad Niemi, Abdul~Hannan Kanji, Katie House,
  Estee~Y Cramer, Johannes Bracher, Andrew Zheng, Teresa~K Yamana, Xinyue
  Xiong, et~al.
\newblock Ensemble forecasts of coronavirus disease 2019 (covid-19) in the us.
\newblock {\em MedRXiv}, 2020.

\bibitem{arokiaraj2020correlation}
Mark~Christopher Arokiaraj.
\newblock Correlation of influenza vaccination and influenza incidence on
  covid-19 severity.
\newblock {\em Available at SSRN 3572814}, 2020.

\bibitem{wang2021association}
Ruitong Wang, Min Liu, and Jue Liu.
\newblock The association between influenza vaccination and covid-19 and its
  outcomes: A systematic review and meta-analysis of observational studies.
\newblock {\em Vaccines}, 9(5):529, 2021.

\bibitem{huang2021universal}
Yi~Huang and Ishanu Chattopadhyay.
\newblock Universal risk phenotype of us counties for flu-like transmission to
  improve county-specific covid-19 incidence forecasts.
\newblock {\em PLoS computational biology}, 17(10):e1009363, 2021.

\bibitem{rodriguez2021steering}
Alexander Rodr{\'\i}guez, Nikhil Muralidhar, Bijaya Adhikari, Anika Tabassum,
  Naren Ramakrishnan, and B~Aditya Prakash.
\newblock Steering a historical disease forecasting model under a pandemic:
  Case of flu and covid-19.
\newblock In {\em Proceedings of the AAAI Conference on Artificial
  Intelligence}, volume~35, pages 4855--4863, 2021.

\bibitem{NYT_COVID}
{The New York Times}.
\newblock Coronavirus (covid-19) data in the united states, 2021.
\newblock \url{https://github.com/nytimes/COVID-19-data}, Last accessed on
  2021-04-03.

\bibitem{CDC_ForecastHub}
Estee~Y Cramer, Yuxin Huang, Yijin Wang, Evan~L Ray, Matthew Cornell, Johannes
  Bracher, Andrea Brennen, Alvaro~J Castro~Rivadeneira, Aaron Gerding, and
  Katie et~al House.
\newblock The united states covid-19 forecast hub dataset.
\newblock {\em medRxiv}, 2021.

\bibitem{JHU_Data}
Ensheng Dong, Hongru Du, and Lauren Gardner.
\newblock An interactive web-based dashboard to track covid-19 in real time.
\newblock {\em Lancet Infect Dis}, 20(5), 2020.

\bibitem{GoogleTrends}
Faq about google trends data.
\newblock
  \url{https://support.google.com/trends/answer/4365533?hl=en&ref_topic=6248052}.
\newblock Accessed: 2021-04-03.

\bibitem{ma2021covid}
Simin Ma and Shihao Yang.
\newblock Covid-19 forecasts using internet search information in the united
  states.
\newblock {\em arXiv preprint arXiv:2106.12160}, 2021.

\bibitem{ARGO}
Shihao Yang, Mauricio Santillana, and S.~C. Kou.
\newblock Accurate estimation of influenza epidemics using google search data
  via argo.
\newblock {\em Proceedings of the National Academy of Sciences},
  112(47):14473--14478, 2015.

\bibitem{UMASS}
Dan Sheldon and Casey Gibson.
\newblock Bayesian seird model, 2020.
\newblock Accessed = 2021-04-03.

\bibitem{MOBS_GLEAM}
Rebecca~K Borchering, C{\'e}cile Viboud, Emily Howerton, Claire~P Smith, Shaun
  Truelove, Michael~C Runge, Nicholas~G Reich, Lucie Contamin, John Levander,
  Jessica Salerno, et~al.
\newblock Modeling of future covid-19 cases, hospitalizations, and deaths, by
  vaccination rates and nonpharmaceutical intervention scenarios—united
  states, april--september 2021.
\newblock {\em Morbidity and Mortality Weekly Report}, 70(19):719, 2021.

\bibitem{LANL_GrowthRate}
L~Castro, G~Fairchild, I~Michaud, and D~Osthus.
\newblock Coffee: Covid-19 forecasts using fast evaluations and estimation,
  2020.

\bibitem{UA_EpiGro}
Joceline Lega.
\newblock Parameter estimation from icc curves.
\newblock {\em arXiv preprint arXiv:2005.08134}, 2020.

\bibitem{epiforecasts_MIT}
S~Abbott, J~Hellewell, RN~Thompson, K~Sherratt, HP~Gibbs, NI~Bosse, JD~Munday,
  S~Meakin, EL~Doughty, JY~Chun, YWD Chan, F~Finger, P~Campbell, A~Endo, CAB
  Pearson, A~Gimma, T~Russell, null null, S~Flasche, AJ~Kucharski, RM~Eggo, and
  S~Funk.
\newblock Estimating the time-varying reproduction number of sars-cov-2 using
  national and subnational case counts [version 1; peer review: awaiting peer
  review].
\newblock {\em Wellcome Open Research}, 5(112), 2020.

\bibitem{USC-SI_kJalpha}
Ajitesh Srivastava, Tianjian Xu, and Viktor~K Prasanna.
\newblock Fast and accurate forecasting of covid-19 deaths using the sikj
  $\alpha$ model.
\newblock {\em arXiv preprint arXiv:2007.05180}, 2020.

\bibitem{JHU_CSSE-DECOM}
Hamada~S Badr, Hongru Du, Maximilian Marshall, Ensheng Dong, Marietta~M Squire,
  and Lauren~M Gardner.
\newblock Association between mobility patterns and covid-19 transmission in
  the usa: a mathematical modelling study.
\newblock {\em The Lancet Infectious Diseases}, 20(11):1247--1254, 2020.

\bibitem{UVA-Ensemble}
Aniruddha Adiga, Lijing Wang, Benjamin Hurt, Akhil~Sai Peddireddy, Przemyslaw
  Porebski, Srinivasan Venkatramanan, Bryan Lewis, and Madhav Marathe.
\newblock All models are useful: Bayesian ensembling for robust high resolution
  covid-19 forecasting.
\newblock {\em medRxiv}, 2021.

\bibitem{CU-Select}
Teresa Yamana, Sen Pei, and Jeffrey Shaman.
\newblock Projection of covid-19 cases and deaths in the us as individual
  states re-open may 4, 2020.
\newblock {\em MedRxiv}, 2020.

\bibitem{COVIDAnalytics-DELPHI}
Michael~Lingzhi Li, Hamza~Tazi Bouardi, Omar~Skali Lami, Thomas~A Trikalinos,
  Nikolaos~K Trichakis, and Dimitris Bertsimas.
\newblock Forecasting covid-19 and analyzing the effect of government
  interventions.
\newblock {\em MedRxiv}, pages 2020--06, 2021.

\bibitem{karlen2020characterizing}
Dean Karlen.
\newblock Characterizing the spread of covid-19.
\newblock {\em arXiv preprint arXiv:2007.07156}, 2020.
\end{thebibliography}
\end{document}